\documentclass[journal,onecolumn]{IEEEtran}

\usepackage{amssymb}
\usepackage{amsfonts}
\usepackage[intlimits,sumlimits]{amsmath}
\usepackage[bbgreekl]{mathbbol}
\usepackage{framed}
\usepackage[boxruled,vlined,longend]{algorithm2e}
\usepackage{rotating}
\usepackage[sans]{dsfont}
\usepackage{yfonts}
\usepackage{pstcol}  
\usepackage{hyperref}

\newtheorem{theorem}{Theorem}

\newtheorem{condition}[theorem]{Condition}

\newtheorem{corollary}[theorem]{Corollary}

\newtheorem{definition}[theorem]{Definition}
\newtheorem{example}[theorem]{Example}

\newtheorem{problem}[theorem]{Problem}
\newtheorem{proposition}[theorem]{Proposition}
\newtheorem{remark}[theorem]{Remark}

\DeclareMathOperator*{\argmin}{argmin}
\DeclareMathOperator*{\arginf}{arginf}

\DeclareMathOperator*{\argmax}{argmax}

\begin{document}

\title{A precise bare simulation approach to 
the minimization of some distances. II. Further Foundations}

\author{Michel~Broniatowski
and Wolfgang~Stummer
\thanks{M. Broniatowski is with the 
LPSM, Sorbonne Universit\'{e}, 4 place Jussieu, 75252 Paris, France.
ORCID 0000-0001-6301-5531.}
\thanks{W. Stummer is with the Department of Mathematics, 
Friedrich-Alexander-Universität Erlangen--N\"{u}rnberg
(FAU),
Cauerstrasse $11$, 91058 Erlangen, Germany; e-mail: stummer@math.fau.de .
ORCID 0000-0002-7831-4558. Corresponding author.}
}

\markboth{\ }
{Broniatowski \& Stummer: 
A bare simulation approach to minimization II}

\maketitle


\begin{abstract} 
The constrained minimization (respectively maximization)
of directed distances and of related generalized entropies
is a fundamental task in information theory as well as in the adjacent fields of 
statistics, machine learning, artificial intelligence, signal processing 
and pattern recognition.
In our previous paper \textquotedblleft A precise bare simulation approach to 
the minimization of some
distances. I. Foundations\textquotedblright, 
we obtained such kind of constrained optima
by a new dimension-free precise bare (pure) simulation method,
provided basically that (i) the underlying directed distance is of 
$f-$divergence type, 
and that (ii) this can be connected to a light-tailed probability distribution in a certain manner. 
In the present paper, we extend 
this approach such that constrained optimization problems of a very huge amount of 
directed distances
and generalized entropies --- and beyond --- can be tackled 
by a newly developed dimension-free \textit{extended bare simulation} method,
for obtaining both optima as well as optimizers. 
Almost no assumptions (like convexity) on the set 
of constraints are needed, within our 
discrete setup of arbitrary dimension, and our method is precise (i.e., converges in the limit).
For instance, we cover constrained optimizations of
\textit{arbitrary} $f-$divergences, Bregman distances, scaled Bregman distances
and weighted $\ell_{r}-$distances. 
The potential for wide-spread applicability
is indicated, too; in particular, we deliver many recent references for uses of
the involved distances/divergences in various different
research fields (which may also serve as an interdisciplinary interface).
\end{abstract}


\IEEEpeerreviewmaketitle
 
\hfill 7th February 2024


\section{Introduction}

\IEEEPARstart{I}{n} our previous paper~\cite{Bro:23a},
we developed a new random-simulation-based approach 
--- called the \textit{(narrow-sense) bare simulation method} ---
to obtain the deterministic 
constrained minimum
$\inf \left\{ \Phi_{\mathbf{P}}(\mathbf{Q}),
\mathbf{Q}\in \mathbf{\Omega}\right\}$
of (generalized versions of) \textit{$\varphi-$divergences}
$\Phi_{\mathbf{P}}(\mathbf{Q})=D_{\varphi}(\mathbf{Q},\mathbf{P})$
(cf. Csiszar~\cite{Csi:63}, Ali \& Silvey~\cite{Ali:66}, Morimoto~\cite{Mori:63}),
where $\mathbf{P} \in 
\mathbb{R}^{K}$,  $\mathbf{\Omega} \subset \mathbb{R}^{K}$
and $\varphi:\mathbb{R} \mapsto \mathbb{R}$ 
is a pregiven convex function 
with certain additional properties 
(namely, $\varphi$ is appropriately linked to an instrumental
probability distribution, under which the crucial auxiliary simulations are performed).
In particular, amongst others, 
in \cite{Bro:23a} 
we covered (generalized versions of) the Kullback-Leibler information distance (relative entropy), 
the (squared) Hellinger distance, the Pearson chi-square divergence and the more general
power divergences (also called Cressie-Read measures \cite{Cre:84} \cite{Rea:88}, 
Tsallis cross-entropies~\cite{Tsa:98},
Amari\textquoteright s alpha-divergences~\cite{Ama:85}), 
and the Jensen-Shannon divergence/distance;
some other important quantities which are closely connected to those $\varphi-$divergences
---
such as e.g. 
the Bhattacharyya coefficient (cf.~\cite{Bha:43},\cite{Bha:46},\cite{Bha:47}),
the Bhattacharyya distance~\cite{Bha:43}
and the more general Renyi divergences~\cite{Renyi:61} (see also 
the comprehensive exposition in van Erven \& Harremo{\"e}s~\cite{VanErv:14}),
as well as the maximization/minimization of some $\varphi-$entropies 
(cf. Burbea \& Rao~\cite{Bur:82}) such as e.g.
the Shannon entropy \cite{Sha:48}, the Renyi entropies \cite{Renyi:61}, 
the Havrda-Charvat entropies \cite{Hav:67}
(also called  Tsallis entropies \cite{Tsa:88})
and of some Euclidean norms 
--- were treated in \cite{Bro:23a}, too.

\vspace{0.2cm}
\noindent
Let us briefly recall the \textit{core steps} of our 
bare-simulation minimization method in \cite{Bro:23a}.
The \textit{first step} is to normalize the vector
\footnote{in this paper, vectors are taken to be row vectors} 
$\mathbf{P}$ into a probability vector
$\mathbf{P}$ \footnote{with a slight abuse of notation; see the main text for a
more comprehensive notation} 
(e.g. the $\varphi-$entropy triggering case $\mathbf{P}=\left( 1,\ldots,1\right)$
is converted into the uniform-probability vector $\mathbf{P} = \left( 1/K,...,1/K\right)$).
The \textit{second step} follows from expressing the function $\varphi$ in form
of the Fenchel-Legendre transform of the 
cumulant (i.e., log moment) generating function of some
random variable $W$; a probabilistic construction based on i.i.d. copies $W_{i}$
of $W$ allows to interpret $\inf \left\{ D_{\varphi }(\mathbf{Q},\mathbf{P}),
\mathbf{Q}\in \mathbf{\Omega }\right\} $ as an asymptotic characteristic for
some explicitly constructable scheme involving both $\mathbf{P}$ and the 
$W_{i}$'s.
The \textit{third and final step} consists in the construction of this
probabilistic scheme, and it differs for the specific problem context.

\vspace{0.2cm}
\noindent 
More detailed, for a deterministic setup where the (transformed) probability vector $\mathbf{P}
=\left( p_{1},\ldots,p_{K}\right)$ is completely known
and $\mathbf{\Omega}$ has non-void interior, 
we construct in \cite{Bro:23a}
the integer part
$n_{i} := \lfloor np_{i} \rfloor$, 
partition the index set $\left\{ 1,\ldots,n\right\} $
into $K$ sets of size $n_{1},\ldots,n_{K}$ and build a $K$-component vector;
\textit{each component} of this vector is an ad hoc weighted empirical mean of the 
$W_{i}$ 's ; up to standard transformations the empirical count of the visits
of this vector in $\mathbf{\Omega }$ approximates the solution of the
optimization problem $\inf \left\{ D_{\varphi }(\mathbf{Q},\mathbf{P}),
\mathbf{Q}\in \mathbf{\Omega }\right\}$. Therefore, the resulting approximation 
can be performed straightforwardly:
the (typically) very complicated minimization task is replaced by a 
much more comfortable 
--- nevertheless convergent ---
random count procedure
which can be based on a fast and accurate 
--- pseudo, true, natural, quantum  
--- random number generator.
In case of a statistical/risk optimizing context, 
one has instead of a \textit{known} 
probability vector $\mathbf{P}$ 
a data-describing 
\textit{sample} $X_{1},\ldots,X_{n}$ of $n$ i.i.d. (and even more general) copies of a discrete random variable $X$ 
with \textit{unknown} distribution (described by an unknown probability vector) $\mathbf{P}$, 
and $\mathbf{\Omega}$ is now a subset of the probability simplex 
$\mathbb{S}^{K}$ in $\mathbb{R}^K$.
For such contexts, in \cite{Bro:23a} we appropriately adapted the above-described method, 
by --- amongst other things --- basically using the corresponding (vectorized)
sample-based 
histogram
as $\mathbf{P}$.

\vspace{0.2cm}
\noindent
Although \cite{Bro:23a} covers a considerable amount of important $\varphi-$divergences,
as shown by numerous solved cases, the question arises
whether the above-described bare-simulation method can be extended 
to achieve the minimum 
$\inf \left\{ \Phi_{\mathbf{P}}(\mathbf{Q}),
\mathbf{Q}\in \mathbf{\Omega}\right\}$
of other important classes 
$\Phi_{\mathbf{P}}(\mathbf{Q})=D(\mathbf{Q},\mathbf{P})$ of 
directed (i.e. non-symmetric) distances, divergences and (dis)similarity measures.
To give the corresponding positive answer, is the main goal of this paper.
Moreover, as opposed to \cite{Bro:23a}, not only the minimum itself
but also the corresponding minimizer(s) 
will be treated as well. Furthermore, we also tackle
the maximum $\sup \left\{ \Phi_{\mathbf{P}}(\mathbf{Q}),
\mathbf{Q}\in \mathbf{\Omega}\right\}$
as well as the corresponding  maximizer(s).

\vspace{0.2cm}
\noindent
Indeed, in this paper we particularly investigate 
the following important 
classes of directed distances (divergences, dissimilarity measures) and connected entropies:

\vspace{0.2cm}

\begin{itemize}

\item general $\varphi-$divergences $\Phi_{\mathbf{P}}(\mathbf{Q}) := D_{\varphi}( \mathbf{Q}, \mathbf{P})$
where $\varphi$ \textit{does not satisfy} the above-mentioned 
Fenchel-Legendre-transform
respresentability. 
For instance, this covers 
the omnipresent $\ell_{1}-$distance (also known as \textit{total variation distance}) 
between $\mathbf{Q}$ and $\mathbf{P}$.
For some comprehensive overviews on $\varphi-$divergences,
the reader is referred to the insightful books of 
e.g.\ Liese \& Vajda~\cite{Lie:87},
Read \& Cressie~\cite{Rea:88},
Vajda~\cite{Vaj:89}, 
Csiszar \& Shields~\cite{Csi:04},
Stummer~\cite{Stu:04a}, 
Pardo~\cite{Par:06}, 
Liese~\& Miescke~\cite{Lie:08},
Basu et al.~\cite{Bas:11},
the survey articles of
e.g. Liese \& Vajda~\cite{Lie:06},  
Vajda \& van der Meulen~\cite{Vaj:10},
Reid \& Williamson~\cite{Reid:11},
Basseville~\cite{Bass:13},
and the references therein.
Moreover, we are interested in the very important special case 
$\mathbf{P}=\left( 1,\ldots,1\right) =: \mathbf{1}$
which leads to the corresponding general
\textit{non-probability-vector extension} of the
\textit{$\varphi-$entropies} 
in the sense of Burbea \& Rao \cite{Bur:82}
(see also Csiszar \cite{Csi:72}, Ben-Bassat \cite{BenB:78}, Ben-Tal \& Teboulle \cite{BenT:86},
Kesavan \& Kapur \cite{Kes:89},
Dacunha-Castelle \& Gamboa \cite{Dac:90},
Teboulle \& Vajda \cite{Teb:93}, 
Gamboa \& Gassiat \cite{Gam:97},
Vajda \& Zvarova \cite{Vaj:07}).
Those general $\varphi-$entropies 
can also be interpreted as an \textquotedblleft  index/degree of (in)equality of
the set $\mathbf{\Omega}$\textquotedblright ,
respectively 
as an \textquotedblleft  index/degree of diversity of
the set $\mathbf{\Omega}$\textquotedblright .
We can also deal with the
corresponding general
\textit{non-probability-vector extension} of the
even wider class of 
$(h,\varphi)-$entropies in the sense of
Salicru et al.~\cite{Sal:93} (see also
e.g. Pardo \cite{Par:06}, Vajda \& Vasek \cite{Vaj:85}).\\

\item 
\textit{separable (ordinary)
Bregman distances} (cf. \cite{Breg:67})
$\Phi_{\mathbf{P}}(\mathbf{Q}) := D_{\varphi}^{OBD}(\mathbf{Q},\mathbf{P})$
with strictly convex function $\varphi:\mathbb{R} \mapsto \mathbb{R}$.
Aside from the vast applications in engineering,
some general connections to probability and statistics are given e.g. 
in Csisz\'ar~\cite{Csi:91},~\cite{Csi:94},~\cite{Csi:95}, 
Pardo \& Vajda~\cite{Par2:97},~\cite{Par2:03}, Stummer \& Vajda~\cite{Stu:12},
and Broniatowski \& Stummer \cite{Bro:19b},\cite{Bro:22}.
Important special cases include 
(i) the omnipresent \textit{squared $\ell_{2}-$distance},
(ii) the more general \textit{density power divergences}
of Basu et al.~\cite{Bas:98} 
(see also e.g. Stummer \& Vajda~\cite{Stu:12}
for a different parameter scaling which covers 
the prominent \textit{Itakura-Saito distance} (in the version of e.g. Banerjee et al. \cite{Ban:05})),
as well as (iii) their rescaled versions
called \textit{beta-divergences} (cf. Eguchi \& Kano \cite{Egu:01}, Mihoko \& Eguchi \cite{Miho:02}).
Another recently launched special case is the \textit{Bregman exponential divergence}
of Mukherjee et al.~\cite{Muk:19} (see also Basak \& Basu~\cite{Basak:22}). \\

\item 
\textit{scaled Bregman distances} (with scaling vector $\mathbf{M}$)
$\Phi_{\mathbf{P}}(\mathbf{Q}) := D_{\varphi,\mathbf{M}}^{SBD}(\mathbf{Q},\mathbf{P})$
of Stummer~\cite{Stu:07} and Stummer \& Vajda~\cite{Stu:12}, 
which cover both $\varphi-$divergences as well as separable (ordinary) Bregman distances.
Further investigations (e.g. on robustness issues) 
can be found in
Ki{\ss}linger \& Stummer~\cite{Kis:13},~\cite{Kis:15a},~\cite{Kis:16},~\cite{Kis:18};
Broniatowski \& Stummer~\cite{Bro:19b} flexibilized/widened 
the involved domains,
and Broniatowski \& Stummer~\cite{Bro:22}
give a comprehensive survey on various different kinds of applications to statistics, 
and to the adjacent fields of
machine learning and artificial intelligence.
Moreover, Stummer \& Ki{\ss}linger~\cite{Stu:17a}
give some structural flexibilizations/generalizations 
of scaled Bregman distances\footnote{(even for non-convex $\varphi$)} 
which cover as \textit{special cases} (the separable versions of) (i) the 
total Bregman divergences of Liu et al.~\cite{Liu8:10},\cite{Liu8:12},  
Vemuri et al.~\cite{Vem:11a}, (ii) its variant given 
in Nock et al.~\cite{Noc:16sB},
as well as (iii) the conformal divergences and the scaled conformal divergences 
of Nock et al. \cite{Noc:16}.
Furthermore, we can also deal with 
the even wider class of the \textit{distances of Broniatowski \& Stummer~\cite{Bro:19b}} (see also
Broniatowski \& Stummer~\cite{Bro:22}), which
cover all the above-mentioned (generalizations of) scaled Bregman distances as special cases.\\

\item 
the very prominent \textit{
non-separable (ordinary) Bregman distances} 
(cf. \cite{Breg:67})
$\Phi_{\mathbf{P}}(\mathbf{Q}) := D_{\boldsymbol{\varphi}}^{gnOBD}(\mathbf{Q},\mathbf{P})$
with strictly convex \textit{multivariate} function 
$\boldsymbol{\varphi}:\mathbb{R}^{K} \mapsto \mathbb{R}$
(as usual, if $\boldsymbol{\varphi}$ is of the separable form 
$\boldsymbol{\varphi}(\mathbf{Q}) := \sum_{k=1}^{K} \varphi(q_{k})$, 
then $D_{\boldsymbol{\varphi}}^{gnOBD}(\mathbf{Q},\mathbf{P})$ collapses to the
separable (ordinary)
Bregman distance $D_{\boldsymbol{\varphi}}^{OBD}(\mathbf{Q},\mathbf{P})$).
A very important special case is the omnipresent (squared) \textit{Mahalanobis distance}
\cite{Maha:36}.\\

\item 
the omnipresent \textit{weighted $\ell_{r}-$distances} 
$\Phi_{\mathbf{P}}(\mathbf{Q}) := ||\mathbf{Q}-\mathbf{P}||_{r,\mathbf{w}}$
where $r \in \, ]0,\infty[$ and $\mathbf{w}$ is a vector of weights.\\

\item \textit{Burbea-Rao divergences} \cite{Bur:82} 
(see also e.g. Pardo \& Vajda~\cite{Par2:97},\cite{Par2:03},
as well as Stummer \& Ki{\ss}linger~\cite{Stu:17a} 
for an imbedding into their scaled-Bregman-distance-flexibilizations).\\

\end{itemize}

\vspace{0.2cm}
\noindent
In the light of the above explanations, the goals of this paper are: 

\vspace{0.2cm}

\begin{enumerate}

\item[(G1)] to \textit{extend} the results of \cite{Bro:23a} on --- narrow sense --- bare-simulation minimization 
$\inf \left\{D_{\varphi}(\mathbf{Q},\mathbf{P}),\mathbf{Q}\in \mathbf{\Omega}\right\}$ 
of $\varphi-$divergences with 
instrumentally linked divergence generator $\varphi$, 
\textit{to} the --- narrow sense --- bare-simulation minimization of
the wider class of scaled Bregman distances 
$D_{\varphi,\mathbf{M}}^{SBD}(\mathbf{Q},\mathbf{P})$
(including separable Bregman distances $D_{\varphi}^{OBD}(\mathbf{Q},\mathbf{P})$)
with the same type of divergence generator $\varphi$;

\item[(G2)] 
to solve --- by means of a newly developed \textit{extension} of our narrow-sense bare-simulation method ---
both constrained \textit{minimization and maximization problems} for the above-mentioned huge range
of directed distances, divergences, \\ 
(dis-)similarity measures, entropies (and beyond);
our new method is precise (i.e., converges in the limit)
and needs almost no assumptions (like convexity) on the set 
$\mathbf{\Omega}$ of constraints of arbitrary finite dimension.

\item[(G3)] to deliver \textit{sharp} estimates of the desired \textit{minima and maxima},
and to derive the left-open estimates of the \textit{minimizers} for the context of \cite{Bro:23a},
as well as the of the minimizers and maximizers for the contexts (G1) and (G2).
 
\end{enumerate}

\vspace{0.2cm}
\noindent
This agenda is achieved in the following way. In the next 
Section \ref{SectMain}, we briefly introduce the principal idea of our new
\textit{extended} bare-simulation optimization paradigm, in addition to
the --- now called \textit{narrow-sense} --- one of \cite{Bro:23a}. 
In order to lay a solid explanatory basis for our new developments, 
we first recall in Section \ref{SectDetNarrow} our main 
results of \cite{Bro:23a} on narrow-sense bare-simulation minimization of
$\varphi-$divergences with instrumentally linked divergence generator $\varphi$,
for \textquotedblleft general\textquotedblright\  constraints sets $\mathbf{\Omega}$ with non-void interior.
For the same type of $\varphi$ and $\mathbf{\Omega}$,
in Section \ref{SectDetNarrow Bregman} we carry out
the above-mentioned goal (G1) on narrow-sense bare-simulation minimization of
separable Bregman distances and even more general
scaled Bregman distances. 
Based on the results of the previous two chapters,
we then achieve in Section \ref{SectDetGeneral}
the goal (G2) and derive four fundamental (non-narrow-sense) bare-simulation
minimization and maximization results on all the above-mentioned general directed distances
$D(\mathbf{Q},\mathbf{P})$ and friends, for the context of $\mathbf{\Omega}$ with non-void interior.
In the next three Sections \ref{SectDetSubsimplex.CASM}, \ref{SectDetSubsimplex.SBD},
\ref{SectDetSubsimplex} we carry out the same program as in the Sections \ref{SectDetNarrow},
\ref{SectDetNarrow Bregman}, \ref{SectDetGeneral},
but for the case that $\mathbf{\Omega}$ contains the side constraint that 
for each member $\mathbf{Q}$ the 
sum of the components equals the same fixed constant $A>0$
(implying that $\mathbf{\Omega}$ has void interior).
Furthermore, in the Sections \ref{SectStochSubsimplex.CASM},
\ref{SectStochSubsimplex.SBD}, \ref{SectStochSubsimplex.General}
we carry out the same program as in the Sections \ref{SectDetNarrow},
\ref{SectDetNarrow Bregman}, \ref{SectDetGeneral},
but for the risk-carrying case that $\mathbf{P}$ --- respectively
some involved parameter --- is unknown
(and $A=1$).
In the Sections \ref{SectEstimators.new.det.nonvoid},
\ref{SectEstimators.new.det.simplex}
and \ref{SubsectEstimators.risk}
we provide corresponding estimators for the
minima, minimizers, maxima and maximizers
of the above-mentioned sections
(cf. Goal (G3)). Finally, all the proofs are given
Appendix \ref{App.A}.


\section{A new minimization paradigm} 
\label{SectMain}

\noindent
We concern with minimization and maximization problems of the following type, where 
$\mathcal{M}$ is a topological space 
and $\mathcal{T}$ is the Borel $\sigma-$field
over a given base on $\mathcal{M}$; e.g. take $\mathcal{M} = \mathbb{R}^{K}$
to be the $K-$dimensional Euclidean space equipped with the Borel $\sigma-$
field $\mathcal{T}$. 

\vspace{0.2cm}

\begin{definition}
\label{brostu5:def.1} 
A measurable function $\Phi: \mathcal{M} \mapsto
\mathbb{R} \cup \{-\infty, \infty\}$ 
and measurable set $\Omega \subset \mathcal{M}$ 
\footnote{
i.e. $\Omega \in \mathcal{T}$} are called 
\textquotedblleft bare-simulation minimizable\textquotedblright\ 
(BS-minimizable) respectively 
\textquotedblleft bare-simulation maximizable\textquotedblright\ 
(BS-maximizable)
if for 
\begin{equation}
\Phi(\Omega ):=\inf_{Q\in\Omega}\left\{ \Phi(Q) \right\} 
\in \, ]-\infty, \infty[
\qquad \textrm{respectively} \qquad 
\Phi(\Omega ):=\sup_{Q\in\Omega}\left\{ \Phi(Q) \right\} 
\in \, ]-\infty, \infty[
\label{brostu5:fo.1}
\end{equation}
there exists a measurable function $G: [0,\infty[ \mapsto 
\mathbb{R}$, a sequence $(f_{n})_{n\in\mathbb{N}}$ of measurable functions 
$f_{n}: \mathcal{M} \mapsto [0,\infty[$ as well as 
a sequence $\left((\mathfrak{X}_{n},\mathcal{A}_{n},\mathbb{\Pi}_{n})\right)_{n \in \mathbb{N}}$ of probability spaces and on them a
sequence $(\xi_{n})_{n\in \mathbb{N}}$ 
\footnote{
in order to emphasize the dependence on $\Phi$, one should use the notations 
$(\xi_{\Phi,n})_{n\in \mathbb{N}}$, $\mathbb{\Pi}_{\Phi,n}$, etc.; this is avoided
for the sake of a better readability.} of $\mathcal{M}-$valued random
variables such that 
\begin{eqnarray}
& &
G\Big(-\lim_{n\rightarrow\infty}\frac{1}{n}\log 
\mathbb{E}_{\mathbb{\Pi}_{n}}\negthinspace \Big[ \, f_{n}(\xi_{n}) \cdot 
\textfrak{1}_{\Omega}(\xi_{n}) \, \Big] 
\Big) =\inf_{Q\in\Omega}\Phi(Q)=\Phi(\Omega)
\label{brostu5:fo.2} \\
&&
\hspace{-2.0cm} \textrm{respectively} \quad
G\Big(-\lim_{n\rightarrow\infty}\frac{1}{n}\log 
\mathbb{E}_{\mathbb{\Pi}_{n}}\negthinspace \Big[ \, f_{n}(\xi_{n}) \cdot 
\textfrak{1}_{\Omega}(\xi_{n}) \, \Big] 
\Big) =\sup_{Q\in\Omega}\Phi(Q) =\Phi(\Omega),
\label{brostu5:fo.2b}
\end{eqnarray}
where $\mathbb{E}_{\mathbb{\Pi}_{n}}[ \, \cdot \, ]$ denotes the expectation with respect to
$\mathbb{\Pi}_{n}$ and $\textfrak{1}_{B}(\cdot)$ denotes the indicator function on the set $B$;
in situations where $\Phi$ is fixed and different $\Omega$\textquoteright s are considered,  
we say that 
\textquotedblleft  $\Phi$ is bare-simulation minimizable (BS-minimizable)
on $\Omega$\textquotedblright\ 
respectively 
\textquotedblleft  $\Phi$ is bare-simulation maximizable (BS-maximizable)
on $\Omega$\textquotedblright.
In case that one can even choose $f_{n}(\cdot) \equiv 1$ ---
and hence 
$\mathbb{E}_{\mathbb{\Pi}_{n}}\negthinspace \big[ \, f_{n}(\xi_{n}) \cdot 
\textfrak{1}_{\Omega}(\xi_{n}) \, \big] = 
\mathbb{\Pi}_{n}\negthinspace \big[ \, \xi_{n} \in \Omega  \, \big] $ 
--- then we speak of 
\textquotedblleft  bare-simulation minimizable/maximizable 
\textit{in the narrow sense}\textquotedblright.
\end{definition}

\vspace{0.2cm}

\begin{remark} 
\label{brostu5:rem.def1}
(a) The above-mentioned Definition \ref{brostu5:def.1} extends the Definition 1 of 
Browniatowski \& Stummer~\cite{Bro:23a} who deal with
the narrow-sense-case $f_{n}(\cdot) \equiv 1$.\\
(b) We could even extend the above-mentioned Definition \ref{brostu5:def.1} to allow for more general
$f_{n}: \mathcal{M} \mapsto \mathbb{R}$ such that\\
$\mathbb{E}_{\mathbb{\Pi}_{n}}\negthinspace \Big[ \, f_{n}(\xi_{n}) \cdot 
\textfrak{1}_{\Omega}(\xi_{n}) \, \Big] \geq 0$ for all large enough $n \in \mathbb{N}$. \\
(c) As usual, we call $\Omega$ the \textit{constraint set} (alternatively used names are e.g.
\textit{choice set} or \textit{search space}).
\end{remark}

\vspace{0.4cm}
\noindent

\noindent The basic idea/incentive of 
this new approach is: if a minimization problem \eqref{brostu5:fo.1} has no
explicit solution and is computationally intractable (or unfeasible) but can
be shown to be BS-minimizable with concretely constructable 
$G$, $(f_{n})_{n\in\mathbb{N}}$, $(\xi_{n})_{n\in 
\mathbb{N}}$ and $(\mathbb{\Pi}_{n})_{n\in \mathbb{N}}$, then one can basically
simulate the log-expectations 
$-\frac{1}{n} \log 
\mathbb{E}_{\mathbb{\Pi}_{n}}\negthinspace \big[ \, f_{n}(\xi_{n}) \cdot 
\textfrak{1}_{\Omega}(\xi_{n}) \, \big]$ 
for large enough integer $n \in \mathbb{N}$ 
to obtain an approximation of the minimum/maximum \eqref{brostu5:fo.1}
without having to evaluate the corresponding (not necessarily unique) 
minimizer/maximizer. As explained in \cite{Bro:23a}, 
this is for instance important
for fast and efficient model search. 
However, in contrast to \cite{Bro:23a}, we show in this paper also how
one can ``nearly synchronously''
achieve an approximation of
the corresponding minimizer(s)/maximizer(s).

\vspace{0.2cm} 
\noindent 
For reasons of transparency, we \textit{start} to demonstrate this approach for the
following important/prominent
class of \textit{deterministic} constrained minimization problems with the following components:

\begin{enumerate}
\item[(i)] 
$\mathcal{M}$ is the $K-$dimensional
Euclidean space $\mathbb{R}^{K}$, i.e. $\Omega$ is a set of vectors $Q$ with 
a number of $K$ components
(where $K$ may be huge, as it is e.g. the case in big data contexts); 

\item[(ii)] $\Phi(\cdot) := \Phi_{P}(\cdot)$ 
depends on some known vector $P$ in $\mathbb{R}^{K}$ 
with $K$ nonnegative components;

\item[(iii)] $\Phi_{P}(\cdot)$ is a \textquotedblleft directed distance\textquotedblright\ (divergence) from $P$ into 
$\Omega$ in the sense of $\Omega \ni Q \mapsto \Phi_{P}(Q)
:= D(Q, P)$, where $D(\cdot,\cdot)$ has the the two properties \textquotedblleft $D(Q, P)
\geq 0$\textquotedblright\ and \textquotedblleft $D(Q, P) = 0$ if and only if $Q=P$\textquotedblright. In particular, 
$D(\cdot,\cdot)$ needs neither satisfy the symmetry $D(Q,P)= D(P,Q)$ nor the
triangular inequality.
\end{enumerate}

\noindent In other words, the left-hand part of (1) together with (i)-(iii) constitutes a
\textit{deterministic}
constrained
distance/divergence-minimization problem; we design a \textquotedblleft
universal\textquotedblright\ method to solve such problems by constructing
appropriate (cf.\eqref{brostu5:fo.2}) $G$, $(f_{n})_{n\in\mathbb{N}}$ and sequences 
$(\xi _{n})_{n\in \mathbb{N}}$ of $\mathbb{R}^{K}-$valued random variables.
The latter will be first constructed ---
for a first insight --- with the help of \textit{narrow-sense methods} 
developed in \cite{Bro:23a}
for directed distances $D(\cdot ,\cdot )$ from 
a large subclass of the 
important omnipresent Csiszar-Ali-Silvey-Morimoto
$\varphi-$divergences (also called $f-$divergences) given in Definition \ref{def div} below.


\section{Deterministic Narrow-Sense Bare-Simulation-Optimization of $\varphi-$divergences}
\label{SectDetNarrow}

\noindent To begin with, concerning the above-mentioned point (i) we take the $K-$
dimensional Euclidean space $\mathcal{M}=\mathbb{R}^{K}$, 
denote from now on --- as usual --- its elements (i.e. vectors) in boldface letters,
and also employ the subsets 
\begin{eqnarray}
&& 
\mathbb{R}_{\ne 0}^{K} := \{\mathbf{Q}:= (q_{1},\ldots,q_{K}) \in \mathbb{R}^K: \,
q_{i} \ne 0 \ \text{for all} \ i=1,\ldots,K \},  
\notag \\
&& 
\mathbb{R}_{> 0}^{K} := \{\mathbf{Q}:= (q_{1},\ldots,q_{K}) \in \mathbb{R}^K: \,
q_{i} > 0 \ \text{for all} \ i=1,\ldots,K \},  \notag \\
&& \mathbb{R}_{\geq 0}^{K} := \{\mathbf{Q}:= (q_{1},\ldots,q_{K}) \in \mathbb{R}^K
: \, q_{i} \geq 0 \ \text{for all} \ i=1,\ldots,K \},  \notag \\
&& \mathbb{R}_{\gneqq 0}^{K} := \mathbb{R}_{\geq 0}^{K}\backslash \{\boldsymbol{0}\} 
 := \{\mathbf{Q}:= (q_{1},\ldots,q_{K}) \in \mathbb{R}_{\geq 0}^{K}
: \, q_{i} \ne 0 \ \text{for some} \ i=1,\ldots,K \},  \notag \\
&& 
\textstyle
\mathbb{S}^{K} := \{\mathbf{Q} := (q_{1},\ldots,q_{K}) \in \mathbb{R}_{\geq 0}^{K}:
\, \sum_{i=1}^{K} q_{i} =1 \} \quad \text{(simplex of probability vectors, probability simplex)},
\notag \\
&& 
\textstyle
\mathbb{S}_{> 0}^{K} := \{\mathbf{Q}:= (q_{1},\ldots,q_{K}) \in \mathbb{R}_{>
0}^{K}: \, \sum_{i=1}^{K} q_{i} =1 \}.  \notag
\end{eqnarray}
Concerning the directed distances $D(\cdot,\cdot)$ in (ii) and (iii), 
as a basis we
\textit{first} deal with the following

\vspace{0.2cm}

\begin{definition}
\label{def div}
(a) \, Let the \textquotedblleft divergence-generator\textquotedblright\ be a 
lower semicontinuous convex function $\varphi: \, ]-\infty,\infty[ \rightarrow [0,\infty]$ 
satisfying $\varphi(1)=0$.
Furthermore, for the effective domain $dom(\varphi) := \{ t \in \mathbb{R} : \varphi(t) < \infty \}$
we assume that its interior $int(dom(\varphi))$ is non-empty
which implies that $int(dom(\varphi)) = \, ]a,b[$ for
some $-\infty \leq a < 1 < b \leq \infty$.
Moreover, we suppose that $\varphi$ is strictly convex
at the point $t=1$ \footnote{
in line with e.g. Liese \& Miescke \cite{Lie:08}, 
here a convex function $\varphi$ 
is called strictly convex at the point $t=1$ if the function $\varphi$ is not
linear in the open interval $]1-\varepsilon,1+\varepsilon[$
for any $\varepsilon > 0$
} (very often in practice, $\varphi$ is strictly convex even
in a non-empty neighborhood $]t_{-}^{sc},t_{+}^{sc}[ \subseteq ]a,b[$ of one
($t_{-}^{sc} < 1 < t_{+}^{sc}$)).
Also, we set $\varphi(a) := \lim_{t \downarrow a} \varphi(t)$ 
and $\varphi(b) := \lim_{t \uparrow b} \varphi(t)$ 
(these limits always exist).
The class of all such functions $\varphi$ will be denoted
by $\widetilde{\Upsilon}(]a,b[)$. A frequent choice is e.g. $]a,b[ \, = \, ]0,\infty[$ or 
$]a,b[ \, = \, ]-\infty,\infty[$.\\
(b) \, For $\varphi \in \widetilde{\Upsilon}(]a,b[)$, 
$\mathbf{P} := (p_{1},\ldots,p_{K}) \in \mathbb{R}_{\geq 0}^{K}$ and $\mathbf{Q} := (q_{1},\ldots,q_{K}) \in \mathbf{\Omega} \subset \mathbb{R}^{K}$, we define the
Csiszar-Ali-Silvey-Morimoto (CASM) $\varphi-$divergence 
\begin{equation}
\Phi_{\mathbf{P}} \left(\mathbf{Q}\right) := D_{\varphi}( \mathbf{Q}, \mathbf{P} ) := 
\sum_{k=1}^{K} p_{k} \cdot
\varphi \left( \frac{q_{k}}{p_{k}}\right) \, \geq 0.  
\label{brostu5:fo.div}
\end{equation}
As usual, in \eqref{brostu5:fo.div} we employ the three conventions that $p
\cdot \varphi \left( \frac{0}{p}\right) = p \cdot \varphi(0) >0$ for all $p > 0$, 
and $0 \cdot \varphi \left( \frac{q}{0}\right) = q \cdot \lim_{x
\rightarrow \infty} \frac{\varphi(x \cdot \textrm{sgn}(q))}{x\cdot \textrm{sgn}(q)} >0$ 
for $q \neq 0$ (employing the sign of $q$), and $0 \cdot
\varphi \left( \frac{0}{0}\right) :=0$. Throughout the paper, we only
consider constellations $(\varphi,\mathbf{P},\mathbf{\Omega})$ 
for which the very mild condition 
\ $\Phi_{\mathbf{P}}(\Omega) := \inf_{\mathbf{Q}\in\mathbf{\Omega}} 
D_{\varphi}( \mathbf{Q}, \mathbf{P} ) \neq \infty$ \ holds.
\end{definition}

\vspace{0.3cm}
\noindent For probability vectors $\mathds{P}$ and $\mathds{Q}$ in 
$\mathbb{S}^{K}$, the 
$\varphi-$divergences $D_{\varphi}( \mathds{Q}, \mathds{P} )$ were introduced by 
Csiszar~\cite{Csi:63}, Ali \& Silvey~\cite{Ali:66} and 
Morimoto~\cite{Mori:63}
(where the first two references even deal with more general probability distributions); 
for some comprehensive overviews 
--- including statistical applications to goodness-of-fit testing and
minimum distance estimation ---
the reader is referred to the insightful 
books the reader is referred to the insightful books of 
e.g.\ Liese \& Vajda~\cite{Lie:87},
Read \& Cressie~\cite{Rea:88},
Vajda~\cite{Vaj:89}, 
Csiszar \& Shields~\cite{Csi:04},
Stummer~\cite{Stu:04a}, 
Pardo~\cite{Par:06}, 
Liese~\& Miescke~\cite{Lie:08},
Basu et al.~\cite{Bas:11}, 
the survey articles of
e.g. Liese \& Vajda~\cite{Lie:06},  
Vajda \& van der Meulen~\cite{Vaj:10},
Reid \& Williamson~\cite{Reid:11},
Basseville~\cite{Bass:13},
and the references therein;
For the setup of $D_{\varphi}( \mathbf{Q}, \mathbf{P} )$ for vectors $\mathbf{P}$, 
$\mathbf{Q}$ with non-negative components
the reader is referred to e.g. Stummer \& Vajda~\cite{Stu:10} 
(who deal with even more general nonnegative measures
and give some statistical as well as information-theoretic applications)
and Gietl \& Reffel~\cite{Gie:17}. 
The case of $\varphi-$divergences for vectors with arbitrary components
can be extracted from e.g. Broniatowski \& Keziou~\cite{Bro:06} 
who actually deal with finite \textit{signed} measures. For a comprehensive technical
treatment, see also Browniatowski \& Stummer~\cite{Bro:19b}.

\vspace{0.2cm} 
\noindent  
As an important special case, we get for the choice $\mathbf{P} := (1, \ldots, 1) := \mathbf{1}$ 
the quantity
\begin{equation}
\Phi_{\mathbf{1}}\negthinspace\left(\mathbf{Q}\right) 
:= D_{\varphi }(\mathbf{Q},\mathbf{1}) = \sum_{k=1}^{K}\varphi (q_{k})     
\label{min Pb one new}
\end{equation}
with $\varphi \in \widetilde{\Upsilon}(]a,b[)$.
As is well known, there are numerous applications of 
$\sum_{k=1}^{K}\varphi (q_{k})$
where $\varphi$ is e.g. interpreted as cost function respectively energy function respectively purpose function.
Furthermore, $\sum_{k=1}^{K}\varphi (q_{k})$ 
can be interpreted as (non-probability extension of a)
$\varphi-$entropy in the sense of Burbea \& Rao \cite{Bur:82}
(see also Csiszar \cite{Csi:72}, Ben-Bassat \cite{BenB:78}, Ben-Tal \& Teboulle \cite{BenT:86},
Kesavan \& Kapur \cite{Kes:89},
Dacunha-Castelle \& Gamboa \cite{Dac:90},
Teboulle \& Vajda \cite{Teb:93}, 
Gamboa \& Gassiat \cite{Gam:97},
Vajda \& Zvarova \cite{Vaj:07}).
Moreover, since $\mathbf{1}$ can be seen 
as a reference vector with (normalized)
equal components, 
$D_{\varphi }(\mathbf{Q},\mathbf{1})$
in \eqref{min Pb one new} 
can be interpreted as an \textquotedblleft  index/degree of (in)equality of
the set $\mathbf{\Omega}$\textquotedblright ,
respectively as an \textquotedblleft  index/degree of diversity of
the set $\mathbf{\Omega}$\textquotedblright .
A comprehensive BS-concerning discussion with
references on the theory and applications of the quantities in 
\eqref{min Pb one new}, 
is given e.g. in Broniatowski \& Stummer \cite{Bro:23a}.

\vspace{0.2cm} 
\noindent 
Returning to the general case, from \eqref{brostu5:fo.div} it is obvious
that in general $D_{\varphi}( \mathbf{Q}, \mathbf{P} ) \ne 
D_{\varphi}(\mathbf{P}, \mathbf{Q})$ (non-symmetry).
Moreover, it is straightforward to deduce that $D_{\varphi}(\mathbf{Q}, \mathbf{P}) = 0$ if
and only if $\mathbf{Q}=\mathbf{P}$ (reflexivity). By appropriate choice of 
$\varphi$, one can get as special cases many very prominent divergences  
which are frequently used in information theory and its applications to e.g.
statistics, artificial intelligence, and machine learning. 

\vspace{0.2cm}
\noindent
For reasons of a more compact representation, we shall henceforth 
assume that $\mathbf{P} := (p_{1},\ldots,p_{K}) \in \mathbb{R}_{> 0}^{K}$,
unless stated otherwise.

\vspace{0.2cm}
\noindent
As a fundamental tool for later purposes, 
let us now briefly explain how the \textit{BS method in the narrow sense} of 
Broniatowski \& Stummer \cite{Bro:23a} can be used to tackle the following
deterministic optimization problems:

\vspace{0.2cm}

\begin{problem}
\label{det Problem}
For pregiven $\varphi \in \widetilde{\Upsilon}(]a,b[)$, 
positive-components vector $\mathbf{P}:=\left( p_{1},..,p_{K}\right) \in \mathbb{R}_{>0}^{K}$
(or from some subset thereof), and subset $\mathbf{\Omega} \subset \mathbb{R}^{K}$
(also denoted in boldface letters, with a slight abuse of notation) 
with regularity properties
\begin{equation}
cl(\mathbf{\Omega} )=cl\left( int\left( \mathbf{\Omega} \right) \right) ,  \qquad int\left( \mathbf{\Omega} \right) \ne \emptyset,
\label{regularity}
\end{equation}
find 
\begin{equation}
\Phi_{\mathbf{P}}(\mathbf{\Omega}) := \inf_{\mathbf{Q}\in \mathbf{\Omega} } D_{\varphi }(\mathbf{Q},\mathbf{P}),  
\label{min Pb}
\end{equation}
provided that 
\begin{equation}
\inf_{\mathbf{Q}\in \mathbf{\Omega} } D_{\varphi }(\mathbf{Q},\mathbf{P}) < \infty 
\label{def fi wrt Omega}
\end{equation}
and that divergence generator $\varphi$  additionally satisfies the following Condition  
\ref{Condition  Fi Tilda in Minimization}.

\begin{condition}
\label{Condition  Fi Tilda in Minimization}
Let $\varphi \in \widetilde{\Upsilon}(]a,b[)$ and $M_{\mathbf{P}} =\sum_{i=1}^{K}p_{i}>0$. 
Then the multiple $\widetilde{\varphi} := M_{\mathbf{P}} \cdot \varphi$
should satisfy the representation 
\begin{equation}
\widetilde{\varphi}(t) = 
\sup_{z \in \mathbb{R}} \Big( z\cdot t - \log \int_{\mathbb{R}} e^{zy} d\widetilde{\mathbb{\bbzeta}}(y) \Big),
\qquad t \in \mathbb{R},  
\label{brostu5:fo.link.var}
\end{equation}
for some probability distribution $\widetilde{\mathbb{\bbzeta}}$ on the real line
such that the function $z \mapsto MGF_{\widetilde{\mathbb{\bbzeta}}}(z) := \int_{\mathbb{R}} e^{zy} d\widetilde{\mathbb{\bbzeta}}(y)$ is finite
on some open interval containing zero.

\end{condition}

\end{problem}

\vspace{0.2cm}
\noindent

\begin{remark} \ 
\label{after represent}
The change from $\varphi$ to $\widetilde{\varphi} := M_{\mathbf{P}} \cdot \varphi$
in Condition \ref{Condition  Fi Tilda in Minimization} stems from the fact that one
can equivalently rewrite \eqref{min Pb} 
such that the vector $\mathbf{P}$ \textquotedblleft turns into\textquotedblright\ 
a probability vector $\widetilde{\mathds{P}}$; the latter will be essentially needed 
for our BS method (cf. Broniatowski \& Stummer \cite{Bro:23a}). Indeed,
one can construct the probability vector 
$\widetilde{\mathds{P}}:=\mathbf{P}/M_{\mathbf{P}}$ and analogously 
$\widetilde{\mathbf{Q}}:=\mathbf{Q}/M_{\mathbf{P}}$ 
(notice that $\widetilde{\mathbf{Q}}$ may be a non-probability vector).
With this, one can equivalently rewrite
\begin{equation}
D_{\varphi }(\mathbf{Q},\mathbf{P})=\sum_{k=1}^{K}p_{k}\cdot \varphi \Big( \frac{q_{k}}{p_{k}}
\Big) =\sum_{k=1}^{K}M_{\mathbf{P}}\cdot \widetilde{p_{k}}\cdot \varphi
\Big( \frac{M_{\mathbf{P}}\cdot \widetilde{q_{k}}}{M_{\mathbf{P}}\cdot \widetilde{p_{k}}}
\Big)=D_{\widetilde{\varphi }}(\widetilde{\mathbf{Q}},\widetilde{\mathds{P}}).
\label{min Pb prob1}
\end{equation}
and thus the solution of \eqref{min Pb} coincides with the one
of the problem of finding
\begin{equation}
\widetilde{\Phi}_{\widetilde{\mathds{P}}}(\widetilde{\mathbf{\Omega}}) := \inf_{\widetilde{\mathbf{Q}}\in \widetilde{\mathbf{\Omega}} }
D_{\widetilde{\varphi} }(\widetilde{\mathbf{Q}},\widetilde{\mathds{P}}),  
\qquad \textrm{with } \widetilde{\mathbf{\Omega}}:=\mathbf{\Omega} /M_{\mathbf{P}}.
\label{min Pb prob2}
\end{equation}

\end{remark} 

\vspace{0.2cm}
\noindent

\begin{remark} \ 
\label{after represent 2}
A comprehensive study on Condition \ref{Condition  Fi Tilda in Minimization}
is given in Section XI of Broniatowski \& Stummer \cite{Bro:23a} as well as 
in Broniatowski \& Stummer \cite{Bro:23b};
numerous explicitly solved cases can be found in Section XII of \cite{Bro:23a}.
In particular, Condition \ref{Condition  Fi Tilda in Minimization}
implies in particular that \\
(i) $\varphi$ is strictly convex 
in a non-empty neighborhood $]t_{-}^{sc},t_{+}^{sc}[ \subseteq ]a,b[$ of one
($t_{-}^{sc} < 1 < t_{+}^{sc}$) 
and affine linear on the rest
$]-\infty,\infty[ \, \backslash \, ]t_{-}^{sc},t_{+}^{sc}[$ in case that
this rest is non-empty, \\
(ii) $\varphi(t) >0$ for all $t \in ]a,b[\backslash{\{1\}}$,  \\
(iii) $\varphi$
is continuously differentiable on $]a,b[$;
accordingly, we denote the corresponding derivative by $\varphi^{\prime}$, \\
(iv) $\varphi^{\prime}(1) = 0$.

\end{remark}

\vspace{0.4cm}

\begin{remark} \ 
\label{after det Problem}
(a) \thinspace\ The purpose of assumption \eqref{regularity} is to get
rid of the $\lim \sup$ type and $\lim \inf$ type results in our
below-mentioned \textquotedblleft bare-simulation\textquotedblright\
approach and to obtain \textit{limit}-statements which motivate our construction. In
practice, it is enough to verify $\mathbf{\Omega }\subseteq cl\left(
int\left( \mathbf{\Omega }\right) \right) $, which is equivalent to 
the left-hand part of \eqref{regularity}. Clearly, any open set 
$\mathbf{\Omega }\subset \mathbb{R}^{K}$ 
satisfies the left-hand part of \eqref{regularity}.
In the subsetup where $\mathbf{\Omega }$ is a closed convex set 
and $int(\mathbf{\Omega })\neq \emptyset $, 
\eqref{regularity} is satisfied and the minimizer $\mathbf{Q}_{min}\in 
\mathbf{\Omega }$ in 
\eqref{min Pb} is attained and even unique. When $\mathbf{\Omega }
$ is open and satisfies \eqref{regularity}, then the
infimum in (\ref{min Pb}) exists but is reached at some generalized
projection of $\mathbf{P}$ on $\mathbf{\Omega }$ 
(see Csiszar
\cite{Csi:84} 
for the Kullback-Leibler-information-distance case of probability measures, 
which extends to any $\varphi-$divergence in our framework).
\\
(b) Without further mentioning,
the regularity assumption \eqref{regularity} is supposed to hold in the \textit{full}
topology. Of course, 
$int\left( \mathbb{S}^{K} \right) = \emptyset$
and thus, for the important probability-vector setup $\mathbf{\Omega} \subset \mathbb{S}^{K}$
the assumption \eqref{regularity} is violated which requires extra refinements
(cf. Section \ref{SectDetSubsimplex.CASM} below).
The same is needed for $\mathbf{\Omega} \subset A \cdot \mathbb{S}^{K}$
for some $A \ne 1$, since obviously $int\left( A \cdot\mathbb{S}^{K} \right) = \emptyset$;
such a context appears naturally e.g. in connection with mass transportation problems 
and with distributed energy management 
(see e.g. Chapter IX of Broniatowski \& Stummer \cite{Bro:23a} and the references therein).
\\
(c) \ 
Our approach is predestined
for \textit{non- or semiparametric} models,
see Broniatowski \& Stummer \cite{Bro:23a} for a detailed discussion.
\\
(d) The Condition \ref{Condition  Fi Tilda in Minimization} implies in particular
that $\varphi$ satisfies
$\int_{\mathbb{R}} y d\widetilde{\mathbb{\bbzeta}}(y) =1$ 
and that $\widetilde{\mathbb{\bbzeta}}$
has light tails;  moreover, $\widetilde{\mathbb{\bbzeta}}$ may depend on $M_{\mathbf{P}}$
in a highly non-trivial way. 
For details --- including also methods for finding $\widetilde{\mathbb{\bbzeta}}$
as well as numerous examples ---
the reader is referred to Broniatowski \& Stummer \cite{Bro:23a}.

\end{remark}

\vspace{0.4cm}
\noindent
Returning to the distance-minimizing Problem \ref{det Problem},
we proceed 
by constructing an appropriate sequence $(\boldsymbol{\xi}_{n})_{n\in \mathbb{N}}$ 
of $\mathbb{R}^{K}-$valued random variables
(cf. \eqref{brostu5:fo.2} in Definition \ref{brostu5:def.1}
and the special case of Remark \ref{brostu5:rem.def1}(a))
as follows:
\ we first transform $\widetilde{\mathds{P}}:=\mathbf{P}/M_{\mathbf{P}}$
having components $(\widetilde{p}_{1}, \ldots, \widetilde{p}_{K})$.
Moreover,
for any $n \in \mathbb{N}$ and 
any $k \in \left\{ 1, \ldots ,K-1\right\}$, let $n_{k}:=\lfloor n \cdot 
\widetilde{p}_{k}\rfloor $ 
(where $\lfloor x \rfloor$ denotes the integer part of $x$)
and $n_{K} := n- \sum_{k=1}^{K-1} n_{k}$;
for this, we assume that $n \in \mathbb{N}$ is large enough, 
namely
$n \geq \max_{k \in \{1, \ldots, K\}} \frac{1}{\widetilde{p}_{k}}$,
such that all the integers $n_{k}$ ($k=1,\ldots,K$) are
non-zero.
Since we assume $\mathbf{P} \in \mathbb{R}_{> 0}^{K}$ 
and thus none of the $\widetilde{p}_{k}$'s is zero, one has 
\begin{equation}
\lim_{n\rightarrow \infty} \frac{n_{k}}{n} = \widetilde{p}_{k}, \qquad k=1,\ldots,K.
\label{fo.freqlim}
\end{equation}
With this at hand,
we decompose the set $\{1, \ldots, n\}$ of all integers from $1$ to $n$
into the following disjoint blocks: $I_{1}^{(n)}:=\left\{
1,\ldots ,n_{1}\right\} $, $I_{2}^{(n)}:=\left\{ n_{1}+1,\ldots
,n_{1}+n_{2}\right\} $, and so on until the last block 
$I_{K}^{(n)} := \{ \sum_{k=1}^{K-1} n_{k} + 1, \ldots, n \}$ which
therefore contains all integers from $n_{1}+ \ldots +n_{K-1}+1$ to $n$.
Due to our construction, $I_{k}^{(n)}$ has $n_{k} \geq 1$ elements (i.e. $card(I_{k}^{(n)}) = n_{k}$
where $card(B)$ denotes the number of elements in a set $B$) 
for all $k \in \{1, \ldots, K\}$
\footnote{
if all $\widetilde{p}_{k}$ ($k=1,\ldots,K$) are rational numbers in $]0,1[$ with 
$\sum_{k=1}^{K} \widetilde{p}_{k} =1$ 
and $N$ is the (always existing) smallest integer such that all
$N \cdot \widetilde{p}_{k}$ ($k=1,\ldots,K$) are integers (i.e. $\in \mathbb{N}$),
then for any multiple $n= m \cdot N$ ($m \in \mathbb{N}$) 
one gets that all $n \cdot \widetilde{p}_{k}$ are integers
and hence $n_{k}= \lfloor n \cdot \widetilde{p}_{k}\rfloor = n \cdot \widetilde{p}_{k}$
($k=1,\ldots,K$)}.
Furthermore, consider a vector 
$\mathbf{\widetilde{W}}:=\left( \widetilde{W}_{1},\ldots ,\widetilde{W}_{n}\right) $ 
where the $\widetilde{W}_{i}$'s are
i.i.d. copies of the random variable $\widetilde{W}$ whose distribution is
associated with the divergence-generator $\widetilde{\varphi }:=
M_{\textbf{P}} \cdot \varphi $ through 
\eqref{brostu5:fo.link.var}, in the sense that 
$\mathbb{\Pi }[\widetilde{W}\in \cdot \,]=\widetilde{\mathbb{\bbzeta}}
[\,\cdot \,]$ on some probability space $(\mathfrak{X},\mathcal{A},\mathbb{\Pi})$. We group the $\widetilde{W}_{i}$'s according to the
above-mentioned blocks and sum them up blockwise, in order to build the
following $K-$ component random vector 
\begin{equation}
\boldsymbol{\xi }_{n}^{\mathbf{\widetilde{W}}}:=\Big(\frac{1}{n}\sum_{i\in
I_{1}^{(n)}}\widetilde{W}_{i},\ldots ,\frac{1}{n}\sum_{i\in I_{K}^{(n)}}
\widetilde{W}_{i}\Big);
\label{Xi_n^W vector}
\end{equation}
notice that the signs of 
its components
may be negative, depending on the
nature of the $\widetilde{W}_{i}$'s; moreover, the expectation of its $k-$th
component converges to $\widetilde{p}_{k}$ as $n$ tends to infinity 
(since the expectation of $\widetilde{W}_{1} $ is $1$), 
whereas the $n-$fold of the corresponding variance converges to $\widetilde{p}_{k}$ times
the variance of $\widetilde{W}_{1}$.

\vspace{0.3cm}
\noindent
For such a context, Broniatowski \& Stummer \cite{Bro:23a} obtain the following assertion\footnote{
with $(\mathfrak{X}_{n},\mathcal{A}_{n},\mathbb{\Pi}_{n})= (\mathfrak{X},\mathcal{A},\mathbb{\Pi})$
for all $n \in \mathbb{N}$} :

\vspace{0.3cm}

\begin{theorem}
\label{brostu5:thm.BSnarrow}
Let $\mathbf{P} \in \mathbb{R}_{> 0}^{K}$, 
$M_{\mathbf{P}}:=\sum_{i=1}^{K}p_{i}>0$,
and suppose that the divergence generator $\varphi$ 
satisfies Condition \ref{Condition  Fi Tilda in Minimization}
above, with $\widetilde{\mathbb{\bbzeta}}$ (cf. \eqref{brostu5:fo.link.var}). 
Additionally, let $\widetilde{W}:=(\widetilde{W}_{i})_{i\in \mathbb{N}}$ be
a sequence of random variables where the $
\widetilde{W}_{i}$'s are i.i.d. copies of the random variable $\widetilde{W}$
whose distribution 
is $\mathbb{\Pi }[\widetilde{W}\in \cdot \,]=\widetilde{\mathbb{\bbzeta}}[\,\cdot \,]$ \footnote{
and thus, $E_{\mathbb{\Pi }}[\widetilde{W}_{i}]=1$ and $\widetilde{W}_{i}$ has light tails}. 
Then, in terms of 
the random vectors $\boldsymbol{\xi }_{n}^{\mathbf{\widetilde{W}}}$
(cf. \eqref{Xi_n^W vector})
there holds 
\begin{equation}
\inf_{\mathbf{Q} \in \mathbf{\Omega }}D_{\varphi }(\mathbf{Q},\mathbf{P})
\ = \ 
-\lim_{n\rightarrow \infty }\frac{1}{n}\log \,\mathbb{\Pi }\negthinspace \left[
\boldsymbol{\xi }_{n}^{\mathbf{\widetilde{W}}}
\in \mathbf{\Omega} /M_{\mathbf{P}}
\right]
\label{LDP Minimization}
\end{equation}
for any $\mathbf{\Omega }\subset \mathbb{R}^{K}$ with regularity properties 
\eqref{regularity} and finiteness property \eqref{def fi wrt Omega}.
In particular, for each $\mathbf{P} \in \mathbb{R}_{> 0}^{K}$ the function 
$\Phi_{\mathbf{P}} \left( \cdot \right) := D_{\varphi}( \cdot , \mathbf{P} )$
(cf. \eqref{brostu5:fo.div}) is bare-simulation minimizable (BS-minimizable)
in the narrow sense (cf. \eqref{brostu5:fo.2} in Definition \ref{brostu5:def.1}
and the special case of Remark \ref{brostu5:rem.def1}(a)) 
on any such $\mathbf{\Omega }\subset \mathbb{R}^{K}$.

\end{theorem}

\vspace{0.2cm} 
\noindent
\begin{remark}
\label{dist of components}
(i)  
For some contexts, we can \textit{explicitly} give the
distribution of each of the independent (non-deterministic parts of the) components
$\Big(\sum_{i\in I_{k}^{(n)}}\widetilde{W}_{i}\Big)_{k=1,\ldots,K}$
of the vector $\boldsymbol{\xi }_{n}^{\mathbf{\widetilde{W}}}$;
this will ease the corresponding concrete simulations.
For corresponding examples, see \cite{Bro:23a}.
\\
(ii)  Let us emphasize that we have assumed $\mathbf{P} \in \mathbb{R}_{> 0}^{K}$
in Theorem \ref{brostu5:thm.BSnarrow} which excludes $\mathbf{P}$
from having zero components. 
However, in cases
where $\lim_{x\rightarrow \infty }\left\vert \frac{\varphi \left(
x \cdot sgn(q)\right) }{x \cdot sgn(q)}\right\vert =+\infty $ for $q\neq 0$, then if 
$p_{k_{0}}=0$ for some $k_{0\text{ }}$ it follows that $q_{k_{0}}=0$,
which proves that $\mathbf{P}\in \mathbb{R}_{>0}^{K}$ imposes no restriction
in Theorem \ref{brostu5:thm.BSnarrow},
since the projection of $\mathbf{P}$ in $\mathbf{\Omega}$ then belongs
to the subspace of $\mathbb{R}^{K}$ generated by the non-null components of $\mathbf{P}$;
such a situation appears e.g. for power divergence generators $\varphi_{\gamma }$ with $\gamma > 2$. 
So there is no loss of generality assuming $\mathbf{P}\in \mathbb{R}_{>0}^{K}$ in this case.\\
(iii) Notice that \eqref{LDP Minimization} even holds when $\mathbf{P}\in \mathbf{\Omega}$
for which the left-hand side becomes zero. 
\end{remark}

\vspace{0.5cm} 
\noindent
Returning to the general context, the limit statement \eqref{LDP Minimization} 
provides the principle for the approximation of the solution of Problem \eqref{min Pb}. Indeed,
by replacing the right-hand side in \eqref{LDP Minimization} by its finite
counterpart, we deduce for given large $n$  
\begin{equation}
- \frac{1}{n}\log \mathbb{\Pi} \negthinspace \left[ \boldsymbol{\xi}_{n}^{\mathbf{\widetilde{W}}}\in 
\mathbf{\Omega}/M_{\mathbf{P}} \right] 
\approx \inf_{Q\in \mathbf{\Omega} }D_{\varphi }(\mathbf{Q},\mathbf{P});
\label{fo.approx.1} 
\end{equation}
it remains to estimate the left-hand side of \eqref{fo.approx.1}.
The latter can be performed either by a \textit{naive estimator} of the
frequency of those replications of $\boldsymbol{\xi }_{n}^{\mathbf{\widetilde{W}}}$  
which hit $\mathbf{\Omega} /M_{\mathbf{P}}$, or more efficiently by some improved estimator; 
for details, the reader is referred to Section X of Broniatowski \& Stummer \cite{Bro:23a}, see   
also the corresponding extensions given in Section 
\ref{SectEstimators.new.det.nonvoid}
below, where the latter also provides e.g. estimates of the \textit{minimizers}).

\vspace{0.3cm}

\begin{remark} \ 
According to 
\eqref{LDP Minimization} of Theorem \ref{brostu5:thm.BSnarrow} as well as \eqref{fo.approx.1},
we can principally tackle 
the (approximative) computation of the minimum value
$\inf_{Q\in \mathbf{\Omega}}
D_{\varphi }(\mathbf{Q},\mathbf{P}) =
\inf_{Q\in \mathbf{\Omega}}
\sum_{k=1}^{K}p_{k}\cdot \varphi \left( \frac{q_{k}}{p_{k}}\right)$
and in particular of
$\inf_{\mathbf{Q} \in \mathbf{\Omega} } \sum_{k=1}^{K}\varphi (q_{k}) = 
\inf_{\mathbf{Q} \in \mathbf{\Omega} } D_{\varphi }(\mathbf{Q},\mathbf{1})  
\  \textrm{(cf. \eqref{min Pb one new})}$
by basically \textit{only employing a fast and accurate 
--- pseudo, true, natural, quantum  
--- random number generator}\footnote{
for exemplary recent references on 
this very active research field on fast and accurate random number generation
see e.g.~\cite{Tuc:13}--\cite{Zheng15:23} },
provided that the constraint set $\mathbf{\Omega}$ 
satisfies the mild assumptions \eqref{regularity} and \eqref{def fi wrt Omega}.
Notice that \eqref{regularity} also covers 
constraint sets $\mathbf{\Omega}$
(of arbitrary dimension $K$)
which are \textit{non-convex} and even \textit{highly disconnected}, and for which
other minimization methods (e.g. pure enumeration, gradient or steepest descent methods, etc.)
may be problematic or intractable.

\end{remark}

\vspace{0.3cm}
\noindent
Returning to the general context, notice that Theorem \ref{brostu5:thm.BSnarrow}
does not cover cases where $\mathbf{\Omega}$ consists of $\mathbf{Q}$ satisfying the
additional constraint $\sum_{i=1}^{K} q_{i} =A$ for some fixed $A>0$
(and thus $int\left( \mathbf{\Omega} \right) = \emptyset$ which violates \eqref{regularity}).
However, such situations can be still handled with an adaption of 
the above-described narrow-sense BS method, see 
Broniatowski \& Stummer~\cite{Bro:23a} and also the beginning of Section \ref{SectDetSubsimplex}
below.

\vspace{0.4cm}
\noindent
To end up this section, let us present some examples of $\varphi-$divergence generators
which satisfy Condition \ref{Condition  Fi Tilda in Minimization} and for which
thus the Theorem \ref{brostu5:thm.BSnarrow} can be applied;
the corresponding essential distributions $\widetilde{\bbzeta}$ can be obtained from 
$\bbzeta$ in the last column of Table 1 below, by employing $M_{\mathbf{P}} \cdot \varphi$ instead of $\varphi$.
For details, construction methods and further examples, the reader is referred to Broniatowski \& Stummer \cite{Bro:23a}
(see also Broniatowski \& Stummer \cite{Bro:23c} for applications to fuzzy sets and basic belief assignments).

\noindent
\begin{example}
\label{brostu5:ex.1}
Let us take the important case of power-divergence generators $\varphi_{\gamma} : \mathbb{R} \mapsto [0,\infty]$
defined by
\begin{eqnarray}
\varphi_{\gamma}(t) \hspace{-0.2cm} &:=& \hspace{-0.2cm}
\begin{cases}
\frac{t^\gamma-\gamma \cdot t+ \gamma - 1}{\gamma \cdot (\gamma-1)}, \hspace{6.0cm} \textrm{if }  \gamma \in \, ]-\infty,0[ \ 
\textrm{and } t \in ]0,\infty[, 
 \\
- \log t + t - 1, \hspace{5.5cm} \textrm{if }  \gamma = 0 \ \textrm{and } t \in \, ]0,\infty[,
\\
\frac{t^\gamma-\gamma \cdot t+ \gamma - 1}{\gamma \cdot (\gamma-1)}, \hspace{6.0cm} \textrm{if }  \gamma \in \, ]0,1[ \ 
\textrm{and } t \in [0,\infty[,
 \\
t \cdot \log t + 1 - t, \hspace{5.5cm} \textrm{if }  \gamma = 1 \ \textrm{and } t \in [0,\infty[,
\\
\frac{t^\gamma-\gamma \cdot t+ \gamma - 1}{\gamma \cdot (\gamma-1)} \cdot \textfrak{1}_{]0,\infty[}(t)
+(\frac{1}{\gamma} - \frac{t}{\gamma-1}) \cdot \textfrak{1}_{]-\infty,0]}(t),
\hspace{1.0cm} \textrm{if }  \gamma \in \, ]1,2[ \ \textrm{and } t \in \, ]-\infty,\infty[,
\\
\frac{(t - 1)^2}{2}, \hspace{6.8cm} \textrm{if }  \gamma = 2 \ 
\textrm{and } t \in \, ]-\infty,\infty[,
\\
\frac{t^\gamma-\gamma \cdot t+ \gamma - 1}{\gamma \cdot (\gamma-1)} \cdot \textfrak{1}_{]0,\infty[}(t)
+(\frac{1}{\gamma} - \frac{t}{\gamma-1}) \cdot \textfrak{1}_{]-\infty,0]}(t),
\hspace{1.0cm} \textrm{if }  \gamma \in \, ]2,\infty[ \ \textrm{and } t \in \, ]-\infty,\infty[,
\\
\infty, \hspace{7.35cm} \textrm{else},
\end{cases}
\label{brostu5:fo.powdivgen} 
\end{eqnarray}
which --- by \eqref{brostu5:fo.div} --- for arbitrary multiplier $\widetilde{c} \in \, ]0,\infty[$ generate (the vector-valued form of) the 
\textit{generalized power divergences} given by
\begin{eqnarray}
D_{\widetilde{c} \cdot \varphi_{\gamma}}(\mathbf{Q},\mathbf{P}) \hspace{-0.2cm} &:=& \hspace{-0.2cm}
\begin{cases}
\widetilde{c} \cdot \Big\{\frac{ \sum\limits_{k=1}^{K} (q_{k})^{\gamma} \cdot (p_{k})^{1-\gamma}}{\gamma \cdot (\gamma-1)}
- \frac{1}{\gamma -1} \cdot \sum\limits_{k=1}^{K} q_{k} + \frac{1}{\gamma} \cdot \sum\limits_{k=1}^{K} p_{k} \Big\}, 
\hspace{2.0cm} \textrm{if }  \gamma \in \, ]-\infty,0[, \  
\mathbf{P} \in \mathbb{R}_{\gneqq 0}^{K} \ \textrm{and } \mathbf{Q} \in \mathbb{R}_{> 0}^{K}, 
 \\
\widetilde{c} \cdot \Big\{ \sum\limits_{k=1}^{K} p_{k} \cdot \log \Big(\frac{p_{k}}{q_{k}} \Big) 
+ \sum\limits_{k=1}^{K} q_{k} - \sum\limits_{k=1}^{K} p_{k} \Big\}, 
\hspace{3.1cm} \textrm{if }  \gamma = 0, \  
\mathbf{P} \in \mathbb{R}_{\gneqq 0}^{K} \ \textrm{and } \mathbf{Q} \in \mathbb{R}_{> 0}^{K},
\\
\widetilde{c} \cdot \Big\{\frac{ \sum\limits_{k=1}^{K} (q_{k})^{\gamma} \cdot (p_{k})^{1-\gamma}}{\gamma \cdot (\gamma-1)}
- \frac{1}{\gamma -1} \cdot \sum\limits_{k=1}^{K} q_{k} + \frac{1}{\gamma} \cdot \sum\limits_{k=1}^{K} p_{k} \Big\}, 
\hspace{2.1cm} \textrm{if }  \gamma \in \, ]0,1[, \  
\mathbf{P} \in \mathbb{R}_{\gneqq 0}^{K} \ \textrm{and } \mathbf{Q} \in \mathbb{R}_{\geq 0}^{K}, 
 \\
\widetilde{c} \cdot \Big\{ \sum\limits_{k=1}^{K} q_{k} \cdot \log \Big(\frac{q_{k}}{p_{k}} \Big)
- \sum\limits_{k=1}^{K} q_{k} + \sum\limits_{k=1}^{K} p_{k} \Big\}, 
\hspace{3.2cm} \textrm{if }  \gamma = 1, \  
\mathbf{P} \in \mathbb{R}_{> 0}^{K} \ \textrm{and } \mathbf{Q} \in \mathbb{R}_{\geq 0}^{K},
\\
\widetilde{c} \cdot \Big\{ \sum\limits_{k=1}^{K} \frac{(q_{k})^{\gamma} \cdot (p_{k})^{1-\gamma}}{\gamma \cdot (\gamma-1)} 
\cdot \textfrak{1}_{[0,\infty[}(q_{k})
- \frac{1}{\gamma -1} \cdot \sum\limits_{k=1}^{K} q_{k} + \frac{1}{\gamma} \cdot \sum\limits_{k=1}^{K} p_{k} \Big\},
\hspace{0.3cm}  \textrm{if }  \gamma \in \, ]1,2[, \  
\mathbf{P} \in \mathbb{R}_{>0}^{K} \ \textrm{and } \mathbf{Q} \in \mathbb{R}^{K}, 
\\
\widetilde{c}\cdot \sum\limits_{k=1}^{K}\frac{  (q_{k}-p_{k})^{2}}{2 \cdot p_{k}} , 
\hspace{7.1cm}  \textrm{if }  \gamma = 2, \  
\mathbf{P} \in \mathbb{R}_{>0}^{K} \ \textrm{and } \mathbf{Q} \in \mathbb{R}^{K},
\\
\widetilde{c} \cdot \Big\{ \sum\limits_{k=1}^{K} \frac{(q_{k})^{\gamma} \cdot (p_{k})^{1-\gamma}}{\gamma \cdot (\gamma-1)} 
\cdot \textfrak{1}_{[0,\infty[}(q_{k})
- \frac{1}{\gamma -1} \cdot \sum\limits_{k=1}^{K} q_{k} + \frac{1}{\gamma} \cdot \sum\limits_{k=1}^{K} p_{k} \Big\},
\hspace{0.4cm}  \textrm{if }  \gamma \in \, ]2,\infty[, \  
\mathbf{P} \in \mathbb{R}_{>0}^{K} \ \textrm{and } \mathbf{Q} \in \mathbb{R}^{K},
\\
\infty, \hspace{9.05cm} \textrm{else};
\end{cases}
\label{brostu5:fo.powdiv.new} 
\end{eqnarray}
notice that one has the straightforward relationship 
$D_{\widetilde{c}\cdot\varphi_{\gamma}}(\cdot ,\cdot )
=\widetilde{c}\cdot D_{\varphi _{\gamma}}(\cdot ,\cdot )$
but $\widetilde{c}$ enters in a non-straightforward way in the construction
of the corresponding simulation distributions $\widetilde{\bbzeta}$, cf. Table 1 below
(for a discussion on the importance of $\widetilde{c}$, the reader is referred to \cite{Bro:23a}).\\
\indent For $\widetilde{c}=1$ and probability vectors $\mathds{Q}$, $\mathds{P}$ in $\mathbb{S}^{K}$ respectively 
in $\mathbb{S}_{>0}^{K}$, the divergences 
\eqref{brostu5:fo.powdiv.new}
simplify considerably, namely to the well-known
\textit{power divergences} $D_{\varphi_{\gamma}}(\mathds{Q},\mathds{P})$
in the scaling of e.g. Liese \& Vajda \cite{Lie:87} 
(in other scalings they are also called
\textit{Rathie \& Kannapan\textquoteright s non-additive directed divergences of order $\gamma$}
\cite{Rat:72}, \textit{Cressie-Read divergences} \cite{Cre:84} \cite{Rea:88}, 
\textit{relative Tsallis entropies or Tsallis cross-entropies} \cite{Tsa:98}
(see also Shiino \cite{Shi:98}), 
\textit{Amari\textquoteright s alpha-divergences} \cite{Ama:85});
for some comprehensive overviews on power divergences 
$D_{\varphi_{\gamma}}(\mathds{Q},\mathds{P})$
--- including statistical applications to goodness-of-fit testing and
minimum distance estimation ---
the reader is referred to the above-mentioned insightful 
books~\cite{Lie:87}--\cite{Lie:08},
the survey articles~\cite{Lie:06},\cite{Vaj:10},
and the references therein.
Prominent and widely used special cases of $D_{\varphi_{\gamma}}(\mathds{Q},\mathds{P})$
are the omnipresent \textit{Kullback-Leibler information divergence (relative entropy)}
where $\gamma=1$, the equally important
\textit{reverse Kullback-Leibler information divergence (reverse relative entropy)}
where $\gamma =0$,
the \textit{Pearson chi-square divergence} ($\gamma=2$), the 
\textit{Neyman chi-square divergence} ($\gamma=-1$),
the \textit{Hellinger divergence} ($\gamma=\frac{1}{2}$,
also called squared Hellinger distance, 
squared Matusita distance \cite{Mat:51} or squared Hellinger-Kakutani metric, 
see e.g. Deza \& Deza \cite{Dez:16}). 
Some exemplary (relatively) recent studies and applications
of power divergences $D_{\varphi_{\gamma}}(\mathds{Q},\mathds{P})$
--- aside from the vast statistical literature (including 
in particular maximum likelihood estimation and Pearson\textquoteright s chi-square test) ---
are cited e.g. in Broniatowski \& Stummer \cite{Bro:23a}.
For $\widetilde{c}=1$ and nonnegative-component vectors $\mathbf{Q}$, $\mathbf{P}$ in 
$\mathbb{R}_{\geq 0}^{K}$ respectively 
$\mathbb{R}_{>0}^{K}$ respectively $\mathbb{R}_{\gneqq 0}^{K}$, the generalized power divergences 
$D_{\varphi_{\gamma}}(\mathbf{Q},\mathbf{P})$ of \eqref{brostu5:fo.powdiv.new}
were treated by 
Stummer \& Vajda \cite{Stu:10} (for even more general probability measures, deriving
e.g. also generalized Pinsker\textquoteright s  inequalities);
for a more general comprehensive technical treatment see also e.g. 
Broniatowski \& Stummer \cite{Bro:19b}.

\vspace{0.3cm}
\noindent
For any fixed $M_{\mathbf{P}}\in \, ]0,\infty[$, Condition \ref{Condition  Fi Tilda in Minimization}
is satisfied for $\varphi := \widetilde{c}\cdot\varphi_{\gamma}$ --- 
and thus the \textit{narrow-sense} BS-minimizability concerning Theorem \ref{brostu5:thm.BSnarrow} can be applied ---
for all $\widetilde{c} \in \, ]0,\infty[$ and all $\gamma \in \mathbb{R}\backslash]1,2[$
(cf. \cite{Bro:23a}). 
(As far as we know at the moment) For the case $\gamma \in \, ]1,2[$ one can not verify
Condition \ref{Condition  Fi Tilda in Minimization}, but BS-minimizability (in the sense of \eqref{brostu5:fo.2} 
with $f_{n}(\cdot) \not\equiv 1$) will be 
shown in Example \ref{brostu5:ex.POWmiss} below.\\
For the latter,
we shall employ the next example, 
which is of interest on its own due to its 
finite-derivative-behaviour at $t = \pm \infty$
(and which will be also helpful for e.g. the study of the total variation distance
in Example \ref{brostu5:ex.TV} below).

\end{example}

\noindent
\begin{example}
\label{brostu5:ex.2}
For any parameter-triple $\alpha,\beta,\widetilde{c} \in \, ]0,\infty[$
we choose 
$]a,b[ \,  :=  \,  ]-\infty, \infty \, [$  and
\begin{equation}
\varphi_{\alpha,\beta,\widetilde{c}}(t) 
:= \widetilde{c} \cdot \alpha \cdot \Big\{
\sqrt{1 + \beta^{2} 
\cdot \Big(\frac{1-t}{\alpha}\Big)^2} \, - \, 1 
+ \log\frac{2 \cdot \Big(
\sqrt{1 + \beta^{2} 
\cdot \Big(\frac{1-t}{\alpha} \Big)^2} \, - \, 1
\Big)
}{
\beta^{2} \cdot \Big(\frac{1-t}{\alpha}\Big)^{2}
}  \Big\}
\ \in [0,\infty[,
\ \  
t \in \, ]-\infty, \infty[ \, .
\label{brostu5:fo.genLap3ba}
\end{equation}
Notice that  
$\varphi_{\alpha,\beta,\widetilde{c}}(1) = 0$, 
$\varphi_{\alpha,\beta,\widetilde{c}}^{\prime}(1) = 0$,
$\varphi_{\alpha,\beta,\widetilde{c}}(-\infty) = \infty$
and $\varphi_{\alpha,\beta,\widetilde{c}}(\infty) = \infty$. 
Moreover,  
$\varphi_{\alpha,\beta,\widetilde{c}}^{\prime}(-\infty) = 
\varphi_{\alpha,\beta,\widetilde{c}}^{\prime}(a) =
-\widetilde{c} \cdot \beta$ and 
$\varphi_{\alpha,\beta,\widetilde{c}}^{\prime}(\infty) = 
\varphi_{\alpha,\beta,\widetilde{c}}^{\prime}(b) =
\widetilde{c} \cdot \beta$.  Furthermore, 
$\varphi_{\alpha,\beta,\widetilde{c}}(\cdot)$ is strictly convex and
smooth (i.e. of $C^{\infty}-$type).
From \eqref{brostu5:fo.genLap3ba},
we construct the corresponding divergence (cf. \eqref{brostu5:fo.div})
\begin{eqnarray}
& & \hspace{-0.7cm}
D_{\varphi_{\alpha,\beta,\widetilde{c}}}(\mathbf{Q},\mathbf{P})
= \sum\limits_{k=1}^{K} p_{k} \cdot 
\varphi_{\alpha,\beta,\widetilde{c}}\Big(\frac{q_{k}}{p_{k}}\Big)
\nonumber \\
& & \hspace{-0.7cm}
= \widetilde{c} \cdot \alpha \cdot  \sum\limits_{k=1}^{K} 
p_{k} \cdot
\Big\{
\sqrt{1 + \beta^{2} 
\cdot \Big(\frac{1-\frac{q_{k}}{p_{k}}}{\alpha}\Big)^2} \, - \, 1 
+ \log\frac{2 \cdot \Big(
\sqrt{1 + \beta^{2} 
\cdot \Big(\frac{1-\frac{q_{k}}{p_{k}}}{\alpha} \Big)^2} \, - \, 1
\Big)
}{
\beta^{2} \cdot \Big(\frac{1-\frac{q_{k}}{p_{k}}}{\alpha}\Big)^{2}
}  \Big\} ,  
\qquad \mathbf{P} \in \mathbb{R}_{> 0}^{K}, \mathbf{Q} \in \mathbb{R}^{K}.
\label{brostu5:fo.genLap4fa}
\end{eqnarray}
For any fixed $M_{\mathbf{P}}\in \, ]0,\infty[$, Condition \ref{Condition  Fi Tilda in Minimization}
is satisfied for $\varphi := \varphi_{\alpha,\beta,\widetilde{c}}$
(with a comfortably computable distribution $\widetilde{\bbzeta}$, cf. Table 1 below)
--- and thus the \textit{narrow-sense} BS-minimizability concerning Theorem \ref{brostu5:thm.BSnarrow} can be applied ---
for all parameter-triples $\alpha,\beta,\widetilde{c} \in \, ]0,\infty[$ (cf. \cite{Bro:23a}).
For the important special case $\alpha=\beta$ the formula \eqref{brostu5:fo.genLap3ba} 
collapses to
\begin{equation}
\varphi_{\beta,\beta,\widetilde{c}}(t) 
= \widetilde{c} \cdot \beta \cdot \Big\{
\sqrt{1 + (1-t)^2} \, - \, 1 
+ \log\frac{2 \cdot \Big(
\sqrt{1 + (1-t)^2} \, - \, 1
\Big)
}{
(1-t)^{2}
}  \Big\}
\ \in [0,\infty[,
\ \  
t \in \, ]-\infty, \infty[ \, ,
\label{brostu5:fo.genLap3ba.equal}
\end{equation}
and \eqref{brostu5:fo.genLap4fa} collapses to
\begin{eqnarray}
& & \hspace{-0.7cm}
D_{\varphi_{\beta,\beta,\widetilde{c}}}(\mathbf{Q},\mathbf{P})
= \sum\limits_{k=1}^{K} p_{k} \cdot 
\varphi_{\beta,\beta,\widetilde{c}}\Big(\frac{q_{k}}{p_{k}}\Big)
\nonumber \\
& & \hspace{-0.7cm}
= \widetilde{c} \cdot \beta \cdot  \sum\limits_{k=1}^{K} 
p_{k} \cdot
\Big\{
\sqrt{1 + \Big(1-\frac{q_{k}}{p_{k}}\Big)^2} \, - \, 1 
+ \log\frac{2 \cdot \Big(
\sqrt{1 + \Big(1-\frac{q_{k}}{p_{k}}\Big)^2} \, - \, 1
\Big)
}{
\Big(1-\frac{q_{k}}{p_{k}}\Big)^{2}
}  \Big\} ,  
\qquad \mathbf{P} \in \mathbb{R}_{> 0}^{K}, \mathbf{Q} \in \mathbb{R}^{K}.
\nonumber
\end{eqnarray}

\end{example}

\vspace{0.1cm}
\noindent
To end this section, let us present Table 1 which gives a compact summary of several important 
solved cases; for details, see Section XII of Broniatowski \& Stummer~\cite{Bro:23a}.
Notice that --- as already explained above --- in the current setup one has to take
in Table 1 the divergence generator $M_{\mathbf{P}} \cdot \varphi$ instead of $\varphi$ (cf. \eqref{brostu5:fo.link.var} 
instead of \eqref{brostu5:fo.link.var.simplex}) and accordingly
$\bbzeta$ turns into $\widetilde{\bbzeta}$.
Here and henceforth, we employ the following notations:

\begin{itemize}
\item $GAM(\alpha,\beta)$ denotes the Gamma distribution 
with rate parameter (inverse scale parameter) $\alpha \in \, ]0,\infty[$ and 
shape parameter $\beta \, ]0,\infty[$ having (Lebesgue-)density
$f(y) := \frac{\alpha^{\beta} \cdot 
y^{\beta-1} \cdot e^{-\alpha \cdot y} }{\Gamma(\beta)} 
\cdot \textfrak{1}_{]0,\infty[}(y)$, 
\  $y \in \mathbb{R}$; 

\item $POI(\kappa)$ denotes the Poisson distribution 
with parameter $\kappa \in \, ]0,\infty[$;

\item Compound $POI(\kappa)$-$GAM(\alpha,\beta)$ denotes the corresponding 
\textquotedblleft  compound Poisson-Gamma distribution\textquotedblright ; 

\item $NOR(\mu,\sigma^2)$ denotes the Normal (i.e. Gaussian) distribution with
mean $\mu \in \mathbb{R}$ and variance $\sigma^2 \in \, ]0,\infty[$;

\item $NB(\tau,p)$ denotes the Negative-Binomial distribution
with parameters $\tau \in \, ]0,\infty[$ and $p \in \, ]0,1[$.

\end{itemize}


\newpage

{
\hspace{-0.9cm}
\tiny
\rotatebox{90}{
\begin{tabular}{|*{12}{l|}} \cline{1-12}
$]a,b[$ &
$\varphi(t)$ for $t \in ]a,b[$ & 
$\mathbf{P} \in$ & 
$\mathbf{Q} \in $  
&
$D_{\varphi}(\mathbf{Q},\mathbf{P})$,  
$P=(p_{k})_{k=1,\ldots K}$ 
& 
$]t_{-}^{sc},t_{+}^{sc}[$ & 
$\varphi(a)$ & 
$\varphi(b)$ & 
$\varphi^{\prime}(a)$ & 
$\varphi^{\prime}(b)$  &  
support of $\mathbb{\bbzeta}$
&
$\mathbb{\bbzeta} =\mathbb{\Pi}[W \in \cdot \, ]$   
\\ 
& 
&
&  
&
$\mathbf{Q}=(q_{k})_{k=1,\ldots K}$  &
&  
&  
& 
& 
& 
& 
cf. \eqref{brostu5:fo.link.var.simplex}
\\\cline{1-12}
$]0,\infty[$ & 
$\widetilde{c} \cdot \varphi_{\gamma}(t) := \widetilde{c} \cdot 
\frac{t^\gamma-\gamma \cdot t+ \gamma - 1}{\gamma \cdot (\gamma-1)}$  & 
 $\mathbb{R}_{\gneqq 0}^{K}$ & 
 $\mathbb{R}_{> 0}^{K}$ & 
$\widetilde{c} \cdot \Big\{\frac{ \sum\limits_{k=1}^{K} (q_{k})^{\gamma} 
\cdot (p_{k})^{1-\gamma}}{\gamma \cdot (\gamma-1)} $ &
$]0,\infty[$ & 
$\infty$ & 
$\infty$ & 
$- \infty$ & 
$\frac{\widetilde{c}}{1-\gamma}$ & 
$]0,\infty[$ & 
dampened stable 
\\  
& 
for $\gamma \in \ ]-\infty,0[$, $\widetilde{c} >0$  &
&  
&
$- \frac{1}{\gamma -1} \cdot \sum\limits_{k=1}^{K} q_{k} + \frac{1}{\gamma} \cdot \sum\limits_{k=1}^{K} p_{k} \Big\}$  
&
&     
&  
&  
&  
&   
&  
distribution on $[0,\infty[$  
\\\cline{1-12}  
$]0,\infty[$ & 
$\widetilde{c} \cdot \varphi_{\gamma}(t)$ & 
 $\mathbb{R}_{\gneqq 0}^{K}$ & 
 $\mathbb{R}_{\geq 0}^{K}$ & 
as above &
$]0,\infty[$ & 
$\frac{\widetilde{c}}{\gamma}$ & 
$\infty$ & 
$- \infty$ & 
$\frac{\widetilde{c}}{1-\gamma}$ & 
$[0,\infty[$ &  
Compound $POI(\frac{\widetilde{c}}{\gamma })$  
\\  
& 
for $\gamma \in ]0,1[ $, $\widetilde{c} >0$  &
&  
&
&
&  
&  
&  
&  
&   
&  
$-$ $GAM(\frac{\widetilde{c}}{1-\gamma } ,\frac{\gamma }{1-\gamma })$
\\\cline{1-12}  
$]\negthinspace \negthinspace - \negthinspace \infty,\infty[$ & 
$\widetilde{c} \cdot \{ \varphi_{\gamma}(t) \cdot \textfrak{1}_{]0,\infty[}(t)$ & 
 $\mathbb{R}_{> 0}^{K}$ & 
 $\mathbb{R}^{K}$ & 
$\widetilde{c} \cdot \Big\{ \sum\limits_{k=1}^{K} \frac{(q_{k})^{\gamma} \cdot (p_{k})^{1-\gamma}}{\gamma \cdot (\gamma-1)} 
\cdot \textfrak{1}_{[0,\infty[}(q_{k})$ &
$]0,\infty[$ & 
$\infty$ & 
$\infty$ & 
$\frac{\widetilde{c}}{1-\gamma}$ & 
$\infty$ & 
$]\negthinspace \negthinspace - \negthinspace \infty,\infty[$ &  
distorted stable distri- 
 \\  
& 
$+(\frac{1}{\gamma} - \frac{t}{\gamma-1}) \cdot \textfrak{1}_{]-\infty,0]}(t) \}$ &
&  
&
$- \frac{1}{\gamma -1} \cdot \sum\limits_{k=1}^{K} q_{k} + \frac{1}{\gamma} \cdot \sum\limits_{k=1}^{K} p_{k} \Big\}$ 
&     
&  
&  
&  
&  
&   
&  
bution on $]\negthinspace \negthinspace - \negthinspace \infty,\infty[$  
\\ 
& 
for $\gamma \in \, ]2,\infty[ $, $\widetilde{c} >0$   &
&  
&
&
&       
&  
&  
&  
&   
&   
\\\cline{1-12}  
$]\negthinspace \negthinspace - \negthinspace \infty,\infty[$ & 
$\widetilde{c} \cdot \varphi_{2}(t) = \widetilde{c} \cdot \frac{(t - 1)^2}{2}$  & 
 $\mathbb{R}_{> 0}^{K}$ & 
 $\mathbb{R}^{K}$ & 
$\widetilde{c}\cdot \sum\limits_{k=1}^{K}\frac{  (q_{k}-p_{k})^{2}}{2 \cdot p_{k}} $ &
$]\negthinspace \negthinspace - \negthinspace \infty,\infty[$ & 
$\infty$ & 
$\infty$ & 
$- \infty$ & 
$\infty$ & 
$]\negthinspace \negthinspace - \negthinspace \infty,\infty[$ & 
$NOR(1, \frac{1}{\widetilde{c}})$ 
\\  
& 
for $\widetilde{c} >0$  &
&  
&
&
&      
&  
&  
&  
&   
&  
\\\cline{1-12}  
$]0,\infty[$ & 
$\widetilde{c} \cdot \varphi_{0}(t) := \widetilde{c} \cdot \{ - \log t $  & 
 $\mathbb{R}_{\gneqq 0}^{K}$ & 
 $\mathbb{R}_{> 0}^{K}$ & 
$\widetilde{c} \cdot \Big\{ \sum\limits_{k=1}^{K} p_{k} \cdot \log \Big(\frac{p_{k}}{q_{k}} \Big) $ &
$]0,\infty[$ & 
$\infty$ & 
$\infty$ & 
$- \infty$ & 
$1$ & 
$]0,\infty[$ & 
$GAM(\widetilde{c} ,\widetilde{c})$ 
\\  
& 
$ + \, t - 1 \}$ \ \ for $\widetilde{c} >0$  &
&  
&
$+ \sum\limits_{k=1}^{K} q_{k} - \sum\limits_{k=1}^{K} p_{k} \Big\}$  &  
&     
&  
&  
&  
&   
&  
\\\cline{1-11}  
$]0,\infty[$ & 
$\widetilde{c} \cdot \varphi_{1}(t) := \widetilde{c} \cdot \{ t \cdot \log t $  & 
 $\mathbb{R}_{> 0}^{K}$ & 
 $\mathbb{R}_{\geq 0}^{K}$ & 
$\widetilde{c} \cdot \Big\{ \sum\limits_{k=1}^{K} q_{k} \cdot \log \Big(\frac{q_{k}}{p_{k}} \Big) $ &
$]0,\infty[$ & 
$1$ & 
$\infty$ & 
$- \infty$ & 
$\infty$ & 
$\{0, \frac{1}{\widetilde{c}},  \ldots \}$ & 
$\frac{1}{\widetilde{c}}-$fold of
$POI(\widetilde{c})$ 
\\  
& 
$ + \, 1 - t \}$ \ \ for $\widetilde{c} >0$  &
&  
&
$- \sum\limits_{k=1}^{K} q_{k} + \sum\limits_{k=1}^{K} p_{k} \Big\}$  & 
&     
&  
&  
&  
&   
$= \frac{1}{\widetilde{c}} \cdot \mathbb{N}_{0}$ &    
\\\cline{1-12}  
$]0,\infty[$& 
$\widetilde{c} \cdot \{ t \cdot \log t  $ & 
 $\mathbb{R}_{\gneqq 0}^{K}$ & 
 $\mathbb{R}_{\gneqq 0}^{K}$ \ \ with & 
$\widetilde{c} \cdot \Big\{ \sum\limits_{k=1}^{K} q_{k} \cdot \log \Big(\frac{2 q_{k}}{q_{k} + p_{k}} \Big) $ &
$]0,\infty[$ & 
$\widetilde{c} \log 2$ & 
$\infty$ & 
$- \infty$ & 
$\widetilde{c} \log 2$ & 
$\{0, \frac{1}{\widetilde{c}}, \ldots \}$ &  
$\frac{1}{\widetilde{c}}-$fold of
$NB(\widetilde{c}, \frac{1}{2})$ 
\\  
& 
$+ \, (t+1) \cdot \log(\frac{2}{t+1}) \}$, $\widetilde{c} >0$  &
& $\mathbf{P} \negthinspace + \negthinspace \mathbf{Q} \negthinspace \in \negthinspace \mathbb{R}_{>0}^{K}$ 
&
$+ \sum\limits_{k=1}^{K} p_{k} \cdot \log \Big(\frac{2 p_{k}}{q_{k} + p_{k}} \Big) \Big\}$
&      
&  
&  
&  
&  
&   
$= \frac{1}{\widetilde{c}} \cdot \mathbb{N}_{0}$ &  
\\\cline{1-12}  
$]\frac{\beta-1}{\beta} \negthinspace , \negthinspace \infty[$ & 
$\widetilde{c} \cdot \frac{(t-1)^{2}}{2(\beta \cdot t +1 -\beta)}$ \ \ 
for 
$\widetilde{c}>0$ & 
$\mathbb{R}_{\gneqq 0}^{K}$ & 
$] \frac{\beta-1}{\beta} \mathbf{P},\infty[$
& 
$\frac{\widetilde{c}}{2} \cdot \sum\limits_{k=1}^{K} 
\frac{(q_{k}-p_{k})^{2}}{\beta \cdot q_{k} + (1 -\beta)\cdot p_{k}}$
&
$]\frac{\beta-1}{\beta},\infty[$ & 
$\infty$ & 
$\infty$ & 
$- \infty$ & 
$\frac{\widetilde{c}}{2\beta}$ & 
$]\frac{\beta-1}{\beta},\infty[$ &  
modified dampened stable 
\\  
$\beta \in \, ]0,1]$ & 
&
&  
component-wise &
&  
&    
&  
&  
&  
&   
&  
distribution  
\\\cline{1-12}  
$]z_{1},z_{2}[$ & 
$\frac{(t-z_{1})}{z_{2}-z_{1}} \cdot \log \big(\frac{(t-z_{1}) \cdot p}{(z_{2}-t) \cdot (1-p)} \big)$ & 
 $\mathbb{R}_{> 0}^{K}$ & 
$[z_{1} \mathbf{P}, z_{2} \mathbf{P}]$ & 
$\sum\limits_{k=1}^{K} \frac{q_{k} - z_{1} \cdot p_{k}}{z_{2} - z_{1}}  \cdot 
\log \Big(\frac{p \cdot (q_{k} - z_{1} \cdot p_{k})}{(1-p) \cdot (z_{2} \cdot p_{k} - q_{k})} \Big)$ &
$]z_{1},z_{2}[$ & 
$\log \frac{1}{p}$ & 
$\log \frac{1}{1-p}$ & 
$- \infty$ & 
$\infty$ & 
$\{z_{1},z_{2}\}$ &  
$\mathbb{\bbzeta}[\{z_{1}\}] = p$, 
\\  
$z_{1} < 1$ & 
$- \log \big(\frac{(z_{2}-z_{1}) \cdot p}{(z_{2}-t)} \big)$  &
&  
comp.-wise &
$- \sum\limits_{k=1}^{K} p_{k} \cdot  
\log \Big(\frac{p \cdot (z_{2}-z_{1}) \cdot  p_{k}}{z_{2} \cdot p_{k} - q_{k}}  \Big)$  & 
&     
&  
&  
&  
&   
&  
$\mathbb{\bbzeta}[\{z_{2}\}] = 1-p$ 
\\ 
$z_{2} >1$ & 
for $p := \frac{z_{2}-1}{z_{2}-z_{1}} \in \, ]0,1[ $ &
&  
&
&
&       
&  
&  
&  
&   
&    
\\\cline{1-12}   
$]\negthinspace \negthinspace - \negthinspace \infty,\infty[$ &  
$\widetilde{c}  \cdot \varphi_{\alpha,\beta_{1},\beta_{2}}(t)  := 
\widetilde{c} \cdot \alpha  \cdot 
\{$ & 
$\mathbb{R}_{\gneqq 0}^{K}$ & 
 $\mathbb{R}^{K}$ &
$\widetilde{c} \cdot \sum\limits_{k=1}^{K} p_{k} \cdot 
\varphi_{\alpha,\beta_{1},\beta_{2}}\Big(\frac{q_{k}}{p_{k}}\Big)$ 
&
$]\negthinspace \negthinspace - \negthinspace \infty,\infty[$ & 
$\infty$ & 
$\infty$ & 
$-\widetilde{c} \beta_{2}$ & 
$\widetilde{c} \beta_{1}$ & 
$]\negthinspace \negthinspace - \negthinspace \infty,\infty[$ &  
law of $\breve{\theta} + Z_{1} - Z_{2}$ 
\\  
&  
$\frac{\sqrt{4 + (\beta_{1} + \beta_{2})^{2} g^2}
 - (\beta_{1} -  \beta_{2}) g - 2}{2}$  &
&  
&
&
&     
&  
&  
&  
&   
&  
$\breve{\theta} := 1 \negthinspace + \negthinspace \alpha \cdot \Big(\frac{1}{\beta_{2}} 
\negthinspace \negthinspace - \negthinspace \negthinspace \frac{1}{\beta_{1}} \Big)$ 
\\ 
& 
$+ \log\frac{\sqrt{4 + (\beta_{1} + \beta_{2})^{2} g^2} - 2}{\beta_{1} \beta_{2} g^{2}} $  \}&
&  
&
&  
&      
&  
&  
&  
&   
&  
$Z_{1}, Z_{2}$ independent,  
\\[0.2cm] 
& 
&
&  
&
&
&        
&  
&  
&  
&   
&  
$Z_{1} \sim GAM(\widetilde{c} \beta_{1},\widetilde{c} \alpha)$, 
\\[0.2cm] 
& 
for $\alpha, \widetilde{c},  \beta_{1}, \beta_{2} >0$,&
&  
&
&        
& 
& 
&  
&  
&   
&  
$Z_{2} \sim GAM(\widetilde{c} \beta_{2},\widetilde{c} \alpha)$  
\\[0.2cm] 
& 
$g := g(t) := \frac{1-t}{\alpha} + \frac{1}{\beta_{2}} - \frac{1}{\beta_{1}}$, &
&  
&
& 
&       
&  
&  
&  
&   
&  
(generalized asymmetric 
\\[0.2cm] 
& 
&
&
&  
&
&        
&  
&  
&  
&   
&  
Laplace  distribution)
\\\cline{1-12}  
$]\negthinspace \negthinspace - \negthinspace \infty,\infty[$ & 
$\widetilde{c} \cdot \varphi_{\alpha,\beta}(t) := \widetilde{c} \cdot \alpha \cdot
\{ \sqrt{1 + \beta^{2} g^2}
$ & 
$\mathbb{R}_{\gneqq 0}^{K}$ & 
 $\mathbb{R}^{K}$ &
$\widetilde{c} \cdot \sum\limits_{k=1}^{K} p_{k} \cdot 
\varphi_{\alpha,\beta}\Big(\frac{q_{k}}{p_{k}}\Big)$ 
&
$]\negthinspace \negthinspace - \negthinspace \infty,\infty[$ & 
$\infty$ & 
$\infty$ & 
$-\widetilde{c} \beta$ & 
$\widetilde{c} \beta$ & 
$]\negthinspace \negthinspace - \negthinspace \infty,\infty[$ &  
as above, but for $\breve{\theta} =1$, 
\\  
& 
$-1 + \log\frac{2 \cdot (\sqrt{1 + \beta^{2} g^2} - 1)}{\beta^{2} g^{2}} \}$  &
&  
&
& 
&      
&  
&  
&  
&   
&  
$Z_{1} \sim GAM(\widetilde{c} \beta,\widetilde{c} \alpha)$, 
\\ 
& 
for $\beta, \alpha, \widetilde{c} >0$ and $g := g(t) := \frac{1-t}{\alpha}$  &
&  
&
&        
& 
& 
&  
&  
&   
&  
$Z_{2} \sim GAM(\widetilde{c} \beta,\widetilde{c} \alpha)$ 
\\\cline{1-12}  
\end{tabular}
}
\rotatebox{90}{
\textbf{Table 1.} \ Selection of concrete examples 
including some of their important features.  
}
}

\newpage


\section{
Deterministic Narrow-Sense Bare-Simulation-Optimization of Bregman divergences}
\label{SectDetNarrow Bregman}

In the previous section, we have recalled/summarized recently achieved 
(cf. Broniatowski \& Stummer \cite{Bro:23a}) \textit{narrow-sense}
bare-simulation minimization results for a special subclass of
--- discrete smooth --- 
CASM $\varphi-$divergences. Let us now present a 
\textit{first generalization} thereof, namely 
\textit{narrow-sense} bare-simulation minimization results for a special subclass of 
--- discrete smooth --- 
\textit{Scaled Bregman Distances} 
of Stummer~\cite{Stu:07} and Stummer \& Vajda~\cite{Stu:12};
this will particularly cover 
narrow-sense bare-simulation minimization results for
(a special subclass of)
\textit{ordinary/classical Bregman Distances}. 

\vspace{0.2cm}
\noindent
To start with, we fix a (scaling) vector $\mathbf{P}$ with strictly positive 
components $p_{k}>0$ as well as a divergence generator
$\varphi \in \widetilde{\Upsilon}(]a,b[)$ having representability \eqref{brostu5:fo.link.var}
(i.e. $\varphi$ satisfies Condition \ref{Condition  Fi Tilda in Minimization}); 
recall from Remark \ref{after represent 2} that
$]t_{-}^{sc},t_{+}^{sc}[$ (covering $t=1$) denotes the interior of its (maximum) domain of strict convexity,
that $\varphi$ is finite and differentiable on $]a,b[$,
that $\varphi(t)=0$ if and only if $t=1$,
and that $\varphi^{\prime}(t)=0$ if and only if $t=1$.  
Moreover, we fix a second vector $\mathbf{Q}^{\ast\ast} \in \mathbb{R}^{K}$ such that 
\begin{equation}
t_{k}^{\ast\ast} :=  \frac{q_{k}^{\ast\ast}}{p_{k}} \in \, ]t_{-}^{sc},t_{+}^{sc}[  
\quad \textrm{for all $k =1,\ldots,K$.} 
\label{brostu5:fo.SBD.qstarstar}
\end{equation}
Notice that this implies $\varphi(t_{k}^{\ast\ast}) < \infty$, 
and $\varphi^{\prime}(t_{k}^{\ast\ast}) \in \, ]-\infty,\infty[$ 
($k =1,\ldots,K$). 
From this, for each $k=1,\ldots,K$ we construct the function 
\begin{equation}
\varphi_{k}(t) \ := \
\varphi(t) - \varphi(t_{k}^{\ast\ast}) - \varphi^{\prime}(t_{k}^{\ast\ast}) \cdot (t-t_{k}^{\ast\ast}),
\hspace{1.0cm} t \in \, \mathbb{R}, 
\label{brostu5:fo.phi_k} 
\end{equation}
whose effective domain is $dom(\varphi_{k}) = dom(\varphi)$ and thus $int(dom(\varphi_{k})) = \, ]a,b[$.
Notice that  $\varphi_{k}(t) \geq 0$
for all $t \in \mathbb{R}$, that $\varphi_{k}(t) = 0$ if and only if $t= t_{k}^{\ast\ast}$,
and that $\varphi_{k}^{\prime}(t) = 0$ if and only if $t= t_{k}^{\ast\ast}$.
Moreover, $]t_{-}^{sc},t_{+}^{sc}[$ is also the interior of the (maximum) domain of strict convexity
of $\varphi_{k}$, and consequently, $\varphi_{k}$ is affine-linear on $]a,t_{-}^{sc}[$ (provided
that this interval is non-empty) and on $]t_{+}^{sc},b[$ (provided
that this interval is non-empty). 
Clearly, we interpret $\varphi_{k}(a) := \lim_{t\downarrow a} 
(\varphi(t) - \varphi(t_{k}^{\ast\ast}) - \varphi^{\prime}(t_{k}^{\ast\ast}) \cdot (t-t_{k}^{\ast\ast}))
= \varphi(a) - \varphi(t_{k}^{\ast\ast}) - \varphi^{\prime}(t_{k}^{\ast\ast}) \cdot (a-t_{k}^{\ast\ast})$
and $\varphi_{k}(b) := \lim_{t\uparrow b} 
(\varphi(t) - \varphi(t_{k}^{\ast\ast}) - \varphi^{\prime}(t_{k}^{\ast\ast}) \cdot (t-t_{k}^{\ast\ast}))
= \varphi(b) - \varphi(t_{k}^{\ast\ast}) - \varphi^{\prime}(t_{k}^{\ast\ast}) \cdot (b-t_{k}^{\ast\ast})$
(where the limits always exist but may be infinite).
By means of these $\varphi_{k}$\textquoteright s, 
under the assumption \eqref{brostu5:fo.SBD.qstarstar} we construct the
--- discrete smooth special cases of --- \textit{scaled Bregman distances} 
(between $\mathbf{Q}$ and $\mathbf{Q}^{\ast\ast}$
\footnote{note that for the finiteness of the divergence we allow $\mathbf{Q}$ to be in a 
domain which is larger or equal to the domain of $\mathbf{Q}^{\ast\ast}$})
of Stummer~\cite{Stu:07} and Stummer \& Vajda~\cite{Stu:12} as
\begin{equation}
D_{\varphi,\mathbf{P}}^{SBD}(\mathbf{Q},\mathbf{Q}^{\ast\ast}) \ := \ 
\sum_{k=1}^{K} p_{k} \cdot
\varphi_{k} \negthinspace\left( \frac{q_{k}}{p_{k}}\right) \ = \ 
\sum\limits_{k=1}^{K} p_{k} \cdot
\left[ \varphi \negthinspace \left( \frac{q_{k}}{p_{k}} \right) -\varphi \negthinspace 
\left( \frac{q_{k}^{\ast\ast}}{p_{k}} \right)
- \varphi^{\prime} \negthinspace
\left( \frac{q_{k}^{\ast\ast}}{p_{k}} \right) \cdot \left(\frac{q_{k}}{p_{k}} - \frac{q_{k}^{\ast\ast}}{p_{k}} \right) 
\right] .
\label{brostu5:fo.SBD.smooth}
\end{equation}

\vspace{0.2cm}
\noindent
Notice that $D_{\varphi,\mathbf{P}}^{SBD}(\mathbf{Q},\mathbf{Q}^{\ast\ast}) < \infty$
if and only if $\frac{q_{k}}{p_{k}} \in  dom(\varphi)$ 
for all $k \in \{1,\ldots,K\}$
(mostly, we deal with cases where $\frac{q_{k}}{p_{k}} \in \, ]t_{-}^{sc},t_{+}^{sc}[$
with eventual involvement of the boundary points). Moreover,
there hold the above-mentioned divergence properties (iii),
i.e. \textquotedblleft $D_{\varphi,\mathbf{P}}^{SBD}(\mathbf{Q},\mathbf{Q}^{\ast\ast})
\geq 0$\textquotedblright\ and \textquotedblleft $D_{\varphi,\mathbf{P}}^{SBD}(\mathbf{Q},\mathbf{Q}^{\ast\ast}) = 0$ if and only if $\mathbf{Q}=\mathbf{Q}^{\ast\ast}$\textquotedblright.
Furthermore, the corresponding CASM $\varphi-$divergence can be recovered as
special case
$D_{\varphi,\mathbf{P}}^{SBD}(\mathbf{Q},\mathbf{P}) = D_{\varphi}(\mathbf{Q},\mathbf{P})$.
Moreover, the particular choice $\mathbf{P} := (1, \ldots, 1) := \mathbf{1}$
leads to  
--- discrete smooth special cases of ---
the omnipresent important class of \textit{separable ordinary/classical (i.e. unscaled) Bregman distances}
$D_{\varphi}^{OBD}(\mathbf{Q},\mathbf{Q}^{\ast\ast}) := 
D_{\varphi,\mathbf{1}}^{SBD}(\mathbf{Q},\mathbf{Q}^{\ast\ast})$
\footnote{where $\varphi$ may even have an affine-linear part, but 
all the components of $\mathbf{Q}^{\ast\ast}$ are in its strictly-convex part}.
Detailed references on the theory and applications of ordinary
Bregman distances respectively on scaled Bregman distances will be given below,
after \eqref{brostu5:fo.scaledBregpow} 
as well as in (D2) respectively (D3)
of the next Section \ref{SectDetGeneral}. 
Notice also that scaled Bregman distances of the form
\eqref{brostu5:fo.SBD.smooth} also appear \textit{in a natural way} in a new context of speed-up simulation,
see Subsection \ref{SectEstimators.new.det.nonvoid.improved.compact} below. 

\vspace{0.2cm}
\noindent
Let us also mention that in 
\eqref{brostu5:fo.phi_k} 
we could also \textit{equivalently} replace the generator 
$\varphi \in \widetilde{\Upsilon}(]a,b[)$ (with \eqref{brostu5:fo.link.var}) by $\breve{\varphi}$
defined through $\breve{\varphi}(t):= \varphi(t) + c_{1} +  c_{2} \cdot t$ 
($t \in dom(\varphi)$) with 
arbitrary $c_{1},c_{2} \in \mathbb{R}$;
in particular, one has $\breve{\varphi}(1) = c_{1} +  c_{2}$ instead
of $\varphi(1) =0$, and $\breve{\varphi}^{\prime}(1) = c_{2}$ instead
of $\varphi(1)^{\prime} =0$. This replacement can be done since
for both $\breve{\varphi}$ and $\varphi$
the respective two intervals $]a,b[$ and $]t_{-}^{sc},t_{+}^{sc}[$ coincide,  
and since $D_{\breve{\varphi},\mathbf{P}}^{SBD}(\mathbf{Q},\mathbf{Q}^{\ast\ast})
= D_{\varphi,\mathbf{P}}^{SBD}(\mathbf{Q},\mathbf{Q}^{\ast\ast})$ by straightforward calculations.
For the sake of brevity, for the rest of this section we stick to $\varphi$ rather than $\breve{\varphi}$.

\vspace{0.2cm}
\begin{remark} \ 
In the analogous spirit of Remark \ref{after represent} --- in terms of
$\widetilde{\varphi} := M_{\mathbf{P}} \cdot \varphi$, 
the probability vector $\widetilde{\mathds{P}}:=\mathbf{P}/M_{\mathbf{P}}$,  
as well as the vectors $\widetilde{\mathbf{Q}}:=\mathbf{Q}/M_{\mathbf{P}}$
and $\widetilde{\mathbf{Q}}^{\ast\ast}:= \mathbf{Q}^{\ast\ast}/M_{\mathbf{P}}$ --- 
one can equivalently rewrite
\eqref{brostu5:fo.SBD.smooth}
as
\begin{equation} 
D_{\varphi,\mathbf{P}}^{SBD}(\mathbf{Q},\mathbf{Q}^{\ast\ast})
=
D_{\widetilde{\varphi},\widetilde{\mathds{P}}}^{SBD}(\widetilde{\mathbf{Q}},\widetilde{\mathbf{Q}}^{\ast\ast})
\label{brostu5:fo.SBD.smooth.equality}
\end{equation}
and thus the two following minimization problems coincide:
\begin{eqnarray}
&& \inf_{\mathbf{Q}\in \mathbf{\Omega} }
D_{\varphi,\mathbf{P}}^{SBD}(\mathbf{Q},\mathbf{Q}^{\ast\ast})
\qquad \textrm{respectively} \qquad
\inf_{\widetilde{\mathbf{Q}}\in \widetilde{\mathbf{\Omega}} }
D_{\widetilde{\varphi},\widetilde{\mathds{P}}}^{SBD}(\widetilde{\mathbf{Q}},\widetilde{\mathbf{Q}}^{\ast\ast})  
\quad \textrm{with } \widetilde{\mathbf{\Omega}}:=\mathbf{\Omega} /M_{\mathbf{P}}.
\nonumber
\end{eqnarray}
Notice that the roles of $\mathbf{P}$ respectively $\widetilde{\mathds{P}}$
are now that of a scaling vector, whereas in \eqref{min Pb} respectively \eqref{min Pb prob2}
of the previous Section \ref{SectDetNarrow} they act as points which are 
projected on the constraint set $\mathbf{\Omega}$ respectively $\widetilde{\mathbf{\Omega}}$.

\end{remark}

\vspace{0.4cm}
\noindent
Below we shall show that the above-mentioned scaled Bregman distances can be BS-minimized 
\textit{in the narrow sense} in their first component. 
For this, recall that we have fixed $\mathbf{P} \in \mathbb{R}_{>0}^{K}$ 
and $\mathbf{Q}^{\ast\ast} \in \mathbb{R}^{K}$ such that
\eqref{brostu5:fo.SBD.qstarstar} holds.
By means of them, and by employing the above-mentioned notations
$n_{k}:=\lfloor n \cdot \widetilde{p}_{k}\rfloor$ ($k \in \left\{ 1, \ldots ,K-1\right\}$),
$n_{K} := n- \sum_{k=1}^{K-1} n_{k}$ (recall \eqref{fo.freqlim}) as well as 
$$I_{1}^{(n)}:=\left\{
1,\ldots ,n_{1}\right\}, \quad I_{2}^{(n)}:=\left\{ n_{1}+1,\ldots
,n_{1}+n_{2}\right\}, \quad \ldots, 
\quad I_{K}^{(n)} := \{ \sum_{k=1}^{K-1} n_{k} + 1, \ldots, n \}, \quad card(I_{k}^{(n)}) = n_{k},
$$
we construct  
the $n-$dimensional vector of random variables (with a slight abuse of notation)
\begin{equation}
\mathbf{\widetilde{V}} := \mathbf{\widetilde{V}}_{n} := \left(\widetilde{V}_{1}, \ldots, \widetilde{V}_{n}  
\right) = \left(\widetilde{V}_{1}, \ldots, \widetilde{V}_{n_{1}}, 
\widetilde{V}_{n_{1}+1}, \ldots, \widetilde{V}_{n_{1}+n_{2}}, \ldots, 
\widetilde{V}_{\sum_{k=1}^{K-1} n_{k} + 1}, \ldots, \widetilde{V}_{n} \right) \, ,
\label{brostu5:V_new}
\end{equation}
as follows: 
independently for each $k \in \{1,\ldots,K\}$ we employ 
$n_{k}$ i.i.d. random variables $\widetilde{V}_{i}$, $i\in I_{k}^{(n)}$, with 
common distribution $\mathbb{\Pi }[\widetilde{V}_{i}\in \cdot \,]=
\widetilde{U}_{k}[\,\cdot \,]$ ($i\in I_{k}^{(n)}$) given by (the $\widetilde{\mathbb{\bbzeta}}-$distortion type)
\begin{equation}
d\widetilde{U}_{k}(v) \ := \ 
\frac{\exp \left(\tau_{k} \cdot v\right)}{MGF_{\widetilde{\mathbb{\bbzeta}}}(\tau_{k})} \, 
d\widetilde{\mathbb{\bbzeta}}(v),
\label{brostu5:Utilde_k_new}
\end{equation}
where $\tau_{k} := \ M_{\mathbf{P}} \cdot \varphi^{\, \prime} \negthinspace
\left(\frac{\widetilde{q}_{k}^{\ast\ast}}{\widetilde{p}_{k}}\right)
= \ M_{\mathbf{P}} \cdot \varphi^{\, \prime} \negthinspace
\left(\frac{q_{k}^{\ast\ast}}{p_{k}}\right)$.
From this, we construct
\begin{equation}
\boldsymbol{\xi }_{n}^{\mathbf{\widetilde{V}}}:=\Big(\frac{1}{n}\sum_{i\in
I_{1}^{(n)}}\widetilde{V}_{i},\ldots ,\frac{1}{n}\sum_{i\in I_{K}^{(n)}}
\widetilde{V}_{i}\Big)
\label{Xi_n^W vector V new2}
\end{equation}
for which (as a consequence of \eqref{brostu5:fo.link.var}) one can prove 
$ \mathbb{E}_{\mathbb{\Pi}}\negthinspace \Big[\boldsymbol{\xi}_{n}^{\mathbf{\widetilde{V}}}\Big]
= \left(
\frac{\lfloor n \cdot \widetilde{p}_{1}\rfloor}{n \cdot \widetilde{p}_{1}} \cdot \widetilde{q}_{1}^{\ast\ast},
\ldots,
\frac{\lfloor n \cdot \widetilde{p}_{K-1}\rfloor}{n \cdot \widetilde{p}_{K-1}} \cdot \widetilde{q}_{K-1}^{\ast\ast},
\frac{n - \sum_{i=1}^{K-1} \lfloor n \cdot \widetilde{p}_{i}\rfloor}{n \cdot \widetilde{p}_{K}} \cdot \widetilde{q}_{K}^{\ast\ast}
\right)
$
(cf. Broniatowski \& Stummer \cite{Bro:23a})
and thus --- due to \eqref{fo.freqlim} ---
$\lim_{n\rightarrow \infty} \mathbb{E}_{\mathbb{\Pi}}\negthinspace \Big[\boldsymbol{\xi}_{n}^{\mathbf{\widetilde{V}}}
\Big]= 
\widetilde{\mathbf{Q}}^{\ast\ast}$
or equivalently  
$\lim_{n\rightarrow \infty}
\mathbb{E}_{\mathbb{\Pi}}\negthinspace \Big[M_{\mathbf{P}} \cdot \boldsymbol{\xi}_{n}^{\mathbf{\widetilde{V}}}
\Big]= \mathbf{Q}^{\ast\ast}$. 

\noindent
\begin{remark}
Notice all the differences to the construction in the previous Section \ref{SectDetNarrow}.
There, instead of the $\mathbf{\widetilde{V}}$ we have employed 
$\mathbf{\widetilde{W}}:=\left( \widetilde{W}_{1},\ldots ,\widetilde{W}_{n}\right) $ 
where the $\widetilde{W}_{i}$'s are
i.i.d. copies of the random variable $\widetilde{W}$ whose 
(block-neutral) distribution is
$\mathbb{\Pi }[\widetilde{W}\in \cdot \,]=\widetilde{\mathbb{\bbzeta}}
[\,\cdot \,]$. We have partitioned these $\mathbf{\widetilde{W}}$
into the same blocks as the $\mathbf{\widetilde{V}}$, and
transformed the former into the random vectors 
$\boldsymbol{\xi }_{n}^{\mathbf{\widetilde{W}}}$
(cf. \eqref{Xi_n^W vector}) which are formally the same
as $\boldsymbol{\xi}_{n}^{\mathbf{\widetilde{V}}}$ (cf. \eqref{Xi_n^W vector V new2})
where $\mathbf{\widetilde{W}}$ is just replaced by $\mathbf{\widetilde{V}}$.
It is straightforward to see that for the special case 
$\mathbf{P} = \mathbf{Q}^{\ast\ast}$ the two constructions coincide
(since then $\tau_{k} = M_{\mathbf{P}} \cdot \varphi^{\, \prime} \negthinspace
\left(1\right) = 0$); this is consistent with the above-mentioned 
collapse-property
$D_{\varphi,\mathbf{P}}^{SBD}(\mathbf{Q},\mathbf{P}) = D_{\varphi}(\mathbf{Q},\mathbf{P})$.

\end{remark}

\vspace{0.2cm}
\noindent
We are now in the position to formulate the following new assertion
on the BS-minimizability of scaled Bregman distances:

\vspace{0.3cm}
\noindent

\begin{theorem}
\label{brostu5:thm.BSnarrow.SBD}
Let $\mathbf{P} \in \mathbb{R}_{> 0}^{K}$, 
$M_{\mathbf{P}}:=\sum_{i=1}^{K}p_{i}>0$, and
suppose that the divergence generator $\varphi$ 
satisfies the above Condition \ref{Condition  Fi Tilda in Minimization},
with $\widetilde{\mathbb{\bbzeta}}$ (cf. \eqref{brostu5:fo.link.var}).  
Additionally, let 
$\mathbf{Q}^{\ast\ast} \in \mathbb{R}^{K}$ such that 
\eqref{brostu5:fo.SBD.qstarstar} holds.
Moreover, we assume that $\mathbf{\Omega}$ satisfies 
the regularity properties \eqref{regularity} as well as
the finiteness property
\begin{equation}
\inf_{\mathbf{Q}\in \mathbf{\Omega} }
D_{\varphi,\mathbf{P}}^{SBD}(\mathbf{Q},\mathbf{Q}^{\ast\ast}) < \infty .
\nonumber
\end{equation}
Furthermore, let $\widetilde{V} := (\mathbf{\widetilde{V}}_{n})_{n \in \mathbb{N}}$ 
be a sequence of random vectors constructed via \eqref{brostu5:V_new} and \eqref{brostu5:Utilde_k_new}. 
Then, in terms of 
the random vectors $\boldsymbol{\xi }_{n}^{\mathbf{\widetilde{V}}}$
(cf. \eqref{Xi_n^W vector V new2})
there holds 
\begin{equation}
\inf_{\mathbf{Q} \in \mathbf{\Omega }}
D_{\varphi,\mathbf{P}}^{SBD}(\mathbf{Q},\mathbf{Q}^{\ast\ast})
\ = \ 
-\lim_{n\rightarrow \infty }\frac{1}{n}\log \,\mathbb{\Pi }\negthinspace \left[
\boldsymbol{\xi}_{n}^{\mathbf{\widetilde{V}}}
\in \mathbf{\Omega} /M_{\mathbf{P}}
\right] \, .
\label{LDP Minimization SBD}
\end{equation}
In particular, for each $\mathbf{P} \in \mathbb{R}_{> 0}^{K}$ 
and each $\mathbf{Q}^{\ast\ast} \in \mathbb{R}^{K}$ with 
\eqref{brostu5:fo.SBD.qstarstar} the function 
$\Phi_{\mathbf{P},\mathbf{Q}^{\ast\ast}} \left( \cdot \right) := 
D_{\varphi,\mathbf{P}}^{SBD}( \cdot,\mathbf{Q}^{\ast\ast})$
(cf. \eqref{brostu5:fo.SBD.smooth})
is bare-simulation minimizable (BS-minimizable)
\textit{in the narrow sense} (cf. \eqref{brostu5:fo.2} in Definition \ref{brostu5:def.1}
and the special case of Remark \ref{brostu5:rem.def1}(a)) 
on any such $\mathbf{\Omega }\subset \mathbb{R}^{K}$.

\end{theorem}

\vspace{0.3cm} 
\noindent
The proof of Theorem \ref{brostu5:thm.BSnarrow.SBD}
will be given in Appendix \ref{App.A}.

\noindent
\begin{remark}
\label{Rem BS narrow SBD}
Analogously to Remark \ref{dist of components}(iii), 
the limit relation \eqref{LDP Minimization SBD}
even holds when $\mathbf{Q}^{\ast\ast} \in \mathbf{\Omega}$
for which the left-hand side becomes zero. 
\end{remark}

\noindent
\begin{remark}
\label{dist of components new}
For some contexts, we can \textit{explicitly} give the
distribution of each of the independent (non-deterministic parts of the) components
$\Big(\sum_{i\in I_{k}^{(n)}}\widetilde{V}_{i}\Big)_{k=1,\ldots,K}$
of the vector $\boldsymbol{\xi }_{n}^{\mathbf{\widetilde{V}}}$
(cf. \cite{Bro:23a} for other purposes than here);
this will ease the corresponding concrete simulations.

\end{remark}

\noindent
Analogously to \eqref{fo.approx.1},
the limit statement 
\eqref{LDP Minimization SBD}
provides the principle for the approximation of the solution of 
the minimization problem 
$\Phi_\mathbf{P}(\mathbf{\Omega}) := \inf_{\mathbf{Q} \in \mathbf{\Omega}}
D_{\varphi,\mathbf{P}}^{SBD}(\mathbf{Q},\mathbf{Q}^{\ast\ast})$.
Namely, by replacing the right-hand side in \eqref{LDP Minimization SBD} by its finite
counterpart, we derive for given large $n$  
\begin{equation}
- \frac{1}{n}\log \mathbb{\Pi} \negthinspace \left[ \boldsymbol{\xi}_{n}^{\mathbf{\widetilde{V}}}\in 
\mathbf{\Omega}/M_{\mathbf{P}} \right] 
\approx \inf_{Q\in \mathbf{\Omega} }
D_{\varphi,\mathbf{P}}^{SBD}(\mathbf{Q},\mathbf{Q}^{\ast\ast});
\label{fo.approx.1.SBD} 
\end{equation}
it remains to estimate the left-hand side of \eqref{fo.approx.1.SBD}
(see Section 
\ref{SectEstimators.new.det.nonvoid}
below, where the latter also provides estimates of the \textit{minimizers}).

\noindent
\begin{example}
Let us take the important case of the
generators $\varphi_{\gamma} : \mathbb{R} \mapsto [0,\infty]$
defined by \eqref{brostu5:fo.powdivgen} in Example \ref{brostu5:ex.1},
which --- by \eqref{brostu5:fo.SBD.smooth} --- for arbitrary multiplier $\widetilde{c} \in \, ]0,\infty[$, 
and arbitrary $\mathbf{P} \in \mathbb{R}_{>0}^{K}$ 
generate (the vector-valued form of) the 
\textit{generalized scaled Bregman power distances} given by
\begin{eqnarray}
D_{\widetilde{c} \cdot \varphi_{\gamma},\mathbf{P}}^{SBD}(\mathbf{Q},\mathbf{Q}^{\ast\ast})
 \hspace{-0.2cm} &:=& \hspace{-0.2cm}
\begin{cases}
\widetilde{c} \cdot \Big\{\frac{ \sum\limits_{k=1}^{K} (q_{k})^{\gamma} \cdot (p_{k})^{1-\gamma}}{\gamma \cdot (\gamma-1)}
+ \frac{ \sum\limits_{k=1}^{K} (q_{k}^{\ast\ast})^{\gamma} \cdot (p_{k})^{1-\gamma}}{\gamma}
- \frac{\sum\limits_{k=1}^{K} q_{k} \cdot  (q_{k}^{\ast\ast})^{\gamma-1} \cdot (p_{k})^{1-\gamma}}{\gamma -1} 
\Big\}, \\
\hspace{7.2cm} \textrm{if }  \gamma \in \, ]-\infty,0[, \  
\mathbf{Q}^{\ast\ast} \in \mathbb{R}_{>0}^{K}
\ \textrm{and } 
\mathbf{Q} \in \mathbb{R}_{> 0}^{K}, 
 \\
\widetilde{c} \cdot \Big\{ - \sum\limits_{k=1}^{K} p_{k} \cdot \log \Big(\frac{q_{k}}{q_{k}^{\ast\ast}} \Big) 
+ \sum\limits_{k=1}^{K} \frac{q_{k} \cdot p_{k}}{q_{k}^{\ast\ast}} - \sum\limits_{k=1}^{K} p_{k} \Big\}, 
\hspace{1.1cm} \textrm{if }  \gamma = 0, \  
\mathbf{Q}^{\ast\ast} \in \mathbb{R}_{>0}^{K} \ \textrm{and } \mathbf{Q} \in \mathbb{R}_{> 0}^{K},
\\
\widetilde{c} \cdot \Big\{\frac{ \sum\limits_{k=1}^{K} (q_{k})^{\gamma} \cdot (p_{k})^{1-\gamma}}{\gamma \cdot (\gamma-1)}
+ \frac{ \sum\limits_{k=1}^{K} (q_{k}^{\ast\ast})^{\gamma} \cdot (p_{k})^{1-\gamma}}{\gamma}
- \frac{\sum\limits_{k=1}^{K} q_{k} \cdot  (q_{k}^{\ast\ast})^{\gamma-1} \cdot (p_{k})^{1-\gamma}}{\gamma -1} 
\Big\}, \\ 
\hspace{7.2cm} \textrm{if }  \gamma \in \, ]0,1[, \  
\mathbf{Q}^{\ast\ast} \in \mathbb{R}_{>0}^{K} \ \textrm{and } \mathbf{Q} \in \mathbb{R}_{\geq 0}^{K}, 
 \\
\widetilde{c} \cdot \Big\{ \sum\limits_{k=1}^{K} q_{k} \cdot \log \Big(\frac{q_{k}}{q_{k}^{\ast\ast}} \Big)
- \sum\limits_{k=1}^{K} q_{k} + \sum\limits_{k=1}^{K} q_{k}^{\ast\ast} \Big\}, 
\hspace{0.75cm} \textrm{if }  \gamma = 1, \  
\mathbf{Q}^{\ast\ast} \in \mathbb{R}_{>0}^{K} \ \textrm{and } \mathbf{Q} \in \mathbb{R}_{\geq 0}^{K},
\\
\widetilde{c} \cdot \Big\{\frac{ \sum\limits_{k=1}^{K} (q_{k})^{\gamma} \cdot (p_{k})^{1-\gamma}
\cdot \textfrak{1}_{[0,\infty[}(q_{k})}{\gamma \cdot (\gamma-1)}
+ \frac{ \sum\limits_{k=1}^{K} (q_{k}^{\ast\ast})^{\gamma} \cdot (p_{k})^{1-\gamma}}{\gamma}
- \frac{\sum\limits_{k=1}^{K} q_{k} \cdot  (q_{k}^{\ast\ast})^{\gamma-1} \cdot (p_{k})^{1-\gamma}}{\gamma -1} 
\Big\}, \\
\hspace{7.2cm}  \textrm{if }  \gamma \in \, ]1,2[, \  
\mathbf{Q}^{\ast\ast} \in \mathbb{R}_{>0}^{K} \ \textrm{and } \mathbf{Q} \in \mathbb{R}^{K}, 
\\
\widetilde{c}\cdot \sum\limits_{k=1}^{K}\frac{  (q_{k}-q_{k}^{\ast\ast})^{2}}{2 \cdot p_{k}} , 
\hspace{4.7cm}  \textrm{if }  \gamma = 2, \  
\mathbf{Q}^{\ast\ast} \in \mathbb{R}^{K} \ \textrm{and } \mathbf{Q} \in \mathbb{R}^{K},
\\
\widetilde{c} \cdot \Big\{\frac{ \sum\limits_{k=1}^{K} (q_{k})^{\gamma} \cdot (p_{k})^{1-\gamma}
\cdot \textfrak{1}_{[0,\infty[}(q_{k})}{\gamma \cdot (\gamma-1)}
+ \frac{ \sum\limits_{k=1}^{K} (q_{k}^{\ast\ast})^{\gamma} \cdot (p_{k})^{1-\gamma}}{\gamma}
- \frac{\sum\limits_{k=1}^{K} q_{k} \cdot  (q_{k}^{\ast\ast})^{\gamma-1} \cdot (p_{k})^{1-\gamma}}{\gamma -1} 
\Big\}, \\
\hspace{7.2cm}  \textrm{if }  \gamma \in \, ]2,\infty[, \  
\mathbf{Q}^{\ast\ast} \in \mathbb{R}_{>0}^{K} \ \textrm{and } \mathbf{Q} \in \mathbb{R}^{K},
\\
\infty, \hspace{6.7cm} \textrm{else};
\end{cases}
\label{brostu5:fo.scaledBregpow} 
\end{eqnarray}
notice that one has the straightforward relationship 
$D_{\widetilde{c} \cdot \varphi_{\gamma},\mathbf{P}}^{SBD}(\cdot ,\cdot ) 
=\widetilde{c}\cdot D_{\varphi_{\gamma},\mathbf{P}}^{SBD}(\cdot ,\cdot ) $. 
The special case of \eqref{brostu5:fo.scaledBregpow} for probability vectors 
 $\mathds{Q} \in \mathbb{S}$, $\mathds{Q}^{\ast\ast} \in \mathbb{S}$
has been introduced in Stummer \& Vajda \cite{Stu:12}
and e.g. applied in Ki{\ss}linger \& Stummer \cite{Kis:13},\cite{Kis:15a},\cite{Kis:16},\cite{Kis:18}
as well as in Roensch \& Stummer \cite{Roe:17},\cite{Roe:19a},\cite{Roe:19b}.
The case of \eqref{brostu5:fo.scaledBregpow} for ``general'' vectors (and even ``general'' measures)
has been first indicated in Ki{\ss}linger \& Stummer \cite{Kis:16}
and fully worked out in Broniatowski \& Stummer \cite{Bro:19b};
for corresponding applications see e.g. Kr{\"o}mer \& Stummer \cite{Kro:19} (to mortality data analytics)
and Broniatowski \& Stummer \cite{Bro:22} (to statistics and the adjacent fields of machine learning and artificial
intelligence).

\vspace{0.2cm}
\noindent
In line with the above-mentioned general imbedding, the particular choice $\mathbf{P} := (1, \ldots, 1) := \mathbf{1}$
leads to --- discrete smooth special cases of ---
the omnipresent important class of \textit{separable ordinary/classical (i.e. unscaled) Bregman power distances}
$D_{\widetilde{c} \cdot \varphi_{\gamma}}^{OBD}(\mathbf{Q},\mathbf{Q}^{\ast\ast}) := 
D_{\widetilde{c} \cdot \varphi_{\gamma},\mathbf{1}}^{SBD}(\mathbf{Q},\mathbf{Q}^{\ast\ast})$
which can be explicitly deduced from \eqref{brostu5:fo.scaledBregpow} by setting 
$p_{k}:=1$ for all $k=1,\ldots,K$. For instance, for the case $\gamma=2$ the omnipresent \textit{
squared weighted $\ell_{2}-$distance} 
$D_{\widetilde{c} \cdot \varphi_{2},\mathbf{P}}^{SBD}(\mathbf{Q},\mathbf{Q}^{\ast\ast})$
reduces to the omnipresent 
(multiple of) \textit{squared $\ell_{2}-$distance 
}
\begin{equation}
0 \leq 
D_{\widetilde{c} \cdot \varphi_{2}}^{OBD}( \mathbf{Q}, \mathbf{Q}^{\ast\ast})
 \ = \  \frac{\widetilde{c}}{2} \cdot \sum_{k=1}^{K} ( \, q_{k} - q_{k}^{\ast\ast} \, )^{2}
\, \, \quad \mathbf{Q} \in \mathbb{R}^{K}, \ \mathbf{Q}^{\ast\ast} \in \mathbb{R}^{K}.
\label{brostu5:fo.sqL2.new}
\end{equation} 

\noindent
In the special case of probability vectors 
$\mathds{Q}^{\ast\ast} \in \mathbb{S}_{> 0}^{K}$ and $\mathds{Q} \in \mathbb{S}_{> 0}^{K}$
together with the special choice $\gamma >0$, the divergences 
$D_{\widetilde{c} \cdot \varphi_{\gamma}}^{OBD}(\mathds{Q},\mathds{Q}^{\ast\ast})$
reduce to the prominent
(finite-discrete-special-cases of) \text{``order$-\gamma$'' 
density power divergences} 
of Basu et al.~\cite{Bas:98}
(also known as Basu-Harris-Hjort-Jones distances) and their rescaled versions
called \textit{beta-divergences} (cf. Eguchi \& Kano \cite{Egu:01}, Mihoko \& Eguchi \cite{Miho:02}).
For general $\gamma \in \mathbb{R}$ and general probability measures 
see e.g. Stummer \& Vajda~\cite{Stu:12}.
The general case of non-probability vectors and measures is treated in
Broniatowski \& Stummer \cite{Bro:19b},\cite{Bro:22}
(see also Hennequin et al. \cite{Henn:11} for finite discrete beta-divergences).\\
As far as further important particular parameter-cases (other than $\gamma=2$) is concerned,
$D_{\widetilde{c} \cdot \varphi_{1}}^{OBD}(\mathbf{Q},\mathbf{Q}^{\ast\ast})$
(with $\gamma=1$) amounts to the \textit{discrete generalized relative entropy}, whereas
$D_{\widetilde{c} \cdot \varphi_{0}}^{OBD}(\mathbf{Q},\mathbf{Q}^{\ast\ast})$
(with $\gamma=0$) leads to the \textit{discrete (e.g. sampled) Itakura-Saito distance}
\cite{Itak:68}.\\
Some exemplary (relatively) recent studies and applications
of (generalized) density power divergences 
$D_{\widetilde{c} \cdot \varphi_{\gamma}}^{OBD}(\mathbf{Q},\mathbf{Q}^{\ast\ast})$ 
(including beta-divergences, Itakura-Saito-divergences, continuous versions)
--- aside from the vast literature on the omnipresent cases $\gamma =1$ and $\gamma=2$ ---
appear e.g. in
Basu et al.~\cite{Bas:15a}, 
Basu et al. \cite{Bas:16g},
Ghosh \& Basu~\cite{Gho:16a},\cite{Gho:16b},
Basu et al. \cite{Bas:17h},
Martin et al. \cite{Martin2:17},
Basu et al. \cite{Bas:18f},
Balakrishnan et al. \cite{Balakri:19},
Balakrishnan et al. \cite{Balakri:20},
Ghosh \& Majumdar \cite{Ghosh:20d},
Leplat et al. \cite{Leplat:20},
Vandecappelle et al. \cite{Vandec:20},
Balakrishnan et al. \cite{Balakri:21},
Basak et al. \cite{Basak:21},
Basu et al. \cite{Bas:21e},
Calvino et al. \cite{Calv:21},
Castilla et al. \cite{Cast:21a},
Castilla et al. \cite{Cast:21b},
Legros et al. \cite{Legros:21},
Pu et al. \cite{Pu11:22},
Ramirez et al. \cite{Rami:22},
Castilla \& Chocano \cite{Cast:23a},
Marmin et al. \cite{Marmin:23},
Saraceno et al. \cite{Sara:23},
Sharma \& Pradhan \cite{Sharma2:23}.

\vspace{0.2cm}
\noindent
Returning to the general setup of generalized scaled Bregman power distances \eqref{brostu5:fo.scaledBregpow},
let us point out that for any fixed $M_{\mathbf{P}}\in \, ]0,\infty[$ 
the Condition \ref{Condition  Fi Tilda in Minimization}
is satisfied for $\varphi := \widetilde{c}\cdot\varphi_{\gamma}$ --- 
and thus the \textit{narrow-sense} BS-minimizability concerning Theorem \ref{brostu5:thm.BSnarrow.SBD} can be applied 
---
for all $\widetilde{c} \in \, ]0,\infty[$ and all $\gamma \in \mathbb{R}\backslash]1,2[$. 
(As far as we know at the moment) For the case $\gamma \in \, ]1,2[$ one can not verify
Condition \ref{Condition  Fi Tilda in Minimization}, but BS-minimizability (in the sense of \eqref{brostu5:fo.2} 
with $f_{n}(\cdot) \not\equiv 1$) will be 
shown in Example \ref{brostu5:ex.POWmiss} below.

\vspace{0.4cm}
\noindent
As indicated in the rows 1 to 6 in the above Table 1 
(with $M_{\mathbf{P}} \cdot \varphi$ instead of $\varphi$), 
the representability \eqref{brostu5:fo.link.var} 
of the power divergence generators $\varphi := \widetilde{c} \cdot \varphi_{\gamma}$
has been completely worked out in Section XII of Broniatowski \& Stummer~\cite{Bro:23a};
there (for other purposes than here) we have also \textit{explicitly} derived the corresponding crucial
block-wise sampling distributions 
(cf. \eqref{brostu5:Utilde_k_new})
$\widetilde{U}_{k}[\,\cdot \,] = \mathbb{\Pi }[ \, \widetilde{V}_{i}\in \cdot \,]$ ($i\in I_{k}^{(n)}$)
and even more comfortably  
the block-sum distributions 
$\widetilde{U}_{k}^{\ast n_{k}}[\,\cdot \,] = \mathbb{\Pi }[ \, \sum_{i\in I_{k}^{(n)}}\widetilde{V}_{i}\in \cdot \,]$,
as follows: 

\begin{itemize}

\item $\gamma \in \, ]-\infty,0[$: \ $\widetilde{U}_{k}^{\ast n_{k}}$
has the (Lebesgue-)density
\[
f_{\widetilde{U}_{k}^{\ast n_{k}}
}(x) \ :=  \ \frac{\exp((\tau_{k} - \frac{\widetilde{c} \cdot M_{\mathbf{P}}}{1-\gamma})\cdot x)}{
\exp\left(n_{k} \cdot \frac{\widetilde{c} \cdot M_{\mathbf{P}}}{\gamma}\cdot
(1+\frac{\gamma-1}{\widetilde{c} \cdot M_{\mathbf{P}}} \cdot \tau_{k})^{\gamma/(\gamma-1)}\right)} 
\cdot f_{\breve{\breve{Z}}}(x) \cdot
\textfrak{1}_{]0,\infty[}(x),
\qquad x \in \mathbb{R}, 
\]
where $\tau_{k} = \widetilde{c} \cdot M_{\mathbf{P}} \cdot
\frac{1-\big(\frac{\widetilde{q}_{k}^{\ast\ast}}{\widetilde{p}_{k}}\big)^{\gamma-1}}{1-\gamma}$
for $\widetilde{q}_{k}^{\ast\ast} >0$, and  
$\breve{\breve{Z}}$ is a random variable with density $f_{\breve{\breve{Z}}}$ of a stable law
with parameter-quadruple
$(\frac{-\gamma }{1-\gamma },1,0,- n_{k} \cdot \frac{(\widetilde{c}\cdot M_{\mathbf{P}})^{1/(1-\gamma )} \cdot (1-\gamma )^{-\gamma /(1-\gamma )}}{\gamma })$
(in terms of the \textquotedblleft form-B notation\textquotedblright\
on p.12 in Zolotarev~\cite{Zol:86});

\item $\gamma =0$: $\widetilde{U}_{k}^{\ast n_{k}}
=GAM\left( \widetilde{c} \cdot M_{\mathbf{P}} - \tau_{k},
n_{k}\cdot \widetilde{c} \cdot M_{\mathbf{P}}\right)$, with 
$\tau_{k}=\widetilde{c} \cdot M_{\mathbf{P}} \cdot ( 1- 
\frac{\widetilde{p}_{k}}{\widetilde{q}_{k}^{\ast\ast}})$
for $\widetilde{q}_{k}^{\ast\ast} >0$;

\item $\gamma \in \, ]0,1[$: $\widetilde{U}_{k}^{\ast n_{k}}$
is Compound $POI( n_{k}\cdot \breve{\theta})-
GAM\big(\frac{\widetilde{c} \cdot M_{\mathbf{P}}}{1-\gamma} - \tau_{k},
\frac{\gamma}{1-\gamma} \big) \big)$ with 
$\breve{\theta}:= \frac{\widetilde{c} \cdot M_{\mathbf{P}}}{\gamma}
\cdot \big(\frac{(\gamma -1) \cdot \tau_{k}}{\widetilde{c} \cdot M_{\mathbf{P}}}
+1\big)^{\gamma/(\gamma -1)}$ and 
$\tau_{k} = \widetilde{c} \cdot M_{\mathbf{P}} \cdot
\frac{
1-\big(\frac{\widetilde{q}_{k}^{\ast\ast}}{\widetilde{p}_{k}}\big)^{\gamma-1}}{1-\gamma}$
for $\widetilde{q}_{k}^{\ast\ast} >0$;

\item $\gamma =1$: $\widetilde{U}_{k}^{\ast n_{k}}$
is the probability distribution 
$\frac{1}{\widetilde{c} \cdot M_{\mathbf{P}}} \cdot 
POI\left(n_{k} \cdot \widetilde{c} \cdot M_{\mathbf{P}} \cdot 
\exp(\frac{\tau _{k}}{\widetilde{c} \cdot M_{\mathbf{P}}})\right) $ 
with support on the lattice 
$\left\{ \frac{j}{\widetilde{c} \cdot M_{\mathbf{P}}}, \, j\in \mathbb{N}_{0}\right\}$,
where $\tau_{k}=\widetilde{c} \cdot \log \left( \frac{\widetilde{q}_{k}^{\ast\ast}}{\widetilde{p}_{k}}\right) $ 
for $\widetilde{q}_{k}^{\ast\ast} >0$;

\item $\gamma =2$: $\widetilde{U}_{k}^{\ast n_{k}}
=NOR(n_{k}\cdot(1+\frac{\tau_{k}}{\widetilde{c} \cdot M_{\mathbf{P}}}),
\frac{n_{k}}{\widetilde{c} \cdot M_{\mathbf{P}}})$ 
with $\tau_{k}=\widetilde{c} \cdot M_{\mathbf{P}} \cdot 
( \frac{\widetilde{q}_{k}^{\ast\ast}}{\widetilde{p}_{k}} - 1 ) $ 
for $\widetilde{q}_{k}^{\ast\ast} \in \mathbb{R}$;

\item $\gamma \in \, ]2,\infty[$:
$\widetilde{U}_{k}^{\ast n_{k}}$
has the (Lebesgue-)density
\[
f_{\widetilde{U}_{k}^{\ast n_{k}}
}(x) \ :=  \ \frac{\exp((\tau_{k} + \frac{\widetilde{c} \cdot M_{\mathbf{P}}}{\gamma-1})\cdot x)}{
\exp\left(n_{k} \cdot \frac{\widetilde{c} \cdot M_{\mathbf{P}}}{\gamma}\cdot
(1+\frac{\gamma-1}{\widetilde{c} \cdot M_{\mathbf{P}}} \cdot \tau_{k})^{\gamma/(\gamma-1)}\right)} 
\cdot f_{\breve{\breve{Z}}}(-x) , 
\qquad x \in \mathbb{R}, 
\]

\noindent
where $\tau_{k} = - \frac{\widetilde{c} \cdot M_{\mathbf{P}}}{\gamma-1} \cdot
\big(1- \big(\frac{\widetilde{q}_{k}^{\ast\ast}}{\widetilde{p}_{k}}\big)^{\gamma-1}
\big)$
for $\widetilde{q}_{k}^{\ast\ast} >0$, and  
$\breve{\breve{Z}}$ is a random variable with density $f_{\breve{\breve{Z}}}$ of a stable law
with parameter-quadruple
$(\frac{\gamma }{\gamma-1},1,0,n_{k} \cdot \frac{(\widetilde{c}\cdot M_{\mathbf{P}})^{1/(1-\gamma )} \cdot (\gamma-1)^{\gamma /(\gamma-1)}}{\gamma })$.

\end{itemize}

\end{example}

\vspace{0.4cm}
\noindent
The explicit block-sum distributions $\widetilde{U}_{k}^{\ast n_{k}}$ for the other 
$\varphi-$generator examples in Table 1 (rows 7 to 11) can also be found in 
Section XII of Broniatowski \& Stummer~\cite{Bro:23a}.
Let us present a new interesting completely worked out case:

\noindent
\begin{example}
For $\beta \in \mathbb{R}\backslash\{0\}$ and $\widetilde{c} \in \, ]0,\infty[$  and
we employ the strictly increasing, smooth function
\begin{eqnarray}
F_{\beta,\widetilde{c}}(t) &:=& 
\frac{2 \widetilde{c}}{\beta} \cdot \left( e^{\beta \cdot t} -  e^{\beta} \right),
\hspace{2.5cm}   t \in \, ]-\infty,\infty[, 
\nonumber 
\end{eqnarray}
which satisfies $F_{\beta,\widetilde{c}}(1)=0$ and which has strictly increasing, smooth
inverse
\begin{eqnarray}
F_{\beta,\widetilde{c}}^{\leftarrow}(x) &:=& 
\frac{1}{\beta} \cdot \log \left( \frac{\beta}{2 \widetilde{c}} \cdot x + e^{\beta} \right),
\hspace{2.5cm}   \textrm{if } \beta \cdot x > -2 \widetilde{c} \cdot e^{\beta}. 
\nonumber 
\end{eqnarray}
By applying Theorem 22 and the corresponding Remark 23(b) of Broniatowski \& Stummer \cite{Bro:23a},
we obtain the divergence generator 
\begin{eqnarray}
\varphi_{\beta,\widetilde{c}}(t)  & := &  
t \cdot F_{\beta,\widetilde{c}}\left(t\right)
-\int\displaylimits_{0}^{F_{\beta,\widetilde{c}}\left(t\right)} F_{\beta,\widetilde{c}}^{-1}(u) \, du
\ = \ t \cdot \frac{2 \widetilde{c}}{\beta} \cdot \left( e^{\beta \cdot t} -  e^{\beta} \right)
- \frac{2 \widetilde{c}}{\beta^2} \cdot \Big\{ [\beta \cdot t \cdot e^{\beta \cdot t}- e^{\beta \cdot t}] 
- [\beta\cdot e^{\beta}- e^{\beta} ] \Big\} 
\nonumber \\
& = & \frac{2 \widetilde{c}}{\beta^2} \cdot \Big\{
e^{\beta \cdot t} - t \cdot \beta \cdot  e^{\beta} + (\beta-1) \cdot e^{\beta}
\Big\},
\qquad t \in \, ]-\infty,\infty[ ;
\label{brostu5:ex.BEDandGNED.3} 
\end{eqnarray}
as well as the cumulant-generating-function-\textit{candidate}
\begin{eqnarray}
 & & \hspace{-1.0cm} \Lambda_{\beta,\widetilde{c}}(z) := 
\int\displaylimits_{0}^{z} F_{\beta,\widetilde{c}}^{\leftarrow}(u) \, du
\ = \ \frac{2 \widetilde{c}}{\beta^2} \cdot \Big\{ 
\Big[(\frac{\beta}{2 \widetilde{c}} \cdot z + e^{\beta}) \cdot 
\log(\frac{\beta}{2 \widetilde{c}} \cdot z + e^{\beta})
- (\frac{\beta}{2 \widetilde{c}} \cdot z + e^{\beta})\Big] 
- \Big[\beta\cdot e^{\beta}- e^{\beta} \Big] \Big\} ,
\ \   \textrm{if } \beta \cdot z > -2 \widetilde{c} \cdot e^{\beta}, 
\quad {\ \ }
\label{brostu5:ex.BEDandGNED.4} 
\end{eqnarray}
which satisfy the representability \eqref{brostu5:fo.link.var.simplex} 
(which is \eqref{brostu5:fo.link.var} with $M_{\mathbf{P}} =1$) below.

\noindent
The function $\varphi_{\beta,\widetilde{c}}$ of \eqref{brostu5:ex.BEDandGNED.3} 
(and thus, equivalently, its linear-part-cleaned sibling
$\check{\varphi}_{\beta,\widetilde{c}}(t) := \frac{2 \widetilde{c}}{\beta^2} \cdot 
e^{\beta \cdot t}$) --- having $]t_{-}^{sc},t_{+}^{sc}[ \, = \, ]-\infty,\infty[$ ---
generates the new scaled Bregman divergence (cf. \eqref{brostu5:fo.SBD.smooth})
\begin{eqnarray}
& & D_{\varphi_{\beta,\widetilde{c}},\mathbf{P}}^{SBD}(\mathbf{Q},\mathbf{Q}^{\ast\ast}) \ = \ 
D_{\check{\varphi}_{\beta,\widetilde{c}},\mathbf{P}}^{SBD}(\mathbf{Q},\mathbf{Q}^{\ast\ast})
= \frac{2 \widetilde{c}}{\beta^2} \cdot 
\sum_{k=1}^{K} \Big\{
p_{k} \cdot \exp\negthinspace\left( \beta \cdot \frac{q_{k}}{p_{k}}\right) 
- \Big(p_{k} + \beta \cdot (q_{k}-q_{k}^{\ast\ast}) \Big)
\cdot \exp\negthinspace\left( \beta \cdot \frac{q_{k}^{\ast\ast}}{p_{k}}\right) 
\Big\},
\nonumber\\
& & \hspace{10.0cm}
\textrm{for }
\mathbf{P} \in \mathbb{R}_{>0}^{K}, \mathbf{Q} \in \mathbb{R}^{K}, \mathbf{Q}^{\ast\ast} \in \mathbb{R}^{K}.
\label{brostu5:ex.BEDandGNED.5} 
\end{eqnarray}
Taking in \eqref{brostu5:ex.BEDandGNED.5} the special case $\mathbf{P} : = \mathbf{1} = (1,\ldots,1)$ 
and probability vectors $\mathbf{Q} := \mathds{Q} \in \mathbb{S}^{K}$, 
$\mathbf{Q}^{\ast\ast} := \mathds{Q}^{\ast\ast} \in \mathbb{S}^{K}$,
one ends up at the
discrete version of the \textit{Bregman exponential divergence} of Mukherjee et al. \cite{Muk:19}; 
the latter has been generalized by Basak \& Basu \cite{Basak:22} 
(in a continuous setting) to the \textit{extended Bregman exponential divergence}
by employing 
$\mathbf{Q} := (\mathds{Q})^{\alpha_{1}}$ and  
$\mathbf{Q}^{\ast\ast} := (\mathds{Q}^{\ast\ast})^{\alpha_{2}}$
(i.e. componentwise $\alpha_{i}-$th powers),
which also arises as a special case of \eqref{brostu5:ex.BEDandGNED.5}
with $\mathbf{P} : = \mathbf{1}$.
Another interesting special case of \eqref{brostu5:ex.BEDandGNED.5}
is to choose $\mathbf{Q} := \mathds{Q} \in \mathbb{S}^{K}$ and  
$\mathbf{Q}^{\ast\ast} := \mathds{Q}^{\ast\ast}\in \mathbb{S}_{>0}^{K}$
together with $\mathbf{P} := \mathds{P} :=\mathds{Q}^{\ast\ast}$;
the outcoming $\phi-$divergence
$D_{\varphi_{\beta,\widetilde{c}},\mathds{P}}^{SBD}(\mathds{Q},\mathds{Q}^{\ast\ast})
= D_{\varphi_{\beta,\widetilde{c}}}(\mathds{Q},\mathds{Q}^{\ast\ast})
= D_{\varphi_{\beta,\widetilde{c}}}(\mathds{Q},\mathds{P})$
is a multiple of the $\beta-$order \textit{generalized negative exponential disparity/divergence GNED}
of Jeong \& Sarkar \cite{Jeo:00} and Bhandari et al. \cite{Bhan:06}
(see also Basu et al. \cite{Bas:11}),
which --- already in the special case $\beta := -1$ 
called \textit{negative exponential disparity/divergence NED}
(cf. Lindsay \cite{Lind:94} and Basu \& Sarkar \cite{Bas:94}) ---
performs good estimation-robustness against both
outliers and inliers. For a more general, comprehensive, systematic treatment of the
robustness-design of scaled Bregman divergences (including ordinary separable Bregman distances
and $\varphi-$divergences), the reader is  e.g. referred to Ki{\ss}linger \& Stummer \cite{Kis:16}.

\vspace{0.3cm}
\noindent
As far as \eqref{brostu5:ex.BEDandGNED.4} is concerned, one can
show that the involved $\Lambda_{\beta,\widetilde{c}}$
is the cumulant generating function of a 
\textquotedblleft distorted stable distribution\textquotedblright\ 
$\mathbb{\bbzeta}[ \, \cdot \,] = \mathbb{\Pi}[W \in \cdot \, ]$
of a random variable $W$,
which can be constructed as follows: 
let $Z$ be an auxiliary random variable 
(having density $f_{Z}$ and support $supp(Z) = ]-\infty,\infty[$) of a stable law
with parameter-quadruple
$(1,1,1,\frac{2\widetilde{c}}{\beta^{2}})$
in terms of the above-mentioned \textquotedblleft form-B notation\textquotedblright ;
by applying a general Laplace-transform result on p.~112 of~\cite{Zol:86} 
we can derive
\begin{eqnarray}
M_{Z}(z) := E_{\mathbb{\Pi}}[\exp(z \cdot Z)] 
= \int_{-\infty}^{\infty} \exp(z \cdot y) \cdot f_{Z}(y) \, dy  \hspace{-0.2cm} &=& \hspace{-0.2cm}
\begin{cases}
\exp\Big(
\frac{2\widetilde{c}}{\beta^{2}} \cdot 
\left[ (-z) \cdot \log (-z) - (-z) \right]
 \Big),  
\quad \textrm{if } \  z \in ]-\infty,0] , \\
\infty, \hspace{5.25cm} \textrm{if } \  z \in \, ]0,\infty[. \\
\end{cases}
\nonumber
\end{eqnarray}
Since $0 \notin int(dom(M_{Z}))$
(and thus, $Z$ does not have light-tails) we have to distort the involved density
in order to extend the effective domain. Accordingly, let $W$ be a random variable having density
\begin{equation}
f_{W}(y)\ :=\ 
\frac{2\widetilde{c}}{|\beta|} \cdot
\frac{\exp \{\frac{2\widetilde{c}\cdot e^{\beta}}{\beta} \cdot y\}}{\exp \{\frac{2\widetilde{c}\cdot 
(\beta-1) \cdot e^{\beta}}{\beta^{2}} \}}
\cdot f_{Z}\Big(- \, \frac{2\widetilde{c}}{\beta} \cdot y \Big), \qquad y \in \, ]-\infty,\infty[.
\label{brostu5:ex.BEDandGNED.7} 
\end{equation}
Then one can straightforwardly deduce from \eqref{brostu5:ex.BEDandGNED.7} that  
$\int_{-\infty}^{\infty} f_{W}(y) \, dy =1$ and that
\begin{eqnarray}
M_{W}(z) &:=& E_{\mathbb{\Pi}}[\exp(z \cdot W)] 
= \int_{-\infty}^{\infty} \exp(z \cdot y) \cdot f_{W}(y) \, dy  
\nonumber \\
&=& 
\begin{cases}
\exp\left(
\frac{2 \widetilde{c}}{\beta^2} \cdot \Big\{ 
(\frac{\beta}{2 \widetilde{c}} \cdot z + e^{\beta}) \cdot 
\log(\frac{\beta}{2 \widetilde{c}} \cdot z + e^{\beta})
- (\frac{\beta}{2 \widetilde{c}} \cdot z + e^{\beta}) 
-  (\beta-1) \cdot e^{\beta} \Big\}
\right),  
\qquad \textrm{if } \  z \cdot \beta \geq - 2\widetilde{c}\cdot e^{\beta}, \\
\infty, \hspace{5.5cm} \textrm{if } \  z \cdot \beta < - 2\widetilde{c}\cdot e^{\beta}, \\
\end{cases}
\label{brostu5:ex.BEDandGNED.8} 
\end{eqnarray}
where the first line in \eqref{brostu5:ex.BEDandGNED.8} 
coincides with the exponentiated right-hand side \eqref{brostu5:ex.BEDandGNED.4}.
Notice that $\mathbb{\bbzeta}$ is an infinitely divisible (cf. Proposition 27 in \cite{Bro:23a})
continuous distribution with density $f_{W}$, and that
$\mathbb{\bbzeta}[ \, ]0,\infty[ \, ] = \mathbb{\Pi}[W > 0]=  
\int_{0}^{\infty} f_{W}(u) \, du \in \, ]0,1[$,
$\mathbb{\bbzeta}[ \, \{0\} \, ] = \mathbb{\Pi}[W = 0]= 0$. 
Having derived the distribution $\mathbb{\bbzeta}[ \, \cdot \,] = \mathbb{\Pi}[W \in \cdot \, ]$
of $W$, by replacing $\widetilde{c}$ with $\widetilde{c} \cdot M_{\mathbf{P}}$
in all the above construction, we end up with the divergence generator 
$\widetilde{\varphi}_{\beta,\widetilde{c}} := M_{\mathbf{P}} \cdot \varphi_{\beta,\widetilde{c}}
= \varphi_{\beta,\widetilde{c}\cdot  M_{\mathbf{P}}}$ and the
distribution 
$\widetilde{\mathbb{\bbzeta}}[\,\cdot \,] = \mathbb{\Pi }[\widetilde{W}\in \cdot \,]$
of a random variable $\widetilde{W}$, such that the representability
\eqref{brostu5:fo.link.var} is satisfied. With those ingredients, we can
apply Theorem \ref{brostu5:thm.BSnarrow} for the narrow-sense
BS-minimization 
$\inf_{\mathbf{Q} \in \mathbf{\Omega}}
D_{\varphi_{\beta,\widetilde{c}},\mathbf{P}}^{SBD}(\mathbf{Q},\mathbf{P})
= \inf_{\mathbf{Q} \in \mathbf{\Omega}}
D_{\varphi_{\beta,\widetilde{c}}}(\mathbf{Q},\mathbf{P})$
of the non-probability version of the above-described
multiple $\beta-$order GNED,
for any $\mathbf{\Omega}\subset \mathbb{R}^{K}$ with regularity properties 
\eqref{regularity} and finiteness property \eqref{def fi wrt Omega}.
For the latter-type constraint sets,
in order to perform the BS-minimization 
$\inf_{\mathbf{Q} \in \mathbf{\Omega}} 
D_{\varphi_{\beta,\widetilde{c}},\mathbf{P}}^{SBD}(\mathbf{Q},\mathbf{Q}^{\ast\ast})$
for the more general scaled Bregman divergences of \eqref{brostu5:ex.BEDandGNED.5},
according to Theorem \ref{brostu5:thm.BSnarrow.SBD}
we have to further derive the corresponding distributions
$\widetilde{U}_{k}[\,\cdot \,] =: \mathbb{\Pi }[\widetilde{V}_{i}\in \cdot \,]$ 
(cf. \eqref{brostu5:Utilde_k_new})
of the $\widetilde{V}_{i}$ ($i\in I_{k}^{(n)}$),
or --- for a better simulation comfort (cf. Remark \ref{dist of components new}) ---
even the block-sum distributions 
$\widetilde{U}_{k}^{\ast n_{k}}[\,\cdot \,] = \mathbb{\Pi }[ \, \sum_{i\in I_{k}^{(n)}}\widetilde{V}_{i}\in \cdot \,]$.
We achieve this by employing \eqref{brostu5:Utilde_k_new} to compute 
the moment generating function of $\sum_{i\in I_{k}^{(n)}}\widetilde{V}_{i}$
as the (due to the assumed independence)
$n_{k}-$th power of the moment generating function of $\widetilde{V}_{i}$.
From this, and with the help of 
$\tau_{k} := M_{\mathbf{P}} \cdot \varphi_{\beta,\widetilde{c}}^{\, \prime}
\negthinspace
\left(\frac{\widetilde{q}_{k}^{\ast\ast}}{\widetilde{p}_{k}}\right)
= \frac{2 \widetilde{c} \cdot M_{\mathbf{P}}}{\beta} \cdot
\left\{ \exp\negthinspace\left( \beta \cdot \frac{\widetilde{q}_{k}^{\ast\ast}}{\widetilde{p}_{k}}\right) 
- \exp(\beta) \right\}$
for arbitrary $\widetilde{q}_{k}^{\ast\ast} \in \mathbb{R}$,
we identify that $\widetilde{U}_{k}^{\ast n_{k}}$
has the (Lebesgue-)density 
\begin{equation}
f_{\widetilde{U}_{k}^{\ast n_{k}}}(y)\ :=\ 
\frac{2\widetilde{c}\cdot M_{\mathbf{P}}}{|\beta|} \cdot
\frac{\exp \left\{
\frac{2 \widetilde{c} \cdot M_{\mathbf{P}}}{\beta} \cdot
\exp\negthinspace\left( \beta \cdot \frac{\widetilde{q}_{k}^{\ast\ast}}{\widetilde{p}_{k}}\right) 
\cdot y
\right\}
}{\exp 
\left\{
\frac{2 \widetilde{c} \cdot M_{\mathbf{P}}\cdot n_{k}}{\beta^{2}} \cdot
\exp\negthinspace\left( \beta \cdot \frac{\widetilde{q}_{k}^{\ast\ast}}{\widetilde{p}_{k}}\right) 
\cdot \left( \beta \cdot \frac{\widetilde{q}_{k}^{\ast\ast}}{\widetilde{p}_{k}} - 1 \right)
\right\}
}
\cdot f_{\breve{\breve{Z}}}\Big(- \, \frac{2\widetilde{c} \cdot M_{\mathbf{P}}}{\beta} \cdot y \Big), \qquad y \in \, ]-\infty,\infty[,
\nonumber 
\end{equation}
where $\breve{\breve{Z}}$ is a random variable with density $f_{\breve{\breve{Z}}}$ of a stable law
with parameter-quadruple $(1,1,1,\frac{2\widetilde{c}\cdot M_{\mathbf{P}}\cdot n_{k}}{\beta^{2}})$.

\end{example}

\vspace{0.3cm}
\noindent


\section{
Bare-Simulation-Method for General Deterministic Divergence-Optimization-Problems}
\label{SectDetGeneral}


\subsection{Further Divergences and Friends}
\label{SectDetGeneral.Friends}

\noindent 
Although the BS-\textit{minimization} results of 

\begin{itemize}
\item the above Section \ref{SectDetNarrow} on $\varphi-$divergences $D_{\varphi}(\mathbf{Q}, \mathbf{P})$ with 
divergence generator $\varphi$ satisfying Condition \ref{Condition  Fi Tilda in Minimization}
--- and thus particularly $\varphi$ is continuously differentiable (cf. Remark \ref{after represent 2}) ---
as well as
\item the above Section \ref{SectDetNarrow Bregman} on 
scaled (including ordinary) Bregman divergences 
$D_{\varphi,\mathbf{M}}^{SBD}(\mathbf{Q},\mathbf{P})$ (with slight re-notation) 
with the same divergence generator $\varphi$ (with possibly affine-linear addition term)

\end{itemize}

\noindent
cover a substantial amount of concrete divergences and related generalized entropies,
(in addition to the desir of \textit{maximization} of the latter two) we would like to go one step beyond and e.g. also minimize and maximize 
the \textit{following} directed distances $\mathbf{Q} \mapsto \Phi_{\mathbf{P}} \left(\mathbf{Q}\right) := D(\mathbf{Q},\mathbf{P})$ and 
connected functions of great importance
in information theory as well as in the adjacent fields of 
statistics, machine learning, artificial intelligence, signal processing 
and pattern recognition:

\begin{enumerate}

\item[(D1)] CASM $\varphi-$divergences $\Phi_{\mathbf{P}}(\mathbf{Q}) := D_{\varphi}( \mathbf{Q}, \mathbf{P})$
which \textit{can not be covered} by the narrow-sense BS minimizability results of Section \ref{SectDetNarrow},
i.e. $\varphi \in \widetilde{\Upsilon}(]a,b[)$ \textit{does not 
satisfy Condition \ref{Condition  Fi Tilda in Minimization}};
for instance, $\varphi$ may be \textit{non-}differentiable 
which is e.g. the case with the choice 
$\varphi_{TV}(t) := |t-1|$ (with $t \in ]a,b[\, = \, ]-\infty,\infty[$)
leading to the very important (discrete special case of the) \textit{total variation distance} 
\begin{equation}
\Phi_{\mathbf{P}}(\mathbf{Q}) \, := \, D_{\varphi_{TV}}(\mathbf{Q}, \mathbf{P}) := 
\sum_{k=1}^{K} p_{k} \cdot
\varphi_{TV} \negthinspace \left( \frac{q_{k}}{p_{k}}\right) 
\ = \  \sum_{k=1}^{K} | \, q_{k} - p_{k} \, |
\, \geq 0 , \quad \mathbf{Q} \in \mathbb{R}^{K}, \ \mathbf{P} \in \mathbb{R}_{\geq 0}^{K}, 
\label{brostu5:fo.TVdiv}
\end{equation}
which is equal to the $\ell_{1}-$distance 
$\Phi_{\mathbf{P}}^{\ell_{1}}(\mathbf{Q}) := || \mathbf{Q} - \mathbf{P}||$ 
between $\mathbf{Q}$ and $\mathbf{P}$ (notice that 
$\Phi_{\mathbf{P}}^{\ell_{1}}(\mathbf{Q})$ can be trivially extended to $\mathbf{P} \in \mathbb{R}^{K}$
which (in case of some strictly negative components $p_{k} <0$) is not a CASM divergence anymore).
As far as literature on $\varphi-$divergences is concerned, e.g. recall the references
given in the paragraph after \eqref{brostu5:fo.div};
some further exemplary (relatively) recent studies and applications
of $D_{\varphi}(\mathbf{Q}, \mathbf{P})$ (partially with 
\eqref{brostu5:fo.TVdiv} as well as 
more general setups)
--- in addition to the vast statistical literature 
(including in particular maximum likelihood estimation and Pearson\textquoteright s chi-square test) 
---
appear e.g. in
Berend et al. \cite{Berend:14},
Jiao et al. \cite{Jiao7:14},
Como \& Fagnani \cite{Como:15},
Han et al. \cite{Han7:15}, 
Sason \cite{Sason:15},
Batsidis et al. \cite{Bats:16},
B{\"o}cherer \& Geiger \cite{Boech:16},
Das \& Kashyap \cite{Das:16},
Alonso-Revenga et al. \cite{Alonso:17},
Keziou \& Regnault \cite{Kez:17},
Liu et al. \cite{LiuJingbo:17},
Tzortzis et al. \cite{Tzortzis:17},
Castilla et al. \cite{Cast:18},
Csisz{\'a}r \& Breuer \cite{Csi:18},
El Gheche et al. \cite{ElGheche:18},
Felipe et al. \cite{Feli:18},
Markatou \& Chen \cite{Mark:18},
Sun et al. \cite{SunYujing:18}, 
Asadi et al. \cite{Asadi:19},
Broniatowski et al. \cite{Bro:19d},
Collet \cite{Coll:19},
Sason \cite{Sas:19},
Yagli \& Cuff \cite{Yagli:19},
Zhao et al. \cite{ZhaoKun:19},
De Ponti \cite{DePonti:20},
Kammerer \& Stummer \cite{Kam:20},
Nishiyama \& Sason \cite{Nish:20},
Nomura \cite{Nom:20},
Rassouli \& G{\"u}nd{\"u}z \cite{Rassouli:20},
Esposito et al. \cite{Espo:21},
Markatou et al. \cite{Mark:21},
Salehkalaibar et al. \cite{Saleh:21},
Stummer \cite{Stu:21},
Tzortzis et al. \cite{Tzortzis:21},
Birrell et al. \cite{Birr:22},
Castilla \& Chocano \cite{Cast:22},
Dixit et al. \cite{Dixit:22},
Hyun et al. \cite{HyunDongwoon:22},
Melbourne et al. \cite{Melb:22},
Peng et al. \cite{Peng4:22},
Tan \& Zhang \cite{Tan11:22},
Zhang et al. \cite{Zhang19:22},
Alba-Fern{\'a}ndez \& Jim{\'e}nez-Gamero \cite{Alba:23},
Baudry et al. \cite{Baudry:23},
Boukeloua \& Keziou \cite{Bouk:23},
Cressie et al. \cite{Cressie:23},
Kateri \cite{Kateri:23},
Manole \& Ramdas \cite{Mano:23},
Markatou \& Liu \cite{Mark:23},
Masiha et al. \cite{Masiha:23},
Miranda et al. \cite{Miranda2:23},
Nielsen \& Okamura \cite{Nielsen:23a},
Perrone \cite{Perr:23},
Velasco-Forero \cite{Vela:23} and
Nielsen \& Okamura \cite{Nielsen:24a}.

\item[(D2)] (discrete special case of) 
\textit{separable \textquotedblleft ordinary/classical\textquotedblright\ Bregman distances} (cf. \cite{Breg:67})
\begin{eqnarray} 
& & 
\Phi_{\mathbf{P}}(\mathbf{Q}) := D_{\varphi}^{OBD}(\mathbf{Q},\mathbf{P}) 
\ := \  \sum_{k=1}^{K} 
\bigg[ \varphi \negthinspace \left( q_{k}\right) -\varphi \negthinspace \left( p_{k} \right)
- \varphi^{\prime} \negthinspace
\left( p_{k} \right) \cdot \left( q_{k} - p_{k} \right) 
\bigg] \, \geq 0 \, ,
\label{brostu5:fo.OBD}
\end{eqnarray}
where (opposed to Section \ref{SectDetNarrow Bregman})
the divergence generator $\varphi$ is \textit{not necessarily in $\widetilde{\Upsilon}(]a,b[)$ 
satisfying additionally the Condition \ref{Condition  Fi Tilda in Minimization}},
but is a lower semicontinuous convex function $\varphi: \, ]-\infty,\infty[ \, \rightarrow \, 
]-\infty,\infty]$ 
which is (continuously) differentiable on $int(dom(\varphi)) = \, ]a,b[$ and strictly convex
on (say, only) some interval $]t_{-}^{sc},t_{+}^{sc}[ \, \subset \, ]a,b[$
\footnote{
recall that the divergence generator $\varphi$ may contain some line parts, see e.g.
$\varphi := \varphi_{\gamma}$ of \eqref{brostu5:fo.powdivgen} with $\gamma \in ]1,2[$,
where $]t_{-}^{sc},t_{+}^{sc}[ \, = ]0,\infty[$ and $]a,b[ = ]-\infty,\infty[$
}
such that $dom(\varphi)$ covers all the involved components $q_{k}$ of all $\mathbf{Q} \in \mathbf{\Omega}$
as well as $]t_{-}^{sc},t_{+}^{sc}[$ covers all $p_{k}$  ($k =1,\ldots,K$);
accordingly, $D_{\varphi}^{OBD}(\mathbf{Q},\mathbf{P})$ in \eqref{brostu5:fo.OBD} is always finite 
and the overall sum can be split into finite ``partial'' sums.
For instance, $]a,b[\, = \, ]0,\infty[$ or $]a,b[ \, = \, ]-\infty,\infty[$.\\ 
For the special case of probability vectors $\mathds{P}$ and $\mathds{Q} \in \mathbb{S}^{K}$, 
$D_{\varphi}^{OBD}(\mathds{Q},\mathds{P})$ are studied e.g. in Csisz\'ar~\cite{Csi:91},~\cite{Csi:94},~\cite{Csi:95}, 
Pardo \& Vajda~\cite{Par2:97},~\cite{Par2:03}, and Stummer \& Vajda~\cite{Stu:12}). 
The general case of non-probability vectors (and even measures) is treated e.g. in
Broniatowski \& Stummer \cite{Bro:19b},\cite{Bro:22}.
Some exemplary (relatively) recent 
applications on separable ordinary Bregman distances (including continuous versions)
--- aside from the literature on the special case of separable ordinary Bregman \textit{power} distances 
already cited after \eqref{brostu5:fo.sqL2.new} ---
appear e.g. in
Jiao et al. \cite{Jiao7:14}, 
Csisz{\'a}r \& Breuer \cite{Csi:18c},
Jana \& Basu \cite{Bas:19h},
Painsky \& Wornell \cite{Painsky:20},
Vial et al. \cite{Vial:21},
Tan \& Zhang \cite{Tan11:22}.

\item[(D3)] 
(discrete special case of) \textit{scaled Bregman distances} 
of Stummer~\cite{Stu:07} and Stummer \& Vajda~\cite{Stu:12} 
\begin{eqnarray}
& & \hspace{-0.2cm}   
\Phi_{\mathbf{P}}(\mathbf{Q}) := D_{\varphi,\mathbf{M}}^{SBD}(\mathbf{Q},\mathbf{P})  
\ := \  \sum_{k=1}^{K} 
\bigg[ \varphi \negthinspace \left( \frac{q_{k}}{m_{k}} \right) -
\varphi \negthinspace \left( \frac{p_{k}}{m_{k}} \right)
- \varphi_{+}^{\prime} \negthinspace
\left( \frac{p_{k}}{m_{k}} \right) \cdot \left(\frac{q_{k}}{m_{k}} - \frac{p_{k}}{m_{k}} \right) 
\bigg] \cdot m_{k}
\ \geq \ 0 \, , 
\label{brostu5:fo.SBD}
\end{eqnarray}
where $\mathbf{M} \in \mathbb{R}_{>0}^{K}$ is a scaling vector with strictly positive 
components $m_{k}>0$ and
(opposed to Section \ref{SectDetNarrow Bregman})
the divergence generator $\varphi$ is \textit{not necessarily in $\widetilde{\Upsilon}(]a,b[)$ 
satisfying additionally the Condition \ref{Condition  Fi Tilda in Minimization}},
but is a
lower semicontinuous convex function $\varphi: \, ]-\infty,\infty[ \, 
\, \rightarrow \, ]-\infty,\infty]$ 
which is strictly convex (with right-hand derivative $\varphi_{+}^{\prime}$) on 
(say, only) some interval 
$]t_{-}^{sc},t_{+}^{sc}[ \, \subset \, int(dom(\varphi)) = \, ]a,b[$
such that $dom(\varphi)$ covers all the involved components $\frac{q_{k}}{m_{k}}$ for all $\mathbf{Q} \in \mathbf{\Omega}$
as well as $]t_{-}^{sc},t_{+}^{sc}[$
covers all $\frac{p_{k}}{m_{k}}$  ($k =1,\ldots,K$);
accordingly, 
$D_{\varphi,\mathbf{M}}^{SBD}(\mathbf{Q},\mathbf{P})$ in \eqref{brostu5:fo.SBD}
is always finite 
and the overall sum can be split into finite ``partial'' sums.
For instance, $]a,b[\, = \, ]0,\infty[$ or $]a,b[ \, = \, ]-\infty,\infty[$.
Notice that \textit{scaled Bregman divergences} 
have been first defined in Stummer~\cite{Stu:07}, Stummer \& Vajda~\cite{Stu:12}
for the context of \textit{probability measures and probability vectors}, see also
Ki{\ss}linger \& Stummer~\cite{Kis:13},~\cite{Kis:15a},~\cite{Kis:16} for the ``purely adaptive'' case 
(i.e. $m_{k} = w\big(q_{k},p_{k}\big)$ for some so-called \textit{scale-connector function} $w(\cdot)$
respectively its continuous version)
and for indications on non-probability measures and non-probability vectors. Later on, 
Broniatowski \& Stummer~\cite{Bro:19b} flexibilized/widened this to \textit{scaled Bregman distances 
of arbitrary functions and vectors}, see also Broniatowski \& Stummer~\cite{Bro:22}
for various different kinds of applications to statistics, and to the adjacent fields of
machine learning and artificial intelligence. 

\item[(D4)] (discrete special case of) the \textit{distances of Broniatowski \& Stummer~\cite{Bro:19b}} (see also
Broniatowski \& Stummer~\cite{Bro:22})
\begin{eqnarray}
& & \hspace{-1.45cm}   
\Phi_{\mathbf{P}}(\mathbf{Q}) := 
D_{\varphi,\mathbf{M_{1}},\mathbf{M_{2}},\mathbf{M_{3}}}^{BSD,c}(\mathbf{Q},\mathbf{P}) 
 :=   \sum_{k=1}^{K} 
\bigg[ \varphi \negthinspace \left( \frac{q_{k}}{m_{1,k}} \right) -\varphi \negthinspace \left( \frac{p_{k}}{m_{2,k}} \right)
- \varphi_{+,c}^{\prime} \negthinspace
\left( \frac{p_{k}}{m_{2,k}} \right) \cdot \left(\frac{q_{k}}{m_{1,k}} - \frac{p_{k}}{m_{2,k}} \right) 
\bigg] \cdot m_{3,k}
\geq 0 , 
\label{brostu5:fo.BSdist}
\end{eqnarray}
where $\mathbf{M}_{1}, \mathbf{M}_{2} \in \mathbb{R}_{>0}^{K}$ are scaling vectors and 
$\mathbf{M}_{3} \in \mathbb{R}_{>0}^{K}$ is an aggregation (weighting) vector
with strictly positive components $m_{i,k} >0$ ($k=1,\ldots,K$, $i=1,2,3$).
Moreover, the divergence generator $\varphi$ is a
lower semicontinuous convex function $\varphi: \, ]-\infty,\infty[ \, 
\, \rightarrow \, ]-\infty,\infty]$  
which is strictly convex (with right-hand derivative $\varphi_{+}^{\prime}$,
left-hand derivative $\varphi_{-}^{\prime}$ and 
intermediate values $\varphi_{+,c}^{\prime}(t) := c \cdot \varphi_{+}^{\prime}(t)
+ (1- c) \cdot \varphi_{-}^{\prime}(t)$ ($c \in [0,1]$)) on 
(say, only) some interval 
$]t_{-}^{sc},t_{+}^{sc}[ \, \subset \, int(dom(\varphi)) = \, ]a,b[$
such that $dom(\varphi)$ covers 
all the involved components $\frac{q_{k}}{m_{1,k}}$ for all $\mathbf{Q} \in \mathbf{\Omega}$
and as well as $]t_{-}^{sc},t_{+}^{sc}[$ covers all $\frac{p_{k}}{m_{2,k}}$  ($k =1,\ldots,K$).
Accordingly, $D_{\varphi,\mathbf{M_{1}},\mathbf{M_{2}},\mathbf{M_{3}}}^{BSD,c}(\mathbf{Q},\mathbf{P})$
in \eqref{brostu5:fo.BSdist}
is always finite 
and the overall sum can be split into finite ``partial'' sums.
Under some mild assumptions,
\cite{Bro:19b} verify the axioms of divergence (i.e. non-negativity and reflexivity). 
Within this general framework, all the above distances/divergences (D1), (D2), (D3) appear as special cases.
If $\varphi$ is also differentiable  (and thus, $\varphi_{+,c}^{\prime}(\cdot)$
becomes the derivative $\varphi^{\prime}(\cdot)$ and we omit the obsolete index $c$),
other important special cases are (the separable versions of)
\begin{enumerate}
\item[(i)] the \textit{total Bregman distances}
of Liu et al.~\cite{Liu8:10},\cite{Liu8:12},  
Vemuri et al.~\cite{Vem:11a} 
(see also e.g. the recent application in 
Lohit \& Kumar \cite{Lohit:23})
which are representable as
$\Phi_{\mathbf{P}}(\mathbf{Q}) := 
D_{\varphi}^{TBD}(\mathbf{Q},\mathbf{P})
:= D_{\varphi,\mathbf{1},\mathbf{1},\mathbf{M}^{to}(\mathbf{P})}^{BSD}(\mathbf{Q},\mathbf{P})$
with constant $m_{1,k} = m_{2,k} = 1$ and 
$m_{3,k} = m_{3} := 
\frac{1}{\sqrt{1+\sum_{i=1}^{K} (\varphi^{\prime} (p_{i}))^2}}$
for all $k=1,\ldots,K$  and $\mathbf{M}^{to}(\mathbf{P})$ is
the vector where each of the $K$ components is equal to $m_{3}$,

\item[(ii)] the variants of (i) with constant $m_{1,k} = m_{1} = H(\mathbf{Q})$,
$m_{2,k} = m_{2} = H(\mathbf{P})$ 
(for some differentiable positive real-valued function $H(\cdot)$) and $m_{3,k}=1$ (cf. Nock et al. \cite{Noc:16sB}), 

\item[(iii)] the \textit{conformal divergences}
of Nock et al.~\cite{Noc:16} which can be represented as
$\Phi_{\mathbf{P}}(\mathbf{Q}) 
:= D_{\varphi,\mathbf{1},\mathbf{1},\mathbf{M}^{conf}(\mathbf{P})}^{BSD}(\mathbf{Q},\mathbf{P})$
where $\mathbf{M}^{conf}(\mathbf{P})$ is the vector where each of the $K$ components is equal to 
$H(\mathbf{P})$ for some technically adequate positive real-valued function $H(\cdot)$, 

\item[(iv)] the \textit{scaled conformal divergences}
of Nock et al.~\cite{Noc:16} which can be represented as\\
$\Phi_{\mathbf{P}}(\mathbf{Q}) 
:= D_{\varphi,m \cdot \mathbf{1},m \cdot \mathbf{1},m\cdot\mathbf{M}^{conf}(\mathbf{P})}^{BSD}(\mathbf{Q},\mathbf{P})$
for some strictly positive constant $m>0$.

\end{enumerate}

A more detailed discussion on (i)--(iv) (and their continuous versions and further generalizations) can be found in 
Stummer \& Ki{\ss}linger \cite{Stu:17a}.

\item[(D5)] 
(discrete special case of the \textit{non-probability extension} of) the \textit{$\varphi-$entropies} 
in the sense of Burbea \& Rao \cite{Bur:82}
(see also Csiszar \cite{Csi:72}, Ben-Bassat \cite{BenB:78}, Ben-Tal \& Teboulle \cite{BenT:86},
Kesavan \& Kapur \cite{Kes:89},
Dacunha-Castelle \& Gamboa \cite{Dac:90},
Teboulle \& Vajda \cite{Teb:93}, 
Gamboa \& Gassiat \cite{Gam:97},
Vajda \& Zvarova \cite{Vaj:07})
\begin{equation}
\Phi_{\mathbf{1}}(\mathbf{Q}) \, := \, \mathcal{E}_{\varphi}( \mathbf{Q}) := 
\sum_{k=1}^{K} \varphi(q_{k}) 
\label{brostu5:fo.phiENT}
\end{equation}
for (say) $\varphi \in \widetilde{\Upsilon}(]a,b[)$
\textit{not necessarily satisfying additionally the Condition \ref{Condition  Fi Tilda in Minimization}}.
Clearly, the constrained maximization (minimization) of $\mathcal{E}_{\varphi}( \cdot)$ corresponds
to a \textit{generalized entropy maximization (minimization)} task.\\
As explained right after \eqref{min Pb one new}, 
there are also numerous other applications of 
$\sum_{k=1}^{K}\varphi (q_{k})$
where $\varphi$ is e.g. a cost function respectively energy function respectively purpose function,
and $\sum_{k=1}^{K}\varphi (q_{k})$
can also be interpreted as an \textquotedblleft  index/degree of (in)equality of
the set $\mathbf{\Omega}$\textquotedblright ,
respectively as an \textquotedblleft  index/degree of diversity of
the set $\mathbf{\Omega}$\textquotedblright .
A corresponding comprehensive BS-concerning discussion 
is given e.g. in Broniatowski \& Stummer \cite{Bro:23a}.\\
A flexibilization of \eqref{brostu5:fo.phiENT} is given by
\begin{equation}
\Phi_{\mathbf{1}}(\mathbf{Q}) \, := \, \mathcal{E}_{\varphi,h}( \mathbf{Q}) :=  
h\Big(\sum_{k=1}^{K}\varphi (q_{k})\Big) \, 
\label{brostu5:fo.hphiENT}
\end{equation}
where $h: \mathcal{H} \, \mapsto \mathbb{R}$
is a continuous strictly increasing (respectively strictly decreasing) function with 
$\mathcal{H} \subseteq [0,\infty[$.
The quantity $h\Big(\sum_{k=1}^{K}\varphi (q_{k})\Big)$
in \eqref{brostu5:fo.hphiENT} can be seen as 
(non-probability extension of an) 
$(h,\varphi)-$entropy in the sense of
Salicru et al.~\cite{Sal:93} (see also
e.g. Pardo \cite{Par:06}, Vajda \& Vasek \cite{Vaj:85},
as well as e.g. Chen et al. \cite{Che:07},
Girardin \& Lhote \cite{Girar:15},
Ren et al. \cite{Ren:15}, 
Girardin et al. \cite{Girar:19}
for exemplary applications).

\item[(D6)] 
Burbea-Rao divergences \cite{Bur:82} (see also e.g. 
Pardo \& Vajda~\cite{Par2:97},\cite{Par2:03})
\begin{equation}
\Phi_{\mathbf{P}}(\mathbf{Q}) := D_{\varphi}^{BR}(\mathbf{Q},\mathbf{P}) 
\ := \  \sum_{k=1}^{K} \Big[ 
\frac{\varphi(q_{k}) + \varphi(p_{k})}{2} \, - \, 
\varphi\Big(\frac{q_{k}+ p_{k}}{2}\Big)  
\Big]
\nonumber
\end{equation}
with some lower semicontinuous convex function $\varphi: \, ]-\infty,\infty[ \, \rightarrow \, 
]-\infty,\infty]$ 
which is strictly convex on $int(dom(\varphi)) = \, ]a,b[$
such that $dom(\varphi)$ covers all the involved components $q_{k}$ of all $\mathbf{Q} \in \mathbf{\Omega}$
as well as 
all $p_{k}$  ($k =1,\ldots,K$).
We can also handle their straightforward generalizations 
(called (separable form of) \textit{skew Burbea-Rao divergences} in e.g. Nielsen \& Boltz \cite{Nielsen:11e}
respectively \textit{skew Jensen divergences} in e.g. Nielsen \& Nock \cite{Nielsen:17e})
\begin{equation}
\Phi_{\mathbf{P}}(\mathbf{Q}) := D_{\varphi,\beta}^{sBR}(\mathbf{Q},\mathbf{P}) 
\ := \  \sum_{k=1}^{K} \Big[ 
\beta \cdot \varphi(q_{k}) + (1-\beta) \cdot \varphi(p_{k}) \, - \, 
\varphi\Big(\beta \cdot q_{k}+ (1-\beta) \cdot p_{k}\Big)  
\Big]
\nonumber
\end{equation}
with some $\beta \in \,  ]0,1[$ (and we can even deal with corresponding non-separable generalizations of the latter);
see also Stummer \& Ki{\ss}linger~\cite{Stu:17a} for obtaining separable skew Burbea-Rao divergences
as a special case of their scaled-Bregman-distance-flexibilizations.

\item[(D7)] (discrete special case of) 
\textit{generally non-separable \textquotedblleft ordinary/classical\textquotedblright\ Bregman distances}
(cf. \cite{Breg:67}) 
\begin{eqnarray} 
& & 
\Phi_{\mathbf{P}}(\mathbf{Q}) := D_{\psi}^{gnOBD}(\mathbf{Q},\mathbf{P}) 
\ := \   \psi \negthinspace \left(\mathbf{Q}\right) - \psi \negthinspace \left(\mathbf{P}\right)
- \nabla\psi \negthinspace
\left( \mathbf{P} \right) \cdot \left( \mathbf{Q} - \mathbf{P} \right)  \, \geq 0 \, ,
\label{brostu5:fo.nOBD}
\end{eqnarray}
where $\psi: \mathbf{B} \mapsto \mathbb{R}$ is strictly convex and continuously differentiable 
(with gradient $\nabla\psi$)
on an open domain $\mathbf{B} \subset \mathbb{R}^{K}$
such that its closure $cl(\mathbf{B})$ covers all the involved $\mathbf{Q} \in \mathbf{\Omega}$
as well as all $\mathbf{P}$. For the separable choice $\psi(\mathbf{Q}) := 
\psi_{\varphi}(\mathbf{Q}) := \sum_{k=1}^{K} \varphi(q_{k})$,
the distance \eqref{brostu5:fo.nOBD} turns into the OBD \eqref{brostu5:fo.OBD}. Moreover, 
for the choice $\psi(\mathbf{X}) := \psi_{MAH,\underline{\underline{A}}}(\mathbf{x}) := \mathbf{X}^{tr} \underline{\underline{A}} \mathbf{X}$
with some positive definite $K \times K-$matrix $\underline{\underline{A}} := (a_{i,j})_{i,j=1,\ldots,K}$,
the distance \eqref{brostu5:fo.nOBD} leads to the prominently
used \textit{squared Mahalanobis distance} \cite{Maha:36}
\begin{equation}
\Phi_{\mathbf{P}}(\mathbf{Q}) := D_{\psi_{MAH,\underline{\underline{A}}}}^{gnOBD}( \mathbf{Q}, \mathbf{P} )
 \ = \ \sum_{k=1}^{K} \sum_{j=1}^{K} a_{i,j} \cdot ( \, q_{k} - p_{k} \, ) \cdot ( \, q_{j} - p_{j} \, )
\,  ;
\label{brostu5:fo.mah}
\end{equation}  
clearly, in the special case where
$\underline{\underline{A}} := \frac{1}{2} \cdot \underline{\underline{I}}$ is half of the unit matrix $\underline{\underline{I}}$,
the squared Mahalanobis distance \eqref{brostu5:fo.mah} simplifies to the
squared $\ell_{2}-$distance \eqref{brostu5:fo.sqL2.new}.

\vspace{0.2cm}
Some exemplary (relatively) recent 
applications of (squared) Mahalanobis distance
can be found e.g. in
Xu et al. \cite{Xu18:13},
Kim et al. \cite{Kim12:14},
Mei et al. \cite{Mei1:16},
Zhang et al. \cite{Zhang33:16},
Li et al. \cite{Li14:17},
Xu et al. \cite{Xu22:17},
Mahony \& Cannon \cite{Mahony:18},
Xu et al. \cite{Xu33:18},
Etherington \cite{Ether:19},
Fitzpatrick \& Dunn \cite{Fitz:19},
Kakavand et al. \cite{Kaka:19},
Sun et al. \cite{Sun20:19},
Bai et al. \cite{Bai:20},
Li et al. \cite{Li22:20},
Naveed \& ur Rehman \cite{Naveed:20},
Wang et al. \cite{Wang31:20},
Winter et al. \cite{Winter:20},
Bartlett et al. \cite{Bartlett:21},
Butterfield et al. \cite{Butter:21},
Chamberland et al. \cite{Chamb:21},
Kang et al. \cite{Kang9:21},
Sato et al. \cite{Sato:21},
Zheng et al. \cite{Zheng8:21},
A et al. \cite{A:22},
Chakraborty et al. \cite{Chakraborty:22},
Chen et al. \cite{Chen31:22},
dos Santos et al. \cite{DosSantos:22},
Guerra et al. \cite{Guerra:22},
Huang et al. \cite{HuangHai:22},
Nomoto et al. \cite{Nomoto:22},
Reichen et al. \cite{Reichen:22},
Sun et al. \cite{SunShuping:22},
Timmermann et al. \cite{Timm:22},
Wauchope et al. \cite{Wauch:22},
Weinberger \cite{Wein:22},
Wen et al. \cite{Wen6:22},
Yang et al. \cite{Yang19:22},
Zhang et al. \cite{Zhang34:22},
Burssens et al. \cite{Burss:23},
Choi et al. \cite{Choi5:23},
Choi et al. \cite{Choi7:23},
Dahlin et al. \cite{Dahlin:23},
Ebrahimi et al. \cite{Ebrahimi:23},
Jeong et al. \cite{Jeong7:23},
Kim et al. \cite{Kim18:23},
Nowakowski et al. \cite{Nowak:23},
Qu et al. \cite{Qu4:23},
Rabby et al. \cite{Rabby:23},
Sarno et al. \cite{Sarno:23},
Tang et al. \cite{Tang18:23},
Tsvieli \& Weinberger \cite{Tsvi:23},
Zhang et al. \cite{Zhang27:23},
Zhou et al. \cite{Zhou13:23}.

\vspace{0.2cm}
Some exemplary (relatively) recent 
applications of generally non-separable (ordinary/classical) Bregman distances
appear e.g. in
Jiao et al. \cite{Jiao7:14},
Varshney \& Varshney \cite{Varshney:14},
Hu et al. \cite{Hu14:15},
Nock et al. \cite{Noc:15f},
Raskutti \& Mukherjee \cite{Rask:15},
Wang et al. \cite{WangHuiwei:15},
He et al. \cite{HeWenwu:17},
Li et al. \cite{Li28:17},
Harremo{\"e}s \cite{Harre:18a},
Xu et al. \cite{Xu22:18},
Halder \cite{Halder:19},
Zhang et al. \cite{ZhangQiping:19},
Shao et al. \cite{ShaoXiaodan:20},
Tembine \cite{Temb:20},
Br{\'e}cheteau et al. \cite{Brech:21},
Lin et al. \cite{Lin17:21},
Yuan et al. \cite{YuanDeming:21},
Azizan et al. \cite{Azizan:22},
Dytso et al. \cite{Dytso:22},
Gruzdeva \& Ushakov \cite{Gruz:22},
Song et al. \cite{Song6:22},
Tan \& Zhang \cite{Tan11:22},
Yu et al. \cite{YuZhan:22},
Cap{\'o} et al. \cite{Capo:23},
Chen et al. \cite{Chen32:23},
Fern{\'a}ndez-Rodriguez \cite{Fern:23},
Hayashi \cite{Haya:23},
Li \& Ralescu \cite{Li33:23},
Xiong et al. \cite{XiongMenghui:23},
Liu et al. \cite{LiuJie:24}.

\item[(D8)] 
\textit{weighted $\ell_{r}-$distances} ($r \in \, ]0,\infty[$)
\begin{equation}
\Phi_{\mathbf{P}}(\mathbf{Q}) \, := \, 
D_{\ell_{r},\mathbf{M}}( \mathbf{Q}, \mathbf{P} ) \, := \, 
\left(\sum_{k=1}^{K} \frac{1}{m_{k}} \cdot | \, q_{k} - p_{k} \, |^{r}
\right)^{1/r}
\, \geq 0 \, 
\label{brostu5:fo.ellrdist}
\end{equation}
where $\mathbf{M}$ is a weighting (scaling) vector with strictly positive 
components $m_{k}>0$.

\end{enumerate}

\vspace{0.3cm}

\begin{remark} \ 
(a) \, 
The above-mentioned contexts of Section \ref{SectDetNarrow}, Section \ref{SectDetNarrow Bregman} 
as well as (D1) to (D8) 
also cover corresponding divergences $D(X,Y)$ (including entropies) between
(possible complex-valued) $M \times N$ matrices $X$ and $Y$, by taking 
$\breve{D}(X,Y) := D(f(X),f(Y))$ where $f$ is an appropriate mapping from 
(a subset of) the space of all (possible complex-valued) $M \times N$ matrices to (a subset of) $\mathbb{R}^{K}$.
For instance, for real-valued $M \times N$ matrices one can employ their \textit{vectorization}
by taking (say) $f(X) := \mathbf{Q} := (q_{1}, \ldots, q_{M \cdot N})$
such that $q_{(i-1)\cdot N +j} := x_{ij}$ ($i=1,\ldots,M$, $j=1,\ldots,N$)
and hence $K:= M \cdot N$;
more flexible versions where $i \in \{1,\ldots,M\}$, $j \in J_{i}$
for some $J_{i} \subseteq \{1,\ldots,N\}$
as well as multidimensional-array/tensor versions 
can be transformed in a similar book-keeping manner, too.\\
(b) \,  Another important special case of (a) is to take
for $K \times K$ Hermitian (e.g. real symmetric) matrices (say) $X$
the function $f(X) = (\lambda_{1}, \ldots, \lambda_{K})$ to be
the $K-$dimensional real-valued vector of its eigenvalues $\lambda_{k}$ 
in e.g. decreasing order. 
Accordingly, $\breve{D}(X,Y) := D(f(X),f(Y))$ measures the dissimilarity between
the vectors of ordered eigenvalues of the two Hermitian matrices $X$ and $Y$.
Depending on the nature of the underlying vector-divergence $D(\cdot,\cdot)$
one may need further restrictions on the involved matrices $X,Y$ in order to
achieve the finiteness $\breve{D}(X,Y) < \infty$.
For instance, if $D(\cdot,\cdot) := D_{\varphi}(\cdot,\cdot)$ is of 
CASM $\varphi-$divergence type (D1),
then \textquotedblleft basically\textquotedblright\ all the ratios of the involved eigenvalues 
should lie in $dom(\varphi)$;
for $dom(\varphi) =[0,\infty[$ 
(respectively $dom(\varphi) = \, ]0,\infty[$)
this is the case for positive semi-definite
(respectively positive definite) Hermitian matrices $X,Y$.
As another example,
one can take a vector-divergence $D(\cdot,\cdot)$ from
the class (D2) of separable ordinary/classical Bregman distances (D2) or
from the class (D7) of generally non-separable ordinary/classical Bregman distances
(see e.g. 
Bauschke \& Borwein \cite{Bau:97}, Dhillon \& Tropp \cite{Dhi:07}
and Kulis et al. \cite{Kulis:09}) which covers in particular
the \textit{von Neumann divergence}, 
the \textit{log-determinant divergence}, the squared Frobenius distance
and the more general \textit{spectral Bregman matrix divergences}.\\
(c) \ According to (a) and (b), all the above-mentioned and below-mentioned BS-optimization
results on vector-divergences $D(\cdot,\cdot)$ carry over to corresponding 
BS-optimization results on matrix-divergences $\breve{D}(\cdot,\cdot)$.
\end{remark}


\subsection{Minimization via Base-Divergence-Method 1}

\vspace{0.2cm}
\noindent
All the above-mentioned contexts (D1) to (D8) share basically the same property that the 
involved function (to be constrained-optimized) 
$\mathbf{Q} \mapsto \Phi_{\mathbf{P}}(\mathbf{Q})$ is continuous. 
For such a context
we obtain the following new fundamental

\noindent
\begin{theorem}
\label{brostu5:thm.Fmin}
Let us arbitrarily fix some $\mathbf{P} \in \mathbb{R}_{> 0}^{K}$, 
$M_{\mathbf{P}}>0$, $\varphi$, $\widetilde{\mathbb{\bbzeta}}$, 
$\widetilde{W}:=(\widetilde{W}_{i})_{i\in \mathbb{N}}$ and 
$\boldsymbol{\xi }_{n}^{\mathbf{\widetilde{W}}}$
(cf. \eqref{Xi_n^W vector})
as in Theorem \ref{brostu5:thm.BSnarrow}; we call the corresponding 
$D_{\varphi}(\cdot,\mathbf{P})$ the \textit{base-divergence} (function). 
\\
(a) Furthermore, suppose that $\mathbf{\Omega}\subset \mathbb{R}^{K}$ is compact
and satisfies the regularity properties \eqref{regularity},
and that $\Phi: \mathbf{\Omega} \mapsto \mathbb{R}$ is a continuous function on $\mathbf{\Omega}$.
Then, there holds 
\begin{equation}
\inf_{\mathbf{Q}\in \mathbf{\Omega}} \Phi(\mathbf{Q})
\ = \ - \, 
\lim_{n\rightarrow \infty }
\frac{1}{n}\log \negthinspace \left( \ 
\mathbb{E}_{\mathbb{\Pi}}\negthinspace \Big[
\exp\negthinspace\Big(n \cdot \Big(
D_{\varphi }\big(M_{\mathbf{P}} \cdot \boldsymbol{\xi }_{n}^{\mathbf{\widetilde{W}}},\mathbf{P}\big) 
- \Phi\big(M_{\mathbf{P}} \cdot \boldsymbol{\xi }_{n}^{\mathbf{\widetilde{W}}}\big)
\Big)
\Big)
\cdot \textfrak{1}_{\mathbf{\Omega}}\big(M_{\mathbf{P}} \cdot \boldsymbol{\xi }_{n}^{\mathbf{\widetilde{W}}}\big)
\, \Big] 
\right)
\, 
\label{brostu5:fo.BSmin.extended}
\end{equation}
and the infimum is attained at some (not necessarily unique) point in $\mathbf{\Omega}$. 
In particular, the function 
$\Phi\left( \cdot \right)$ is bare-simulation minimizable (BS-minimizable)
on $\mathbf{\Omega}$ (cf. \eqref{brostu5:fo.2} in Definition \ref{brostu5:def.1}).\\
(b) If $\mathbf{\Omega}\subset \mathbb{R}^{K}$ is not necessarily compact
but satisfies the regularity properties 
\eqref{regularity} and the finiteness property \eqref{def fi wrt Omega},
and $\Phi: \mathbf{\Omega } \mapsto \mathbb{R}$ is a continuous function
which satisfies the lower-bound condition
\begin{equation}
\textrm{there exists a constant $c_{1} \in \mathbb{R}$ such that for all $\mathbf{Q} \in \mathbf{\Omega}$ there holds} 
\quad
\Phi(\mathbf{Q}) \geq  D_{\varphi }(\mathbf{Q},\mathbf{P}) - c_{1} 
 \, ,
\label{brostu5:fo.phibound}
\end{equation}
then the representation/convergence \eqref{brostu5:fo.BSmin.extended} 
--- and hence the corresponding BS-minimizability --- still holds,
but the infimum may not necessarily be attained/reached at some point in $\mathbf{\Omega}$.

\end{theorem}

\vspace{0.3cm}
\noindent
The proof of Theorem \ref{brostu5:thm.Fmin} will be given in Appendix \ref{App.A} below.

\vspace{0.2cm}
\noindent

\begin{remark} 
\label{brostu5:rem.thm.Fmin}
(i) In Theorem \ref{brostu5:thm.Fmin}(b),
one gets \eqref{brostu5:fo.BSmin.extended} with exponent 
$D_{\varphi }\big(M_{\mathbf{P}} \cdot \boldsymbol{\xi }_{n}^{\mathbf{\widetilde{W}}},\mathbf{P}\big) 
- \Phi\big(M_{\mathbf{P}} \cdot \boldsymbol{\xi }_{n}^{\mathbf{\widetilde{W}}}\big) \leq c_{1}$,
which turns into an 
\textit{exponential dampening} in case of $c_{1} < 0$.
Examples for the applicability of \eqref{brostu5:fo.phibound} and Theorem \ref{brostu5:thm.Fmin}(b)
will be given right below.
\\
(ii) In Theorem \ref{brostu5:thm.Fmin} we have allowed for the special case that 
$\mathbf{P}$ can be in $\mathbf{\Omega}$ (and thus, 
$\inf_{\mathbf{Q} \in \mathbf{\Omega }}D_{\varphi }(\mathbf{Q},\mathbf{P})=0$ 
of Remark \ref{dist of components}(iii) applies); however, in such a situation
one gets
$\inf_{\mathbf{Q} \in \mathbf{\Omega }} \Phi(\mathbf{Q}) \ne 0$ in general.\\
(iii) As indicated above, in Theorem \ref{brostu5:thm.Fmin} the function $\Phi(\cdot)$ can be e.g. of the form 
$\Phi(\cdot) := \Phi_{\breve{\mathbf{P}}}(\cdot) := D_{\breve{\varphi}}(\cdot,\breve{\mathbf{P}})$
where the pregiven $\breve{\mathbf{P}}$ and $\breve{\varphi}$ do not necessarily coincide 
with the $\mathbf{P}$ and $\varphi$ of the base-divergence 
$D_{\varphi }(\cdot,\mathbf{P})$ employed in \eqref{brostu5:fo.BSmin.extended}.

\end{remark}

\vspace{0.4cm} 
\noindent
Analogously to \eqref{fo.approx.1},
the limit statement \eqref{brostu5:fo.BSmin.extended}
provides the principle for the approximation of the solution of 
the minimization problem 
$\Phi(\mathbf{\Omega}) := \inf_{\mathbf{Q} \in \mathbf{\Omega}}
\Phi(\mathbf{Q})$. This can be achieved by replacing the 
right-hand side in \eqref{brostu5:fo.BSmin.extended} by its finite
counterpart, from which we obtain for given large $n$  
\begin{equation}
- \frac{1}{n}\log \negthinspace \left( \ 
\mathbb{E}_{\mathbb{\Pi}}\negthinspace \Big[
\exp\negthinspace\Big(n \cdot \Big(
D_{\varphi }\big(M_{\mathbf{P}} \cdot \boldsymbol{\xi }_{n}^{\mathbf{\widetilde{W}}},\mathbf{P}\big) 
- \Phi\big(M_{\mathbf{P}} \cdot \boldsymbol{\xi }_{n}^{\mathbf{\widetilde{W}}}\big)
\Big)
\Big)
\cdot \textfrak{1}_{\mathbf{\Omega}}\big(M_{\mathbf{P}} \cdot \boldsymbol{\xi }_{n}^{\mathbf{\widetilde{W}}}\big)
\, \Big] 
\right) 
\ \approx \ \inf_{Q\in \mathbf{\Omega} }
\Phi(\mathbf{Q});
\label{fo.approx.1.extended} 
\end{equation}
it remains to estimate the left-hand side of \eqref{fo.approx.1.extended}
(see Section \ref{SectEstimators.new.det.nonvoid}
below, where the latter also provides estimates of the \textit{minimizers}).

\vspace{0.3cm}
\noindent
Clearly, Theorem \ref{brostu5:thm.Fmin}(a) can be applied to obtain
$\inf_{\mathbf{Q}\in \mathbf{\Omega}} \Phi(\mathbf{Q})$
for all the directed distances/divergences $\Phi(\cdot) := \Phi_{\mathbf{P}}(\cdot)$ and friends 
given in (D1) to (D8), which are therefore all BS-minimizable on compact $\mathbf{\Omega}$ with 
regularity \eqref{regularity}.
As far as the application of Theorem \ref{brostu5:thm.Fmin}(b) to the case (D1)
of CASM $\varphi-$divergences is concerned, we derive the following

\noindent
\begin{corollary}
\label{brostu5:cor.Fmin.PhiDiv}
Let us arbitrarily fix some $\mathbf{P} \in \mathbb{R}_{> 0}^{K}$, 
$M_{\mathbf{P}}>0$, $\varphi$, $\widetilde{\mathbb{\bbzeta}}$, 
$\widetilde{W}:=(\widetilde{W}_{i})_{i\in \mathbb{N}}$
and $\boldsymbol{\xi }_{n}^{\mathbf{\widetilde{W}}}$
(cf. \eqref{Xi_n^W vector})
as in Theorem \ref{brostu5:thm.BSnarrow}.\\
If $\breve{\varphi} \in \widetilde{\Upsilon}(]a,b[)$ satisfies
the lower-bound condition
\begin{equation}
\textrm{there exists a constant $\breve{c} \in \, ]0,\infty[$ such that for all $t \in \, ]a,b[$ there holds} 
\ \
\breve{c} \cdot \breve{\varphi} (t) \geq  \varphi(t) 
\ \  \textrm{with equality if and only if $t=1$}
 \, ,
\label{brostu5:fo.phibrevebound}
\end{equation}
then there holds 
\begin{equation}
\inf_{\mathbf{Q}\in \mathbf{\Omega}} 
D_{\breve{c} \cdot \breve{\varphi}}(\mathbf{Q},\mathbf{P})
\ = \ - \, 
\lim_{n\rightarrow \infty }\frac{1}{n}\log \negthinspace \left( \ 
\mathbb{E}_{\mathbb{\Pi}}\negthinspace \Big[
\exp\negthinspace\Big(n \cdot \Big(
D_{\varphi}\big(M_{\mathbf{P}} \cdot \boldsymbol{\xi }_{n}^{\mathbf{\widetilde{W}}},\mathbf{P}\big) 
- 
D_{\breve{c} \cdot \breve{\varphi}}\big(M_{\mathbf{P}} \cdot \boldsymbol{\xi }_{n}^{\mathbf{\widetilde{W}}},\mathbf{P}\big)
\Big)
\Big)
\cdot \textfrak{1}_{\mathbf{\Omega}}\big(M_{\mathbf{P}} \cdot \boldsymbol{\xi }_{n}^{\mathbf{\widetilde{W}}}\big)
\, \Big] 
\right)
\, 
\label{brostu5:fo.BSmin.extended.breve}
\end{equation}
for all $\mathbf{\Omega}\subset \mathbb{R}^{K}$ satisfying the regularity properties 
\eqref{regularity} and the finiteness property \eqref{def fi wrt Omega}.
In particular, the CASM divergence 
$\Phi_{\mathbf{P}}(\cdot) \, := \, D_{\breve{c} \cdot \breve{\varphi}}( \cdot, \mathbf{P} )$
is BS-minimizable (on any such $\mathbf{\Omega}$).
The exponent in \eqref{brostu5:fo.BSmin.extended.breve} satisfies
$D_{\varphi}\big(M_{\mathbf{P}} \cdot \boldsymbol{\xi }_{n}^{\mathbf{\widetilde{W}}},\mathbf{P}\big) 
- 
D_{\breve{c} \cdot \breve{\varphi}}\big(M_{\mathbf{P}} \cdot \boldsymbol{\xi }_{n}^{\mathbf{\widetilde{W}}},\mathbf{P}\big) \leq 0$
 \footnote{
in other words, the exponential-dampening
concerning Remark \ref{brostu5:rem.thm.Fmin} is applicable with $c_{1} = 0$.
}
with equality if and only if $M_{\mathbf{P}} \cdot \boldsymbol{\xi }_{n}^{\mathbf{\widetilde{W}}} = \mathbf{P}$
(where typically the latter happens at most very occasionally).
\end{corollary}

\vspace{0.4cm}
\noindent
The assertion of Corollary \ref{brostu5:cor.Fmin.PhiDiv} follows immediately from Theorem \ref{brostu5:thm.Fmin}(b)
and the fact that \eqref{brostu5:fo.phibrevebound} implies
\begin{equation}
D_{\breve{c} \cdot \breve{\varphi}}(\mathbf{Q},\mathbf{P}) \geq  D_{\varphi}(\mathbf{Q},\mathbf{P}) 
\quad \textrm{for all $\mathbf{Q} \in \mathbf{\Omega}$} 
\nonumber
\end{equation}
(with equality if and only if $\mathbf{Q} = \mathbf{P}$).

\vspace{0.1cm}
\noindent
\begin{remark} 
In Corollary \ref{brostu5:cor.Fmin.PhiDiv}, the assumed \textit{narrow-sense} BS-minimizability
of $D_{\varphi}( \cdot, \mathbf{P} )$
is transformed into the (non-narrow-sense) BS-minimizability  
of $D_{\breve{c} \cdot \breve{\varphi}}( \cdot, \mathbf{P} )$.\\

\end{remark}

\begin{example}
\label{brostu5:ex.TV}
Let us show how Corollary \ref{brostu5:cor.Fmin.PhiDiv} can be used
to tackle the BS-minimizability of the very important total variation distance ($\ell_{1}-$distance)
\begin{equation}
\Phi_{\mathbf{P}}(\mathbf{Q}) \, := \, 
D_{\breve{\varphi}}( \mathbf{Q}, \mathbf{P} ) :=
D_{\varphi_{TV}}( \mathbf{Q}, \mathbf{P} ) := 
\sum_{k=1}^{K} p_{k} \cdot
\varphi_{TV} \negthinspace \left( \frac{q_{k}}{p_{k}}\right) 
\ = \  \sum_{k=1}^{K} | \, q_{k} - p_{k} \, |
\, \geq 0 \, 
\qquad \textrm{(cf. \eqref{brostu5:fo.TVdiv})}
\nonumber
\end{equation}
where $\breve{\varphi}(t) := \varphi_{TV}(t) := |t-1|$ for all $t \in ]a,b[\, = \, ]-\infty,\infty[$.
For this, as generator of the base-divergence we take  $\varphi:=\varphi_{\alpha,\beta,\widetilde{c}}$ of Example \ref{brostu5:ex.2},
with (for simplicity) parameters $\widetilde{c} :=1$ and $\alpha = \beta \in \, ]0,1[$. 
From $\varphi_{\beta,\beta,1}(1) = 0 = \varphi_{TV}(1)$, $\varphi_{\beta,\beta,1}^{\prime}(1) = 0$,
$\varphi_{\beta,\beta,1}^{\prime}(-\infty) = - \beta > \varphi_{TV,-}^{\prime}(1) = -1$, 
$\varphi_{\beta,\beta,1}^{\prime}(\infty) = \beta < \varphi_{TV,+}^{\prime}(1) = 1$ 
and the strict convexity of $\varphi_{\beta,\beta,1}(\cdot)$ (and thus the strict increasingness of 
$\varphi_{\beta,\beta,1}^{\prime}$), 
one can easily see that $\varphi_{\beta,\beta,1}(t) \leq |t-1|$ for all $t\in\mathbb{R}$,
with equality if and only if $t=1$. Hence, \eqref{brostu5:fo.phibrevebound} holds with 
$\breve{c} := 1$.
To proceed, let us arbitrarily fix some $\mathbf{P} \in \mathbb{R}_{> 0}^{K}$.
The transformed generator $\widetilde{\varphi_{\beta,\beta,1}} := M_{\mathbf{P}} \cdot \varphi_{\beta,\beta,1}
= \varphi_{\beta,\beta,M_{\mathbf{P}}}$ satisfies 
Condition \ref{Condition  Fi Tilda in Minimization} (i.e. \eqref{brostu5:fo.link.var})
with $\widetilde{\mathbb{\bbzeta}}$ such that 
the $\widetilde{W}_{i}$'s are i.i.d. copies of the random variable $\widetilde{W}$ of the form
$\widetilde{W} := 1 + Z_{1} - Z_{2}$, where
$Z_{1}$ and $Z_{2}$ are auxiliary random variables which are 
independent and $GAM(M_{\mathbf{P}} \cdot \beta,M_{\mathbf{P}} \cdot \beta)-$distributed 
(cf. Section XII of Broniatowski \& Stummer \cite{Bro:23a}, see also Table 1).
With these choices, \eqref{brostu5:fo.BSmin.extended.breve} specializes to
\begin{equation}
\inf_{\mathbf{Q}\in \mathbf{\Omega}} D_{\varphi_{TV}}( \mathbf{Q}, \mathbf{P} )
\ = \ - \, 
\lim_{n\rightarrow \infty }\frac{1}{n}\log \negthinspace \left( \ 
\mathbb{E}_{\mathbb{\Pi}}\negthinspace \Big[
\exp\negthinspace\Big(n \cdot \Big(
D_{\varphi_{\beta,\beta,1}}\big(M_{\mathbf{P}} \cdot \boldsymbol{\xi }_{n}^{\mathbf{\widetilde{W}}},\mathbf{P}\big) 
- 
D_{\varphi_{TV}}\big(M_{\mathbf{P}} \cdot \boldsymbol{\xi }_{n}^{\mathbf{\widetilde{W}}},\mathbf{P}\big)
\Big)
\Big)
\cdot \textfrak{1}_{\mathbf{\Omega}}\big(M_{\mathbf{P}} \cdot \boldsymbol{\xi }_{n}^{\mathbf{\widetilde{W}}}\big)
\, \Big] 
\right)
\, 
\label{brostu5:fo.BSmin.extended.TV}
\end{equation}
for all $\mathbf{\Omega}\subset \mathbb{R}^{K}$ satisfying the regularity properties 
\eqref{regularity} and the finiteness property \eqref{def fi wrt Omega}.
In particular, the total variation distance ($\ell_{1}-$distance) 
$\Phi_{\mathbf{P}}(\cdot) \, := \, D_{\varphi_{TV}}( \cdot, \mathbf{P} )$
is BS-minimizable (on any such $\mathbf{\Omega}$).
As an exemplary application, let us take $\mathbf{P} := \mathbf{1}$
and \
\begin{equation}
\mathbf{\Omega} \subseteqq \mathbf{\Omega}_{0} := \Big\{ \mathbf{Q} \in \mathbb{R}^{K}: \,   
\sum_{i=1}^{d} \Big(y_{i} + \sum_{k=1}^{K} x_{i,k} - \sum_{k=1}^{K} x_{i,k} \cdot q_{k} \Big)^{2} \leq \varepsilon
\, \Big\}
\label{brostu5:fo.OmegaRegression}
\end{equation}
with pregiven $y_{i} \in \mathbb{R}$ and $x_{i,k} \in \mathbb{R}$ ($i=1,\ldots,d$, $k=1,\ldots,K)$
--- where the integer $d$ is (say) much smaller than $K$ ---
and pregiven $\varepsilon \in \, ]0,\infty[$.
Non-degeneration assumptions on the entries $(x_{i,k})_{i=1,\ldots,d;k=1,\ldots,K}$
have to be considered so that $\mathbf{\Omega}$ (e.g. $\mathbf{\Omega}_{0}$)
satisfies the regularity properties 
\eqref{regularity} and the corresponding finiteness property \eqref{def fi wrt Omega}.

\noindent
Accordingly, with our BS method \eqref{brostu5:fo.BSmin.extended.TV} we can approximate/tackle the corresponding 
minimimum $\inf_{\mathbf{Q}\in \mathbf{\Omega}} 
D_{\varphi_{TV}}( \mathbf{Q}, \mathbf{1})$
which with $\widetilde{q}_{k} := q_{k} - 1$ transforms equivalently  into
the constrained $\ell_{1}-$norm minimum
\begin{eqnarray}
&& 
\inf_{\widetilde{\mathbf{Q}}} \, 
||  \widetilde{\mathbf{Q}} ||_{1}
\label{brostu5:fo.BPD1}
\\
&&
\textrm{such that}
\nonumber
\\
&& \sum_{i=1}^{d} \Big(y_{i} - \sum_{k=1}^{K} x_{i,k} \cdot \widetilde{q}_{k} \Big)^{2} \leq \varepsilon,
\label{brostu5:fo.BPD2}
\\
&&  \textrm{eventually  (cf. $\subsetneqq$ in \eqref{brostu5:fo.OmegaRegression})
with further (i.e. side) constraints on $\mathbf{Q}$}.
\label{brostu5:fo.BPD3}
\end{eqnarray}
In a multiple-linear-regression context where $\varepsilon$ is a pregiven small error
and each data observation $y_{i}$ is of the form
$y_{i} = \widetilde{q}_{1} \cdot x_{i,1} + \cdots + \widetilde{q}_{K} \cdot x_{i,K} + \eta_{i}$
with deterministic explanatory variables $x_{i,k}$,
parameters $\widetilde{q}_{k}$ and i.i.d. homoscedastic/standard Gaussian noise $\eta_{i}$,
the problem
\eqref{brostu5:fo.BPD1} to \eqref{brostu5:fo.BPD3} corresponds to
the (minimum value of a formulation of the) well-known \textit{basis pursuit denoising problem} 
(cf. Donoho et al. \cite{Donoho:06a}, 
see also e.g. 
Cand{\`e}s et al. \cite{Can:06},
Lustig et al. \cite{Lustig:07},
Cand\`es \cite{Can:08a}, 
Cand\`es et al. \cite{Can:08b}, 
Goldstein \& Osher \cite{Goldst:09},
Zhang et al. \cite{Zhang21:14},
Edgar et al. \cite{Edgar:19})
--- with eventual further constraints.
As a side remark, let us mention the \textit{direct} application of Theorem \ref{brostu5:thm.Fmin}(a)
to the continuous weighted $\ell_{r}-$distance ($r \in \, ]0,\infty[$)
\begin{equation}
\Phi_{\mathbf{P}}(\mathbf{Q}) \, := \, 
D_{\ell_{r},\mathbf{M}}( \mathbf{Q}, \mathbf{P} ) \, := \, 
\left(\sum_{k=1}^{K} \frac{1}{m_{k}} \cdot | \, q_{k} - p_{k} \, |^{r}
\right)^{1/r}
\, \geq 0 \, , \qquad \mathbf{M} \in \mathbb{R}_{> 0}^{K},
\qquad \qquad \textrm{(cf. \eqref{brostu5:fo.ellrdist})}
\nonumber
\end{equation}
with the special constellation $\varphi = \varphi_{\beta,\beta,1}$, 
$\mathbf{P} := \mathbf{1}$ and compact $\mathbf{\Omega}_{0}$;
by proceeding analogously as above, we can tackle by our BS-method the
constrained weighted $\ell_{r}-$norm minimum
\begin{eqnarray}
&& 
\inf_{\widetilde{\mathbf{Q}}} \, 
\left(\sum_{k=1}^{K} \frac{1}{m_{k}} \cdot | \, \widetilde{q}_{k} \, |^{r}
\right)^{1/r}
\label{brostu5:fo.weigth.lr.min.1}
\\
&&
\textrm{such that the constraint \eqref{brostu5:fo.BPD2} is satisfied.}
\label{brostu5:fo.weigth.lr.min.2}
\end{eqnarray}
For the subcase $m_{k}=1$ for all $k=1,\ldots,K$,
the problem \eqref{brostu5:fo.weigth.lr.min.1},\eqref{brostu5:fo.weigth.lr.min.2} has been
tackled by other methods e.g. in 
Foucart \& Lai \cite{Foucart:09}
and 
Liu et al. \cite{Liu8:15}, whereas e.g.
Bruckstein et al. \cite{Bruck:09} deal with general $m_{k} >0$ together with $r=1$;
notice that for $m_{k}=1$ the minimization problem 
\eqref{brostu5:fo.weigth.lr.min.1},\eqref{brostu5:fo.weigth.lr.min.2}
is a \textit{relaxation} of the \textquotedblleft  pure\textquotedblright\ 
sparseness optimization (dimension reduction) problem
(with non-continuous objective function)
\begin{eqnarray}
&& 
\inf_{\widetilde{\mathbf{Q}}} \, 
||  \widetilde{\mathbf{Q}} ||_{0}
\nonumber
\\
&&
\textrm{such that the constraint \eqref{brostu5:fo.BPD2} is satisfied,}
\nonumber
\end{eqnarray}
since
$$\lim_{r \rightarrow 0} \left(\sum_{k=1}^{K} | \, \widetilde{q}_{k} \, |^{r}
\right)^{1/r} \ = \ ||  \widetilde{\mathbf{Q}} ||_{0}
\ =: \ card(\{ k \in \{1,\ldots,K\} : \widetilde{q}_{k} \ne 0\}).  
$$
\end{example}

\vspace{0.2cm}
\noindent
\begin{remark} 
As a side effect of the above considerations (in slightly more general form), 
for any parameter-triple $\alpha,\beta,\widetilde{c} \in \, ]0,\infty[$ with 
$\widetilde{c} \cdot \beta <1$ we can show the new divergence inequality
\begin{equation}
\textrm{for all $\mathbf{Q} \in \mathbb{R}^{K}$ and $\mathbf{P} \in \mathbb{R}_{>0}^{K}$ there holds} 
\ \ 
D_{\varphi_{\alpha,\beta,\widetilde{c}}}( \mathbf{Q}, \mathbf{P} )  \leq  D_{\varphi_{TV}}( \mathbf{Q}, \mathbf{P} ) 
\quad \textrm{with equality if and only if $\mathbf{Q}=\mathbf{P}$}  \, ,
\nonumber
\end{equation}
since $\varphi_{\alpha,\beta,\widetilde{c}}(t) \leq |t-1|$ for all $t\in\mathbb{R}$,
with equality if and only if $t=1$.

\end{remark}

\vspace{0.4cm}
\noindent
\begin{example}
\label{brostu5:ex.POWmiss}
Let us now show how Corollary \ref{brostu5:cor.Fmin.PhiDiv} can be used
to tackle the BS-minimizability of the generalized power divergences
$\Phi(\cdot) := \Phi_{\mathbf{P}}(\cdot) := D_{\widetilde{c} \cdot \varphi_{\gamma}}(\cdot,\mathbf{P})$
(cf. \eqref{brostu5:fo.powdiv.new}) for the missing case $\gamma \in \, ]1,2[$ (and $\widetilde{c} \in \, ]0,\infty[$) 
in Example \ref{brostu5:ex.1}.
For this, as generator of the base-divergence we take 
$\varphi:=\varphi_{\alpha,\beta,\frac{\widetilde{c}}{\gamma}}$ of Example \ref{brostu5:ex.2},
with (for simplicity) parameters $\alpha = \beta \in \, ]0,1[$.
In Appendix \ref{App.A} we shall prove the bound
\begin{equation}
\textrm{for all $\gamma \in \, ]1,\infty[$, $\beta \in \, ]0,1[$,
$\widetilde{c} \in \, ]0,\infty[$ and $t \in \mathbb{R}$ there holds} 
\ \varphi_{\beta,\beta,\frac{\widetilde{c}}{\gamma}}(t) \, \leq \, \widetilde{c} \cdot \varphi_{\gamma}(t)
\quad \textrm{with equality if and only if $t=1$}  \, .
\label{brostu5:fo.boundPOWdivmiss}
\end{equation}
Thus, by confining ourselves to $\gamma \in \, ]1,2[$ we can proceed analogously to Example \ref{brostu5:ex.TV}, with the difference that 
$\widetilde{\varphi_{\alpha,\beta,\frac{\widetilde{c}}{\gamma}}} := M_{\mathbf{P}} \cdot \varphi_{\alpha,\beta,\frac{\widetilde{c}}{\gamma}}
= \varphi_{\beta,\beta,\frac{M_{\mathbf{P}} \cdot \widetilde{c}}{\gamma}}$
and hence
the $\widetilde{W}_{i}$'s are i.i.d. copies of the random variable $\widetilde{W}$ of the form
$\widetilde{W} := 1 + Z_{1} - Z_{2}$, where
$Z_{1}$ and $Z_{2}$ are auxiliary random variables which are independent and 
$GAM(\frac{M_{\mathbf{P}} \cdot \widetilde{c}}{\gamma} \cdot \beta,\frac{M_{\mathbf{P}} \cdot \widetilde{c}}{\gamma} \cdot \beta)-$distributed 
(cf. Section XII of Broniatowski \& Stummer \cite{Bro:23a}, see also Table 1). 
With these choices, \eqref{brostu5:fo.BSmin.extended.breve} specializes to
\begin{equation}
\inf_{\mathbf{Q}\in \mathbf{\Omega}} 
D_{\widetilde{c} \cdot \varphi_{\gamma}}( \mathbf{Q}, \mathbf{P} )
\ = \ - \, 
\lim_{n\rightarrow \infty }\frac{1}{n}\log \negthinspace \left( \ 
\mathbb{E}_{\mathbb{\Pi}}\negthinspace \Big[
\exp\negthinspace\Big(n \cdot \Big(
D_{\varphi_{\beta,\beta,\frac{\widetilde{c}}{\gamma}}}\big(M_{\mathbf{P}} \cdot \boldsymbol{\xi }_{n}^{\mathbf{\widetilde{W}}},\mathbf{P}\big) 
- 
D_{\widetilde{c} \cdot \varphi_{\gamma}}\big(M_{\mathbf{P}} \cdot \boldsymbol{\xi }_{n}^{\mathbf{\widetilde{W}}},\mathbf{P}\big)
\Big)
\Big)
\cdot \textfrak{1}_{\mathbf{\Omega}}\big(M_{\mathbf{P}} \cdot \boldsymbol{\xi }_{n}^{\mathbf{\widetilde{W}}}\big)
\, \Big] 
\right)
\, 
\nonumber
\end{equation}
for all $\mathbf{\Omega}\subset \mathbb{R}^{K}$ satisfying the regularity properties 
\eqref{regularity} and the finiteness property \eqref{def fi wrt Omega}.
In particular, for $\gamma \in \, ]1,2[$ and $\widetilde{c} \in \, ]0,\infty[$
the generalized power divergence
$\Phi_{\mathbf{P}}(\cdot) \, := \, D_{\widetilde{c} \cdot \varphi_{\gamma}}( \cdot, \mathbf{P} )$
is BS-minimizable (on any such $\mathbf{\Omega}$).
This contrasts to Example \ref{brostu5:ex.1}, where 
$D_{\widetilde{c} \cdot \varphi_{\gamma}}( \cdot, \mathbf{P} )$
is BS-minimizable \textit{in the narrow sense} 
for all the other cases
$\gamma \in \mathbb{R}\backslash]1,2[$ and $\widetilde{c} \in \, ]0,\infty[$
(cf. Broniatowski \& Stummer \cite{Bro:23a}).

\end{example}

\vspace{0.1cm}
\noindent
\begin{remark} 
\label{brostu5:rem.ex.POWmiss}
(a)
As a side effect of the above considerations (in slightly more general form), 
for all parameters $\widetilde{c} \in \, ]0,\infty[$, $\beta \in \, ]0,1[$ 
and $\gamma \in \, ]1,\infty[$ we have shown the new divergence inequality
\begin{equation}
\textrm{for all $\mathbf{Q} \in \mathbb{R}^{K}$ and $\mathbf{P} \in \mathbb{R}_{>0}^{K}$ there holds} 
\ \ 
D_{\varphi_{\beta,\beta,\frac{\widetilde{c}}{\gamma}}}( \mathbf{Q}, \mathbf{P} )  \leq   D_{\widetilde{c} \cdot \varphi_{\gamma}}( \mathbf{Q}, \mathbf{P} ) 
\quad \textrm{with equality if and only if $\mathbf{Q}=\mathbf{P}$}  \, .
\nonumber
\end{equation}
(b) For the restricted cases where $\mathbf{\Omega} \subset \mathbb{R}_{>0}^{K}$ and $\gamma \in \, ]1,2[$,
one can e.g. alternatively use as the base-divergence-generator 
$\varphi := \frac{\widetilde{c}}{\gamma} \cdot \varphi_{1}$  
and employ the well-known bound (cf. e.g. Liese \& Vajda \cite{Lie:87})
\begin{equation}
\textrm{for all $\gamma \in \, ]1,\infty[$, 
$\widetilde{c} \in \, ]0,\infty[$ and $t \in \, [0,\infty[$ there holds}  
\ \frac{\widetilde{c}}{\gamma} \cdot \varphi_{1}(t) \, \leq \, \widetilde{c} \cdot \varphi_{\gamma}(t)
\quad \textrm{with equality if and only if $t=1$}  \, ,
\label{brostu5:fo.boundPOWdivmiss.altern1}
\end{equation}
and hence
\begin{equation}
\textrm{for all $\mathbf{Q} \in \mathbb{R}_{\geq 0}^{K}$ and $\mathbf{P} \in \mathbb{R}_{>0}^{K}$ there holds} 
\ \ 
D_{\frac{\widetilde{c}}{\gamma} \cdot \varphi_{1}}( \mathbf{Q}, \mathbf{P} )  \leq  D_{\widetilde{c} \cdot \varphi_{\gamma}}( \mathbf{Q}, \mathbf{P} ) 
\quad \textrm{with equality if and only if $\mathbf{Q}=\mathbf{P}$}  \, .
\nonumber
\end{equation}
Thus, we can proceed analogously to Example \ref{brostu5:ex.TV}, with the difference that 
$\widetilde{\frac{\widetilde{c}}{\gamma} \cdot \varphi_{1}} := 
M_{\mathbf{P}} \cdot \frac{\widetilde{c}}{\gamma} \cdot \varphi_{1}$
and hence
the $\widetilde{W}_{i}$'s are i.i.d. copies of the random variable $\widetilde{W}$ of the form
$\widetilde{W} = \frac{\gamma}{\widetilde{c}\cdot M_{\mathbf{P}}} \cdot Z$ for a Poisson $POI(\frac{\widetilde{c}\cdot M_{\mathbf{P}}}{\gamma})-$distributed
random variable $Z$ (cf. Section XII of Broniatowski \& Stummer \cite{Bro:23a}, see also Table 1). 
With these choices, \eqref{brostu5:fo.BSmin.extended.breve} specializes accordingly.\\
(c) As a side effect, we can even show the new divergence bounds
\begin{equation}
\textrm{for all $\gamma \in \, ]1,\infty[$, 
$\beta \in \, ]0,\frac{8}{5}]$,
$\widetilde{c} \in \, ]0,\infty[$ and $t \in \, [0,\infty[$ there holds} 
\ \varphi_{\beta,\beta,\frac{\widetilde{c}}{\gamma}}(t) \, \leq \, \frac{\widetilde{c}}{\gamma} \cdot \varphi_{1}(t)
\quad \textrm{with equality if and only if $t=1$} ,
\label{brostu5:fo.boundPOWdivmiss.altern2}
\end{equation}
and hence 
\begin{equation}
\textrm{for all $\mathbf{Q} \in \mathbb{R}_{\geq 0}^{K}$ and $\mathbf{P} \in \mathbb{R}_{>0}^{K}$ there holds} 
\ \ 
D_{\varphi_{\beta,\beta,\frac{\widetilde{c}}{\gamma}}}( \mathbf{Q}, \mathbf{P} )  \leq  
 D_{\frac{\widetilde{c}}{\gamma} \cdot \varphi_{1}}( \mathbf{Q}, \mathbf{P} ) 
\quad \textrm{with equality if and only if $\mathbf{Q}=\mathbf{P}$}  \, ;
\nonumber
\end{equation}
see Appendix \ref{App.A} for a proof.
\end{remark}


\subsection{Minimization via Base-Divergence-Method 2}

\vspace{0.2cm}
\noindent
Due to our investigations in Section \ref{SectDetNarrow Bregman},
as an alternative to the new Theorem \ref{brostu5:thm.Fmin}
we can also derive the following new assertions,
by using a different base-divergence (function):

\noindent
\begin{theorem}
\label{brostu5:thm.Fmin.SBD}
Let us arbitrarily fix some $\mathbf{P} \in \mathbb{R}_{> 0}^{K}$, 
$M_{\mathbf{P}}>0$, $\mathbf{Q}^{\ast\ast}$,
$\varphi$,
$\widetilde{\mathbb{\bbzeta}}$, 
$\mathbf{\Omega}$, 
$\widetilde{V}$ 
and $\boldsymbol{\xi }_{n}^{\mathbf{\widetilde{V}}}$
(cf. \eqref{Xi_n^W vector V new2})
as in Theorem \ref{brostu5:thm.BSnarrow.SBD}; we call the corresponding 
$D_{\varphi,\mathbf{P}}^{SBD}(\cdot,\mathbf{Q}^{\ast\ast})$ (cf. \eqref{brostu5:fo.SBD.smooth})
the \textit{base-SBD-divergence} (function). 
\\
(a) Furthermore, suppose that $\mathbf{\Omega}\subset \mathbb{R}^{K}$ is also compact
and that $\Phi: \mathbf{\Omega} \mapsto \mathbb{R}$ is a continuous function on $\mathbf{\Omega}$.
Then, there holds 
\begin{equation}
\inf_{\mathbf{Q}\in \mathbf{\Omega}} \Phi(\mathbf{Q})
\ = \ - \, 
\lim_{n\rightarrow \infty }\frac{1}{n}\log \negthinspace \left( \ 
\mathbb{E}_{\mathbb{\Pi}}\negthinspace \Big[
\exp\negthinspace\Big(n \cdot \Big(
D_{\varphi,\mathbf{P}}^{SBD}\negthinspace\left(M_{\mathbf{P}} 
\cdot \boldsymbol{\xi }_{n}^{\mathbf{\widetilde{V}}},\mathbf{Q}^{\ast\ast}\right)  
- \Phi\big(M_{\mathbf{P}} \cdot \boldsymbol{\xi }_{n}^{\mathbf{\widetilde{V}}}\big)
\Big)
\Big)
\cdot \textfrak{1}_{\mathbf{\Omega}}\big(M_{\mathbf{P}} \cdot \boldsymbol{\xi }_{n}^{\mathbf{\widetilde{V}}}\big)
\, \Big] 
\right)
\, 
\label{brostu5:fo.BSmin.extended.SBD}
\end{equation}
and the infimum is attained at some (not necessarily unique) point in $\mathbf{\Omega}$. 
In particular, the function 
$\Phi\left( \cdot \right)$ is bare-simulation minimizable (BS-minimizable)
on $\mathbf{\Omega}$ (cf. \eqref{brostu5:fo.2} in Definition \ref{brostu5:def.1}),
\textit{in terms of the SBD method}.\\
(b) If $\mathbf{\Omega}\subset \mathbb{R}^{K}$ is not necessarily compact
and $\Phi: \mathbf{\Omega } \mapsto \mathbb{R}$ is a continuous function
which satisfies the lower-bound condition
\begin{equation}
\textrm{there exists a constant $c_{1} \in \mathbb{R}$ such that for all $\mathbf{Q} \in \mathbf{\Omega}$ there holds} 
\quad
\Phi(\mathbf{Q}) \geq 
D_{\varphi,\mathbf{P}}^{SBD}(\mathbf{Q},\mathbf{Q}^{\ast\ast}) - c_{1} 
 \, ,
\nonumber
\end{equation}
then the representation/convergence \eqref{brostu5:fo.BSmin.extended.SBD} 
--- and hence the corresponding BS-minimizability --- still holds,
but the infimum may not necessarily be attained/reached at some point in $\mathbf{\Omega}$.

\end{theorem}

\vspace{0.4cm}
\noindent
The proof of Theorem \ref{brostu5:thm.Fmin.SBD} is given in Appendix \ref{App.A} below.

\vspace{0.4cm}
\noindent

\begin{remark} 
(a) In Theorem \ref{brostu5:thm.Fmin.SBD} we have allowed for the special case that 
$\mathbf{Q}^{\ast\ast}$ can be in $\mathbf{\Omega}$ (and thus, 
$\inf_{\mathbf{Q} \in \mathbf{\Omega }}
D_{\varphi,\mathbf{P}}^{SBD}(\mathbf{Q},\mathbf{Q}^{\ast\ast})=0$ 
of Remark \ref{Rem BS narrow SBD} applies); however, in such a situation
one gets
$\inf_{\mathbf{Q} \in \mathbf{\Omega }} \Phi(\mathbf{Q}) \ne 0$ in general.
Choosing (if possible due to according freedom) $\mathbf{Q}^{\ast\ast} \in \mathbf{\Omega}$,
turns out to be useful for important improvements of the corresponding estimators of
$\inf_{\mathbf{Q}\in \mathbf{\Omega}} \Phi(\mathbf{Q})$ and their corresponding minimizers.
\\
(b) For the special case $\Phi(\cdot) := D_{\varphi,\mathbf{P}}^{SBD}(\cdot,\mathbf{Q}^{\ast\ast})$,
the exponent in \eqref{brostu5:fo.BSmin.extended.SBD} becomes zero and then
Theorem \ref{brostu5:thm.Fmin.SBD}(b) collapses to Theorem \ref{brostu5:thm.BSnarrow.SBD}.

\end{remark}

\vspace{0.5cm} 
\noindent
Alternatively to \eqref{fo.approx.1.extended},
the limit statement \eqref{brostu5:fo.BSmin.extended.SBD}
provides another principle for the approximation of the solution of 
the minimization problem 
$\Phi(\mathbf{\Omega}) 
:= \inf_{\mathbf{Q} \in \mathbf{\Omega}} \Phi(\mathbf{Q})
$.
Indeed, by replacing the right-hand side in \eqref{brostu5:fo.BSmin.extended.SBD} by its finite
counterpart, we get for given large $n$  
\begin{equation}
- \frac{1}{n}\log \negthinspace \left( \ 
\mathbb{E}_{\mathbb{\Pi}}\negthinspace \Big[
\exp\negthinspace\Big(n \cdot \Big(
D_{\varphi,\mathbf{P}}^{SBD}\negthinspace\left(M_{\mathbf{P}} 
\cdot \boldsymbol{\xi }_{n}^{\mathbf{\widetilde{V}}},\mathbf{Q}^{\ast\ast}\right) - \Phi\big(M_{\mathbf{P}} \cdot \boldsymbol{\xi }_{n}^{\mathbf{\widetilde{V}}}\big)
\Big)
\Big)
\cdot \textfrak{1}_{\mathbf{\Omega}}\big(M_{\mathbf{P}} \cdot \boldsymbol{\xi }_{n}^{\mathbf{\widetilde{V}}}\big)
\, \Big] 
\right) 
\ \approx \ \inf_{Q\in \mathbf{\Omega} }
\Phi(\mathbf{Q});
\label{fo.approx.1.extended.SBD} 
\end{equation}
it remains to estimate the left-hand side of \eqref{fo.approx.1.extended.SBD}
(see Section \ref{SectEstimators.new.det.nonvoid}
below, where the latter also provides estimates of the \textit{minimizers}).


\subsection{Maximization via Base-Divergence-Method 1}

\vspace{0.2cm}
\noindent
In the previous two subsections, for the above-mentioned contexts (D1) to (D8) --- and beyond --- 
we have only dealt with BS-minimizability so far.
For their \textit{BS-maximizability} we obtain the following new fundamental

\vspace{0.2cm}
\noindent

\begin{theorem}
\label{brostu5:thm.Fmax}
Let us arbitrarily fix some $\mathbf{P} \in \mathbb{R}_{> 0}^{K}$, 
$M_{\mathbf{P}}>0$, 
$\varphi$,  
$\widetilde{\mathbb{\bbzeta}}$, 
$\widetilde{W}:=(\widetilde{W}_{i})_{i\in \mathbb{N}}$
and $\boldsymbol{\xi }_{n}^{\mathbf{\widetilde{W}}}$
(cf. \eqref{Xi_n^W vector})
as in Theorem \ref{brostu5:thm.BSnarrow}; recall that we have named 
$D_{\varphi}(\cdot,\mathbf{P})$ as the corresponding \textit{base-divergence} (function). 
\\
(a) Furthermore, suppose that $\mathbf{\Omega}\subset \mathbb{R}^{K}$ is compact
and satisfies the regularity properties \eqref{regularity},
and that $\Phi: \mathbf{\Omega } \mapsto \mathbb{R}$ is a continuous function on $\mathbf{\Omega}$.
Then, there holds 
\begin{equation}
\sup_{\mathbf{Q}\in \mathbf{\Omega}} \Phi(\mathbf{Q})
\ = \  
\lim_{n\rightarrow \infty }\frac{1}{n}\log \negthinspace \left( \ 
\mathbb{E}_{\mathbb{\Pi}}\negthinspace \Big[
\exp\negthinspace\Big(n \cdot \Big(
D_{\varphi }\big(M_{\mathbf{P}} \cdot \boldsymbol{\xi }_{n}^{\mathbf{\widetilde{W}}},\mathbf{P}\big) 
+ \Phi\big(M_{\mathbf{P}} \cdot \boldsymbol{\xi }_{n}^{\mathbf{\widetilde{W}}}\big)
\Big)
\Big)
\cdot \textfrak{1}_{\mathbf{\Omega}}\big(M_{\mathbf{P}} \cdot \boldsymbol{\xi }_{n}^{\mathbf{\widetilde{W}}}\big)
\, \Big] 
\right)
\, 
\label{brostu5:fo.BSmax.extended}
\end{equation}
and the supremum is attained at some (not necessarily unique) point in $\mathbf{\Omega}$. 
In particular, the function 
$\Phi\left( \cdot \right)$ is bare-simulation maximizable (BS-maximizable)
on $\mathbf{\Omega}$ (cf. \eqref{brostu5:fo.2b} in Definition \ref{brostu5:def.1}).\\
(b) If $\mathbf{\Omega}\subset \mathbb{R}^{K}$ is not necessarily compact
but satisfies only the regularity properties \eqref{regularity}, 
and $\Phi: \mathbf{\Omega } \mapsto \mathbb{R}$ is a continuous function
which satisfies the upper-bound condition
\begin{equation}
\textrm{there exists a constant $c_{1} \in \mathbb{R}$ such that for all $\mathbf{Q} \in \mathbf{\Omega}$ there holds} 
\quad
\Phi(\mathbf{Q}) \leq  c_{1} - D_{\varphi }(\mathbf{Q},\mathbf{P})   
 \, ,
\nonumber
\end{equation}
then the representation/convergence \eqref{brostu5:fo.BSmax.extended} 
--- and hence the corresponding BS-maximizability --- still holds,
but the supremum may not necessarily be attained/reached at some point in $\mathbf{\Omega}$.

\end{theorem}

\vspace{0.4cm}
\noindent
The proof of Theorem \ref{brostu5:thm.Fmax} will be given in Appendix \ref{App.A} below.

\vspace{0.4cm}
\noindent

\begin{remark} 
(i) In Theorem \ref{brostu5:thm.Fmax}(b),
one gets \eqref{brostu5:fo.BSmax.extended} with exponent 
$D_{\varphi}\big(M_{\mathbf{P}} \cdot \boldsymbol{\xi }_{n}^{\mathbf{\widetilde{W}}},\mathbf{P}\big) 
+ \Phi\big(M_{\mathbf{P}} \cdot \boldsymbol{\xi }_{n}^{\mathbf{\widetilde{W}}}\big) \leq c_{1}$,
which turns into an 
\textit{exponential dampening}
in case of $c_{1} \leq 0$.\\
(ii) In Theorem \ref{brostu5:thm.Fmax} we have allowed for the special case that 
$\mathbf{P}$ can be in $\mathbf{\Omega}$.

\end{remark}

\vspace{0.4cm} 
\noindent
Analogously to \eqref{fo.approx.1.extended},
the limit statement \eqref{brostu5:fo.BSmax.extended}
provides the principle for the approximation of the solution of 
the maximization problem 
$\Phi(\mathbf{\Omega}) 
:= \sup_{\mathbf{Q} \in \mathbf{\Omega}}
\Phi(\mathbf{Q})$.
This can be achieved by replacing the right-hand side in \eqref{brostu5:fo.BSmax.extended} by its finite
counterpart, from which we obtain for given large $n$  
\begin{equation}
\frac{1}{n}\log \negthinspace \left( \ 
\mathbb{E}_{\mathbb{\Pi}}\negthinspace \Big[
\exp\negthinspace\Big(n \cdot \Big(
D_{\varphi }\big(M_{\mathbf{P}} \cdot \boldsymbol{\xi }_{n}^{\mathbf{\widetilde{W}}},\mathbf{P}\big) 
+ \Phi\big(M_{\mathbf{P}} \cdot \boldsymbol{\xi }_{n}^{\mathbf{\widetilde{W}}}\big)
\Big)
\Big)
\cdot \textfrak{1}_{\mathbf{\Omega}}\big(M_{\mathbf{P}} \cdot \boldsymbol{\xi }_{n}^{\mathbf{\widetilde{W}}}\big)
\, \Big] 
\right) 
\ \approx \ \sup_{Q\in \mathbf{\Omega} }
\Phi(\mathbf{Q});
\label{fo.approx.1.extended.max} 
\end{equation}
it remains to estimate the left-hand side of \eqref{fo.approx.1.extended.max}
(see Section \ref{SectEstimators.new.det.nonvoid}
below, where the latter also provides estimates of the \textit{maximizers}).


\subsection{Maximization via Base-Divergence-Method 2}

\noindent
Due to our investigations in Section \ref{SectDetNarrow Bregman},
as an alternative to the new Theorem \ref{brostu5:thm.Fmax}
we can also derive the following new assertions, by employing
a different base-divergence (function):

\noindent
\begin{theorem}
\label{brostu5:thm.Fmax.SBD}
Let us arbitrarily fix some $\mathbf{P} \in \mathbb{R}_{> 0}^{K}$, 
$M_{\mathbf{P}}>0$, $\mathbf{Q}^{\ast\ast}$,
$\varphi$,
$\widetilde{\mathbb{\bbzeta}}$, 
$\mathbf{\Omega}$, 
$\widetilde{V}$ 
and $\boldsymbol{\xi }_{n}^{\mathbf{\widetilde{V}}}$
(cf. \eqref{Xi_n^W vector V new2})
as in Theorem \ref{brostu5:thm.BSnarrow.SBD}; recall that we have named 
$D_{\varphi,\mathbf{P}}^{SBD}(\cdot,\mathbf{Q}^{\ast\ast})$ 
as the corresponding \textit{base-SBD-divergence} (function). 
\\
(a) Furthermore, suppose that $\mathbf{\Omega}\subset \mathbb{R}^{K}$ is also compact
and that $\Phi: \mathbf{\Omega} \mapsto \mathbb{R}$ is a continuous function on $\mathbf{\Omega}$.
Then, there holds 
\begin{equation}
\sup_{\mathbf{Q}\in \mathbf{\Omega}} \Phi(\mathbf{Q})
\ = \  
\lim_{n\rightarrow \infty }\frac{1}{n}\log \negthinspace \left( \ 
\mathbb{E}_{\mathbb{\Pi}}\negthinspace \Big[
\exp\negthinspace\Big(n \cdot \Big(
D_{\varphi,\mathbf{P}}^{SBD}\negthinspace\left(M_{\mathbf{P}} 
\cdot \boldsymbol{\xi }_{n}^{\mathbf{\widetilde{V}}},\mathbf{Q}^{\ast\ast}\right)  
+ \Phi\big(M_{\mathbf{P}} \cdot \boldsymbol{\xi }_{n}^{\mathbf{\widetilde{V}}}\big)
\Big)
\Big)
\cdot \textfrak{1}_{\mathbf{\Omega}}\big(M_{\mathbf{P}} \cdot \boldsymbol{\xi }_{n}^{\mathbf{\widetilde{V}}}\big)
\, \Big] 
\right)
\, 
\label{brostu5:fo.BSmax.extended.SBD}
\end{equation}
and the supremum is attained at some (not necessarily unique) point in $\mathbf{\Omega}$. 
In particular, the function 
$\Phi\left( \cdot \right)$ is bare-simulation maximizable (BS-maximizable)
on $\mathbf{\Omega}$ (cf. \eqref{brostu5:fo.2b} in Definition \ref{brostu5:def.1}),
\textit{in terms of the SBD method}.\\
(b) If $\mathbf{\Omega}\subset \mathbb{R}^{K}$ is not necessarily compact
and $\Phi: \mathbf{\Omega } \mapsto \mathbb{R}$ is a continuous function
which satisfies the upper-bound condition
\begin{equation}
\textrm{there exists a constant $c_{1} \in \mathbb{R}$ such that for all $\mathbf{Q} \in \mathbf{\Omega}$ there holds} 
\quad
\Phi(\mathbf{Q}) \leq 
c_{1}  - D_{\varphi,\mathbf{P}}^{SBD}(\mathbf{Q},\mathbf{Q}^{\ast\ast}) 
 \, ,
\nonumber
\end{equation}
then the representation/convergence \eqref{brostu5:fo.BSmax.extended.SBD} 
--- and hence the corresponding BS-maximizability --- still holds,
but the supremum may not necessarily be attained/reached at some point in $\mathbf{\Omega}$.

\end{theorem}

\vspace{0.4cm}
\noindent
The proof of Theorem \ref{brostu5:thm.Fmax.SBD} is given in Appendix \ref{App.A} below.

\noindent
\begin{remark} 
In Theorem \ref{brostu5:thm.Fmax.SBD} we have allowed for the special case that 
$\mathbf{Q}^{\ast\ast}$ can be in $\mathbf{\Omega}$ (and thus, 
$\inf_{\mathbf{Q} \in \mathbf{\Omega }}
D_{\varphi,\mathbf{P}}^{SBD}(\mathbf{Q},\mathbf{Q}^{\ast\ast})=0$ 
of Remark \ref{Rem BS narrow SBD} applies).

\end{remark}

\vspace{0.3cm} 
\noindent
Analogously to \eqref{fo.approx.1.extended.max},
the limit statement \eqref{brostu5:fo.BSmax.extended.SBD}
provides the principle for the approximation of the solution of 
the maximization problem 
$\Phi(\mathbf{\Omega}) := \sup_{\mathbf{Q} \in \mathbf{\Omega}}
\Phi(\mathbf{Q})$.
Indeed, by replacing the right-hand side in \eqref{brostu5:fo.BSmax.extended.SBD} by its finite
counterpart, we derive for given large $n$  
\begin{equation}
\frac{1}{n}\log \negthinspace \left( \ 
\mathbb{E}_{\mathbb{\Pi}}\negthinspace \Big[
\exp\negthinspace\Big(n \cdot \Big(
D_{\varphi,\mathbf{P}}^{SBD}\negthinspace\left(M_{\mathbf{P}} 
\cdot \boldsymbol{\xi }_{n}^{\mathbf{\widetilde{V}}},\mathbf{Q}^{\ast\ast}\right) + \Phi\big(M_{\mathbf{P}} \cdot \boldsymbol{\xi }_{n}^{\mathbf{\widetilde{V}}}\big)
\Big)
\Big)
\cdot \textfrak{1}_{\mathbf{\Omega}}\big(M_{\mathbf{P}} \cdot \boldsymbol{\xi }_{n}^{\mathbf{\widetilde{V}}}\big)
\, \Big] 
\right) 
\ \approx \ \sup_{Q\in \mathbf{\Omega} }
\Phi(\mathbf{Q});
\label{fo.approx.1.extended.max.SBD} 
\end{equation}
it remains to estimate the left-hand side of \eqref{fo.approx.1.extended.max.SBD}
(see Section \ref{SectEstimators.new.det.nonvoid}
below, where the latter also provides estimates of the \textit{maximizers}).


\section{Deterministic Narrow-Sense Bare-Simulation-Optimization of $\varphi-$divergences
with Constant-Component-Sum Side Constraint }
\label{SectDetSubsimplex.CASM}

\vspace{0.2cm}
\noindent
Recall that we have denoted by $\mathbb{S}^{K} := \{\mathbf{Q} := (q_{1},\ldots,q_{K}) \in \mathbb{R}_{\geq 0}^{K}:
\, \sum_{i=1}^{K} q_{i} =1 \}$ the simplex of probability vectors (probability simplex)
and its interior by $\mathbb{S}_{> 0}^{K} := \{\mathbf{Q}:= (q_{1},\ldots,q_{K}) \in \mathbb{R}_{>
0}^{K}: \, \sum_{i=1}^{K} q_{i} =1 \}$.
For better emphasis, for elements of these two sets we use the symbols 
$\mathds{Q}$,$\mathds{P}$ instead of $\mathbf{Q}$,$\mathbf{P}$, etc., but for their components
we still use our notation $q_{k}$,$p_{k}$. Moreover, subsets of $\mathbb{S}^{K}$
or $\mathbb{S}_{> 0}^{K}$ will be denoted by 
\textrm{$\boldsymbol{\Omega}$\hspace{-0.23cm}$\boldsymbol{\Omega}$} instead of $\mathbf{\Omega}$ etc.
In this section, we work with constraint sets of the form 
$A \cdot \textrm{$\boldsymbol{\Omega}$\hspace{-0.23cm}$\boldsymbol{\Omega}$}$
for some arbitrary $A \in \,]0,\infty[$ (sometimes even $A \in \mathbb{R}\backslash\{0\}$) which satisfy 
$int\left( A \cdot \textrm{$\boldsymbol{\Omega}$\hspace{-0.23cm}$\boldsymbol{\Omega}$} \right) = \emptyset$
(cf. Remark \ref{after det Problem}(b)) and thus need extra refinements. In more detail, we deal with

\vspace{0.2cm}
\noindent

\begin{problem}
\label{det Problem simplex}
For pregiven 
$\varphi \in \widetilde{\Upsilon}(]a,b[)$, 
positive-components probability vector $\mathds{P}:=\left( p_{1},..,p_{K}\right) \in \mathbb{S}_{> 0}^{K}$,
and subset $A \cdot \textrm{$\boldsymbol{\Omega}$\hspace{-0.23cm}$\boldsymbol{\Omega}$} \subset A \cdot \mathbb{S}^{K}$
with regularity properties --- in the relative topology (!!) ---
\begin{equation}
cl(A \cdot \textrm{$\boldsymbol{\Omega}$\hspace{-0.23cm}$\boldsymbol{\Omega}$})=
cl\left( int\left( A \cdot \textrm{$\boldsymbol{\Omega}$\hspace{-0.23cm}$\boldsymbol{\Omega}$} \right) \right) ,  
\qquad int\left( A \cdot \textrm{$\boldsymbol{\Omega}$\hspace{-0.23cm}$\boldsymbol{\Omega}$} \right) \ne \emptyset,
\label{regularity simplex}
\end{equation}
find 
\begin{equation}
\Phi_{\mathds{P}}(A \cdot \textrm{$\boldsymbol{\Omega}$\hspace{-0.23cm}$\boldsymbol{\Omega}$}) := 
\inf_{\mathbf{Q}\in A \cdot \textrm{$\boldsymbol{\Omega}$\hspace{-0.19cm}$\boldsymbol{\Omega}$}} 
D_{\varphi }(\mathbf{Q},\mathds{P}),  
\nonumber
\end{equation}
provided that 
\begin{equation}
\inf_{\mathbf{Q}\in A \cdot \textrm{$\boldsymbol{\Omega}$\hspace{-0.19cm}$\boldsymbol{\Omega}$} } 
D_{\varphi }(\mathbf{Q},\mathds{P}) < \infty 
\label{def fi wrt Omega simplex}
\end{equation}
and that divergence generator $\varphi$  additionally satisfies the following Condition  
\ref{Condition  Fi Tilda in Minimization simplex}.

\begin{condition}
\label{Condition  Fi Tilda in Minimization simplex}
Let $\varphi \in \widetilde{\Upsilon}(]a,b[)$ and 
satisfy the representation 
\begin{equation}
\varphi(t) = 
\sup_{z \in \mathbb{R}} \Big( z\cdot t - \log \int_{\mathbb{R}} e^{zy} d\mathbb{\bbzeta}(y) \Big),
\qquad t \in \mathbb{R},  
\label{brostu5:fo.link.var.simplex}
\end{equation}
for some probability distribution $\mathbb{\bbzeta}$ on the real line
such that the function $z \mapsto MGF_{\mathbb{\bbzeta}}(z) := \int_{\mathbb{R}} e^{zy} d\mathbb{\bbzeta}(y)$ is finite
on some open interval containing zero.

\end{condition}

\end{problem}

\vspace{0.2cm}
\noindent
\begin{remark} \ 
Since here $M_{\mathds{P}} =\sum_{i=1}^{K} p_{i} = 1$ and hence (in the notation of the previous Sections \ref{SectDetNarrow} and \ref{SectDetGeneral})
$\widetilde{\mathds{P}}:=\mathds{P}/M_{\mathds{P}} = \mathds{P}$,
the Condition  \ref{Condition  Fi Tilda in Minimization} collapses to Condition \ref{Condition  Fi Tilda in Minimization simplex}.
\end{remark} 

\vspace{0.3cm}
\noindent
For the directed-distance-minimization Problem \ref{det Problem simplex},
we proceed (mostly analogously to Section \ref{SectDetNarrow} above)
by constructing an appropriate sequence $(\boldsymbol{\xi}_{n})_{n\in \mathbb{N}}$ 
of $\mathbb{R}^{K}-$valued random variables
(cf. \eqref{brostu5:fo.2} in Definition \ref{brostu5:def.1}
and the special case of Remark \ref{brostu5:rem.def1}(a))
as follows:
for any $n \in \mathbb{N}$ and 
any $k \in \left\{ 1, \ldots ,K-1\right\}$, let $n_{k}:=\lfloor n \cdot 
p_{k}\rfloor $ 
(where $\lfloor x \rfloor$ denotes the integer part of $x$)
and $n_{K} := n- \sum_{k=1}^{K-1} n_{k}$;
for this, we assume that $n \in \mathbb{N}$ is large enough, 
namely
$n \geq \max_{k \in \{1, \ldots, K\}} \frac{1}{p_{k}}$,
such that all the integers $n_{k}$ ($k=1,\ldots,K$) are
non-zero.
Since we assume $\mathbf{P} \in \mathbb{R}_{> 0}^{K}$ 
and thus none of the $p_{k}$'s is zero, one has 
\begin{equation}
\lim_{n\rightarrow \infty} \frac{n_{k}}{n} = p_{k}, \qquad k=1,\ldots,K.
\nonumber
\end{equation}
With this at hand,
we decompose the set $\{1, \ldots, n\}$ of all integers from $1$ to $n$
into the following disjoint blocks: $I_{1}^{(n)}:=\left\{
1,\ldots ,n_{1}\right\} $, $I_{2}^{(n)}:=\left\{ n_{1}+1,\ldots
,n_{1}+n_{2}\right\} $, and so on until the last block 
$I_{K}^{(n)} := \{ \sum_{k=1}^{K-1} n_{k} + 1, \ldots, n \}$ which
therefore contains all integers from $n_{1}+ \ldots +n_{K-1}+1$ to $n$.
Due to our construction, $I_{k}^{(n)}$ has $n_{k} \geq 1$ elements (i.e. $card(I_{k}^{(n)}) = n_{k}$)
for all $k \in \{1, \ldots, K\}$
\footnote{
if all $p_{k}$ ($k=1,\ldots,K$) are rational numbers in $]0,1[$ with 
$\sum_{k=1}^{K} p_{k} =1$ 
and $N$ is the (always existing) smallest integer such that all
$N \cdot p_{k}$ ($k=1,\ldots,K$) are integers (i.e. $\in \mathbb{N}$),
then for any multiple $n= m \cdot N$ ($m \in \mathbb{N}$) 
one gets that all $n \cdot p$ are integers
and hence $n_{k}= \lfloor n \cdot p_{k}\rfloor = n \cdot p_{k}$
($k=1,\ldots,K$)}.
Furthermore, consider a vector $\mathbf{W}:=\left( W_{1},\ldots ,W_{n}\right) $ where the $W_{i}$'s are
i.i.d. copies of the random variable $W$ whose distribution is
associated with the divergence-generator $\varphi$ through  
\eqref{brostu5:fo.link.var.simplex},
in the sense that 
$\mathbb{\Pi }[W\in \cdot \,]= \mathbb{\bbzeta}
[\,\cdot \,]$. We group the $W_{i}$'s according to the
above-mentioned blocks and sum them up blockwise, in order to build the
following $K-$ component random vector 
(instead of $\boldsymbol{\xi }_{n}^{\mathbf{\widetilde{W}}}$ in \eqref{Xi_n^W vector})
\begin{eqnarray}
\boldsymbol{\xi}_{n}^{w\mathbf{W}} &:=&
\begin{cases}
\left(\frac{\sum_{i \in I_{1}^{(n)}}W_{i}}{\sum_{k=1}^{K}\sum_{i \in I_{k}^{(n)}}W_{i}},
\ldots, \frac{\sum_{i \in I_{K}^{(n)}}W_{i}}{\sum_{k=1}^{K}\sum_{i \in I_{k}^{(n)}}W_{i}} \right) ,
\qquad \textrm{if } \sum_{j=1}^{n} W_{j} \ne 0, \\
\ (\infty, \ldots, \infty) =: \boldsymbol{\infty}, \hspace{4.0cm} \textrm{if } \sum_{j=1}^{n} W_{j} = 0.
\end{cases}
\label{brostu5:fo.norweiemp.vec.det} 
\end{eqnarray}

\vspace{0.2cm}
\noindent

\begin{remark} 
(i)  
(Concerning e.g. computer-program command availability) 
In case of $\sum_{j=1}^{n}W_{j}=0$, in \eqref{brostu5:fo.norweiemp.vec.det} 
we may equivalently assign to 
$\boldsymbol{\xi}_{n}^{w\mathbf{W}}$ any vector outside of
$\boldsymbol{\Omega}$\hspace{-0.23cm}$\boldsymbol{\Omega}$
instead of $\boldsymbol{\infty}$.\\
(ii) \ By construction, in case of $\sum_{j=1}^{n} W_{j} \ne 0$, the sum of the random 
$K$ vector components of \eqref{brostu5:fo.norweiemp.vec.det} is now automatically equal to 1, 
but --- as (depending on $\varphi$) the $W_{i}$\textquoteright s may take both positive and negative values --- 
these random components may be negative with probability strictly 
greater than zero (respectively nonnegative with probability strictly less than 1).
However, $\mathbb{\Pi} [\boldsymbol{\xi}_{n}^{w\mathbf{W}}\in \mathbb{S}_{>0}^{K}]>0$ 
since all the (identically distributed) random variables $W_{i}$ have expectation 1 
(as a consequence of the assumed representability
\eqref{brostu5:fo.link.var.simplex};
in case of $\mathbb{\Pi}[W_{1}>0]=1$ one has even 
$\mathbb{\Pi}[\boldsymbol{\xi}_{n}^{w\mathbf{W}}\in \mathbb{S}_{>0}^{K}]=1$.
Summing up things, the probability $\mathbb{\Pi}[\boldsymbol{\xi }_{n}^{w\mathbf{W}}
\in \boldsymbol{\Omega}$\hspace{-0.23cm}$\boldsymbol{\Omega}]
= \mathbb{\Pi}[A \cdot \boldsymbol{\xi }_{n}^{w\mathbf{W}} \in 
A \cdot 
\boldsymbol{\Omega}$\hspace{-0.23cm}$\boldsymbol{\Omega}]$
is strictly positive and finite at least for large n,
whenever $\Phi_{\mathds{P}}(\boldsymbol{\Omega}$\hspace{-0.23cm}$\boldsymbol{\Omega})
= \inf_{\mathds{Q}\in \textrm{$\boldsymbol{\Omega}$\hspace{-0.19cm}$\boldsymbol{\Omega}$} }
\ D_{\varphi }(\mathds{Q},\mathds{P})$ is finite.
\end{remark}

\vspace{0.3cm}
\noindent
With the above-mentioned ingredients, we have proven in Theorem 12 of 
Broniatowski \& Stummer \cite{Bro:23a} (an even more general, conditional-expectations-involving
version of) the following assertion:

\vspace{0.2cm}

\begin{theorem}
\label{brostu3:thm.divnormW.new.det} 
Let $\mathds{P} \in \mathbb{S}_{> 0}^{K}$, 
and suppose that the divergence generator $\varphi$ satisfies Condition 
\ref{Condition  Fi Tilda in Minimization simplex}.
Additionally, let $(W_{i})_{i \in  \mathbb{N}}$  be  a family
of independent and identically distributed $\mathbb{R}-$valued random variables
with probability distribution $\mathbb{\bbzeta}[ \cdot \, ] := \mathbb{\Pi}[W_{1} \in \cdot \, ]$
being connected with the divergence generator $\varphi \in \Upsilon(]a,b[)$ via the representability
\eqref{brostu5:fo.link.var.simplex}.
Then there holds (in terms of \eqref{brostu5:fo.norweiemp.vec.det})
\begin{align}
& \inf_{\mathds{Q}\in \textrm{$\boldsymbol{\Omega}$\hspace{-0.19cm}$\boldsymbol{\Omega}$} }
\ \inf_{m\neq 0} 
D_{\varphi }(m\cdot \mathds{Q},\mathds{P})  = 
\inf_{m\neq 0}\ \inf_{\mathds{Q}\in \textrm{$\boldsymbol{\Omega}$\hspace{-0.19cm}$\boldsymbol{\Omega}$} }
D_{\varphi }(m\cdot \mathds{Q},\mathds{P})  
= -\lim_{n\rightarrow \infty }\frac{1}{n}\log \, 
\mathbb{\Pi}\negthinspace \left[\boldsymbol{\xi}_{n}^{w\mathbf{W}}\in 
\textrm{$\boldsymbol{\Omega}$\hspace{-0.23cm}$\boldsymbol{\Omega}$}\right]
\label{LDP Normalized Vec BS2}
\end{align}
for all sets $\boldsymbol{\Omega}$\hspace{-0.23cm}$\boldsymbol{\Omega}$ satisfying the 
regularity properties \eqref{regularity simplex} 
\textit{in the relative topology}
and the finiteness property 
\eqref{def fi wrt Omega simplex}
(the latter two with $A=1$).
\end{theorem}

\vspace{0.2cm}
\noindent
From this and with the help of \eqref{min Pb prob1}, one gets immediately 

\vspace{0.2cm}

\begin{corollary}
\label{brostu3:cor.divnormW.new.det} 
Let $A \in \mathbb{R} \backslash\{0\}$, 
$\mathbf{P} \in \mathbb{R}_{> 0}^{K}$ with $M_{\mathbf{P}} : =\sum_{i=1}^{K}p_{i}>0$,
and suppose that the divergence generator $\varphi$ satisfies Condition 
\ref{Condition  Fi Tilda in Minimization}.
Additionally, let $(\widetilde{W}_{i})_{i \in  \mathbb{N}}$ be a family
of independent and identically distributed $\mathbb{R}-$valued random variables
with probability distribution $\widetilde{\mathbb{\bbzeta}}[ \cdot \, ] := 
\mathbb{\Pi}[\widetilde{W}_{1} \in \cdot \, ]$
being connected with the divergence generator $\varphi \in \Upsilon(]a,b[)$ via the representability
\eqref{brostu5:fo.link.var}.
Then there holds (in terms of \eqref{brostu5:fo.norweiemp.vec.det} with 
$\widetilde{W}$ instead of $W$)
\begin{align}
& \inf_{\mathbf{Q}\in A \cdot \textrm{$\boldsymbol{\Omega}$\hspace{-0.19cm}$\boldsymbol{\Omega}$} }
\ \inf_{m\neq 0} 
D_{\varphi}(m\cdot \mathbf{Q},\mathbf{P})  = 
\inf_{m\neq 0}\ \inf_{\mathbf{Q}\in A \cdot \textrm{$\boldsymbol{\Omega}$\hspace{-0.19cm}$\boldsymbol{\Omega}$} }
D_{\varphi}(m\cdot \mathbf{Q},\mathbf{P})  
= -\lim_{n\rightarrow \infty }\frac{1}{n}\log \, 
\mathbb{\Pi}\negthinspace \left[\boldsymbol{\xi}_{n}^{w\mathbf{\widetilde{W}}}\in 
\textrm{$\boldsymbol{\Omega}$\hspace{-0.23cm}$\boldsymbol{\Omega}$}\right]
\label{LDP Normalized Vec BS2 nonnormalized}
\end{align}
for all sets $\boldsymbol{\Omega}$\hspace{-0.23cm}$\boldsymbol{\Omega}$ satisfying the 
regularity properties \eqref{regularity simplex} \textit{in the relative topology}
and the \eqref{def fi wrt Omega simplex}.

\end{corollary}

\vspace{0.2cm} 
\noindent
\begin{remark}
\label{m-minimization CASM}
(a) As shown in \cite{Bro:23a}, the involved 
\textquotedblleft  inner\textquotedblright\ $m-$minimizations on the
left-hand side of \eqref{LDP Normalized Vec BS2} 
(and analogously, of \eqref{LDP Normalized Vec BS2 nonnormalized}) can be solved explicitly in
the important special case of the power divergences \eqref{brostu5:fo.powdiv.new}.\\
(b) Even more, we have worked out in \cite{Bro:23a} that the outcomes of (a) can be rewritten in terms of invertible
functions $F(\cdot)$ of the divergences $D_{\widetilde{c} \cdot \varphi_{\gamma}}(\mathds{Q},\mathds{P})$
(and analogously of $D_{\widetilde{c} \cdot \varphi_{\gamma}}(\mathbf{Q},\mathbf{P})$). 

\end{remark}

\vspace{0.3cm} 
\noindent
Let us illuminate the details of the previous Remark \ref{m-minimization CASM}. 
The required representability \eqref{brostu5:fo.link.var.simplex} 
is satisfied for all (multiple of) the generators
$\varphi(\cdot) := \widetilde{c} \cdot \varphi_{\gamma}(\cdot)$ of \eqref{brostu5:fo.powdivgen}
with $\widetilde{c} \in \, ]0,\infty[$ and $\gamma \in \mathbb{R}\backslash\, ]1,2[$
(cf. Broniatowski \& Stummer~\cite{Bro:23a}); the corresponding crucial
probability laws $\bbzeta$ can be found in Table 1.
The corresponding generalized power divergences
$D_{\widetilde{c} \cdot \varphi_{\gamma}}(\mathbf{Q},\mathbf{P})$ given by \eqref{brostu5:fo.powdiv.new} 
will be used as tools to derive our base-divergences for 
our new fundamental Theorem \ref{brostu5:thm.Fmin.simplex} below.
In order to obtain this, for fixed $A \in \, ]0,\infty[$ and 
$\mathds{P} \in \mathbb{S}_{>0}^{K}$
we employ the following auxiliary notations:

\vspace{0.1cm}

\begin{enumerate}

\item[(S1)] 
$\widetilde{\mathcal{M}}_{\gamma} := A \cdot \mathbb{S}_{>0}^{K}$
for $\gamma \in \, ]-\infty ,0]$, 
respectively, $\widetilde{\mathcal{M}}_{\gamma} := A \cdot \mathbb{S}^{K}$ for 
$\gamma \in \, ]0,1] \cup [2,\infty[$. 

\item[(S2)] the \textit{modified $\gamma-$order Hellinger integral of $\mathbf{Q}$ and $\mathds{P}$}
given by
\begin{equation}
\hspace{-0.9cm}
0 < H_{\gamma}(\mathbf{Q},\mathds{P}) := \sum\displaylimits_{k=1}^{K} (q_{k})^{\gamma} \cdot (p_{k})^{1-\gamma}
\ = \ 1 + \gamma \cdot (A-1) +
\gamma \cdot (\gamma-1) \cdot D_{\varphi_{\gamma}}(\mathbf{Q}, \mathds{P}), 
\quad \gamma \in \, ]-\infty ,0[ \, \cup \, ]0,1[ \, 
\cup \, [2,\infty[, \ \mathbf{Q}\in \widetilde{\mathcal{M}}_{\gamma};
\nonumber
\end{equation}

\item[(S3)] for $\gamma \in \, ]-\infty ,0[ \, \cup \, ]0,1[ \, 
\cup \, [2,\infty[$
and $\widetilde{c} \in \, ]0,\infty[$ we define the function
$F_{\gamma,\widetilde{c},A}: \, 
]-\infty,\infty[ \, \mapsto \, ]-\infty,\infty]$ by
\begin{eqnarray}
F_{\gamma,\widetilde{c},A}(x) &:=&
\begin{cases}
\frac{\widetilde{c}}{\gamma }
\cdot \big\{ 1- A^{\gamma/(\gamma-1)} \cdot 
\big[ 1+ \gamma \cdot (A-1) + \frac{\gamma \cdot \left( \gamma -1\right) }{\widetilde{c}} \cdot x \big]
^{-1/\left( \gamma -1\right) }\big\} ,\\
\hspace{1.65cm} 
\textrm{if } \gamma \in \, ]-\infty,0[ \, \cup \, ]0,1[ \,
\textrm{ and } 1+ \gamma \cdot (A-1) + \frac{\gamma \cdot \left( \gamma -1\right) }{\widetilde{c}} \cdot x \geq 0, \\
\hspace{1.65cm}
\textrm{or if } \gamma \in \, [2,\infty[ \,
\ \textrm{ and } \ 1+ \gamma \cdot (A-1) + \frac{\gamma \cdot \left( \gamma -1\right) }{\widetilde{c}} \cdot x > 0,
\\
\ \infty, \hspace{1.0cm} \textrm{if $\gamma \in \, [2,\infty[$ and } 
1+ \gamma \cdot (A-1) + \frac{\gamma \cdot \left( \gamma -1\right) }{\widetilde{c}} \cdot x \leq 0,\\
\ \infty, \hspace{1.0cm} \textrm{if $\gamma \in \, ]-\infty ,0[ \, \cup \, ]0,1[$ and } 
1+ \gamma \cdot (A-1) + \frac{\gamma \cdot \left( \gamma -1\right) }{\widetilde{c}} \cdot x < 0.
\end{cases}
\label{brostu5:fo.divpow.F}
\end{eqnarray}
Notice that in the constellations of (S2) only the first line of \eqref{brostu5:fo.divpow.F} 
is relevant, and we can even show
$F_{\gamma,\widetilde{c},A}\Big(D_{\widetilde{c} \cdot \varphi_{\gamma}}(\mathbf{Q},\mathds{P})\Big)
\in \, [0,\infty[$ 
for all $\mathbf{Q} \in \widetilde{\mathcal{M}}_{\gamma}$ 
(cf. Broniatowski \& Stummer~\cite{Bro:23a}). 
Moreover, the function 
$x \mapsto F_{\gamma,\widetilde{c},A}(x)$ is 
on its effective domain $dom(F_{\gamma,\widetilde{c},A}) := \{x: F_{\gamma,\widetilde{c},A}(x) \in \, ]-\infty,\infty[\}$
strictly increasing and has the strictly increasing inverse function
\begin{equation}
F_{\gamma,\widetilde{c},A}^{\leftarrow}(z) :=
\frac{\widetilde{c}}{\gamma \cdot (\gamma-1)}
\cdot \left\{
A^{\gamma} \cdot 
\left[ 1 - \frac{\gamma}{\widetilde{c}} \cdot z \right]^{-\left( \gamma -1\right)}
- 1 - \gamma \cdot (A-1) 
\right\} ,
\qquad \textrm{for all $z \in \mathbb{R}$ such that } \gamma \cdot z  \leq \widetilde{c}. 
\nonumber
\end{equation}

\item[(S4)] for $\gamma = 1$ 
we employ for $\widetilde{c} \in \, ]0,\infty[$ the strictly increasing function
$F_{1,\widetilde{c},A}: \, 
]-\infty,\infty[ 
\, \mapsto \, ]-\infty,\infty[$ defined by
\begin{equation}
F_{1,\widetilde{c},A}(x) := 
\widetilde{c} \cdot \Big\{ 1- A \cdot 
\exp\Big( \frac{1}{A} - 1 - \frac{x}{A \cdot \widetilde{c}} \Big)
\Big\} .
\label{brostu5:fo.divpow.F.KL}
\end{equation}
Notice that $F_{1,\widetilde{c},A}\Big(D_{\widetilde{c} \cdot \varphi_{1}}(\mathbf{Q},\mathds{P})\Big)
\in \, [0,\infty[$ for all $\mathbf{Q} \in \widetilde{\mathcal{M}}_{1}$
(cf. Broniatowski \& Stummer~\cite{Bro:23a}). The corresponding inverse is given by
\begin{equation}
F_{1,\widetilde{c},A}^{\leftarrow}(z) :=
\widetilde{c} \cdot \left\{
1 - A - A \cdot \left[\log\left(1-\frac{z}{\widetilde{c}}\right)  - \log A \right] 
\right\} ,
\qquad \textrm{for all } z  \in \, ]-\infty, \widetilde{c}[. 
\nonumber
\end{equation}

\item[(S5)] for $\gamma = 0$ 
we employ for $\widetilde{c} \in \, ]0,\infty[$ the strictly increasing function
$F_{0,\widetilde{c},A}: \, 
]-\infty,\infty[ 
\, \mapsto \, ]-\infty,\infty[$ defined by
\begin{equation}
F_{0,\widetilde{c},A}(x) := 
\widetilde{c} \cdot (1 - A + \log A) + x .
\nonumber
\end{equation}
Notice that $F_{0,\widetilde{c},A}\Big(D_{\widetilde{c} \cdot \varphi_{1}}(\mathbf{Q},\mathds{P})\Big)
\in \, [0,\infty[$ for all $\mathbf{Q} \in \widetilde{\mathcal{M}}_{0}$
(cf. Broniatowski \& Stummer~\cite{Bro:23a}). The corresponding inverse is given by
\begin{equation}
F_{0,\widetilde{c},A}^{\leftarrow}(z) := z - \widetilde{c} \cdot (1 - A + \log A) ,
\qquad \textrm{for all } z \in \mathbb{R}. 
\nonumber
\end{equation}

\end{enumerate}

\vspace{0.3cm}

\begin{remark}
\label{rem.FasDivergence}
For the special case $A=1$, 
$D(\cdot,\cdot) := F_{\gamma,\widetilde{c},1}\Big(D_{\widetilde{c} \cdot \varphi_{\gamma}}(\cdot,\cdot)\Big)$
is a divergence on $\textrm{$\boldsymbol{\Omega}$\hspace{-0.23cm}$\boldsymbol{\Omega}$}$
(i.e. for all $\mathds{Q},\mathds{P} \in \textrm{$\boldsymbol{\Omega}$\hspace{-0.23cm}$\boldsymbol{\Omega}$}$
there holds
$F_{\gamma,\widetilde{c},1}\Big(D_{\widetilde{c} \cdot \varphi_{\gamma}}(\mathds{Q},\mathds{P})\Big) \geq 0$ 
with equality if and only if $\mathds{Q}=\mathds{P}$).
Moreover, due to the above-mentioned strict increasingness of $F_{\gamma,\widetilde{c},1}$,
the (not necessarily unique) minimizer respectively maximizer of $D(\cdot,\mathds{P})$ and 
$D_{\widetilde{c} \cdot \varphi_{\gamma}}(\cdot,\mathds{P})$
on (say) compact $\textrm{$\boldsymbol{\Omega}$\hspace{-0.23cm}$\boldsymbol{\Omega}$}$ coincide.
\end{remark}

\vspace{0.4cm}
\noindent  
For such a context, Broniatowski \& Stummer \cite{Bro:23a} (cf. Theorem 12 \footnote{
which coincides with the above Theorem \ref{brostu3:thm.divnormW.new.det}}, 
Formula (39) and Lemma 14 therein) obtain the following assertion
on BS-minimizability in the narrow sense:

\noindent
\begin{theorem}
\label{brostu5:thm.divnormW.new} 
Let $\mathds{P} \in \mathbb{S}_{> 0}^{K}$, 
$\widetilde{c} \in \, ]0,\infty[$, $\gamma \in \mathbb{R}\backslash\, ]1,2[$
and $A \in \, ]0,\infty[$
be arbitrary but fixed.
Moreover, let $(W_{i})_{i \in  \mathbb{N}}$  be  a family
of independent and identically distributed $\mathbb{R}-$valued random variables
with probability distribution $\mathbb{\bbzeta}[ \cdot \, ] := \mathbb{\Pi}[W_{1} \in \cdot \, ]$
being connected with the divergence generator 
$\varphi :=  \widetilde{c} \cdot \varphi_{\gamma}(\cdot) \in \widetilde{\Upsilon}(]a,b[)$ 
via the representability \eqref{brostu5:fo.link.var.simplex}. \\
(a) Then there holds 
\begin{eqnarray}
& & \inf_{\mathbf{Q}\in A \cdot \textrm{$\boldsymbol{\Omega}$\hspace{-0.19cm}$\boldsymbol{\Omega}$} }
F_{\gamma,\widetilde{c},A}\Big(D_{\widetilde{c} \cdot \varphi_{\gamma}}(\mathbf{Q},\mathds{P})\Big)
= -\lim_{n\rightarrow \infty }\frac{1}{n}\log \, 
\mathbb{\Pi}\negthinspace \left[\boldsymbol{\xi}_{n}^{w\mathbf{W}}\in 
\textrm{$\boldsymbol{\Omega}$\hspace{-0.23cm}$\boldsymbol{\Omega}$}\right]
\label{LDP Normalized Vec simplex}
\\
& & 
= \inf_{\mathds{Q}\in \textrm{$\boldsymbol{\Omega}$\hspace{-0.19cm}$\boldsymbol{\Omega}$} }
\ \inf_{m\neq 0} 
D_{\widetilde{c} \cdot \varphi_{\gamma}}(m\cdot \mathds{Q},\mathds{P})
= \inf_{\mathbf{Q}\in A \cdot \textrm{$\boldsymbol{\Omega}$\hspace{-0.19cm}$\boldsymbol{\Omega}$} }
\ \inf_{\widetilde{m}\neq 0} 
D_{\widetilde{c} \cdot \varphi_{\gamma}}(\widetilde{m}\cdot \mathbf{Q},\mathds{P})
= -\lim_{n\rightarrow \infty }\frac{1}{n}\log \, 
\mathbb{\Pi}\negthinspace \left[A \cdot \boldsymbol{\xi}_{n}^{w\mathbf{W}}\in 
A \cdot \textrm{$\boldsymbol{\Omega}$\hspace{-0.23cm}$\boldsymbol{\Omega}$}\right]
\nonumber
\end{eqnarray}
for all sets $A \cdot \boldsymbol{\Omega}$\hspace{-0.23cm}$\boldsymbol{\Omega} \subset \widetilde{\mathcal{M}}_{\gamma}$ satisfying the 
regularity properties \eqref{regularity simplex} 
\textit{in the relative topology}. 
In particular, 
for each $\mathds{P} \in \mathbb{S}_{>0}^{K}$
the function 
$\Phi_{\mathds{P}}(\cdot) := F_{\gamma,\widetilde{c},A}\Big(D_{\widetilde{c} \cdot \varphi_{\gamma}}(\cdot,\mathds{P})\Big)$
is bare-simulation minimizable (BS-minimizable)
in the narrow sense 
(cf. \eqref{brostu5:fo.2} in Definition \ref{brostu5:def.1}
and the special case of Remark \ref{brostu5:rem.def1}(a)) 
on all sets 
$A \cdot \textrm{$\boldsymbol{\Omega}$\hspace{-0.23cm}$\boldsymbol{\Omega}$} \subset \widetilde{\mathcal{M}}_{\gamma}$ 
satisfying 
\eqref{regularity simplex}
\textit{in the relative topology}.\\
(b) Moreover, there holds 
\begin{align}
& \inf_{\mathbf{Q}\in A \cdot \textrm{$\boldsymbol{\Omega}$\hspace{-0.19cm}$\boldsymbol{\Omega}$} }
D_{\widetilde{c} \cdot \varphi_{\gamma}}(\mathbf{Q},\mathds{P})
= F_{\gamma,\widetilde{c},A}^{\leftarrow}\Big(
-\lim_{n\rightarrow \infty }\frac{1}{n}\log \, 
\mathbb{\Pi}\negthinspace \left[\boldsymbol{\xi}_{n}^{w\mathbf{W}}\in 
\textrm{$\boldsymbol{\Omega}$\hspace{-0.23cm}$\boldsymbol{\Omega}$}\right]\Big)
\label{LDP Normalized Vec simplex 2}
\end{align}
for all sets $A \cdot \boldsymbol{\Omega}$\hspace{-0.23cm}$\boldsymbol{\Omega} \subset \widetilde{\mathcal{M}}_{\gamma}$ satisfying the 
regularity properties \eqref{regularity simplex} 
\textit{in the relative topology}.
In particular, 
for each $\mathds{P} \in \mathbb{S}_{>0}^{K}$
the function 
$\Phi_{\mathds{P}}(\cdot) := D_{\widetilde{c} \cdot \varphi_{\gamma}}(\cdot,\mathds{P})$
is bare-simulation minimizable (BS-minimizable)
in the narrow sense on all sets 
$A \cdot \textrm{$\boldsymbol{\Omega}$\hspace{-0.23cm}$\boldsymbol{\Omega}$} \subset \widetilde{\mathcal{M}}_{\gamma}$ 
satisfying 
\eqref{regularity simplex}
\textit{in the relative topology}.  

\end{theorem}

\noindent
\begin{remark}
(a) \, By straightforward rescaling (cf. Corollary \ref{brostu3:cor.divnormW.new.det} above), 
for $\mathbf{P} \in \mathbb{R}_{> 0}^{K}$ with 
$M_{\mathbf{P}} :=\sum_{i=1}^{K}p_{i}$, 
$\widetilde{c} \in \, ]0,\infty[$, $\gamma \in \mathbb{R}\backslash\, ]1,2[$
and $A \in \, ]0,\infty[$
one can generalize \eqref{LDP Normalized Vec simplex 2}
to \begin{align}
& \inf_{\mathbf{Q}\in A \cdot \textrm{$\boldsymbol{\Omega}$\hspace{-0.19cm}$\boldsymbol{\Omega}$} }
D_{\widetilde{c} \cdot \varphi_{\gamma}}(\mathbf{Q},\mathbf{P})
= F_{\gamma,\breve{c},\breve{A}}^{\leftarrow}\Big(
-\lim_{n\rightarrow \infty }\frac{1}{n}\log \, 
\mathbb{\Pi}\negthinspace \left[\boldsymbol{\xi}_{n}^{w\widetilde{\mathbf{W}}}\in 
\textrm{$\boldsymbol{\Omega}$\hspace{-0.23cm}$\boldsymbol{\Omega}$}\right]\Big)
\label{LDP Normalized Vec simplex 2 var}
\end{align}
for all sets $A \cdot \boldsymbol{\Omega}$\hspace{-0.23cm}$\boldsymbol{\Omega} \subset \widetilde{\mathcal{M}}_{\gamma}$ satisfying the 
regularity properties \eqref{regularity simplex} 
\textit{in the relative topology}; here, 
$\breve{c}:= M_{\mathbf{P}} \cdot \widetilde{c}$, 
$\breve{A}:= \frac{A}{M_{\mathbf{P}}}$, and
$(\widetilde{W}_{i})_{i \in  \mathbb{N}}$  is a family
of independent and identically distributed $\mathbb{R}-$valued random variables
with probability distribution $\widetilde{\mathbb{\bbzeta}}[ \cdot \, ] := \mathbb{\Pi}[\widetilde{W}_{1} \in \cdot \, ]$
being connected with the divergence generator $\varphi :=  \widetilde{c} \cdot \varphi_{\gamma}(\cdot) 
$ 
via the representability 
\eqref{brostu5:fo.link.var}.\\
(b) \, 
For the case $\gamma =2$ in Theorem \ref{brostu5:thm.divnormW.new} 
as well as in (a) of this remark, one can even take $A \in \mathbb{R}\backslash\{0\}$
instead of $A \in \, ]0,\infty[$.\\ 
(c) \ For applications of Theorem \ref{brostu5:thm.divnormW.new}
to fuzzy sets and basic belief assignments, see Broniatowski \& Stummer \cite{Bro:23c}.

\end{remark}

\vspace{0.4cm} 
\noindent
Analogously to \eqref{LDP Minimization},
the limit statement \eqref{LDP Normalized Vec simplex 2 var}
provides the principle for the approximation of the solution of 
the minimization problem 
$\Phi_{\mathbf{P}}(\mathbf{\Omega}) := 
\inf_{\mathbf{Q}\in A \cdot \textrm{$\boldsymbol{\Omega}$\hspace{-0.19cm}$\boldsymbol{\Omega}$} }
D_{\widetilde{c} \cdot \varphi_{\gamma}}(\mathbf{Q},\mathbf{P})$.
Indeed,
by replacing the right-hand side in \eqref{LDP Normalized Vec simplex 2 var} by its finite
counterpart, we deduce for given large $n$  
\begin{equation}
F_{\gamma,\breve{c},\breve{A}}^{\leftarrow}\Big(
- \frac{1}{n}\log \, 
\mathbb{\Pi}\negthinspace \left[\boldsymbol{\xi}_{n}^{w\widetilde{\mathbf{W}}}\in 
\textrm{$\boldsymbol{\Omega}$\hspace{-0.23cm}$\boldsymbol{\Omega}$}\right]\Big)
\ \approx \ \inf_{\mathbf{Q}\in A \cdot \textrm{$\boldsymbol{\Omega}$\hspace{-0.19cm}$\boldsymbol{\Omega}$} }
D_{\widetilde{c} \cdot \varphi_{\gamma}}(\mathbf{Q},\mathbf{P});
\label{fo.approx.1.simplex} 
\end{equation}
it remains to estimate the left-hand side of \eqref{fo.approx.1.simplex}
(see Section \ref{SectEstimators.new.det.simplex}
below, where the latter also provides estimates of the \textit{minimizers}).


\section{
Deterministic Narrow-Sense Bare-Simulation-Optimization of Bregman Distances
with Constant-Component-Sum Side Constraint}
\label{SectDetSubsimplex.SBD}

In the previous Section \ref{SectDetSubsimplex.CASM}, we have recalled/summarized recently achieved 
(cf. Broniatowski \& Stummer \cite{Bro:23a}) \textit{narrow-sense}
bare-simulation minimization results 
on CASM $\varphi-$divergences $\mathbf{Q} \mapsto D_{\varphi}(\mathbf{Q},\mathbf{P})$
under the \textit{additional constraints} that 
all the components $q_{k}$ are nonnegative/strictly positive
and that $\sum_{k=1}^{K} q_{k} =A$ for some (say) $A >0$,
where we have mostly concentrated on the explicitly solvable subcase 
$\varphi = \widetilde{c} \cdot \varphi_{\gamma}$ ($\widetilde{c} \in ]0,\infty[$, 
$\gamma \in \mathbb{R}\backslash]1,2[$). On the other hand, independently of any constraints, 
we have seen in Section \ref{SectDetNarrow Bregman} that the $\varphi-$divergences 
generalize to the scaled Bregman distances (cf. \eqref{brostu5:fo.SBD.smooth}) 
$\mathbf{Q} \mapsto D_{\varphi,\mathbf{P}}^{SBD}(\mathbf{Q},\mathbf{Q}^{\ast\ast})$
(recall that $D_{\varphi,\mathbf{P}}^{SBD}(\mathbf{Q},\mathbf{P})
= D_{\varphi}(\mathbf{Q},\mathbf{P})$). In the following, we work out
how the results of Section \ref{SectDetSubsimplex.CASM} generalize
to the bare-simulation minimization of these scaled Bregman distances 
under the above-mentioned additional constraints. 
We first achieve

\vspace{0.3cm}

\begin{theorem}
\label{brostu3:thm.divnormW.new.det.SBD} 
Let $\mathds{P} \in \mathbb{S}_{> 0}^{K}$, 
and suppose that the divergence generator $\varphi$ satisfies Condition 
\ref{Condition  Fi Tilda in Minimization simplex}.
Additionally, let $\mathbf{Q}^{\ast\ast} \in \mathbb{R}^{K}$ such that 
\eqref{brostu5:fo.SBD.qstarstar} holds.
Moreover, we assume that $\boldsymbol{\Omega}$\hspace{-0.23cm}$\boldsymbol{\Omega} \subset \mathbb{S}^{K}$ 
satisfies the regularity properties \eqref{regularity simplex} in the relative topology as well as
the finiteness property
\begin{equation}
\inf_{\mathds{Q}\in \textrm{$\boldsymbol{\Omega}$\hspace{-0.19cm}$\boldsymbol{\Omega}$}}
D_{\varphi,\mathds{P}}^{SBD}(\mathds{Q},\mathbf{Q}^{\ast\ast}) < \infty .
\label{def fi wrt Omega Bregman simplex}
\end{equation}
Additionally, let $V := (\mathbf{V}_{n})_{n \in \mathbb{N}}$ be a sequence of random vectors 
constructed via \eqref{brostu5:V_new} and \eqref{brostu5:Utilde_k_new} (where we
write $V$ instead of $\widetilde{V}$, since $M_{\mathds{P}}=1$ and thus $\widetilde{p}_{k}=
p_{k}/M_{\mathds{P}} = p_{k}$ as well as $\widetilde{\mathbb{\bbzeta}}=\mathbb{\bbzeta}$).
Then, in terms of 
the random vectors $\boldsymbol{\xi}_{n}^{w\mathbf{V}}$ given by
\begin{eqnarray}
\boldsymbol{\xi}_{n}^{w\mathbf{V}} &:=&
\begin{cases}
\left(\frac{\sum_{i \in I_{1}^{(n)}}V_{i}}{\sum_{k=1}^{K}\sum_{i \in I_{k}^{(n)}}V_{i}},
\ldots, \frac{\sum_{i \in I_{K}^{(n)}}V_{i}}{\sum_{k=1}^{K}\sum_{i \in I_{k}^{(n)}}V_{i}} \right) ,
\qquad \textrm{if } \sum_{j=1}^{n} V_{j} \ne 0, \\
\ (\infty, \ldots, \infty) =: \boldsymbol{\infty}, \hspace{3.8cm} \textrm{if } \sum_{j=1}^{n} V_{j} = 0,
\end{cases}
\label{brostu5:fo.norweiemp.vec.det.SBD} 
\end{eqnarray}
there holds 
\begin{align}
& \inf_{\mathds{Q}\in \textrm{$\boldsymbol{\Omega}$\hspace{-0.19cm}$\boldsymbol{\Omega}$} }
\ \inf_{m\neq 0} 
D_{\varphi,\mathds{P}}^{SBD}(m \cdot \mathds{Q},\mathbf{Q}^{\ast\ast}) 
= 
\inf_{m\neq 0}\ \inf_{\mathds{Q}\in \textrm{$\boldsymbol{\Omega}$\hspace{-0.19cm}$\boldsymbol{\Omega}$} }
D_{\varphi,\mathds{P}}^{SBD}(m \cdot \mathds{Q},\mathbf{Q}^{\ast\ast})  
= -\lim_{n\rightarrow \infty }\frac{1}{n}\log \, 
\mathbb{\Pi}\negthinspace \left[\boldsymbol{\xi}_{n}^{w\mathbf{V}}\in 
\textrm{$\boldsymbol{\Omega}$\hspace{-0.23cm}$\boldsymbol{\Omega}$}\right].
\label{LDP Normalized Vec BS2 SBD}
\end{align}
\end{theorem}

\vspace{0.3cm}
\noindent
The proof of Theorem \ref{brostu3:thm.divnormW.new.det.SBD}
will be given in Appendix \ref{App.A} below.

\vspace{0.3cm}
\noindent
From Theorem \ref{brostu3:thm.divnormW.new.det.SBD} and with the help of 
\eqref{brostu5:fo.SBD.smooth.equality}, one gets immediately 

\vspace{0.2cm}

\begin{corollary}
\label{brostu3:cor.divnormW.new.det.SBD} 
Let $A \in \mathbb{R}$, 
$\mathbf{P} \in \mathbb{R}_{> 0}^{K}$ with $M_{\mathbf{P}} : =\sum_{i=1}^{K}p_{i}>0$,
and suppose that the divergence generator $\varphi$ satisfies Condition 
\ref{Condition  Fi Tilda in Minimization}.
Additionally, let $\mathbf{Q}^{\ast\ast} \in \mathbb{R}^{K}$ such that 
\eqref{brostu5:fo.SBD.qstarstar} holds.
Moreover, we assume that $\boldsymbol{\Omega}$\hspace{-0.23cm}$\boldsymbol{\Omega} \subset \mathbb{S}^{K}$ 
satisfies the regularity properties \eqref{regularity simplex} in the relative topology as well as
the finiteness property
\begin{equation}
\inf_{\mathbf{Q}\in A \cdot \textrm{$\boldsymbol{\Omega}$\hspace{-0.19cm}$\boldsymbol{\Omega}$}}
D_{\varphi,\mathbf{P}}^{SBD}(\mathbf{Q},\mathbf{Q}^{\ast\ast}) < \infty .
\nonumber
\end{equation}
Additionally, let $\widetilde{V} := (\mathbf{\widetilde{V}}_{n})_{n \in \mathbb{N}}$ be a sequence of random vectors 
constructed via \eqref{brostu5:V_new} and \eqref{brostu5:Utilde_k_new}.
Then there holds (in terms of \eqref{brostu5:fo.norweiemp.vec.det.SBD} with 
$\widetilde{V}$ instead of $V$)
\begin{align}
& \inf_{\mathbf{Q}\in A \cdot \textrm{$\boldsymbol{\Omega}$\hspace{-0.19cm}$\boldsymbol{\Omega}$} }
\ \inf_{m\neq 0} 
D_{\varphi,\mathbf{P}}^{SBD}(m \cdot \mathbf{Q},\mathbf{Q}^{\ast\ast}) 
= \inf_{m\neq 0}\ \inf_{\mathbf{Q}\in A \cdot \textrm{$\boldsymbol{\Omega}$\hspace{-0.19cm}$\boldsymbol{\Omega}$} }
D_{\varphi,\mathbf{P}}^{SBD}(m \cdot \mathbf{Q},\mathbf{Q}^{\ast\ast}) 
= -\lim_{n\rightarrow \infty }\frac{1}{n}\log \, 
\mathbb{\Pi}\negthinspace \left[\boldsymbol{\xi}_{n}^{w\mathbf{\widetilde{V}}}\in 
\textrm{$\boldsymbol{\Omega}$\hspace{-0.23cm}$\boldsymbol{\Omega}$}\right] .
\label{LDP Normalized Vec BS2 nonnormalized SBD 2}
\end{align}

\end{corollary}

\vspace{0.2cm} 
\noindent
\begin{remark}
\label{m-minimization SBD}
(a) \textit{Analogously} to Remark \ref{m-minimization CASM}(a), the involved 
\textquotedblleft  inner\textquotedblright\ $m-$minimizations on the
left-hand side of \eqref{LDP Normalized Vec BS2 SBD} 
(and analogously, of \eqref{LDP Normalized Vec BS2 nonnormalized SBD 2}) can be solved explicitly in
the important special case of the scaled Bregman power distances \eqref{brostu5:fo.scaledBregpow}.\\
(b) \textit{Contrary} to Remark \ref{m-minimization CASM}(b),
the outcomes of (a) can generally \textit{NOT} be rewritten in terms of invertible
functions $F(\cdot)$ of the divergences 
$D_{\widetilde{c} \cdot \varphi_{\gamma},\mathds{P}}^{SBD}(\mathds{Q},\mathbf{Q}^{\ast\ast})$
(and analogously of 
$D_{\widetilde{c} \cdot \varphi_{\gamma},\mathbf{P}}^{SBD}(\mathbf{Q},\mathbf{Q}^{\ast\ast})$). 
This means that we can generally \textit{NOT} use our \textit{narrow-sense} bare-simulation procedure
for the minimization of 
$\mathds{Q} \mapsto D_{\widetilde{c} \cdot \varphi_{\gamma},\mathds{P}}^{SBD}(\mathds{Q},\mathbf{Q}^{\ast\ast})
=:\Phi_{\mathds{P}}(\mathds{Q})$
(respectively $\mathbf{Q} \mapsto D_{\widetilde{c} \cdot \varphi_{\gamma},\mathbf{P}}^{SBD}(\mathbf{Q},\mathbf{Q}^{\ast\ast})
=:\Phi_{\mathbf{P}}(\mathbf{Q})$). However, we are able to employ our bare-simulation procedure in the (non-narrow-)sense of \eqref{brostu5:fo.2}
for these problems (by applying the below-mentioned Theorem \ref{brostu5:thm.Fmin.simplex}
or alternatively, Theorem \ref{brostu5:thm.Fmin.simplex.SBD} and Remark \ref{rem.original}.\\
(c) \, For the special case $\mathds{P} = \mathds{Q}^{\ast\ast}$
(respectively $\mathbf{P} = \mathbf{Q}^{\ast\ast}$),
the Theorem \ref{brostu3:thm.divnormW.new.det.SBD} collapses to Theorem \ref{brostu3:thm.divnormW.new.det}
(respectively, Corollary \ref{brostu3:cor.divnormW.new.det.SBD} collapses to
Corollary \ref{brostu3:cor.divnormW.new.det}).\\
(d) \ Notice that the quantity 
$\breve{D}_{\varphi,\mathbf{P}}^{SBD}(\mathbf{Q},\mathbf{Q}^{\ast\ast})
:= \inf_{m\neq 0} D_{\varphi,\mathbf{P}}^{SBD}(m \cdot \mathbf{Q},\mathbf{Q}^{\ast\ast})$
satisfies the axioms of a divergence, that is,
$\breve{D}_{\varphi,\mathbf{P}}^{SBD}(\mathbf{Q},\mathbf{Q}^{\ast\ast}) \geq 0$, as well as 
$\breve{D}_{\varphi,\mathbf{P}}^{SBD}(\mathbf{Q},\mathbf{Q}^{\ast\ast}) = 0$ if
and only if $\mathbb{Q} = \mathbf{Q}^{\ast\ast}$ (reflexivity). 
Hence, \eqref{LDP Normalized Vec BS2 nonnormalized SBD 2} reflects a corresponding
narrow-sense bare-simulation for this.

\end{remark}

\vspace{0.3cm}
\noindent
In the following, we illuminate the details of the previous Remark \ref{m-minimization SBD}(a),(b).
The required representability \eqref{brostu5:fo.link.var.simplex} respectively
\eqref{brostu5:fo.link.var}
is satisfied for all the generators
$\varphi(\cdot) := \widetilde{c} \cdot \varphi_{\gamma}(\cdot)$ of \eqref{brostu5:fo.powdivgen}
with $\widetilde{c} \in \, ]0,\infty[$ and $\gamma \in \mathbb{R}\backslash\, ]1,2[$
(cf. Broniatowski \& Stummer~\cite{Bro:23a}); the corresponding crucial
probability laws $\bbzeta$ (respectively $\widetilde{\bbzeta}$ by taking
$\widetilde{c} \cdot M_{\mathbf{P}}$ instead of $\widetilde{c}$)
can be found in Table 1.
Moreover, we fix $A \in \, ]0,\infty[$, $\mathbf{P} \in \mathbb{R}_{>0}^{K}$
and employ the following auxiliary notations:

\vspace{0.2cm}
\noindent

\begin{enumerate}

\item[(T1)]
recall from (S1) that $\widetilde{\mathcal{M}}_{\gamma} := A \cdot \mathbb{S}_{>0}^{K}$
for $\gamma \in \, ]-\infty ,0]$, 
respectively, $\widetilde{\mathcal{M}}_{\gamma} := A \cdot \mathbb{S}^{K}$ for 
$\gamma \in \, ]0,1] \cup [2,\infty[$ and also set 
$\widetilde{\mathcal{M}}_{\gamma} := A \cdot \mathbb{S}^{K}$ for 
$\gamma \in \, ]1,2[$ (the latter will be mainly employed for the purposes of the next theorem only),
$\widetilde{\mathcal{N}}_{\gamma} := \mathbb{R}_{>0}^{K}$
for $\gamma \in \, \mathbb{R}^{K}\backslash\{2\}$, 
respectively, $\widetilde{\mathcal{N}}_{2} := \mathbb{R}$
(for consistency of $\mathbf{Q}^{\ast\ast}$ in \eqref{brostu5:fo.scaledBregpow}
and $\mathbf{Q}^{\ast\ast}$ in \eqref{brostu5:fo.triplepowersum}).

\item[(T2)] the \textit{modified (discrete) $\gamma-$order Hellinger integral of $\mathbf{Q}$ and $\mathbf{P}$}
given by 
\begin{equation}
0 < H_{\gamma}(\mathbf{Q},\mathbf{P}) := \sum\displaylimits_{k=1}^{K} (q_{k})^{\gamma} \cdot (p_{k})^{1-\gamma},
\qquad \gamma \in \mathbb{R}\backslash\{0,1\}, \ \mathbf{Q}\in \widetilde{\mathcal{M}}_{\gamma};
\label{brostu5:fo.divpow.hellinger1.extended}
\end{equation}

\item[(T3)] the \textit{modified (discrete) $\gamma-$order Hellinger integral of $\mathbf{Q}^{\ast\ast}$
and $\mathbf{P}$} given by 
\begin{equation}
H_{\gamma}(\mathbf{Q}^{\ast\ast},\mathbf{P}) = 
\sum\displaylimits_{k=1}^{K} (q_{k}^{\ast\ast})^{\gamma} \cdot (p_{k})^{1-\gamma}, 
\qquad \gamma \in \mathbb{R}\backslash\{0,1\}, \ \mathbf{Q}^{\ast\ast}\in \widetilde{\mathcal{N}}_{\gamma};
\label{brostu5:fo.divpow.hellinger1.extended2}
\end{equation}

\item[(T4)] the \textit{$\gamma-$order triple power sum of $\mathbf{Q}$, $\mathbf{Q}^{\ast\ast}$ and $\mathbf{P}$}
given by 
\begin{equation}
T_{\gamma}(\mathbf{Q},\mathbf{Q}^{\ast\ast},\mathbf{P}) := 
\sum\displaylimits_{k=1}^{K} q_{k} \cdot  (q_{k}^{\ast\ast})^{\gamma-1} \cdot (p_{k})^{1-\gamma}, 
\qquad \gamma \in \mathbb{R}\backslash\{1\}, \ \mathbf{Q}\in \mathbb{R}^{K}, \ 
\mathbf{Q}^{\ast\ast} \in \widetilde{\mathcal{N}}_{\gamma};
\label{brostu5:fo.triplepowersum}
\end{equation}

\item[(T5)] 
the \textit{modified Kullback-Leibler information distance (modified relative entropy)} given by
\begin{eqnarray}
I(\mathbf{Q},\mathbf{Q}^{\ast\ast}) :=  
\sum\displaylimits_{k=1}^{K} q_{k} \cdot \log\left( \frac{q_{k}}{q_{k}^{\ast\ast}} \right),
\qquad \mathbf{Q} \in \widetilde{\mathcal{M}}_{1} = A \cdot \mathbb{S}^{K}, \
\mathbf{Q}^{\ast\ast} \in \widetilde{\mathcal{N}}_{1} = \mathbb{R}_{>0}^{K};
\label{brostu3:fo.divpow.Kull1.SBD}
\end{eqnarray}

\item[(T6)] the \textit{logarithmic $0-$order triple power sum of 
$\mathbf{Q}$, $\mathbf{Q}^{\ast\ast}$ and $\mathbf{P}$}
given by 
\begin{equation}
\breve{T}_{0}(\mathbf{Q},\mathbf{Q}^{\ast\ast},\mathbf{P}) := 
- \sum\displaylimits_{k=1}^{K} p_{k} \cdot \log\left( \frac{q_{k}}{q_{k}^{\ast\ast}} \right),
\qquad
\mathbf{Q}\in \mathbb{R}_{>0}^{K}, \ 
\mathbf{Q}^{\ast\ast} \in \widetilde{\mathcal{N}}_{0} = \mathbb{R}_{>0}^{K}.
\label{brostu5:fo.triplepowersum.log}
\end{equation}

\end{enumerate}

\vspace{0.3cm}
\noindent
In terms of (T1) to (T6),   
we obtain the following assertions: 

\vspace{0.2cm}

\begin{theorem}
\label{theorem inner min Bregman power general}
\ Let $A \in \, ]0,\infty[$ and $\mathbf{P} \in \mathbb{R}_{>0}^{K}$ be arbitrarily fixed.\\ 
(a) Let  $\widetilde{c}\in \, ]0,\infty[$ be arbitrary,
$\gamma \in \, ]-\infty,0[ \, \cup \, ]0,1[ \, \cup \, ]1,\infty[$,  
$\mathbf{Q}\in \widetilde{\mathcal{M}}_{\gamma}$
and $\mathbf{Q}^{\ast\ast} \in\mathbb{R}_{>0}^{K}$.
Then one has \\[-0.4cm]
\begin{eqnarray}
\inf_{m\neq 0} D_{\widetilde{c} \cdot \varphi_{\gamma},\mathbf{P}}^{SBD}(m \cdot \mathbf{Q},\mathbf{Q}^{\ast\ast})
& = &  \frac{\widetilde{c}}{\gamma }
\cdot \left[ H_{\gamma}(\mathbf{Q}^{\ast\ast},\mathbf{P}) - 
T_{\gamma}(\mathbf{Q},\mathbf{Q}^{\ast\ast},\mathbf{P})^{\gamma/(\gamma-1)} \cdot 
H_{\gamma}(\mathbf{Q},\mathbf{P})^{-1/\left( \gamma -1\right) }\right]
\nonumber\\
&=:& \breve{D}_{\widetilde{c} \cdot \varphi_{\gamma},\mathbf{P}}^{SBD}( \mathbf{Q},\mathbf{Q}^{\ast\ast}),
\label{brostu3:fo.676b.SBD}
\end{eqnarray}
and consequently (cf. \eqref{LDP Normalized Vec BS2 nonnormalized SBD 2})
for any $\gamma \in \, ]-\infty,0[ \, \cup \, ]0,1[ \, \cup \, [2,\infty[$
and any subset 
$A \cdot \boldsymbol{\Omega}$\hspace{-0.23cm}$\boldsymbol{\Omega} \subset \widetilde{\mathcal{M}}_{\gamma}$ 
with \eqref{regularity simplex} one gets
\begin{align}
& \inf_{\mathbf{Q}\in A \cdot \textrm{$\boldsymbol{\Omega}$\hspace{-0.19cm}$\boldsymbol{\Omega}$} }
\frac{\widetilde{c}}{\gamma }
\cdot \left[ H_{\gamma}(\mathbf{Q}^{\ast\ast},\mathbf{P}) - 
T_{\gamma}(\mathbf{Q},\mathbf{Q}^{\ast\ast},\mathbf{P})^{\gamma/(\gamma-1)} \cdot 
H_{\gamma}(\mathbf{Q},\mathbf{P})^{-1/\left( \gamma -1\right) }\right]
\nonumber \\
& = \inf_{\mathbf{Q}\in A \cdot \textrm{$\boldsymbol{\Omega}$\hspace{-0.19cm}$\boldsymbol{\Omega}$} }
\breve{D}_{\widetilde{c} \cdot \varphi_{\gamma},\mathbf{P}}^{SBD}( \mathbf{Q},\mathbf{Q}^{\ast\ast})
= -\lim_{n\rightarrow \infty }\frac{1}{n}\log \, 
\mathbb{\Pi}\negthinspace \left[\boldsymbol{\xi}_{n}^{w\mathbf{\widetilde{V}}}\in 
\textrm{$\boldsymbol{\Omega}$\hspace{-0.23cm}$\boldsymbol{\Omega}$}\right] .
\label{LDP Normalized Vec BS2 nonnormalized SBD 2 simplex}
\end{align}
For the case $\gamma=2$ one can even allow for $A<0$ and $\mathbf{Q}^{\ast\ast} \in\mathbb{R}^{K}$.
\\[0.1cm]
\noindent (b) For any $\widetilde{c}>0$, $\mathbf{Q} \in 
\widetilde{\mathcal{M}}_{1} = A \cdot \mathbb{S}^{K}$
and $\mathbf{Q}^{\ast\ast} \in\mathbb{R}_{>0}^{K}$ 
 one gets 
\begin{eqnarray}
\inf_{m\neq 0} D_{\widetilde{c} \cdot \varphi_{1},\mathbf{P}}^{SBD}(m \cdot \mathbf{Q},\mathbf{Q}^{\ast\ast})
&=& 
\widetilde{c}\cdot \Big[ M_{\mathbf{Q}^{\ast\ast}} - A \cdot \exp
\Big( -\frac{1}{A}\cdot I(\mathbf{Q},\mathbf{Q}^{\ast\ast})\Big) \Big] 
=: \breve{D}_{\widetilde{c} \cdot \varphi_{1},\mathbf{P}}^{SBD}( \mathbf{Q},\mathbf{Q}^{\ast\ast}) 
\label{brostu3:fo.677a.SBD}
\end{eqnarray}

\vspace{-0.2cm}
\noindent
(which exceptionally does not depend on $\mathbf{P}$) and consequently 
(cf. \eqref{LDP Normalized Vec BS2 nonnormalized SBD 2}) for any subset 
$A \cdot \boldsymbol{\Omega}$\hspace{-0.23cm}$\boldsymbol{\Omega} \subset 
\widetilde{\mathcal{M}}_{1} = A \cdot \mathbb{S}^{K}$ with \eqref{regularity simplex} one has
\begin{align}
& \inf_{\mathbf{Q}\in A \cdot \textrm{$\boldsymbol{\Omega}$\hspace{-0.19cm}$\boldsymbol{\Omega}$} }
\widetilde{c}\cdot \Big[ M_{\mathbf{Q}^{\ast\ast}} - A \cdot \exp
\Big( -\frac{1}{A}\cdot I(\mathbf{Q},\mathbf{Q}^{\ast\ast})\Big) \Big] 
\nonumber \\
& = \inf_{\mathbf{Q}\in A \cdot \textrm{$\boldsymbol{\Omega}$\hspace{-0.19cm}$\boldsymbol{\Omega}$} }
\breve{D}_{\widetilde{c} \cdot \varphi_{1},\mathbf{P}}^{SBD}( \mathbf{Q},\mathbf{Q}^{\ast\ast})
= -\lim_{n\rightarrow \infty }\frac{1}{n}\log \, 
\mathbb{\Pi}\negthinspace \left[\boldsymbol{\xi}_{n}^{w\mathbf{\widetilde{V}}}\in 
\textrm{$\boldsymbol{\Omega}$\hspace{-0.23cm}$\boldsymbol{\Omega}$}\right] .
\label{LDP Normalized Vec BS2 nonnormalized SBD 2 simplex KL}
\end{align}
\noindent (c) For any $\widetilde{c}>0$, 
$\mathbf{Q} \in \widetilde{\mathcal{M}}_{0} = A \cdot \mathbb{S}_{>0}^{K}$
and $\mathbf{Q}^{\ast\ast} \in\mathbb{R}_{>0}^{K}$ 
we obtain
\begin{align}
&\inf_{m\neq 0} D_{\widetilde{c} \cdot \varphi_{0},\mathbf{P}}^{SBD}(m \cdot \mathbf{Q},\mathbf{Q}^{\ast\ast})
= \ \widetilde{c} \cdot \Big[ 
M_{\mathbf{P}} \cdot \log\Big(T_{0}(\mathbf{Q},\mathbf{Q}^{\ast\ast},\mathbf{P})\Big)
+ \breve{T}_{0}(\mathbf{Q},\mathbf{Q}^{\ast\ast},\mathbf{P})
-  M_{\mathbf{P}} \cdot \log( M_{\mathbf{P}}) \Big]
\ =: \ \breve{D}_{\widetilde{c} \cdot \varphi_{0},\mathbf{P}}^{SBD}( \mathbf{Q},\mathbf{Q}^{\ast\ast})
\label{brostu3:fo.678.SBD}
\end{align}
and consequently (cf. \eqref{LDP Normalized Vec BS2 nonnormalized SBD 2}) for any set subset
$A \cdot \boldsymbol{\Omega}$\hspace{-0.23cm}$\boldsymbol{\Omega} \subset 
\widetilde{\mathcal{M}}_{0} = A \cdot \mathbb{S}_{>0}^{K}$ 
with \eqref{regularity simplex} one gets
\begin{align}
& \inf_{\mathbf{Q}\in A \cdot \textrm{$\boldsymbol{\Omega}$\hspace{-0.19cm}$\boldsymbol{\Omega}$} }
\widetilde{c} \cdot \Big[ 
M_{\mathbf{P}} \cdot \log\Big(T_{0}(\mathbf{Q},\mathbf{Q}^{\ast\ast},\mathbf{P})\Big)
+ \breve{T}_{0}(\mathbf{Q},\mathbf{Q}^{\ast\ast},\mathbf{P})
-  M_{\mathbf{P}} \cdot \log( M_{\mathbf{P}}) \Big]
\nonumber \\
& = \inf_{\mathbf{Q}\in A \cdot \textrm{$\boldsymbol{\Omega}$\hspace{-0.19cm}$\boldsymbol{\Omega}$} }
\breve{D}_{\widetilde{c} \cdot \varphi_{0},\mathbf{P}}^{SBD}( \mathbf{Q},\mathbf{Q}^{\ast\ast})
= -\lim_{n\rightarrow \infty }\frac{1}{n}\log \, 
\mathbb{\Pi}\negthinspace \left[\boldsymbol{\xi}_{n}^{w\mathbf{\widetilde{V}}}\in 
\textrm{$\boldsymbol{\Omega}$\hspace{-0.23cm}$\boldsymbol{\Omega}$}\right] .
\label{LDP Normalized Vec BS2 nonnormalized SBD 2 simplex RKL}
\end{align}

\end{theorem}

\vspace{0.2cm}
\begin{remark} \ 
\label{after theorem inner min Bregman power general}
In accordance with Remark \ref{m-minimization SBD}(d), the quantity
$\breve{D}_{\widetilde{c} \cdot \varphi_{\gamma},\mathbf{P}}^{SBD}( \mathbf{Q},\mathbf{Q}^{\ast\ast})$
in Theorem \ref{theorem inner min Bregman power general}
can be regarded as (new class of) divergences between $\mathbf{Q}$ and $\mathbf{Q}^{\ast\ast}$.
Accordingly, we call 
$\breve{D}_{\widetilde{c} \cdot \varphi_{\gamma},\mathbf{P}}^{SBD}( \mathbf{Q},\mathbf{Q}^{\ast\ast})$
the \textit{inner-minimization-scaled-Bregman-distance} (in short, innmin-SBD),
which is BS-minimized --- in a narrow sense ---
by \eqref{LDP Normalized Vec BS2 nonnormalized SBD 2 simplex}
respectively \eqref{LDP Normalized Vec BS2 nonnormalized SBD 2 simplex KL}
respectively \eqref{LDP Normalized Vec BS2 nonnormalized SBD 2 simplex RKL}.
The BS-minimization --- in a wide sense --- of the
``original'' scaled Bregman divergence
$\mathbf{Q} \mapsto D_{\widetilde{c} \cdot \varphi_{\gamma},\mathbf{P}}^{SBD}( \mathbf{Q},\mathbf{Q}^{\ast\ast})$
will be treated in Remark \ref{rem.original} below.

\end{remark}

\vspace{0.4cm} 
\noindent
The proof of Theorem \ref{theorem inner min Bregman power general}
will be given in Appendix \ref{App.A}.

\vspace{0.5cm} 
\noindent
Analogously to \eqref{fo.approx.1.simplex},
the limit statements 
\eqref{LDP Normalized Vec BS2 nonnormalized SBD 2 simplex},
\eqref{LDP Normalized Vec BS2 nonnormalized SBD 2 simplex KL}
and \eqref{LDP Normalized Vec BS2 nonnormalized SBD 2 simplex RKL},
provide the principle for the approximation of the solution of 
the divergence minimization problem 
$\Phi_{\mathbf{P}}(\mathbf{\Omega}) := 
\inf_{\mathbf{Q}\in A \cdot \textrm{$\boldsymbol{\Omega}$\hspace{-0.19cm}$\boldsymbol{\Omega}$} }
\breve{D}_{\widetilde{c} \cdot \varphi_{\gamma},\mathbf{P}}^{SBD}( \mathbf{Q},\mathbf{Q}^{\ast\ast})$.
Indeed, by replacing the right-hand side in those
by their finite counterparts,
 we deduce for given large $n$  
\begin{equation}
- \frac{1}{n}\log \, 
\mathbb{\Pi}\negthinspace \left[\boldsymbol{\xi}_{n}^{w\widetilde{\mathbf{V}}}\in 
\textrm{$\boldsymbol{\Omega}$\hspace{-0.23cm}$\boldsymbol{\Omega}$}\right]
\ \approx \ \inf_{\mathbf{Q}\in A \cdot \textrm{$\boldsymbol{\Omega}$\hspace{-0.19cm}$\boldsymbol{\Omega}$} }
\breve{D}_{\widetilde{c} \cdot \varphi_{\gamma},\mathbf{P}}^{SBD}( \mathbf{Q},\mathbf{Q}^{\ast\ast});
\label{fo.approx.1.simplex.innminSBD} 
\end{equation}
it remains to estimate the left-hand side of \eqref{fo.approx.1.simplex}
(see Section \ref{SectEstimators.new.det.simplex}
below, where the latter also provides estimates of the \textit{minimizers}).

\vspace{0.3cm}
\noindent
Theorem \ref{theorem inner min Bregman power general} establishes 
the part (a) of our above-mentioned Remark \ref{m-minimization SBD}.
As far as the corresponding part (b) is concerned, it is clear 
that the involved functions
$\mathbf{Q} \mapsto \breve{D}_{\widetilde{c} \cdot \varphi_{\gamma},\mathbf{P}}^{SBD}( \mathbf{Q},\mathbf{Q}^{\ast\ast})$
can generally \textit{NOT} be rewritten in terms of $\mathbf{Q}-$independent invertible
functions $\breve{F}(\cdot)$ of the divergences 
$D_{\widetilde{c} \cdot \varphi_{\gamma},\mathbf{P}}^{SBD}(\mathbf{Q},\mathbf{Q}^{\ast\ast})$. 
This means that we can generally \textit{NOT} employ our \textit{narrow-sense} bare-simulation procedure
for the minimization of
$\mathbf{Q} \mapsto D_{\widetilde{c} \cdot \varphi_{\gamma},\mathbf{P}}^{SBD}(\mathbf{Q},\mathbf{Q}^{\ast\ast})$.
However, there are a few important exceptions given in the following 

\noindent
\begin{theorem}
\label{brostu5:thm.divnormW.new.SBD} 
Let $\mathbf{P} \in \mathbb{R}_{> 0}^{K}$, $\widetilde{c} \in \, ]0,\infty[$, 
$\gamma \in \mathbb{R}$
and $A \in \, ]0,\infty[$ be arbitrary but fixed.
Moreover, let $(W_{i})_{i \in  \mathbb{N}}$  be  a family
of independent and identically distributed $\mathbb{R}-$valued random variables
with probability distribution $\mathbb{\bbzeta}[ \cdot \, ] := \mathbb{\Pi}[W_{1} \in \cdot \, ]$
being connected with the divergence generator $\varphi :=  \widetilde{c} \cdot \varphi_{\gamma}(\cdot) \in \widetilde{\Upsilon}(]a,b[)$ 
via the representability \eqref{brostu5:fo.link.var.simplex}. \\
(a) If $\gamma \in \mathbb{R}\backslash\{1\}$, then for all $C \in ]0,\infty[$, $\widetilde{c} \in ]0,\infty[$ 
and $\mathbf{Q} \in \widetilde{\mathcal{M}}_{\gamma}$
there holds in the special subsetup $\mathbf{Q}^{\ast\ast} := C \cdot \mathbf{P}$
the representation
\begin{equation}
\breve{D}_{\widetilde{c} \cdot \varphi_{\gamma},\mathbf{P}}^{SBD}( \mathbf{Q},C \cdot \mathbf{P})
= \breve{F}_{\gamma,\widetilde{c},A,M_{\mathbf{P}},C}\Big(D_{\widetilde{c} 
\cdot \varphi_{\gamma},\mathbf{P}}^{SBD}(\mathbf{Q}, C \cdot \mathbf{P})
\Big)
\label{brostu5:fo.SBDrepres}
\end{equation}
with function
$\breve{F}_{\gamma,\widetilde{c},A,M_{\mathbf{P}},C}: \, 
]-\infty,\infty[ \, \mapsto \, ]-\infty,\infty]$ given by
\begin{eqnarray}
\hspace{-0.5cm}
\breve{F}_{\gamma,\widetilde{c},A,M_{\mathbf{P}},C}(x) &:=&
\begin{cases}
\frac{\widetilde{c} \cdot C^{\gamma}}{\gamma }
\cdot \big\{ M_{\mathbf{P}} - A^{\gamma/(\gamma-1)} \cdot 
\big[ C^{\gamma} \cdot  M_{\mathbf{P}}
+ \gamma \cdot C^{\gamma-1} \cdot (A- C \cdot  M_{\mathbf{P}}) 
+ \frac{\gamma \cdot \left( \gamma -1\right) }{\widetilde{c}} \cdot x \big]
^{-1/\left( \gamma -1\right) }\big\} ,\\
\hspace{1.0cm}
\textrm{if } \gamma \in \, ]-\infty,0[ \, \cup \, ]0,1[ \,
\ \textrm{and } \ C^{\gamma} \cdot  M_{\mathbf{P}}
+ \gamma \cdot C^{\gamma-1} \cdot (A- C \cdot  M_{\mathbf{P}}) 
+ \frac{\gamma \cdot \left( \gamma -1\right) }{\widetilde{c}} \cdot x \geq 0, \\
\hspace{1.0cm}
\textrm{or if } \gamma \in \, ]1,2[ \, \cup \, ]2,\infty[ \,
\ \textrm{and } \ C^{\gamma} \cdot  M_{\mathbf{P}}
+ \gamma \cdot C^{\gamma-1} \cdot (A- C \cdot  M_{\mathbf{P}}) 
+ \frac{\gamma \cdot \left( \gamma -1\right) }{\widetilde{c}} \cdot x > 0, \\
\frac{\widetilde{c} \cdot C^{2}}{2}
\cdot \big\{ 
M_{\mathbf{P}} - A^{2} \cdot 
\big[ C^{2} \cdot  M_{\mathbf{P}}
+ 2 \cdot C \cdot (A- C \cdot  M_{\mathbf{P}}) 
+ \frac{2}{\widetilde{c}} \cdot x 
\big]
^{-1}
\big\} ,\\
\hspace{6.7cm}
\textrm{if } \gamma =2 
\ \textrm{and } \ x \in \, ]\frac{\widetilde{c} \cdot C}{2} \cdot (C \cdot  M_{\mathbf{P}} -2 A), \infty[,\\
\widetilde{c} \cdot \Big(M_{\mathbf{P}} - \frac{A}{C} + M_{\mathbf{P}} \cdot \log\Big(\frac{A}{C}\Big) 
- M_{\mathbf{P}} \cdot \log (M_{\mathbf{P}})\Big) + x,
\hspace{1.0cm}
\textrm{if } \gamma =0 \ \textrm{and } \ x \in \, ]-\infty,\infty[,\\
\infty, \hspace{0.5cm}
\textrm{if } \gamma \in \, ]-\infty,0[ \, \cup \, ]0,1[ \,
\ \textrm{and } \ C^{\gamma} \cdot  M_{\mathbf{P}}
+ \gamma \cdot C^{\gamma-1} \cdot (A- C \cdot  M_{\mathbf{P}}) 
+ \frac{\gamma \cdot \left( \gamma -1\right) }{\widetilde{c}} \cdot x < 0, \\
\hspace{1.0cm}
\textrm{or if } \gamma \in \, ]1,2[ \, \cup \, ]2,\infty[ \,
\ \textrm{and } \ C^{\gamma} \cdot  M_{\mathbf{P}}
+ \gamma \cdot C^{\gamma-1} \cdot (A- C \cdot  M_{\mathbf{P}}) 
+ \frac{\gamma \cdot \left( \gamma -1\right) }{\widetilde{c}} \cdot x \leq 0,\\ 
\hspace{1.0cm}
\textrm{or if } \gamma =2 
\ \textrm{and } \ x \in \, ]-\infty,\frac{\widetilde{c} \cdot C}{2} \cdot (C \cdot  M_{\mathbf{P}} -2 A)],
\end{cases}
\label{brostu5:fo.divpow.F.SBD}
\end{eqnarray}
which on its effective domain 
$dom(\breve{F}_{\gamma,\widetilde{c},A,M_{\mathbf{P}},C})$
is strictly increasing with strictly increasing inverse function
\begin{eqnarray}
\hspace{-0.5cm}
\breve{F}_{\gamma,\widetilde{c},A,M_{\mathbf{P}},C}^{\leftarrow}(z) &:=&
\begin{cases}
\frac{\widetilde{c}}{\gamma \cdot (\gamma-1)}
\cdot \left\{
A^{\gamma} \cdot 
\left[ M_{\mathbf{P}} - \frac{\gamma}{\widetilde{c} \cdot C^{\gamma}} 
\cdot z \right]^{-\left( \gamma -1\right)}
- C^{\gamma} \cdot M_{\mathbf{P}} - \gamma \cdot C^{\gamma-1} \cdot (A-C \cdot M_{\mathbf{P}}) 
\right\} , \\
\hspace{6.1cm}
\textrm{if } \gamma \in \, ]-\infty,0[ \, \cup \, ]0,1[ \,
\ \textrm{and } \ \gamma \cdot z \leq \widetilde{c} \cdot C^{\gamma} \cdot M_{\mathbf{P}}, \\
\hspace{6.1cm}
\textrm{or if } \gamma \in \, ]1,2[ \, \cup \, ]2,\infty[ \,
\ \textrm{and } \ \gamma \cdot z < \widetilde{c} \cdot C^{\gamma} \cdot M_{\mathbf{P}}, \\
\frac{\widetilde{c}}{2}
\cdot \left\{
A^{2} \cdot 
\left[ M_{\mathbf{P}} - \frac{2}{\widetilde{c} \cdot C^{2}} 
\cdot z \right]^{-1}
- C^{2} \cdot M_{\mathbf{P}} - 2 \cdot C \cdot (A-C \cdot M_{\mathbf{P}}) 
\right\} , \\
\hspace{6.1cm}
\textrm{if } \gamma =2
\ \textrm{and } \ 2 \cdot z < \widetilde{c} \cdot C^{2} \cdot M_{\mathbf{P}}, \\
z - \widetilde{c} \cdot \Big(M_{\mathbf{P}} - \frac{A}{C} + M_{\mathbf{P}} \cdot \log\Big(\frac{A}{C}\Big) 
- M_{\mathbf{P}} \cdot \log (M_{\mathbf{P}})\Big),
\hspace{1.0cm}
\textrm{if } \gamma =0 \ \textrm{and } \ z \in \, ]-\infty,\infty[.\\
\end{cases}
\label{brostu5:fo.divpow.F.SBD.inv}
\end{eqnarray}
Consequently (cf. \eqref{LDP Normalized Vec BS2 nonnormalized SBD 2 simplex},
\eqref{LDP Normalized Vec BS2 nonnormalized SBD 2 simplex RKL}) 
for any $\gamma \in \mathbb{R}\backslash[1,2[$ and any subset
$A \cdot \boldsymbol{\Omega}$\hspace{-0.23cm}$\boldsymbol{\Omega} \subset 
\widetilde{\mathcal{M}}_{\gamma}$ 
with \eqref{regularity simplex} one gets
\begin{align}
& \inf_{\mathbf{Q}\in A \cdot \textrm{$\boldsymbol{\Omega}$\hspace{-0.19cm}$\boldsymbol{\Omega}$} }
\breve{F}_{\gamma,\widetilde{c},A,M_{\mathbf{P}},C}\Big(
D_{\widetilde{c} \cdot \varphi_{\gamma},\mathbf{P}}^{SBD}(\mathbf{Q}, C \cdot \mathbf{P})
\Big)
= -\lim_{n\rightarrow \infty }\frac{1}{n}\log \, 
\mathbb{\Pi}\negthinspace \left[\boldsymbol{\xi}_{n}^{w\mathbf{\widetilde{V}}}\in 
\textrm{$\boldsymbol{\Omega}$\hspace{-0.23cm}$\boldsymbol{\Omega}$}\right] 
\nonumber
\end{align}
and
\begin{align}
& \inf_{\mathbf{Q}\in A \cdot \textrm{$\boldsymbol{\Omega}$\hspace{-0.19cm}$\boldsymbol{\Omega}$} }
D_{\widetilde{c} \cdot \varphi_{\gamma},\mathbf{P}}^{SBD}(\mathbf{Q}, C \cdot \mathbf{P})
= \breve{F}_{\gamma,\widetilde{c},A,M_{\mathbf{P}},C}^{\leftarrow}\Big(
-\lim_{n\rightarrow \infty }\frac{1}{n}\log \, 
\mathbb{\Pi}\negthinspace \left[\boldsymbol{\xi}_{n}^{w\mathbf{\widetilde{V}}}\in 
\textrm{$\boldsymbol{\Omega}$\hspace{-0.23cm}$\boldsymbol{\Omega}$}\right]\Big) .
\label{LDP Normalized Vec simplex 2 SBD}
\end{align}
In particular, the functions 
$\Phi_{\mathbf{P}}(\cdot) :=
\breve{F}_{\gamma,\widetilde{c},A,M_{\mathbf{P}},C}\Big(
D_{\widetilde{c} \cdot \varphi_{\gamma},\mathbf{P}}^{SBD}(\cdot, C \cdot \mathbf{P})
\Big)$
and
$\Phi_{\mathbf{P}}(\cdot) :=
D_{\widetilde{c} \cdot \varphi_{\gamma},\mathbf{P}}^{SBD}(\cdot, C \cdot \mathbf{P})$
are bare-simulation minimizable (BS-minimizable)
in the narrow sense 
(cf. \eqref{brostu5:fo.2} in Definition \ref{brostu5:def.1}
and the special case of Remark \ref{brostu5:rem.def1}(a)) 
on all sets 
$A \cdot \textrm{$\boldsymbol{\Omega}$\hspace{-0.23cm}$\boldsymbol{\Omega}$} \subset \widetilde{\mathcal{M}}_{\gamma}$ 
satisfying 
\eqref{regularity simplex}
\textit{in the relative topology}.\\
(b) If $\gamma=1$, then for all $\widetilde{c} \in ]0,\infty[$, 
$\mathbf{Q}^{\ast\ast} \in \mathbb{R}_{>0}^{K}$
 and $\mathbf{Q} \in \widetilde{\mathcal{M}}_{1} = A \cdot \mathbb{S}^{K}$
there holds the representation
\begin{equation}
\breve{D}_{\widetilde{c} \cdot \varphi_{1},\mathbf{P}}^{SBD}( \mathbf{Q},\mathbf{Q}^{\ast\ast})
= \breve{F}_{1,\widetilde{c},A,M_{\mathbf{Q}^{\ast\ast}}}\Big(D_{\widetilde{c} 
\cdot \varphi_{1},\mathbf{P}}^{SBD}(\mathbf{Q}, \mathbf{Q}^{\ast\ast})
\Big)
\label{brostu5:fo.SBDrepres.gamma1}
\end{equation}
with strictly increasing function
$\breve{F}_{1,\widetilde{c},A,M_{\mathbf{Q}^{\ast\ast}}}: \, 
]-\infty,\infty[ \, \mapsto \, ]-\infty,\infty[$ given by
\begin{equation}
\breve{F}_{1,\widetilde{c},A,M_{\mathbf{Q}^{\ast\ast}}}(x) \ := \ 
\widetilde{c} \cdot \Big\{ M_{\mathbf{Q}^{\ast\ast}} - A \cdot 
\exp\Big( \frac{M_{\mathbf{Q}^{\ast\ast}}}{A} - 1 - \frac{x}{A \cdot \widetilde{c}} \Big)
\Big\}, \qquad x \in \, ]-\infty,\infty[,
\label{brostu5:fo.divpow.F.SBD.gamma1}
\end{equation}
having strictly increasing inverse function
\begin{equation}
\breve{F}_{1,\widetilde{c},A,M_{\mathbf{Q}^{\ast\ast}}}^{\leftarrow}(z) \ := \ 
\widetilde{c} \cdot \left\{
M_{\mathbf{Q}^{\ast\ast}} - A - A \cdot \left[\log\left(M_{\mathbf{Q}^{\ast\ast}}
-\frac{z}{\widetilde{c}}\right)  - \log A \right] 
\right\} ,
\qquad \textrm{for all } z  \in \, ]-\infty, \widetilde{c} \cdot M_{\mathbf{Q}^{\ast\ast}}[.
\label{brostu5:fo.SBDrepres.gamma1.inverse}
\end{equation}
Consequently (cf. \eqref{LDP Normalized Vec BS2 nonnormalized SBD 2 simplex KL}) for any set subset
$A \cdot \boldsymbol{\Omega}$\hspace{-0.23cm}$\boldsymbol{\Omega} \subset 
\widetilde{\mathcal{M}}_{1}$ 
with \eqref{regularity simplex} one gets
\begin{align}
& \inf_{\mathbf{Q}\in A \cdot \textrm{$\boldsymbol{\Omega}$\hspace{-0.19cm}$\boldsymbol{\Omega}$} }
\breve{F}_{1,\widetilde{c},A,M_{\mathbf{Q}^{\ast\ast}}}\Big(
D_{\widetilde{c} \cdot \varphi_{1},\mathbf{P}}^{SBD}(\mathbf{Q}, \mathbf{Q}^{\ast\ast})
\Big)
= -\lim_{n\rightarrow \infty }\frac{1}{n}\log \, 
\mathbb{\Pi}\negthinspace \left[\boldsymbol{\xi}_{n}^{w\mathbf{\widetilde{V}}}\in 
\textrm{$\boldsymbol{\Omega}$\hspace{-0.23cm}$\boldsymbol{\Omega}$}\right] 
\nonumber
\end{align}
and
\begin{align}
& \inf_{\mathbf{Q}\in A \cdot \textrm{$\boldsymbol{\Omega}$\hspace{-0.19cm}$\boldsymbol{\Omega}$} }
D_{\widetilde{c} \cdot \varphi_{1},\mathbf{P}}^{SBD}(\mathbf{Q}, \mathbf{Q}^{\ast\ast})
= \breve{F}_{1,\widetilde{c},A,M_{\mathbf{Q}^{\ast\ast}}}^{\leftarrow}\Big(
-\lim_{n\rightarrow \infty }\frac{1}{n}\log \, 
\mathbb{\Pi}\negthinspace \left[\boldsymbol{\xi}_{n}^{w\mathbf{\widetilde{V}}}\in 
\textrm{$\boldsymbol{\Omega}$\hspace{-0.23cm}$\boldsymbol{\Omega}$}\right]\Big) .
\label{LDP Normalized Vec simplex 2 SBD gamma1}
\end{align}
In particular, the functions 
$\Phi_{\mathbf{P}}(\cdot) :=
\breve{F}_{1,\widetilde{c},A,M_{\mathbf{Q}^{\ast\ast}}}\Big(
D_{\widetilde{c} \cdot \varphi_{1},\mathbf{P}}^{SBD}(\cdot, \mathbf{Q}^{\ast\ast})
\Big)$
and
$\Phi_{\mathbf{P}}(\cdot) :=
D_{\widetilde{c} \cdot \varphi_{1},\mathbf{P}}^{SBD}(\cdot, \mathbf{Q}^{\ast\ast})$
are bare-simulation minimizable (BS-minimizable)
in the narrow sense 
(cf. \eqref{brostu5:fo.2} in Definition \ref{brostu5:def.1}
and the special case of Remark \ref{brostu5:rem.def1}(a)) 
on all sets 
$A \cdot \textrm{$\boldsymbol{\Omega}$\hspace{-0.23cm}$\boldsymbol{\Omega}$} \subset \widetilde{\mathcal{M}}_{1}$ 
satisfying 
\eqref{regularity simplex}
\textit{in the relative topology}.

\end{theorem}

\vspace{0.4cm}
\noindent
The assertions of Theorem \ref{brostu5:thm.divnormW.new.SBD} 
follow by straightforward caluclations from Theorem \ref{theorem inner min Bregman power general}.
For the special case $C=1$, Theorem \ref{brostu5:thm.divnormW.new.SBD} collapses to
Theorem \ref{brostu5:thm.divnormW.new}.

\vspace{0.5cm} 
\noindent
Analogously to \eqref{fo.approx.1.simplex},
the limit statements 
\eqref{LDP Normalized Vec simplex 2 SBD}
and \eqref{LDP Normalized Vec simplex 2 SBD gamma1}
provide the principle for the approximation of the solution of the
minimization problems 
$\Phi_{\mathbf{P}}(\mathbf{\Omega}) := 
\inf_{\mathbf{Q}\in A \cdot \textrm{$\boldsymbol{\Omega}$\hspace{-0.19cm}$\boldsymbol{\Omega}$} }
D_{\widetilde{c} \cdot \varphi_{\gamma},\mathbf{P}}^{SBD}(\mathbf{Q}, C \cdot \mathbf{P})$
(for $\gamma \in \mathbb{R}\backslash[1,2[$)
and 
$\inf_{\mathbf{Q}\in A \cdot \textrm{$\boldsymbol{\Omega}$\hspace{-0.19cm}$\boldsymbol{\Omega}$} }
D_{\widetilde{c} \cdot \varphi_{1},\mathbf{P}}^{SBD}(\mathbf{Q}, \mathbf{Q}^{\ast\ast})$
(for $\gamma=1$).
Indeed, we replace the right-hand side in those
by their finite counterparts,
and accordingly obtain for given large $n$  
\begin{eqnarray}
& & \breve{F}_{\gamma,\widetilde{c},A,M_{\mathbf{P}},C}^{\leftarrow}\Big(
-\frac{1}{n}\log \, 
\mathbb{\Pi}\negthinspace \left[\boldsymbol{\xi}_{n}^{w\mathbf{\widetilde{V}}}\in 
\textrm{$\boldsymbol{\Omega}$\hspace{-0.23cm}$\boldsymbol{\Omega}$}\right]\Big)
\ \approx \ 
\inf_{\mathbf{Q}\in A \cdot \textrm{$\boldsymbol{\Omega}$\hspace{-0.19cm}$\boldsymbol{\Omega}$} }
D_{\widetilde{c} \cdot \varphi_{\gamma},\mathbf{P}}^{SBD}(\mathbf{Q}, C \cdot \mathbf{P}),
\qquad \textrm{for } \gamma \in \mathbb{R}\backslash[1,2[,
\label{fo.approx.1.simplex.SBD.1} \\
& &
\breve{F}_{1,\widetilde{c},A,M_{\mathbf{Q}^{\ast\ast}}}^{\leftarrow}\Big(
- \frac{1}{n}\log \, 
\mathbb{\Pi}\negthinspace \left[\boldsymbol{\xi}_{n}^{w\mathbf{\widetilde{V}}}\in 
\textrm{$\boldsymbol{\Omega}$\hspace{-0.23cm}$\boldsymbol{\Omega}$}\right]\Big)
\ \approx \ 
\inf_{\mathbf{Q}\in A \cdot \textrm{$\boldsymbol{\Omega}$\hspace{-0.19cm}$\boldsymbol{\Omega}$} }
D_{\widetilde{c} \cdot \varphi_{1},\mathbf{P}}^{SBD}(\mathbf{Q}, \mathbf{Q}^{\ast\ast}),
\hspace{1.0cm} \textrm{for } \gamma =1;
\label{fo.approx.1.simplex.SBD.2}
\end{eqnarray}
it remains to estimate the left-hand sides of \eqref{fo.approx.1.simplex.SBD.1}
and \eqref{fo.approx.1.simplex.SBD.2}
(see Section \ref{SectEstimators.new.det.simplex}
below, where the latter also provides estimates of the \textit{minimizers}).

 
\section{
Bare-Simulation-Method for General Deterministic Divergence-Optimization-Problems with 
Constant-Component-Sum Side Constraint}
\label{SectDetSubsimplex}


\subsection{Minimization via Base-Divergence-Method 1}

\vspace{0.2cm}
\noindent
Recall that we are interested in the constrained optimization of the \textit{continuous} functions 
$\mathbf{Q} \mapsto \Phi_{\mathbf{P}}(\mathbf{Q})$
in the above-mentioned cases (D1) to (D8) of Subsection \ref{SectDetGeneral.Friends}, and beyond. 
Notice that --- 
on constraint sets of the form $A \cdot \textrm{$\boldsymbol{\Omega}$\hspace{-0.23cm}$\boldsymbol{\Omega}$}$ 
with \eqref{regularity simplex} 
(to be treated in this section) ---
the class (D1) of
CASM $\varphi-$divergences $\Phi_{\mathbf{P}}(\mathbf{Q}) := D_{\varphi}( \mathbf{Q}, \mathbf{P})$
which \textit{can not be covered} by the \textit{narrow-sense} BS minimizability results of Section 
\ref{SectDetSubsimplex.CASM} (namely seemingly all $\varphi-$divergences which are not generalized 
power divergences $D_{\widetilde{c} \cdot \varphi_{\gamma}}(\mathbf{Q},\mathbf{P})$,
cf. \eqref{brostu5:fo.powdiv.new})
is now \textit{much larger} than the class (D1) of 
CASM $\varphi-$divergences 
which \textit{can not be covered} by the narrow-sense BS minimizability results of Section 
\ref{SectDetNarrow} dealing with constraint sets of the form $\mathbf{\Omega}$ with \eqref{regularity}.
One corresponding example is the prominent \textit{Jensen-Shannon divergence} 
(being also called symmetrized and normalized Kullback-Leibler information distance, 
symmetrized and normalized relative entropy, capacitory discrimination),
see Broniatowski \& Stummer \cite{Bro:23a} for details.

\vspace{0.3cm}
\noindent
However, for such cases --- and beyond --- we can apply the 
following new fundamental \textit{non-narrow-sense} BS-minimizability:

\vspace{0.3cm}
\noindent

\begin{theorem}
\label{brostu5:thm.Fmin.simplex}
Let us arbitrarily fix some
$\mathds{P} \in \mathbb{S}_{> 0}^{K}$, 
$\widetilde{c} \in \, ]0,\infty[$, $A \in \, ]0,\infty[$, $\varphi_{\gamma}$ 
with $\gamma \in \mathbb{R}\backslash\, ]1,2[$,
$\mathbb{\bbzeta}$, $W:=(W_{i})_{i\in \mathbb{N}}$
and $\boldsymbol{\xi}_{n}^{w\mathbf{W}}$
(cf. \eqref{brostu5:fo.norweiemp.vec.det})
as in Theorem \ref{brostu5:thm.divnormW.new}; moreover, recall $\widetilde{\mathcal{M}}_{\gamma}$ from (T1). 
\\
(a) Furthermore, suppose that $A \cdot \textrm{$\boldsymbol{\Omega}$\hspace{-0.23cm}$\boldsymbol{\Omega}$} 
\subset \widetilde{\mathcal{M}}_{\gamma}$ is compact
and satisfies the regularity properties \eqref{regularity simplex} in the relative topology,
and that $\Phi: A \cdot \textrm{$\boldsymbol{\Omega}$\hspace{-0.23cm}$\boldsymbol{\Omega}$} 
\mapsto \mathbb{R}$ is a continuous function on $A \cdot \textrm{$\boldsymbol{\Omega}$\hspace{-0.23cm}$\boldsymbol{\Omega}$}$.
Then, there holds 
\begin{equation}
\inf_{\mathbf{Q}\in A \cdot \textrm{$\boldsymbol{\Omega}$\hspace{-0.19cm}$\boldsymbol{\Omega}$}} \Phi(\mathbf{Q})
\ = \ - \, 
\lim_{n\rightarrow \infty }\frac{1}{n}\log \negthinspace \left( \ 
\mathbb{E}_{\mathbb{\Pi}}\negthinspace \Big[
\exp\negthinspace\Big(n \cdot \Big(
F_{\gamma,\widetilde{c},A}\Big(D_{\widetilde{c} \cdot \varphi_{\gamma}}(A \cdot \boldsymbol{\xi}_{n}^{w\mathbf{W}},\mathds{P})\Big)
- \Phi\big(A \cdot \boldsymbol{\xi}_{n}^{w\mathbf{W}}\big)
\Big)
\Big)
\cdot \textfrak{1}_{
\textrm{$\boldsymbol{\Omega}$\hspace{-0.19cm}$\boldsymbol{\Omega}$}}\big(\boldsymbol{\xi}_{n}^{w\mathbf{W}}\big)
\, \Big] 
\right)
\, 
\label{brostu5:fo.BSmin.extended.simplex}
\end{equation}
and the infimum is attained at some (not necessarily unique) point in 
$A \cdot \textrm{$\boldsymbol{\Omega}$\hspace{-0.23cm}$\boldsymbol{\Omega}$}$. 
In particular, the function 
$\Phi\left( \cdot \right)$ is bare-simulation minimizable (BS-minimizable)
on $A \cdot \textrm{$\boldsymbol{\Omega}$\hspace{-0.23cm}$\boldsymbol{\Omega}$}$
 (cf. \eqref{brostu5:fo.2} in Definition \ref{brostu5:def.1}).\\
(b) If $A \cdot \textrm{$\boldsymbol{\Omega}$\hspace{-0.23cm}$\boldsymbol{\Omega}$} \subset \widetilde{\mathcal{M}}_{\gamma}$ 
is not necessarily compact
but satisfies the regularity properties 
\eqref{regularity simplex} in the relative topology and the finiteness property \eqref{def fi wrt Omega simplex}
with $\varphi := \widetilde{c} \cdot \varphi_{\gamma}$,
and $\Phi: A \cdot \textrm{$\boldsymbol{\Omega}$\hspace{-0.23cm}$\boldsymbol{\Omega}$} \mapsto \mathbb{R}$ is a continuous function
which satisfies the lower-bound condition
\begin{equation}
\textrm{there exists a constant $c_{1} \in \mathbb{R}$ such that for all $\mathbf{Q} \in 
A \cdot \textrm{$\boldsymbol{\Omega}$\hspace{-0.23cm}$\boldsymbol{\Omega}$}$ there holds} 
\quad
\Phi(\mathbf{Q}) \geq  
F_{\gamma,\widetilde{c},A}\Big(D_{\widetilde{c} \cdot \varphi_{\gamma}}(\mathbf{Q},\mathds{P})\Big) - c_{1} 
 \, ,
\label{brostu5:fo.phibound.simplex}
\end{equation}
then the representation/convergence \eqref{brostu5:fo.BSmin.extended.simplex} 
--- and hence the corresponding BS-minimizability --- still holds,
but the infimum may not necessarily be attained/reached at some point in 
$A \cdot \textrm{$\boldsymbol{\Omega}$\hspace{-0.23cm}$\boldsymbol{\Omega}$}$.

\end{theorem}

\vspace{0.4cm}
\noindent
The proof of Theorem \ref{brostu5:thm.Fmin.simplex} will be given in Appendix \ref{App.A} below.
Clearly, Theorem \ref{brostu5:thm.Fmin.simplex}(a) can be applied to obtain
$\inf_{\mathbf{Q}\in A \cdot \textrm{$\boldsymbol{\Omega}$\hspace{-0.19cm}$\boldsymbol{\Omega}$}} \Phi(\mathbf{Q})$
for all the directed distances/divergences $\Phi(\cdot) := \Phi_{\mathbf{\breve{P}}}(\cdot)$ 
(where $\mathbf{\breve{P}}$ needs not coincide with $\mathds{P}$) and friends 
given in (D1) to (D8) of Subsection \ref{SectDetGeneral.Friends}, which are therefore all BS-minimizable on compact 
$A \cdot \textrm{$\boldsymbol{\Omega}$\hspace{-0.23cm}$\boldsymbol{\Omega}$} \subset \widetilde{\mathcal{M}}_{\gamma}$ with 
regularity \eqref{regularity simplex} in the relative topology.

\vspace{0.5cm} 
\noindent
Analogously to 
\eqref{fo.approx.1.simplex}, 
the limit statement 
\eqref{brostu5:fo.BSmin.extended.simplex}
provides the principle for the approximation of the solution of 
the minimization problem 
$\Phi(\mathbf{\Omega}) := 
\inf_{\mathbf{Q}\in A \cdot \textrm{$\boldsymbol{\Omega}$\hspace{-0.19cm}$\boldsymbol{\Omega}$} }
\Phi(\mathbf{Q})$.
This can be achieved by replacing the right-hand side in \eqref{brostu5:fo.BSmin.extended.simplex}
 by its finite
counterpart, from which we obtain for given large $n$  
\begin{equation}
- \frac{1}{n}\log \negthinspace \left( \ 
\mathbb{E}_{\mathbb{\Pi}}\negthinspace \Big[
\exp\negthinspace\Big(n \cdot \Big(
F_{\gamma,\widetilde{c},A}\Big(D_{\widetilde{c} \cdot \varphi_{\gamma}}(A \cdot \boldsymbol{\xi}_{n}^{w\mathbf{W}},\mathds{P})\Big)
- \Phi\big(A \cdot \boldsymbol{\xi}_{n}^{w\mathbf{W}}\big)
\Big)
\Big)
\cdot \textfrak{1}_{
\textrm{$\boldsymbol{\Omega}$\hspace{-0.19cm}$\boldsymbol{\Omega}$}}\big(\boldsymbol{\xi}_{n}^{w\mathbf{W}}\big)
\, \Big] 
\right)
\ \approx \ \inf_{Q\in \mathbf{\Omega} }
\Phi(\mathbf{Q});
\label{fo.approx.1.extended.simplex} 
\end{equation}
it remains to estimate the left-hand side of \eqref{fo.approx.1.extended.simplex} 
(see Section 
\ref{SectEstimators.new.det.simplex}
below, where the latter also provides estimates of the \textit{minimizers}).

\noindent
\begin{example}
As a continuation of Example \ref{brostu5:ex.POWmiss}
in connection with Remark \ref{brostu5:rem.ex.POWmiss}(b),
let us how Theorem \ref{brostu5:thm.Fmin.simplex}
can be used to tackle the BS-minimizability --- 
on $A \cdot \textrm{$\boldsymbol{\Omega}$\hspace{-0.23cm}$\boldsymbol{\Omega}$}$ 
--- of the generalized power divergences
$\Phi(\cdot) := \Phi_{\mathds{P}}(\cdot) := D_{\widetilde{c} \cdot \varphi_{\gamma}}(\cdot,\mathds{P})$
(cf. \eqref{brostu5:fo.powdiv.new}) for the missing case $\gamma \in \, ]1,2[$ (and $\widetilde{c} \in \, ]0,\infty[$).
Indeed, by employing \eqref{brostu5:fo.divpow.F.KL} it is easy to see that
\begin{eqnarray}
& & \hspace{-0.7cm}
\textrm{for all $x \in \, [0,\infty[$, $A \in \, ]0,\infty[$, $\widetilde{c} \in \, ]0,\infty[$,
$\gamma \in \, ]1,2[$ there holds} 
\ x \, \geq \, 
F_{1,\frac{\widetilde{c}}{\gamma},A}(x) = 
\frac{\widetilde{c}}{\gamma} \cdot \Big\{ 1- A \cdot 
\exp\Big( \frac{1}{A} - 1 - \frac{x}{A \cdot \frac{\widetilde{c}}{\gamma}} \Big)
\Big\} \qquad {\ }
\label{brostu5:fo.bound.100a}
\\
& & \hspace{8.8cm}
\textrm{with equality if and only if $x= \frac{\widetilde{c}}{\gamma} \cdot (1-A) \geq 0$} 
\nonumber
\end{eqnarray}
(and even for a wider range of $\gamma$). By using \eqref{brostu5:fo.boundPOWdivmiss.altern1}
and \eqref{brostu5:fo.bound.100a} we obtain
\begin{equation}
\textrm{for all $\mathbf{Q} \in 
A \cdot \textrm{$\boldsymbol{\Omega}$\hspace{-0.23cm}$\boldsymbol{\Omega}$}$ 
and all $\gamma \in \, ]1,2[$ there holds} 
\quad
\Phi_{\mathds{P}}(\mathbf{Q}) := D_{\widetilde{c} \cdot \varphi_{\gamma}}(\mathbf{Q},\mathds{P}) \ \geq \ 
D_{\frac{\widetilde{c}}{\gamma} \cdot \varphi_{1}}(\mathbf{Q},\mathds{P}) 
\ \geq \ 
F_{1,\frac{\widetilde{c}}{\gamma},A}\Big(D_{\frac{\widetilde{c}}{\gamma} \cdot \varphi_{1}}(\mathbf{Q},\mathds{P})\Big)  
 \, ;
\label{brostu5:fo.phibound.simplex.pow}
\end{equation}
notice that the inequalities in \eqref{brostu5:fo.phibound.simplex.pow}
turn into equalities if $q_{k} = p_{k}$ for all $k \in \{1,\ldots,K\}$ (and hence, $A=1$)
but also e.g. if $q_{K}=0$ and $q_{k} = p_{k}$ for all $k \in \{1,\ldots,K-1\}$ (and hence, $A < 1$).
According to \eqref{brostu5:fo.phibound.simplex.pow}, the bound \eqref{brostu5:fo.phibound.simplex} 
is satisfied with $c_{1}=0$.
Thus, we can proceed analogously to Remark \ref{brostu5:rem.ex.POWmiss}(b), 
and choose the $W_{i}$'s to be i.i.d. copies of the random variable $W$ of the form
$W = \frac{\gamma}{\widetilde{c}} \cdot Z$ for a Poisson $POI(\frac{\widetilde{c}}{\gamma})-$distributed
random variable $Z$ (cf. Broniatowski \& Stummer \cite{Bro:23a}, see also Table 1). 
With these choices, \eqref{brostu5:fo.BSmin.extended.simplex} specializes to
\begin{equation}
\inf_{\mathbf{Q}\in A \cdot \textrm{$\boldsymbol{\Omega}$\hspace{-0.19cm}$\boldsymbol{\Omega}$}} 
D_{\widetilde{c} \cdot \varphi_{\gamma}}(\mathbf{Q},\mathds{P})
\ = \ - \, 
\lim_{n\rightarrow \infty }\frac{1}{n}\log \negthinspace \left( \ 
\mathbb{E}_{\mathbb{\Pi}}\negthinspace \Big[
\exp\negthinspace\Big(n \cdot \Big(
F_{1,\frac{\widetilde{c}}{\gamma},A}\Big(D_{\frac{\widetilde{c}}{\gamma} \cdot \varphi_{1}}(A \cdot \boldsymbol{\xi}_{n}^{w\mathbf{W}},\mathds{P})\Big)
- D_{\widetilde{c} \cdot \varphi_{\gamma}}(A \cdot \boldsymbol{\xi}_{n}^{w\mathbf{W}},\mathds{P})
\Big)
\Big)
\cdot \textfrak{1}_{
\textrm{$\boldsymbol{\Omega}$\hspace{-0.19cm}$\boldsymbol{\Omega}$}}\big(\boldsymbol{\xi}_{n}^{w\mathbf{W}}\big)
\, \Big] 
\right)
\, 
\nonumber
\end{equation}
for all $A \cdot \textrm{$\boldsymbol{\Omega}$\hspace{-0.23cm}$\boldsymbol{\Omega}$} \subset \widetilde{\mathcal{M}}_{\gamma}$ 
satisfying the regularity properties 
\eqref{regularity simplex} in the relative topology and the finiteness property \eqref{def fi wrt Omega simplex}
with $\varphi := \frac{\widetilde{c}}{\gamma} \cdot \varphi_{1}$.

\end{example}


\subsection{Maximization via Base-Divergence-Method 1}

\vspace{0.2cm}
\noindent
For the \textit{non-narrow-sense BS-maximizability} we obtain the following new fundamental

\vspace{0.2cm}
\noindent

\begin{theorem}
\label{brostu5:thm.Fmax.simplex}
Let us arbitrarily fix some
$\mathds{P} \in \mathbb{S}_{> 0}^{K}$, 
$\widetilde{c} \in \, ]0,\infty[$, $A \in \, ]0,\infty[$, $\varphi_{\gamma}$ 
with $\gamma \in \mathbb{R}\backslash\, ]1,2[$,
$\mathbb{\bbzeta}$, 
$W:=(W_{i})_{i\in \mathbb{N}}$
and $\boldsymbol{\xi}_{n}^{w\mathbf{W}}$
(cf. \eqref{brostu5:fo.norweiemp.vec.det})
as in Theorem \ref{brostu5:thm.divnormW.new}.
\\
(a) Furthermore, suppose that $A \cdot \textrm{$\boldsymbol{\Omega}$\hspace{-0.23cm}$\boldsymbol{\Omega}$} \subset \widetilde{\mathcal{M}}_{\gamma}$ is compact
and satisfies the regularity properties \eqref{regularity simplex} in the relative topology,
and that $\Phi: A \cdot \textrm{$\boldsymbol{\Omega}$\hspace{-0.23cm}$\boldsymbol{\Omega}$} 
\mapsto \mathbb{R}$ is a continuous function on $A \cdot \textrm{$\boldsymbol{\Omega}$\hspace{-0.23cm}$\boldsymbol{\Omega}$}$.
Then, there holds 
\begin{equation}
\sup_{\mathbf{Q}\in A \cdot \textrm{$\boldsymbol{\Omega}$\hspace{-0.19cm}$\boldsymbol{\Omega}$}} \Phi(\mathbf{Q})
\ = \ 
\lim_{n\rightarrow \infty }\frac{1}{n}\log \negthinspace \left( \ 
\mathbb{E}_{\mathbb{\Pi}}\negthinspace \Big[
\exp\negthinspace\Big(n \cdot \Big(
F_{\gamma,\widetilde{c},A}\Big(D_{\widetilde{c} \cdot \varphi_{\gamma}}(A \cdot \boldsymbol{\xi}_{n}^{w\mathbf{W}},\mathds{P})\Big)
+ \Phi\big(A \cdot \boldsymbol{\xi}_{n}^{w\mathbf{W}}\big)
\Big)
\Big)
\cdot \textfrak{1}_{
\textrm{$\boldsymbol{\Omega}$\hspace{-0.19cm}$\boldsymbol{\Omega}$}}\big(\boldsymbol{\xi}_{n}^{w\mathbf{W}}\big)
\, \Big] 
\right)
\, 
\label{brostu5:fo.BSmax.extended.simplex}
\end{equation}
and the supremum is attained at some (not necessarily unique) point in 
$A \cdot \textrm{$\boldsymbol{\Omega}$\hspace{-0.23cm}$\boldsymbol{\Omega}$}$. 
In particular, the function 
$\Phi\left( \cdot \right)$ is bare-simulation maximizable (BS-maximizable)
on $A \cdot \textrm{$\boldsymbol{\Omega}$\hspace{-0.23cm}$\boldsymbol{\Omega}$}$
 (cf. \eqref{brostu5:fo.2b} in Definition \ref{brostu5:def.1}).\\
(b) If $A \cdot \textrm{$\boldsymbol{\Omega}$\hspace{-0.23cm}$\boldsymbol{\Omega}$} \subset \widetilde{\mathcal{M}}_{\gamma}$ 
is not necessarily compact
but satisfies the regularity properties 
\eqref{regularity simplex} in the relative topology 
and $\Phi: A \cdot \textrm{$\boldsymbol{\Omega}$\hspace{-0.23cm}$\boldsymbol{\Omega}$} 
\mapsto \mathbb{R}$ is a continuous function
which satisfies the upper-bound condition
\begin{equation}
\textrm{there exists a constant $c_{1} \in \mathbb{R}$ such that for all $\mathbf{Q} \in 
A \cdot \textrm{$\boldsymbol{\Omega}$\hspace{-0.23cm}$\boldsymbol{\Omega}$}$ there holds} 
\quad
\Phi(\mathbf{Q}) \leq c_{1} -    
F_{\gamma,\widetilde{c},A}\Big(D_{\widetilde{c} \cdot \varphi_{\gamma}}(\mathbf{Q},\mathds{P})\Big) 
 \, ,
\nonumber
\end{equation}
then the representation/convergence \eqref{brostu5:fo.BSmax.extended.simplex} 
--- and hence the corresponding BS-maximizability --- still holds,
but the supremum may not necessarily be attained/reached at some point in 
$A \cdot \textrm{$\boldsymbol{\Omega}$\hspace{-0.23cm}$\boldsymbol{\Omega}$}$.

\end{theorem}

\vspace{0.4cm}
\noindent
The proof of Theorem \ref{brostu5:thm.Fmax.simplex} will be given in Appendix \ref{App.A} below.

\vspace{0.3cm} 
\noindent
Analogously to 
\eqref{fo.approx.1.extended.simplex},
the limit statement 
\eqref{brostu5:fo.BSmax.extended.simplex}
provides the principle for the approximation of the solution of 
the maximization problem 
$\Phi(\mathbf{\Omega}) := 
\sup_{\mathbf{Q}\in A \cdot \textrm{$\boldsymbol{\Omega}$\hspace{-0.19cm}$\boldsymbol{\Omega}$} }
\Phi(\mathbf{Q})$.
This can be achieved by replacing the right-hand side in \eqref{brostu5:fo.BSmax.extended.simplex}
by its finite
counterpart, from which we obtain for given large $n$  
\begin{equation}
\frac{1}{n}\log \negthinspace \left( \ 
\mathbb{E}_{\mathbb{\Pi}}\negthinspace \Big[
\exp\negthinspace\Big(n \cdot \Big(
F_{\gamma,\widetilde{c},A}\Big(D_{\widetilde{c} \cdot \varphi_{\gamma}}(A \cdot \boldsymbol{\xi}_{n}^{w\mathbf{W}},\mathds{P})\Big)
+ \Phi\big(A \cdot \boldsymbol{\xi}_{n}^{w\mathbf{W}}\big)
\Big)
\Big)
\cdot \textfrak{1}_{
\textrm{$\boldsymbol{\Omega}$\hspace{-0.19cm}$\boldsymbol{\Omega}$}}\big(\boldsymbol{\xi}_{n}^{w\mathbf{W}}\big)
\, \Big] 
\right)
\ \approx \ \sup_{Q\in \mathbf{\Omega} }
\Phi(\mathbf{Q});
\label{fo.approx.1.extended.simplex.max} 
\end{equation}
it remains to estimate the left-hand side of \eqref{fo.approx.1.extended.simplex.max} 
(see Section \ref{SectEstimators.new.det.simplex}
below, where the latter also provides estimates of the \textit{maximizers}).


\subsection{Minimization via Base-Divergence-Method 2}

\vspace{0.2cm}
\noindent
Due to our investigations in Section \ref{SectDetSubsimplex.SBD},
as an alternative to the new Theorem \ref{brostu5:thm.Fmin.simplex}
we can also derive the following 
new assertions, by switching the involved base-divergences:

\vspace{0.2cm}
\noindent

\begin{theorem}
\label{brostu5:thm.Fmin.simplex.SBD}
Let us arbitrarily fix some $\mathbf{P} \in \mathbb{R}_{> 0}^{K}$ 
(recall 
$M_{\mathbf{P}}:=\sum_{i=1}^{K}p_{i}>0$), 
$\mathbf{Q}^{\ast\ast}$,
$\widetilde{c} \in \, ]0,\infty[$, $A \in \, ]0,\infty[$, $\varphi_{\gamma}$ 
with $\gamma \in \mathbb{R}\backslash\, ]1,2[$,
$\widetilde{V}$ and 
$\boldsymbol{\xi}_{n}^{w\mathbf{\widetilde{V}}}$
(cf. \eqref{brostu5:fo.norweiemp.vec.det.SBD} with $\widetilde{V}$ instead of $V$)
as in Theorem \ref{theorem inner min Bregman power general}; we call the corresponding 
$\breve{D}_{\widetilde{c} \cdot \varphi_{\gamma},\mathbf{P}}^{SBD}(\cdot,\mathbf{Q}^{\ast\ast})$
(cf. \eqref{brostu3:fo.676b.SBD},\eqref{brostu3:fo.677a.SBD},\eqref{brostu3:fo.678.SBD})
the \textit{base-innmin-SBD-divergence} (function). 
\\
(a) Furthermore, suppose that 
$A \cdot \textrm{$\boldsymbol{\Omega}$\hspace{-0.23cm}$\boldsymbol{\Omega}$} \subset \widetilde{\mathcal{M}}_{\gamma}$ is compact and satisfies the regularity properties \eqref{regularity simplex} in the relative topology,
and that $\Phi: A \cdot \textrm{$\boldsymbol{\Omega}$\hspace{-0.23cm}$\boldsymbol{\Omega}$} 
\mapsto \mathbb{R}$ is a continuous function on $A \cdot \textrm{$\boldsymbol{\Omega}$\hspace{-0.23cm}$\boldsymbol{\Omega}$}$.
Then, there holds 
\begin{equation}
\inf_{\mathbf{Q}\in A \cdot \textrm{$\boldsymbol{\Omega}$\hspace{-0.19cm}$\boldsymbol{\Omega}$}} \Phi(\mathbf{Q})
\ = \ - \, 
\lim_{n\rightarrow \infty }\frac{1}{n}\log \negthinspace \left( \ 
\mathbb{E}_{\mathbb{\Pi}}\negthinspace \Big[
\exp\negthinspace\Big(n \cdot \Big(
\breve{D}_{\widetilde{c} \cdot \varphi_{\gamma},\mathbf{P}}^{SBD}\negthinspace\left(A 
\cdot \boldsymbol{\xi}_{n}^{w\mathbf{\widetilde{V}}},\mathbf{Q}^{\ast\ast}\right) 
- \Phi\big(A \cdot \boldsymbol{\xi}_{n}^{w\mathbf{\widetilde{V}}}\big)
\Big)
\Big)
\cdot \textfrak{1}_{
\textrm{$\boldsymbol{\Omega}$\hspace{-0.19cm}$\boldsymbol{\Omega}$}}\big(\boldsymbol{\xi}_{n}^{w\mathbf{\widetilde{V}}}\big)
\, \Big] 
\right)
\, 
\label{brostu5:fo.BSmin.extended.simplex.SBD}
\end{equation}
and the infimum is attained at some (not necessarily unique) point in 
$A \cdot \textrm{$\boldsymbol{\Omega}$\hspace{-0.23cm}$\boldsymbol{\Omega}$}$. 
In particular, the function 
$\Phi\left( \cdot \right)$ is bare-simulation minimizable (BS-minimizable)
on $A \cdot \textrm{$\boldsymbol{\Omega}$\hspace{-0.23cm}$\boldsymbol{\Omega}$}$
 (cf. \eqref{brostu5:fo.2} in Definition \ref{brostu5:def.1}),
\textit{in terms of the SBD method}.\\
(b) If $A \cdot \textrm{$\boldsymbol{\Omega}$\hspace{-0.23cm}$\boldsymbol{\Omega}$} \subset \widetilde{\mathcal{M}}_{\gamma}$ 
is not necessarily compact
but satisfies the regularity properties 
\eqref{regularity simplex} in the relative topology 
with $\varphi := \widetilde{c} \cdot \varphi_{\gamma}$,
and $\Phi: A \cdot \textrm{$\boldsymbol{\Omega}$\hspace{-0.23cm}$\boldsymbol{\Omega}$} 
\mapsto \mathbb{R}$ is a continuous function
which satisfies the lower-bound condition
\begin{equation}
\textrm{there exists a constant $c_{1} \in \mathbb{R}$ such that for all $\mathbf{Q} \in 
A \cdot \textrm{$\boldsymbol{\Omega}$\hspace{-0.23cm}$\boldsymbol{\Omega}$}$ there holds} 
\quad
\Phi(\mathbf{Q}) \geq  
\breve{D}_{\widetilde{c} \cdot 
\varphi_{\gamma},\mathbf{P}}^{SBD}\negthinspace\left(\mathbf{Q},\mathbf{Q}^{\ast\ast}\right)
- c_{1} \, ,
\nonumber
\end{equation}
then the representation/convergence 
\eqref{brostu5:fo.BSmin.extended.simplex.SBD}
--- and hence the corresponding BS-minimizability --- still holds,
but the infimum may not necessarily be attained/reached at some point in 
$A \cdot \textrm{$\boldsymbol{\Omega}$\hspace{-0.23cm}$\boldsymbol{\Omega}$}$.

\end{theorem}

\vspace{0.3cm}

\begin{remark} \ 
(a) For the case $\gamma \in \mathbb{R}\backslash\{1\}$
and the special subsetup $\mathbf{Q}^{\ast\ast} := C \cdot \mathbf{P}$
(cf. Theorem \ref{brostu5:thm.divnormW.new.SBD}(a)),
in Theorem \ref{brostu5:thm.Fmin.simplex.SBD} one can equivalently replace 
$\breve{D}_{\widetilde{c} \cdot \varphi_{\gamma},\mathbf{P}}^{SBD}( \cdot,C \cdot \mathbf{P})$
by $\breve{F}_{\gamma,\widetilde{c},A,M_{\mathbf{P}},C}\Big(D_{\widetilde{c} 
\cdot \varphi_{\gamma},\mathbf{P}}^{SBD}(\cdot, C \cdot \mathbf{P})
\Big)$ 
(cf. \eqref{brostu5:fo.SBDrepres},\eqref{brostu5:fo.divpow.F.SBD}).
By additionally taking $C=1$ and $\mathbf{P} := \mathds{P}$ 
to be a probability vector, one collapsedly ends up with 
the corresponding assertions of Theorem \ref{brostu5:thm.Fmin.simplex}.\\
(b) For the case $\gamma =1$,
in Theorem \ref{brostu5:thm.Fmin.simplex.SBD} one can equivalently replace 
$\breve{D}_{\widetilde{c} \cdot \varphi_{1},\mathbf{P}}^{SBD}( \cdot,\mathbf{Q}^{\ast\ast})$ 
by $\breve{F}_{1,\widetilde{c},A,M_{\mathbf{Q}^{\ast\ast}}}\Big(D_{\widetilde{c} 
\cdot \varphi_{1},\mathbf{P}}^{SBD}(\cdot, \mathbf{Q}^{\ast\ast})\Big)$
(cf. \eqref{brostu5:fo.SBDrepres.gamma1},\eqref{brostu5:fo.divpow.F.SBD.gamma1}).

\end{remark} 

\vspace{0.4cm}

\begin{remark} \ 
\label{rem.original}
The BS-minimization --- in a wide sense --- 
$\inf_{\mathbf{Q}\in A \cdot \textrm{$\boldsymbol{\Omega}$\hspace{-0.19cm}$\boldsymbol{\Omega}$}} \Phi(\mathbf{Q})$
of the scaled Bregman divergence
$\Phi(\mathbf{Q}) := D_{\widetilde{c} \cdot \varphi_{\gamma},\mathbf{P}}^{SBD}( \mathbf{Q},\mathbf{Q}^{\ast\ast})$
follows immediately as special case of Theorem \ref{brostu5:thm.Fmin.simplex.SBD}. 
Notice that in \eqref{brostu5:fo.BSmin.extended.simplex.SBD}
the corresponding exponential part
$ \breve{D}_{\widetilde{c} \cdot \varphi_{\gamma},\mathbf{P}}^{SBD}\negthinspace\left(A 
\cdot \boldsymbol{\xi}_{n}^{w\mathbf{\widetilde{V}}},\mathbf{Q}^{\ast\ast}\right) 
- \Phi\big(A \cdot \boldsymbol{\xi}_{n}^{w\mathbf{\widetilde{V}}}\big)$
is generally non-zero.

\end{remark} 

\vspace{0.4cm}
\noindent
The proof of Theorem \ref{brostu5:thm.Fmin.simplex.SBD} will be given in Appendix \ref{App.A} below.
Clearly, Theorem \ref{brostu5:thm.Fmin.simplex.SBD}(a) can be applied to obtain
$\inf_{\mathbf{Q}\in A \cdot \textrm{$\boldsymbol{\Omega}$\hspace{-0.19cm}$\boldsymbol{\Omega}$}} \Phi(\mathbf{Q})$
for all the directed distances/divergences $\Phi(\cdot) := \Phi_{\mathbf{P}}(\cdot)$ and friends 
given in (D1) to (D8) of Subsection \ref{SectDetGeneral.Friends}, which are therefore all BS-minimizable on compact 
$A \cdot \textrm{$\boldsymbol{\Omega}$\hspace{-0.23cm}$\boldsymbol{\Omega}$} \subset \widetilde{\mathcal{M}}_{\gamma}$ with 
regularity \eqref{regularity simplex} in the relative topology.

\vspace{0.3cm} 
\noindent
Analogously to 
\eqref{fo.approx.1.extended.simplex}, 
the limit statement 
\eqref{brostu5:fo.BSmin.extended.simplex.SBD}
provides the principle for the approximation of the solution of 
the minimization problem 
$\Phi(\mathbf{\Omega}) := 
\inf_{\mathbf{Q}\in A \cdot \textrm{$\boldsymbol{\Omega}$\hspace{-0.19cm}$\boldsymbol{\Omega}$} }
\Phi(\mathbf{Q})$. 
This can be achieved by replacing the right-hand side in \eqref{brostu5:fo.BSmin.extended.simplex.SBD}
 by its finite
counterpart, from which we obtain for given large $n$  
\begin{equation}
- \frac{1}{n}\log \negthinspace \left( \ 
\mathbb{E}_{\mathbb{\Pi}}\negthinspace \Big[
\exp\negthinspace\Big(n \cdot \Big(
\breve{D}_{\widetilde{c} \cdot \varphi_{\gamma},\mathbf{P}}^{SBD}\negthinspace\left(A 
\cdot \boldsymbol{\xi}_{n}^{w\mathbf{\widetilde{V}}},\mathbf{Q}^{\ast\ast}\right)
- \Phi\big(A \cdot \boldsymbol{\xi}_{n}^{w\mathbf{\widetilde{V}}}\big)
\Big)
\Big)
\cdot \textfrak{1}_{
\textrm{$\boldsymbol{\Omega}$\hspace{-0.19cm}$\boldsymbol{\Omega}$}}\big(\boldsymbol{\xi}_{n}^{w\mathbf{\widetilde{V}}}\big)
\, \Big] 
\right)
\ \approx \ \inf_{Q\in \mathbf{\Omega} }
\Phi(\mathbf{Q});
\label{fo.approx.1.extended.simplex.SBD} 
\end{equation}
it remains to estimate the left-hand side of \eqref{fo.approx.1.extended.simplex.SBD}  
(see Section \ref{SectEstimators.new.det.simplex}
below, where the latter also provides estimates of the \textit{minimizers}).


\subsection{Maximization via Base-Divergence-Method 2}

\vspace{0.2cm}
\noindent
For the \textit{non-narrow-sense BS-maximizability} via innmin-SBDs as base divergences, 
we obtain the following new fundamental

\vspace{0.2cm}

\begin{theorem}
\label{brostu5:thm.Fmax.simplex.SBD}
Let us arbitrarily fix some $\mathbf{P} \in \mathbb{R}_{> 0}^{K}$ 
(recall 
$M_{\mathbf{P}}:=\sum_{i=1}^{K}p_{i}>0$), 
$\mathbf{Q}^{\ast\ast}$,
$\widetilde{c} \in \, ]0,\infty[$, $A \in \, ]0,\infty[$, $\varphi_{\gamma}$ 
with $\gamma \in \mathbb{R}\backslash\, ]1,2[$,
$\widetilde{V}$ and 
$\boldsymbol{\xi}_{n}^{w\mathbf{\widetilde{V}}}$
(cf. \eqref{brostu5:fo.norweiemp.vec.det.SBD} with $\widetilde{V}$ instead of $V$)
as in Theorem \ref{theorem inner min Bregman power general}; recall that we have named the corresponding 
$\breve{D}_{\widetilde{c} \cdot \varphi_{\gamma},\mathbf{P}}^{SBD}(\cdot,\mathbf{Q}^{\ast\ast})$
(cf. \eqref{brostu3:fo.676b.SBD},\eqref{brostu3:fo.677a.SBD},\eqref{brostu3:fo.678.SBD})
the \textit{base-innmin-SBD-divergence} (function). 
\\
(a) Furthermore, suppose that 
$A \cdot \textrm{$\boldsymbol{\Omega}$\hspace{-0.23cm}$\boldsymbol{\Omega}$} \subset \widetilde{\mathcal{M}}_{\gamma}$ is compact and satisfies the regularity properties \eqref{regularity simplex} in the relative topology,
and that $\Phi: A \cdot \textrm{$\boldsymbol{\Omega}$\hspace{-0.23cm}$\boldsymbol{\Omega}$} 
\mapsto \mathbb{R}$ is a continuous function on $A \cdot \textrm{$\boldsymbol{\Omega}$\hspace{-0.23cm}$\boldsymbol{\Omega}$}$.
Then, there holds 
\begin{equation}
\sup_{\mathbf{Q}\in A \cdot \textrm{$\boldsymbol{\Omega}$\hspace{-0.19cm}$\boldsymbol{\Omega}$}} \Phi(\mathbf{Q})
\ = \ 
\lim_{n\rightarrow \infty }\frac{1}{n}\log \negthinspace \left( \ 
\mathbb{E}_{\mathbb{\Pi}}\negthinspace \Big[
\exp\negthinspace\Big(n \cdot \Big(
\breve{D}_{\widetilde{c} \cdot \varphi_{\gamma},\mathbf{P}}^{SBD}\negthinspace\left(A 
\cdot \boldsymbol{\xi}_{n}^{w\mathbf{\widetilde{V}}},\mathbf{Q}^{\ast\ast}\right) 
+ \Phi\big(A \cdot \boldsymbol{\xi}_{n}^{w\mathbf{\widetilde{V}}}\big)
\Big)
\Big)
\cdot \textfrak{1}_{
\textrm{$\boldsymbol{\Omega}$\hspace{-0.19cm}$\boldsymbol{\Omega}$}}\big(\boldsymbol{\xi}_{n}^{w\mathbf{\widetilde{V}}}\big)
\, \Big] 
\right)
\, 
\label{brostu5:fo.BSmax.extended.simplex.SBD}
\end{equation}
and the supremum is attained at some (not necessarily unique) point in 
$A \cdot \textrm{$\boldsymbol{\Omega}$\hspace{-0.23cm}$\boldsymbol{\Omega}$}$. 
In particular, the function 
$\Phi\left( \cdot \right)$ is bare-simulation maximizable (BS-maximizable)
on $A \cdot \textrm{$\boldsymbol{\Omega}$\hspace{-0.23cm}$\boldsymbol{\Omega}$}$
 (cf. \eqref{brostu5:fo.2b} in Definition \ref{brostu5:def.1}),
\textit{in terms of the SBD method}.\\
(b) If $A \cdot \textrm{$\boldsymbol{\Omega}$\hspace{-0.23cm}$\boldsymbol{\Omega}$} \subset \widetilde{\mathcal{M}}_{\gamma}$ 
is not necessarily compact
but satisfies the regularity properties 
\eqref{regularity simplex} in the relative topology, 
and $\Phi: A \cdot \textrm{$\boldsymbol{\Omega}$\hspace{-0.23cm}$\boldsymbol{\Omega}$} 
\mapsto \mathbb{R}$ is a continuous function which satisfies the lower-bound condition
\begin{equation}
\textrm{there exists a constant $c_{1} \in \mathbb{R}$ such that for all $\mathbf{Q} \in 
A \cdot \textrm{$\boldsymbol{\Omega}$\hspace{-0.23cm}$\boldsymbol{\Omega}$}$ there holds} 
\quad
\Phi(\mathbf{Q}) \leq  c_{1} -
\breve{D}_{\widetilde{c} \cdot 
\varphi_{\gamma},\mathbf{P}}^{SBD}\negthinspace\left(\mathbf{Q},\mathbf{Q}^{\ast\ast}\right) \, ,
\nonumber
\end{equation}
then the representation/convergence 
\eqref{brostu5:fo.BSmax.extended.simplex.SBD}
--- and hence the corresponding BS-minimizability --- still holds,
but the supremum may not necessarily be attained/reached at some point in 
$A \cdot \textrm{$\boldsymbol{\Omega}$\hspace{-0.23cm}$\boldsymbol{\Omega}$}$.

\end{theorem}

\vspace{0.3cm}

\begin{remark} \ 
(a) For the case $\gamma \in \mathbb{R}\backslash\{1\}$
and the special subsetup $\mathbf{Q}^{\ast\ast} := C \cdot \mathbf{P}$
(cf. Theorem \ref{brostu5:thm.divnormW.new.SBD}(a)),
in Theorem \ref{brostu5:thm.Fmax.simplex.SBD} one can equivalently replace 
$\breve{D}_{\widetilde{c} \cdot \varphi_{\gamma},\mathbf{P}}^{SBD}( \cdot,C \cdot \mathbf{P})$
by $\breve{F}_{\gamma,\widetilde{c},A,M_{\mathbf{P}},C}\Big(D_{\widetilde{c} 
\cdot \varphi_{\gamma},\mathbf{P}}^{SBD}(\cdot, C \cdot \mathbf{P})
\Big)$ 
(cf. \eqref{brostu5:fo.SBDrepres},\eqref{brostu5:fo.divpow.F.SBD}).
By additionally taking $C=1$ and $\mathbf{P} := \mathds{P}$ 
to be a probability vector, one collapsedly ends up with 
the corresponding assertions of Theorem \ref{brostu5:thm.Fmax.simplex}.\\
(b) For the case $\gamma =1$,
in Theorem \ref{brostu5:thm.Fmax.simplex.SBD} one can equivalently replace 
$\breve{D}_{\widetilde{c} \cdot \varphi_{1},\mathbf{P}}^{SBD}( \cdot,\mathbf{Q}^{\ast\ast})$ 
by $\breve{F}_{1,\widetilde{c},A,M_{\mathbf{Q}^{\ast\ast}}}\Big(D_{\widetilde{c} 
\cdot \varphi_{1},\mathbf{P}}^{SBD}(\cdot, \mathbf{Q}^{\ast\ast})\Big)$
(cf. \eqref{brostu5:fo.SBDrepres.gamma1},\eqref{brostu5:fo.divpow.F.SBD.gamma1}).

\end{remark} 

\vspace{0.4cm}
\noindent
The proof of Theorem \ref{brostu5:thm.Fmax.simplex.SBD} will be given in Appendix 
\ref{App.A} below. Certainly, Theorem \ref{brostu5:thm.Fmax.simplex.SBD}(a) can be applied to obtain
$\sup_{\mathbf{Q}\in A \cdot \textrm{$\boldsymbol{\Omega}$\hspace{-0.19cm}$\boldsymbol{\Omega}$}} \Phi(\mathbf{Q})$
for all the directed distances/divergences $\Phi(\cdot) := \Phi_{\mathbf{\breve{P}}}(\cdot)$ 
(where $\mathbf{\breve{P}}$ needs not coincide with $\mathbf{P}$) and friends 
given in (D1) to (D8) of Subsection \ref{SectDetGeneral.Friends}, which are therefore all BS-maximizable on compact 
$A \cdot \textrm{$\boldsymbol{\Omega}$\hspace{-0.23cm}$\boldsymbol{\Omega}$} 
\subset \widetilde{\mathcal{M}}_{\gamma}$ with 
regularity properties \eqref{regularity simplex} in the relative topology.

\vspace{0.3cm} 
\noindent
Analogously to 
\eqref{fo.approx.1.extended.simplex.max}, 
the limit statement \eqref{brostu5:fo.BSmax.extended.simplex.SBD}
provides the principle for the approximation of the solution of 
the maximization problem 
$\Phi(\mathbf{\Omega}) := 
\sup_{\mathbf{Q}\in A \cdot \textrm{$\boldsymbol{\Omega}$\hspace{-0.19cm}$\boldsymbol{\Omega}$} }
\Phi(\mathbf{Q})$.
This can be achieved by replacing the right-hand side in \eqref{brostu5:fo.BSmax.extended.simplex.SBD}
 by its finite
counterpart, from which we obtain for given large $n$  
\begin{equation}
\frac{1}{n}\log \negthinspace \left( \ 
\mathbb{E}_{\mathbb{\Pi}}\negthinspace \Big[
\exp\negthinspace\Big(n \cdot \Big(
\breve{D}_{\widetilde{c} \cdot \varphi_{\gamma},\mathbf{P}}^{SBD}\negthinspace\left(A 
\cdot \boldsymbol{\xi}_{n}^{w\mathbf{\widetilde{V}}},\mathbf{Q}^{\ast\ast}\right)
+ \Phi\big(A \cdot \boldsymbol{\xi}_{n}^{w\mathbf{\widetilde{V}}}\big)
\Big)
\Big)
\cdot \textfrak{1}_{
\textrm{$\boldsymbol{\Omega}$\hspace{-0.19cm}$\boldsymbol{\Omega}$}}\big(\boldsymbol{\xi}_{n}^{w\mathbf{\widetilde{V}}}\big)
\, \Big] 
\right)
\ \approx \ \sup_{Q\in \mathbf{\Omega} }
\Phi(\mathbf{Q});
\label{fo.approx.1.extended.simplex.SBD.max} 
\end{equation}
it remains to estimate the left-hand side of \eqref{fo.approx.1.extended.simplex.SBD.max}  
(see Section \ref{SectEstimators.new.det.simplex}
below, where the latter also provides estimates of the \textit{maximizers}).


\section{Narrow-Sense Bare-Simulation-Minimization of $\varphi-$divergences under risk}
\label{SectStochSubsimplex.CASM}


\subsection{The statistical view}
\label{SectStochSubsimplex.CASM.stat}

\vspace{0.2cm}
\noindent
In contrast to the previous Sections 
\ref{SectDetSubsimplex.CASM},\ref{SectDetSubsimplex.SBD},\ref{SectDetSubsimplex}, 
we now work out our BS method for the important setup where basically $\mathds{P}$ is a 
\textit{random} (unknown) element of the (open) simplex $\mathbb{S}_{>0}^{K}$ 
of zeros-free $K-$component probability (frequency) vectors 
and $\textrm{$\boldsymbol{\Omega}$\hspace{-0.23cm}$\boldsymbol{\Omega}$} \subset \mathbb{S}^{K}$ 
(i.e. $A=1$). Its importance stems from the fact that
in the statistics of discrete data --- and in the adjacent research fields of
information theory, artificial intelligence and machine learning --- one
often encounters the following \textit{minimum distance estimation (MDE)
problem} which is often also named as \textit{estimation of the asymptotic risk}:

\vspace{0.2cm}

\begin{enumerate}

\item[(MDE1)] for index $i\in \mathbb{N}$, let the generation of the $i-$th
(uncertainty-prone) data point be represented by the random variable $X_{i}$
which takes values in the discrete set $\mathcal{Y}:=\left\{ d_{1},\cdots ,d_{K}\right\}$ 
of $K$ distinct values \textquotedblleft of any
kind\textquotedblright. It is assumed that there exists a --- true unknown --- probability
distribution $\mathbb{P}[\cdot \,]$ on $\mathcal{Y}$ which is the a.s. limit 
(as $n$ tends to infinity) of
the empirical distributions $\mathbb{P}_{n}^{emp}(\mathbf{X}_{1}^{n})$ defined by the collection 
$\mathbf{X}_{1}^{n} := \left( X_{1},\ldots,X_{n}\right)$, i.e.
\begin{equation}
\lim_{n\rightarrow \infty }\mathbb{P}_{n}^{emp}(\mathbf{X}_{1}^{n}):=\lim_{n\rightarrow \infty }
\frac{1}{n}\sum_{i=1}^{n}\delta _{X_{i}}=\mathbb{P} \qquad \text{a.s.}
\label{cv emp measure X to P}
\end{equation}
where $\delta _{y}$ denotes the one-point distribution (Dirac mass) at point 
$y$ 
\footnote{
notice that $\mathbb{P}_{n}^{emp}(\mathbf{X}_{1}^{n})$ a probability distribution 
(on the data space
$\mathcal{Y}$), which is random due to its dependence on the $X_{i}$\textquoteright s}. 
We assume that none of the entries of $\mathbb{P}$ bears zero mass
so that $\mathbb{P}$ is identified with a point in the interior of $\mathbb{S}^{K}$ 
(see below).
The underlying probability space (say, $(\mathfrak{X},\mathcal{A},\mathbb{\Pi})$) 
where the above a.s. convergence holds, pertains 
to the random generation of the sequence $\left( X_{i}\right)_{i\in\mathbb{N}}$, 
of which we do not need to know but for \eqref{cv emp measure X to P}. 
Examples include the i.i.d. case (where the $X_{i}$\textquoteright s
are independent and have common distribution $\mathbb{P}$), 
ergodic Markov chains on $\mathcal{Y}$ with
stationary distribution $\mathbb{P}$, 
more globally autoregressive chains
with stationary distribution $\mathbb{P}$, etc.

\vspace{0.2cm}
\noindent

\begin{remark} \ 
(a) Let us briefly discuss our assumption
\eqref{cv emp measure X to P}
(resp. its vector form \eqref{cv emp measure X to P vector} below)
on the limit behavior of the empirical distribution 
of the sample $\mathbf{X}_{1}^{n} = \left(X_{1},\ldots,X_{n}\right)$ 
as $n$ tends to infinity.
In the \textquotedblleft  basic\textquotedblright\ statistical context, 
the sample $\mathbf{X}_{1}^{n}$
consists of i.i.d. replications of a generic random variable $X$ with 
probability distribution $\mathds{P}$.
However, our approach captures many other sampling schemes, where the 
distribution $\mathds{P}$ is defined implicitly through 
\eqref{cv emp measure X to P}
for which we aim at some estimate of $\Phi_{\mathds{P}}\left( 
\mathbb{\Omega} \right)$
of a family $\mathbb{\Omega}$ of probability distributions on $\mathcal{Y}$.
Sometimes the sequence of samples 
may stem from a \textquotedblleft  triangular\textquotedblright\  array 
$\left(
X_{1,n},..,X_{k_{n},n}\right)_{n\in \mathbb{N}} $ with $k_{n}\rightarrow \infty$ (as $n$ tends to infinity)
and \eqref{cv emp measure X to P} is substituted by 
\vspace{-0.2cm}
\[
\lim_{n\rightarrow \infty }\frac{1}{k_{n}}\sum_{j=1}^{k_{n}}\delta_{X_{j,n}} =
\mathds{P} \text{ \ a.s.} 
\]

\vspace{-0.2cm}
\noindent
which does not alter the results of this paper by any means.\\
(b) As another alternative to \eqref{cv emp measure X to P}
(resp. its vector form \eqref{cv emp measure X to P vector} below), 
in the following we could also (verbatim) employ any other sequence
$\mathbb{P}_{n}(\mathbf{X}_{1}^{n})$ of
probability-distribution-valued function(al)s for which 
$lim_{n\rightarrow \infty }\mathbb{P}_{n}(\mathbf{X}_{1}^{n})=\mathbb{P}$ a.s.

\end{remark} 

\vspace{0.2cm}
\noindent

\item[(MDE2)] given a \textit{model} $\mathbb{\Omega}$, i.e. a 
family $\mathbb{\Omega}$  
of probability distributions $\mathbb{Q}$ on $\mathcal{Y}$
each of which serves as a potential description of
the underlying (unknown) data-generating mechanism $\mathbb{P}$, one would like to find
the \textit{minimum directed distance/divergence} (cf. (iii) in Section \ref{SectMain})

\begin{equation}
\Phi_{\mathbb{P}}(\mathbb{\Omega}) := \inf_{\mathbb{Q}\in\mathbb{\Omega}} D( \mathbb{Q}, \mathbb{P} )
\label{inf proba new}
\end{equation}
which quantifies the \textit{adequacy} \footnote{
also called \textit{misspecification error}}
of the model
$\mathbb{\Omega}$ for modelling the true unknown data-generating mechanism $\mathbb{P}$, 
\textit{via} the minimal distance/dissimilarity of $\mathbb{\Omega}$ to $\mathbb{P}$;
a lower $\Phi_{\mathbb{P}}-$value means a better adequacy
(in the sense of a lower departure between the model and the truth,
cf. Lindsay \cite{Lind:04}, Lindsay et al. \cite{Lind:08},
Markatou \& Chen \cite{Mark:18}, 
Markatou \& Sofikitou \cite{Mark:19}). 
Hence, especially in the context of \textit{model selection} within 
complicated big-data contexts, for the \textit{search of
appropriate models} $\mathbb{\Omega}$ and model elements/members therein,
the (fast and efficient) computation of $\Phi_{\mathbb{P}}(\mathbb{\Omega})$ constitutes
a decisive first step, since
if the latter is \textquotedblleft  too large\textquotedblright\ (respectively 
\textquotedblleft  much larger than\textquotedblright\ 
$\Phi_{\mathbb{P}}(\overline{\mathbb{\Omega}})$ for some competing
model $\overline{\mathbb{\Omega}}$), then the model 
$\mathbb{\Omega}$ is \textquotedblleft  not adequate enough\textquotedblright\ (respectively
\textquotedblleft  much less adequate than\textquotedblright\ 
 $\overline{\mathbb{\Omega}}$);
in such a situation, the effort of computing the (not necessarily unique) 
best model
element/member 
$\arg \inf_{\mathbb{Q}\in\mathbb{\Omega}} D( \mathbb{Q}, \mathbb{P} )$
within the model $\mathbb{\Omega}$ 
is \textquotedblleft  not very useful\textquotedblright\ and is thus
a \textquotedblleft waste of computational time\textquotedblright.
Because of such considerations, we concentrate \textit{first}
on finding the infimum \eqref{inf proba new}; 
finding the corresponding \textit{minimizer(s)} 
$\arginf_{\mathbb{Q}\in\mathbb{\Omega}} D( \mathbb{Q}, \mathbb{P})$ ---
which can be interpreted as the (not necessarily unique)
model member $\mathbb{Q}\in\mathbb{\Omega}$ which
\textit{most adequately} describes the true unknown data-generating mechanism $\mathbb{P}$ ---
will be treated \textit{later} in Section \ref{SubsectEstimators.risk}.

\end{enumerate}

\vspace{0.3cm}
\noindent
Since $int(\mathbb{\Omega})$ is required to be a non-empty set 
(in the relative topology)
in the space of probability distributions on $\mathcal{Y}$, the present procedure is fitted for
semi-parametric models $\mathbb{\Omega}$, 
e.g. defined through
moment conditions (as extensions of the Empirical Likelihood paradigm, see
e.g. Broniatowski \& Keziou~\cite{Bro:12}), 
or through L-moment conditions
(i.e. moment conditions pertaining to quantile measures, 
see Broniatowski \& Decurninge \cite{Bro:16}), 
or even more involved non-parametric models
where the geometry of $\mathbb{\Omega }$ does not allow for ad-hoc
procedures.
In such setups, there is typically no closed form of
the divergence with respect to any probability distribution available.

\vspace{0.2cm}
\noindent
The measurement or the estimation of $\Phi _{\mathbb{P}}(\mathbb{\Omega})$ is a tool
for the choice of pertinent putative models $\mathbb{\Omega}$ among a class of
specifications.
The case when  $\Phi _{\mathbb{P}}(\mathbb{\Omega })>0$ \,  is interesting in its own, 
since it is quite common in engineering modelling to argue in
favor of misspecified models (or (non-void) neighborhoods of such models
for sake of robustness issues), due to quest for conservatism; the choice
between them is a widely open field e.g. in the practice of reliability.

\vspace{0.2cm}
\noindent
An estimate of $\Phi_{\mathbb{P}}(\mathbb{\Omega})$ can be used as a statistics for some
test of fit, and indeed the likelihood ratio test adapted to some semi-parametric models
has been generalized to the divergence setting (see Broniatowski \& Keziou~\cite{Bro:12}). 
The statement of the limit distributions of our estimate, under
the model and under misspecification, is postponed to future work.
In the following, we compute/approximate \eqref{inf proba new} --- and beyond --- by our 
\textit{bare simulation (BS)} method,
by appropriately adapting our methods of the previous Sections 
\ref{SectDetSubsimplex.CASM},\ref{SectDetSubsimplex.SBD},\ref{SectDetSubsimplex}.
To achieve this, let us first remark that, as usual, each probability
distribution (probability measure) $\mathbb{P}$ on $\mathcal{Y}=\left\{d_{1},\ldots,d_{K}\right\}$
can be uniquely identified with the (row) vector $\mathds{P} := (p_{1}, \ldots, p_{K}) \in \mathbb{S}^{K}$ 
of the corresponding probability masses (frequencies) 
$p_{k} = \mathbb{P}[\{ d_{k} \}]$ via 
$\mathbb{P}[B] = \sum_{k=1}^{K} p_{k} \cdot \textfrak{1}_{B}(d_{k}) $ for each $B \subset \mathcal{Y}$,
where $\textfrak{1}_{B}(\cdot)$ denotes the indicator function on the set $B$.
In particular, the probability distribution $\mathbb{P}$ in (MDE1) can be identified with 
$(p_{1}, \ldots, p_{K})$ in terms of $p_{k} = \mathbb{P}[\{ d_{k} \}]$
(which in the i.i.d. case turns into $p_{k} = \mathbb{\Pi}[X_{1} = d_{k}]$).
Along this line, the family
$\mathbb{\Omega}$ of probability distributions in (MDE2)
can be identified with a subset $\boldsymbol{\Omega}$\hspace{-0.23cm}$\boldsymbol{\Omega} 
 \subset  \mathbb{S}^{K}$ of probability vectors 
(viz. of vectors of probability masses).
Analogously, each finite nonnegative measure $Q$ on $\mathcal{Y}$
can be uniquely identified with a vector $\mathbf{Q} := (q_{1}, \ldots, q_{K}) \in \mathbb{R}_{\geq 0}^{K}$,
and each finite signed measure $Q$ with a vector 
$\mathbf{Q} := (q_{1}, \ldots, q_{K}) \in \mathbb{R}^{K}$.
The corresponding divergences between distributions/measures are then, as usual, defined 
through the divergences between their respective masses/frequencies:

\begin{equation}
D(Q,\mathbb{P}) : = D(\mathbf{Q},\mathds{P}).
\nonumber
\end{equation}

\vspace{0.2cm}
\noindent
In particular, $\mathbb{P}_{n}^{emp}(\mathbf{X}_{1}^{n})$ can be identified with 
the vector $\mathds{P}_{n}^{emp}(\mathbf{X}_{1}^{n}) := 
(p_{n,1}^{emp}(\mathbf{X}_{1}^{n}), \ldots, p_{n,K}^{emp}(\mathbf{X}_{1}^{n}))$ where
\begin{equation}
p_{n,k}^{emp}(\mathbf{X}_{1}^{n}) \ := \ 
\frac{1}{n} \cdot n_{k}(\mathbf{X}_{1}^{n}) 
\ := \ \frac{1}{n} \cdot card(\bigl\{ i \in \{ 1, \ldots, n\}:  \ X_{i} = d_{k} \bigr\})
\ =: \ \frac{1}{n} \cdot card(I_{k}^{(n)}(\mathbf{X}_{1}^{n})) , \quad k \in \{1, \ldots, K\},
\label{I^(n)_k for stat case}
\end{equation}
and accordingly the required limit behaviour \eqref{cv emp measure X to P}
is equivalent to the vector-convergence
\begin{equation}
\lim_{n\rightarrow \infty } \Big( \frac{n_{1}(\mathbf{X}_{1}^{n})}{n}, 
\ldots, \frac{n_{K}(\mathbf{X}_{1}^{n})}{n} \Big) 
= (p_{1}, \ldots, p_{K}) 
\qquad 
\textrm{a.s.}
\label{cv emp measure X to P vector}
\end{equation}
As usual, $\mathds{P}_{n}^{emp}(\mathbf{X}_{1}^{n})$ can be interpreted as the
\textit{vectorized histogram of the sample $\left( X_{1},\ldots,X_{n}\right)$}.

\noindent
Notice that, in contrast to 
the above Section \ref{SectDetSubsimplex},
the sets $I_{k}^{(n)}(\mathbf{X}_{1}^{n})
$ of indexes introduced in \eqref{I^(n)_k for stat case} 
and their numbers $n_{k}(\mathbf{X}_{1}^{n}) = card(I_{k}^{(n)})$ of elements are now
\textit{random} (due to their dependence on the $X_{i}$\textquoteright s).
Of course, $M_{\mathds{P}_{n}^{emp}(\mathbf{X}_{1}^{n})}=1$.
In a \textit{batch procedure}, when 
inference is done once/after the sample  $\mathbf{X}_{1}^{n} = \left(X_{1},\ldots,X_{n}\right)$ is observed, 
we may reorder this sample by putting the $n_{1}$ sample points $X_{i}$ which are equal to $d_{1}$ 
in the first places, and so on; accordingly one ends up with index sets $I_{k}^{(n)}(\mathbf{X}_{1}^{n})$ 
which are completely analogue to the ones used in the Sections \ref{SectDetSubsimplex.CASM},\ref{SectDetSubsimplex.SBD},\ref{SectDetSubsimplex}. 
When the \textit{online acquisition} of the data $X_{i}$\textquoteright s
is required, then we usually do not reorder the sample, and the 
$I_{k}^{(n)}(\mathbf{X}_{1}^{n})$\textquoteright s do
generally not consist in consecutive indexes, which does not make any change with
respect to the resulting construction nor to the estimator.
Let us first concentrate on the well-known CASM $\varphi-$divergences
$D(\cdot,\cdot) := D_{\varphi}(\cdot,\cdot)$ --- in the probability-vector form context ---
and solve the following

\vspace{0.2cm}
\noindent

\begin{problem}
\label{det Problem simplex risk}
For pregiven $\varphi \in \widetilde{\Upsilon}(]a,b[)$, 
unknown zeros-free probability vector $\mathds{P}:=\left( p_{1},..,p_{K}\right) \in \mathbb{S}_{> 0}^{K}$
satisfying \eqref{cv emp measure X to P vector},
and subset $\textrm{$\boldsymbol{\Omega}$\hspace{-0.23cm}$\boldsymbol{\Omega}$} \subset \mathbb{S}^{K}$
with regularity properties --- in the relative topology (!!) ---
\begin{equation}
cl(\textrm{$\boldsymbol{\Omega}$\hspace{-0.23cm}$\boldsymbol{\Omega}$})=
cl\left( int\left(\textrm{$\boldsymbol{\Omega}$\hspace{-0.23cm}$\boldsymbol{\Omega}$} \right) \right) ,  
\qquad int\left(\textrm{$\boldsymbol{\Omega}$\hspace{-0.23cm}$\boldsymbol{\Omega}$} \right) \ne \emptyset
\qquad \textrm{(cf. \eqref{regularity simplex})},
\nonumber
\end{equation}
find 
\begin{equation}
\Phi_{\mathds{P}}(\textrm{$\boldsymbol{\Omega}$\hspace{-0.23cm}$\boldsymbol{\Omega}$}) := 
\inf_{\mathds{Q}\in \textrm{$\boldsymbol{\Omega}$\hspace{-0.19cm}$\boldsymbol{\Omega}$}} 
D_{\varphi }(\mathds{Q},\mathds{P}),  
\nonumber
\end{equation}
provided that 
\begin{equation}
\inf_{\mathds{Q}\in \textrm{$\boldsymbol{\Omega}$\hspace{-0.19cm}$\boldsymbol{\Omega}$} } 
D_{\varphi }(\mathds{Q},\mathds{P}) < \infty 
\nonumber
\end{equation}
and that the divergence generator $\varphi$  additionally satisfies the Condition  
\ref{Condition  Fi Tilda in Minimization simplex}
(i.e. the representability \eqref{brostu5:fo.link.var.simplex} 
holds with probability distribution $\mathbb{\bbzeta}$).

\end{problem}

\vspace{0.2cm}
\noindent
To tackle Problem \ref{det Problem simplex risk}, we proceed analogously to Section \ref{SectDetSubsimplex.CASM}
and employ
a family of random variables $(W_{i})_{i \in  \mathbb{N}}$  
of independent and identically distributed $\mathbb{R}-$valued random variables
with probability distribution $\mathbb{\bbzeta}[ \cdot \, ] := \mathbb{\Pi}[W_{1} \in \cdot \, ]$
--- being connected with the divergence generator $\varphi \in \widetilde{\Upsilon}(]a,b[)$ via the representability
\eqref{brostu5:fo.link.var.simplex} --- 
such that $(W_{i})_{i \in  \mathbb{N}}$ is independent of $(X_{i})_{i \in  \mathbb{N}}$
\footnote{
notice that all the $W_{i}$ and $X_{i}$ live on the \textit{same} underlying probability space $(\mathfrak{X},\mathcal{A},\mathbb{\Pi})$
}. Moreover, we use
\begin{eqnarray}
\boldsymbol{\xi}_{n,\mathbf{X}}^{w\mathbf{W}} &:=&
\begin{cases}
\left(\frac{\sum_{i \in I_{1}^{(n)}(\mathbf{X}_{1}^{n})}W_{i}}{\sum_{k=1}^{K}
\sum_{i \in I_{k}^{(n)}(\mathbf{X}_{1}^{n})}W_{i}},
\ldots, \frac{\sum_{i \in I_{K}^{(n)}(\mathbf{X}_{1}^{n})}W_{i}}{\sum_{k=1}^{K}
\sum_{i \in I_{k}^{(n)}(\mathbf{X}_{1}^{n})}W_{i}} \right) ,
\qquad \textrm{if } \sum_{j=1}^{n} W_{j} \ne 0, \\
\ (\infty, \ldots, \infty) =: \boldsymbol{\infty}, \hspace{5.1cm} \textrm{if } \sum_{j=1}^{n} W_{j} = 0.
\end{cases}
\label{brostu5:fo.norweiemp.vec.risk} 
\end{eqnarray}
Notice that the right-hand side of \eqref{brostu5:fo.norweiemp.vec.risk} 
structurally coincides
with the right-hand side of \eqref{brostu5:fo.norweiemp.vec.det}, however 
--- as explained above --- the construction 
of the involved partition $I_{k}^{(n)}(\mathbf{X}_{1}^{n})$ differs 
from the non-random $I_{k}^{(n)}$
which we also indicate with (respective) different indexing on the corresponding
left-hand sides.

\vspace{0.3cm}
\noindent
With the above-mentioned ingredients, we have proven (cf. Theorem 12 of \cite{Bro:23a})
a more general, conditional-expectations-involving
version of Theorem \ref{brostu3:thm.divnormW.new.det}; from this, we have proceeded analogously
to the derivation of the Theorem \ref{brostu5:thm.divnormW.new}.
Indeed, as mentioned above, the required representability \eqref{brostu5:fo.link.var.simplex} 
is satisfied for all (multiple of) the generators
$\widetilde{c} \cdot \varphi_{\gamma}(\cdot)$ of \eqref{brostu5:fo.powdivgen}
with $\widetilde{c} \in \, ]0,\infty[$ and $\gamma \in \mathbb{R}\backslash\, ]1,2[$.
In terms of the auxiliary notations in (S1) to (S5) with $A:=1$
and the \textit{conditional} distributions
$\mathbb{\Pi}_{n}[\, \cdot \, ] := \mathbb{\Pi}_{\mathbf{X}_{1}^{n}}[\, \cdot \, ]  := \mathbb{\Pi}[ \, \cdot \, | \, 
X_{1}, \ldots, X_{n} ]$
\footnote{
\label{foot.condprob}
recall from basic probability theory that for any event $B \in \mathcal{A}$
the conditional probability
$\mathbb{\Pi}_{\mathbf{X}_{1}^{n}}[\, B \, ]  := \mathbb{\Pi}[ \, B \, | \, 
X_{1}, \ldots, X_{n} ]$
can in particular be rewritten as $h(X_{1}, \ldots, X_{n})$ for some ($B-$dependent) measurable function 
$h: \mathcal{Y}^{n} \mapsto \mathbb{R}$; accordingly, 
for observed \textit{concrete data}
$\mathbf{x}_{1}^{n} := \left( x_{1},\ldots,x_{n}\right)$
(which are, as usual, interpreted as realizations of 
the random variables $\mathbf{X}_{1}^{n} = \left( X_{1},\ldots,X_{n}\right)$)
one takes
$\mathbb{\Pi}_{\mathbf{x}_{1}^{n}}[\, B \, ] := 
\mathbb{\Pi}[ \, B \, | \, X_{1}=x_{1}, \ldots, X_{n}=x_{n} ]
:= h(x_{1}, \ldots, x_{n})$.
The corresponding expectations will be denoted by
$\mathbb{E}_{\mathbb{\Pi}_{\mathbf{X}_{1}^{n}}}[\, \cdot \, ]
:= \mathbb{E}_{\mathbb{\Pi}}[ \, \cdot \, | \, 
X_{1}, \ldots, X_{n} ]$ \, respectively \, 
$\mathbb{E}_{\mathbb{\Pi}_{\mathbf{x}_{1}^{n}}}[\, \cdot \, ]
:= \mathbb{E}_{\mathbb{\Pi}}[ \, \cdot \, | \, 
x_{1}, \ldots, x_{n} ]$. 
},
we have obtained in in Broniatowski \& Stummer \cite{Bro:23a} the following assertion
on BS-minimizability in the narrow sense:

\vspace{0.2cm}
\noindent

\begin{theorem}
\label{brostu5:thm.divnormW.new.risk} 
Suppose that $(X_{i})_{i\in \mathbb{N}}$ is a sequence of random variables
with values in $\mathcal{Y}:=\left\{ d_{1},\cdots ,d_{K}\right\}$ 
such that \eqref{cv emp measure X to P vector}
holds for some probability vector $\mathds{P} \in \mathbb{S}_{> 0}^{K}$.
Furthermore, we arbitrarily fix $\widetilde{c} \in \, ]0,\infty[$ and $\gamma \in \mathbb{R}\backslash\, ]1,2[$.
Moreover, let $(W_{i})_{i \in  \mathbb{N}}$  be  a family
of independent and identically distributed $\mathbb{R}-$valued random variables
with probability distribution $\mathbb{\bbzeta}[ \cdot \, ] := \mathbb{\Pi}[W_{1} \in \cdot \, ]$
being connected with the divergence generator $\widetilde{c} \cdot \varphi_{\gamma} \in \Upsilon(]a,b[)$ 
via the representability
\eqref{brostu5:fo.link.var.simplex},
such that $(W_{i})_{i \in  \mathbb{N}}$ is independent of $(X_{i})_{i \in  \mathbb{N}}$.\\
(a) Then there holds 
\begin{align}
& \inf_{\mathds{Q}\in \textrm{$\boldsymbol{\Omega}$\hspace{-0.19cm}$\boldsymbol{\Omega}$} }
F_{\gamma,\widetilde{c},1}\Big(D_{\widetilde{c} \cdot \varphi_{\gamma}}(\mathds{Q},\mathds{P})\Big)  
= -\lim_{n\rightarrow \infty }\frac{1}{n}\log \, 
\mathbb{\Pi}_{\mathbf{X}_{1}^{n}}\negthinspace \left[\boldsymbol{\xi}_{n,\mathbf{X}}^{w\mathbf{W}}\in 
\textrm{$\boldsymbol{\Omega}$\hspace{-0.23cm}$\boldsymbol{\Omega}$}\right]
\nonumber
\end{align}
for all sets $\boldsymbol{\Omega}$\hspace{-0.23cm}$\boldsymbol{\Omega} \subset \widetilde{\mathcal{M}}_{\gamma}$ satisfying the regularity properties 
\eqref{regularity simplex} \textit{in the relative topology}.
In particular, 
for each such $\mathds{P} \in \mathbb{S}_{>0}^{K}$ with \eqref{cv emp measure X to P vector}
the function 
$\Phi_{\mathds{P}}(\cdot) := 
F_{\gamma,\widetilde{c},1}\Big(D_{\widetilde{c} \cdot \varphi_{\gamma}}(\cdot,\mathds{P})\Big)$
is bare-simulation minimizable (BS-minimizable)
in the narrow sense 
(cf. \eqref{brostu5:fo.2} in Definition \ref{brostu5:def.1}
and the special case of Remark \ref{brostu5:rem.def1}(a)) 
on all sets 
$\textrm{$\boldsymbol{\Omega}$\hspace{-0.23cm}$\boldsymbol{\Omega}$} \subset \widetilde{\mathcal{M}}_{\gamma}$ 
satisfying \eqref{regularity simplex} \textit{in the relative topology}.\\
(b) Moreover, there holds 
\begin{align}
& \inf_{\mathds{Q}\in \textrm{$\boldsymbol{\Omega}$\hspace{-0.19cm}$\boldsymbol{\Omega}$} }
D_{\widetilde{c} \cdot \varphi_{\gamma}}(\mathds{Q},\mathds{P})
= F_{\gamma,\widetilde{c},1}^{\leftarrow}\Big(
-\lim_{n\rightarrow \infty }\frac{1}{n}\log \, 
\mathbb{\Pi}_{\mathbf{X}_{1}^{n}}\negthinspace \left[\boldsymbol{\xi}_{n,\mathbf{X}}^{w\mathbf{W}}\in 
\textrm{$\boldsymbol{\Omega}$\hspace{-0.23cm}$\boldsymbol{\Omega}$}\right]\Big)
= \lim_{n\rightarrow \infty } \, F_{\gamma,\widetilde{c},1}^{\leftarrow}\Big(
-\frac{1}{n}\log \, 
\mathbb{\Pi}_{\mathbf{X}_{1}^{n}}\negthinspace \left[\boldsymbol{\xi}_{n,\mathbf{X}}^{w\mathbf{W}}\in 
\textrm{$\boldsymbol{\Omega}$\hspace{-0.23cm}$\boldsymbol{\Omega}$}\right]\Big)
\label{LDP Normalized Vec simplex 2 risk}
\end{align}
for all sets $\boldsymbol{\Omega}$\hspace{-0.23cm}$\boldsymbol{\Omega} \subset \widetilde{\mathcal{M}}_{\gamma}$ satisfying the regularity properties \eqref{regularity simplex} \textit{in the relative topology}.
In particular, 
for each such $\mathds{P} \in \mathbb{S}_{>0}^{K}$ with \eqref{cv emp measure X to P vector}
the function 
$\Phi_{\mathds{P}}(\cdot) := D_{\widetilde{c} \cdot \varphi_{\gamma}}(\cdot,\mathds{P})$
is bare-simulation minimizable (BS-minimizable)
in the narrow sense on all sets 
$\textrm{$\boldsymbol{\Omega}$\hspace{-0.23cm}$\boldsymbol{\Omega}$} \subset \widetilde{\mathcal{M}}_{\gamma}$ 
satisfying \eqref{regularity simplex} \textit{in the relative topology}.  

\end{theorem}

\vspace{0.2cm}
\noindent

\begin{remark} \ 
For an equivalent version of Theorem \ref{brostu5:thm.divnormW.new.risk} in terms of 
probability \textit{distributions} $\mathbb{Q} \in \mathbb{\Omega}$
rather than probability \textit{vectors} $\mathds{Q} \in \textrm{$\boldsymbol{\Omega}$\hspace{-0.23cm}$\boldsymbol{\Omega}$}$,
the reader is referred to
Broniatowski \& Stummer \cite{Bro:23a} (cf. Theorem 12, Formula (39) and Lemma 14 therein).
\end{remark} 

\vspace{0.3cm} 
\noindent
The limit statement \eqref{LDP Normalized Vec simplex 2 risk}
provides the principle for the \textit{estimation} 
--- based on the random sample $\mathbf{X}_{1}^{n} = \left(X_{1},\ldots,X_{n}\right)$ ---
of the solution of the minimization problem
\begin{equation}
\Phi_{\mathds{P}}(\textrm{$\boldsymbol{\Omega}$\hspace{-0.23cm}$\boldsymbol{\Omega}$}) := 
\inf_{\mathds{Q}\in \textrm{$\boldsymbol{\Omega}$\hspace{-0.19cm}$\boldsymbol{\Omega}$} }
D_{\widetilde{c} \cdot \varphi_{\gamma}}(\mathds{Q},\mathds{P})
\nonumber
\end{equation}
which quantifies the adequacy (misspecification error)
of the model
$\textrm{$\boldsymbol{\Omega}$\hspace{-0.23cm}$\boldsymbol{\Omega}$}$ 
for modelling the true unknown data-generating mechanism $\mathbb{P}$. 
Indeed, by replacing the right-hand side in 
\eqref{LDP Normalized Vec simplex 2 risk} by its finite
counterpart, we deduce for given large $n$  
(with a slight abuse of notation)
\begin{equation}
\underline{M}_{n}^{BS,PD}\Big(X_{1},\ldots,X_{n}\Big) := 
\underline{M}_{n}^{BS,PD}\Big(\mathds{P}_{n}^{emp}(\mathbf{X}_{1}^{n})\Big) := 
 F_{\gamma,\widetilde{c},1}^{\leftarrow}\Big(
-\frac{1}{n}\log \, 
\mathbb{\Pi}_{\mathbf{X}_{1}^{n}}\negthinspace \left[\boldsymbol{\xi}_{n,\mathbf{X}}^{w\mathbf{W}}\in 
\textrm{$\boldsymbol{\Omega}$\hspace{-0.23cm}$\boldsymbol{\Omega}$}\right]\Big)
\ \approx \ 
\inf_{\mathds{Q}\in \textrm{$\boldsymbol{\Omega}$\hspace{-0.19cm}$\boldsymbol{\Omega}$} }
D_{\widetilde{c} \cdot \varphi_{\gamma}}(\mathds{Q},\mathds{P}),
\label{fo.approx.1.risk} 
\end{equation}
where $PD$ is the abbreviation for \textit{power divergences}; it remains to estimate the involved conditional probability
$\mathbb{\Pi}_{\mathbf{X}_{1}^{n}}\negthinspace \left[\boldsymbol{\xi}_{n,\mathbf{X}}^{w\mathbf{W}}\in 
\textrm{$\boldsymbol{\Omega}$\hspace{-0.23cm}$\boldsymbol{\Omega}$}\right]$
in \eqref{fo.approx.1.risk}.
The latter can be performed either by a \textit{naive estimator} of the
frequency of those replications of $\boldsymbol{\xi}_{n,\mathbf{X}}^{\mathbf{W}}$  
which hit $\textrm{$\boldsymbol{\Omega}$\hspace{-0.23cm}$\boldsymbol{\Omega}$}$, 
or more efficiently by some improved estimator; 
for details, the reader is referred to Section X of Broniatowski \& Stummer \cite{Bro:23a}, see   
also the corresponding extensions given in Section \ref{SubsectEstimators.risk}
below, where the latter also provides e.g. estimates of the \textit{minimizers}).


\subsection{The pure data-analytic view on the risk case}
\label{SectStochSubsimplex.CASM.pure}

\noindent
Let us begin with the remark that --- by construction\footnote{
$\mathds{P}_{n}^{emp}(\mathbf{X}_{1}^{n})$ only enters in the construction
of the partition $I_{k}^{(n)}(\mathbf{X}_{1}^{n})$ ($k=1,\ldots,K$)
} --- the above-mentioned sample-based function (statistical functional)
$\underline{M}_{n}^{BS,PD}\Big(X_{1},\ldots,X_{n}\Big)$
\textit{does not} coincide with the sample-based function
(with slight abuse of notation)

\begin{equation}
\underline{FM}_{n}^{BS,PD}\Big(X_{1},\ldots,X_{n}\Big) := 
\underline{FM}_{n}^{BS,PD}\Big(\mathds{P}_{n}^{emp}(\mathbf{X}_{1}^{n})\Big) := 
\inf_{\mathds{Q}\in \textrm{$\boldsymbol{\Omega}$\hspace{-0.19cm}$\boldsymbol{\Omega}$} }
D_{\widetilde{c} \cdot \varphi_{\gamma}}\Big(\mathds{Q},\mathds{P}_{n}^{emp}(\mathbf{X}_{1}^{n})\Big)
\nonumber
\end{equation}
which quantifies the adequacy (misspecification error)
of the model
$\textrm{$\boldsymbol{\Omega}$\hspace{-0.23cm}$\boldsymbol{\Omega}$}$ for describing --- 
the vectorized histogram of ---
the \textit{sample} $X_{1},\ldots,X_{n}$; the corresponding minimizer 
corresponds to the prominent (non-parametric) \textit{sample-based minimum (power) divergence
estimator}. For more details on the latter, see Section \ref{SubsectEstimators.risk} 
(and in particular, Remark \ref{rem.FasDivergence.estimator.risk}) below.

\vspace{0.2cm}
\noindent
Now suppose that we have observed some \textit{concrete data}
$\mathbf{x}_{1}^{n} := \left( x_{1},\ldots,x_{n}\right)$
(which are, as usual, interpreted as realizations of 
the random variables $\mathbf{X}_{1}^{n} = \left( X_{1},\ldots,X_{n}\right)$).
Moreover, let us summarize the data by the corresponding vectorized histogram
$\mathds{P}_{n}^{emp}(\mathbf{x}_{1}^{n}) \in \mathbb{S}_{>0}^{K}$
(or, more generally, any data-dependent vector
$\mathds{P}_{n}(\mathbf{x}_{1}^{n}) \in \mathbb{S}_{>0}^{K}$)
so that (for large enough $n$) each data value $d_{k}$ ($k=1,\ldots,K$) is observed at least once.
Independently of the nature of the true unknown original data-generating meachanism
$\mathds{P}$ (and thus, independently on the validity of \eqref{cv emp measure X to P vector}),
let us find --- as usual --- 
the adequacy (misspecification error)
of the model
$\textrm{$\boldsymbol{\Omega}$\hspace{-0.23cm}$\boldsymbol{\Omega}$}$ for describing --- 
the vectorized histogram of ---
the \textit{concrete data} $x_{1},\ldots,x_{n}$; in other words, 
in the following let us solve
the \textit{deterministic} minimization problem
(also called estimation of the \textit{empirical} risk)
\begin{equation}
\underline{FM}_{n}^{BS,PD}\Big(x_{1},\ldots,x_{n}\Big) = 
\underline{FM}_{n}^{BS,PD}\Big(\mathds{P}_{n}^{emp}(\mathbf{x}_{1}^{n})\Big) = 
\inf_{\mathds{Q}\in \textrm{$\boldsymbol{\Omega}$\hspace{-0.19cm}$\boldsymbol{\Omega}$} }
D_{\widetilde{c} \cdot \varphi_{\gamma}}\Big(\mathds{Q},\mathds{P}_{n}^{emp}(\mathbf{x}_{1}^{n})\Big);
\label{fo.rem.unequal.statistics.powerdiv.det} 
\end{equation}
the corresponding minimizer, i.e. the \textit{data-based} minimum (power) divergence
estimator, will be given in Section \ref{SubsectEstimators.risk} below.

\vspace{0.2cm}
\noindent
The deterministic minimization problem \eqref{fo.rem.unequal.statistics.powerdiv.det} 
can be treated via our BS-method of the above Section \ref{SectDetSubsimplex.CASM},
by taking $\mathbf{P} := \mathds{P}_{n}^{emp}(\mathbf{x}_{1}^{n})$
and thus $A=M_{\mathds{P}_{n}^{emp}(\mathbf{x}_{1}^{n})}=1$ (and accordingly, we
omit the tildes for the involved variables). Since the index $n$
now plays a completely different role as in Section \ref{SectDetSubsimplex.CASM},
we introduce a \textit{new additional index} $m \in \mathbb{N}$ which will be the
analogue of the approximation-step-indicating index $n$ used in Section \ref{SectDetSubsimplex.CASM}.
Accordingly, for any $m \in \mathbb{N}$ and 
any $k \in \left\{ 1, \ldots ,K-1\right\}$, let $m_{k}(\mathbf{x}_{1}^{n}):=\lfloor m \cdot 
p_{n,k}^{emp}(\mathbf{x}_{1}^{n})\rfloor $ 
and $m_{K}(\mathbf{x}_{1}^{n}) := m- \sum_{k=1}^{K-1} m_{k}(\mathbf{x}_{1}^{n})$;
for this, we assume that $m \in \mathbb{N}$ is large enough, 
namely
$m \geq \max_{k \in \{1, \ldots, K\}} \frac{1}{p_{n,k}^{emp}(\mathbf{x}_{1}^{n})}$,
such that all the integers $m_{k}(\mathbf{x}_{1}^{n})$ ($k=1,\ldots,K$) are
non-zero (recall that we have $p_{n,k}^{emp}(\mathbf{x}_{1}^{n}) >0$
for large enough $n \in \mathbb{N}$). Clearly, one gets

\begin{equation}
\lim_{m\rightarrow \infty} \frac{m_{k}(\mathbf{x}_{1}^{n})}{m} 
= p_{n,k}^{emp}(\mathbf{x}_{1}^{n}), \qquad k=1,\ldots,K.
\label{fo.freqlim.simplex.pnemp}
\end{equation}
With this at hand,
we decompose the set $\{1, \ldots, m\}$ of all integers from $1$ to $m$
into the disjoint blocks $I_{1}^{(m)}(\mathbf{x}_{1}^{n}):=\left\{
1,\ldots ,m_{1}(\mathbf{x}_{1}^{n})\right\}$, 
$I_{2}^{(m)}(\mathbf{x}_{1}^{n}):=\left\{ m_{1}(\mathbf{x}_{1}^{n})+1,\ldots
,m_{1}(\mathbf{x}_{1}^{n})+m_{2}(\mathbf{x}_{1}^{n})\right\} $, and so on until the last block \\
$I_{K}^{(m)}(\mathbf{x}_{1}^{n}) := \{ \sum_{k=1}^{K-1} m_{k}(\mathbf{x}_{1}^{n}) + 1, \ldots, m \}$. 
Thus, $I_{k}^{(m)}(\mathbf{x}_{1}^{n})$ has $m_{k}(\mathbf{x}_{1}^{n}) \geq 1$ elements. 
From this, we construct

\begin{eqnarray}
\boldsymbol{\xi}_{n,m,\mathbf{x}}^{w\mathbf{W}} &:=&
\begin{cases}
\left(\frac{\sum_{i \in I_{1}^{(m)}(\mathbf{X}_{1}^{n})} W_{i}}{\sum_{k=1}^{K}
\sum_{i \in I_{k}^{(m)}(\mathbf{X}_{1}^{n})} W_{i}},
\ldots, \frac{\sum_{i \in I_{K}^{(m)}(\mathbf{X}_{1}^{n})} W_{i}}{\sum_{k=1}^{K}
\sum_{i \in I_{k}^{(m)}(\mathbf{X}_{1}^{n})} W_{i}} \right) ,
\qquad \textrm{if } \sum_{j=1}^{m} W_{j} \ne 0, \\
\ (\infty, \ldots, \infty) =: \boldsymbol{\infty}, \hspace{5.1cm} \textrm{if } \sum_{j=1}^{m} W_{j} = 0.
\end{cases}
\label{brostu5:fo.norweiemp.vec.risk.pnemp} 
\end{eqnarray}

\vspace{0.3cm}

\begin{remark} 
\label{brostu5:rem.pnemp}
Alternatively to the above procedure, we could 
also introduce a \textit{new additional index} 
(with a slight abuse of notation)
$m \in \mathbb{N}$ such that
$m\cdot n$ (rather than $m$ itself)
will be the analogue of the approximation-step-indicating index $n$ used in Section \ref{SectDetSubsimplex.CASM}.
Indeed, by construction,
all the $p_{n,k}^{emp}(\mathbf{x}_{1}^{n})$ are
rational numbers and $n$ is the smallest integer such that all
$n \cdot p_{n,k}^{emp}(\mathbf{x}_{1}^{n}) =  n_{k}(\mathbf{x}_{1}^{n})$ ($k=1,\ldots,K$) are 
integers (which for large enough data size $n$ are non-zero); 
consequently, also all $m \cdot n \cdot p_{n,k}^{emp}(\mathbf{x}_{1}^{n}) =  
m \cdot n_{k}(\mathbf{x}_{1}^{n})$ are (non-zero) integers.
With this, we replace the partitions $I_{k}^{(n)}$ (having $n_{k}$ elements, cf. \eqref{I^(n)_k for stat case})
by the new \textquotedblleft $m-$fold blown-up\textquotedblright\
partitions $A_{k}^{(m \cdot n)}(\mathbf{x}_{1}^{n})$ of size $m_{k}(\mathbf{x}_{1}^{n}):= m \cdot n_{k}(\mathbf{x}_{1}^{n})$
and thus, $\frac{m_{k}(\mathbf{x}_{1}^{n})}{m \cdot n}= p_{n,k}^{emp}(\mathbf{x}_{1}^{n}) 
$, \, $k \in \{1,\ldots,K\}$
(which means that here the appropriately adapted condition \eqref{fo.freqlim.simplex.pnemp}
holds even pointwise rather than only in the limit).
More detailed, 
in the above-mentioned \textit{batch procedure}  
we may reorder such that 
$A_{1}^{(m\cdot n)}(\mathbf{x}_{1}^{n}) := \{1, \ldots, m \cdot n_{1}(\mathbf{x}_{1}^{n}) \}$,
$A_{2}^{(m\cdot n)}(\mathbf{x}_{1}^{n}) := \{m \cdot n_{1}(\mathbf{x}_{1}^{n}) + 1, \ldots, 
m \cdot (n_{1}(\mathbf{x}_{1}^{n}) + n_{2}(\mathbf{x}_{1}^{n})) \}$,
\ldots, $A_{K}^{(m \cdot n)}(\mathbf{x}_{1}^{n}) := 
\{m \cdot \sum_{i=1}^{K-1} n_{i}(\mathbf{x}_{1}^{n}) + 1, \ldots, 
m \cdot \sum_{i=1}^{K} n_{i} (\mathbf{x}_{1}^{n})\}$,
such that $\bigcup_{k=1}^{K} A_{k}^{(m\cdot n)}(\mathbf{x}_{1}^{n}) = \{1, \ldots, m \cdot n\}$.
When the above-mentioned \textit{online acquisition} of the data $x_{i}$\textquoteright s
is required, then we usually do not reorder but after the appearance of a data point (say) 
$x_{n+1}=d_{k}$
in the $(n+1)-$th concrete data acquisition
we immediately add $m$ \textit{new} (e.g. the next available) indices to the old index set 
$A_{k}^{(m\cdot n)}(\mathbf{x}_{1}^{n})$ to end up with
the new index subset $A_{k}^{(m\cdot (n+1))}(\mathbf{x}_{1}^{n+1})$
(and keep all the other index subsets the same,
i.e. $A_{j}^{(m\cdot (n+1))}(\mathbf{x}_{1}^{n+1}) := A_{j}^{(m\cdot n)}(\mathbf{x}_{1}^{n})$
for $j \ne k$). Accordingly,  
the $A_{k}^{(m\cdot n)}(\mathbf{x}_{1}^{n})$'s 
do generally not consist in consecutive indexes, which does not make any change with
respect to the resulting construction. 
With all this in hand, in the following investigations we can alternatively employ
\begin{eqnarray}
\boldsymbol{\xi}_{n,m,\mathbf{x}}^{w\mathbf{W}} &:=&
\begin{cases}
\left(\frac{\sum_{i \in A_{1}^{(m\cdot n)}(\mathbf{x}_{1}^{n})}W_{i}}{\sum_{k=1}^{K}
\sum_{i \in A_{k}^{(m\cdot n)}(\mathbf{x}_{1}^{n})}W_{i}},
\ldots, \frac{\sum_{i \in A_{K}^{(m\cdot n)}(\mathbf{x}_{1}^{n})}W_{i}}{\sum_{k=1}^{K}
\sum_{i \in A_{k}^{(m\cdot n)}(\mathbf{x}_{1}^{n})}W_{i}} \right) ,
\qquad \textrm{if } \sum_{j=1}^{m \cdot n} W_{j} \ne 0, \\
\ (\infty, \ldots, \infty) =: \boldsymbol{\infty}, \hspace{5.7cm} \textrm{if } \sum_{j=1}^{m \cdot n} W_{j} = 0,
\end{cases}
\label{brostu5:fo.norweiemp.vec.risk.pnemp.var}
\end{eqnarray}
\textit{instead} of \eqref{brostu5:fo.norweiemp.vec.risk.pnemp} together with 
appropriately using $m \cdot n$ instead of $m$ for the corresponding rates and limit-takings;
such a construction has been employed in Broniatowski \& Stummer \cite{Bro:23a}. 

\end{remark}

\vspace{0.4cm}
\noindent
In terms of \eqref{brostu5:fo.norweiemp.vec.risk.pnemp} 
(with fixed $n$), we can apply Theorem \ref{brostu5:thm.divnormW.new}(b)
to end up with 
\begin{align}
& 
\underline{FM}_{n}^{BS,PD}\Big(x_{1},\ldots,x_{n}\Big) = 
\underline{FM}_{n}^{BS,PD}\Big(\mathds{P}_{n}^{emp}(\mathbf{x}_{1}^{n})\Big) 
= \inf_{\mathbf{Q}\in \textrm{$\boldsymbol{\Omega}$\hspace{-0.19cm}$\boldsymbol{\Omega}$} }
D_{\widetilde{c} \cdot \varphi_{\gamma}}(\mathbf{Q},\mathds{P}_{n}^{emp}(\mathbf{x}_{1}^{n}))
= F_{\gamma,\widetilde{c},1}^{\leftarrow}\Big(
-\lim_{m\rightarrow \infty } \frac{1}{m} \log \, 
\mathbb{\Pi}_{\mathbf{x}_{1}^{n}}
\negthinspace \left[\boldsymbol{\xi}_{n,m,\mathbf{x}}^{w\mathbf{W}} \in 
\textrm{$\boldsymbol{\Omega}$\hspace{-0.23cm}$\boldsymbol{\Omega}$}\right]\Big)
\label{LDP Normalized Vec simplex 2 risk det}
\end{align}
for all sets $\boldsymbol{\Omega}$\hspace{-0.23cm}$\boldsymbol{\Omega} \subset \widetilde{\mathcal{M}}_{\gamma}$ 
(with $A=1$) satisfying the 
regularity properties \eqref{regularity simplex} 
\textit{in the relative topology};
here, in \eqref{LDP Normalized Vec simplex 2 risk det} we have used the
abbreviation 
$\mathbb{\Pi}_{\mathbf{x}_{1}^{n}}[\, \cdot \, ]  := \mathbb{\Pi}[ \, \cdot \, | \, 
X_{1}=x_{1}, \ldots, X_{n}=x_{n} ]$ (recall footnote \ref{foot.condprob}).


\subsection{The combined view}
\label{SectStochSubsimplex.CASM.comb}

\vspace{0.2cm}
\noindent
Of course, one may also want to combine the statistical view
(cf. Subsection \ref{SectStochSubsimplex.CASM.stat})
with the pure data-analytic view on the risk case
(cf. Subsection \ref{SectStochSubsimplex.CASM.pure}).
This amounts to replace the concrete (deterministic) data
$\mathbf{x}_{1}^{n} := (x_{1}, \ldots, x_{n})$ with
the (random) sample $\mathbf{X}_{1}^{n} := (X_{1}, \ldots, X_{n})$
assuming \eqref{cv emp measure X to P}.
Then the following
coherence/consistency result
\begin{align}
& 
\Phi_{\mathds{P}}(\textrm{$\boldsymbol{\Omega}$\hspace{-0.23cm}$\boldsymbol{\Omega}$}) = 
\inf_{\mathds{Q}\in \textrm{$\boldsymbol{\Omega}$\hspace{-0.19cm}$\boldsymbol{\Omega}$} }
D_{\widetilde{c} \cdot \varphi_{\gamma}}(\mathds{Q},\mathds{P}) =
\lim_{n\rightarrow \infty }
\underline{FM}_{n}^{BS,PD}\Big(X_{1},\ldots,X_{n}\Big)  
= \lim_{n\rightarrow \infty} \lim_{m\rightarrow \infty}
F_{\gamma,\widetilde{c},1}^{\leftarrow}\Big(
-\frac{1}{m} \log \, 
\mathbb{\Pi}_{\mathbf{X}_{1}^{n}} \negthinspace \left[\boldsymbol{\xi}_{n,m,\mathbf{X}}^{w\mathbf{W}} \in 
\textrm{$\boldsymbol{\Omega}$\hspace{-0.23cm}$\boldsymbol{\Omega}$}\right]\Big)
\qquad a.s.
\label{LDP Normalized Vec simplex 2 risk det coh}
\end{align} 
should hold ---
under appropriate analytic conditions ---
for all compact sets $\boldsymbol{\Omega}$\hspace{-0.23cm}$\boldsymbol{\Omega} \subset \widetilde{\mathcal{M}}_{\gamma}$ 
(with $A=1$) 
satisfying \eqref{regularity simplex} and \eqref{def fi wrt Omega simplex}. 
Indeed, by denoting 
\begin{equation}
\mathbf{R}_{n} := \mathbf{R}_{n}(\mathbf{X}_{1}^{n}) := \mathds{P}_{n}^{emp}(\mathbf{X}_{1}^{n}),
\quad
\Phi_{\mathbf{R}_{n}}(\mathds{Q}) := 
D_{\widetilde{c} \cdot \varphi_{\gamma}}(\mathbf{Q},\mathbf{R}_{n}),
\quad
\mathbf{R} := \mathds{P},
\quad
\Phi_{\mathbf{R}}(\mathds{Q}) := 
D_{\widetilde{c} \cdot \varphi_{\gamma}}(\mathbf{Q},\mathbf{R}),
\nonumber
\end{equation}
the desired result \eqref{LDP Normalized Vec simplex 2 risk det coh} follows 
from \eqref{LDP Normalized Vec simplex 2 risk det} 
together with
\begin{equation}
\Phi_{\mathbf{R}_{n}}(\cdot) \ \textrm{converges (as $n$ tends to $\infty$) a.s. to} \ 
\Phi_{\mathbf{R}}(\cdot) \ \textrm{uniformly on}
\ \textrm{$\boldsymbol{\Omega}$\hspace{-0.23cm}$\boldsymbol{\Omega}$} \ 
\ \textrm{whenever} \ \mathbf{R}_{n} \ 
\textrm{converges (as $n$ tends to $\infty$) a.s. to} \ \mathbf{R}.
\label{consistency1}
\end{equation} 
As far as the latter (which will be also used in other contexts below) is concerned,
in the case where the minimizer-set 
$\mathcal{A} := 
\argmin_{\mathds{Q}\in \textrm{$\boldsymbol{\Omega}$\hspace{-0.19cm}$\boldsymbol{\Omega}$}}
\Phi_{\mathbf{R}}(\mathds{Q})$ basically consists of isolated points,
the a.s. convergence of the family of minimizers of $\Phi_{\mathbf{R}_{n}}$
towards $\mathcal{A}$ can be proved under adequate analytical conditions
(see e.g. Theorem 5.7 of Van der Vaart \cite{Vaa:98}). 
The latter can, for instance, be tackled by appropriately carrying over  
the parametric-case-concerning \textit{grand consistency theorem}
(i.e. Theorem 2.1) of Kuchibhotla \& Basu~\cite{Kuc:17}
to our non-parametric set-up with compact $\textrm{$\boldsymbol{\Omega}$\hspace{-0.23cm}$\boldsymbol{\Omega}$}$
(e.g. for the divergence generator 
$\varphi_{\alpha,\beta,\widetilde{c}}$
of Example \ref{brostu5:ex.2},
which satisfies 
$\varphi_{\alpha,\beta,\widetilde{c}}(0) + 
\varphi_{\alpha,\beta,\widetilde{c}}^{\prime}(\infty) < \infty$
and thus their Assumption (C2)).


\section{Narrow-Sense Bare-Simulation-Minimization of 
innmin-Bregman divergences under risk}
\label{SectStochSubsimplex.SBD}

\vspace{0.2cm}
\noindent
For scaled Bregman distances we first derive the following fundamental
risk-case-extension of Theorem \ref{brostu3:thm.divnormW.new.det.SBD}:

\vspace{0.3cm}

\begin{theorem}
\label{brostu3:thm.divnormW.new.det.SBD.risk} 
Suppose that $(X_{i})_{i\in \mathbb{N}}$ is a sequence of random variables
with values in $\mathcal{Y}:=\left\{ d_{1},\cdots ,d_{K}\right\}$ 
such that \eqref{cv emp measure X to P vector}
holds for some probability vector $\mathds{P} \in \mathbb{S}_{> 0}^{K}$.
Moreover, assume that the divergence generator $\varphi$ satisfies Condition 
\ref{Condition  Fi Tilda in Minimization simplex} and
let $\mathbf{Q}^{\ast\ast} \in \mathbb{R}^{K}$ be such that 
\eqref{brostu5:fo.SBD.qstarstar} holds.
Additionally, we suppose that $\boldsymbol{\Omega}$\hspace{-0.23cm}$\boldsymbol{\Omega} \subset \mathbb{S}^{K}$ 
satisfies the regularity properties \eqref{regularity simplex} in the relative topology as well as
the finiteness property
$\inf_{\mathds{Q}\in \textrm{$\boldsymbol{\Omega}$\hspace{-0.19cm}$\boldsymbol{\Omega}$}}
D_{\varphi,\mathds{P}}^{SBD}(\mathds{Q},\mathbf{Q}^{\ast\ast}) < \infty$
(cf. \eqref{def fi wrt Omega Bregman simplex}).
Moreover, let $V := (\mathbf{V}_{i})_{i \in \mathbb{N}}$ be a sequence of random vectors 
which are independent of $(X_{i})_{i\in \mathbb{N}}$ \footnote{
notice that all the $V_{i}$ and $X_{i}$ live on the \textit{same} underlying probability space $(\mathfrak{X},\mathcal{A},\mathbb{\Pi})$
}
and which are 
constructed by \eqref{brostu5:V_new} and \eqref{brostu5:Utilde_k_new} (where we
write $V$ instead of $\widetilde{V}$, since $M_{\mathds{P}}=1$ and thus $\widetilde{p}_{k}=
p_{k}/M_{\mathds{P}} = p_{k}$ as well as $\widetilde{\mathbb{\bbzeta}}=\mathbb{\bbzeta}$).
Moreover, let 
$I_{k}^{(n)}(\mathbf{X}_{1}^{n})
:= \bigl\{ i \in \{ 1, \ldots, n\}:  \ X_{i} = d_{k} \bigr\}$
($k=1,\ldots,K$) 
be the partitions constructed in \eqref{I^(n)_k for stat case}.
Then, in terms of 
the random vectors $\boldsymbol{\xi}_{n,\mathbf{X}}^{w\mathbf{V}}$ given by
\begin{eqnarray}
\boldsymbol{\xi}_{n,\mathbf{X}}^{w\mathbf{V}} &:=&
\begin{cases}
\left(\frac{\sum_{i \in I_{1}^{(n)}(\mathbf{X}_{1}^{n})}V_{i}}{\sum_{k=1}^{K}
\sum_{i \in I_{k}^{(n)}(\mathbf{X}_{1}^{n})}V_{i}},
\ldots, \frac{\sum_{i \in I_{K}^{(n)}(\mathbf{X}_{1}^{n})}V_{i}}{\sum_{k=1}^{K}
\sum_{i \in I_{k}^{(n)}(\mathbf{X}_{1}^{n})}V_{i}} \right) ,
\qquad \textrm{if } \sum_{j=1}^{n} V_{j} \ne 0, \\
\ (\infty, \ldots, \infty) =: \boldsymbol{\infty}, \hspace{5.0cm} \textrm{if } \sum_{j=1}^{n} V_{j} = 0,
\end{cases}
\label{brostu5:fo.norweiemp.vec.det.SBD.risk} 
\end{eqnarray}
there holds 
\begin{align}
& \inf_{\mathds{Q}\in \textrm{$\boldsymbol{\Omega}$\hspace{-0.19cm}$\boldsymbol{\Omega}$} }
\ \inf_{m\neq 0} 
D_{\varphi,\mathds{P}}^{SBD}(m \cdot \mathds{Q},\mathbf{Q}^{\ast\ast}) 
= 
\inf_{m\neq 0}\ \inf_{\mathds{Q}\in \textrm{$\boldsymbol{\Omega}$\hspace{-0.19cm}$\boldsymbol{\Omega}$} }
D_{\varphi,\mathds{P}}^{SBD}(m \cdot \mathds{Q},\mathbf{Q}^{\ast\ast})  
= -\lim_{n\rightarrow \infty }\frac{1}{n}\log \, 
\mathbb{\Pi}_{\mathbf{X}_{1}^{n}}\negthinspace \negthinspace\left[\boldsymbol{\xi}_{n,\mathbf{X}}^{w\mathbf{V}}\in 
\textrm{$\boldsymbol{\Omega}$\hspace{-0.23cm}$\boldsymbol{\Omega}$}\right].
\label{LDP Normalized Vec BS2 SBD risk}
\end{align}
\end{theorem}

\vspace{0.3cm}
\noindent
The proof of Theorem \ref{brostu3:thm.divnormW.new.det.SBD.risk}
will be given in Appendix \ref{App.A} below.

\vspace{0.3cm}
\noindent
By combining Theorem \ref{brostu3:thm.divnormW.new.det.SBD.risk}
with \eqref{brostu3:fo.676b.SBD}, \eqref{brostu3:fo.677a.SBD}
and \eqref{brostu3:fo.678.SBD}
we straightforwardly deduce the following

\vspace{0.3cm}
\noindent

\begin{theorem}
\label{brostu5:thm.divnormW.new.risk.SBD} 
Let $\mathds{P} \in \mathbb{S}_{> 0}^{K}$, $(X_{i})_{i\in \mathbb{N}}$, 
$V := (\mathbf{V}_{i})_{i \in \mathbb{N}}$, $\boldsymbol{\xi}_{n,\mathbf{X}}^{w\mathbf{V}}$
as in Theorem \ref{brostu3:thm.divnormW.new.det.SBD.risk},
for the special choice $\varphi := \widetilde{c} \cdot \varphi_{\gamma}$ 
(cf. \eqref{brostu5:fo.powdivgen}) with some arbitrarily fixed
$\widetilde{c} \in \, ]0,\infty[$, $\gamma \in \mathbb{R}\backslash\, ]1,2[$
together with some arbitrarily fixed
$\mathbf{Q}^{\ast\ast} \in \widetilde{\mathcal{N}}_{\gamma}$ (cf. (T1)) implying 
\eqref{brostu5:fo.SBD.qstarstar}. Then there holds 
\begin{align}
& \inf_{\mathds{Q}\in \textrm{$\boldsymbol{\Omega}$\hspace{-0.19cm}$\boldsymbol{\Omega}$} }
\breve{D}_{\widetilde{c} \cdot \varphi_{\gamma},\mathds{P}}^{SBD}( \mathds{Q},\mathbf{Q}^{\ast\ast}) 
= -\lim_{n\rightarrow \infty }\frac{1}{n}\log \, 
\mathbb{\Pi}_{\mathbf{X}_{1}^{n}}\negthinspace \left[\boldsymbol{\xi}_{n,\mathbf{X}}^{w\mathbf{V}}\in 
\textrm{$\boldsymbol{\Omega}$\hspace{-0.23cm}$\boldsymbol{\Omega}$}\right]
\label{LDP Normalized Vec simplex risk SBD}
\end{align}
for all sets $\boldsymbol{\Omega}$\hspace{-0.23cm}$\boldsymbol{\Omega} \subset \widetilde{\mathcal{M}}_{\gamma}$ 
(cf. (T1)) satisfying the 
regularity properties 
\eqref{regularity simplex} \textit{in the relative topology}.
In particular, 
for each such $\mathds{P} \in \mathbb{S}_{>0}^{K}$ with \eqref{cv emp measure X to P vector}
the function 
$\Phi_{\mathds{P}}(\cdot) := 
\breve{D}_{\widetilde{c} \cdot \varphi_{\gamma},\mathds{P}}^{SBD}( \cdot,\mathbf{Q}^{\ast\ast})$
is bare-simulation minimizable (BS-minimizable)
in the narrow sense 
(cf. \eqref{brostu5:fo.2} in Definition \ref{brostu5:def.1}
and the special case of Remark \ref{brostu5:rem.def1}(a)) 
on all sets 
$\textrm{$\boldsymbol{\Omega}$\hspace{-0.23cm}$\boldsymbol{\Omega}$} \subset \widetilde{\mathcal{M}}_{\gamma}$ 
satisfying 
\eqref{regularity simplex} 
\textit{in the relative topology}.\\

\end{theorem}

\vspace{0.2cm}
\noindent
Recall from Remark \ref{after theorem inner min Bregman power general} that
$\breve{D}_{\widetilde{c} \cdot \varphi_{\gamma},\mathds{P}}^{SBD}( \mathds{Q},\mathbf{Q}^{\ast\ast})$
is a divergence (which we have called innmin-SBD). Accordingly, 
based on the limit statement \eqref{LDP Normalized Vec simplex risk SBD}
(rather than \eqref{LDP Normalized Vec simplex 2 risk}) we can follow the spirit of the previous
Section \ref{SectStochSubsimplex.CASM}
to estimate --- based on random scaling-samples $\mathbf{X}_{1}^{n} = \left(X_{1},\ldots,X_{n}\right)$ 
satisfying \eqref{cv emp measure X to P} ---
the directed distance (divergence) from $\mathbf{Q}^{\ast\ast}$ to the model
$\textrm{$\boldsymbol{\Omega}$\hspace{-0.23cm}$\boldsymbol{\Omega}$}$ 
under a true unknown ``scaling-data-generating'' mechanism $\mathbb{P}$,
\textit{given by}

\begin{equation}
\Phi_{\mathds{P}}(\textrm{$\boldsymbol{\Omega}$\hspace{-0.23cm}$\boldsymbol{\Omega}$}) := 
\inf_{\mathds{Q}\in \textrm{$\boldsymbol{\Omega}$\hspace{-0.19cm}$\boldsymbol{\Omega}$} }
\breve{D}_{\widetilde{c} \cdot \varphi_{\gamma},\mathds{P}}^{SBD}( \mathds{Q},\mathbf{Q}^{\ast\ast}),
\nonumber
\end{equation}
via (with a slight abuse of notation)
\begin{equation}
\underline{M}_{n}^{BS,SBD}\Big(X_{1},\ldots,X_{n}\Big) := 
\underline{M}_{n}^{BS,SBD}\Big(\mathds{P}_{n}^{emp}(\mathbf{X}_{1}^{n})\Big) := 
-\frac{1}{n}\log \, 
\mathbb{\Pi}_{\mathbf{X}_{1}^{n}}\negthinspace \left[\boldsymbol{\xi}_{n,\mathbf{X}}^{w\mathbf{V}}\in 
\textrm{$\boldsymbol{\Omega}$\hspace{-0.23cm}$\boldsymbol{\Omega}$}\right]
\ \approx \ 
\inf_{\mathds{Q}\in \textrm{$\boldsymbol{\Omega}$\hspace{-0.19cm}$\boldsymbol{\Omega}$} }
\breve{D}_{\widetilde{c} \cdot \varphi_{\gamma},\mathds{P}}^{SBD}( \mathds{Q},\mathbf{Q}^{\ast\ast})
\label{fo.approx.1.risk.innminSBD} 
\end{equation}
together with an estimate the involved conditional probability
$\mathbb{\Pi}_{\mathbf{X}_{1}^{n}}\negthinspace \left[\boldsymbol{\xi}_{n,\mathbf{X}}^{w\mathbf{V}}\in 
\textrm{$\boldsymbol{\Omega}$\hspace{-0.23cm}$\boldsymbol{\Omega}$}\right]$
in \eqref{fo.approx.1.risk.innminSBD} (for given large $n$).
Moreover, in the spirit of Subsection \ref{SectStochSubsimplex.CASM.pure}
we can approximate  
the above-mentioned randomly-scaled directed distance
--- in terms of the vectorized histogram of 
the \textit{concrete scaling-data} $x_{1},\ldots,x_{n}$ --- by
\begin{equation}
\underline{FM}_{n}^{BS,SBD}\Big(x_{1},\ldots,x_{n}\Big) = 
\underline{FM}_{n}^{BS,SBD}\Big(\mathds{P}_{n}^{emp}(\mathbf{x}_{1}^{n})\Big) := 
\inf_{\mathds{Q}\in \textrm{$\boldsymbol{\Omega}$\hspace{-0.19cm}$\boldsymbol{\Omega}$} }
\breve{D}_{\widetilde{c} \cdot \varphi_{\gamma},\mathds{P}_{n}^{emp}(\mathbf{x}_{1}^{n})}^{SBD}( 
\mathds{Q},\mathbf{Q}^{\ast\ast}),
\label{fo.rem.unequal.statistics.innminSBD.det} 
\end{equation}
via an appropriate application of Theorem \ref{theorem inner min Bregman power general}
--- by substituting $\mathbf{P}$ by $\mathds{P}_{n}^{emp}(\mathbf{x}_{1}^{n})$
and $\mathbb{\Pi}[\, \cdot \, ]$ by $\mathbb{\Pi}_{\mathbf{x}_{1}^{n}}[\, \cdot \, ]$ ---
and end up with 

\begin{align}
& 
\underline{FM}_{n}^{BS,SBD}\Big(x_{1},\ldots,x_{n}\Big)   
=  -\lim_{m\rightarrow \infty }
\frac{1}{m} \log \, 
\mathbb{\Pi}_{\mathbf{x}_{1}^{n}}\negthinspace \left[\boldsymbol{\xi}_{n,m,\mathbf{x}}^{w\mathbf{V}} \in 
\textrm{$\boldsymbol{\Omega}$\hspace{-0.23cm}$\boldsymbol{\Omega}$}\right]
\label{LDP Normalized Vec simplex 2 risk det innminSBD}
\end{align}
for all sets $\boldsymbol{\Omega}$\hspace{-0.23cm}$\boldsymbol{\Omega} \subset \widetilde{\mathcal{M}}_{\gamma}$ 
(with $A=1$) satisfying the 
regularity properties \eqref{regularity simplex} 
\textit{in the relative topology}. 
Here, in \eqref{LDP Normalized Vec simplex 2 risk det innminSBD} we have employed

\begin{eqnarray}
\boldsymbol{\xi}_{n,m,\mathbf{x}}^{w\mathbf{V}} &:=&
\begin{cases}
\left(\frac{\sum_{i \in I_{1}^{(m)}(\mathbf{X}_{1}^{n})} V_{i}}{\sum_{k=1}^{K}
\sum_{i \in I_{k}^{(m)}(\mathbf{X}_{1}^{n})} V_{i}},
\ldots, \frac{\sum_{i \in I_{K}^{(m)}(\mathbf{X}_{1}^{n})} V_{i}}{\sum_{k=1}^{K}
\sum_{i \in I_{k}^{(m)}(\mathbf{X}_{1}^{n})} V_{i}} \right) ,
\qquad \textrm{if } \sum_{j=1}^{m} V_{j} \ne 0, \\
\ (\infty, \ldots, \infty) =: \boldsymbol{\infty}, \hspace{5.1cm} \textrm{if } \sum_{j=1}^{m} V_{j} = 0.
\end{cases}
\label{brostu5:fo.norweiemp.vec.risk.pnemp.innminSBD} 
\end{eqnarray}
where the partitions $I_{k}^{(m)}(\mathbf{x}_{1}^{n})$ are the same as in 
\eqref{brostu5:fo.norweiemp.vec.risk.pnemp}.

\vspace{0.3cm}

\begin{remark} 
In analogy with Remark \ref{brostu5:rem.pnemp},
alternatively to \eqref{brostu5:fo.norweiemp.vec.risk.pnemp.innminSBD} 
we can use
\begin{eqnarray}
\boldsymbol{\xi}_{n,m,\mathbf{x}}^{w\mathbf{V}} &:=&
\begin{cases}
\left(\frac{\sum_{i \in A_{1}^{(m\cdot n)}(\mathbf{x}_{1}^{n})}V_{i}}{\sum_{k=1}^{K}
\sum_{i \in A_{k}^{(m\cdot n)}(\mathbf{x}_{1}^{n})}V_{i}},
\ldots, \frac{\sum_{i \in A_{K}^{(m\cdot n)}(\mathbf{x}_{1}^{n})}V_{i}}{\sum_{k=1}^{K}
\sum_{i \in A_{k}^{(m\cdot n)}(\mathbf{x}_{1}^{n})}V_{i}} \right) ,
\qquad \textrm{if } \sum_{j=1}^{m \cdot n} V_{j} \ne 0, \\
\ (\infty, \ldots, \infty) =: \boldsymbol{\infty}, \hspace{5.5cm} \textrm{if } \sum_{j=1}^{m \cdot n} V_{j} = 0,
\end{cases}
\label{brostu5:fo.norweiemp.vec.risk.pnemp.var.innminSBD}
\end{eqnarray}
(where the partitions $A_{k}^{(m\cdot n)}(\mathbf{x}_{1}^{n})$ are the same as in 
\eqref{brostu5:fo.norweiemp.vec.risk.pnemp.var})
to derive
\begin{align}
& 
\underline{FM}_{n}^{BS,SBD}\Big(x_{1},\ldots,x_{n}\Big)  
= -\lim_{m\rightarrow \infty }\frac{1}{m \cdot n}\log \, 
\mathbb{\Pi}_{\mathbf{x}_{1}^{n}}\negthinspace \left[\boldsymbol{\xi}_{n,m,\mathbf{x}}^{w\mathbf{V}} \in 
\textrm{$\boldsymbol{\Omega}$\hspace{-0.23cm}$\boldsymbol{\Omega}$}\right] .
\nonumber
\end{align}

\end{remark}

\vspace{0.4cm}
\noindent
In the spirit of Subsection \ref{SectStochSubsimplex.CASM.comb},
by combining 
\eqref{fo.rem.unequal.statistics.innminSBD.det},
\eqref{LDP Normalized Vec simplex 2 risk det innminSBD}
and \eqref{consistency1}
with  
\begin{equation}
\mathbf{R}_{n} := \mathbf{R}_{n}(\mathbf{X}_{1}^{n}) := \mathds{P}_{n}^{emp}(\mathbf{X}_{1}^{n}),
\quad
\Phi_{\mathbf{R}_{n}}(\mathds{Q}) := 
\breve{D}_{\widetilde{c} \cdot \varphi_{\gamma},\mathbf{R}_{n}}^{SBD}( 
\mathds{Q},\mathbf{Q}^{\ast\ast}),
\quad
\mathbf{R} := \mathds{P},
\quad
\Phi_{\mathbf{R}}(\mathds{Q}) := 
\breve{D}_{\widetilde{c} \cdot \varphi_{\gamma},\mathbf{R}}^{SBD}( 
\mathds{Q},\mathbf{Q}^{\ast\ast}),
\nonumber
\end{equation}
it is possible --- under appropriate analytic conditions ---
to obtain the following coherence/consistency result
\begin{align}
& 
\Phi_{\mathds{P}}(\textrm{$\boldsymbol{\Omega}$\hspace{-0.23cm}$\boldsymbol{\Omega}$}) = 
\inf_{\mathds{Q}\in \textrm{$\boldsymbol{\Omega}$\hspace{-0.19cm}$\boldsymbol{\Omega}$} }
\breve{D}_{\widetilde{c} \cdot \varphi_{\gamma},\mathds{P}}^{SBD}( \mathds{Q},\mathbf{Q}^{\ast\ast})
= \lim_{n\rightarrow \infty }
\underline{FM}_{n}^{BS,SBD}\Big(X_{1},\ldots,X_{n}\Big) 
= - \lim_{n\rightarrow \infty} \lim_{m\rightarrow \infty} \frac{1}{m}
\log \, 
\mathbb{\Pi}_{\mathbf{X}_{1}^{n}} \negthinspace \left[\boldsymbol{\xi}_{n,m,\mathbf{X}}^{w\mathbf{V}} \in 
\textrm{$\boldsymbol{\Omega}$\hspace{-0.23cm}$\boldsymbol{\Omega}$}\right]
\qquad a.s.
\nonumber
\end{align}
for all sets $\boldsymbol{\Omega}$\hspace{-0.23cm}$\boldsymbol{\Omega} \subset \widetilde{\mathcal{M}}_{\gamma}$ 
(with $A=1$) satisfying the 
regularity properties \eqref{regularity simplex} 
\textit{in the relative topology}.  

\vspace{0.3cm}
\noindent
Due to the lack of representability (cf. the explanations 
right after \eqref{fo.approx.1.simplex.innminSBD})
we \textit{generally} can not give analogous results for the \textit{original}
scaled Bregman distance 
$D_{\widetilde{c} \cdot \varphi_{\gamma},\mathds{P}}^{SBD}( \mathds{Q},\mathbf{Q}^{\ast\ast})$. 
The latter can be achieved, however, in the subsetup $\gamma=1$ as well as in the subsetup
where $\mathbf{Q}^{\ast\ast} := C \cdot \mathbf{P}$ and $\gamma \in \mathbb{R}\backslash[1,2[$.
To start with, 
by combining Theorem \ref{brostu5:thm.divnormW.new.risk.SBD} 
with \eqref{brostu5:fo.SBDrepres}, \eqref{brostu5:fo.divpow.F.SBD.inv},
\eqref{brostu5:fo.SBDrepres.gamma1} and \eqref{brostu5:fo.SBDrepres.gamma1.inverse}
for the special choice $A=1$,
we can straightforwardly derive the following 

\vspace{0.2cm}
\noindent
\begin{theorem}
\label{brostu5:thm.divnormW.new.SBD.risk} 
Let $\mathds{P} \in \mathbb{S}_{> 0}^{K}$, $(X_{i})_{i\in \mathbb{N}}$, 
$V := (\mathbf{V}_{i})_{i \in \mathbb{N}}$, $\boldsymbol{\xi}_{n,\mathbf{X}}^{w\mathbf{V}}$
as in Theorem \ref{brostu3:thm.divnormW.new.det.SBD.risk},
for the special choice $\varphi := \widetilde{c} \cdot \varphi_{\gamma}$ 
(cf. \eqref{brostu5:fo.powdivgen}) with some arbitrarily fixed
$\widetilde{c} \in \, ]0,\infty[$, $\gamma \in \mathbb{R}\backslash\, ]1,2[$.
Then the following holds for any $C>0$:\\
(a) In case of $\gamma \in \mathbb{R}\backslash[1,2[$, for any subset
$\boldsymbol{\Omega}$\hspace{-0.23cm}$\boldsymbol{\Omega} \subset 
\widetilde{\mathcal{M}}_{\gamma}$ (with $A=1$)
with \eqref{regularity simplex} one gets
\begin{align}
& \inf_{\mathds{Q}\in \textrm{$\boldsymbol{\Omega}$\hspace{-0.19cm}$\boldsymbol{\Omega}$} }
\breve{F}_{\gamma,\widetilde{c},1,1,C}\Big(
D_{\widetilde{c} \cdot \varphi_{\gamma},\mathds{P}}^{SBD}(\mathds{Q}, C \cdot \mathds{P})
\Big)
= -\lim_{n\rightarrow \infty }\frac{1}{n}\log \, 
\mathbb{\Pi}\negthinspace \left[\boldsymbol{\xi}_{n,\mathbf{X}}^{w\mathbf{V}}\in 
\textrm{$\boldsymbol{\Omega}$\hspace{-0.23cm}$\boldsymbol{\Omega}$}\right] 
\nonumber
\end{align}
and (equivalently)
\begin{align}
& \inf_{\mathds{Q}\in \textrm{$\boldsymbol{\Omega}$\hspace{-0.19cm}$\boldsymbol{\Omega}$} }
D_{\widetilde{c} \cdot \varphi_{\gamma},\mathds{P}}^{SBD}(\mathds{Q}, C \cdot \mathds{P})
= \breve{F}_{\gamma,\widetilde{c},1,1,C}^{\leftarrow}\Big(
-\lim_{n\rightarrow \infty }\frac{1}{n}\log \, 
\mathbb{\Pi}\negthinspace \left[\boldsymbol{\xi}_{n}^{w\mathbf{V}}\in 
\textrm{$\boldsymbol{\Omega}$\hspace{-0.23cm}$\boldsymbol{\Omega}$}\right]\Big) .
\label{LDP Normalized Vec simplex 2 SBD risk}
\end{align}
In particular, the functions 
$\Phi_{\mathbf{P}}(\cdot) :=
\breve{F}_{\gamma,\widetilde{c},1,1,C}\Big(
D_{\widetilde{c} \cdot \varphi_{\gamma},\mathds{P}}^{SBD}(\cdot, C \cdot \mathds{P})
\Big)$
and 
$\Phi_{\mathbf{P}}(\cdot) :=
D_{\widetilde{c} \cdot \varphi_{\gamma},\mathds{P}}^{SBD}(\cdot, C \cdot \mathds{P})$
are bare-simulation minimizable (BS-minimizable)
in the narrow sense 
(cf. \eqref{brostu5:fo.2} in Definition \ref{brostu5:def.1}
and the special case of Remark \ref{brostu5:rem.def1}(a)) 
on all sets 
$\textrm{$\boldsymbol{\Omega}$\hspace{-0.23cm}$\boldsymbol{\Omega}$} \subset \widetilde{\mathcal{M}}_{\gamma}$ 
(with $A=1$) satisfying 
\eqref{regularity simplex}
\textit{in the relative topology}.\\
(b) If $\gamma=1$, then for any $\mathbf{Q}^{\ast\ast} \in \mathbb{R}_{>0}^{K}$
and any subset
$\boldsymbol{\Omega}$\hspace{-0.23cm}$\boldsymbol{\Omega} \subset 
\mathbb{S}^{K}$
with \eqref{regularity simplex} one gets
\begin{align}
& \inf_{\mathds{Q}\in \textrm{$\boldsymbol{\Omega}$\hspace{-0.19cm}$\boldsymbol{\Omega}$} }
\breve{F}_{1,\widetilde{c},1,M_{\mathbf{Q}^{\ast\ast}}}\Big(
D_{\widetilde{c} \cdot \varphi_{1},\mathds{P}}^{SBD}(\mathds{Q}, \mathbf{Q}^{\ast\ast})
\Big)
= -\lim_{n\rightarrow \infty }\frac{1}{n}\log \, 
\mathbb{\Pi}\negthinspace \left[\boldsymbol{\xi}_{n,\mathbf{X}}^{w\mathbf{V}}\in 
\textrm{$\boldsymbol{\Omega}$\hspace{-0.23cm}$\boldsymbol{\Omega}$}\right] 
\nonumber
\end{align}
and (equivalently)
\begin{align}
& \inf_{\mathds{Q}\in \textrm{$\boldsymbol{\Omega}$\hspace{-0.19cm}$\boldsymbol{\Omega}$} }
D_{\widetilde{c} \cdot \varphi_{1},\mathds{P}}^{SBD}(\mathds{Q}, \mathbf{Q}^{\ast\ast}) 
= \breve{F}_{1,\widetilde{c},1,M_{\mathbf{Q}^{\ast\ast}}}^{\leftarrow}\Big(
-\lim_{n\rightarrow \infty }\frac{1}{n}\log \, 
\mathbb{\Pi}\negthinspace \left[\boldsymbol{\xi}_{n,\mathbf{X}}^{w\mathbf{V}}\in 
\textrm{$\boldsymbol{\Omega}$\hspace{-0.23cm}$\boldsymbol{\Omega}$}\right]\Big) .
\label{LDP Normalized Vec simplex 2 SBD gamma1 risk}
\end{align}
In particular, the functions 
$\Phi_{\mathds{P}}(\cdot) :=
\breve{F}_{1,\widetilde{c},A,M_{\mathbf{Q}^{\ast\ast}}}\Big(
D_{\widetilde{c} \cdot \varphi_{1},\mathds{P}}^{SBD}(\cdot, \mathbf{Q}^{\ast\ast})
\Big)$
and
$\Phi_{\mathds{P}}(\cdot) :=
D_{\widetilde{c} \cdot \varphi_{1},\mathds{P}}^{SBD}(\cdot, \mathbf{Q}^{\ast\ast})$
are bare-simulation minimizable (BS-minimizable)
in the narrow sense 
(cf. \eqref{brostu5:fo.2} in Definition \ref{brostu5:def.1}
and the special case of Remark \ref{brostu5:rem.def1}(a)) 
on all sets 
$\textrm{$\boldsymbol{\Omega}$\hspace{-0.23cm}$\boldsymbol{\Omega}$} \subset \mathbb{S}^{K}$ 
satisfying 
\eqref{regularity simplex}
\textit{in the relative topology}.

\end{theorem}

\vspace{0.4cm}
\noindent
For the special case $C=1$ --- respectively $\mathbf{Q}^{\ast\ast} = \mathds{P}$ if $\gamma=1$ ---
Theorem \ref{brostu5:thm.divnormW.new.SBD.risk} collapses to Theorem \ref{brostu5:thm.divnormW.new}.

\vspace{0.3cm}
\noindent

\begin{remark} \ 
Analogously as in Theorem 12 of Broniatowski \& Stummer \cite{Bro:23a}, we can also give equivalent versions of 
the three Theorems \ref{brostu3:thm.divnormW.new.det.SBD.risk},
\ref{brostu5:thm.divnormW.new.risk.SBD} and \ref{brostu5:thm.divnormW.new.SBD.risk}
(and the proofs thereof) in terms of probability \textit{distributions} $\mathbb{Q} \in \mathbb{\Omega}$ and $\mathbb{P}$
as well as measure $Q^{\ast\ast}$,
rather than probability \textit{vectors} 
$\mathds{Q} \in \textrm{$\boldsymbol{\Omega}$\hspace{-0.23cm}$\boldsymbol{\Omega}$}$
and $\mathds{P}$ as well as vector $\mathbf{Q}^{\ast\ast}$;
for the sake of brevity, we omit the corresponding explicitness. 

\end{remark}

\vspace{0.3cm}
\noindent
Now, analogously to the above investigations, we employ the 
limit statements \eqref{LDP Normalized Vec simplex 2 SBD risk}
and \eqref{LDP Normalized Vec simplex 2 SBD gamma1 risk}
(rather than \eqref{LDP Normalized Vec simplex risk SBD}) 
and estimate 
--- based on random scaling-samples $\mathbf{X}_{1}^{n} = \left(X_{1},\ldots,X_{n}\right)$ 
satisfying \eqref{cv emp measure X to P} ---
the directed distance (divergence) from $\mathbf{Q}^{\ast\ast}$ to the model
$\textrm{$\boldsymbol{\Omega}$\hspace{-0.23cm}$\boldsymbol{\Omega}$}$ 
under a true unknown ``scaling-data-generating'' mechanism $\mathbb{P}$,
\textit{given by}
\begin{eqnarray}
& &
\Phi_{\mathds{P}}(\textrm{$\boldsymbol{\Omega}$\hspace{-0.23cm}$\boldsymbol{\Omega}$}) := 
\inf_{\mathds{Q}\in \textrm{$\boldsymbol{\Omega}$\hspace{-0.19cm}$\boldsymbol{\Omega}$} }
D_{\widetilde{c} \cdot \varphi_{\gamma},\mathds{P}}^{SBD}( \mathds{Q},C \cdot \mathds{P}),
\quad \textrm{for } \gamma \in \mathbb{R}\backslash[1,2[,
\nonumber\\
& & 
\Phi_{\mathds{P}}(\textrm{$\boldsymbol{\Omega}$\hspace{-0.23cm}$\boldsymbol{\Omega}$}) := 
\inf_{\mathds{Q}\in \textrm{$\boldsymbol{\Omega}$\hspace{-0.19cm}$\boldsymbol{\Omega}$} }
D_{\widetilde{c} \cdot \varphi_{1},\mathds{P}}^{SBD}(\mathds{Q}, \mathbf{Q}^{\ast\ast}),
\hspace{0.5cm} \textrm{for } \gamma =1,
\nonumber
\end{eqnarray}
via (with a slight abuse of notation)
\begin{eqnarray}
& &  \hspace{-0.7cm}
\underline{M}_{n}^{BS,SBD}\Big(X_{1},\ldots,X_{n}\Big) := 
\underline{M}_{n}^{BS,SBD}\Big(\mathds{P}_{n}^{emp}(\mathbf{X}_{1}^{n})\Big) := 
\breve{F}_{\gamma,\widetilde{c},1,1,C}^{\leftarrow}\Big(
-\frac{1}{n}\log \, 
\mathbb{\Pi}_{\mathbf{X}_{1}^{n}}\negthinspace \left[\boldsymbol{\xi}_{n,\mathbf{X}}^{w\mathbf{V}}\in 
\textrm{$\boldsymbol{\Omega}$\hspace{-0.23cm}$\boldsymbol{\Omega}$}\right]\Big)
\ \approx \ 
\inf_{\mathds{Q}\in \textrm{$\boldsymbol{\Omega}$\hspace{-0.19cm}$\boldsymbol{\Omega}$} }
D_{\widetilde{c} \cdot \varphi_{\gamma},\mathds{P}}^{SBD}( \mathds{Q},C \cdot \mathds{P}),
\nonumber \\
& & \hspace{13.7cm} \textrm{for } \gamma \in \mathbb{R}\backslash[1,2[,
\label{fo.approx.1.risk.SBD.1} \\
& & \hspace{-0.7cm}
\underline{M}_{n}^{BS,SBD}\Big(X_{1},\ldots,X_{n}\Big) := 
\underline{M}_{n}^{BS,SBD}\Big(\mathds{P}_{n}^{emp}(\mathbf{X}_{1}^{n})\Big) := 
\breve{F}_{1,\widetilde{c},1,M_{\mathbf{Q}^{\ast\ast}}}^{\leftarrow}\Big(
-\frac{1}{n}\log \, 
\mathbb{\Pi}_{\mathbf{X}_{1}^{n}}\negthinspace \left[\boldsymbol{\xi}_{n,\mathbf{X}}^{w\mathbf{V}}\in 
\textrm{$\boldsymbol{\Omega}$\hspace{-0.23cm}$\boldsymbol{\Omega}$}\right]\Big)
\ \approx \ 
\inf_{\mathds{Q}\in \textrm{$\boldsymbol{\Omega}$\hspace{-0.19cm}$\boldsymbol{\Omega}$} }
D_{\widetilde{c} \cdot \varphi_{1},\mathds{P}}^{SBD}( \mathds{Q},\mathbf{Q}^{\ast\ast}),
\nonumber \\
& & \hspace{13.7cm} \textrm{for } \gamma =1,
\label{fo.approx.1.risk.SBD.2} 
\end{eqnarray}
together with an estimate of the involved conditional probability
$\mathbb{\Pi}_{\mathbf{X}_{1}^{n}}\negthinspace \left[\boldsymbol{\xi}_{n,\mathbf{X}}^{w\mathbf{V}}\in 
\textrm{$\boldsymbol{\Omega}$\hspace{-0.23cm}$\boldsymbol{\Omega}$}\right]$
in \eqref{fo.approx.1.risk.SBD.1} and \eqref{fo.approx.1.risk.SBD.2} (for given large $n$).
Moreover, in the spirit of Subsection \ref{SectStochSubsimplex.CASM.pure}
we can approximate  
the above-mentioned randomly-scaled directed distance
--- in terms of the vectorized histogram of 
the \textit{concrete scaling-data} $x_{1},\ldots,x_{n}$ --- by
\begin{eqnarray}
& & \hspace{-1.7cm}
\underline{FM}_{n}^{BS,SBD}\Big(x_{1},\ldots,x_{n}\Big) = 
\underline{FM}_{n}^{BS,SBD}\Big(\mathds{P}_{n}^{emp}(\mathbf{x}_{1}^{n})\Big) := 
\inf_{\mathds{Q}\in \textrm{$\boldsymbol{\Omega}$\hspace{-0.19cm}$\boldsymbol{\Omega}$} }
D_{\widetilde{c} \cdot \varphi_{\gamma},\mathds{P}_{n}^{emp}(\mathbf{x}_{1}^{n})}^{SBD}( 
\mathds{Q},C \cdot \mathds{P}_{n}^{emp}(\mathbf{x}_{1}^{n})),
\quad \textrm{for } \gamma \in \mathbb{R}\backslash[1,2[,
\nonumber\\
& & \hspace{-1.7cm}
\underline{FM}_{n}^{BS,SBD}\Big(x_{1},\ldots,x_{n}\Big) = 
\underline{FM}_{n}^{BS,SBD}\Big(\mathds{P}_{n}^{emp}(\mathbf{x}_{1}^{n})\Big) := 
\inf_{\mathds{Q}\in \textrm{$\boldsymbol{\Omega}$\hspace{-0.19cm}$\boldsymbol{\Omega}$} }
D_{\widetilde{c} \cdot \varphi_{1},\mathds{P}_{n}^{emp}(\mathbf{x}_{1}^{n})}^{SBD}( 
\mathds{Q},\mathbf{Q}^{\ast\ast}),
\hspace{1.8cm} \textrm{for } \gamma =1,
\nonumber
\end{eqnarray}
via an appropriate application of Theorem \ref{brostu5:thm.divnormW.new.SBD},
and end up with 
\begin{align}
& 
\underline{FM}_{n}^{BS,SBD}\Big(x_{1},\ldots,x_{n}\Big) 
= \breve{F}_{\gamma,\widetilde{c},1,1,C}^{\leftarrow}\Big(
-\lim_{m\rightarrow \infty }
\frac{1}{m} \log \, 
\mathbb{\Pi}_{\mathbf{x}_{1}^{n}}\negthinspace \left[\boldsymbol{\xi}_{n,m,\mathbf{x}}^{w\mathbf{V}} \in 
\textrm{$\boldsymbol{\Omega}$\hspace{-0.23cm}$\boldsymbol{\Omega}$}\right] \Big),
\hspace{1.0cm} \textrm{for } \gamma \in \mathbb{R}\backslash[1,2[,
\nonumber\\
& \underline{FM}_{n}^{BS,SBD}\Big(x_{1},\ldots,x_{n}\Big)  
= \breve{F}_{1,\widetilde{c},1,M_{\mathbf{Q}^{\ast\ast}}}^{\leftarrow}\Big(
-\lim_{m\rightarrow \infty }
\frac{1}{m} \log \, 
\mathbb{\Pi}_{\mathbf{x}_{1}^{n}}\negthinspace \left[\boldsymbol{\xi}_{n,m,\mathbf{x}}^{w\mathbf{V}} \in 
\textrm{$\boldsymbol{\Omega}$\hspace{-0.23cm}$\boldsymbol{\Omega}$}\right] \Big),
\qquad \textrm{for } \gamma =1,
\nonumber
\end{align}
for all sets $\boldsymbol{\Omega}$\hspace{-0.23cm}$\boldsymbol{\Omega} \subset \widetilde{\mathcal{M}}_{\gamma}$ 
(with $A=1$) satisfying the 
regularity properties \eqref{regularity simplex} 
\textit{in the relative topology}. 

\vspace{0.2cm}
\noindent
Finally, following the spirit of Subsection \ref{SectStochSubsimplex.CASM.comb}
in an analogous fashion as above,
it is possible --- under appropriate analytic conditions ---
to obtain the following coherence/consistency result
\begin{eqnarray}
& & \hspace{-1.0cm}
\Phi_{\mathds{P}}(\textrm{$\boldsymbol{\Omega}$\hspace{-0.23cm}$\boldsymbol{\Omega}$}) = 
\inf_{\mathds{Q}\in \textrm{$\boldsymbol{\Omega}$\hspace{-0.19cm}$\boldsymbol{\Omega}$} }
\breve{D}_{\widetilde{c} \cdot \varphi_{\gamma},\mathds{P}}^{SBD}( \mathds{Q},C \cdot \mathds{P})
= \lim_{n\rightarrow \infty }
\underline{FM}_{n}^{BS,SBD}\Big(X_{1},\ldots,X_{n}\Big) 
= \breve{F}_{\gamma,\widetilde{c},1,1,C}^{\leftarrow}\Big(
- \lim_{n\rightarrow \infty} \lim_{m\rightarrow \infty}
\frac{1}{m} \log \, 
\mathbb{\Pi}_{\mathbf{X}_{1}^{n}} \negthinspace \left[\boldsymbol{\xi}_{n,m,\mathbf{X}}^{w\mathbf{V}} \in 
\textrm{$\boldsymbol{\Omega}$\hspace{-0.23cm}$\boldsymbol{\Omega}$}\right]\Big)
\ a.s.
\nonumber \\
& & \hspace{13.7cm} 
\textrm{for } \gamma \in \mathbb{R}\backslash[1,2[,
\nonumber\\
& & \hspace{-1.0cm}
\Phi_{\mathds{P}}(\textrm{$\boldsymbol{\Omega}$\hspace{-0.23cm}$\boldsymbol{\Omega}$}) = 
\inf_{\mathds{Q}\in \textrm{$\boldsymbol{\Omega}$\hspace{-0.19cm}$\boldsymbol{\Omega}$} }
\breve{D}_{\widetilde{c} \cdot \varphi_{1},\mathds{P}}^{SBD}( \mathds{Q},\mathbf{Q}^{\ast\ast})
= \lim_{n\rightarrow \infty }
\underline{FM}_{n}^{BS,SBD}\Big(X_{1},\ldots,X_{n}\Big) 
= 
\breve{F}_{1,\widetilde{c},1,M_{\mathbf{Q}^{\ast\ast}}}^{\leftarrow}\Big(
- \lim_{n\rightarrow \infty} \lim_{m\rightarrow \infty}
\frac{1}{m} \log \, 
\mathbb{\Pi}_{\mathbf{X}_{1}^{n}} \negthinspace \left[\boldsymbol{\xi}_{n,m,\mathbf{X}}^{w\mathbf{V}} \in 
\textrm{$\boldsymbol{\Omega}$\hspace{-0.23cm}$\boldsymbol{\Omega}$}\right]\Big)
\ a.s.
\nonumber \\
& & \hspace{13.7cm} 
\textrm{for } \gamma =1,
\nonumber
\end{eqnarray}
for all sets $\boldsymbol{\Omega}$\hspace{-0.23cm}$\boldsymbol{\Omega} \subset \widetilde{\mathcal{M}}_{\gamma}$ 
(with $A=1$) satisfying the 
regularity properties \eqref{regularity simplex} 
\textit{in the relative topology}.


\section{Bare-Simulation-Method For General Divergence-Optimization-Problems Under Risk}
\label{SectStochSubsimplex.General}


\subsection{Divergences under Risk and Friends}
\label{SectDetGeneral.BDM1.min.risk.friends}

\vspace{0.1cm}
\noindent
In the two previous Sections \ref{SectStochSubsimplex.CASM} and \ref{SectStochSubsimplex.SBD}
we have dealt --- under risk (i.e. with unknown $\mathds{P}$) ---
with \textit{narrow-sense} bare-simulation \textit{minimization}
of $\varphi-$divergences 
$\Phi_{\mathds{P}}(\mathds{Q}) := 
D_{\varphi}(\mathds{Q},\mathds{P})$,  
innmin scaled Bregman distances
$\Phi_{\mathds{P}}(\mathds{Q}) := 
\breve{D}_{\varphi,\mathds{P}}^{SBD}( \mathds{Q},\mathbf{Q}^{\ast\ast})$
and scaled Bregman distances 
$\Phi_{\mathds{P}}(\mathds{Q}) := 
D_{\varphi,\mathds{P}}^{SBD}( \mathds{Q},\mathbf{Q}^{\ast\ast})$
where (except for Theorem \ref{brostu3:thm.divnormW.new.det.SBD.risk}) as divergence generator $\varphi$ 
we have taken the power functions 
$\varphi=\widetilde{c} \cdot \varphi_{\gamma}$ (cf. \eqref{brostu5:fo.powdivgen} 
with $\gamma \in \mathbb{R}\backslash[1,2[$)
(and for Bregman distances restrictively $\mathbf{Q}^{\ast\ast} = C \cdot \mathds{P}$ in case of $\gamma \ne 1$);
recall the different role of $\mathds{P}$ (``point to be projected'' vs. scaling) in these two contexts.

\vspace{0.3cm}
\noindent
Also recall that we are interested in the (generally, non-narrow sense) bare-simulation constrained 
\textit{minimization and maximization} of the \textit{continuous} functions 
$\mathbf{Q} \mapsto \Phi_{\mathbf{P}}(\mathbf{Q})$
--- given in Subsection \ref{SectDetGeneral.Friends}, and beyond. 
In fact, the there-involved functions
(\textbf{supposedly on a compact 
$\textrm{$\boldsymbol{\Omega}$\hspace{-0.23cm}$\boldsymbol{\Omega}$} \subset \mathbb{S}^{K}$
in this section}) 
$\mathds{Q} \mapsto \Phi_{\mathbf{R}}(\mathds{Q})$
partially carry even another (respectively, more general) vector-valued ``parameter'' $\mathbf{R}$
which in this section we \textit{consider as unknown}.
For instance, in (D1) and (D2) we have the only
possibility $\Phi_{\mathbf{R}}(\mathds{Q}) := D_{\varphi}( \mathds{Q}, \mathbf{R})$
and $\Phi_{\mathbf{R}}(\mathds{Q}) := D_{\varphi}^{OBD}(\mathds{Q},\mathbf{R})$
with $\mathbf{R} := \mathbf{P}$,
whereas in (D3) we have three possibilities of an unknown ``parameter'' $\mathbf{R}$:
$\Phi_{\mathbf{R}}(\mathds{Q}) : = 
D_{\varphi,\mathbf{M}}^{SBD}(\mathds{Q},\mathbf{R})$
with $\mathbf{R} := \mathbf{P}$,
$\Phi_{\mathbf{R}}(\mathds{Q}) : = 
D_{\varphi,\mathbf{R}}^{SBD}(\mathds{Q},\mathbf{P})$
with $\mathbf{R} := \mathbf{M}$\footnote{e.g.
take $\varphi(t) := 2 \cdot \varphi_{2}(t) = (t - 1)^2$
(cf. \eqref{brostu5:fo.powdivgen}) to end up with
the squared weighted $\ell_{2}-$distance
$G:=D_{2\varphi_{2},\mathbf{R}}^{SBD}(\mathds{Q},\mathbf{P})
= \sum_{k=1}^{K} \frac{1}{r_{k}} \cdot | \, q_{k} - p_{k} \, |^{2}
= \left(D_{\ell_{2},\mathbf{R}}( \mathds{Q}, \mathbf{P} )\right)^{2}$ 
(cf. \eqref{brostu5:fo.ellrdist}),
and choose $q_{k}>0$, $p_{k}>0$ to be two alternative fixed (i.e. non-random) cash-flows at
(say) some future time $t_k >0$ as well as $r_{k}$ to be a random (e.g. 
random-term-structure-of-interest-rates connected) discount factor
from $t_{k}$ to $t_{0} := 0$ (e.g. $t_{0} \mathrel{\widehat{=}} 12.00$ hours tomorrow);
accordingly, $G$ is connected with the aggregated dissimilarity of the corresponding 
(interest-rate-prone) present-values at $t_{0}$.
},
and $\Phi_{\mathbf{R}}(\mathds{Q}) : = 
D_{\varphi,\mathbf{R}_{1}}^{SBD}(\mathds{Q},\mathbf{R}_{2})$
with 2K-dimensional concatenation $\mathbf{R}=(\mathbf{R}_{1},\mathbf{R}_{2})
=(\mathbf{M},\mathbf{P})$.
In (D4) we have even more possibilities, unless we deal e.g. with the interesting
special case of the total Bregman distance 
$D_{\varphi}^{TBD}(\mathds{Q},\mathbf{R})$
with $\mathbf{R}:= \mathbf{P}$.
The cases (D6), (D7) and (D8) can take unknown parameters $\mathbf{R}$ 
in an analogous way, whereas (D5) has no parameter.

\vspace{0.3cm}
\noindent
Returning to the general context,
in case that $\mathbf{R}$ is \textit{known}, 
we can simply apply the two
Theorems \ref{brostu5:thm.Fmin.simplex} and \ref{brostu5:thm.Fmax.simplex} 
(or alternatively, \ref{brostu5:thm.Fmin.simplex.SBD} and \ref{brostu5:thm.Fmax.simplex.SBD})
with the special subsetup $A=1$
to $\Phi(\cdot) := \Phi_{\mathbf{R}}(\cdot)$. In case that $\mathbf{R}$ is \textit{unknown}
(i.e. under risk),
we have to adapt the above-mentioned procedures of the 
Sections \ref{SectStochSubsimplex.CASM} respectively \ref{SectStochSubsimplex.SBD}.
For this, we assume that for index $i\in \mathbb{N}$ the generation of the $i-$th
(uncertainty-prone) data point is represented by the random variable $Y_{i}$
\footnote{
on the underlying probability space $(\mathfrak{X},\mathcal{A},\mathbb{\Pi})$
}
which takes values in the discrete set $\widetilde{\mathcal{Y}}:=
\left\{ \widetilde{d}_{1},\cdots ,\widetilde{d}_{K}\right\}$ 
of $K$ distinct values \textquotedblleft of any
kind\textquotedblright \.{ }.\ It is assumed that
--- with the notation  $\mathbf{Y}_{1}^{n} := (Y_{1}, \ldots, Y_{n})$ ---
there is a sequence
$(\mathbf{R}_{n}(\mathbf{Y}_{1}^{n}))_{n\in\mathbb{N}}$ of (say)
vector-valued function(al)s $\mathbf{R}_{n}(\mathbf{Y}_{1}^{n})$ for which 
\begin{equation}
\lim_{n\rightarrow \infty }\mathbf{R}_{n}(\mathbf{Y}_{1}^{n})=\mathbf{R} \quad \textrm{a.s.}
\label{R-convergence}
\end{equation}


\subsection{Minimization via Base-Divergence-Method 1}
\label{SectDetGeneral.BDM1.min.risk}

\vspace{0.2cm}
\noindent
We first choose the ``transformed'' power divergences 
$F_{\gamma,\widetilde{c},1}(D_{\widetilde{c} \cdot \varphi_{\gamma}}(\cdot,\cdot))$
as the involved base-divergences.
In analogy with the considerations of the Subsection \ref{SectStochSubsimplex.CASM.pure},
let us examine the pure data-analytic view on the risk case.
Accordingly, we can approximate 
--- based on concrete data 
$\mathbf{y}_{1}^{n} = \left(y_{1},\ldots,y_{n}\right)$
from the samples $\mathbf{Y}_{1}^{n} = \left(Y_{1},\ldots,Y_{n}\right)$ 
satisfying \eqref{R-convergence} ---
the \textit{unknown-parameter-carrying function}
\begin{equation}
\Phi_{\mathbf{R}}(\textrm{$\boldsymbol{\Omega}$\hspace{-0.23cm}$\boldsymbol{\Omega}$}) := 
\inf_{\mathds{Q}\in \textrm{$\boldsymbol{\Omega}$\hspace{-0.19cm}$\boldsymbol{\Omega}$} }
\Phi_{\mathbf{R}}(\mathds{Q})
\nonumber
\end{equation}
by the corresponding empirical (i.e. concrete-data-dependent) minimum
\begin{equation}
\underline{FM}_{n}^{BS}\Big(y_{1},\ldots,y_{n}\Big) = 
\underline{FM}_{n}^{BS}\Big(\mathbf{R}_{n}(\mathbf{y}_{1}^{n})\Big) := 
\inf_{\mathds{Q}\in \textrm{$\boldsymbol{\Omega}$\hspace{-0.19cm}$\boldsymbol{\Omega}$}} 
\Phi_{\mathbf{R}_{n}(\mathbf{y}_{1}^{n})}(\mathds{Q}),
\label{min.risk.4} 
\end{equation}
via an appropriate application of Theorem \ref{brostu5:thm.Fmin.simplex},
and end up with

\begin{align}
& 
\underline{FM}_{n}^{BS}\Big(y_{1},\ldots,y_{n}\Big)  = 
-  \lim_{m\rightarrow \infty } \frac{1}{m}
\log \negthinspace \left( \ 
\mathbb{E}_{\mathbb{\Pi}_{\mathbf{y}_{1}^{n}}}
\negthinspace \Big[
\exp\negthinspace\Big(
m \cdot \Big(
F_{\gamma,\widetilde{c},1}\Big(D_{\widetilde{c} \cdot 
\varphi_{\gamma}}(\boldsymbol{\xi}_{m}^{w\mathbf{W}},\mathds{P}^{aux})\Big)
- \Phi_{\mathbf{R}_{n}(\mathbf{y}_{1}^{n})}\big(\boldsymbol{\xi}_{m}^{w\mathbf{W}}\big)
\Big)
\Big)
\cdot \textfrak{1}_{
\textrm{$\boldsymbol{\Omega}$\hspace{-0.19cm}$\boldsymbol{\Omega}$}}\big(\boldsymbol{\xi}_{m}^{w\mathbf{W}}\big)
\, \Big] 
\right)
\label{min.risk.5a.new}
\end{align}
for all sets $\boldsymbol{\Omega}$\hspace{-0.23cm}$\boldsymbol{\Omega} \subset \widetilde{\mathcal{M}}_{\gamma}$ 
(with $A=1$) satisfying the 
regularity properties \eqref{regularity simplex} 
\textit{in the relative topology}. 
Here, in \eqref{min.risk.5a.new} we have employed the random vectors
$\boldsymbol{\xi}_{m}^{w\mathbf{W}}$ constructed from an auxiliary deterministic
probability vector $\mathds{P}^{aux}  \in \mathbb{S}_{> 0}^{K}$ by ---
cf. \eqref{brostu5:fo.norweiemp.vec.det} with $m$ instead of $n$
and blocks $I_{k}^{(m)}$ of sizes $m_{k} :=\lfloor m \cdot 
p_{k}^{aux}\rfloor$ ($k=1,\ldots,K-1$) and $m_{K} := m- \sum_{k=1}^{K-1} m_{k}$
instead of $n_{k}$, $n_{K}$ ---
\begin{eqnarray}
\boldsymbol{\xi}_{m}^{w\mathbf{W}} &:=&
\begin{cases}
\left(\frac{\sum_{i \in I_{1}^{(m)}}W_{i}}{\sum_{k=1}^{K}\sum_{i \in I_{k}^{(m)}}W_{i}},
\ldots, \frac{\sum_{i \in I_{K}^{(m)}}W_{i}}{\sum_{k=1}^{K}\sum_{i \in I_{k}^{(m)}}W_{i}} \right) ,
\qquad \textrm{if } \sum_{j=1}^{m} W_{j} \ne 0, \\
\ (\infty, \ldots, \infty) =: \boldsymbol{\infty}, \hspace{4.0cm} \textrm{if } \sum_{j=1}^{m} W_{j} = 0;
\end{cases}
\label{brostu5:fo.norweiemp.vec.det.m} 
\end{eqnarray}
moreover, $(W_{i})_{i \in \mathbb{N}}$ and $(Y_{i})_{i \in \mathbb{N}}$ 
are supposed to be independent\footnote{
notice that all the $Y_{i}$ and $W_{i}$ live on the \textit{same} underlying probability space $(\mathfrak{X},\mathcal{A},\mathbb{\Pi})$
}. In \eqref{min.risk.5a.new} we have also used
the \textit{conditional} expectations
$\mathbb{E}_{\mathbb{\Pi}_{\mathbf{y}_{1}^{n}}} [\, \cdot \, ] := 
\mathbb{E}_{\mathbb{\Pi}}[ \, \cdot \, | \, Y_{1}=y_{1}, \ldots, Y_{n}=y_{n} ]$
(cf. footnote \ref{foot.condprob}).

\vspace{0.3cm}

\begin{remark} 
\label{brostu5:rem.Paux}
The involved $\mathds{P}^{aux}$ may also depend on the deterministic data sequence
$\mathbf{y}_{1}^{n}$;
for the special case $\mathds{P}^{aux} := \mathds{P}_{n}^{emp}(\mathbf{y}_{1}^{n})$,
we obtain $\boldsymbol{\xi}_{m}^{w\mathbf{W}} = \boldsymbol{\xi}_{n,m,\mathbf{y}}^{w\mathbf{W}}$
(cf. \eqref{brostu5:fo.norweiemp.vec.risk.pnemp} respectively \eqref{brostu5:fo.norweiemp.vec.risk.pnemp.var}).

\end{remark} 

\vspace{0.3cm}
\noindent
Based on \eqref{min.risk.5a.new}, one can (for fixed $n$)
approximate the corresponding empirical minimum
$\inf_{\mathds{Q}\in \textrm{$\boldsymbol{\Omega}$\hspace{-0.19cm}$\boldsymbol{\Omega}$}} 
\Phi_{\mathbf{R}_{n}(\mathbf{y}_{1}^{n})}(\mathds{Q})$ by
\begin{align}
& 
-  \frac{1}{m}
\log \negthinspace \left( \ 
\mathbb{E}_{\mathbb{\Pi}_{\mathbf{y}_{1}^{n}}}
\negthinspace \Big[
\exp\negthinspace\Big(
m \cdot \Big(
F_{\gamma,\widetilde{c},1}\Big(D_{\widetilde{c} \cdot 
\varphi_{\gamma}}(\boldsymbol{\xi}_{m}^{w\mathbf{W}},\mathds{P}^{aux})\Big)
- \Phi_{\mathbf{R}_{n}(\mathbf{y}_{1}^{n})}\big(\boldsymbol{\xi}_{m}^{w\mathbf{W}}\big)
\Big)
\Big)
\cdot \textfrak{1}_{
\textrm{$\boldsymbol{\Omega}$\hspace{-0.19cm}$\boldsymbol{\Omega}$}}\big(\boldsymbol{\xi}_{m}^{w\mathbf{W}}\big)
\, \Big] 
\right)
\ \approx \ \inf_{\mathds{Q}\in \textrm{$\boldsymbol{\Omega}$\hspace{-0.19cm}$\boldsymbol{\Omega}$}} 
\Phi_{\mathbf{R}_{n}(\mathbf{y}_{1}^{n})}(\mathds{Q})
\nonumber
\end{align}
for large $m$.

\vspace{0.3cm}
\noindent
Clearly, from \eqref{min.risk.5a.new} we can immediately derive
for the random sample $\mathbf{Y}_{1}^{n} = (Y_{1},\ldots,Y_{n})$
\begin{align}
& 
\underline{FM}_{n}^{BS}\Big(Y_{1},\ldots,Y_{n}\Big)  = 
-  \lim_{m\rightarrow \infty } 
\frac{1}{m}
\log \negthinspace \left( \ 
\mathbb{E}_{\mathbb{\Pi}_{\mathbf{Y}_{1}^{n}}
}
\negthinspace \Big[
\exp\negthinspace\Big(
m \cdot \Big(
F_{\gamma,\widetilde{c},1}\Big(D_{\widetilde{c} \cdot 
\varphi_{\gamma}}(\boldsymbol{\xi}_{m}^{w\mathbf{W}},\mathds{P}^{aux})\Big)
- \Phi_{\mathbf{R}_{n}(\mathbf{Y}_{1}^{n})}\big(\boldsymbol{\xi}_{m}^{w\mathbf{W}}\big)
\Big)
\Big)
\cdot \textfrak{1}_{
\textrm{$\boldsymbol{\Omega}$\hspace{-0.19cm}$\boldsymbol{\Omega}$}}\big(\boldsymbol{\xi}_{m}^{w\mathbf{W}}\big)
\, \Big] 
\right)
\label{min.risk.5a.new.sample}
\end{align}
for all sets $\boldsymbol{\Omega}$\hspace{-0.23cm}$\boldsymbol{\Omega} \subset \widetilde{\mathcal{M}}_{\gamma}$ 
(with $A=1$) satisfying the 
regularity properties \eqref{regularity simplex} 
\textit{in the relative topology};
also the Remark \ref{brostu5:rem.Paux} carries over correspondingly. 
Here, in \eqref{min.risk.5a.new.sample} we have employed
the \textit{conditional} expectations
$\mathbb{E}_{\mathbb{\Pi}_{\mathbf{Y}_{1}^{n}}}[\, \cdot \, ]  := 
\mathbb{E}_{\mathbb{\Pi}}[ \, \cdot \, | \, 
Y_{1}, \ldots, Y_{n} ]$
(see also footnote \ref{foot.condprob}).

\vspace{0.3cm}
\noindent
Based on \eqref{min.risk.5a.new.sample}, one can (for fixed $n$)
approximate the corresponding sample-dependent minimum
$\inf_{\mathds{Q}\in \textrm{$\boldsymbol{\Omega}$\hspace{-0.19cm}$\boldsymbol{\Omega}$}} 
\Phi_{\mathbf{R}_{n}(\mathbf{Y}_{1}^{n})}(\mathds{Q})$ 
by
\begin{align}
& 
-  \frac{1}{m}
\log \negthinspace \left( \ 
\mathbb{E}_{\mathbb{\Pi}_{\mathbf{Y}_{1}^{n}}
}
\negthinspace \Big[
\exp\negthinspace\Big(
m \cdot \Big(
F_{\gamma,\widetilde{c},1}\Big(D_{\widetilde{c} \cdot 
\varphi_{\gamma}}(\boldsymbol{\xi}_{m}^{w\mathbf{W}},\mathds{P}^{aux})\Big)
- \Phi_{\mathbf{R}_{n}(\mathbf{Y}_{1}^{n})}\big(\boldsymbol{\xi}_{m}^{w\mathbf{W}}\big)
\Big)
\Big)
\cdot \textfrak{1}_{
\textrm{$\boldsymbol{\Omega}$\hspace{-0.19cm}$\boldsymbol{\Omega}$}}\big(\boldsymbol{\xi}_{m}^{w\mathbf{W}}\big)
\, \Big] 
\right)
\ \approx \ \inf_{\mathds{Q}\in \textrm{$\boldsymbol{\Omega}$\hspace{-0.19cm}$\boldsymbol{\Omega}$}} 
\Phi_{\mathbf{R}_{n}(\mathbf{Y}_{1}^{n})}(\mathds{Q})
\label{min.risk.5b.sample}
\end{align}
for large $m$.

\vspace{0.3cm}
\noindent
In the spirit of Subsection \ref{SectStochSubsimplex.CASM.comb},
by combining \eqref{min.risk.4},\eqref{min.risk.5a.new.sample},
\eqref{R-convergence} and \eqref{consistency1}
with  $\mathbf{R}_{n} := \mathbf{R}_{n}(\mathbf{Y}_{1}^{n})$, 
it is possible --- under appropriate analytic conditions ---
to obtain the following coherence/consistency result
\begin{align}
& 
\Phi_{\mathbf{R}}(\textrm{$\boldsymbol{\Omega}$\hspace{-0.23cm}$\boldsymbol{\Omega}$}) := 
\inf_{\mathds{Q}\in \textrm{$\boldsymbol{\Omega}$\hspace{-0.19cm}$\boldsymbol{\Omega}$} }
\Phi_{\mathbf{R}}(\mathds{Q})
= \lim_{n\rightarrow \infty }
\underline{FM}_{n}^{BS}\Big(Y_{1},\ldots,Y_{n}\Big) 
\nonumber\\
&
\ = \ - \, 
\lim_{n\rightarrow \infty } \lim_{m\rightarrow \infty } 
\frac{1}{m} \log \negthinspace \left( \ 
\mathbb{E}_{\mathbb{\Pi}_{\mathbf{Y}_{1}^{n}}
}
\negthinspace \Big[
\exp\negthinspace\Big(
m \cdot \Big(
F_{\gamma,\widetilde{c},1}\Big(D_{\widetilde{c} \cdot 
\varphi_{\gamma}}(\boldsymbol{\xi}_{m}^{w\mathbf{W}},\mathds{P}^{aux})\Big)
- \Phi_{\mathbf{R}_{n}(\mathbf{Y}_{1}^{n})}\big(\boldsymbol{\xi}_{m}^{w\mathbf{W}}\big)
\Big)
\Big)
\cdot \textfrak{1}_{
\textrm{$\boldsymbol{\Omega}$\hspace{-0.19cm}$\boldsymbol{\Omega}$}}\big(\boldsymbol{\xi}_{m}^{w\mathbf{W}}\big)
\, \Big] 
\right) 
\quad a.s. 
\, 
\nonumber
\end{align}
for all compact sets $\boldsymbol{\Omega}$\hspace{-0.23cm}$\boldsymbol{\Omega} \subset \widetilde{\mathcal{M}}_{\gamma}$ 
(with $A=1$) satisfying the 
regularity properties \eqref{regularity simplex} 
\textit{in the relative topology}. 

\vspace{0.3cm}
\noindent
Clearly, all the above considerations can be applied to tackle
$\inf_{\mathds{Q}\in \textrm{$\boldsymbol{\Omega}$\hspace{-0.19cm}$\boldsymbol{\Omega}$}} \Phi(\mathbf{Q})$
for all the directed distances/divergences $\Phi(\cdot) := \Phi_{\mathbf{R}}(\cdot)$ and friends 
discussed in the above Subsection \ref{SectDetGeneral.BDM1.min.risk.friends}.


\subsection{Maximization via Base-Divergence-Method 1}

\vspace{0.2cm}
\noindent
For the \textit{non-narrow-sense} BS-maximizability under risk,
one can proceed analogously to the previous Subsection
\ref{SectDetGeneral.BDM1.min.risk},
and approximate 
--- based on concrete data 
$\mathbf{y}_{1}^{n} = \left(y_{1},\ldots,y_{n}\right)$
from the samples $\mathbf{Y}_{1}^{n} = \left(Y_{1},\ldots,Y_{n}\right)$ 
satisfying \eqref{R-convergence} ---
the \textit{unknown-parameter-carrying function}
\begin{equation}
\Phi_{\mathbf{R}}(\textrm{$\boldsymbol{\Omega}$\hspace{-0.23cm}$\boldsymbol{\Omega}$}) := 
\sup_{\mathds{Q}\in \textrm{$\boldsymbol{\Omega}$\hspace{-0.19cm}$\boldsymbol{\Omega}$} }
\Phi_{\mathbf{R}}(\mathds{Q})
\label{max.risk.2}
\end{equation}
by the empirical (i.e. concrete-data-dependent) maximum given by
\begin{equation}
\overline{FM}_{n}^{BS}\Big(y_{1},\ldots,y_{n}\Big) = 
\overline{FM}_{n}^{BS}\Big(\mathbf{R}_{n}(\mathbf{y}_{1}^{n})\Big) := 
\sup_{\mathds{Q}\in \textrm{$\boldsymbol{\Omega}$\hspace{-0.19cm}$\boldsymbol{\Omega}$}} 
\Phi_{\mathbf{R}_{n}(\mathbf{y}_{1}^{n})}(\mathds{Q}),
\label{max.risk.4} 
\end{equation}
via an appropriate application of Theorem \ref{brostu5:thm.Fmax.simplex},
and end up with 
\begin{align}
& 
\overline{FM}_{n}^{BS}\Big(y_{1},\ldots,y_{n}\Big)  = 
\lim_{m\rightarrow \infty } 
\frac{1}{m}
\log \negthinspace \left( \ 
\mathbb{E}_{\mathbb{\Pi}_{\mathbf{y}_{1}^{n}}}\negthinspace \Big[
\exp\negthinspace\Big(
m \cdot \Big(
F_{\gamma,\widetilde{c},1}\Big(D_{\widetilde{c} \cdot 
\varphi_{\gamma}}(\boldsymbol{\xi}_{m}^{w\mathbf{W}},\mathds{P}^{aux})\Big)
+ \Phi_{\mathbf{R}_{n}(\mathbf{y}_{1}^{n})}\big(\boldsymbol{\xi}_{m}^{w\mathbf{W}}\big)
\Big)
\Big)
\cdot \textfrak{1}_{
\textrm{$\boldsymbol{\Omega}$\hspace{-0.19cm}$\boldsymbol{\Omega}$}}\big(\boldsymbol{\xi}_{m}^{w\mathbf{W}}\big)
\, \Big] 
\right)
\label{max.risk.5a}
\end{align}
for all sets $\boldsymbol{\Omega}$\hspace{-0.23cm}$\boldsymbol{\Omega} \subset \widetilde{\mathcal{M}}_{\gamma}$ 
(with $A=1$) satisfying the 
regularity properties \eqref{regularity simplex} 
\textit{in the relative topology};
in accordance with Remark \ref{brostu5:rem.Paux},
the involved $\mathds{P}^{aux}$ may also depend on the deterministic data sequence
$\mathbf{y}_{1}^{n}$;
for the special case $\mathds{P}^{aux} := \mathds{P}_{n}^{emp}(\mathbf{y}_{1}^{n})$,
we obtain $\boldsymbol{\xi}_{m}^{w\mathbf{W}} = \boldsymbol{\xi}_{n,m,\mathbf{y}}^{w\mathbf{W}}$
(cf. \eqref{brostu5:fo.norweiemp.vec.risk.pnemp} respectively \eqref{brostu5:fo.norweiemp.vec.risk.pnemp.var}).
Based on \eqref{max.risk.5a}, one can
approximate (for fixed $n$) the corresponding empirical maximum by
\begin{align}
& 
\frac{1}{m} \log \negthinspace \left( \ 
\mathbb{E}_{\mathbb{\Pi}_{\mathbf{y}_{1}^{n}}}\negthinspace \Big[
\exp\negthinspace\Big(
m \cdot \Big(
F_{\gamma,\widetilde{c},1}\Big(D_{\widetilde{c} \cdot 
\varphi_{\gamma}}(\boldsymbol{\xi}_{m}^{w\mathbf{W}},\mathds{P}^{aux})\Big)
+ \Phi_{\mathbf{R}_{n}(\mathbf{y}_{1}^{n})}\big(\boldsymbol{\xi}_{m}^{w\mathbf{W}}\big)
\Big)
\Big)
\cdot \textfrak{1}_{
\textrm{$\boldsymbol{\Omega}$\hspace{-0.19cm}$\boldsymbol{\Omega}$}}\big(\boldsymbol{\xi}_{m}^{w\mathbf{W}}\big)
\, \Big] 
\right)
\ \approx \ \sup_{\mathds{Q}\in \textrm{$\boldsymbol{\Omega}$\hspace{-0.19cm}$\boldsymbol{\Omega}$}} 
\Phi_{\mathbf{R}_{n}(\mathbf{y}_{1}^{n})}(\mathds{Q})
\nonumber
\end{align}
for large $m$.
Clearly, from \eqref{max.risk.5a} we can immediately derive
for the random sample $\mathbf{Y}_{1}^{n} = (Y_{1},\ldots,Y_{n})$
\begin{align}
& 
\overline{FM}_{n}^{BS}\Big(Y_{1},\ldots,Y_{n}\Big)  = 
\lim_{m\rightarrow \infty } 
\frac{1}{m}
\log \negthinspace \left( \ 
\mathbb{E}_{\mathbb{\Pi}_{\mathbf{Y}_{1}^{n}}
}
\negthinspace \Big[
\exp\negthinspace\Big(
m \cdot \Big(
F_{\gamma,\widetilde{c},1}\Big(D_{\widetilde{c} \cdot 
\varphi_{\gamma}}(\boldsymbol{\xi}_{m}^{w\mathbf{W}},\mathds{P}^{aux})\Big)
+ \Phi_{\mathbf{R}_{n}(\mathbf{Y}_{1}^{n})}\big(\boldsymbol{\xi}_{m}^{w\mathbf{W}}\big)
\Big)
\Big)
\cdot \textfrak{1}_{
\textrm{$\boldsymbol{\Omega}$\hspace{-0.19cm}$\boldsymbol{\Omega}$}}\big(\boldsymbol{\xi}_{m}^{w\mathbf{W}}\big)
\, \Big] 
\right)
\label{max.risk.5a.new.sample}
\end{align}
for all sets $\boldsymbol{\Omega}$\hspace{-0.23cm}$\boldsymbol{\Omega} \subset \widetilde{\mathcal{M}}_{\gamma}$ 
(with $A=1$) satisfying the 
regularity properties \eqref{regularity simplex} 
\textit{in the relative topology};
also the Remark \ref{brostu5:rem.Paux} carries over correspondingly. 

\vspace{0.3cm}
\noindent
Based on \eqref{max.risk.5a.new.sample}, one can
approximate (for fixed $n$) the corresponding 
sample-dependent maximum
$\sup_{\mathds{Q}\in \textrm{$\boldsymbol{\Omega}$\hspace{-0.19cm}$\boldsymbol{\Omega}$}} 
\Phi_{\mathbf{R}_{n}(\mathbf{Y}_{1}^{n})}(\mathds{Q})$ by
\begin{align}
& 
\frac{1}{m} \log \negthinspace \left( \ 
\mathbb{E}_{\mathbb{\Pi}_{\mathbf{Y}_{1}^{n}}
}
\negthinspace \Big[
\exp\negthinspace\Big(
m \cdot \Big(
F_{\gamma,\widetilde{c},1}\Big(D_{\widetilde{c} \cdot 
\varphi_{\gamma}}(\boldsymbol{\xi}_{m}^{w\mathbf{W}},\mathds{P}^{aux})\Big)
+ \Phi_{\mathbf{R}_{n}(\mathbf{Y}_{1}^{n})}\big(\boldsymbol{\xi}_{m}^{w\mathbf{W}}\big)
\Big)
\Big)
\cdot \textfrak{1}_{
\textrm{$\boldsymbol{\Omega}$\hspace{-0.19cm}$\boldsymbol{\Omega}$}}\big(\boldsymbol{\xi}_{m}^{w\mathbf{W}}\big)
\, \Big] 
\right)
\ \approx \ \sup_{\mathds{Q}\in \textrm{$\boldsymbol{\Omega}$\hspace{-0.19cm}$\boldsymbol{\Omega}$}} 
\Phi_{\mathbf{R}_{n}(\mathbf{Y}_{1}^{n})}(\mathds{Q})
\label{max.risk.5b.sample}
\end{align}
for large $m$.

\vspace{0.3cm}
\noindent
In the spirit of Subsection \ref{SectStochSubsimplex.CASM.comb},
by combining 
\eqref{max.risk.4}, \eqref{max.risk.5a.new.sample},
\eqref{R-convergence} and \eqref{consistency1}
with  $\mathbf{R}_{n} := \mathbf{R}_{n}(\mathbf{Y}_{1}^{n})$, 
it is possible --- under appropriate analytic conditions ---
to obtain the following coherence/consistency result
\begin{align}
& 
\Phi_{\mathbf{R}}(\textrm{$\boldsymbol{\Omega}$\hspace{-0.23cm}$\boldsymbol{\Omega}$}) := 
\sup_{\mathds{Q}\in \textrm{$\boldsymbol{\Omega}$\hspace{-0.19cm}$\boldsymbol{\Omega}$} }
\Phi_{\mathbf{R}}(\mathds{Q})
= \lim_{n\rightarrow \infty }
\overline{FM}_{n}^{BS}\Big(Y_{1},\ldots,Y_{n}\Big) 
\nonumber\\
&
\ = \ \lim_{n\rightarrow \infty } \lim_{m\rightarrow \infty } 
\frac{1}{m}
\log \negthinspace \left( \ 
\mathbb{E}_{\mathbb{\Pi}_{\mathbf{Y}_{1}^{n}}
}
\negthinspace \Big[
\exp\negthinspace\Big(
m \cdot \Big(
F_{\gamma,\widetilde{c},1}\Big(D_{\widetilde{c} \cdot 
\varphi_{\gamma}}(\boldsymbol{\xi}_{m}^{w\mathbf{W}},\mathds{P}^{aux})\Big)
+ \Phi_{\mathbf{R}_{n}(\mathbf{Y}_{1}^{n})}\big(\boldsymbol{\xi}_{m}^{w\mathbf{W}}\big)
\Big)
\Big)
\cdot \textfrak{1}_{
\textrm{$\boldsymbol{\Omega}$\hspace{-0.19cm}$\boldsymbol{\Omega}$}}\big(\boldsymbol{\xi}_{m}^{w\mathbf{W}}\big)
\, \Big] 
\right) 
\quad a.s. 
\, 
\nonumber
\end{align}
for all compact sets $\boldsymbol{\Omega}$\hspace{-0.23cm}$\boldsymbol{\Omega} \subset \widetilde{\mathcal{M}}_{\gamma}$ 
(with $A=1$) satisfying the 
regularity properties \eqref{regularity simplex} 
\textit{in the relative topology}. 

\vspace{0.3cm}
\noindent
Clearly, all the above considerations can be applied to tackle
$\sup_{\mathds{Q}\in \textrm{$\boldsymbol{\Omega}$\hspace{-0.19cm}$\boldsymbol{\Omega}$}} \Phi(\mathbf{Q})$
for all the directed distances/divergences $\Phi(\cdot) := \Phi_{\mathbf{R}}(\cdot)$ and friends 
discussed in the above Subsection \ref{SectDetGeneral.BDM1.min.risk.friends}.


\subsection{Minimization via Base-Divergence-Method 2}
\label{SectDetGeneral.BDM2.min.risk}

\vspace{0.2cm}
\noindent
For the \textit{non-narrow-sense} BS-minimizability under risk,
one proceed analogously to the Subsection
\ref{SectDetGeneral.BDM1.min.risk},
but now alternatively choose the
innmin scaled Bregman power divergences 
$\breve{D}_{\widetilde{c} \cdot 
\varphi_{\gamma},\cdot}^{SBD}(\cdot,\mathbf{Q}^{\ast\ast})$
as involved base-divergences. Firstly, we approximate 
--- based on the \textit{concrete data} $\mathbf{y}_{1}^{n} :=(y_{1},\ldots,y_{n})$ 
and $\mathbf{R}_{n}(\mathbf{y}_{1}^{n})$ ---
the corresponding empirical (i.e. concrete-data-dependent) minimum given by
\begin{equation}
\underline{FM}_{n}^{BS}\Big(y_{1},\ldots,y_{n}\Big) = 
\underline{FM}_{n}^{BS}\Big(\mathbf{R}_{n}(\mathbf{y}_{1}^{n})\Big) := 
\inf_{\mathds{Q}\in \textrm{$\boldsymbol{\Omega}$\hspace{-0.19cm}$\boldsymbol{\Omega}$}} 
\Phi_{\mathbf{R}_{n}(\mathbf{y}_{1}^{n})}(\mathds{Q})
\qquad \textrm{(cf. \eqref{min.risk.4})}
\nonumber
\end{equation}
via an appropriate application of Theorem \ref{brostu5:thm.Fmin.simplex.SBD},
and end up with 
\begin{align}
& 
\underline{FM}_{n}^{BS}\Big(y_{1},\ldots,y_{n}\Big)  = 
-  \lim_{m\rightarrow \infty } 
\frac{1}{m}\log \negthinspace \left( \ 
\mathbb{E}_{\mathbb{\Pi}_{\mathbf{y}_{1}^{n}}}\negthinspace \Big[
\exp\negthinspace\Big(
m \cdot \Big(
\breve{D}_{\widetilde{c} \cdot 
\varphi_{\gamma},\mathds{P}^{aux}}^{SBD}(\boldsymbol{\xi}_{m}^{w\mathbf{V}},\mathbf{Q}^{\ast\ast}) 
- \Phi_{\mathbf{R}_{n}(\mathbf{y}_{1}^{n})}\big(\boldsymbol{\xi}_{m}^{w\mathbf{V}}\big)
\Big)
\Big)
\cdot \textfrak{1}_{
\textrm{$\boldsymbol{\Omega}$\hspace{-0.19cm}$\boldsymbol{\Omega}$}}\big(\boldsymbol{\xi}_{m}^{w\mathbf{V}}\big)
\, \Big] 
\right)
\label{min.risk.5a.SBD}
\end{align}
for all sets $\boldsymbol{\Omega}$\hspace{-0.23cm}$\boldsymbol{\Omega} \subset \widetilde{\mathcal{M}}_{\gamma}$ 
(with $A=1$) satisfying the 
regularity properties \eqref{regularity simplex} 
\textit{in the relative topology}. 
Here, in \eqref{min.risk.5a.SBD} we have employed the random vectors
$\boldsymbol{\xi}_{m}^{w\mathbf{V}}$ constructed from an auxiliary deterministic
probability vector $\mathds{P}^{aux}  \in \mathbb{S}_{> 0}^{K}$ by ---
cf. \eqref{brostu5:fo.norweiemp.vec.det.SBD} 
with $m$ instead of $n$
and blocks $I_{k}^{(m)}$ of sizes $m_{k} :=\lfloor m \cdot 
p_{k}^{aux}\rfloor$ ($k=1,\ldots,K-1$) and $m_{K} := m- \sum_{k=1}^{K-1} m_{k}$
instead of $n_{k}$, $n_{K}$ ---
\begin{eqnarray}
\boldsymbol{\xi}_{m}^{w\mathbf{V}} &:=&
\begin{cases}
\left(\frac{\sum_{i \in I_{1}^{(m)}}V_{i}}{\sum_{k=1}^{K}\sum_{i \in I_{k}^{(m)}}V_{i}},
\ldots, \frac{\sum_{i \in I_{K}^{(m)}}V_{i}}{\sum_{k=1}^{K}\sum_{i \in I_{k}^{(m)}}V_{i}} \right) ,
\qquad \textrm{if } \sum_{j=1}^{m} V_{j} \ne 0, \\
\ (\infty, \ldots, \infty) =: \boldsymbol{\infty}, \hspace{4.0cm} \textrm{if } \sum_{j=1}^{m} V_{j} = 0;
\end{cases}
\label{brostu5:fo.norweiemp.vec.det.m.SBD} 
\end{eqnarray}
moreover, $(V_{i})_{i \in \mathbb{N}}$ and $(Y_{i})_{i \in \mathbb{N}}$ 
are supposed to be independent\footnote{
notice that all the $Y_{i}$ and $V_{i}$ and live on the \textit{same} underlying probability space $(\mathfrak{X},\mathcal{A},\mathbb{\Pi})$
}.
In accordance with Remark \ref{brostu5:rem.Paux},
the involved $\mathds{P}^{aux}$ may also depend on the deterministic data sequence
$\mathbf{y}_{1}^{n}$;
for the special case $\mathds{P}^{aux} := \mathds{P}_{n}^{emp}(\mathbf{y}_{1}^{n})$ ,
we obtain $\boldsymbol{\xi}_{m}^{w\mathbf{V}} = \boldsymbol{\xi}_{n,m,\mathbf{y}}^{w\mathbf{V}}$
(cf. \eqref{brostu5:fo.norweiemp.vec.risk.pnemp.innminSBD}
respectively \eqref{brostu5:fo.norweiemp.vec.risk.pnemp.var.innminSBD}). 
Based on \eqref{min.risk.5a.SBD}, one can (for fixed $n$)
approximate the corresponding empirical minimum by
\begin{align}
& 
- \frac{1}{m}
\log \negthinspace \left( \ 
\mathbb{E}_{\mathbb{\Pi}_{\mathbf{y}_{1}^{n}}}\negthinspace \Big[
\exp\negthinspace\Big(
m \cdot \Big(
\breve{D}_{\widetilde{c} \cdot 
\varphi_{\gamma},\mathds{P}^{aux}}^{SBD}(\boldsymbol{\xi}_{m}^{w\mathbf{V}},\mathbf{Q}^{\ast\ast}) 
- \Phi_{\mathbf{R}_{n}(\mathbf{y}_{1}^{n})}\big(\boldsymbol{\xi}_{m}^{w\mathbf{V}}\big)
\Big) \Big)
\cdot \textfrak{1}_{
\textrm{$\boldsymbol{\Omega}$\hspace{-0.19cm}$\boldsymbol{\Omega}$}}\big(\boldsymbol{\xi}_{m}^{w\mathbf{V}}\big)
\, \Big] 
\right)
\ \approx \ \inf_{\mathds{Q}\in \textrm{$\boldsymbol{\Omega}$\hspace{-0.19cm}$\boldsymbol{\Omega}$}} 
\Phi_{\mathbf{R}_{n}(\mathbf{y}_{1}^{n})}(\mathds{Q})
\nonumber
\end{align}
for large $m$.
Certainly, from \eqref{min.risk.5a.SBD} we can straightforwardly deduce
for the random sample $\mathbf{Y}_{1}^{n} = (Y_{1},\ldots,Y_{n})$
\begin{align}
& 
\underline{FM}_{n}^{BS}\Big(Y_{1},\ldots,Y_{n}\Big)  = 
-  \lim_{m\rightarrow \infty } 
\frac{1}{m}\log \negthinspace \left( \ 
\mathbb{E}_{\mathbb{\Pi}_{\mathbf{Y}_{1}^{n}}}\negthinspace \Big[
\exp\negthinspace\Big(
m \cdot \Big(
\breve{D}_{\widetilde{c} \cdot 
\varphi_{\gamma},\mathds{P}^{aux}}^{SBD}(\boldsymbol{\xi}_{m}^{w\mathbf{V}},\mathbf{Q}^{\ast\ast}) 
- \Phi_{\mathbf{R}_{n}(\mathbf{Y}_{1}^{n})}\big(\boldsymbol{\xi}_{m}^{w\mathbf{V}}\big)
\Big)
\Big)
\cdot \textfrak{1}_{
\textrm{$\boldsymbol{\Omega}$\hspace{-0.19cm}$\boldsymbol{\Omega}$}}\big(\boldsymbol{\xi}_{m}^{w\mathbf{V}}\big)
\, \Big] 
\right)
\label{min.risk.5a.SBD.sample}
\end{align}
for all sets $\boldsymbol{\Omega}$\hspace{-0.23cm}$\boldsymbol{\Omega} \subset \widetilde{\mathcal{M}}_{\gamma}$ 
(with $A=1$) satisfying the 
regularity properties \eqref{regularity simplex} 
\textit{in the relative topology}; also the Remark \ref{brostu5:rem.Paux} carries over correspondingly.

\vspace{0.3cm}
\noindent
Based on \eqref{min.risk.5a.SBD.sample}, one can (for fixed $n$)
approximate the corresponding sample-dependent minimum by
\begin{align}
& 
- \frac{1}{m}\log \negthinspace \left( \ 
\mathbb{E}_{\mathbb{\Pi}_{\mathbf{Y}_{1}^{n}}}\negthinspace \Big[
\exp\negthinspace\Big(
m \cdot \Big(
\breve{D}_{\widetilde{c} \cdot 
\varphi_{\gamma},\mathds{P}^{aux}}^{SBD}(\boldsymbol{\xi}_{m}^{w\mathbf{V}},\mathbf{Q}^{\ast\ast}) 
- \Phi_{\mathbf{R}_{n}(\mathbf{Y}_{1}^{n})}\big(\boldsymbol{\xi}_{m}^{w\mathbf{V}}\big)
\Big)
\Big)
\cdot \textfrak{1}_{
\textrm{$\boldsymbol{\Omega}$\hspace{-0.19cm}$\boldsymbol{\Omega}$}}\big(\boldsymbol{\xi}_{m}^{w\mathbf{V}}\big)
\, \Big] 
\right)
\ \approx \ \inf_{\mathds{Q}\in \textrm{$\boldsymbol{\Omega}$\hspace{-0.19cm}$\boldsymbol{\Omega}$}} 
\Phi_{\mathbf{R}_{n}(\mathbf{Y}_{1}^{n})}(\mathds{Q})
\label{min.risk.5b.SBD.sample}
\end{align}
for large $m$.

\vspace{0.2cm}
\noindent
Finally, in the spirit of Subsection \ref{SectStochSubsimplex.CASM.comb},
by combining 
\eqref{min.risk.4}, \eqref{min.risk.5a.SBD.sample},
\eqref{R-convergence} and \eqref{consistency1}
with  $\mathbf{R}_{n} := \mathbf{R}_{n}(\mathbf{Y}_{1}^{n})$, 
it is possible --- under appropriate analytic conditions ---
to obtain the following coherence/consistency result
\begin{align}
& 
\Phi_{\mathbf{R}}(\textrm{$\boldsymbol{\Omega}$\hspace{-0.23cm}$\boldsymbol{\Omega}$}) := 
\inf_{\mathds{Q}\in \textrm{$\boldsymbol{\Omega}$\hspace{-0.19cm}$\boldsymbol{\Omega}$} }
\Phi_{\mathbf{R}}(\mathds{Q})
= \lim_{n\rightarrow \infty }
\underline{FM}_{n}^{BS}\Big(Y_{1},\ldots,Y_{n}\Big) 
\nonumber\\
&
\ = \ - \, 
\lim_{n\rightarrow \infty } \lim_{m\rightarrow \infty } 
\frac{1}{m}
\log \negthinspace \left( \ 
\mathbb{E}_{\mathbb{\Pi}_{\mathbf{Y}_{1}^{n}}}\negthinspace \Big[
\exp\negthinspace\Big(
m \cdot \Big(
\breve{D}_{\widetilde{c} \cdot 
\varphi_{\gamma},\mathds{P}^{aux}}^{SBD}(\boldsymbol{\xi}_{m}^{w\mathbf{V}},\mathbf{Q}^{\ast\ast}) 
- \Phi_{\mathbf{R}_{n}(\mathbf{Y}_{1}^{n})}\big(\boldsymbol{\xi}_{m}^{w\mathbf{V}}\big)
\Big)
\Big)
\cdot \textfrak{1}_{
\textrm{$\boldsymbol{\Omega}$\hspace{-0.19cm}$\boldsymbol{\Omega}$}}\big(\boldsymbol{\xi}_{m}^{w\mathbf{V}}\big)
\, \Big] 
\right)
\, \quad a.s.
\nonumber
\end{align}
for all sets $\boldsymbol{\Omega}$\hspace{-0.23cm}$\boldsymbol{\Omega} \subset \widetilde{\mathcal{M}}_{\gamma}$ 
(with $A=1$) satisfying the 
regularity properties \eqref{regularity simplex} 
\textit{in the relative topology}. 

\vspace{0.3cm}
\noindent
Clearly, all the above considerations can be applied to tackle
$\inf_{\mathds{Q}\in \textrm{$\boldsymbol{\Omega}$\hspace{-0.19cm}$\boldsymbol{\Omega}$}} \Phi(\mathbf{Q})$
for all the directed distances/divergences $\Phi(\cdot) := \Phi_{\mathbf{R}}(\cdot)$ and friends 
discussed in the above Subsection \ref{SectDetGeneral.BDM1.min.risk.friends}.


\subsection{Maximization via Base-Divergence-Method 2}

\vspace{0.2cm}
\noindent
For the \textit{non-narrow-sense} BS-maximizability under risk,
one can proceed analogously to the previous Subsection
\ref{SectDetGeneral.BDM2.min.risk},
and approximate --- based on concrete data 
$\mathbf{y}_{1}^{n} = \left(y_{1},\ldots,y_{n}\right)$
from the samples $\mathbf{Y}_{1}^{n} = \left(Y_{1},\ldots,Y_{n}\right)$ 
satisfying \eqref{R-convergence} ---
the \textit{unknown-parameter-carrying function}
\begin{equation}
\Phi_{\mathbf{R}}(\textrm{$\boldsymbol{\Omega}$\hspace{-0.23cm}$\boldsymbol{\Omega}$}) := 
\sup_{\mathds{Q}\in \textrm{$\boldsymbol{\Omega}$\hspace{-0.19cm}$\boldsymbol{\Omega}$} }
\Phi_{\mathbf{R}}(\mathds{Q})
\qquad \textrm{(cf. \eqref{max.risk.2})}
\nonumber
\end{equation}
by the empirical (i.e. concrete-data-dependent) maximum
\begin{equation}
\overline{FM}_{n}^{BS}\Big(y_{1},\ldots,y_{n}\Big) = 
\overline{FM}_{n}^{BS}\Big(\mathbf{R}_{n}(\mathbf{y}_{1}^{n})\Big) := 
\sup_{\mathds{Q}\in \textrm{$\boldsymbol{\Omega}$\hspace{-0.19cm}$\boldsymbol{\Omega}$}} 
\Phi_{\mathbf{R}_{n}(\mathbf{y}_{1}^{n})}(\mathds{Q})
\qquad \textrm{(cf. \eqref{max.risk.4})}
\nonumber
\end{equation}
via an appropriate application of Theorem \ref{brostu5:thm.Fmax.simplex.SBD},
and end up with 
\begin{align}
& 
\overline{FM}_{n}^{BS}\Big(y_{1},\ldots,y_{n}\Big)  = 
\lim_{m\rightarrow \infty } 
\frac{1}{m}
\log \negthinspace \left( \ 
\mathbb{E}_{\mathbb{\Pi}_{\mathbf{y}_{1}^{n}}}\negthinspace \Big[
\exp\negthinspace\Big(
m \cdot \Big(
\breve{D}_{\widetilde{c} \cdot 
\varphi_{\gamma},\mathds{P}^{aux}}^{SBD}(\boldsymbol{\xi}_{m}^{w\mathbf{V}},\mathbf{Q}^{\ast\ast})  
+ \Phi_{\mathbf{R}_{n}(\mathbf{y}_{1}^{n})}\big(\boldsymbol{\xi}_{m}^{w\mathbf{V}}\big)
\Big)
\Big)
\cdot \textfrak{1}_{
\textrm{$\boldsymbol{\Omega}$\hspace{-0.19cm}$\boldsymbol{\Omega}$}}\big(\boldsymbol{\xi}_{m}^{w\mathbf{V}}\big)
\, \Big] 
\right)
\label{max.risk.5a.SBD}
\end{align}
for all sets $\boldsymbol{\Omega}$\hspace{-0.23cm}$\boldsymbol{\Omega} \subset \widetilde{\mathcal{M}}_{\gamma}$ 
(with $A=1$) satisfying the 
regularity properties \eqref{regularity simplex} 
\textit{in the relative topology};
in accordance with Remark \ref{brostu5:rem.Paux},
the involved $\mathds{P}^{aux}$ may also depend on the deterministic data sequence
$\mathbf{y}_{1}^{n}$;
for the special case $\mathds{P}^{aux} := \mathds{P}_{n}^{emp}(\mathbf{y}_{1}^{n})$,
we obtain $\boldsymbol{\xi}_{m}^{w\mathbf{V}} = \boldsymbol{\xi}_{n,m,\mathbf{y}}^{w\mathbf{V}}$
(cf. \eqref{brostu5:fo.norweiemp.vec.risk.pnemp.innminSBD}
respectively \eqref{brostu5:fo.norweiemp.vec.risk.pnemp.var.innminSBD}).
Based on \eqref{max.risk.5a.SBD}, one can (for fixed $n$)
approximate the corresponding empirical maximum by
\begin{align}
& 
\frac{1}{m} \log \negthinspace \left( \ 
\mathbb{E}_{\mathbb{\Pi}_{\mathbf{y}_{1}^{n}}}\negthinspace \Big[
\exp\negthinspace\Big(
m \cdot \Big(
\breve{D}_{\widetilde{c} \cdot 
\varphi_{\gamma},\mathds{P}^{aux}}^{SBD}(\boldsymbol{\xi}_{m}^{w\mathbf{V}},\mathbf{Q}^{\ast\ast}) 
+ \Phi_{\mathbf{R}_{n}(\mathbf{y}_{1}^{n})}\big(\boldsymbol{\xi}_{m}^{w\mathbf{V}}\big)
\Big)
\Big)
\cdot \textfrak{1}_{
\textrm{$\boldsymbol{\Omega}$\hspace{-0.19cm}$\boldsymbol{\Omega}$}}\big(\boldsymbol{\xi}_{m}^{w\mathbf{V}}\big)
\, \Big] 
\right)
\ \approx \ \sup_{\mathds{Q}\in \textrm{$\boldsymbol{\Omega}$\hspace{-0.19cm}$\boldsymbol{\Omega}$}} 
\Phi_{\mathbf{R}_{n}(\mathbf{y}_{1}^{n})}(\mathds{Q})
\nonumber
\end{align}
for large $m$. Clearly, from \eqref{max.risk.5a.SBD} we can immediately derive
for the random sample $\mathbf{Y}_{1}^{n} = (Y_{1},\ldots,Y_{n})$
\begin{align}
& 
\overline{FM}_{n}^{BS}\Big(Y_{1},\ldots,Y_{n}\Big)  = 
\lim_{m\rightarrow \infty } 
\frac{1}{m}\log \negthinspace \left( \ 
\mathbb{E}_{\mathbb{\Pi}_{\mathbf{Y}_{1}^{n}}}\negthinspace \Big[
\exp\negthinspace\Big(
m \cdot \Big(
\breve{D}_{\widetilde{c} \cdot 
\varphi_{\gamma},\mathds{P}^{aux}}^{SBD}(\boldsymbol{\xi}_{m}^{w\mathbf{V}},\mathbf{Q}^{\ast\ast}) 
+ \Phi_{\mathbf{R}_{n}(\mathbf{Y}_{1}^{n})}\big(\boldsymbol{\xi}_{m}^{w\mathbf{V}}\big)
\Big)
\Big)
\cdot \textfrak{1}_{
\textrm{$\boldsymbol{\Omega}$\hspace{-0.19cm}$\boldsymbol{\Omega}$}}\big(\boldsymbol{\xi}_{m}^{w\mathbf{V}}\big)
\, \Big] 
\right)
\label{max.risk.5a.SBD.sample}
\end{align}
for all sets $\boldsymbol{\Omega}$\hspace{-0.23cm}$\boldsymbol{\Omega} \subset \widetilde{\mathcal{M}}_{\gamma}$ 
(with $A=1$) satisfying the 
regularity properties \eqref{regularity simplex} 
\textit{in the relative topology}; also the Remark \ref{brostu5:rem.Paux} carries over correspondingly.

\vspace{0.3cm}
\noindent
Based on \eqref{max.risk.5a.SBD.sample}, one can (for fixed $n$)
approximate the corresponding sample-dependent maximum by
\begin{align}
& 
\frac{1}{m}\log \negthinspace \left( \ 
\mathbb{E}_{\mathbb{\Pi}_{\mathbf{Y}_{1}^{n}}}\negthinspace \Big[
\exp\negthinspace\Big(
m \cdot \Big(
\breve{D}_{\widetilde{c} \cdot 
\varphi_{\gamma},\mathds{P}^{aux}}^{SBD}(\boldsymbol{\xi}_{m}^{w\mathbf{V}},\mathbf{Q}^{\ast\ast}) 
+ \Phi_{\mathbf{R}_{n}(\mathbf{Y}_{1}^{n})}\big(\boldsymbol{\xi}_{m}^{w\mathbf{V}}\big)
\Big)
\Big)
\cdot \textfrak{1}_{
\textrm{$\boldsymbol{\Omega}$\hspace{-0.19cm}$\boldsymbol{\Omega}$}}\big(\boldsymbol{\xi}_{m}^{w\mathbf{V}}\big)
\, \Big] 
\right)
\ \approx \ \sup_{\mathds{Q}\in \textrm{$\boldsymbol{\Omega}$\hspace{-0.19cm}$\boldsymbol{\Omega}$}} 
\Phi_{\mathbf{R}_{n}(\mathbf{Y}_{1}^{n})}(\mathds{Q})
\label{max.risk.5b.SBD.sample}
\end{align}
for large $m$.

\vspace{0.3cm}
\noindent
In the spirit of Subsection \ref{SectStochSubsimplex.CASM.comb},
by combining \eqref{max.risk.4}, \eqref{max.risk.5a.SBD.sample},
\eqref{R-convergence} and \eqref{consistency1}
with  $\mathbf{R}_{n} := \mathbf{R}_{n}(\mathbf{Y}_{1}^{n})$, 
it is possible --- under appropriate analytic conditions ---
to obtain the following coherence/consistency result
\begin{align}
& 
\Phi_{\mathbf{R}}(\textrm{$\boldsymbol{\Omega}$\hspace{-0.23cm}$\boldsymbol{\Omega}$}) := 
\sup_{\mathds{Q}\in \textrm{$\boldsymbol{\Omega}$\hspace{-0.19cm}$\boldsymbol{\Omega}$} }
\Phi_{\mathbf{R}}(\mathds{Q})
= \lim_{n\rightarrow \infty }
\overline{FM}_{n}^{BS}\Big(Y_{1},\ldots,Y_{n}\Big) 
\nonumber\\
& \ = \ 
\lim_{n\rightarrow \infty} \lim_{m\rightarrow \infty} 
\frac{1}{m} \log \negthinspace \left( \ 
\mathbb{E}_{\mathbb{\Pi}_{\mathbf{Y}_{1}^{n}}}\negthinspace \Big[
\exp\negthinspace\Big(
m \cdot \Big(
\breve{D}_{\widetilde{c} \cdot 
\varphi_{\gamma},\mathds{P}^{aux}}^{SBD}(\boldsymbol{\xi}_{m}^{w\mathbf{V}},\mathbf{Q}^{\ast\ast})
+ \Phi_{\mathbf{R}_{n}(\mathbf{Y}_{1}^{n})}\big(\boldsymbol{\xi}_{m}^{w\mathbf{V}}\big)
\Big)
\Big)
\cdot \textfrak{1}_{
\textrm{$\boldsymbol{\Omega}$\hspace{-0.19cm}$\boldsymbol{\Omega}$}}\big(\boldsymbol{\xi}_{m}^{w\mathbf{V}}\big)
\, \Big] 
\right)
\, \quad a.s.
\nonumber
\end{align}
for all sets $\boldsymbol{\Omega}$\hspace{-0.23cm}$\boldsymbol{\Omega} \subset \widetilde{\mathcal{M}}_{\gamma}$ 
(with $A=1$) satisfying the 
regularity properties \eqref{regularity simplex}. 

\vspace{0.3cm}
\noindent
Clearly, all the above considerations can be applied to tackle
$\sup_{\mathds{Q}\in \textrm{$\boldsymbol{\Omega}$\hspace{-0.19cm}$\boldsymbol{\Omega}$}} \Phi(\mathbf{Q})$
for all the directed distances/divergences $\Phi(\cdot) := \Phi_{\mathbf{R}}(\cdot)$ and friends 
discussed in the above Subsection \ref{SectDetGeneral.BDM1.min.risk.friends}.


\section{Bare-Simulation Estimators For General Deterministic Divergence-Optimization-Problems}
\label{SectEstimators.new.det.nonvoid}

\vspace{0.2cm}
Recall that we are interested in the constrained optimization of the \textit{continuous} 
distance-connected functions 
$\mathbf{\Omega} \ni \mathbf{Q} \mapsto \Phi(\mathbf{Q})$
in the above-mentioned cases (D1) to (D8) of Subsection \ref{SectDetGeneral.Friends}, and beyond
(e.g. $\Phi(\cdot) := D_{\breve{\varphi}}(\cdot,\breve{\mathbf{P}})$
may be a $\breve{\varphi}-$divergence with pregiven $\breve{\mathbf{P}}$ and $\breve{\varphi}$, cf.
Remark \ref{brostu5:rem.thm.Fmin}(iii)). 
In this section, we consider constraint sets $\mathbf{\Omega} \in \mathbb{R}^{K}$ 
with regularity properties \eqref{regularity} 
such that
the function $\Phi(\cdot)$ possesses a (not necessarily unique) minimizer; 
for this, we construct 
\textit{naive estimators} as well as 
\textit{speed-up estimators} (approximations) of the \textit{minimum value}
$\inf_{\mathbf{Q}\in \mathbf{\Omega}}\Phi (\mathbf{Q}) 
= \min_{\mathbf{Q}\in \mathbf{\Omega}} \Phi (\mathbf{Q})$ and of the corresponding
(set of) \textit{minimizers} $\arg \inf_{\mathbf{Q}\in \mathbf{\Omega }}\Phi (\mathbf{Q}) 
=\arg \min_{\mathbf{Q}\in \mathbf{\Omega}} \Phi (\mathbf{Q})$.
The case of maximum values and maximizers will be treated analogously. 
We mainly focus on compact sets $\mathbf{\Omega}$ but also discuss some relaxations thereof.


\subsection{Naive estimators of min and argmin --- Base-Divergence-Method 1, compact case}
\label{SectEstimators.new.det.nonvoid.meth1.compact.min}

\vspace{0.2cm}
\noindent
Recall that from Theorem \ref{brostu5:thm.Fmin}(a)
we obtain for any continuous function $\Phi: \mathbf{\Omega} \mapsto \mathbb{R}$ on a compact
set $\mathbf{\Omega}\subset \mathbb{R}^{K}$ with \eqref{regularity} the assertion 
\vspace{0.2cm}
\noindent
\begin{equation}
\min_{\mathbf{Q}\in \mathbf{\Omega}} \Phi(\mathbf{Q})
\ = \ 
- \, 
\lim_{n\rightarrow \infty }\frac{1}{n}\log \negthinspace \left( \ 
\mathbb{E}_{\mathbb{\Pi}}\negthinspace \Big[
\exp\negthinspace\Big(n \cdot \Big(
D_{\varphi }\big(M_{\mathbf{P}} \cdot \boldsymbol{\xi }_{n}^{\mathbf{\widetilde{W}}},\mathbf{P}\big) 
- \Phi\big(M_{\mathbf{P}} \cdot \boldsymbol{\xi }_{n}^{\mathbf{\widetilde{W}}}\big)
\Big)
\Big)
\cdot \textfrak{1}_{\mathbf{\Omega}}\big(M_{\mathbf{P}} \cdot \boldsymbol{\xi }_{n}^{\mathbf{\widetilde{W}}}\big)
\, \Big]
\right) \, ,
\label{brostu5:fo.BSmin.extended.min}
\end{equation}
where $M_{\mathbf{P}} =\sum_{i=1}^{K}p_{i}>0$ and 
\begin{equation}
\boldsymbol{\xi }_{n}^{\mathbf{\widetilde{W}}}:=\Big(\frac{1}{n}\sum_{i\in
I_{1}^{(n)}}\widetilde{W}_{i},\ldots ,\frac{1}{n}\sum_{i\in I_{K}^{(n)}}
\widetilde{W}_{i}\Big)
\qquad \textrm{(cf. \eqref{Xi_n^W vector})}
\nonumber
\end{equation}
is constructed from a sequence $\widetilde{W}:=(\widetilde{W}_{i})_{i\in \mathbb{N}}$ of random variables, 
where the $\widetilde{W}_{i}$'s are i.i.d. copies of the random variable $\widetilde{W}$
whose distribution 
is $\mathbb{\Pi }[\widetilde{W}\in \cdot \,]=\widetilde{\mathbb{\bbzeta}}[\,\cdot \,]$ 
being attributed to the divergence generator  
$\widetilde{\varphi} := M_{\mathbf{P}} \cdot \varphi \in \widetilde{\Upsilon}(]a,b[)$
by the representability \eqref{brostu5:fo.link.var}. 
Within such a set-up, we thus obtain for large $n \in \mathbb{N}$ (cf. \eqref{brostu5:fo.BSmin.extended.min})
the approximation
\begin{equation}
\Phi(\mathbf{\Omega}) := \min_{\mathbf{Q}\in \mathbf{\Omega}} \Phi(\mathbf{Q})
\ \approx \ - \, 
\frac{1}{n}\log \negthinspace \left( \ 
\mathbb{E}_{\mathbb{\Pi}}\negthinspace \Big[
\exp\negthinspace\Big(n \cdot \Big(
D_{\varphi }\big(M_{\mathbf{P}} \cdot \boldsymbol{\xi }_{n}^{\mathbf{\widetilde{W}}},\mathbf{P}\big) 
- \Phi\big(M_{\mathbf{P}} \cdot \boldsymbol{\xi }_{n}^{\mathbf{\widetilde{W}}}\big)
\Big)
\Big)
\cdot \textfrak{1}_{\mathbf{\Omega}}\big(M_{\mathbf{P}} \cdot \boldsymbol{\xi }_{n}^{\mathbf{\widetilde{W}}}\big)
\, \Big] 
\right)
\,  ,
\label{brostu5:fo.BSmin.extended.approx}
\end{equation}
and hence for getting an estimator of the minimum value $\Phi(\mathbf{\Omega})$
one can estimate the right-hand side of \eqref{brostu5:fo.BSmin.extended.approx}.
To achieve this, for the rest of this section we assume 
that $n$ is chosen such that all 
$n \cdot \frac{p_{k}}{M_{\mathbf{P}}}$ are integers
(and hence, $n = \sum_{k=1}^{K} n_{k}$ with $n_{k} = n \cdot \frac{p_{k}}{M_{\mathbf{P}}}$) --- the remaining case works analogously.
Extending the lines of Broniatowski \& Stummer \cite{Bro:23a} (who deal with the narrow-sense
BS minimizability), a corresponding \textit{naive (crude) estimator} can be constructed by 
\begin{equation}
\widehat{\Phi(\mathbf{\Omega})}_{n,L}^{naive,1} \ := \ 
- \frac{1}{n}\log \frac{1}{L}\sum_{\ell =1}^{L} 
\exp\negthinspace\Big(n \cdot \Big(
D_{\varphi }\big(M_{\mathbf{P}} \cdot \boldsymbol{\xi }_{n}^{\mathbf{\widetilde{W}}^{(\ell)}},\mathbf{P}\big) 
- \Phi\big(M_{\mathbf{P}} \cdot \boldsymbol{\xi }_{n}^{\mathbf{\widetilde{W}}^{(\ell)}}\big)
\Big)
\Big)
\cdot
\mathbf{1}_{\mathbf{\Omega}}
\Big(M_{\mathbf{P}} \cdot \boldsymbol{\xi}_{n}^{\mathbf{\widetilde{W}}^{(\ell)}} \Big)
\, ,
\label{brostu5:fo.BSmin.extended.naive.estim}
\end{equation}
where we simulate independently $L$ copies 
$\mathbf{\widetilde{W}}^{(1)},\ldots,\mathbf{\widetilde{W}}^{(L)}$ of the vector
 $\mathbf{\widetilde{W}}:=\left( \widetilde{W}_{1},\ldots,\widetilde{W}_{n}\right) $ 
with independent entries under $\widetilde{\mathbb{\bbzeta}}$, 
and compute each of $\boldsymbol{\xi}_{n}^{\mathbf{\widetilde{W}}^{(1)}}, \ldots,
\boldsymbol{\xi}_{n}^{\mathbf{\widetilde{W}}^{(L)}}$ according to \eqref{Xi_n^W vector}.
Clearly, with the help of the strong law of large numbers we get with 
$\widehat{\Phi(\mathbf{\Omega})}_{n,\infty}^{naive,1}: = 
\lim_{L\rightarrow \infty} \widehat{\Phi(\mathbf{\Omega})}_{n,L}^{naive,1}$
the following assertion:

\vspace{0.2cm}

\begin{proposition}
\begin{equation}
\lim_{n\rightarrow \infty} \widehat{\Phi(\mathbf{\Omega})}_{n,\infty}^{naive,1}
\ = \ 
\lim_{n\rightarrow \infty} \lim_{L\rightarrow \infty} 
\widehat{\Phi(\mathbf{\Omega})}_{n,L}^{naive,1}
\ = \ \Phi(\mathbf{\Omega}) \qquad \textrm{a.s.}
\label{brostu5:fo.BSmin.extended.naive.estim.lim}
\end{equation}
\end{proposition}

\vspace{0.2cm}
\noindent
As the corresponding \textit{very natural naive (crude) estimator} 
of the minimizer-set 
$\mathcal{Q}^{\ast} := \argmin_{\mathbf{Q} \in \mathbf{\Omega }} \Phi(\mathbf{Q})$, 
we take 
\begin{equation}
\widehat{\argmin_{\mathbf{Q} \in \mathbf{\Omega}} \Phi(\mathbf{Q})}_{n,L}^{naive,1}
:= \argmin_{\boldsymbol{\nu} \in \mathcal{W}_{n,L}} \Phi(\boldsymbol{\nu})
\label{brostu5:fo.1555}
\end{equation}
where $\mathcal{W}_{n,L} := \{M_{\mathbf{P}} \cdot \boldsymbol{\xi}_{n}^{\mathbf{\widetilde{W}}^{(\ell)}}:
\ell \in \{1,\ldots,L\} \,  \} \cap \mathbf{\Omega}$. 
In other words, as a corresponding \textit{naive (crude) estimator} 
of the (not necessarily unique) element $\mathbf{Q}^{\ast}$ of the minimizer-set 
$\mathcal{Q}^{\ast} := \argmin_{\mathbf{Q} \in \mathbf{\Omega}} \Phi(\mathbf{Q})$, 
we take the (not necessarily unique) element 
$M_{\mathbf{P}} \cdot \boldsymbol{\xi}_{n}^{\mathbf{\widetilde{W}}^{L,\ast}}$
of the set $\{M_{\mathbf{P}} \cdot \boldsymbol{\xi}_{n}^{\mathbf{\widetilde{W}}^{(\ell)}}:
\ell \in \{1,\ldots,L\} \,  \} $ such that 
$M_{\mathbf{P}} \cdot \boldsymbol{\xi}_{n}^{\mathbf{\widetilde{W}}^{L,\ast}} \in \mathbf{\Omega}$ 
and 
\begin{equation}
\Phi(M_{\mathbf{P}} \cdot \boldsymbol{\xi}_{n}^{\mathbf{\widetilde{W}}^{L,\ast}}) 
\ \leq \Phi(M_{\mathbf{P}} \cdot \boldsymbol{\xi}_{n}^{\mathbf{\widetilde{W}}^{(\ell)}})   
\qquad \textrm{for all $\ell =1,\ldots,L$ for which 
$M_{\mathbf{P}} \cdot \boldsymbol{\xi}_{n}^{\mathbf{\widetilde{W}}^{(\ell)}}$ 
belongs to $\mathbf{\Omega }$.} 
\nonumber
\end{equation}
In short, $M_{\mathbf{P}} \cdot \boldsymbol{\xi}_{n}^{\mathbf{\widetilde{W}}^{L,\ast}}$
minimizes $\Phi(\cdot)$ amongst all values 
$M_{\mathbf{P}} \cdot \boldsymbol{\xi}_{n}^{\mathbf{\widetilde{W}}^{(\ell)}}$
at hand which fall into $\mathbf{\Omega}$. 
For large enough $n \in \mathbb{N}$ and $L \in \mathbb{N}$, 
such $M_{\mathbf{P}} \cdot \boldsymbol{\xi}_{n}^{\mathbf{\widetilde{W}}^{L,\ast}}$ exists
since $\mathbf{\Omega}$ has non-void interior, by assumption \eqref{regularity}.
We prove that if $L$ and $n$ tend to infinity, then 
$M_{\mathbf{P}} \cdot \boldsymbol{\xi}_{n}^{\mathbf{\widetilde{W}}^{L,\ast}}$
concentrates to the above-mentioned set $\mathcal{Q}^{\ast}$ 
of minimizers of $\Phi(\cdot)$ on $\mathbf{\Omega}$.
As usual in similar procedures, $L$ is assumed to be large enough in order to justify 
some approximation for fixed $n$, typically the substitution of empirical means by expectations,
since $L$ is at disposal.

\noindent
Next we show that $M_{\mathbf{P}} \cdot \boldsymbol{\xi}_{n}^{\mathbf{\widetilde{W}}^{L,\ast}}$ is
a proxy minimizer of $\Phi(\cdot)$ on $\mathbf{\Omega}$, by proving the following

\vspace{0.2cm}

\begin{proposition}
\label{brostu5:prop.generaldeterministic.minimizer.naive}
There holds
\begin{equation}
\min_{\mathbf{Q}\in \mathbf{\Omega}} \Phi(\mathbf{Q}) 
\ \leq \ 
\Phi \left( M_{\mathbf{P}} \cdot \boldsymbol{\xi}_{n}^{\mathbf{\widetilde{W}}^{L,\ast}} \right)
\ \leq \ 
\widehat{\Phi(\mathbf{\Omega})}_{n,\infty}^{naive,1}
\ + \ o_{\mathbb{\Pi}}(1)
\label{minim}
\end{equation}
where $o_{\mathbb{\Pi}}(1)$ goes to $0$ as $L\rightarrow \infty$ and $n\rightarrow \infty$ under the
distribution $\mathbb{\Pi}$
(recall that $\mathbb{\Pi }[(\widetilde{W}_{1},\ldots,\widetilde{W}_{n}) \in \cdot \,]=
\widetilde{\mathbb{\bbzeta}}^{\otimes n}[\,\cdot \,]$).
\end{proposition}

\vspace{0.3cm}
\noindent
The proof of Proposition \ref{brostu5:prop.generaldeterministic.minimizer.naive}
is given in Appendix \ref{App.A} below.

\vspace{0.3cm}

\begin{remark}
\label{brostu5:rem.116}
In the current set-up of compact $\mathbf{\Omega}$ with \eqref{regularity}, 
by taking the special case $\Phi(\mathbf{Q}) := D_{\varphi}(\mathbf{Q},\mathbf{P})$ 
we obtain the naive BS-estimator $\widehat{D_{\varphi}(\mathbf{\Omega},\mathbf{P})}_{n,L}^{naive,1}$ 
of the minimum value $\min_{\mathbf{Q}\in \mathbf{\Omega}} D_{\varphi}(\mathbf{Q},\mathbf{P})$
given in Broniatowski \& Stummer~\cite{Bro:23a}.
In the latter paper, we had left the \textit{open gap} of constructing a naive BS-estimator of the 
corresponding minimizer $\argmin_{\mathbf{Q}\in \mathbf{\Omega}} D_{\varphi}(\mathbf{Q},\mathbf{P})$,
which we have now \textit{filled/resolved} by taking the corresponding special case 
$\widehat{\argmin_{\mathbf{Q} \in \mathbf{\Omega }} D_{\varphi}(\mathbf{Q},\mathbf{P})}_{n,L}^{naive,1}$
of \eqref{brostu5:fo.1555} (for a speed-up version thereof see
$\widehat{\argmin_{\mathbf{Q} \in \mathbf{\Omega }} D_{\varphi}(\mathbf{Q},\mathbf{P})}_{n,L}^{speedup,1}$
of \eqref{brostu5:fo.1555.improved} below).
\end{remark}

\vspace{0.3cm}

\begin{proposition}
In the above set-up, one has
\begin{equation}
\lim_{n\rightarrow \infty} \lim_{L\rightarrow \infty} 
\Phi\Big( \widehat{\argmin_{\mathbf{Q} \in \mathbf{\Omega}} \Phi(\mathbf{Q})}_{n,L}^{naive,1} \Big)
\ = \ \Phi(\mathbf{\Omega}) \qquad \textrm{a.s.},
\nonumber
\end{equation}
and thus the quantity 
$\Phi\Big( \widehat{\argmin_{\mathbf{Q} \in \mathbf{\Omega}} \Phi(\mathbf{Q})}_{n,L}^{naive,1} \Big)$
is a natural alternative to the estimate
$\widehat{\Phi(\mathbf{\Omega})}_{n,L}^{naive,1}$ given in \eqref{brostu5:fo.BSmin.extended.naive.estim}.

\end{proposition}


\subsection{Naive estimators of max and argmax --- Base-Divergence-Method 1, compact case}
\label{SectEstimators.new.det.nonvoid.meth1.compact.max}

\vspace{0.2cm}
\noindent
In the set-up of compact $\mathbf{\Omega}$ with \eqref{regularity}, we can treat
the maximizing problem completely analogously to the method in 
the previous Subsection \ref{SectEstimators.new.det.nonvoid.meth1.compact.min}.
Indeed, by employing Theorem \ref{brostu5:thm.Fmax}(a) instead of Theorem \ref{brostu5:thm.Fmin}(a)
we construct the naive BS-estimator 
\begin{equation}
\widehat{\Phi(\mathbf{\Omega})}_{n,L}^{naive,1} \ := \ 
\frac{1}{n}\log \frac{1}{L}\sum_{\ell =1}^{L} 
\exp\negthinspace\Big(n \cdot \Big(
D_{\varphi }\big(M_{\mathbf{P}} \cdot \boldsymbol{\xi }_{n}^{\mathbf{\widetilde{W}}^{(\ell)}},\mathbf{P}\big) 
+ \Phi\big(M_{\mathbf{P}} \cdot \boldsymbol{\xi }_{n}^{\mathbf{\widetilde{W}}^{(\ell)}}\big)
\Big)
\Big)
\cdot
\mathbf{1}_{\mathbf{\Omega}}
\Big(M_{\mathbf{P}} \cdot \boldsymbol{\xi}_{n}^{\mathbf{\widetilde{W}}^{(\ell)}} \Big)
\, ,
\label{brostu5:fo.BSmax.extended.naive.estim}
\end{equation}
of the maximum value $\Phi(\mathbf{\Omega}) := \max_{\mathbf{Q}\in \mathbf{\Omega}} \Phi(\mathbf{Q})$;
for this, we get
\begin{equation}
\lim_{n\rightarrow \infty} \widehat{\Phi(\mathbf{\Omega})}_{n,\infty}^{naive,1}
\ = \ 
\lim_{n\rightarrow \infty} \lim_{L\rightarrow \infty} 
\widehat{\Phi(\mathbf{\Omega})}_{n,L}^{naive,1}
\ = \ \Phi(\mathbf{\Omega}) \qquad \textrm{a.s.}
\nonumber
\end{equation}
instead of \eqref{brostu5:fo.BSmin.extended.naive.estim.lim} (notice the different definition of the involved quantities).
As the corresponding very natural naive BS-estimator of the 
maximizer $\argmax_{\mathbf{Q} \in \mathbf{\Omega }} \Phi(\mathbf{Q})$ 
we take 
\begin{equation}
\widehat{\argmax_{\mathbf{Q} \in \mathbf{\Omega }} \Phi(\mathbf{Q})}_{n,L}^{naive,1}
:= \argmax_{\boldsymbol{\nu} \in \mathcal{W}_{n,L}} \Phi(\boldsymbol{\nu})
\nonumber
\end{equation}
instead of \eqref{brostu5:fo.1555}; in short, 
we take any $M_{\mathbf{P}} \cdot \boldsymbol{\xi}_{n}^{\mathbf{\widetilde{W}}^{L,\ast}}$
which maximizes $\Phi(\cdot)$ amongst all values 
$M_{\mathbf{P}} \cdot \boldsymbol{\xi}_{n}^{\mathbf{\widetilde{W}}^{(\ell)}}$
at hand which fall into $\mathbf{\Omega}$. For this, we obtain --- instead of \eqref{minim} ---
the assertion
\begin{equation}
\max_{\mathbf{Q}\in \mathbf{\Omega}} \Phi(\mathbf{Q}) 
\ \geq \ 
\Phi \left( M_{\mathbf{P}} \cdot \boldsymbol{\xi}_{n}^{\mathbf{\widetilde{W}}^{L,\ast}} \right)
\ \geq \ 
\widehat{\Phi(\mathbf{\Omega})}_{n,\infty}^{naive,1}
\ - \ o_{\mathbb{\Pi}}(1)
\nonumber
\end{equation}
where $o_{\mathbb{\Pi}}(1)$ goes to $0$ as $L\rightarrow \infty$ and $n\rightarrow \infty$ under the
distribution $\mathbb{\Pi}$.
Hence, as $L$ and $n$ tend to infinity,  
$M_{\mathbf{P}} \cdot \boldsymbol{\xi}_{n}^{\mathbf{\widetilde{W}}^{L,\ast}}$
concentrates to the set of maximizers of $\Phi(\cdot)$ on $\mathbf{\Omega}$.

\vspace{0.3cm}

\begin{proposition}
In the above set-up, one has
\begin{equation}
\lim_{n\rightarrow \infty} \lim_{L\rightarrow \infty} 
\Phi\Big( \widehat{\argmax_{\mathbf{Q} \in \mathbf{\Omega}} \Phi(\mathbf{Q})}_{n,L}^{naive,1} \Big)
\ = \ \Phi(\mathbf{\Omega}) \qquad \textrm{a.s.},
\nonumber
\end{equation}
and thus the quantity 
$\Phi\Big( \widehat{\argmax_{\mathbf{Q} \in \mathbf{\Omega}} \Phi(\mathbf{Q})}_{n,L}^{naive,1} \Big)$
is a natural alternative to the estimate
$\widehat{\Phi(\mathbf{\Omega})}_{n,L}^{naive,1}$ given in \eqref{brostu5:fo.BSmax.extended.naive.estim}.

\end{proposition}


\subsection{Naive estimators --- Base-Divergence-Method 1, non-compact case}
\label{SectEstimators.new.det.nonvoid.meth1.noncompact}

\vspace{0.2cm}
\noindent
Suppose that we are in the set-up of Theorem \ref{brostu5:thm.Fmin}(b),
which particularly means that $\mathbf{\Omega}\subset \mathbb{R}^{K}$ is not necessarily compact
but satisfies \eqref{regularity} and \eqref{def fi wrt Omega},
and that $\Phi: \mathbf{\Omega} \mapsto \mathbb{R}$ is a continuous function
which satisfies the lower-bound condition \eqref{brostu5:fo.phibound}
(notice that \eqref{brostu5:fo.phibound} trivially holds for the special case 
$\Phi(\mathbf{Q}) := D_{\varphi}(\mathbf{Q},\mathbf{P})$,
or if $\mathbf{\Omega}$ is bounded but not necessarily closed).
Additionally, let us now assume that the minimum value is achieved,
i.e. $\inf_{\mathbf{Q}\in \mathbf{\Omega}} \Phi(\mathbf{Q}) = 
\min_{\mathbf{Q}\in \mathbf{\Omega}} \Phi(\mathbf{Q}) = \Phi(\mathbf{Q}_{min})$
for some (not necessarily unique) point $\mathbf{Q}_{min} \in \mathbf{\Omega}$,
and that the corresponding set 
$\mathcal{Q}^{\ast} := \argmin_{\mathbf{Q} \in \mathbf{\Omega}} \Phi(\mathbf{Q})$
of minimizers is covered by a compact set $\mathbf{B} \in \mathbb{R}^{K}$
(e.g. we take $\mathbf{B}:= cl(\mathbf{\Omega})$ 
in case that $\mathbf{\Omega}$ is bounded but not necessarily closed). 
In such a context, we can proceed as in
Subsection \ref{SectEstimators.new.det.nonvoid.meth1.compact.min} 
by replacing $\mathbf{\Omega}$ with $\mathbf{\Omega} \cap \mathbf{B}$.
The analogous procedure applies to the maximization problem,
by proceeding as in Subsection \ref{SectEstimators.new.det.nonvoid.meth1.compact.max} 
by replacing $\mathbf{\Omega}$ with $\mathbf{\Omega} \cap \mathbf{B}$.


\subsection{Naive estimators of min and argmin --- Base-Divergence-Method 2, compact case}
\label{SectEstimators.new.det.nonvoid.meth2.compact.min}

\vspace{0.2cm}
\noindent
In the above Subsections 
\ref{SectEstimators.new.det.nonvoid.meth1.compact.min},
\ref{SectEstimators.new.det.nonvoid.meth1.compact.max},
\ref{SectEstimators.new.det.nonvoid.meth1.noncompact},
in order to optimize $\Phi(\cdot)$
we have chosen an appropriate $\varphi-$divergence 
$D_{\varphi}(M_{\mathbf{P}} 
\cdot \boldsymbol{\xi}_{n}^{\mathbf{\widetilde{W}}},\mathbf{P})$ as the so-called \textit{base divergence}
(cf. Theorem \ref{brostu5:thm.Fmin} and Theorem \ref{brostu5:thm.Fmax});
we have referred to this choice as \textit{Base-Divergence-Method 1}.
However, as can be seen from the alternative Theorem \ref{brostu5:thm.Fmin.SBD}
and Theorem \ref{brostu5:thm.Fmax.SBD}, 
for the optimization of the same $\Phi(\cdot)$ we can also choose 
--- as Base-Divergence-Method 2 --- as \textit{base divergence} an appropriate scaled Bregman distance
$D_{\varphi,\mathbf{P}}^{SBD}(M_{\mathbf{P}} 
\cdot \boldsymbol{\xi }_{n}^{\mathbf{\widetilde{V}}},\mathbf{Q}^{\ast\ast})$ (cf. \eqref{brostu5:fo.SBD.smooth}),
where $\mathbf{Q}^{\ast\ast} \in \mathbb{R}^{K}$ 
\textit{NEED NOT} be in $\mathbf{\Omega}$; for instance, if the goal
is to optimize $\mathbf{Q} \mapsto \Phi(\mathbf{Q}) := D_{\varphi,\mathbf{P}}^{SBD}(\mathbf{Q},\mathbf{Q}^{\ast\ast})$ 
for application-context-specific $\mathbf{Q}^{\ast\ast} \notin \mathbf{\Omega}$, the use of the
base-divergence
$D_{\varphi,\mathbf{P}}^{SBD}(M_{\mathbf{P}} 
\cdot \boldsymbol{\xi }_{n}^{\mathbf{\widetilde{V}}},\mathbf{Q}^{\ast\ast})$ appears naturally
(as a side remark, recall that separable ordinary Bregman distances are subsumed as 
special cases $D_{\varphi}^{OBD}(\mathbf{Q},\mathbf{Q}^{\ast\ast}) := 
D_{\varphi,\mathbf{P}}^{SBD}(\mathbf{Q},\mathbf{Q}^{\ast\ast})$  
with $\mathbf{P} := \mathbf{1} = (1,\ldots,1)$).

\vspace{0.3cm}

\begin{remark}
As will be seen in Subsection \ref{SectEstimators.new.det.nonvoid.improved.compact} 
below, if we can freely choose 
$\mathbf{Q}^{\ast\ast} \in int(\mathbf{\Omega})$
in an appropriate way, then we shall end up with \textit{improved/speed-up estimator versions} of the above-mentioned
\textit{naive extimators} of Base-Divergence-Method 1.

\end{remark}

\vspace{0.3cm}
\noindent
However, let us start with the general case of any fixed $\mathbf{Q}^{\ast\ast} \in \mathbb{R}^{K}$ 
and any fixed $\mathbf{P} \in \mathbb{R}_{>0}^{K}$ 
(with $M_{\mathbf{P}} =\sum_{i=1}^{K}p_{i}>0$)
satisfying
\begin{equation}
t_{k}^{\ast\ast} :=  \frac{q_{k}^{\ast\ast}}{p_{k}} \in \, ]t_{-}^{sc},t_{+}^{sc}[  
\quad \textrm{for all $k =1,\ldots,K$} 
\qquad \textrm{(cf. \eqref{brostu5:fo.SBD.qstarstar})}.
\nonumber
\end{equation}
In such a setup, recall from Theorem \ref{brostu5:thm.Fmin.SBD}(a) that
we have obtained for any continuous function $\Phi: \mathbf{\Omega} \mapsto \mathbb{R}$ on a compact
set $\mathbf{\Omega}\subset \mathbb{R}^{K}$ with \eqref{regularity} the assertion 
\begin{equation}
\min_{\mathbf{Q}\in \mathbf{\Omega}} \Phi(\mathbf{Q})
\ = \ - \, 
\lim_{n\rightarrow \infty }\frac{1}{n}\log \negthinspace \left( \ 
\mathbb{E}_{\mathbb{\Pi}}\negthinspace \Big[
\exp\negthinspace\Big(n \cdot \Big(
D_{\varphi,\mathbf{P}}^{SBD}\negthinspace\left(M_{\mathbf{P}} 
\cdot \boldsymbol{\xi }_{n}^{\mathbf{\widetilde{V}}},\mathbf{Q}^{\ast\ast}\right)  
- \Phi\big(M_{\mathbf{P}} \cdot \boldsymbol{\xi }_{n}^{\mathbf{\widetilde{V}}}\big)
\Big)
\Big)
\cdot \textfrak{1}_{\mathbf{\Omega}}\big(M_{\mathbf{P}} \cdot \boldsymbol{\xi }_{n}^{\mathbf{\widetilde{V}}}\big)
\, \Big] 
\right)
\, 
\label{brostu5:fo.BSmin.extended.min.SBD.new}
\end{equation}
by the following construction: in terms of the notations
$n_{k}:=\lfloor n \cdot \widetilde{p}_{k}\rfloor$ ($k \in \left\{ 1, \ldots ,K-1\right\}$),
$n_{K} := n- \sum_{k=1}^{K-1} n_{k}$
(recall \eqref{fo.freqlim}), 
$$I_{1}^{(n)}:=\left\{
1,\ldots ,n_{1}\right\}, \quad I_{2}^{(n)}:=\left\{ n_{1}+1,\ldots
,n_{1}+n_{2}\right\}, \quad \ldots, 
\quad I_{K}^{(n)} := \{ \sum_{k=1}^{K-1} n_{k} + 1, \ldots, n \}, \quad card(I_{k}^{(n)}) = n_{k},
$$
\begin{equation}
\mathbf{\widetilde{V}} := \mathbf{\widetilde{V}}_{n} := \left(\widetilde{V}_{1}, \ldots, \widetilde{V}_{n}  
\right) = \left(\widetilde{V}_{1}, \ldots, \widetilde{V}_{n_{1}}, 
\widetilde{V}_{n_{1}+1}, \ldots, \widetilde{V}_{n_{1}+n_{2}}, \ldots, 
\widetilde{V}_{\sum_{k=1}^{K-1} n_{k} + 1}, \ldots, \widetilde{V}_{n} \right) \, 
\qquad \textrm{(cf. \eqref{brostu5:V_new})},
\nonumber
\end{equation}
\begin{equation}
\boldsymbol{\xi}_{n}^{\mathbf{\widetilde{V}}}:=\Big(\frac{1}{n}\sum_{i\in
I_{1}^{(n)}}\widetilde{V}_{i},\ldots ,\frac{1}{n}\sum_{i\in I_{K}^{(n)}}
\widetilde{V}_{i}\Big)
\qquad \textrm{(cf. \eqref{Xi_n^W vector V new2})},
\nonumber
\end{equation}
we have employed ---
independently for each $k \in \{1,\ldots,K\}$ ---
$n_{k}$ i.i.d. random variables $\widetilde{V}_{i}$, $i\in I_{k}^{(n)}$, with 
common distribution $\mathbb{\Pi }[\widetilde{V}_{i}\in \cdot \,]=
\widetilde{U}_{k}[\,\cdot \,]$ ($i\in I_{k}^{(n)}$) given by 
\begin{equation}
d\widetilde{U}_{k}(v) \ := \ 
\frac{\exp \left(\tau_{k} \cdot v\right)}{MGF_{\widetilde{\mathbb{\bbzeta}}}(\tau_{k})} \, 
d\widetilde{\mathbb{\bbzeta}}(v),
\qquad \textrm{(cf. \eqref{brostu5:Utilde_k_new})}
\nonumber
\end{equation}
with $\tau_{k} := \ M_{\mathbf{P}} \cdot \varphi^{\, \prime} \negthinspace
\left(\frac{\widetilde{q}_{k}^{\ast\ast}}{\widetilde{p}_{k}}\right)
= \ M_{\mathbf{P}} \cdot \varphi^{\, \prime} \negthinspace
\left(\frac{q_{k}^{\ast\ast}}{p_{k}}\right)$.
Recall (cf. the lines right below \eqref{Xi_n^W vector V new2}) that\\
$ \mathbb{E}_{\mathbb{\Pi}}\negthinspace \Big[\boldsymbol{\xi}_{n}^{\mathbf{\widetilde{V}}}\Big]
= \left(
\frac{\lfloor n \cdot \widetilde{p}_{1}\rfloor}{n \cdot \widetilde{p}_{1}} \cdot \widetilde{q}_{1}^{\ast\ast},
\ldots,
\frac{\lfloor n \cdot \widetilde{p}_{K-1}\rfloor}{n \cdot \widetilde{p}_{K-1}} \cdot \widetilde{q}_{K-1}^{\ast\ast},
\frac{n - \sum_{i=1}^{K-1} \lfloor n \cdot \widetilde{p}_{i}\rfloor}{n \cdot \widetilde{p}_{K}} \cdot \widetilde{q}_{K}^{\ast\ast}
\right)
$
and thus
$\lim_{n\rightarrow \infty} \mathbb{E}_{\mathbb{\Pi}}\negthinspace \Big[\boldsymbol{\xi}_{n}^{\mathbf{\widetilde{V}}}
\Big]= 
\widetilde{\mathbf{Q}}^{\ast\ast}$
or equivalently \\ 
$\lim_{n\rightarrow \infty}
\mathbb{E}_{\mathbb{\Pi}}\negthinspace \Big[M_{\mathbf{P}} \cdot \boldsymbol{\xi}_{n}^{\mathbf{\widetilde{V}}}
\Big]= \mathbf{Q}^{\ast\ast}$.
Within such a set-up, we thus obtain for large $n \in \mathbb{N}$ 
(cf. \eqref{brostu5:fo.BSmin.extended.min.SBD.new}) 
the approximation
\begin{equation}
\Phi(\mathbf{\Omega}) := \min_{\mathbf{Q}\in \mathbf{\Omega}} \Phi(\mathbf{Q})
\ \approx \ - \, 
\frac{1}{n}\log \negthinspace \left( \ 
\mathbb{E}_{\mathbb{\Pi}}\negthinspace \Big[
\exp\negthinspace\Big(n \cdot \Big(
D_{\varphi,\mathbf{P}}^{SBD}\negthinspace\left(M_{\mathbf{P}} 
\cdot \boldsymbol{\xi }_{n}^{\mathbf{\widetilde{V}}},\mathbf{Q}^{\ast\ast}\right)  
- \Phi\big(M_{\mathbf{P}} \cdot \boldsymbol{\xi }_{n}^{\mathbf{\widetilde{V}}}\big)
\Big)
\Big)
\cdot \textfrak{1}_{\mathbf{\Omega}}\big(M_{\mathbf{P}} \cdot \boldsymbol{\xi }_{n}^{\mathbf{\widetilde{V}}}\big)
\, \Big] 
\right)
\, 
\label{brostu5:fo.BSmin.extended.min.SBD.approx.new}
\end{equation}
and hence for getting an estimator of the minimum value $\Phi(\mathbf{\Omega})$
one can estimate the right-hand side of \eqref{brostu5:fo.BSmin.extended.min.SBD.approx.new}.
To achieve this, for the rest of this section we assume 
that $n$ is chosen such that all 
$n \cdot \frac{p_{k}}{M_{\mathbf{P}}}$ are integers
(and hence, $n = \sum_{k=1}^{K} n_{k}$ with $n_{k} = n \cdot \frac{p_{k}}{M_{\mathbf{P}}}$) --- the remaining case works analogously. With this, we construct the corresponding \textit{naive estimator} 
for the minimum value $\min_{\mathbf{Q}\in \mathbf{\Omega}} \Phi(\mathbf{Q})$ as 
\begin{equation}
\widehat{\Phi(\mathbf{\Omega})}_{n,L}^{naive,2}\ := \ 
- \frac{1}{n}\log \frac{1}{L}\sum_{\ell =1}^{L} 
\exp\negthinspace\Big(n \cdot \Big(
D_{\varphi,\mathbf{P}}^{SBD}\negthinspace\left(M_{\mathbf{P}} 
\cdot \boldsymbol{\xi }_{n}^{\mathbf{\widetilde{V}}^{(\ell)}},\mathbf{Q}^{\ast\ast}\right) 
- \Phi\big(M_{\mathbf{P}} \cdot \boldsymbol{\xi }_{n}^{\mathbf{\widetilde{V}}^{(\ell)}}\big)
\Big)
\Big)
\cdot \textfrak{1}_{\mathbf{\Omega}}\big(M_{\mathbf{P}} \cdot \boldsymbol{\xi}_{n}^{\mathbf{\widetilde{V}}^{(\ell)}}\big)
\, ,
\label{brostu5:fo.BSmin.extended.improved.estim.new}
\end{equation}
where we simulate independently $L$ copies 
$\mathbf{\widetilde{V}}^{(1)},\ldots,\mathbf{\widetilde{V}}^{(L)}$ of the random vector
$\mathbf{\widetilde{V}}:=\left( \widetilde{V}_{1},\ldots,\widetilde{V}_{n}\right)$ 
constructed by the above-mentioned method, 
and compute each of $\boldsymbol{\xi}_{n}^{\mathbf{\widetilde{V}}^{(1)}}, \ldots,
\boldsymbol{\xi}_{n}^{\mathbf{\widetilde{V}}^{(L)}}$ according to 
\eqref{Xi_n^W vector V new2}. Clearly, with the help of the strong law of large numbers we get with 
$\widehat{\Phi(\mathbf{\Omega})}_{n,\infty}^{naive,2}: = 
\lim_{L\rightarrow \infty} \widehat{\Phi(\mathbf{\Omega})}_{n,L}^{naive,2}$
the following assertion:

\vspace{0.2cm}

\begin{proposition}
Let the assumptions of Theorem \ref{brostu5:thm.Fmin.SBD}(a) be satisfied. Then one has
\begin{equation}
\lim_{n\rightarrow \infty} 
\widehat{\Phi(\mathbf{\Omega})}_{n,\infty}^{naive,2}
\ = \ 
\lim_{n\rightarrow \infty} \lim_{L\rightarrow \infty} 
\widehat{\Phi(\mathbf{\Omega})}_{n,L}^{naive,2}
\ = \ \Phi(\mathbf{\Omega}) \qquad \textrm{a.s.}
\label{brostu5:fo.BSmin.extended.improved.estim.lim.new}
\end{equation}
\end{proposition}

\vspace{0.2cm}
\noindent
As the corresponding \textit{very natural naive BS-estimator of the minimizer-set} 
$\mathcal{Q}^{\ast} := \argmin_{\mathbf{Q} \in \mathbf{\Omega }} \Phi(\mathbf{Q})$, 
we take 
\begin{equation}
\widehat{\argmin_{\mathbf{Q} \in \mathbf{\Omega }} 
\Phi(\mathbf{Q})}_{n,L}^{naive,2}
:= \argmin_{\boldsymbol{\nu} \in \mathcal{V}_{n,L}} \Phi(\boldsymbol{\nu})
\label{brostu5:fo.1555.improved.new}
\end{equation}
where $\mathcal{V}_{n,L} := \{M_{\mathbf{P}} \cdot \boldsymbol{\xi}_{n}^{\mathbf{\widetilde{V}}^{(\ell)}}:
\ell \in \{1,\ldots,L\} \,  \} \cap \mathbf{\Omega}$. 
In short, we take as minimizer-estimate
the (not necessarily unique) element 
$M_{\mathbf{P}} \cdot \boldsymbol{\xi}_{n}^{\mathbf{\widetilde{V}}^{L,\ast}}$
which minimizes $\Phi(\cdot)$ amongst all values 
$M_{\mathbf{P}} \cdot \boldsymbol{\xi}_{n}^{\mathbf{\widetilde{V}}^{(\ell)}}$
at hand which fall into $\mathbf{\Omega}$. 
We prove that if $L$ and $n$ tend to infinity, then 
$M_{\mathbf{P}} \cdot \boldsymbol{\xi}_{n}^{\mathbf{\widetilde{V}}^{L,\ast}}$
concentrates to the above-mentioned set $\mathcal{Q}^{\ast}$ 
of minimizers of $\Phi(\cdot)$ on $\mathbf{\Omega}$.
Indeed, we show that $M_{\mathbf{P}} \cdot \boldsymbol{\xi}_{n}^{\mathbf{\widetilde{V}}^{L,\ast}}$ is
a proxy minimizer of $\Phi(\cdot)$ on $\mathbf{\Omega}$, by proving the following

\vspace{0.2cm}

\begin{proposition}
\label{brostu5:prop.generaldeterministic.minimizer.improved.new}
There holds
\begin{equation}
\min_{\mathbf{Q}\in \mathbf{\Omega}} \Phi(\mathbf{Q}) 
\ \leq \ 
\Phi \left( M_{\mathbf{P}} \cdot \boldsymbol{\xi}_{n}^{\mathbf{\widetilde{V}}^{L,\ast}} \right)
\ \leq \ \widehat{\Phi(\mathbf{\Omega})}_{n,\infty}^{naive,2}
\ + \ o_{\mathbb{\Pi}}(1)
\label{minim.improved.new}
\end{equation}
where $o_{\mathbb{\Pi}}(1)$ goes to $0$ as 
$L\rightarrow \infty$ and $n\rightarrow \infty$ under the
distribution $\mathbb{\Pi}$
(recall that 
$\mathbb{\Pi}[(\widetilde{V}_{1},\ldots,\widetilde{V}_{n}) \in \cdot \,]=
\bigotimes\limits_{k=1}^{K}\widetilde{U}_{k}^{\otimes n_{k}}[\,\cdot \,]$)
\end{proposition}

\vspace{0.4cm}
\noindent
The proof of Proposition \ref{brostu5:prop.generaldeterministic.minimizer.improved.new}
works completely analogous to that of Proposition \ref{brostu5:prop.generaldeterministic.minimizer.naive}.

\vspace{0.3cm}

\begin{remark}
In the current set-up of compact $\mathbf{\Omega}$ with \eqref{regularity}, 
by taking the special case $\Phi(\mathbf{Q}) := 
D_{\varphi,\mathbf{P}}^{SBD}\negthinspace\left(\mathbf{Q},\mathbf{Q}^{\ast\ast}\right)$
we obtain the naive BS-estimator 
$\widehat{D_{\varphi,\mathbf{P}}^{SBD}\negthinspace\left(\mathbf{\Omega},\mathbf{Q}^{\ast\ast}\right)}_{n,L}^{naive,2}$ 
of the minimum value $\min_{\mathbf{Q}\in \mathbf{\Omega}} 
D_{\varphi,\mathbf{P}}^{SBD}\negthinspace\left(\mathbf{Q},\mathbf{Q}^{\ast\ast}\right)$, as
well as the corresponding naive BS-estimator 
$\widehat{\argmin_{\mathbf{Q} \in \mathbf{\Omega }} 
D_{\varphi,\mathbf{P}}^{SBD}\negthinspace\left(\mathbf{Q},\mathbf{Q}^{\ast\ast}\right)
}_{n,L}^{naive,2}$
of the minimizer(s);
recall that the fixed pregiven $\mathbf{Q}^{\ast\ast}$ NEED NOT be in $\mathbf{\Omega}$
(e.g. in some application-specific contexts).
For the special case $\mathbf{P} :=  \mathbf{1} = (1,\ldots, 1)$, this leads
to the corresponding estimators for the minimum and the minimizer(s)
of the separable ordinary Bregman distance
$\mathbf{Q} \mapsto D_{\varphi}^{OBD}\negthinspace\left(\mathbf{Q},\mathbf{Q}^{\ast\ast}\right)
= D_{\varphi,\mathbf{1}}^{SBD}\negthinspace\left(\mathbf{Q},\mathbf{Q}^{\ast\ast}\right)$.

\end{remark}

\vspace{0.3cm}

\begin{proposition}
In the above set-up, one has
\begin{equation}
\lim_{n\rightarrow \infty} \lim_{L\rightarrow \infty} 
\Phi\Big( \widehat{\argmin_{\mathbf{Q} \in \mathbf{\Omega}} \Phi(\mathbf{Q})}_{n,L}^{naive,2} \Big)
\ = \ \Phi(\mathbf{\Omega}) \qquad \textrm{a.s.},
\nonumber
\end{equation}
and thus the quantity 
$\Phi\Big( \widehat{\argmin_{\mathbf{Q} \in \mathbf{\Omega}} \Phi(\mathbf{Q})}_{n,L}^{naive,2} \Big)$
is a natural alternative to the estimate
$\widehat{\Phi(\mathbf{\Omega})}_{n,L}^{naive,2}$ given in 
\eqref{brostu5:fo.BSmin.extended.improved.estim.new}.

\end{proposition}


\subsection{Naive estimators of max and argmax --- Base-Divergence-Method 2, compact case}
\label{SectEstimators.new.det.nonvoid.meth2.compact.max}

\vspace{0.2cm}
\noindent
\vspace{0.2cm}
\noindent
In the set-up of compact $\mathbf{\Omega}$ with \eqref{regularity}, we can treat
the maximizing problem completely analogously to the method in 
the previous Subsection \ref{SectEstimators.new.det.nonvoid.meth2.compact.min}.
Indeed, by employing Theorem \ref{brostu5:thm.Fmax.SBD}(a) instead of Theorem \ref{brostu5:thm.Fmin.SBD}(a)
we construct the naive BS-estimator 
\begin{equation}
\widehat{\Phi(\mathbf{\Omega})}_{n,L}^{naive,2}\ := \ 
\frac{1}{n}\log \frac{1}{L}\sum_{\ell =1}^{L} 
\exp\negthinspace\Big(n \cdot \Big(
D_{\varphi,\mathbf{P}}^{SBD}\negthinspace\left(M_{\mathbf{P}} 
\cdot \boldsymbol{\xi }_{n}^{\mathbf{\widetilde{V}}^{(\ell)}},\mathbf{Q}^{\ast\ast}\right) 
+ \Phi\big(M_{\mathbf{P}} \cdot \boldsymbol{\xi }_{n}^{\mathbf{\widetilde{V}}^{(\ell)}}\big)
\Big)
\Big)
\cdot \textfrak{1}_{\mathbf{\Omega}}\big(M_{\mathbf{P}} \cdot \boldsymbol{\xi}_{n}^{\mathbf{\widetilde{V}}^{(\ell)}}\big)
\, ,
\label{brostu5:fo.BSmax.extended.improved.estim.new}
\end{equation}
of the maximum value $\Phi(\mathbf{\Omega}) := \max_{\mathbf{Q}\in \mathbf{\Omega}} \Phi(\mathbf{Q})$;
for this, we get
\begin{equation}
\lim_{n\rightarrow \infty} \widehat{\Phi(\mathbf{\Omega})}_{n,\infty}^{naive,2}
\ = \ 
\lim_{n\rightarrow \infty} \lim_{L\rightarrow \infty} 
\widehat{\Phi(\mathbf{\Omega})}_{n,L}^{naive,2}
\ = \ \Phi(\mathbf{\Omega}) \qquad \textrm{a.s.}
\nonumber
\end{equation}
instead of 
\eqref{brostu5:fo.BSmin.extended.improved.estim.lim.new}
(notice the different definition of the involved quantities).
As the corresponding very natural naive BS-estimator of the 
maximizer $\argmax_{\mathbf{Q} \in \mathbf{\Omega }} \Phi(\mathbf{Q})$
we take 
\begin{equation}
\widehat{\argmax_{\mathbf{Q} \in \mathbf{\Omega }} \Phi(\mathbf{Q})}_{n,L}^{naive,2}
:= \argmax_{\boldsymbol{\nu} \in \mathcal{V}_{n,L}} \Phi(\boldsymbol{\nu})
\nonumber
\end{equation}
instead of 
\eqref{brostu5:fo.1555.improved.new}; in short, 
we take any $M_{\mathbf{P}} \cdot \boldsymbol{\xi}_{n}^{\mathbf{\widetilde{V}}^{L,\ast}}$
which maximizes $\Phi(\cdot)$ amongst all values 
$M_{\mathbf{P}} \cdot \boldsymbol{\xi}_{n}^{\mathbf{\widetilde{V}}^{(\ell)}}$
at hand which fall into $\mathbf{\Omega}$. For this, we obtain --- instead of 
\eqref{minim.improved.new} ---
the assertion
\begin{equation}
\max_{\mathbf{Q}\in \mathbf{\Omega}} \Phi(\mathbf{Q}) 
\ \geq \ 
\Phi \left( M_{\mathbf{P}} \cdot \boldsymbol{\xi}_{n}^{\mathbf{\widetilde{V}}^{L,\ast}} \right)
\ \geq \ 
\widehat{\Phi(\mathbf{\Omega})}_{n,\infty}^{naive,2}
\ - \ o_{\mathbb{\Pi}}(1)
\nonumber
\end{equation}
where $o_{\mathbb{\Pi}}(1)$ goes to $0$ as $L\rightarrow \infty$ and $n\rightarrow \infty$ under the
distribution $\mathbb{\Pi}$.
Hence, as $L$ and $n$ tend to infinity,  
$M_{\mathbf{P}} \cdot \boldsymbol{\xi}_{n}^{\mathbf{\widetilde{V}}^{L,\ast}}$
concentrates to the set of maximizers of $\Phi(\cdot)$ on $\mathbf{\Omega}$.

\vspace{0.3cm}

\begin{proposition}
In the above set-up, one has
\begin{equation}
\lim_{n\rightarrow \infty} \lim_{L\rightarrow \infty} 
\Phi\Big( \widehat{\argmax_{\mathbf{Q} \in \mathbf{\Omega}} \Phi(\mathbf{Q})}_{n,L}^{naive,2} \Big)
\ = \ \Phi(\mathbf{\Omega}) \qquad \textrm{a.s.},
\nonumber
\end{equation}
and thus the quantity 
$\Phi\Big( \widehat{\argmax_{\mathbf{Q} \in \mathbf{\Omega}} \Phi(\mathbf{Q})}_{n,L}^{naive,2} \Big)$
is a natural alternative to the estimate
$\widehat{\Phi(\mathbf{\Omega})}_{n,L}^{naive,2}$ given in 
\eqref{brostu5:fo.BSmax.extended.improved.estim.new}.

\end{proposition}


\subsection{Naive estimators --- Base-Divergence-Method 2, non-compact case}

\vspace{0.2cm}
\noindent
We can proceed analogously to the Base-Divergence-Method-1-treating 
Subsection \ref{SectEstimators.new.det.nonvoid.meth1.noncompact}
(e.g. by employing Theorem \ref{brostu5:thm.Fmin.SBD}(b)).


\subsection{Improved/Speed-up estimators 
of min and argmin --- compact case}
\label{SectEstimators.new.det.nonvoid.improved.compact}

\vspace{0.2cm}
\noindent
\textit{G.1) \, The underlying method}

\vspace{0.2cm}
\noindent
For the sake of brevity, we deal here only with improving/speeding-up
the naive estimators given in Subsection \ref{SectEstimators.new.det.nonvoid.meth1.compact.min}
which concerns about estimating the min and argmin of $\Phi(\cdot)$ on a compact constraint set $\mathbf{\Omega}$
with \eqref{regularity}, 
via Base-Divergence-Method 1, by choosing the
$\varphi-$divergence $D_{\varphi}(\cdot,\mathbf{P})$ as the so-called \textit{base divergence}.
To start with, recall that
\begin{equation}
\min_{\mathbf{Q}\in \mathbf{\Omega}} \Phi(\mathbf{Q})
\ = \ 
- \, 
\lim_{n\rightarrow \infty }\frac{1}{n}\log \negthinspace \left( \ 
\mathbb{E}_{\mathbb{\Pi}}\negthinspace \Big[
\exp\negthinspace\Big(n \cdot \Big(
D_{\varphi }\big(M_{\mathbf{P}} \cdot \boldsymbol{\xi }_{n}^{\mathbf{\widetilde{W}}},\mathbf{P}\big) 
- \Phi\big(M_{\mathbf{P}} \cdot \boldsymbol{\xi }_{n}^{\mathbf{\widetilde{W}}}\big)
\Big)
\Big)
\cdot \textfrak{1}_{\mathbf{\Omega}}\big(M_{\mathbf{P}} \cdot \boldsymbol{\xi }_{n}^{\mathbf{\widetilde{W}}}\big)
\, \Big]
\right) \, 
\qquad \textrm{(cf. \eqref{brostu5:fo.BSmin.extended.min})},
\nonumber
\end{equation}
which for large $n \in \mathbb{N}$ leads to the approximation 
\begin{equation}
\Phi(\mathbf{\Omega}) := \min_{\mathbf{Q}\in \mathbf{\Omega}} \Phi(\mathbf{Q})
\ \approx \ - \, 
\frac{1}{n}\log \negthinspace \left( \ 
\mathbb{E}_{\mathbb{\Pi}}\negthinspace \Big[
\exp\negthinspace\Big(n \cdot \Big(
D_{\varphi }\big(M_{\mathbf{P}} \cdot \boldsymbol{\xi }_{n}^{\mathbf{\widetilde{W}}},\mathbf{P}\big) 
- \Phi\big(M_{\mathbf{P}} \cdot \boldsymbol{\xi }_{n}^{\mathbf{\widetilde{W}}}\big)
\Big)
\Big)
\cdot \textfrak{1}_{\mathbf{\Omega}}\big(M_{\mathbf{P}} \cdot \boldsymbol{\xi }_{n}^{\mathbf{\widetilde{W}}}\big)
\, \Big] 
\right)
\,  
\qquad 
\textrm{(cf. \eqref{brostu5:fo.BSmin.extended.approx})}.
\nonumber
\end{equation}
Notice that by construction of $\boldsymbol{\xi }_{n}^{\mathbf{\widetilde{W}}}$ (cf. \eqref{Xi_n^W vector}) one has
$$ \mathbb{E}_{\mathbb{\Pi}}\negthinspace \Big[M_{\mathbf{P}} \cdot \boldsymbol{\xi}_{n}^{\mathbf{\widetilde{W}}}\Big]
= \left(M_{\mathbf{P}} \cdot \frac{\lfloor n \cdot \widetilde{p}_{1}\rfloor}{n},
\ldots,
M_{\mathbf{P}} \cdot \frac{\lfloor n \cdot \widetilde{p}_{K-1}\rfloor}{n},
M_{\mathbf{P}} \cdot \frac{n- \sum_{i=1}^{K-1} \lfloor n \cdot \widetilde{p}_{i}\rfloor}{n}
\right)
$$ 
and $\lim_{n\rightarrow \infty} \mathbb{E}_{\mathbb{\Pi}}\negthinspace 
\Big[M_{\mathbf{P}} \cdot\boldsymbol{\xi}_{n}^{\mathbf{\widetilde{W}}}
\Big]= \mathbf{P}$. Since typically $\mathbf{P} \notin \mathbf{\Omega}$,
the method of Subsection \ref{SectEstimators.new.det.nonvoid.meth1.compact.min}
can be time costly (relative to the employed computer-power), since 
the \textit{runtime} which is required in order to obtain some accuracy while estimating
--- via $\widehat{\Phi(\mathbf{\Omega})}_{n,L}^{naive,1}$ given in \eqref{brostu5:fo.BSmin.extended.naive.estim} ---
the expectation (cf. \eqref{brostu5:fo.BSmin.extended.approx})
\begin{equation}
\mathbb{E}_{\mathbb{\Pi}}\negthinspace \Big[
\exp\negthinspace\Big(n \cdot \Big(
D_{\varphi }\big(M_{\mathbf{P}} \cdot \boldsymbol{\xi }_{n}^{\mathbf{\widetilde{W}}},\mathbf{P}\big) 
- \Phi\big(M_{\mathbf{P}} \cdot \boldsymbol{\xi }_{n}^{\mathbf{\widetilde{W}}}\big)
\Big)
\Big)
\cdot \textfrak{1}_{\mathbf{\Omega}}\big(M_{\mathbf{P}} \cdot \boldsymbol{\xi }_{n}^{\mathbf{\widetilde{W}}}\big)
\, \Big]
\nonumber
\end{equation}
depends on the hit rate of the set $\mathbf{\Omega}$ by the
replicates/copies 
$M_{\mathbf{P}} \cdot \boldsymbol{\xi}_{n}^{\mathbf{\widetilde{W}}^{(\ell)}}$
of the random vector $M_{\mathbf{P}} \cdot \boldsymbol{\xi}_{n}^{\mathbf{\widetilde{W}}}$.
We thus may substitute 
$M_{\mathbf{P}} \cdot \boldsymbol{\xi}_{n}^{\mathbf{\widetilde{W}}^{(\ell)}}$ 
by the replicates/copies $M_{\mathbf{P}} \cdot \boldsymbol{\xi}_{n}^{\mathbf{\widetilde{V}}^{(\ell)}}$
of some new random vector 
$M_{\mathbf{P}} \cdot \boldsymbol{\xi}_{n}^{\mathbf{\widetilde{V}}}$
whose visits in $\mathbf{\Omega}$ will be much more frequent. 
For instance, we may have the goal that the new replication mean
$\mathbb{E}_{\mathbb{\Pi}}\negthinspace \Big[M_{\mathbf{P}} 
\cdot \boldsymbol{\xi}_{n}^{\mathbf{\widetilde{V}}^{(\ell)}} \Big]$ 
should be equal to some point $\mathbf{Q}^{\ast}$ which lies in the interior $int(\mathbf{\Omega})$
of $\mathbf{\Omega}$; for pregiven $\mathbf{P} \in \mathbb{R}_{>0}^{K}$, such a vector
$\mathbf{Q}^{\ast}$ 
may be either pregiven (e.g. by the nature of the application context)
or it may be simulatively achieved by e.g. proxy method $1$ or proxy method $2$
of Subsection X-A of Broniatowski \& Stummer \cite{Bro:23a}.

\vspace{0.2cm}
\noindent
One \textit{comfortable} way to achieve from 
$\mathbf{Q}^{\ast} \in int(\mathbf{\Omega})$ the above-mentioned desired construction 
$M_{\mathbf{P}} \cdot\boldsymbol{\xi}_{n}^{\mathbf{\widetilde{V}}^{(\ell)}}$,
is to employ 
\eqref{brostu5:V_new}, \eqref{brostu5:Utilde_k_new} and \eqref{Xi_n^W vector V new2}
with the special choice $\mathbf{Q}^{\ast\ast}:=\mathbf{Q}^{\ast}$
(under the assumption \eqref{brostu5:fo.SBD.qstarstar}),
and accordingly obtain from Theorem \ref{brostu5:thm.Fmin.SBD}(a) 
the limit assertion 
\begin{equation}
\min_{\mathbf{Q}\in \mathbf{\Omega}} \Phi(\mathbf{Q})
\ = \ - \, 
\lim_{n\rightarrow \infty }\frac{1}{n}\log \negthinspace \left( \ 
\mathbb{E}_{\mathbb{\Pi}}\negthinspace \Big[
\exp\negthinspace\Big(n \cdot \Big(
D_{\varphi,\mathbf{P}}^{SBD}\negthinspace\left(M_{\mathbf{P}} 
\cdot \boldsymbol{\xi }_{n}^{\mathbf{\widetilde{V}}},\mathbf{Q}^{\ast}\right)  
- \Phi\big(M_{\mathbf{P}} \cdot \boldsymbol{\xi }_{n}^{\mathbf{\widetilde{V}}}\big)
\Big)
\Big)
\cdot \textfrak{1}_{\mathbf{\Omega}}\big(M_{\mathbf{P}} \cdot \boldsymbol{\xi }_{n}^{\mathbf{\widetilde{V}}}\big)
\, \Big] 
\right) .
\, 
\label{brostu5:fo.BSmin.extended.min.SBD.onestar}
\end{equation}
Notice that
(cf. the lines right below \eqref{Xi_n^W vector V new2} with $\mathbf{Q}^{\ast\ast}:=\mathbf{Q}^{\ast}$)
\begin{eqnarray}
\mathbb{E}_{\mathbb{\Pi}}\negthinspace \Big[M_{\mathbf{P}} \cdot \boldsymbol{\xi}_{n}^{\mathbf{\widetilde{V}}}\Big]
& = & \left(M_{\mathbf{P}} \cdot
\frac{\lfloor n \cdot \widetilde{p}_{1}\rfloor}{n \cdot \widetilde{p}_{1}} \cdot \widetilde{q}_{1}^{\ast},
\ldots,
M_{\mathbf{P}} \cdot
\frac{\lfloor n \cdot \widetilde{p}_{K-1}\rfloor}{n \cdot \widetilde{p}_{K-1}} \cdot \widetilde{q}_{K-1}^{\ast},
M_{\mathbf{P}} \cdot \frac{n - \sum_{i=1}^{K-1} 
\lfloor n \cdot \widetilde{p}_{i}\rfloor}{n \cdot \widetilde{p}_{K}} \cdot \widetilde{q}_{K}^{\ast}
\right)
\nonumber\\
& = & \left(
\frac{\lfloor n \cdot \widetilde{p}_{1}\rfloor}{n \cdot \widetilde{p}_{1}} \cdot q_{1}^{\ast},
\ldots,
\frac{\lfloor n \cdot \widetilde{p}_{K-1}\rfloor}{n \cdot \widetilde{p}_{K-1}} \cdot q_{K-1}^{\ast},
\frac{n - \sum_{i=1}^{K-1} 
\lfloor n \cdot \widetilde{p}_{i}\rfloor}{n \cdot \widetilde{p}_{K}} \cdot q_{K}^{\ast}
\right)
\nonumber
\end{eqnarray} 
which lies in 
$int(\mathbf{\Omega})$ for large enough $n$;
moreover, $\lim_{n\rightarrow \infty} M_{\mathbf{P}} \cdot\boldsymbol{\xi}_{n}^{\mathbf{\widetilde{V}}}
= \mathbf{Q}^{\ast}$ a.s. and thus, for the hit rate we get\\
$\lim_{n\rightarrow \infty} \textfrak{1}_{\mathbf{\Omega}}\big(M_{\mathbf{P}} \cdot \boldsymbol{\xi }_{n}^{\mathbf{\widetilde{V}}}\big) =1$ a.s.
Within such a set-up, we thus obtain for large $n \in \mathbb{N}$ the approximation
\begin{equation}
\Phi(\mathbf{\Omega}) := \min_{\mathbf{Q}\in \mathbf{\Omega}} \Phi(\mathbf{Q})
\ \approx \ - \, 
\frac{1}{n}\log \negthinspace \left( \ 
\mathbb{E}_{\mathbb{\Pi}}\negthinspace \Big[
\exp\negthinspace\Big(n \cdot \Big(
D_{\varphi,\mathbf{P}}^{SBD}\negthinspace\left(M_{\mathbf{P}} 
\cdot \boldsymbol{\xi }_{n}^{\mathbf{\widetilde{V}}},\mathbf{Q}^{\ast}\right)  
- \Phi\big(M_{\mathbf{P}} \cdot \boldsymbol{\xi }_{n}^{\mathbf{\widetilde{V}}}\big)
\Big)
\Big)
\cdot \textfrak{1}_{\mathbf{\Omega}}\big(M_{\mathbf{P}} \cdot \boldsymbol{\xi }_{n}^{\mathbf{\widetilde{V}}}\big)
\, \Big] 
\right)
\, 
\label{brostu5:fo.BSmin.extended.min.SBD.approx.onestar}
\end{equation}
and hence for getting an estimator of the minimum value $\Phi(\mathbf{\Omega})$
one can estimate the involved expectation
\begin{equation}
\mathbb{E}_{\mathbb{\Pi}}\negthinspace \Big[
\exp\negthinspace\Big(n \cdot \Big(
D_{\varphi,\mathbf{P}}^{SBD}\negthinspace\left(M_{\mathbf{P}} 
\cdot \boldsymbol{\xi }_{n}^{\mathbf{\widetilde{V}}},\mathbf{Q}^{\ast}\right)  
- \Phi\big(M_{\mathbf{P}} \cdot \boldsymbol{\xi }_{n}^{\mathbf{\widetilde{V}}}\big)
\Big)
\Big)
\cdot \textfrak{1}_{\mathbf{\Omega}}\big(M_{\mathbf{P}} \cdot \boldsymbol{\xi }_{n}^{\mathbf{\widetilde{V}}}\big)
\, \Big] 
\, 
\nonumber
\end{equation}
with higher hit rate than above. Indeed, we proceed analogously to
Subsection \ref{SectEstimators.new.det.nonvoid.meth2.compact.min}
(with $\mathbf{Q}^{\ast\ast} := \mathbf{Q}^{\ast}$),
assume that $n$ is chosen such that all 
$n \cdot \frac{p_{k}}{M_{\mathbf{P}}}$ are integers
(and hence, $n = \sum_{k=1}^{K} n_{k}$ with $n_{k} = n \cdot \frac{p_{k}}{M_{\mathbf{P}}}$; 
the remaining case works analogously)
and construct the corresponding \textit{speed-up BS-estimator}
for the minimum value $\min_{\mathbf{Q}\in \mathbf{\Omega}} \Phi(\mathbf{Q})$ as 
\begin{equation}
\widehat{\Phi(\mathbf{\Omega})}_{n,L}^{speedup,1} \ := \ 
- \frac{1}{n}\log \frac{1}{L}\sum_{\ell =1}^{L} 
\exp\negthinspace\Big(n \cdot \Big(
D_{\varphi,\mathbf{P}}^{SBD}\negthinspace\left(M_{\mathbf{P}} 
\cdot \boldsymbol{\xi }_{n}^{\mathbf{\widetilde{V}}^{(\ell)}},\mathbf{Q}^{\ast}\right) 
- \Phi\big(M_{\mathbf{P}} \cdot \boldsymbol{\xi }_{n}^{\mathbf{\widetilde{V}}^{(\ell)}}\big)
\Big)
\Big)
\cdot \textfrak{1}_{\mathbf{\Omega}}\big(M_{\mathbf{P}} \cdot \boldsymbol{\xi}_{n}^{\mathbf{\widetilde{V}}^{(\ell)}}\big)
\, ;
\label{brostu5:fo.BSmin.extended.improved.estim}
\end{equation}
notice the difference to the 
\textit{naive estimator} $\widehat{\Phi(\mathbf{\Omega})}_{n,L}^{naive,2}$
for the minimum value $\min_{\mathbf{Q}\in \mathbf{\Omega}} \Phi(\mathbf{Q})$
given in \eqref{brostu5:fo.BSmin.extended.improved.estim.new}:
due to the different nature of $\mathbf{Q}^{\ast}$ and $\mathbf{Q}^{\ast\ast}$,
the construction of $M_{\mathbf{P}} \cdot \boldsymbol{\xi}_{n}^{\mathbf{\widetilde{V}}^{(\ell)}}$ in 
\eqref{brostu5:fo.BSmin.extended.improved.estim} typically hits
$\mathbf{\Omega}$ \textit{more often}
than the construction of $M_{\mathbf{P}} \cdot \boldsymbol{\xi}_{n}^{\mathbf{\widetilde{V}}^{(\ell)}}$
in \eqref{brostu5:fo.BSmin.extended.improved.estim.new} does.
From the corresponding special case of Subsection \ref{SectEstimators.new.det.nonvoid.meth2.compact.min}
we obtain with 
$\widehat{\Phi(\mathbf{\Omega})}_{n,\infty}^{speedup,1}: = 
\lim_{L\rightarrow \infty} \widehat{\Phi(\mathbf{\Omega})}_{n,L}^{speedup,1}$
the following assertion:

\vspace{0.2cm}

\begin{proposition}
\label{brostu5:prop.generaldeterministic.minimum.improved}
Let the assumptions of Theorem \ref{brostu5:thm.Fmin.SBD}(a) be satisfied (with 
$\mathbf{Q}^{\ast}$ instead of $\mathbf{Q}^{\ast\ast}$). Then one has
\begin{equation}
\lim_{n\rightarrow \infty} \widehat{\Phi(\mathbf{\Omega})}_{n,\infty}^{speedup,1}
\ = \ 
\lim_{n\rightarrow \infty} \lim_{L\rightarrow \infty} 
\widehat{\Phi(\mathbf{\Omega})}_{n,L}^{speedup,1}
\ = \ \Phi(\mathbf{\Omega}) \qquad \textrm{a.s.}
\nonumber
\end{equation}
\end{proposition}

\vspace{0.2cm}
\noindent
As the corresponding \textit{very natural speed-up BS-estimator 
of the minimizer-set} 
$\mathcal{Q}^{\ast} := \argmin_{\mathbf{Q} \in \mathbf{\Omega }} \Phi(\mathbf{Q})$, 
we take 
\begin{equation}
\widehat{\argmin_{\mathbf{Q} \in \mathbf{\Omega }} \Phi(\mathbf{Q})}_{n,L}^{speedup,1}
:= \argmin_{\boldsymbol{\nu} \in \mathcal{V}_{n,L}} \Phi(\boldsymbol{\nu})
\label{brostu5:fo.1555.improved}
\end{equation}
where $\mathcal{V}_{n,L} := \{M_{\mathbf{P}} \cdot \boldsymbol{\xi}_{n}^{\mathbf{\widetilde{V}}^{(\ell)}}:
\ell \in \{1,\ldots,L\} \,  \} \cap \mathbf{\Omega}$. 
In short, we take as minimizer-estimate
the (not necessarily unique) element 
$M_{\mathbf{P}} \cdot \boldsymbol{\xi}_{n}^{\mathbf{\widetilde{V}}^{L,\ast}}$
which minimizes $\Phi(\cdot)$ amongst all values 
$M_{\mathbf{P}} \cdot \boldsymbol{\xi}_{n}^{\mathbf{\widetilde{V}}^{(\ell)}}$
at hand which fall into $\mathbf{\Omega}$. 
From the corresponding special case of Subsection \ref{SectEstimators.new.det.nonvoid.meth2.compact.min}
we derive that --- as $L$ and $n$ tend to infinity --- 
$M_{\mathbf{P}} \cdot \boldsymbol{\xi}_{n}^{\mathbf{\widetilde{V}}^{L,\ast}}$
concentrates to the above-mentioned set $\mathcal{Q}^{\ast}$ 
of minimizers of $\Phi(\cdot)$ on $\mathbf{\Omega}$;
indeed, $M_{\mathbf{P}} \cdot \boldsymbol{\xi}_{n}^{\mathbf{\widetilde{V}}^{L,\ast}}$ is
a proxy minimizer of $\Phi(\cdot)$ on $\mathbf{\Omega}$, due to the following

\vspace{0.2cm}

\begin{proposition}
\label{brostu5:prop.generaldeterministic.minimizer.improved}
There holds
\begin{equation}
\min_{\mathbf{Q}\in \mathbf{\Omega}} \Phi(\mathbf{Q}) 
\ \leq \ 
\Phi \left( M_{\mathbf{P}} \cdot \boldsymbol{\xi}_{n}^{\mathbf{\widetilde{V}}^{L,\ast}} \right)
\ \leq \ 
\widehat{\Phi(\mathbf{\Omega})}_{n,\infty}^{speedup,1}
\ + \ o_{\mathbb{\Pi}}(1)
\nonumber
\end{equation}
where $o_{\mathbb{\Pi}}(1)$ goes to $0$ as 
$L\rightarrow \infty$ and $n\rightarrow \infty$ under the
distribution $\mathbb{\Pi}$
(recall that 
$\mathbb{\Pi}[(\widetilde{V}_{1},\ldots,\widetilde{V}_{n}) \in \cdot \,]=
\bigotimes\limits_{k=1}^{K}\widetilde{U}_{k}^{\otimes n_{k}}[\,\cdot \,]$).
\end{proposition}

\vspace{0.3cm}

\begin{proposition}
In the above set-up, one has
\begin{equation}
\lim_{n\rightarrow \infty} \lim_{L\rightarrow \infty} 
\Phi\Big( \widehat{\argmin_{\mathbf{Q} \in \mathbf{\Omega}} \Phi(\mathbf{Q})}_{n,L}^{speedup,1} \Big)
\ = \ \Phi(\mathbf{\Omega}) \qquad \textrm{a.s.},
\nonumber
\end{equation}
and thus the quantity 
$\Phi\Big( \widehat{\argmin_{\mathbf{Q} \in \mathbf{\Omega}} \Phi(\mathbf{Q})}_{n,L}^{speedup,1} \Big)$
is a natural alternative to the estimate
$\widehat{\Phi(\mathbf{\Omega})}_{n,L}^{speedup,1}$ given in 
\eqref{brostu5:fo.BSmin.extended.improved.estim}.

\end{proposition}

\vspace{0.4cm}
\noindent
Analogously to the above investigations, 
by applying the results of Subsection \ref{SectEstimators.new.det.nonvoid.meth2.compact.max}
to the special choice $\mathbf{Q}^{\ast\ast}:= \mathbf{Q}^{\ast} \in int(\mathbf{\Omega})$
we can also derive improved/speed-up estimators 
$\widehat{\Phi(\mathbf{\Omega})}_{n,L}^{speedup,1}$ and
$\Phi\Big( \widehat{\argmax_{\mathbf{Q} \in \mathbf{\Omega}} \Phi(\mathbf{Q})}_{n,L}^{speedup,1} \Big)$
for $\Phi(\mathbf{\Omega}) := \max_{\mathbf{Q}\in \mathbf{\Omega}} \Phi(\mathbf{Q})$,
as well as 
$\widehat{\argmax_{\mathbf{Q} \in \mathbf{\Omega}} \Phi(\mathbf{Q})}_{n,L}^{speedup,1}$
for $\argmax_{\mathbf{Q}\in \mathbf{\Omega}} \Phi(\mathbf{Q})$.
For the sake of brevity, the details are omitted.


\vspace{0.5cm}
\noindent
\textit{G.2) \,  
Connections to importance sampling}

\vspace{0.2cm}
\noindent
Usually, the importance-sampling paradigm basically 
aims for the more accurate simulative estimation of an (say) integral
$\int_{\breve{\mathbf{\Omega}}} \breve{g}(\breve{\mathbf{y}})  \, \mathrm{d}\breve{\mathbb{\bbzeta}}(\breve{\mathbf{y}})$
represented as
$$\int_{\breve{\mathbf{\Omega}}} \breve{g}(\breve{\mathbf{y}})  \, \mathrm{d}\breve{\mathbb{\bbzeta}}(\breve{\mathbf{y}}) = 
\int_{\mathbb{R}^{K}} \textfrak{1}_{\breve{\mathbf{\Omega}}}(\breve{\mathbf{y}}) \cdot  \breve{g}(\breve{\mathbf{y}})  
\, \mathrm{d}\breve{\mathbb{\bbzeta}}(\breve{\mathbf{y}}) 
= \mathbb{E}_{\mathbb{\Pi}}\negthinspace \Big[
\, \breve{g}(\breve{\mathbf{Y}})  \cdot  \textfrak{1}_{\breve{\mathbf{\Omega}}}(\breve{\mathbf{Y}})
\, \Big] 
$$ 
(with distribution $\mathbb{\Pi}[\breve{\mathbf{Y}} \in \cdot \,]=\breve{\mathbb{\bbzeta}}[\,\cdot \,]$),
by simulatively estimating the rewritten integral
$$ 
\int_{\breve{\mathbf{\Omega}}} \breve{g}(\breve{\mathbf{y}})  \, \mathrm{d}\breve{\mathbb{\bbzeta}}(\breve{\mathbf{y}}) 
= \int_{\mathbb{R}^{K}} 
\textfrak{1}_{\breve{\mathbf{\Omega}}}(\breve{\mathbf{z}})
\cdot 
\breve{g}(\breve{\mathbf{z}})  
\cdot 
\frac{\mathrm{d}\breve{\mathbb{\bbzeta}}}{\mathrm{d}\breve{\mathbb{S}}}(\breve{\mathbf{z}}) 
\, \mathrm{d}\breve{\mathbb{S}}(\breve{\mathbf{z}}) 
= \mathbb{E}_{\mathbb{\Pi}}\negthinspace \Big[ \, 
\frac{\mathrm{d}\breve{\mathbb{\bbzeta}}}{\mathrm{d}\breve{\mathbb{S}}}(\breve{\mathbf{Z}}) \cdot
\, \breve{g}(\breve{\mathbf{Z}})  \cdot  \textfrak{1}_{\breve{\mathbf{\Omega}}}(\breve{\mathbf{Z}})
\, \Big] 
$$
(with distribution $\mathbb{\Pi}[\breve{\mathbf{Z}} \in \cdot \,]=
\breve{\mathbb{S}}[\,\cdot \,]$), where $\breve{\mathbb{S}}$ and thus $\breve{\mathbf{Z}}$
are constructed in a (goal-depending) appropriate manner.
To establish some connection with our \textit{BS-minimization} context,
recall that in the above Subsection \ref{SectEstimators.new.det.nonvoid.meth1.compact.min}
we have employed
$\mathbb{\Pi}[\mathbf{\widetilde{W}} \in \cdot \,] 
= \mathbb{\Pi}[(\widetilde{W}_{1},\ldots,\widetilde{W}_{n}) \in \cdot \,]=
\widetilde{\mathbb{\bbzeta}}^{\otimes n}[\,\cdot \,]$ 
and obtained
for large $n \in \mathbb{N}$ 
the approximation
\begin{eqnarray}
\Phi(\mathbf{\Omega}) := \min_{\mathbf{Q}\in \mathbf{\Omega}} \Phi(\mathbf{Q})
&\approx& 
- \, \frac{1}{n}\log \negthinspace \left( \ 
\mathbb{E}_{\mathbb{\Pi}}\negthinspace \Big[
\exp\negthinspace\Big(n \cdot \Big(
D_{\varphi }\big(M_{\mathbf{P}} \cdot \boldsymbol{\xi }_{n}^{\mathbf{\widetilde{W}}},\mathbf{P}\big) 
- \Phi\big(M_{\mathbf{P}} \cdot \boldsymbol{\xi }_{n}^{\mathbf{\widetilde{W}}}\big)
\Big)
\Big)
\cdot \textfrak{1}_{\mathbf{\Omega}}\big(M_{\mathbf{P}} \cdot \boldsymbol{\xi }_{n}^{\mathbf{\widetilde{W}}}\big)
\, \Big] 
\right)
\quad \textrm{(cf. \ref{brostu5:fo.BSmin.extended.approx})}
\nonumber
\\
&=& 
- \, \frac{1}{n}\log \negthinspace \left( \ 
\int_{\mathbb{R}^{K}}
\textfrak{1}_{\mathbf{\Omega}}\big(M_{\mathbf{P}} \cdot \boldsymbol{\xi }_{n}^{\mathbf{y}}\big) \cdot
\exp\negthinspace\Big(n \cdot \Big(
D_{\varphi }\big(M_{\mathbf{P}} \cdot \boldsymbol{\xi}_{n}^{\mathbf{y}},\mathbf{P}\big) 
- \Phi\big(M_{\mathbf{P}} \cdot \boldsymbol{\xi}_{n}^{\mathbf{y}}\big)
\Big)
\Big)
\, 
\, \mathrm{d}\widetilde{\mathbb{\bbzeta}}^{\otimes n}(\mathbf{y})
\right)
\label{brostu5:fo.BSmin.extended.approx2}
\end{eqnarray}
where 
$
\boldsymbol{\xi}_{n}^{\mathbf{y}}:=\Big(\frac{1}{n}\sum_{i\in
I_{1}^{(n)}}y_{i},\ldots ,\frac{1}{n}\sum_{i\in I_{K}^{(n)}}
y_{i}\Big)
$ 
(analogously to \eqref{Xi_n^W vector}). We can apply the above-mentioned general importance-sampling
procedure to the \textit{very special case} (amongst many other alternative possibilities)
$$
\breve{\mathbf{\Omega}} := \mathbf{\Omega},
\quad \breve{\mathbf{Y}} := M_{\mathbf{P}} \cdot \boldsymbol{\xi }_{n}^{\mathbf{\widetilde{W}}}, \quad
\breve{g}(\breve{\mathbf{Y}}) :=
\exp\negthinspace\Big(n \cdot \Big(
D_{\varphi }\big(\breve{\mathbf{Y}},\mathbf{P}\big) - \Phi\big(\breve{\mathbf{Y}}\big)
\Big) \Big), \quad
\breve{\mathbb{\bbzeta}}[\,\cdot \,] := \mathbb{\Pi}[\breve{\mathbf{Y}} \in \cdot \,],
\quad \breve{\mathbf{Z}} :=  M_{\mathbf{P}} \cdot \boldsymbol{\xi }_{n}^{\mathbf{\widetilde{V}}},
\quad \breve{\mathbb{S}}[\,\cdot \,] := \mathbb{\Pi}[\breve{\mathbf{Z}} \in \cdot \,],
$$
where we employ --- with the same choice $\mathbf{Q}^{\ast\ast} := \mathbf{Q}^{\ast} \in int(\mathbf{\Omega})$
as in the previous Subsection \ref{SectEstimators.new.det.nonvoid.improved.compact}.1) ---
the random vector $\mathbf{\widetilde{V}}$ of \eqref{brostu5:V_new}
having distribution 
$\mathbb{\Pi}[\mathbf{\widetilde{V}} \in \cdot \,] 
= \mathbb{\Pi}[(\widetilde{V}_{1},\ldots,\widetilde{V}_{n}) \in \cdot \,]=
\bigotimes\limits_{k=1}^{K}\widetilde{U}_{k}^{\otimes n_{k}}[\,\cdot \,]$, and
$\boldsymbol{\xi}_{n}^{\mathbf{\widetilde{V}}}$ of \eqref{Xi_n^W vector V new2};
accordingly, by rewriting \eqref{brostu5:fo.BSmin.extended.approx}
and \eqref{brostu5:fo.BSmin.extended.approx2} 
we end up with the approximation
\begin{eqnarray}
& &\hspace{-0.7cm}
\Phi(\mathbf{\Omega}) := \min_{\mathbf{Q}\in \mathbf{\Omega}} \Phi(\mathbf{Q})
\nonumber \\
&\approx& 
- \, \frac{1}{n}\log \negthinspace \left( \ 
\mathbb{E}_{\mathbb{\Pi}}\negthinspace \Big[ \, 
\textfrak{1}_{\mathbf{\Omega}}\big(M_{\mathbf{P}} \cdot \boldsymbol{\xi }_{n}^{\mathbf{\widetilde{V}}}\big)
\cdot
\exp\negthinspace\Big(n \cdot \Big(
D_{\varphi }\big(M_{\mathbf{P}} \cdot \boldsymbol{\xi }_{n}^{\mathbf{\widetilde{V}}},\mathbf{P}\big) 
- \Phi\big(M_{\mathbf{P}} \cdot \boldsymbol{\xi }_{n}^{\mathbf{\widetilde{V}}}\big)
\Big)
\Big)
\cdot \frac{\mathrm{d}\breve{\mathbb{\bbzeta}}}{\mathrm{d}\breve{\mathbb{S}}}
\negthinspace\left(M_{\mathbf{P}} \cdot \boldsymbol{\xi}_{n}^{\mathbf{\widetilde{V}}}\right)\, 
\Big] 
\right)
\label{brostu5:fo.BSmin.extended.approx3}
\\
&=& 
- \, \frac{1}{n}\log \negthinspace \left( \ 
\int_{\mathbb{R}^{K}}
\textfrak{1}_{\mathbf{\Omega}}\big(M_{\mathbf{P}} \cdot \boldsymbol{\xi}_{n}^{\mathbf{z}}\big) \cdot
\exp\negthinspace\Big(n \cdot \Big(
D_{\varphi }\big(M_{\mathbf{P}} \cdot \boldsymbol{\xi}_{n}^{\mathbf{z}},\mathbf{P}\big) 
- \Phi\big(M_{\mathbf{P}} \cdot \boldsymbol{\xi}_{n}^{\mathbf{z}}\big)
\Big)
\Big)
\cdot \frac{\mathrm{d}\breve{\mathbb{\bbzeta}}}{\mathrm{d}\breve{\mathbb{S}}}
\Big(M_{\mathbf{P}} \cdot \boldsymbol{\xi}_{n}^{\mathbf{z}}\Big)\, 
\, \mathrm{d}\bigotimes\limits_{k=1}^{K}\widetilde{U}_{k}^{\otimes n_{k}}(z)
\right) .
\nonumber
\end{eqnarray}
Notice that the above-mentioned underlying importance-sampling goal is 
in our current context the improved hit rate of the set $\mathbf{\Omega}$ 
(and this is achieved by construction).
At this point, one may wonder about the connection between the two alternative
approximations \eqref{brostu5:fo.BSmin.extended.approx3}
and
\begin{equation}
\Phi(\mathbf{\Omega}) := \min_{\mathbf{Q}\in \mathbf{\Omega}} \Phi(\mathbf{Q})
\ \approx \ - \, 
\frac{1}{n}\log \negthinspace \left( \ 
\mathbb{E}_{\mathbb{\Pi}}\negthinspace \Big[
\exp\negthinspace\Big(n \cdot \Big(
D_{\varphi,\mathbf{P}}^{SBD}\negthinspace\left(M_{\mathbf{P}} 
\cdot \boldsymbol{\xi }_{n}^{\mathbf{\widetilde{V}}},\mathbf{Q}^{\ast}\right)  
- \Phi\big(M_{\mathbf{P}} \cdot \boldsymbol{\xi }_{n}^{\mathbf{\widetilde{V}}}\big)
\Big)
\Big)
\cdot \textfrak{1}_{\mathbf{\Omega}}\big(M_{\mathbf{P}} \cdot \boldsymbol{\xi }_{n}^{\mathbf{\widetilde{V}}}\big)
\, \Big] 
\right)
\, 
\textrm{(cf. 
\eqref{brostu5:fo.BSmin.extended.min.SBD.approx.onestar}).}
\nonumber
\end{equation}
The corresponding answer is that they \textit{essentially coincide}:

\hspace{0.2cm}
\begin{proposition}
\label{brostu5:prop.coincide}
Under the assumptions of the Subsections \ref{SectEstimators.new.det.nonvoid.meth1.compact.min}
and \ref{SectEstimators.new.det.nonvoid.improved.compact}.1),
if $n \in \mathbb{N}$ is fixed and $\frac{n_{k}}{n} = \widetilde{p}_{k}$ for all $k=1,\ldots,K$, 
then the term in formula \eqref{brostu5:fo.BSmin.extended.approx3} 
is equal to the term in formula \eqref{brostu5:fo.BSmin.extended.min.SBD.approx.onestar}.
\end{proposition}

\vspace{0.2cm} 
\noindent
\begin{remark}
Concerning the above Proposition
\ref{brostu5:prop.coincide},
recall that 
$\lim_{n\rightarrow \infty} \frac{n_{k}}{n} = \widetilde{p}_{k}$
(cf. \eqref{fo.freqlim});
moreover, if all $\widetilde{p}_{k}$ ($k=1,\ldots,K$) are rational numbers in $]0,1[$ with 
$\sum_{k=1}^{K} \widetilde{p}_{k} =1$ 
and $N$ is the (always existing) smallest integer such that all
$N \cdot \widetilde{p}_{k}$ ($k=1,\ldots,K$) are integers (i.e. $\in \mathbb{N}$),
then for any multiple $n= \ell \cdot N$ ($\ell \in \mathbb{N}$) 
one gets that all $n_{k} = n \cdot \widetilde{p}_{k}$ are integers
and that $card(I_{K}^{(n)}) = n_{K}$.
\end{remark}

\vspace{0.4cm} 
\noindent
The proof of Proposition \ref{brostu5:prop.coincide}
will be given in Appendix \ref{App.A}.

\vspace{0.4cm}
\noindent
For the corresponding \textit{importance-sampling estimate} of
the approximation \eqref{brostu5:fo.BSmin.extended.approx3},
we take
\begin{eqnarray}
& & \widehat{\Phi(\mathbf{\Omega})}_{n,L}^{IS,1} \ := \ 
- \frac{1}{n}\log \frac{1}{L}\sum_{\ell =1}^{L} 
\textfrak{1}_{\mathbf{\Omega}}\big(M_{\mathbf{P}} \cdot \boldsymbol{\xi }_{n}^{\mathbf{\widetilde{V}}^{(\ell)}}\big)
\cdot
\exp\negthinspace\Big(n \cdot \Big(
D_{\varphi }\big(M_{\mathbf{P}} \cdot \boldsymbol{\xi }_{n}^{\mathbf{\widetilde{V}}^{(\ell)}},\mathbf{P}\big) 
- \Phi\big(M_{\mathbf{P}} \cdot \boldsymbol{\xi }_{n}^{\mathbf{\widetilde{V}}^{(\ell)}}\big)
\Big)
\Big)
\cdot \frac{\mathrm{d}\breve{\mathbb{\bbzeta}}}{\mathrm{d}\breve{\mathbb{S}}}
\negthinspace\left(M_{\mathbf{P}} \cdot \boldsymbol{\xi}_{n}^{\mathbf{\widetilde{V}}^{(\ell)}}\right)\, 
\nonumber\\
& & \ = \ 
- \frac{1}{n}\log \frac{1}{L}\sum_{\ell =1}^{L} 
\textfrak{1}_{\mathbf{\Omega}}\big(M_{\mathbf{P}} \cdot \boldsymbol{\xi }_{n}^{\mathbf{\widetilde{V}}^{(\ell)}}\big)
\cdot
\exp\negthinspace\Big(n \cdot \Big(
D_{\varphi }\big(M_{\mathbf{P}} \cdot \boldsymbol{\xi }_{n}^{\mathbf{\widetilde{V}}^{(\ell)}},\mathbf{P}\big) 
- \Phi\big(M_{\mathbf{P}} \cdot \boldsymbol{\xi }_{n}^{\mathbf{\widetilde{V}}^{(\ell)}}\big)
\Big)
\Big)
\cdot 
\prod_{k=1}^{K} IS_{k}(\mathbf{\widetilde{V}}_{k}^{(\ell)})
\label{brostu5:fo.BSmin.extended.approx15}
\end{eqnarray}
where we simulate independently $L$ copies 
$\mathbf{\widetilde{V}}^{(1)},\ldots,\mathbf{\widetilde{V}}^{(L)}$ of the random vector
$\mathbf{\widetilde{V}}:=\left( \widetilde{V}_{1},\ldots,\widetilde{V}_{n}\right)$ 
constructed by the above-mentioned method, 
and compute each of $\boldsymbol{\xi}_{n}^{\mathbf{\widetilde{V}}^{(1)}}, \ldots,
\boldsymbol{\xi}_{n}^{\mathbf{\widetilde{V}}^{(L)}}$ according to 
\eqref{Xi_n^W vector V new2}; notice that in \eqref{brostu5:fo.BSmin.extended.approx15}
we have employed (the $k-$block importance-sampling factor) 
$\widetilde{IS}_{k}(z_{1},\ldots,z_{n_{k}}) 
\ := \  \exp\Big(n_{k} \cdot \Lambda_{\widetilde{\mathbb{\bbzeta}}}(\tau_{k}) \, - \, \tau_{k} 
\cdot \sum_{i=1}^{n_{k}} z_{i} \Big)$
(cf. Subsection X-A of Broniatowski \& Stummer \cite{Bro:23a}).

\vspace{0.4cm}
\noindent
As can be seen straightforwardly from the above calculations,
due to \eqref{fo.freqlim} one gets even the limit relation
\begin{eqnarray}
& & \hspace{-0.7cm}
\Phi(\mathbf{\Omega}) := \min_{\mathbf{Q}\in \mathbf{\Omega}} \Phi(\mathbf{Q})
\nonumber \\
&=& 
- \, \lim_{n\rightarrow \infty} \frac{1}{n}\log \negthinspace \left( \ 
\mathbb{E}_{\mathbb{\Pi}}\negthinspace \Big[ \, 
\textfrak{1}_{\mathbf{\Omega}}\big(M_{\mathbf{P}} \cdot \boldsymbol{\xi }_{n}^{\mathbf{\widetilde{V}}}\big)
\cdot
\exp\negthinspace\Big(n \cdot \Big(
D_{\varphi }\big(M_{\mathbf{P}} \cdot \boldsymbol{\xi }_{n}^{\mathbf{\widetilde{V}}},\mathbf{P}\big) 
- \Phi\big(M_{\mathbf{P}} \cdot \boldsymbol{\xi }_{n}^{\mathbf{\widetilde{V}}}\big)
\Big)
\Big)
\cdot \frac{\mathrm{d}\breve{\mathbb{\bbzeta}}}{\mathrm{d}\breve{\mathbb{S}}}
\negthinspace\left(M_{\mathbf{P}} \cdot \boldsymbol{\xi}_{n}^{\mathbf{\widetilde{V}}}\right)\, 
\Big] 
\right)
\label{brostu5:fo.BSmin.extended.limit1}
\\
&=& 
\ - \, 
\lim_{n\rightarrow \infty }\frac{1}{n}\log \negthinspace \left( \ 
\mathbb{E}_{\mathbb{\Pi}}\negthinspace \Big[
\, \textfrak{1}_{\mathbf{\Omega}}\big(M_{\mathbf{P}} \cdot \boldsymbol{\xi }_{n}^{\mathbf{\widetilde{V}}}\big)
\cdot
\exp\negthinspace\Big(n \cdot \Big(
D_{\varphi,\mathbf{P}}^{SBD}\negthinspace\left(M_{\mathbf{P}} 
\cdot \boldsymbol{\xi }_{n}^{\mathbf{\widetilde{V}}},\mathbf{Q}^{\ast}\right)  
- \Phi\big(M_{\mathbf{P}} \cdot \boldsymbol{\xi }_{n}^{\mathbf{\widetilde{V}}}\big)
\Big)
\Big)
\, \Big] 
\right)
\, .
\qquad \textrm{(cf. 
\eqref{brostu5:fo.BSmin.extended.min.SBD.onestar})}
\nonumber
\end{eqnarray}

\vspace{0.2cm}
\noindent
Summing up things, for an \textit{improved} estimation of the \textit{minimum value} 
$\min_{\mathbf{Q}\in \mathbf{\Omega}} \Phi (\mathbf{Q})$ 
the two above-described approaches are \textit{asymptotically} equal. 
However, notice that for the --- quite comfortable ---
essential convergence proofs of the Propositions
\ref{brostu5:prop.generaldeterministic.minimum.improved} and 
\ref{brostu5:prop.generaldeterministic.minimizer.improved}  (including the treatment of estimates
of \textit{minimizers}) 
we rely heavily on the use of our newly developed Theorems \ref{brostu5:thm.BSnarrow.SBD}
and \ref{brostu5:thm.Fmin.SBD}, which put us in position to further proceed 
completely analogously to the case of naive estimators.

\vspace{0.4cm}
\noindent
To end up this section, let us mention that as the corresponding \textit{very natural importance-sampling BS-estimator 
of the minimizer-set} 
$\mathcal{Q}^{\ast} := \argmin_{\mathbf{Q} \in \mathbf{\Omega }} \Phi(\mathbf{Q})$, 
one can take 
\begin{equation}
\widehat{\argmin_{\mathbf{Q} \in \mathbf{\Omega }} \Phi(\mathbf{Q})}_{n,L}^{IS,1}
:= \argmin_{\boldsymbol{\nu} \in \mathcal{V}_{n,L}} \Phi(\boldsymbol{\nu})
\nonumber
\end{equation}
where $\mathcal{V}_{n,L} := \{M_{\mathbf{P}} \cdot \boldsymbol{\xi}_{n}^{\mathbf{\widetilde{V}}^{(\ell)}}:
\ell \in \{1,\ldots,L\} \,  \} \cap \mathbf{\Omega}$,
for which we can prove --- due to the equality of \eqref{brostu5:fo.BSmin.extended.limit1} 
and \eqref{brostu5:fo.BSmin.extended.min.SBD.onestar}
--- that the limit behaviour of $\widehat{\argmin_{\mathbf{Q} \in \mathbf{\Omega}} \Phi(\mathbf{Q})}_{n,L}^{speedup}$ 
carries over. Hence, also the above-mentioned Remark  \ref{brostu5:rem.116} carries over.


\section{Bare-Simulation Estimators For General Deterministic Divergence-Optimization-Problems
with Constant-Component-Sum Side Constraint}
\label{SectEstimators.new.det.simplex}

\vspace{0.2cm}
Recall that we are interested in the constrained optimization of the \textit{continuous} 
distance-connected functions 
$\mathbf{\Omega} \ni \mathbf{Q} \mapsto \Phi(\mathbf{Q})$
in the above-mentioned cases (D1) to (D8) of Subsection \ref{SectDetGeneral.Friends}, and beyond
(e.g. $\Phi(\cdot) := D_{\breve{\varphi}}(\cdot,\breve{\mathbf{P}})$
may be a $\breve{\varphi}-$divergence with pregiven $\breve{\mathbf{P}}$ and $\breve{\varphi}$, cf.
Remark \ref{brostu5:rem.thm.Fmin}(iii)). 
Contrary to the previous Section \ref{SectEstimators.new.det.nonvoid},
we now involve constraint sets $A \cdot \textrm{$\boldsymbol{\Omega}$\hspace{-0.23cm}$\boldsymbol{\Omega}$}$
where $A \in \, ]0,\infty[$ is a pregiven constant and 
$\textrm{$\boldsymbol{\Omega}$\hspace{-0.23cm}$\boldsymbol{\Omega}$} \in \mathbb{S}^{K}$
(respectively $\mathbb{S}_{>0}^{K}$);
this means that we employ the side constraints $q_{k} \geq 0$ (respectively, $q_{k} > 0$) for all $k=1,\ldots,K$ 
as well as $\sum_{k=1}^{K} q_{k} = A$.
Clearly, one has
$int\left( A \cdot \textrm{$\boldsymbol{\Omega}$\hspace{-0.23cm}$\boldsymbol{\Omega}$} \right) = \emptyset$
in the \textit{full} topology 
(cf. Remark \ref{after det Problem}(b)), and thus we can not apply the estimator-results 
of Section \ref{SectEstimators.new.det.nonvoid} but we need some extra refinements.
Those will be worked out in the following,
for the context that $A \cdot \textrm{$\boldsymbol{\Omega}$\hspace{-0.23cm}$\boldsymbol{\Omega}$}$
satisfies the regularity properties \eqref{regularity simplex} --- in the \textit{relative} topology (!!) ---
such that
the function $\Phi(\cdot)$ possesses a (not necessarily unique) minimizer; 
for this, we construct 
\textit{naive estimators} as well as 
\textit{speed-up estimators} (approximations) of the \textit{minimum value}
$\inf_{\mathbf{Q}\in A \cdot \textrm{$\boldsymbol{\Omega}$\hspace{-0.19cm}$\boldsymbol{\Omega}$}}\Phi (\mathbf{Q}) 
= \min_{\mathbf{Q}\in A \cdot \textrm{$\boldsymbol{\Omega}$\hspace{-0.19cm}$\boldsymbol{\Omega}$}} \Phi (\mathbf{Q})$ and of the corresponding
(set of) \textit{minimizers} $\arg \inf_{\mathbf{Q}\in A \cdot 
\textrm{$\boldsymbol{\Omega}$\hspace{-0.19cm}$\boldsymbol{\Omega}$}}\Phi (\mathbf{Q}) 
=\arg \min_{\mathbf{Q}\in A \cdot \textrm{$\boldsymbol{\Omega}$\hspace{-0.19cm}$\boldsymbol{\Omega}$}} \Phi (\mathbf{Q})$.
The case of maximum values and maximizers will be treated analogously. 
We mainly focus on compact sets $A \cdot \textrm{$\boldsymbol{\Omega}$\hspace{-0.23cm}$\boldsymbol{\Omega}$}$ 
(in the relative toplogy) but also discuss some relaxations thereof.


\subsection{Naive estimators of min and argmin --- Base-Divergence-Method 1, compact case}
\label{SectEstimators.new.det.simplex.meth1.compact.min}

\vspace{0.2cm}
\noindent
Recall that from Theorem 
\ref{brostu5:thm.Fmin.simplex}(a)
we obtain for any continuous function $\Phi: A \cdot \textrm{$\boldsymbol{\Omega}$\hspace{-0.23cm}$\boldsymbol{\Omega}$} \mapsto \mathbb{R}$ on a compact
set $A \cdot \textrm{$\boldsymbol{\Omega}$\hspace{-0.23cm}$\boldsymbol{\Omega}$}
\subset \widetilde{\mathcal{M}}_{\gamma}$ 
with \eqref{regularity simplex} the assertion 
\begin{equation}
\min_{\mathbf{Q}\in A \cdot \textrm{$\boldsymbol{\Omega}$\hspace{-0.19cm}$\boldsymbol{\Omega}$}} \Phi(\mathbf{Q})
\ = \ - \, 
\lim_{n\rightarrow \infty }\frac{1}{n}\log \negthinspace \left( \ 
\mathbb{E}_{\mathbb{\Pi}}\negthinspace \Big[
\exp\negthinspace\Big(n \cdot \Big(
F_{\gamma,\widetilde{c},A}\Big(D_{\widetilde{c} \cdot \varphi_{\gamma}}(A \cdot \boldsymbol{\xi}_{n}^{w\mathbf{W}},\mathds{P})\Big)
- \Phi\big(A \cdot \boldsymbol{\xi}_{n}^{w\mathbf{W}}\big)
\Big)
\Big)
\cdot \textfrak{1}_{
\textrm{$\boldsymbol{\Omega}$\hspace{-0.19cm}$\boldsymbol{\Omega}$}}\big(\boldsymbol{\xi}_{n}^{w\mathbf{W}}\big)
\, \Big] 
\right)
\,  \, ,
\label{brostu5:fo.BSmin.extended.min.simplex}
\end{equation}
where $M_{\mathbf{P}} =\sum_{i=1}^{K}p_{i}>0$, $\gamma \in \mathbb{R}\backslash\, ]1,2[$,
$\widetilde{\mathcal{M}}_{\gamma}$ as defined in (S1), and 
\begin{eqnarray}
\boldsymbol{\xi}_{n}^{w\mathbf{W}} &:=&
\begin{cases}
\left(\frac{\sum_{i \in I_{1}^{(n)}}W_{i}}{\sum_{k=1}^{K}\sum_{i \in I_{k}^{(n)}}W_{i}},
\ldots, \frac{\sum_{i \in I_{K}^{(n)}}W_{i}}{\sum_{k=1}^{K}\sum_{i \in I_{k}^{(n)}}W_{i}} \right) ,
\qquad \textrm{if } \sum_{j=1}^{n} W_{j} \ne 0, \\
\ (\infty, \ldots, \infty) =: \boldsymbol{\infty}, \hspace{4.0cm} \textrm{if } \sum_{j=1}^{n} W_{j} = 0,
\end{cases}
\qquad 
\textrm{(cf. \eqref{brostu5:fo.norweiemp.vec.det})}
\nonumber
\end{eqnarray}
is constructed from a sequence $W:=(W_{i})_{i\in \mathbb{N}}$ of random variables, 
where the $W_{i}$'s are i.i.d. copies of the random variable $W$
whose distribution 
is $\mathbb{\Pi }[W\in \cdot \,]=\mathbb{\bbzeta}[\,\cdot \,]$ 
being attributed to the power divergence generator  
$\varphi := \widetilde{c} \cdot \varphi_{\gamma}$ 
by the representability \eqref{brostu5:fo.link.var.simplex}. 
Within such a set-up, we thus obtain for large $n \in \mathbb{N}$ 
(cf. \eqref{brostu5:fo.BSmin.extended.min.simplex})
the approximation
\begin{equation}
\Phi(A \cdot \textrm{$\boldsymbol{\Omega}$\hspace{-0.23cm}$\boldsymbol{\Omega}$}) := 
\min_{\mathbf{Q}\in A \cdot \textrm{$\boldsymbol{\Omega}$\hspace{-0.19cm}$\boldsymbol{\Omega}$}} \Phi(\mathbf{Q})
\ \approx \ - \, 
\frac{1}{n}\log \negthinspace \left( \ 
\mathbb{E}_{\mathbb{\Pi}}\negthinspace \Big[
\exp\negthinspace\Big(n \cdot \Big(
F_{\gamma,\widetilde{c},A}\Big(D_{\widetilde{c} \cdot \varphi_{\gamma}}(A \cdot \boldsymbol{\xi}_{n}^{w\mathbf{W}},\mathds{P})\Big)
- \Phi\big(A \cdot \boldsymbol{\xi}_{n}^{w\mathbf{W}}\big)
\Big)
\Big)
\cdot \textfrak{1}_{
\textrm{$\boldsymbol{\Omega}$\hspace{-0.19cm}$\boldsymbol{\Omega}$}}\big(\boldsymbol{\xi}_{n}^{w\mathbf{W}}\big)
\, \Big] 
\right)
\,  \, ,
\label{brostu5:fo.BSmin.extended.approx.simplex}
\end{equation}
and hence for getting an estimator of the minimum value 
$\Phi(A \cdot \textrm{$\boldsymbol{\Omega}$\hspace{-0.23cm}$\boldsymbol{\Omega}$})$
one can estimate the right-hand side of \eqref{brostu5:fo.BSmin.extended.approx.simplex}.
To achieve this, for the rest of this section we assume 
that $n$ is chosen such that all 
$n \cdot \frac{p_{k}}{M_{\mathbf{P}}}$ are integers
(and hence, $n = \sum_{k=1}^{K} n_{k}$ with $n_{k} = n \cdot \frac{p_{k}}{M_{\mathbf{P}}}$) --- 
the remaining case works analogously.
As above, a corresponding \textit{naive (crude) estimator} can be constructed by 
\begin{equation}
\widehat{\Phi(A \cdot \textrm{$\boldsymbol{\Omega}$\hspace{-0.23cm}$\boldsymbol{\Omega}$})}_{n,L}^{naive,1} \ := \ 
- \frac{1}{n}\log \frac{1}{L}\sum_{\ell =1}^{L} 
\exp\negthinspace\Big(n \cdot \Big(
F_{\gamma,\widetilde{c},A}\Big(D_{\widetilde{c} \cdot \varphi_{\gamma}}(A \cdot 
\boldsymbol{\xi}_{n}^{w\mathbf{W}^{(\ell)}},\mathds{P})\Big)
- \Phi\big(A \cdot \boldsymbol{\xi}_{n}^{w\mathbf{W}^{(\ell)}}\big)
\Big)
\Big)
\cdot \textfrak{1}_{
\textrm{$\boldsymbol{\Omega}$\hspace{-0.19cm}$\boldsymbol{\Omega}$}}\big(\boldsymbol{\xi}_{n}^{w\mathbf{W}^{(\ell)}}\big)
\, ,
\label{brostu5:fo.BSmin.extended.naive.estim.simplex}
\end{equation}
where we simulate independently $L$ copies 
$\mathbf{W}^{(1)},\ldots,\mathbf{W}^{(L)}$ of the vector
 $\mathbf{W}:=\left( W_{1},\ldots,W_{n}\right) $ 
with independent entries under $\mathbb{\bbzeta}$, 
and compute each of $\boldsymbol{\xi}_{n}^{w\mathbf{W}^{(1)}}, \ldots,
\boldsymbol{\xi}_{n}^{w\mathbf{W}^{(L)}}$ according to 
\eqref{brostu5:fo.norweiemp.vec.det}.
Clearly, with the help of the strong law of large numbers we get with 
$\widehat{\Phi(A \cdot 
\textrm{$\boldsymbol{\Omega}$\hspace{-0.23cm}$\boldsymbol{\Omega}$})}_{n,\infty}^{naive,1}: = 
\lim_{L\rightarrow \infty} \widehat{\Phi(A \cdot 
\textrm{$\boldsymbol{\Omega}$\hspace{-0.23cm}$\boldsymbol{\Omega}$})}_{n,L}^{naive,1}$
the following assertion:

\vspace{0.2cm}

\begin{proposition}
\begin{equation}
\lim_{n\rightarrow \infty} 
\widehat{\Phi(A \cdot \textrm{$\boldsymbol{\Omega}$\hspace{-0.23cm}$\boldsymbol{\Omega}$})}_{n,\infty}^{naive,1}
\ = \ 
\lim_{n\rightarrow \infty} \lim_{L\rightarrow \infty} 
\widehat{\Phi(A \cdot \textrm{$\boldsymbol{\Omega}$\hspace{-0.23cm}$\boldsymbol{\Omega}$})}_{n,L}^{naive,1}
\ = \ \Phi(A \cdot \textrm{$\boldsymbol{\Omega}$\hspace{-0.23cm}$\boldsymbol{\Omega}$}) \qquad \textrm{a.s.}
\label{brostu5:fo.BSmin.extended.naive.estim.lim.simplex}
\end{equation}
\end{proposition}

\vspace{0.2cm}
\noindent
As the corresponding \textit{very natural naive (crude) estimator}  
of the minimizer-set 
$\mathcal{Q}^{\ast} := \argmin_{\mathbf{Q} \in 
A \cdot \textrm{$\boldsymbol{\Omega}$\hspace{-0.19cm}$\boldsymbol{\Omega}$}} \Phi(\mathbf{Q})$, 
we take 
\begin{equation}
\widehat{\argmin_{\mathbf{Q} \in 
A \cdot \textrm{$\boldsymbol{\Omega}$\hspace{-0.19cm}$\boldsymbol{\Omega}$}} \Phi(\mathbf{Q})}_{n,L}^{naive,1}
:= \argmin_{\boldsymbol{\nu} \in \mathcal{W}_{n,L}} \Phi(\boldsymbol{\nu})
\label{brostu5:fo.1555.simplex}
\end{equation}
where $\mathcal{W}_{n,L} := \{A \cdot \boldsymbol{\xi}_{n}^{w\mathbf{W}^{(\ell)}}:
\ell \in \{1,\ldots,L\} \,  \} \cap A \cdot \textrm{$\boldsymbol{\Omega}$\hspace{-0.23cm}$\boldsymbol{\Omega}$}$. 
In other words, as a corresponding \textit{naive (crude) estimator} 
of the (not necessarily unique) element $\mathbf{Q}^{\ast}$ of the minimizer-set 
$\mathcal{Q}^{\ast} := \argmin_{\mathbf{Q} \in 
A \cdot \textrm{$\boldsymbol{\Omega}$\hspace{-0.19cm}$\boldsymbol{\Omega}$}} \Phi(\mathbf{Q})$, 
we take the (not necessarily unique) element 
$A \cdot \boldsymbol{\xi}_{n}^{w\mathbf{W}^{L,\ast}}$
of the set $\{A \cdot \boldsymbol{\xi}_{n}^{w\mathbf{W}^{(\ell)}}:
\ell \in \{1,\ldots,L\} \,  \} $ such that 
$A \cdot \boldsymbol{\xi}_{n}^{w\mathbf{W}^{L,\ast}} \in 
A \cdot \textrm{$\boldsymbol{\Omega}$\hspace{-0.23cm}$\boldsymbol{\Omega}$}$ and 
\begin{equation}
\Phi(A \cdot \boldsymbol{\xi}_{n}^{w\mathbf{W}^{L,\ast}}) 
\ \leq \Phi(A \cdot \boldsymbol{\xi}_{n}^{w\mathbf{W}^{(\ell)}})   
\qquad \textrm{for all $\ell =1,\ldots,L$ for which 
$A \cdot \boldsymbol{\xi}_{n}^{w\mathbf{W}^{(\ell)}}$ 
belongs to $A \cdot \textrm{$\boldsymbol{\Omega}$\hspace{-0.23cm}$\boldsymbol{\Omega}$}$.} 
\nonumber
\end{equation}
In short, $A \cdot \boldsymbol{\xi}_{n}^{w\mathbf{W}^{L,\ast}}$
minimizes $\Phi(\cdot)$ amongst all values 
$A \cdot \boldsymbol{\xi}_{n}^{w\mathbf{W}^{(\ell)}}$
at hand which fall into $A \cdot \textrm{$\boldsymbol{\Omega}$\hspace{-0.23cm}$\boldsymbol{\Omega}$}$. 
For large enough $n \in \mathbb{N}$ and $L \in \mathbb{N}$, 
such $A \cdot \boldsymbol{\xi}_{n}^{w\mathbf{\widetilde{W}}^{L,\ast}}$ exists
since $A \cdot \textrm{$\boldsymbol{\Omega}$\hspace{-0.23cm}$\boldsymbol{\Omega}$}$ 
has non-void interior in the relative topology, by assumption 
\eqref{regularity simplex}.
We prove that if $L$ and $n$ tend to infinity, then 
$A \cdot \boldsymbol{\xi}_{n}^{w\mathbf{W}^{L,\ast}}$
concentrates to the above-mentioned set $\mathcal{Q}^{\ast}$ 
of minimizers of $\Phi(\cdot)$ on $A \cdot \textrm{$\boldsymbol{\Omega}$\hspace{-0.23cm}$\boldsymbol{\Omega}$}$.
As usual in similar procedures, $L$ is assumed to be large enough in order to justify 
some approximation for fixed $n$, typically the substitution of empirical means by expectations,
since $L$ is at disposal.

\noindent
Next we derive that $A \cdot \boldsymbol{\xi}_{n}^{w\mathbf{W}^{L,\ast}}$ is
a proxy minimizer of $\Phi(\cdot)$ on $A \cdot \textrm{$\boldsymbol{\Omega}$\hspace{-0.23cm}$\boldsymbol{\Omega}$}$, manifested by the following

\vspace{0.2cm}

\begin{proposition}
\label{brostu5:prop.generaldeterministic.minimizer.naive.simplex}
There holds
\begin{equation}
\min_{\mathbf{Q}\in 
A \cdot \textrm{$\boldsymbol{\Omega}$\hspace{-0.19cm}$\boldsymbol{\Omega}$}} \Phi(\mathbf{Q}) 
\ \leq \ 
\Phi \left( A \cdot \boldsymbol{\xi}_{n}^{w\mathbf{W}^{L,\ast}} \right)
\ \leq \ 
\widehat{\Phi(A \cdot \textrm{$\boldsymbol{\Omega}$\hspace{-0.23cm}$\boldsymbol{\Omega}$})}_{n,\infty}^{naive,1}
\ + \ o_{\mathbb{\Pi}}(1)
\label{minim simplex}
\end{equation}
where $o_{\mathbb{\Pi}}(1)$ goes to $0$ as $L\rightarrow \infty$ and $n\rightarrow \infty$ under the
distribution $\mathbb{\Pi}$
(recall that $\mathbb{\Pi }[(W_{1},\ldots,W_{n}) \in \cdot \,]=
\mathbb{\bbzeta}^{\otimes n}[\,\cdot \,]$).
\end{proposition}

\vspace{0.3cm}
\noindent
The proof of Proposition \ref{brostu5:prop.generaldeterministic.minimizer.naive.simplex}
works analogously to the proof of
Proposition \ref{brostu5:prop.generaldeterministic.minimizer.naive},
by replacing Theorem \ref{brostu5:thm.Fmin}(a) with Theorem \ref{brostu5:thm.Fmin.simplex}(a).

\vspace{0.3cm}

\begin{remark}
In the current set-up of compact 
$A \cdot \textrm{$\boldsymbol{\Omega}$\hspace{-0.23cm}$\boldsymbol{\Omega}$}$ with \eqref{regularity simplex}, 
by taking the special case $\Phi(\mathbf{Q}) := D_{\varphi}(\mathbf{Q},\mathds{P})$ 
we obtain the naive BS-estimator 
$\widehat{D_{\varphi}(A \cdot \textrm{$\boldsymbol{\Omega}$\hspace{-0.23cm}$\boldsymbol{\Omega}$},\mathds{P})}_{n,L}^{naive,1}$ 
of the minimum value $\min_{\mathbf{Q}\in 
A \cdot \textrm{$\boldsymbol{\Omega}$\hspace{-0.19cm}$\boldsymbol{\Omega}$}} D_{\varphi}(\mathbf{Q},\mathds{P})$.
\end{remark}

\vspace{0.3cm}

\begin{proposition}
In the above set-up, one has
\begin{equation}
\lim_{n\rightarrow \infty} \lim_{L\rightarrow \infty} 
\Phi\Big( \widehat{\argmin_{\mathbf{Q} \in 
A \cdot \textrm{$\boldsymbol{\Omega}$\hspace{-0.19cm}$\boldsymbol{\Omega}$}} \Phi(\mathbf{Q})}_{n,L}^{naive,1} \Big)
\ = \ \Phi(A \cdot \textrm{$\boldsymbol{\Omega}$\hspace{-0.23cm}$\boldsymbol{\Omega}$}) \qquad \textrm{a.s.},
\nonumber
\end{equation}
and thus the quantity 
$\Phi\Big( \widehat{\argmin_{\mathbf{Q} \in 
A \cdot \textrm{$\boldsymbol{\Omega}$\hspace{-0.19cm}$\boldsymbol{\Omega}$}} \Phi(\mathbf{Q})}_{n,L}^{naive,1} \Big)$
is a natural alternative to the estimate
$\widehat{\Phi(A \cdot \textrm{$\boldsymbol{\Omega}$\hspace{-0.23cm}$\boldsymbol{\Omega}$})}_{n,L}^{naive,1}$ 
given in 
\eqref{brostu5:fo.BSmin.extended.naive.estim.simplex}.

\end{proposition}

\vspace{0.3cm}

\begin{remark}
By applying the above-mentioned results (with $A=1$)
to the special \textit{divergence} (cf. Remark \ref{rem.FasDivergence})
 $\Phi(\mathbf{Q}) := F_{\gamma,\widetilde{c},1}\Big(D_{\widetilde{c} \cdot \varphi_{\gamma}}(\mathbf{Q},\mathds{P})\Big)$,
we obtain for $\Phi(\textrm{$\boldsymbol{\Omega}$\hspace{-0.23cm}$\boldsymbol{\Omega}$})
:= \min_{\mathbf{Q} \in 
\textrm{$\boldsymbol{\Omega}$\hspace{-0.19cm}$\boldsymbol{\Omega}$}} \Phi(\mathbf{Q}) $
the estimator 
$
\widehat{\Phi(\textrm{$\boldsymbol{\Omega}$\hspace{-0.23cm}$\boldsymbol{\Omega}$})}_{n,L}^{naive,1} \ := \ 
- \frac{1}{n}\log \frac{1}{L}\sum_{\ell =1}^{L} 
\textfrak{1}_{
\textrm{$\boldsymbol{\Omega}$\hspace{-0.19cm}$\boldsymbol{\Omega}$}}\big(\boldsymbol{\xi}_{n}^{w\mathbf{W}^{(\ell)}}\big)
$
as well as the alternative estimator
$\Phi\Big( \widehat{\argmin_{\mathbf{Q} \in 
\textrm{$\boldsymbol{\Omega}$\hspace{-0.19cm}$\boldsymbol{\Omega}$}} \Phi(\mathbf{Q})}_{n,L}^{naive,1} \Big)$.
The involved estimator 
$\widehat{\argmin_{\mathbf{Q} \in 
\textrm{$\boldsymbol{\Omega}$\hspace{-0.19cm}$\boldsymbol{\Omega}$}} \Phi(\mathbf{Q})}_{n,L}^{naive,1}$
of the minimizer set $\mathcal{Q}^{\ast} := \argmin_{\mathbf{Q} \in 
\textrm{$\boldsymbol{\Omega}$\hspace{-0.19cm}$\boldsymbol{\Omega}$}} \Phi(\mathbf{Q})$
is --- at the same time (due to the strict increasingness of $F_{\gamma,\widetilde{c},1}$)  ---
also the estimator of the minimizer set $\mathcal{Q}^{\ast} := \argmin_{\mathbf{Q} \in 
\textrm{$\boldsymbol{\Omega}$\hspace{-0.19cm}$\boldsymbol{\Omega}$}} 
D_{\widetilde{c} \cdot \varphi_{\gamma}}(\mathbf{Q},\mathds{P})$;
giving the latter had been left as an \textit{open gap} in Broniatowski \& Stummer~\cite{Bro:23a},
which we have now \textit{filled/resolved}. Moreover, 
$D_{\widetilde{c} \cdot \varphi_{\gamma}}(
\widehat{\argmin_{\mathbf{Q} \in 
\textrm{$\boldsymbol{\Omega}$\hspace{-0.19cm}$\boldsymbol{\Omega}$}} \Phi(\mathbf{Q})}_{n,L}^{naive,1}
,\mathds{P})$
is an estimator of $\min_{\mathbf{Q} \in 
\textrm{$\boldsymbol{\Omega}$\hspace{-0.19cm}$\boldsymbol{\Omega}$}} 
D_{\widetilde{c} \cdot \varphi_{\gamma}}(\mathbf{Q},\mathds{P})$
which serves an alternative to the one given in Broniatowski \& Stummer~\cite{Bro:23a}.

\end{remark}


\subsection{Naive estimators of max and argmax --- Base-Divergence-Method 1, compact case}
\label{SectEstimators.new.det.simplex.meth1.compact.max}

\vspace{0.2cm}
\noindent
In the set-up of compact $A \cdot \textrm{$\boldsymbol{\Omega}$\hspace{-0.23cm}$\boldsymbol{\Omega}$}$ 
with \eqref{regularity simplex}, we can handle
the maximizing problem completely analogously to the method in the previous Subsection 
\ref{SectEstimators.new.det.simplex.meth1.compact.min}.
Indeed, by applying Theorem \ref{brostu5:thm.Fmax.simplex}(a) instead of 
Theorem \ref{brostu5:thm.Fmin.simplex}(a)
we construct the naive BS-estimator 
\begin{equation}
\widehat{\Phi(A \cdot \textrm{$\boldsymbol{\Omega}$\hspace{-0.23cm}$\boldsymbol{\Omega}$})}_{n,L}^{naive,1} \ := \ 
\frac{1}{n}\log \frac{1}{L}\sum_{\ell =1}^{L} 
\exp\negthinspace\Big(n \cdot \Big(
F_{\gamma,\widetilde{c},A}\Big(D_{\widetilde{c} \cdot \varphi_{\gamma}}(A \cdot 
\boldsymbol{\xi}_{n}^{w\mathbf{W}^{(\ell)}},\mathds{P})\Big)
+ \Phi\big(A \cdot \boldsymbol{\xi}_{n}^{w\mathbf{W}^{(\ell)}}\big)
\Big)
\Big)
\cdot \textfrak{1}_{
\textrm{$\boldsymbol{\Omega}$\hspace{-0.19cm}$\boldsymbol{\Omega}$}}\big(\boldsymbol{\xi}_{n}^{w\mathbf{W}^{(\ell)}}\big)
\, ,
\label{brostu5:fo.BSmax.simplex.naive.estim}
\end{equation}
of the maximum value $\Phi(A \cdot \textrm{$\boldsymbol{\Omega}$\hspace{-0.23cm}$\boldsymbol{\Omega}$}) 
:= \max_{\mathbf{Q}\in A \cdot \textrm{$\boldsymbol{\Omega}$\hspace{-0.19cm}$\boldsymbol{\Omega}$}} 
\Phi(\mathbf{Q})$;
for this, we get
\begin{equation}
\lim_{n\rightarrow \infty} 
\widehat{\Phi(A \cdot \textrm{$\boldsymbol{\Omega}$\hspace{-0.23cm}$\boldsymbol{\Omega}$})}_{n,\infty}^{naive,1}
\ = \ 
\lim_{n\rightarrow \infty} \lim_{L\rightarrow \infty} 
\widehat{\Phi(A \cdot \textrm{$\boldsymbol{\Omega}$\hspace{-0.23cm}$\boldsymbol{\Omega}$})}_{n,L}^{naive,1}
\ = \ \Phi(\mathbf{\Omega}) \qquad \textrm{a.s.}
\nonumber
\end{equation}
instead of \eqref{brostu5:fo.BSmin.extended.naive.estim.lim.simplex}
(notice the different definition of the involved quantities).
As the corresponding very natural naive BS-estimator of the 
maximizer $\argmax_{\mathbf{Q}\in A \cdot \textrm{$\boldsymbol{\Omega}$\hspace{-0.19cm}$\boldsymbol{\Omega}$}} 
\Phi(\mathbf{Q})$ 
we take 
\begin{equation}
\widehat{\argmax_{\mathbf{Q} \in 
A \cdot \textrm{$\boldsymbol{\Omega}$\hspace{-0.19cm}$\boldsymbol{\Omega}$}} \Phi(\mathbf{Q})}_{n,L}^{naive,1}
:= \argmax_{\boldsymbol{\nu} \in \mathcal{W}_{n,L}} \Phi(\boldsymbol{\nu})
\nonumber
\end{equation}
instead of 
\eqref{brostu5:fo.1555.simplex}; 
in short, 
we take any $A \cdot \boldsymbol{\xi}_{n}^{w\mathbf{W}^{L,\ast}}$
which maximizes $\Phi(\cdot)$ amongst all values 
$A \cdot \boldsymbol{\xi}_{n}^{w\mathbf{W}^{(\ell)}}$
at hand which fall into $A \cdot \textrm{$\boldsymbol{\Omega}$\hspace{-0.23cm}$\boldsymbol{\Omega}$}$. 
For this, we obtain --- instead of \eqref{minim simplex} ---
the assertion
\begin{equation}
\max_{\mathbf{Q}\in 
A \cdot \textrm{$\boldsymbol{\Omega}$\hspace{-0.19cm}$\boldsymbol{\Omega}$}
} \Phi(\mathbf{Q}) 
\ \geq \ 
\Phi \left( A \cdot \boldsymbol{\xi}_{n}^{w\mathbf{W}^{L,\ast}} \right)
\ \geq \ 
\widehat{\Phi(A \cdot \textrm{$\boldsymbol{\Omega}$\hspace{-0.23cm}$\boldsymbol{\Omega}$})}_{n,\infty}^{naive,1}
\ - \ o_{\mathbb{\Pi}}(1)
\nonumber
\end{equation}
where $o_{\mathbb{\Pi}}(1)$ goes to $0$ as $L\rightarrow \infty$ and $n\rightarrow \infty$ under the
distribution $\mathbb{\Pi}$.
Hence, as $L$ and $n$ tend to infinity,  
$A \cdot \boldsymbol{\xi}_{n}^{w\mathbf{W}^{L,\ast}}$
concentrates to the set of maximizers of 
$\Phi(\cdot)$ on $A \cdot \textrm{$\boldsymbol{\Omega}$\hspace{-0.23cm}$\boldsymbol{\Omega}$}$.

\vspace{0.2cm}

\begin{proposition}
In the above set-up, one has
\begin{equation}
\lim_{n\rightarrow \infty} \lim_{L\rightarrow \infty} 
\Phi\Big( \widehat{\argmax_{\mathbf{Q} \in 
A \cdot \textrm{$\boldsymbol{\Omega}$\hspace{-0.19cm}$\boldsymbol{\Omega}$}} \Phi(\mathbf{Q})}_{n,L}^{naive,1} \Big)
\ = \ \Phi(A \cdot \textrm{$\boldsymbol{\Omega}$\hspace{-0.23cm}$\boldsymbol{\Omega}$}) \qquad \textrm{a.s.},
\nonumber
\end{equation}
and thus the quantity 
$\Phi\Big( \widehat{\argmax_{\mathbf{Q} \in 
A \cdot \textrm{$\boldsymbol{\Omega}$\hspace{-0.19cm}$\boldsymbol{\Omega}$}} \Phi(\mathbf{Q})}_{n,L}^{naive,1} \Big)$
is a natural alternative to the estimate
$\widehat{\Phi(A \cdot \textrm{$\boldsymbol{\Omega}$\hspace{-0.23cm}$\boldsymbol{\Omega}$})}_{n,L}^{naive,1}$ 
given in 
\eqref{brostu5:fo.BSmax.simplex.naive.estim}.

\end{proposition}


\subsection{Naive estimators --- Base-Divergence-Method 1, non-compact case}
\label{SectEstimators.new.det.simplex.meth1.noncompact}

\vspace{0.2cm}
\noindent
Suppose that we are in the set-up of Theorem \ref{brostu5:thm.Fmin.simplex}(b),
which particularly means that $A \cdot \textrm{$\boldsymbol{\Omega}$\hspace{-0.23cm}$\boldsymbol{\Omega}$}$ 
is not necessarily compact
but satisfies \eqref{regularity simplex} and \eqref{def fi wrt Omega simplex},
and that $\Phi: A \cdot \textrm{$\boldsymbol{\Omega}$\hspace{-0.23cm}$\boldsymbol{\Omega}$} \mapsto \mathbb{R}$ 
is a continuous function
which satisfies the lower-bound condition \eqref{brostu5:fo.phibound.simplex}
(notice that \eqref{brostu5:fo.phibound.simplex} trivially holds if 
$A \cdot \textrm{$\boldsymbol{\Omega}$\hspace{-0.23cm}$\boldsymbol{\Omega}$}$ 
is bounded but not necessarily closed).
Additionally, let us now assume that the minimum value is achieved,
i.e. $\inf_{\mathbf{Q}\in A \cdot \textrm{$\boldsymbol{\Omega}$\hspace{-0.19cm}$\boldsymbol{\Omega}$}} 
\Phi(\mathbf{Q}) = 
\min_{\mathbf{Q}\in A \cdot \textrm{$\boldsymbol{\Omega}$\hspace{-0.19cm}$\boldsymbol{\Omega}$}} 
\Phi(\mathbf{Q}) = \Phi(\mathbf{Q}_{min})$
for some (not necessarily unique) point $\mathbf{Q}_{min} 
\in A \cdot \textrm{$\boldsymbol{\Omega}$\hspace{-0.23cm}$\boldsymbol{\Omega}$}$,
and that the corresponding set 
$\mathcal{Q}^{\ast} := \argmin_{\mathbf{Q} \in
A \cdot \textrm{$\boldsymbol{\Omega}$\hspace{-0.19cm}$\boldsymbol{\Omega}$}} \Phi(\mathbf{Q})$
of minimizers is covered by a compact set $\mathbf{B} \in A \cdot \mathbb{S}^{K}$
(e.g. we take $\mathbf{B}:= cl(A \cdot \textrm{$\boldsymbol{\Omega}$\hspace{-0.23cm}$\boldsymbol{\Omega}$})$ 
in case that $A \cdot \textrm{$\boldsymbol{\Omega}$\hspace{-0.23cm}$\boldsymbol{\Omega}$}$ 
is bounded but not necessarily closed). 
In such a context, we can proceed as in
Subsection \ref{SectEstimators.new.det.simplex.meth1.compact.min} 
by replacing $A \cdot \textrm{$\boldsymbol{\Omega}$\hspace{-0.23cm}$\boldsymbol{\Omega}$}$ 
with $A \cdot \textrm{$\boldsymbol{\Omega}$\hspace{-0.23cm}$\boldsymbol{\Omega}$} \cap \mathbf{B}$.
The analogous procedure applies to the maximization problem,
by proceeding as in
Subsection \ref{SectEstimators.new.det.simplex.meth1.compact.max} 
by replacing $A \cdot \textrm{$\boldsymbol{\Omega}$\hspace{-0.23cm}$\boldsymbol{\Omega}$}$ with 
$A \cdot \textrm{$\boldsymbol{\Omega}$\hspace{-0.23cm}$\boldsymbol{\Omega}$} \cap \mathbf{B}$.


\subsection{Naive estimators of min and argmin --- Base-Divergence-Method 2, compact case}
\label{SectEstimators.new.det.simplex.meth2.compact.min}

\vspace{0.2cm}
\noindent
In the above Subsections 
\ref{SectEstimators.new.det.simplex.meth1.compact.min},
\ref{SectEstimators.new.det.simplex.meth1.compact.max},
\ref{SectEstimators.new.det.simplex.meth1.noncompact}
we have chosen the divergence
$F_{\gamma,\widetilde{c},A}\Big(D_{\widetilde{c} \cdot \varphi_{\gamma}}(\cdot,\mathds{P})\Big)$
as the \textit{base divergence}
(cf. Theorem \ref{brostu5:thm.Fmin.simplex} and Theorem \ref{brostu5:thm.Fmax.simplex});
we have referred to this choice as \textit{Base-Divergence-Method 1}.
However, as can be seen from the alternative Theorem \ref{brostu5:thm.Fmin.simplex.SBD}
and Theorem \ref{brostu5:thm.Fmax.simplex.SBD}, we can also choose 
--- as Base-Divergence-Method 2 --- as \textit{base divergence} 
an appropriate ``innmin scaled Bregman distance''
$\breve{D}_{\widetilde{c} \cdot 
\varphi_{\gamma},\mathbf{P}}^{SBD}\negthinspace\left(\cdot,\mathbf{Q}^{\ast\ast}\right)$ 
(cf. \eqref{brostu3:fo.676b.SBD},\eqref{brostu3:fo.677a.SBD},\eqref{brostu3:fo.678.SBD}),
where $\mathbf{Q}^{\ast\ast} \in \mathbb{R}_{>0}^{K}$ 
\textit{NEED NOT} be in $A \cdot \textrm{$\boldsymbol{\Omega}$\hspace{-0.23cm}$\boldsymbol{\Omega}$}$.

\vspace{0.2cm}
\noindent
Let us start with any fixed $\mathbf{Q}^{\ast\ast} \in \mathbb{R}_{>0}^{K}$ 
and any fixed $\mathbf{P} \in \mathbb{R}_{>0}^{K}$ 
(with $M_{\mathbf{P}} =\sum_{i=1}^{K}p_{i}>0$) satisfying
\begin{equation}
t_{k}^{\ast\ast} :=  \frac{q_{k}^{\ast\ast}}{p_{k}} \in \, ]t_{-}^{sc},t_{+}^{sc}[  
\quad \textrm{for all $k =1,\ldots,K$} 
\qquad \textrm{(cf. \eqref{brostu5:fo.SBD.qstarstar})}.
\nonumber
\end{equation}
In such a setup, recall from Theorem \ref{brostu5:thm.Fmin.simplex.SBD}(a) that
we have obtained for any continuous function $\Phi: \mathbf{\Omega} \mapsto \mathbb{R}$ on a compact
set $A \cdot \textrm{$\boldsymbol{\Omega}$\hspace{-0.23cm}$\boldsymbol{\Omega}$}
\subset \widetilde{\mathcal{M}}_{\gamma}$ ($\gamma \in \mathbb{R}\backslash\, ]1,2[$)
with \eqref{regularity simplex} the assertion 
\begin{equation}
\min_{\mathbf{Q}\in A \cdot \textrm{$\boldsymbol{\Omega}$\hspace{-0.19cm}$\boldsymbol{\Omega}$}} 
\Phi(\mathbf{Q})
\ = \ - \, 
\lim_{n\rightarrow \infty }\frac{1}{n}\log \negthinspace \left( \ 
\mathbb{E}_{\mathbb{\Pi}}\negthinspace \Big[
\exp\negthinspace\Big(n \cdot \Big(
\breve{D}_{\widetilde{c} \cdot \varphi_{\gamma},\mathbf{P}}^{SBD}\negthinspace\left(A 
\cdot \boldsymbol{\xi}_{n}^{w\mathbf{\widetilde{V}}},\mathbf{Q}^{\ast\ast}\right)  
- \Phi\big(A \cdot \boldsymbol{\xi}_{n}^{w\mathbf{\widetilde{V}}}\big)
\Big)
\Big)
\cdot \textfrak{1}_{\textrm{$\boldsymbol{\Omega}$\hspace{-0.19cm}$\boldsymbol{\Omega}$}}
\big(\boldsymbol{\xi}_{n}^{w\mathbf{\widetilde{V}}}\big)
\, \Big] 
\right)
\, ,
\label{brostu5:fo.BSmin.extended.min.SBD.new.simplex}
\end{equation}
where $\boldsymbol{\xi}_{n}^{w\mathbf{\widetilde{V}}}$ is as in
\eqref{brostu5:fo.norweiemp.vec.det.SBD} with $\widetilde{V}$ instead of $V$
(where $\widetilde{V}$ is constructed via \eqref{brostu5:V_new} and \eqref{brostu5:Utilde_k_new}
with the special choice $\widetilde{\varphi}(t) := M_{\mathbf{P}} \cdot
\widetilde{c} \cdot \varphi_{\gamma}(t)$).
Within such a set-up, we thus obtain for large $n \in \mathbb{N}$ 
(cf. \eqref{brostu5:fo.BSmin.extended.min.SBD.new.simplex}) 
the approximation
\begin{equation}
\Phi(A \cdot \textrm{$\boldsymbol{\Omega}$\hspace{-0.23cm}$\boldsymbol{\Omega}$}) := 
\min_{\mathbf{Q}\in A \cdot \textrm{$\boldsymbol{\Omega}$\hspace{-0.19cm}$\boldsymbol{\Omega}$}} 
\Phi(\mathbf{Q})
\ \approx \ - \, 
\frac{1}{n}\log \negthinspace \left( \ 
\mathbb{E}_{\mathbb{\Pi}}\negthinspace \Big[
\exp\negthinspace\Big(n \cdot \Big(
\breve{D}_{\widetilde{c} \cdot \varphi_{\gamma},\mathbf{P}}^{SBD}\negthinspace\left(A 
\cdot \boldsymbol{\xi }_{n}^{w\mathbf{\widetilde{V}}},\mathbf{Q}^{\ast\ast}\right)  
- \Phi\big(A \cdot \boldsymbol{\xi }_{n}^{w\mathbf{\widetilde{V}}}\big)
\Big)
\Big)
\cdot \textfrak{1}_{\textrm{$\boldsymbol{\Omega}$\hspace{-0.19cm}$\boldsymbol{\Omega}$}
}
\big(\boldsymbol{\xi }_{n}^{w\mathbf{\widetilde{V}}}\big)
\, \Big] 
\right)
\, 
\label{brostu5:fo.BSmin.extended.min.SBD.approx.new.simplex}
\end{equation}
and hencefor getting an estimator of the minimum value 
$\Phi(A \cdot \textrm{$\boldsymbol{\Omega}$\hspace{-0.23cm}$\boldsymbol{\Omega}$})$
one can estimate the right-hand side of \eqref{brostu5:fo.BSmin.extended.min.SBD.approx.new.simplex}.
To achieve this, for the rest of this section we assume 
that $n$ is chosen such that all 
$n \cdot \frac{p_{k}}{M_{\mathbf{P}}}$ are integers
(and hence, $n = \sum_{k=1}^{K} n_{k}$ with $n_{k} = n \cdot \frac{p_{k}}{M_{\mathbf{P}}}$) --- the remaining case works analogously. With this, we construct the 
corresponding \textit{naive estimator} 
for the minimum value 
$\min_{\mathbf{Q}\in A \cdot \textrm{$\boldsymbol{\Omega}$\hspace{-0.19cm}$\boldsymbol{\Omega}$}} 
\Phi(\mathbf{Q})$ as 
\begin{equation}
\widehat{\Phi(A \cdot \textrm{$\boldsymbol{\Omega}$\hspace{-0.23cm}$\boldsymbol{\Omega}$})}_{n,L}^{naive,2}\ := \ 
- \frac{1}{n}\log \frac{1}{L}\sum_{\ell =1}^{L} 
\exp\negthinspace\Big(n \cdot \Big(
\breve{D}_{\widetilde{c} \cdot \varphi_{\gamma},\mathbf{P}}^{SBD}\negthinspace\left(A 
\cdot \boldsymbol{\xi}_{n}^{w\mathbf{\widetilde{V}}^{(\ell)}},\mathbf{Q}^{\ast\ast}\right) 
- \Phi\big(A \cdot \boldsymbol{\xi}_{n}^{w\mathbf{\widetilde{V}}^{(\ell)}}\big)
\Big)
\Big)
\cdot \textfrak{1}_{\textrm{$\boldsymbol{\Omega}$\hspace{-0.19cm}$\boldsymbol{\Omega}$}}
\big(\boldsymbol{\xi}_{n}^{w\mathbf{\widetilde{V}}^{(\ell)}}\big)
\, ,
\label{brostu5:fo.BSmin.extended.improved.estim.new.simplex}
\end{equation}
where we simulate independently $L$ copies 
$\mathbf{\widetilde{V}}^{(1)},\ldots,\mathbf{\widetilde{V}}^{(L)}$ of the random vector
$\mathbf{\widetilde{V}}:=\left( \widetilde{V}_{1},\ldots,\widetilde{V}_{n}\right)$ ,
and compute each of $\boldsymbol{\xi}_{n}^{w\mathbf{\widetilde{V}}^{(1)}}, \ldots,
\boldsymbol{\xi}_{n}^{w\mathbf{\widetilde{V}}^{(L)}}$ according to 
\eqref{brostu5:fo.norweiemp.vec.det.SBD} (with $\widetilde{V}$ instead of $V$).
Clearly, with the help of the strong law of large numbers we get with 
$\widehat{\Phi(A \cdot \textrm{$\boldsymbol{\Omega}$\hspace{-0.23cm}$\boldsymbol{\Omega}$}
)}_{n,\infty}^{naive,2}: = 
\lim_{L\rightarrow \infty} \widehat{\Phi(A \cdot \textrm{$\boldsymbol{\Omega}$\hspace{-0.23cm}$\boldsymbol{\Omega}$}
)}_{n,L}^{naive,2}$
the following assertion:

\vspace{0.2cm}

\begin{proposition}
Let the assumptions of Theorem \ref{brostu5:thm.Fmin.simplex.SBD}(a) be satisfied. Then one has
\begin{equation}
\lim_{n\rightarrow \infty} 
\widehat{\Phi(A \cdot \textrm{$\boldsymbol{\Omega}$\hspace{-0.23cm}$\boldsymbol{\Omega}$}
)}_{n,\infty}^{naive,2}
\ = \ 
\lim_{n\rightarrow \infty} \lim_{L\rightarrow \infty} 
\widehat{\Phi(A \cdot \textrm{$\boldsymbol{\Omega}$\hspace{-0.23cm}$\boldsymbol{\Omega}$}
)}_{n,L}^{naive,2}
\ = \ \Phi(A \cdot \textrm{$\boldsymbol{\Omega}$\hspace{-0.23cm}$\boldsymbol{\Omega}$}
) \qquad \textrm{a.s.}
\label{brostu5:fo.BSmin.extended.improved.estim.lim.new.simplex}
\end{equation}
\end{proposition}

\vspace{0.2cm}
\noindent
As the corresponding \textit{very natural naive BS-estimator 
of the minimizer-set} 
$\mathcal{Q}^{\ast} := \argmin_{\mathbf{Q} \in 
A \cdot \textrm{$\boldsymbol{\Omega}$\hspace{-0.19cm}$\boldsymbol{\Omega}$}
} \Phi(\mathbf{Q})$, 
we take 
\begin{equation}
\widehat{\argmin_{\mathbf{Q} \in 
A \cdot \textrm{$\boldsymbol{\Omega}$\hspace{-0.19cm}$\boldsymbol{\Omega}$}} 
\Phi(\mathbf{Q})}_{n,L}^{naive,2}
:= \argmin_{\boldsymbol{\nu} \in \mathcal{V}_{n,L}} \Phi(\boldsymbol{\nu})
\label{brostu5:fo.1555.improved.new.simplex}
\end{equation}
where $\mathcal{V}_{n,L} := \{A \cdot \boldsymbol{\xi}_{n}^{w\mathbf{\widetilde{V}}^{(\ell)}}:
\ell \in \{1,\ldots,L\} \,  \} \cap A \cdot \textrm{$\boldsymbol{\Omega}$\hspace{-0.23cm}$\boldsymbol{\Omega}$}$. 
In short, we take as minimizer-estimate
the (not necessarily unique) element 
$A \cdot \boldsymbol{\xi}_{n}^{w\mathbf{\widetilde{V}}^{L,\ast}}$
which minimizes $\Phi(\cdot)$ amongst all values 
$A \cdot \boldsymbol{\xi}_{n}^{w\mathbf{\widetilde{V}}^{(\ell)}}$
at hand which fall into $A \cdot \textrm{$\boldsymbol{\Omega}$\hspace{-0.23cm}$\boldsymbol{\Omega}$}$. 
We prove that if $L$ and $n$ tend to infinity, then 
$A \cdot \boldsymbol{\xi}_{n}^{\mathbf{w\widetilde{V}}^{L,\ast}}$
concentrates to the above-mentioned set $\mathcal{Q}^{\ast}$ 
of minimizers of $\Phi(\cdot)$ on $A \cdot \textrm{$\boldsymbol{\Omega}$\hspace{-0.23cm}$\boldsymbol{\Omega}$}$.
Indeed, we show that $A \cdot \boldsymbol{\xi}_{n}^{w\mathbf{\widetilde{V}}^{L,\ast}}$ is
a proxy minimizer of $\Phi(\cdot)$ on $A \cdot \textrm{$\boldsymbol{\Omega}$\hspace{-0.23cm}$\boldsymbol{\Omega}$}$, by proving the following

\vspace{0.2cm}

\begin{proposition}
\label{brostu5:prop.generaldeterministic.minimizer.improved.new.simplex}
There holds
\begin{equation}
\min_{\mathbf{Q}\in A \cdot \textrm{$\boldsymbol{\Omega}$\hspace{-0.19cm}$\boldsymbol{\Omega}$}} \Phi(\mathbf{Q}) 
\ \leq \ 
\Phi \left( A \cdot \boldsymbol{\xi}_{n}^{w\mathbf{\widetilde{V}}^{L,\ast}} \right)
\ \leq \ 
\widehat{\Phi(A \cdot \textrm{$\boldsymbol{\Omega}$\hspace{-0.23cm}$\boldsymbol{\Omega}$})}_{n,\infty}^{naive,2}
\ + \ o_{\mathbb{\Pi}}(1)
\label{minim.improved.new.simplex}
\end{equation}
where $o_{\mathbb{\Pi}}(1)$ goes to $0$ as 
$L\rightarrow \infty$ and $n\rightarrow \infty$ under the
distribution $\mathbb{\Pi}$
(recall that $\mathbb{\Pi}[(\widetilde{V}_{1},\ldots,\widetilde{V}_{n}) \in \cdot \,]=
\bigotimes\limits_{k=1}^{K}\widetilde{U}_{k}^{\otimes n_{k}}[\,\cdot \,]$)
\end{proposition}

\vspace{0.4cm}
\noindent
The proof of Proposition \ref{brostu5:prop.generaldeterministic.minimizer.improved.new.simplex}
works analogously to the proof of
Proposition \ref{brostu5:prop.generaldeterministic.minimizer.naive},
by replacing Theorem \ref{brostu5:thm.Fmin}(a) with Theorem \ref{brostu5:thm.Fmin.simplex.SBD}(a).

\vspace{0.4cm}

\begin{proposition}
In the above set-up, one has
\begin{equation}
\lim_{n\rightarrow \infty} \lim_{L\rightarrow \infty} 
\Phi\Big( \widehat{\argmin_{\mathbf{Q} \in 
A \cdot \textrm{$\boldsymbol{\Omega}$\hspace{-0.19cm}$\boldsymbol{\Omega}$}} \Phi(\mathbf{Q})}_{n,L}^{naive,2} \Big)
\ = \ \Phi(A \cdot \textrm{$\boldsymbol{\Omega}$\hspace{-0.23cm}$\boldsymbol{\Omega}$}) \qquad \textrm{a.s.},
\nonumber
\end{equation}
and thus the quantity 
$\Phi\Big( \widehat{\argmin_{\mathbf{Q} \in 
A \cdot \textrm{$\boldsymbol{\Omega}$\hspace{-0.19cm}$\boldsymbol{\Omega}$}} \Phi(\mathbf{Q})}_{n,L}^{naive,2} \Big)$
is a natural alternative to the estimate
$\widehat{\Phi(A \cdot \textrm{$\boldsymbol{\Omega}$\hspace{-0.23cm}$\boldsymbol{\Omega}$})}_{n,L}^{naive,2}$ 
given in 
\eqref{brostu5:fo.BSmin.extended.improved.estim.new.simplex}.

\end{proposition}


\subsection{Naive estimators of max and argmax --- Base-Divergence-Method 2, compact case}
\label{SectEstimators.new.det.simplex.meth2.compact.max}

\vspace{0.2cm}
\noindent
\vspace{0.2cm}
\noindent
In the set-up of compact constraint set 
$A \cdot \textrm{$\boldsymbol{\Omega}$\hspace{-0.23cm}$\boldsymbol{\Omega}$}
\subset \widetilde{\mathcal{M}}_{\gamma}$ ($\gamma \in \mathbb{R}\backslash\, ]1,2[$)
with \eqref{regularity simplex}, we can treat
the maximizing problem completely analogously to the method in 
the previous Subsection \ref{SectEstimators.new.det.simplex.meth2.compact.min}.
Indeed, by employing Theorem \ref{brostu5:thm.Fmax.simplex.SBD}(a) instead of 
Theorem \ref{brostu5:thm.Fmin.simplex.SBD}(a)
we construct the naive BS-estimator 
\begin{equation}
\widehat{\Phi(A \cdot \textrm{$\boldsymbol{\Omega}$\hspace{-0.23cm}$\boldsymbol{\Omega}$})}_{n,L}^{naive,2}\ := \ 
\frac{1}{n}\log \frac{1}{L}\sum_{\ell =1}^{L} 
\exp\negthinspace\Big(n \cdot \Big(
\breve{D}_{\widetilde{c} \cdot \varphi_{\gamma},\mathbf{P}}^{SBD}\negthinspace\left(A 
\cdot \boldsymbol{\xi}_{n}^{w\mathbf{\widetilde{V}}^{(\ell)}},\mathbf{Q}^{\ast\ast}\right) 
+ \Phi\big(A \cdot \boldsymbol{\xi }_{n}^{\mathbf{w\widetilde{V}}^{(\ell)}}\big)
\Big)
\Big)
\cdot \textfrak{1}_{\textrm{$\boldsymbol{\Omega}$\hspace{-0.19cm}$\boldsymbol{\Omega}$}}
\big(\boldsymbol{\xi}_{n}^{w\mathbf{\widetilde{V}}^{(\ell)}}\big)
\, ,
\label{brostu5:fo.BSmax.extended.improved.estim.new.simplex}
\end{equation}
of the maximum value $\Phi(A \cdot \textrm{$\boldsymbol{\Omega}$\hspace{-0.23cm}$\boldsymbol{\Omega}$})
 := \max_{\mathbf{Q}\in A \cdot \textrm{$\boldsymbol{\Omega}$\hspace{-0.19cm}$\boldsymbol{\Omega}$}} 
\Phi(\mathbf{Q})$;
for this, we get
\begin{equation}
\lim_{n\rightarrow \infty} 
\widehat{\Phi(A \cdot \textrm{$\boldsymbol{\Omega}$\hspace{-0.23cm}$\boldsymbol{\Omega}$})}_{n,\infty}^{naive,2}
\ = \ 
\lim_{n\rightarrow \infty} \lim_{L\rightarrow \infty} 
\widehat{\Phi(A \cdot \textrm{$\boldsymbol{\Omega}$\hspace{-0.23cm}$\boldsymbol{\Omega}$})}_{n,L}^{naive,2}
\ = \ \Phi(A \cdot \textrm{$\boldsymbol{\Omega}$\hspace{-0.23cm}$\boldsymbol{\Omega}$}) \qquad \textrm{a.s.}
\nonumber
\end{equation}
instead of 
\eqref{brostu5:fo.BSmin.extended.improved.estim.lim.new.simplex}
(notice the different definition of the involved quantities).
As the corresponding very natural naive BS-estimator of the 
maximizer $\argmax_{\mathbf{Q} \in A \cdot 
\textrm{$\boldsymbol{\Omega}$\hspace{-0.23cm}$\boldsymbol{\Omega}$}} \Phi(\mathbf{Q})$
we take 
\begin{equation}
\widehat{\argmax_{\mathbf{Q} \in A \cdot \textrm{$\boldsymbol{\Omega}$\hspace{-0.23cm}$\boldsymbol{\Omega}$}} 
\Phi(\mathbf{Q})}_{n,L}^{naive,2}
:= \argmax_{\boldsymbol{\nu} \in \mathcal{V}_{n,L}} \Phi(\boldsymbol{\nu})
\nonumber
\end{equation}
instead of 
\eqref{brostu5:fo.1555.improved.new.simplex}; in short, 
we take any $A \cdot \boldsymbol{\xi}_{n}^{\mathbf{w\widetilde{V}}^{L,\ast}}$
which maximizes $\Phi(\cdot)$ amongst all values 
$A \cdot \boldsymbol{\xi}_{n}^{w\mathbf{\widetilde{V}}^{(\ell)}}$
at hand which fall into $A \cdot \textrm{$\boldsymbol{\Omega}$\hspace{-0.23cm}$\boldsymbol{\Omega}$}$. 
For this, we obtain --- instead of 
\eqref{minim.improved.new.simplex} ---
the assertion
\begin{equation}
\max_{\mathbf{Q}\in A \cdot \textrm{$\boldsymbol{\Omega}$\hspace{-0.19cm}$\boldsymbol{\Omega}$}} \Phi(\mathbf{Q}) 
\ \geq \ 
\Phi \left( A \cdot \boldsymbol{\xi}_{n}^{w\mathbf{\widetilde{V}}^{L,\ast}} \right)
\ \geq \ 
\widehat{\Phi(A \cdot \textrm{$\boldsymbol{\Omega}$\hspace{-0.23cm}$\boldsymbol{\Omega}$})}_{n,\infty}^{naive,2}
\ - \ o_{\mathbb{\Pi}}(1)
\nonumber
\end{equation}
where $o_{\mathbb{\Pi}}(1)$ goes to $0$ as $L\rightarrow \infty$ and $n\rightarrow \infty$ under the
distribution $\mathbb{\Pi}$.
Hence, as $L$ and $n$ tend to infinity,  
$A \cdot \boldsymbol{\xi}_{n}^{w\mathbf{\widetilde{V}}^{L,\ast}}$
concentrates to the set of maximizers of 
$\Phi(\cdot)$ on $A \cdot \textrm{$\boldsymbol{\Omega}$\hspace{-0.23cm}$\boldsymbol{\Omega}$}$.

\vspace{0.4cm}

\begin{proposition}
In the above set-up, one has
\begin{equation}
\lim_{n\rightarrow \infty} \lim_{L\rightarrow \infty} 
\Phi\Big( \widehat{\argmax_{\mathbf{Q} \in 
A \cdot \textrm{$\boldsymbol{\Omega}$\hspace{-0.19cm}$\boldsymbol{\Omega}$}} \Phi(\mathbf{Q})}_{n,L}^{naive,2} \Big)
\ = \ \Phi(A \cdot \textrm{$\boldsymbol{\Omega}$\hspace{-0.23cm}$\boldsymbol{\Omega}$}) \qquad \textrm{a.s.},
\nonumber
\end{equation}
and thus the quantity 
$\Phi\Big( \widehat{\argmax_{\mathbf{Q} \in 
A \cdot \textrm{$\boldsymbol{\Omega}$\hspace{-0.19cm}$\boldsymbol{\Omega}$}} \Phi(\mathbf{Q})}_{n,L}^{naive,2} \Big)$
is a natural alternative to the estimate
$\widehat{\Phi(A \cdot \textrm{$\boldsymbol{\Omega}$\hspace{-0.23cm}$\boldsymbol{\Omega}$})}_{n,L}^{naive,2}$ 
given in \eqref{brostu5:fo.BSmax.extended.improved.estim.new.simplex}.

\end{proposition}


\subsection{Naive estimators --- Base-Divergence-Method 2, non-compact case}

\vspace{0.2cm}
\noindent
We can proceed analogously to the above Base-Divergence-Method-1-treating 
Subsection \ref{SectEstimators.new.det.simplex.meth1.noncompact}
(e.g. by employing Theorem 
\ref{brostu5:thm.Fmin.simplex.SBD}(b)).


\subsection{Improved/Speed-up estimators 
of min, argmin, max and argmax --- compact case}
\label{SectEstimators.new.det.simplex.improved.compact}

\vspace{0.2cm}
\noindent
Recall from Subsection \ref{SectEstimators.new.det.nonvoid.improved.compact}.1)
that --- for the case of compact constraint set $\mathbf{\Omega}$ with \eqref{regularity} ---
we have constructed improved/speed-up estimators of 
$\Phi(\mathbf{\Omega}) := \min_{\mathbf{Q}\in \mathbf{\Omega}} \Phi(\mathbf{Q})$
and the corresponding minimizers $\argmin_{\mathbf{Q}\in \mathbf{\Omega}} \Phi(\mathbf{Q})$
by switching from the base-divergence 
$D_{\varphi}\big(M_{\mathbf{P}} \cdot \boldsymbol{\xi}_{n}^{\mathbf{\widetilde{W}}},\mathbf{P}\big)$
to the different base-divergence 
$D_{\varphi,\mathbf{P}}^{SBD}\negthinspace\left(M_{\mathbf{P}} 
\cdot \boldsymbol{\xi}_{n}^{\mathbf{\widetilde{V}}},\mathbf{Q}^{\ast}\right)$
where $\mathbf{Q}^{\ast}$ is chosen to be in $int(\mathbf{\Omega})$
such that by construction the involved random vectors $\boldsymbol{\xi}_{n}^{\mathbf{\widetilde{V}}}$ 
satisfy 
$\mathbb{E}_{\mathbb{\Pi}}\negthinspace \Big[M_{\mathbf{P}} \cdot \boldsymbol{\xi}_{n}^{\mathbf{\widetilde{V}}}\Big]
\in int(\mathbf{\Omega})$ for all large enough $n$
and even $\lim_{n\rightarrow \infty} M_{\mathbf{P}} \cdot\boldsymbol{\xi}_{n}^{\mathbf{\widetilde{V}}}
= \mathbf{Q}^{\ast}$ a.s. 
For the current set-up of compact constraint sets 
$A \cdot \textrm{$\boldsymbol{\Omega}$\hspace{-0.23cm}$\boldsymbol{\Omega}$}$
--- with fixed constant-component-sum $A \in \, ]0,\infty[$  
and $\textrm{$\boldsymbol{\Omega}$\hspace{-0.23cm}$\boldsymbol{\Omega}$} \in \mathbb{S}^{K}$
(respectively $\mathbb{S}_{>0}^{K}$)  ---
with regularity assumptions \eqref{regularity simplex}, we can proceed similarly.
Indeed, for achieving improved/speed-up estimators 
of the \textit{minimum value}
$\inf_{\mathbf{Q}\in A \cdot \textrm{$\boldsymbol{\Omega}$\hspace{-0.19cm}$\boldsymbol{\Omega}$}}\Phi (\mathbf{Q}) 
= \min_{\mathbf{Q}\in A \cdot \textrm{$\boldsymbol{\Omega}$\hspace{-0.19cm}$\boldsymbol{\Omega}$}} \Phi (\mathbf{Q})$ and of the corresponding
(set of) \textit{minimizers} $\arg \inf_{\mathbf{Q}\in A \cdot 
\textrm{$\boldsymbol{\Omega}$\hspace{-0.19cm}$\boldsymbol{\Omega}$}}\Phi (\mathbf{Q}) 
=\arg \min_{\mathbf{Q}\in A \cdot \textrm{$\boldsymbol{\Omega}$\hspace{-0.19cm}$\boldsymbol{\Omega}$}} \Phi (\mathbf{Q})$,
we switch (say) from the
base-divergence
$F_{\gamma,\widetilde{c},A}\Big(D_{\widetilde{c} 
\cdot \varphi_{\gamma}}(A \cdot \boldsymbol{\xi}_{n}^{w\mathbf{W}},\mathds{P})\Big)$
to the different base-divergence 
$\breve{D}_{\widetilde{c} \cdot \varphi_{\gamma},\mathbf{P}}^{SBD}\negthinspace\left(A 
\cdot \boldsymbol{\xi}_{n}^{w\mathbf{\widetilde{V}}},\mathbf{Q}^{\ast\ast}\right)$
where $\mathbf{Q}^{\ast\ast}$
should be chosen in an appropriate way, 
e.g. as follows:
suppose that we have at hand some point $\mathbf{Q}^{\ast} \in 
int(A \cdot \textrm{$\boldsymbol{\Omega}$\hspace{-0.23cm}$\boldsymbol{\Omega}$})$;
for pregiven $\mathbf{P} \in \mathbb{R}_{>0}^{K}$, such a vector
$\mathbf{Q}^{\ast}$ may be either pregiven (e.g. by the nature of the application context)
or it may be simulatively achieved by e.g. proxy method $1$ or proxy method $2$
of Subsection X-A of Broniatowski \& Stummer \cite{Bro:23a}.
From this, we construct $\mathbf{Q}^{\ast\ast} := \mathbf{Q}^{\ast}$.
Also recall \eqref{brostu5:fo.SBD.smooth.equality} with the setting
$\widetilde{\varphi} := M_{\mathbf{P}} \cdot \varphi$, 
$\widetilde{\mathds{P}}:=\mathbf{P}/M_{\mathbf{P}}$,  
$\widetilde{\mathbf{Q}}:=\mathbf{Q}/M_{\mathbf{P}}$
and $\widetilde{\mathbf{Q}}^{\ast\ast}:= \mathbf{Q}^{\ast\ast}/M_{\mathbf{P}}$.
Clearly, $\frac{\widetilde{q}_{k}^{\ast\ast}}{\widetilde{p}_{k}} 
= \frac{q_{k}^{\ast\ast}}{p_{k}} = 
\frac{q_{k}^{\ast}}{p_{k}}
\in \, ]t_{-}^{sc},t_{+}^{sc}[$ 
for all $k =1,\ldots,K$
(recall from the first six rows in Table 1 
that $]t_{-}^{sc},t_{+}^{sc}[ \, = \, ]0,\infty[$ 
for $\gamma \in \mathbb{R}\backslash\{2\}$
and $]t_{-}^{sc},t_{+}^{sc}[\, = \, ]-\infty,\infty[$ for $\gamma =2$)
and thus \eqref{brostu5:fo.SBD.qstarstar} is satisfied.
Accordingly, by the law of large numbers, the unweighted random vector $\boldsymbol{\xi}_{n}^{\mathbf{\widetilde{V}}}$
(cf. \eqref{Xi_n^W vector V new2}) 
performs the a.s. convergence
\begin{eqnarray}
\boldsymbol{\xi}_{n}^{\mathbf{\widetilde{V}}}
&:=& \Big(
\frac{1}{n} \cdot \sum_{i\in I_{1}^{(n)}}\widetilde{V}_{i} , \, 
\ldots , \, 
\frac{1}{n} \cdot \sum_{i\in I_{K-1}^{(n)}}\widetilde{V}_{i}, \, 
\frac{1}{n} \cdot \sum_{i\in I_{K}^{(n)}}\widetilde{V}_{i}\Big)
\nonumber \\
&=& \Big(
\frac{\lfloor n \cdot \widetilde{p}_{1}\rfloor}{n} \cdot \frac{1}{n_{1}} \cdot \sum_{i\in I_{1}^{(n)}}\widetilde{V}_{i}, \,
\ldots , \,
\frac{\lfloor n \cdot \widetilde{p}_{K-1}\rfloor}{n} \cdot \frac{1}{n_{K-1}} \cdot \sum_{i\in I_{K-1}^{(n)}}\widetilde{V}_{i}, \,
\frac{n-\sum_{k=1}^{K-1} \lfloor n \cdot \widetilde{p}_{k}\rfloor}{n} \frac{1}{n_{K}} 
\cdot \sum_{i\in I_{K}^{(n)}} \widetilde{V}_{i}\Big)
\nonumber\\
&\underset{n\rightarrow\infty}{\overset{a.s.}{\longrightarrow}}& 
\Big(\widetilde{q}_{1}^{\ast\ast}, \,  \ldots, \widetilde{q}_{K-1}^{\ast\ast}, \, 
\widetilde{q}_{K}^{\ast\ast}\Big) = \mathbf{\widetilde{Q}}^{\ast\ast} ,
\nonumber
\end{eqnarray}
where we have particularly used the fact that within each block $I_{k}^{(n)}$ all the $\widetilde{V}_{i}$ are independent
and identically distributed with
mean $E_{\mathbb{\Pi}}[\widetilde{V}_{i}]=\frac{\widetilde{q}_{k}^{\ast\ast}}{\widetilde{p}_{k}}$ and finite variance.
Hence, the corresponding total sum of components 
$M_{\boldsymbol{\xi}_{n}^{\mathbf{\widetilde{V}}}} = \frac{1}{n} \sum_{i=1}^{n}\widetilde{V}_{i}$ converges a.s.
(as $n \rightarrow \infty$)
to $M_{\mathbf{\widetilde{Q}}^{\ast\ast}} = \sum_{k=1}^{K} \widetilde{q}_{k}^{\ast\ast} = 
\frac{A}{M_{\mathbf{P}}}$.
Moreover, --- since all $\widetilde{q}_{k}^{\ast\ast}$ are strictly positive ---
the indicator $\textfrak{1}_{\mathbb{R}\backslash\{0\}}(\sum_{i=1}^{n}\widetilde{V}_{i})$ converges a.s. 
(as $n \rightarrow \infty$) to $1$
and, thus, $\textfrak{1}_{\{0\}}(\sum_{i=1}^{n}\widetilde{V}_{i})$ converges a.s. 
(as $n \rightarrow \infty$) to zero.
Consequently,\\
 $\boldsymbol{\xi}_{n}^{w\mathbf{\widetilde{V}}} =
\frac{\boldsymbol{\xi}_{n}^{\mathbf{\widetilde{V}}}}{\frac{1}{n} \sum_{i=1}^{n}\widetilde{V}_{i}}
\, \cdot \, \textfrak{1}_{\mathbb{R}\backslash\{0\}}(\sum_{i=1}^{n}\widetilde{V}_{i})
+ \boldsymbol{\infty} \, \cdot \,  \textfrak{1}_{\{0\}}(\sum_{i=1}^{n}\widetilde{V}_{i})$
converges a.s. (as $n \rightarrow \infty$) to $\frac{M_{\mathbf{P}}}{A} \cdot \mathbf{\widetilde{Q}}^{\ast\ast}
= \frac{1}{A} \cdot \mathbf{Q}^{\ast\ast}$.
Hence, we have derived the following

\vspace{0.2cm}

\begin{proposition}
\begin{equation}
\lim_{n\rightarrow \infty} 
\mathbb{\Pi}[A \cdot \boldsymbol{\xi}_{n}^{w\mathbf{\widetilde{V}}} \in 
A \cdot \textrm{$\boldsymbol{\Omega}$\hspace{-0.23cm}$\boldsymbol{\Omega}$}]
\ = \ 1 .
\nonumber
\end{equation}

\end{proposition}

\vspace{0.3cm}
\noindent
In other words, with the specific deliberate choice 
$\mathbf{Q}^{\ast\ast} := \mathbf{Q}^{\ast}
\in int(A \cdot \textrm{$\boldsymbol{\Omega}$\hspace{-0.23cm}$\boldsymbol{\Omega}$})$, 
we obtain --- for large enough approximation step $n$ --- the desired good hit-rate
needed for the construction of the improved/speed-up estimators.
For the latter we can now proceed, indeed, 
analogously to Subsection \ref{SectEstimators.new.det.nonvoid.improved.compact}.1),
by appropriately applying the results of
the Subsections \ref{SectEstimators.new.det.simplex.meth2.compact.min}
and \ref{SectEstimators.new.det.simplex.meth2.compact.max}
(instead of Subsections \ref{SectEstimators.new.det.nonvoid.meth2.compact.min} 
and \ref{SectEstimators.new.det.nonvoid.meth2.compact.max}).
For the sake of brevity, the details are omitted.


\section{Bare-Simulation Estimators For General Divergence-Optimization-Problems
Under Risk}
\label{SubsectEstimators.risk}

\vspace{0.2cm}
\noindent
Recall from Subsection \ref{SectDetGeneral.BDM1.min.risk.friends}
that we are interested in the constrained optimization of the \textit{continuous} 
distance-connected functions 
$\textrm{$\boldsymbol{\Omega}$\hspace{-0.23cm}$\boldsymbol{\Omega}$}
\ni \mathds{Q} \mapsto 
\Phi_{\mathbf{R}}(\mathds{Q})$
where $\mathbf{R}$ is an \textit{unknown} (say) vector-valued ``parameter'' 
(i.e. under risk). In continuation of 
Section \ref{SectStochSubsimplex.General}
we now give \textit{estimators}
of both the minimum value 
$\min_{\mathds{Q}\in \textrm{$\boldsymbol{\Omega}$\hspace{-0.19cm}$\boldsymbol{\Omega}$}} \Phi_{\mathbf{R}}(\mathbf{Q})$
and the corresponding minimizer(s) 
$\argmin_{\mathds{Q}\in \textrm{$\boldsymbol{\Omega}$\hspace{-0.19cm}$\boldsymbol{\Omega}$}} \Phi_{\mathbf{R}}(\mathbf{Q})$.
The case of maximum values and maximizers will be treated analogously. 
For the sake of brevity,
we only focus on compact constraint sets $\textrm{$\boldsymbol{\Omega}$\hspace{-0.23cm}$\boldsymbol{\Omega}$}
\subset \mathbb{S}^{K}$.


\subsection{Naive estimators of min and argmin --- Base-Divergence-Method 1, compact case}
\label{SectEstimators.new.det.risk.meth1.compact.min}

\vspace{0.2cm}
\noindent
Let the constraint set $\textrm{$\boldsymbol{\Omega}$\hspace{-0.23cm}$\boldsymbol{\Omega}$}$
be compact --- with \eqref{regularity simplex} ---
in the relative topology, and thus the minimum value of $\Phi_{\mathbf{R}}(\cdot)$
is achieved at some (not necessarily unique) point 
in $\textrm{$\boldsymbol{\Omega}$\hspace{-0.23cm}$\boldsymbol{\Omega}$}$. 
Applied to such a situation, \eqref{min.risk.4} and 
\eqref{min.risk.5a.new.sample}
give --- for the random sample $\mathbf{Y}_{1}^{n} = (Y_{1},\ldots,Y_{n})$
and the corresponding sequence
$(\mathbf{R}_{n}(\mathbf{Y}_{1}^{n}))_{n\in\mathbb{N}}$ of (say)
vector-valued function(al)s $\mathbf{R}_{n}(\mathbf{Y}_{1}^{n})$ for which 
$\lim_{n\rightarrow \infty }\mathbf{R}_{n}(\mathbf{Y}_{1}^{n})=\mathbf{R} \ \textrm{a.s.}$
(cf. \eqref{R-convergence})
the \textit{sample-dependent minimum}
\begin{align}
& \Phi_{\mathbf{R}_{n}(\mathbf{Y}_{1}^{n})}(\textrm{$\boldsymbol{\Omega}$\hspace{-0.23cm}$\boldsymbol{\Omega}$}) := 
\min_{\mathds{Q}\in \textrm{$\boldsymbol{\Omega}$\hspace{-0.19cm}$\boldsymbol{\Omega}$}} 
\Phi_{\mathbf{R}_{n}(\mathbf{Y}_{1}^{n})}(\mathds{Q}) 
\nonumber \\
& \ = \ -  \lim_{m\rightarrow \infty}
\frac{1}{m} \log \negthinspace \left( \ 
\mathbb{E}_{\mathbb{\Pi}_{\mathbf{Y}_{1}^{n}}
}
\negthinspace \Big[
\exp\negthinspace\Big(
m \cdot \Big(
F_{\gamma,\widetilde{c},1}\Big(D_{\widetilde{c} \cdot 
\varphi_{\gamma}}(\boldsymbol{\xi}_{m}^{w\mathbf{W}},\mathds{P}^{aux})\Big)
- \Phi_{\mathbf{R}_{n}(\mathbf{Y}_{1}^{n})}\big(\boldsymbol{\xi}_{m}^{w\mathbf{W}}\big)
\Big)
\Big)
\cdot \textfrak{1}_{
\textrm{$\boldsymbol{\Omega}$\hspace{-0.19cm}$\boldsymbol{\Omega}$}}\big(\boldsymbol{\xi}_{m}^{w\mathbf{W}}\big)
\, \Big] 
\right),
\ \label{min.risk.5b.sample.var.2}
\end{align}
where 
\begin{eqnarray}
\boldsymbol{\xi}_{m}^{w\mathbf{W}} &:=&
\begin{cases}
\left(\frac{\sum_{i \in I_{1}^{(m)}}W_{i}}{\sum_{k=1}^{K}\sum_{i \in I_{k}^{(m)}}W_{i}},
\ldots, \frac{\sum_{i \in I_{K}^{(m)}}W_{i}}{\sum_{k=1}^{K}\sum_{i \in I_{k}^{(m)}}W_{i}} \right) ,
\qquad \textrm{if } \sum_{j=1}^{m} W_{j} \ne 0, \\
\ (\infty, \ldots, \infty) =: \boldsymbol{\infty}, \hspace{4.0cm} \textrm{if } \sum_{j=1}^{m} W_{j} = 0,
\end{cases}
\qquad 
\textrm{(cf. \eqref{brostu5:fo.norweiemp.vec.det.m})}
\nonumber 
\end{eqnarray}
is the random vector constructed from 
(i) an auxiliary deterministic
probability vector $\mathds{P}^{aux}  \in \mathbb{S}_{> 0}^{K}$ inducing 
index blocks $I_{k}^{(m)}$ of sizes $m_{k} :=\lfloor m \cdot 
p_{k}^{aux}\rfloor$ ($k=1,\ldots,K-1$) and $m_{K} := m- \sum_{k=1}^{K-1} m_{k}$,
and (ii) from $W_{i}$'s which are i.i.d. copies of the random variable $W$
whose distribution 
is $\mathbb{\Pi }[W\in \cdot \,]=\mathbb{\bbzeta}[\,\cdot \,]$ 
being attributed to the power divergence generator  
$\varphi := \widetilde{c} \cdot \varphi_{\gamma}$ 
($\gamma \in \mathbb{R}\backslash\, ]1,2[$)
by the representability \eqref{brostu5:fo.link.var.simplex};
moreover, $(W_{i})_{i \in \mathbb{N}}$ and $(Y_{i})_{i \in \mathbb{N}}$ 
are supposed to be independent.
Recall from Remark \ref{brostu5:rem.Paux} that
the involved $\mathds{P}^{aux}$ may also depend on the 
sample $\mathbf{Y}_{1}^{n}$;
for the special case $\mathds{P}^{aux} := \mathds{P}_{n}^{emp}(\mathbf{Y}_{1}^{n})$,
we obtain $\boldsymbol{\xi}_{m}^{w\mathbf{W}} = \boldsymbol{\xi}_{n,m,\mathbf{Y}}^{w\mathbf{W}}$
(cf. \eqref{brostu5:fo.norweiemp.vec.risk.pnemp} respectively 
\eqref{brostu5:fo.norweiemp.vec.risk.pnemp.var}).

\vspace{0.2cm}
\noindent
From \eqref{min.risk.5b.sample.var.2}, we obtain for large $m$
the minimal-empirical-risk approximation (cf. \eqref{min.risk.5b.sample})
\begin{align}
& \Phi_{\mathbf{R}_{n}(\mathbf{Y}_{1}^{n})}(\textrm{$\boldsymbol{\Omega}$\hspace{-0.23cm}$\boldsymbol{\Omega}$}) = 
\min_{\mathds{Q}\in \textrm{$\boldsymbol{\Omega}$\hspace{-0.19cm}$\boldsymbol{\Omega}$}} 
\Phi_{\mathbf{R}_{n}(\mathbf{Y}_{1}^{n})}(\mathds{Q}) 
\nonumber \\
& \ \approx \ -  
\frac{1}{m} \log \negthinspace \left( \ 
\mathbb{E}_{\mathbb{\Pi}_{\mathbf{Y}_{1}^{n}}
}
\negthinspace \Big[
\exp\negthinspace\Big(
m \cdot \Big(
F_{\gamma,\widetilde{c},1}\Big(D_{\widetilde{c} \cdot 
\varphi_{\gamma}}(\boldsymbol{\xi}_{m}^{w\mathbf{W}},\mathds{P}^{aux})\Big)
- \Phi_{\mathbf{R}_{n}(\mathbf{Y}_{1}^{n})}\big(\boldsymbol{\xi}_{m}^{w\mathbf{W}}\big)
\Big)
\Big)
\cdot \textfrak{1}_{
\textrm{$\boldsymbol{\Omega}$\hspace{-0.19cm}$\boldsymbol{\Omega}$}}\big(\boldsymbol{\xi}_{m}^{w\mathbf{W}}\big)
\, \Big] 
\right). 
\label{min.risk.5b.sample.var}
\end{align}

\vspace{0.2cm}
\noindent
Hence for getting an estimator of the minimal empirical risk 
$\Phi_{\mathbf{R}_{n}(\mathbf{Y}_{1}^{n})}(\textrm{$\boldsymbol{\Omega}$\hspace{-0.23cm}$\boldsymbol{\Omega}$})
= \min_{\mathds{Q}\in \textrm{$\boldsymbol{\Omega}$\hspace{-0.19cm}$\boldsymbol{\Omega}$}} 
\Phi_{\mathbf{R}_{n}(\mathbf{Y}_{1}^{n})}(\mathds{Q})$
one can estimate the right-hand side of \eqref{min.risk.5b.sample.var}.

\vspace{0.2cm}
\noindent
To achieve this, for the rest of this section we assume 
that $m$ is chosen such that all 
$m \cdot p_{k}^{aux}$ are integers
(and hence, $m = \sum_{k=1}^{K} m_{k}$ with $m_{k} = m \cdot p_{k}^{aux}$) --- 
the remaining case works analogously.
Following the lines of Subsection \ref{SectEstimators.new.det.simplex.meth1.compact.min}
(with $A=1$), a corresponding 
\textit{naive (crude) estimator} can be constructed by 
\begin{equation}
\widehat{
\Phi_{\mathbf{R}_{n}(\mathbf{Y}_{1}^{n})}(\textrm{$\boldsymbol{\Omega}$\hspace{-0.23cm}$\boldsymbol{\Omega}$})
}_{m,L}^{naive,1} \ := \ 
- \frac{1}{m}\log \frac{1}{L}\sum_{\ell =1}^{L} 
\exp\negthinspace\Big(m \cdot \Big(
F_{\gamma,\widetilde{c},1}\Big(D_{\widetilde{c} \cdot \varphi_{\gamma}}( 
\boldsymbol{\xi}_{m}^{w\mathbf{W}^{(\ell)}},\mathds{P}^{aux})\Big)
- \Phi_{\mathbf{R}_{n}(\mathbf{Y}_{1}^{n})}\big(\boldsymbol{\xi}_{m}^{w\mathbf{W}^{(\ell)}}\big)
\Big)
\Big)
\cdot \textfrak{1}_{
\textrm{$\boldsymbol{\Omega}$\hspace{-0.19cm}$\boldsymbol{\Omega}$}}\big(\boldsymbol{\xi}_{m}^{w\mathbf{W}^{(\ell)}}\big)
\, ,
\label{brostu5:fo.BSmin.extended.naive.estim.risk}
\end{equation}
where we (independently of $\mathbf{Y}_{1}^{n}$) simulate
independently $L$ copies 
$\mathbf{W}^{(1)},\ldots,\mathbf{W}^{(L)}$ of the vector
 $\mathbf{W}:=\left( W_{1},\ldots,W_{m}\right) $ 
with independent entries under $\mathbb{\bbzeta}$, 
and compute each of $\boldsymbol{\xi}_{m}^{w\mathbf{W}^{(1)}}, \ldots,
\boldsymbol{\xi}_{m}^{w\mathbf{W}^{(L)}}$ according to 
\eqref{brostu5:fo.norweiemp.vec.det.m}.
Clearly, with the help the strong law of large numbers we get with 
$\widehat{\Phi_{\mathbf{R}_{n}(\mathbf{Y}_{1}^{n})}( 
\textrm{$\boldsymbol{\Omega}$\hspace{-0.23cm}$\boldsymbol{\Omega}$})}_{m,\infty}^{naive,1}: = 
\lim_{L\rightarrow \infty} \widehat{\Phi_{\mathbf{R}_{n}(\mathbf{Y}_{1}^{n})}( 
\textrm{$\boldsymbol{\Omega}$\hspace{-0.23cm}$\boldsymbol{\Omega}$})}_{m,L}^{naive,1}$
the following assertion:

\vspace{0.2cm}

\begin{proposition}
\begin{equation}
\lim_{m\rightarrow \infty} 
\widehat{\Phi_{\mathbf{R}_{n}(\mathbf{Y}_{1}^{n})}(\textrm{$\boldsymbol{\Omega}$\hspace{-0.23cm}$\boldsymbol{\Omega}$})}_{m,\infty}^{naive,1}
\ = \ 
\lim_{m\rightarrow \infty} \lim_{L\rightarrow \infty} 
\widehat{
\Phi_{\mathbf{R}_{n}(\mathbf{Y}_{1}^{n})}(\textrm{$\boldsymbol{\Omega}$\hspace{-0.23cm}$\boldsymbol{\Omega}$})
}_{m,L}^{naive,1}
\ = \ \Phi_{\mathbf{R}_{n}(\mathbf{Y}_{1}^{n})}(\textrm{$\boldsymbol{\Omega}$\hspace{-0.23cm}$\boldsymbol{\Omega}$}) \qquad \textrm{a.s.}
\label{brostu5:fo.BSmin.extended.naive.estim.lim.risk}
\end{equation}
\end{proposition}

\vspace{0.2cm}
\noindent
As the corresponding \textit{very natural naive (crude) estimator} 
of the minimizer-set 
$\mathcal{Q}^{\ast} := \argmin_{\mathds{Q} \in 
\textrm{$\boldsymbol{\Omega}$\hspace{-0.19cm}$\boldsymbol{\Omega}$}} 
\Phi_{\mathbf{R}_{n}(\mathbf{Y}_{1}^{n})}(\mathds{Q})$, 
we take 
\begin{equation}
\widehat{
\argmin_{\mathds{Q}\in \textrm{$\boldsymbol{\Omega}$\hspace{-0.19cm}$\boldsymbol{\Omega}$}} 
\Phi_{\mathbf{R}_{n}(\mathbf{Y}_{1}^{n})}(\mathds{Q})
}_{m,L}^{naive,1}
:= \argmin_{\bbnu \in \mathcal{W}_{m,L}} \Phi_{\mathbf{R}_{n}(\mathbf{Y}_{1}^{n})}(\bbnu)
\label{brostu5:fo.1555.risk}
\end{equation}
where $\mathcal{W}_{m,L} := \{\boldsymbol{\xi}_{m}^{w\mathbf{W}^{(\ell)}}:
\ell \in \{1,\ldots,L\} \,  \} \cap \textrm{$\boldsymbol{\Omega}$\hspace{-0.23cm}$\boldsymbol{\Omega}$}$. 
In other words, as a corresponding \textit{naive (crude) estimator} 
of the (not necessarily unique) element $\mathds{Q}^{\ast}$ of the minimizer-set 
$\mathcal{Q}^{\ast} := \argmin_{\mathds{Q} \in 
\textrm{$\boldsymbol{\Omega}$\hspace{-0.19cm}$\boldsymbol{\Omega}$}} 
\Phi_{\mathbf{R}_{n}(\mathbf{Y}_{1}^{n})}(\mathds{Q})$, 
we take the (not necessarily unique) element 
$\boldsymbol{\xi}_{m}^{w\mathbf{W}^{L,\ast}}$
of the set $\{\boldsymbol{\xi}_{m}^{w\mathbf{W}^{(\ell)}}:
\ell \in \{1,\ldots,L\} \,  \} $ such that 
$\boldsymbol{\xi}_{m}^{w\mathbf{W}^{L,\ast}} \in 
\textrm{$\boldsymbol{\Omega}$\hspace{-0.23cm}$\boldsymbol{\Omega}$}$ and 
\begin{equation}
\Phi_{\mathbf{R}_{n}(\mathbf{Y}_{1}^{n})}(\boldsymbol{\xi}_{m}^{w\mathbf{W}^{L,\ast}}) 
\ \leq \Phi_{\mathbf{R}_{n}(\mathbf{Y}_{1}^{n})}(\boldsymbol{\xi}_{m}^{w\mathbf{W}^{(\ell)}})   
\qquad \textrm{for all $\ell =1,\ldots,L$ for which 
$\boldsymbol{\xi}_{m}^{w\mathbf{W}^{(\ell)}}$ 
belongs to $\textrm{$\boldsymbol{\Omega}$\hspace{-0.23cm}$\boldsymbol{\Omega}$}$.} 
\nonumber
\end{equation}
In short, $\boldsymbol{\xi}_{m}^{w\mathbf{W}^{L,\ast}}$
minimizes $\Phi_{\mathbf{R}_{n}(\mathbf{Y}_{1}^{n})}(\cdot)$ amongst all values 
$\boldsymbol{\xi}_{m}^{w\mathbf{W}^{(\ell)}}$
at hand which fall into $\textrm{$\boldsymbol{\Omega}$\hspace{-0.23cm}$\boldsymbol{\Omega}$}$. 
For large enough $m \in \mathbb{N}$ and $L \in \mathbb{N}$, 
such $\boldsymbol{\xi}_{m}^{w\mathbf{\widetilde{W}}^{L,\ast}}$ exists
since $\textrm{$\boldsymbol{\Omega}$\hspace{-0.23cm}$\boldsymbol{\Omega}$}$ 
has non-void interior in the relative topology, by assumption 
\eqref{regularity simplex}.
We prove that if $L$ and $m$ tend to infinity, then 
$\boldsymbol{\xi}_{m}^{w\mathbf{W}^{L,\ast}}$
concentrates to the above-mentioned set $\mathcal{Q}^{\ast}$ 
of minimizers of $\Phi_{\mathbf{R}_{n}(\mathbf{Y}_{1}^{n})}(\cdot)$ on 
$\textrm{$\boldsymbol{\Omega}$\hspace{-0.23cm}$\boldsymbol{\Omega}$}$.
As usual in similar procedures, $L$ is assumed to be large enough in order to justify 
some approximation for fixed $m$, typically the substitution of empirical means by expectations,
since $L$ is at disposal.

\vspace{0.2cm}
\noindent
Next we derive that $\boldsymbol{\xi}_{m}^{w\mathbf{W}^{L,\ast}}$ is
a proxy minimizer of $\Phi_{\mathbf{R}_{n}(\mathbf{Y}_{1}^{n})}(\cdot)$ 
on $\textrm{$\boldsymbol{\Omega}$\hspace{-0.23cm}$\boldsymbol{\Omega}$}$, manifested by the following

\vspace{0.2cm}

\begin{proposition}
\label{brostu5:prop.generaldeterministic.minimizer.naive.risk}
There holds
\begin{equation}
\min_{\mathbf{Q}\in 
\textrm{$\boldsymbol{\Omega}$\hspace{-0.19cm}$\boldsymbol{\Omega}$}} 
\Phi_{\mathbf{R}_{n}(\mathbf{Y}_{1}^{n})}(\mathbf{Q}) 
\ \leq \ 
\Phi_{\mathbf{R}_{n}(\mathbf{Y}_{1}^{n})} \left( \boldsymbol{\xi}_{m}^{w\mathbf{W}^{L,\ast}} \right)
\ \leq \ 
\widehat{\Phi_{\mathbf{R}_{n}(\mathbf{Y}_{1}^{n})}(\textrm{$\boldsymbol{\Omega}$\hspace{-0.23cm}$\boldsymbol{\Omega}$})}_{m,\infty}^{naive,1}
\ + \ o_{\mathbb{\Pi}}(1)
\label{minim risk}
\end{equation}
where $o_{\mathbb{\Pi}}(1)$ goes to $0$ as $L\rightarrow \infty$ and $m\rightarrow \infty$ under the
distribution $\mathbb{\Pi}$
(recall that $\mathbb{\Pi }[(W_{1},\ldots,W_{m}) \in \cdot \,]=
\mathbb{\bbzeta}^{\otimes m}[\,\cdot \,]$).
\end{proposition}

\vspace{0.4cm}
\noindent
The proof of Proposition \ref{brostu5:prop.generaldeterministic.minimizer.naive.risk}
works analogously to the proof of Proposition 
\ref{brostu5:prop.generaldeterministic.minimizer.naive},
by replacing Theorem \ref{brostu5:thm.Fmin}(a) with \eqref{min.risk.5b.sample.var.2}.

\vspace{0.4cm}

\begin{remark}
In the current set-up of compact 
$\textrm{$\boldsymbol{\Omega}$\hspace{-0.23cm}$\boldsymbol{\Omega}$}$ with \eqref{regularity simplex}, 
by taking the special case $\Phi_{\mathbf{R}_{n}(\mathbf{Y}_{1}^{n})}(\mathds{Q}) 
:= D_{\varphi}(\mathds{Q},\mathbf{R}_{n}(\mathbf{Y}_{1}^{n}))$ 
with $\mathbf{R}_{n}(\mathbf{Y}_{1}^{n}) := \mathds{P}_{n}^{emp}(\mathbf{Y}_{1}^{n})$
we obtain the naive BS-estimator 
$\widehat{D_{\varphi}(\textrm{$\boldsymbol{\Omega}$\hspace{-0.23cm}$\boldsymbol{\Omega}$},\mathds{P}_{n}^{emp}(\mathbf{Y}_{1}^{n}))}_{n,L}^{naive,1}$ 
of the minimum value $\min_{\mathds{Q}\in 
\textrm{$\boldsymbol{\Omega}$\hspace{-0.19cm}$\boldsymbol{\Omega}$}} 
D_{\varphi}(\mathds{Q},\mathds{P}_{n}^{emp}(\mathbf{Y}_{1}^{n}))$.

\end{remark}

\vspace{0.3cm}

\begin{proposition}
In the above set-up, one has
\begin{equation}
\lim_{m\rightarrow \infty} \lim_{L\rightarrow \infty} 
\Phi_{\mathbf{R}_{n}(\mathbf{Y}_{1}^{n})}\Big( \widehat{\argmin_{\mathds{Q} \in 
\textrm{$\boldsymbol{\Omega}$\hspace{-0.19cm}$\boldsymbol{\Omega}$}} 
\Phi_{\mathbf{R}_{n}(\mathbf{Y}_{1}^{n})}(\mathds{Q})}_{m,L}^{naive,1} \Big)
\ = \ \Phi_{\mathbf{R}_{n}(\mathbf{Y}_{1}^{n})}(\textrm{$\boldsymbol{\Omega}$\hspace{-0.23cm}$\boldsymbol{\Omega}$}) 
\qquad \textrm{a.s.},
\nonumber
\end{equation}
and thus the quantity 
$\Phi_{\mathbf{R}_{n}(\mathbf{Y}_{1}^{n})}\Big( \widehat{\argmin_{\mathds{Q} \in 
\textrm{$\boldsymbol{\Omega}$\hspace{-0.19cm}$\boldsymbol{\Omega}$}} 
\Phi_{\mathbf{R}_{n}(\mathbf{Y}_{1}^{n})}(\mathbf{Q})}_{m,L}^{naive,1} \Big)$
is a natural alternative to the estimate
$\widehat{\Phi_{\mathbf{R}_{n}(\mathbf{Y}_{1}^{n})}(\textrm{$\boldsymbol{\Omega}$\hspace{-0.23cm}$\boldsymbol{\Omega}$})}_{m,L}^{naive,1}$ 
given in 
\eqref{brostu5:fo.BSmin.extended.naive.estim.risk}.

\end{proposition}

\vspace{0.3cm}

\begin{remark}
\label{rem.FasDivergence.estimator.risk}
(a) By applying the above-mentioned results 
to the special \textit{divergence} (cf. Remark \ref{rem.FasDivergence})\\
 $\Phi_{\mathds{P}_{n}^{emp}(\mathbf{Y}_{1}^{n})}(\mathds{Q}) := F_{\gamma,\widetilde{c},1}\Big(D_{\widetilde{c} \cdot \varphi_{\gamma}}(\mathds{Q},\mathds{P}_{n}^{emp}(\mathbf{Y}_{1}^{n}))\Big)$,
we obtain for $\Phi_{\mathds{P}_{n}^{emp}(\mathbf{Y}_{1}^{n})}(\textrm{$\boldsymbol{\Omega}$\hspace{-0.23cm}$\boldsymbol{\Omega}$})
:= \min_{\mathds{Q} \in 
\textrm{$\boldsymbol{\Omega}$\hspace{-0.19cm}$\boldsymbol{\Omega}$}} 
\Phi_{\mathds{P}_{n}^{emp}(\mathbf{Y}_{1}^{n})}(\mathds{Q}) $
the estimator 
$
\widehat{\Phi_{\mathds{P}_{n}^{emp}(\mathbf{Y}_{1}^{n})}(\textrm{$\boldsymbol{\Omega}$\hspace{-0.23cm}$\boldsymbol{\Omega}$})}_{m,L}^{naive,1} \ := \ 
- \frac{1}{m}\log \frac{1}{L}\sum_{\ell =1}^{L} 
\textfrak{1}_{
\textrm{$\boldsymbol{\Omega}$\hspace{-0.19cm}$\boldsymbol{\Omega}$}}\big(\boldsymbol{\xi}_{m}^{w\mathbf{W}^{(\ell)}}\big)
$
as well as the alternative estimator\\
$\Phi_{\mathds{P}_{n}^{emp}(\mathbf{Y}_{1}^{n})}\Big( \widehat{\argmin_{\mathds{Q} \in 
\textrm{$\boldsymbol{\Omega}$\hspace{-0.19cm}$\boldsymbol{\Omega}$}} 
\Phi_{\mathds{P}_{n}^{emp}(\mathbf{Y}_{1}^{n})}(\mathds{Q})}_{m,L}^{naive,1} \Big)$.
The involved estimator 
$\widehat{\argmin_{\mathds{Q} \in 
\textrm{$\boldsymbol{\Omega}$\hspace{-0.19cm}$\boldsymbol{\Omega}$}} 
\Phi_{\mathds{P}_{n}^{emp}(\mathbf{Y}_{1}^{n})}(\mathds{Q})}_{m,L}^{naive,1}$
of the minimizer set $\mathcal{Q}^{\ast} := \argmin_{\mathds{Q} \in 
\textrm{$\boldsymbol{\Omega}$\hspace{-0.19cm}$\boldsymbol{\Omega}$}} 
\Phi_{\mathds{P}_{n}^{emp}(\mathbf{Y}_{1}^{n})}(\mathds{Q})$
is --- at the same time (due to the strict increasingness of $F_{\gamma,\widetilde{c},1}$)  ---
also the estimator of the minimizer set $\mathcal{Q}^{\ast} := \argmin_{\mathds{Q} \in 
\textrm{$\boldsymbol{\Omega}$\hspace{-0.19cm}$\boldsymbol{\Omega}$}} 
D_{\widetilde{c} \cdot \varphi_{\gamma}}(\mathds{Q},\mathds{P}_{n}^{emp}(\mathbf{Y}_{1}^{n}))$;
giving the latter had been left as an \textit{open gap} in Broniatowski \& Stummer~\cite{Bro:23a},
which we have now \textit{filled/resolved}. 
By the way, these newly developed minimizer(s) 
are the \textit{non-parametric} analogues of the very prominent
\textit{parametric} minimum-$\varphi-$divergence estimator(s)
(e.g. the omnipresent \textit{maximum-likelihood estimator(s)}
corresponds to the minimum-Kullback-Leibler-information-distance estimator, i.e. 
$\varphi := \varphi_{1}$ given in \eqref{brostu5:fo.powdivgen}).\\
(b) By construction,
$D_{\widetilde{c} \cdot \varphi_{\gamma}}(
\widehat{\argmin_{\mathds{Q} \in 
\textrm{$\boldsymbol{\Omega}$\hspace{-0.19cm}$\boldsymbol{\Omega}$}} 
\Phi_{\mathds{P}_{n}^{emp}(\mathbf{Y}_{1}^{n})}(\mathds{Q})}_{m,L}^{naive,1}
,\mathds{P})$
is an estimator of $\min_{\mathds{Q} \in 
\textrm{$\boldsymbol{\Omega}$\hspace{-0.19cm}$\boldsymbol{\Omega}$}} 
D_{\widetilde{c} \cdot \varphi_{\gamma}}(\mathds{Q},\mathds{P}_{n}^{emp}(\mathbf{Y}_{1}^{n}))$
which serves an alternative to the one given in Broniatowski \& Stummer~\cite{Bro:23a}.

\end{remark}


\subsection{Naive estimators of max and argmax --- Base-Divergence-Method 1, compact case}
\label{SectEstimators.new.det.risk.meth1.compact.max}

\vspace{0.2cm}
\noindent
In the set-up of compact $\textrm{$\boldsymbol{\Omega}$\hspace{-0.23cm}$\boldsymbol{\Omega}$}$ 
--- with \eqref{regularity simplex} --- in the relative topology, we can handle
the maximizing problem completely analogously to the method in the previous Subsection 
\ref{SectEstimators.new.det.risk.meth1.compact.min}.
In fact, instead of \eqref{min.risk.4} and \eqref{min.risk.5a.new.sample}
(leading to \eqref{min.risk.5b.sample}) we employ 
\eqref{max.risk.4} and 
\eqref{max.risk.5a.new.sample}
(leading to \eqref{max.risk.5b.sample})
--- for the random sample $\mathbf{Y}_{1}^{n} = (Y_{1},\ldots,Y_{n})$ ---
to obtain (for large $m$) the following approximation of the sample-dependent maximum
\begin{align}
& \Phi_{\mathbf{R}_{n}(\mathbf{Y}_{1}^{n})}(\textrm{$\boldsymbol{\Omega}$\hspace{-0.23cm}$\boldsymbol{\Omega}$}) := 
\max_{\mathds{Q}\in \textrm{$\boldsymbol{\Omega}$\hspace{-0.19cm}$\boldsymbol{\Omega}$}} 
\Phi_{\mathbf{R}_{n}(\mathbf{Y}_{1}^{n})}(\mathds{Q}) 
\nonumber \\
& \ \approx \  
\frac{1}{m} \log \negthinspace \left( \ 
\mathbb{E}_{\mathbb{\Pi}_{\mathbf{Y}_{1}^{n}}
}
\negthinspace \Big[
\exp\negthinspace\Big(
m \cdot \Big(
F_{\gamma,\widetilde{c},1}\Big(D_{\widetilde{c} \cdot 
\varphi_{\gamma}}(\boldsymbol{\xi}_{m}^{w\mathbf{W}},\mathds{P}^{aux})\Big)
+ \Phi_{\mathbf{R}_{n}(\mathbf{Y}_{1}^{n})}\big(\boldsymbol{\xi}_{m}^{w\mathbf{W}}\big)
\Big)
\Big)
\cdot \textfrak{1}_{
\textrm{$\boldsymbol{\Omega}$\hspace{-0.19cm}$\boldsymbol{\Omega}$}}\big(\boldsymbol{\xi}_{m}^{w\mathbf{W}}\big)
\, \Big] 
\right),
\nonumber
\end{align}
from which we construct the corresponding \textit{naive BS-estimator} as
\begin{equation}
\widehat{
\Phi_{\mathbf{R}_{n}(\mathbf{Y}_{1}^{n})}(\textrm{$\boldsymbol{\Omega}$\hspace{-0.23cm}$\boldsymbol{\Omega}$})
}_{m,L}^{naive,1} \ := \ 
\frac{1}{m}\log \frac{1}{L}\sum_{\ell =1}^{L} 
\exp\negthinspace\Big(m \cdot \Big(
F_{\gamma,\widetilde{c},1}\Big(D_{\widetilde{c} \cdot \varphi_{\gamma}}( 
\boldsymbol{\xi}_{m}^{w\mathbf{W}^{(\ell)}},\mathds{P}^{aux})\Big)
+ \Phi_{\mathbf{R}_{n}(\mathbf{Y}_{1}^{n})}\big(\boldsymbol{\xi}_{m}^{w\mathbf{W}^{(\ell)}}\big)
\Big)
\Big)
\cdot \textfrak{1}_{
\textrm{$\boldsymbol{\Omega}$\hspace{-0.19cm}$\boldsymbol{\Omega}$}}\big(\boldsymbol{\xi}_{m}^{w\mathbf{W}^{(\ell)}}\big)
\, ;
\label{brostu5:fo.BSmax.extended.naive.estim.risk}
\end{equation}
for this, we get
\begin{equation}
\lim_{m\rightarrow \infty} 
\widehat{\Phi_{\mathbf{R}_{n}(\mathbf{Y}_{1}^{n})}(\textrm{$\boldsymbol{\Omega}$\hspace{-0.23cm}$\boldsymbol{\Omega}$})}_{m,\infty}^{naive,1}
\ = \ 
\lim_{m\rightarrow \infty} \lim_{L\rightarrow \infty} 
\widehat{
\Phi_{\mathbf{R}_{n}(\mathbf{Y}_{1}^{n})}(\textrm{$\boldsymbol{\Omega}$\hspace{-0.23cm}$\boldsymbol{\Omega}$})
}_{m,L}^{naive,1}
\ = \ \Phi_{\mathbf{R}_{n}(\mathbf{Y}_{1}^{n})}(\textrm{$\boldsymbol{\Omega}$\hspace{-0.23cm}$\boldsymbol{\Omega}$}) \qquad \textrm{a.s.}
\nonumber
\end{equation}
instead of \eqref{brostu5:fo.BSmin.extended.naive.estim.lim.risk}
(notice the different definition of the involved quantities).
As the corresponding very natural naive BS-estimator of the 
maximizer(-set) \, 
$\argmax_{\mathds{Q} \in 
\textrm{$\boldsymbol{\Omega}$\hspace{-0.19cm}$\boldsymbol{\Omega}$}} 
\Phi_{\mathbf{R}_{n}(\mathbf{Y}_{1}^{n})}(\mathds{Q})$ \, 
we take 
\begin{equation}
\widehat{
\argmax_{\mathds{Q}\in \textrm{$\boldsymbol{\Omega}$\hspace{-0.19cm}$\boldsymbol{\Omega}$}} 
\Phi_{\mathbf{R}_{n}(\mathbf{Y}_{1}^{n})}(\mathds{Q})
}_{m,L}^{naive,1}
:= \argmax_{\bbnu \in \mathcal{W}_{m,L}} \Phi_{\mathbf{R}_{n}(\mathbf{Y}_{1}^{n})}(\bbnu)
\nonumber
\end{equation}
instead of \eqref{brostu5:fo.1555.risk};
in short, 
we take any $\boldsymbol{\xi}_{m}^{w\mathbf{W}^{L,\ast}}$
which maximizes $\Phi_{\mathbf{R}_{n}(\mathbf{Y}_{1}^{n})}(\cdot)$ amongst all values 
$\boldsymbol{\xi}_{m}^{w\mathbf{W}^{(\ell)}}$
at hand which fall into $\textrm{$\boldsymbol{\Omega}$\hspace{-0.23cm}$\boldsymbol{\Omega}$}$. 
For this, we obtain --- instead of \eqref{minim risk} ---
the assertion
\begin{equation}
\max_{\mathbf{Q}\in 
\textrm{$\boldsymbol{\Omega}$\hspace{-0.19cm}$\boldsymbol{\Omega}$}} 
\Phi_{\mathbf{R}_{n}(\mathbf{Y}_{1}^{n})}(\mathbf{Q}) 
\ \geq \ 
\Phi_{\mathbf{R}_{n}(\mathbf{Y}_{1}^{n})} \left( \boldsymbol{\xi}_{m}^{w\mathbf{W}^{L,\ast}} \right)
\ \geq \ 
\widehat{\Phi_{\mathbf{R}_{n}(\mathbf{Y}_{1}^{n})}(\textrm{$\boldsymbol{\Omega}$\hspace{-0.23cm}$\boldsymbol{\Omega}$})}_{m,\infty}^{naive,1}
\ - \ o_{\mathbb{\Pi}}(1)
\nonumber
\end{equation}
where $o_{\mathbb{\Pi}}(1)$ goes to $0$ as $L\rightarrow \infty$ and $m\rightarrow \infty$ under the
distribution $\mathbb{\Pi}$.
Hence, as $L$ and $m$ tend to infinity,  
$\boldsymbol{\xi}_{m}^{w\mathbf{W}^{L,\ast}}$ concentrates to the 
set of maximizers of $\Phi(\cdot)$ on $\textrm{$\boldsymbol{\Omega}$\hspace{-0.23cm}$\boldsymbol{\Omega}$}$.

\vspace{0.2cm}

\begin{proposition}
In the above set-up, one has
\begin{equation}
\lim_{m\rightarrow \infty} \lim_{L\rightarrow \infty} 
\Phi_{\mathbf{R}_{n}(\mathbf{Y}_{1}^{n})}\Big( \widehat{\argmax_{\mathds{Q} \in 
\textrm{$\boldsymbol{\Omega}$\hspace{-0.19cm}$\boldsymbol{\Omega}$}} 
\Phi_{\mathbf{R}_{n}(\mathbf{Y}_{1}^{n})}(\mathds{Q})}_{m,L}^{naive,1} \Big)
\ = \ \Phi_{\mathbf{R}_{n}(\mathbf{Y}_{1}^{n})}(\textrm{$\boldsymbol{\Omega}$\hspace{-0.23cm}$\boldsymbol{\Omega}$}) 
\qquad \textrm{a.s.},
\nonumber
\end{equation}
and thus the quantity 
$\Phi_{\mathbf{R}_{n}(\mathbf{Y}_{1}^{n})}\Big( \widehat{\argmax_{\mathds{Q} \in 
\textrm{$\boldsymbol{\Omega}$\hspace{-0.19cm}$\boldsymbol{\Omega}$}} 
\Phi_{\mathbf{R}_{n}(\mathbf{Y}_{1}^{n})}(\mathbf{Q})}_{m,L}^{naive,1} \Big)$
is a natural alternative to the estimate
$\widehat{\Phi_{\mathbf{R}_{n}(\mathbf{Y}_{1}^{n})}(\textrm{$\boldsymbol{\Omega}$\hspace{-0.23cm}$\boldsymbol{\Omega}$})}_{m,L}^{naive,1}$ 
given in 
\eqref{brostu5:fo.BSmax.extended.naive.estim.risk}. 

\end{proposition}


\subsection{Naive estimators of min and argmin --- Base-Divergence-Method 2, compact case}
\label{SectEstimators.new.det.risk.meth2.compact.min}

\vspace{0.2cm}
\noindent
Relying on Subsection \ref{SectDetGeneral.BDM1.min.risk}, in the above Subsections 
\ref{SectEstimators.new.det.risk.meth1.compact.min} and
\ref{SectEstimators.new.det.risk.meth1.compact.max}
we have chosen the divergence
$F_{\gamma,\widetilde{c},1}(D_{\widetilde{c} \cdot \varphi_{\gamma}}(\cdot,\mathds{P}^{aux}))$
as the \textit{base divergence};
we have referred to this choice as \textit{Base-Divergence-Method 1}.
However, relying on the alternative Subsection 
\ref{SectDetGeneral.BDM2.min.risk}, we can also choose 
--- as Base-Divergence-Method 2 --- as \textit{base divergence} 
the ``innmin scaled Bregman distance''
$\breve{D}_{\widetilde{c} \cdot 
\varphi_{\gamma},\mathds{P}^{aux}}^{SBD}\negthinspace\left(\cdot,\mathbf{Q}^{\ast\ast}\right)$ 
(cf. \eqref{brostu3:fo.676b.SBD},\eqref{brostu3:fo.677a.SBD},\eqref{brostu3:fo.678.SBD}),
where $\mathbf{Q}^{\ast\ast} \in \mathbb{R}_{>0}^{K}$ 
\textit{NEED NOT} be in $\textrm{$\boldsymbol{\Omega}$\hspace{-0.23cm}$\boldsymbol{\Omega}$}$
(e.g. $\mathbf{Q}^{\ast\ast} \in \mathbb{S}_{>0}^{K} \backslash
\textrm{$\boldsymbol{\Omega}$\hspace{-0.23cm}$\boldsymbol{\Omega}$}$).

\vspace{0.2cm}

\vspace{0.2cm}
\noindent
Let us start with any fixed $\mathbf{Q}^{\ast\ast} \in \mathbb{R}_{>0}^{K}$ 
and any fixed $\mathds{P}^{aux} \in \mathbb{S}_{>0}^{K}$ satisfying
\begin{equation}
t_{k}^{\ast\ast} :=  \frac{q_{k}^{\ast\ast}}{p_{k}^{aux}} \in \, ]t_{-}^{sc},t_{+}^{sc}[  
\quad \textrm{for all $k =1,\ldots,K$} 
\qquad \textrm{(cf. \eqref{brostu5:fo.SBD.qstarstar})}.
\nonumber
\end{equation}
In such a setup, recall from \eqref{min.risk.5b.SBD.sample}
that for all compact sets 
$\boldsymbol{\Omega}$\hspace{-0.23cm}$\boldsymbol{\Omega} \subset \widetilde{\mathcal{M}}_{\gamma}$ 
(with $A=1$ and $\gamma \in \mathbb{R}\backslash\, ]1,2[$) satisfying the 
regularity properties \eqref{regularity simplex} 
\textit{in the relative topology} one gets the approximation (for large $m$)
\begin{align}
& \Phi_{\mathbf{R}_{n}(\mathbf{Y}_{1}^{n})}(\textrm{$\boldsymbol{\Omega}$\hspace{-0.23cm}$\boldsymbol{\Omega}$}) := 
\min_{\mathds{Q}\in \textrm{$\boldsymbol{\Omega}$\hspace{-0.19cm}$\boldsymbol{\Omega}$}} 
\Phi_{\mathbf{R}_{n}(\mathbf{Y}_{1}^{n})}(\mathds{Q}) 
\nonumber \\
& \ \approx \ - 
\frac{1}{m} \log \negthinspace \left( \ 
\mathbb{E}_{\mathbb{\Pi}_{\mathbf{Y}_{1}^{n}}}\negthinspace \Big[
\exp\negthinspace\Big(
m \cdot \Big(
\breve{D}_{\widetilde{c} \cdot 
\varphi_{\gamma},\mathds{P}^{aux}}^{SBD}(\boldsymbol{\xi}_{m}^{w\mathbf{V}},\mathbf{Q}^{\ast\ast}) 
- \Phi_{\mathbf{R}_{n}(\mathbf{Y}_{1}^{n})}\big(\boldsymbol{\xi}_{m}^{w\mathbf{V}}\big)
\Big)
\Big)
\cdot \textfrak{1}_{
\textrm{$\boldsymbol{\Omega}$\hspace{-0.19cm}$\boldsymbol{\Omega}$}}\big(\boldsymbol{\xi}_{m}^{w\mathbf{V}}\big)
\, \Big] 
\right),
\label{min.risk.5b.SBD.sample.var}
\end{align}
where $\boldsymbol{\xi}_{m}^{w\mathbf{V}}$ is as in
\eqref{brostu5:fo.norweiemp.vec.det.m.SBD}. 
Also recall that --- in accordance with Remark \ref{brostu5:rem.Paux} ---
the involved $\mathds{P}^{aux}$ may also depend on the 
sample $\mathbf{Y}_{1}^{n}$;
for the special case $\mathds{P}^{aux} := \mathds{P}_{n}^{emp}(\mathbf{Y}_{1}^{n})$
we obtain $\boldsymbol{\xi}_{m}^{w\mathbf{V}} = \boldsymbol{\xi}_{n,m,\mathbf{Y}}^{w\mathbf{V}}$
(cf. \eqref{brostu5:fo.norweiemp.vec.risk.pnemp.innminSBD}
respectively \eqref{brostu5:fo.norweiemp.vec.risk.pnemp.var.innminSBD}, with $\mathbf{Y}$
instead of $\mathbf{y}$).

\vspace{0.3cm}
\noindent
Accordingly, for getting an estimator of the 
sample-dependent minimum
$\Phi_{\mathbf{R}_{n}(\mathbf{Y}_{1}^{n})}(\textrm{$\boldsymbol{\Omega}$\hspace{-0.23cm}$\boldsymbol{\Omega}$})
= \min_{\mathds{Q}\in \textrm{$\boldsymbol{\Omega}$\hspace{-0.19cm}$\boldsymbol{\Omega}$}} 
\Phi_{\mathbf{R}_{n}(\mathbf{Y}_{1}^{n})}(\mathds{Q})$
one can estimate the right-hand side of 
\eqref{min.risk.5b.SBD.sample.var}.

\vspace{0.2cm}
\noindent
To achieve this, for the rest of this section we assume 
that $m$ is chosen such that all 
$m \cdot p_{k}^{aux}$ are integers
(and hence, $m = \sum_{k=1}^{K} m_{k}$ with $m_{k} = m \cdot p_{k}^{aux}$) --- 
the remaining case works analogously. Following the lines of Subsection 
\ref{SectEstimators.new.det.simplex.meth2.compact.min}
(with $A=1$), a corresponding 
\textit{naive (crude) estimator} can be constructed by 

\begin{equation}
\widehat{
\Phi_{\mathbf{R}_{n}(\mathbf{Y}_{1}^{n})}(\textrm{$\boldsymbol{\Omega}$\hspace{-0.23cm}$\boldsymbol{\Omega}$})
}_{m,L}^{naive,2} \ := \ 
- \frac{1}{m}\log \frac{1}{L}\sum_{\ell =1}^{L} 
\exp\negthinspace\Big(m \cdot \Big(
\breve{D}_{\widetilde{c} \cdot \varphi_{\gamma},\mathds{P}^{aux}}^{SBD}\negthinspace\left(
\boldsymbol{\xi}_{m}^{w\mathbf{V}^{(\ell)}},\mathbf{Q}^{\ast\ast}\right) 
\Big)
- \Phi_{\mathbf{R}_{n}(\mathbf{Y}_{1}^{n})}\big(\boldsymbol{\xi}_{m}^{w\mathbf{V}^{(\ell)}}\big)
\Big)
\Big)
\cdot \textfrak{1}_{
\textrm{$\boldsymbol{\Omega}$\hspace{-0.19cm}$\boldsymbol{\Omega}$}}\big(\boldsymbol{\xi}_{m}^{w\mathbf{V}^{(\ell)}}\big)
\, ,
\label{brostu5:fo.BSmin.extended.improved.estim.risk}
\end{equation}
where we simulate independently $L$ copies 
$\mathbf{V}^{(1)},\ldots,\mathbf{V}^{(L)}$ of the random vector
$\mathbf{V}:=\left( V_{1},\ldots,V_{m}\right)$,
and compute each of $\boldsymbol{\xi}_{m}^{w\mathbf{V}^{(1)}}, \ldots,
\boldsymbol{\xi}_{m}^{w\mathbf{V}^{(L)}}$ according to 
\eqref{brostu5:fo.norweiemp.vec.det.SBD} (with $m$ instead of $n$).
Clearly, with the help of the strong law of large numbers we get with 
$\widehat{\Phi_{\mathbf{R}_{n}(\mathbf{Y}_{1}^{n})}( 
\textrm{$\boldsymbol{\Omega}$\hspace{-0.23cm}$\boldsymbol{\Omega}$})}_{m,\infty}^{naive,2}: = 
\lim_{L\rightarrow \infty} \widehat{\Phi_{\mathbf{R}_{n}(\mathbf{Y}_{1}^{n})}( 
\textrm{$\boldsymbol{\Omega}$\hspace{-0.23cm}$\boldsymbol{\Omega}$})}_{m,L}^{naive,2}$
the following assertion:

\vspace{0.2cm}

\begin{proposition}
\begin{equation}
\lim_{m\rightarrow \infty} 
\widehat{\Phi_{\mathbf{R}_{n}(\mathbf{Y}_{1}^{n})}(\textrm{$\boldsymbol{\Omega}$\hspace{-0.23cm}$\boldsymbol{\Omega}$})}_{m,\infty}^{naive,2}
\ = \ 
\lim_{m\rightarrow \infty} \lim_{L\rightarrow \infty} 
\widehat{
\Phi_{\mathbf{R}_{n}(\mathbf{Y}_{1}^{n})}(\textrm{$\boldsymbol{\Omega}$\hspace{-0.23cm}$\boldsymbol{\Omega}$})
}_{m,L}^{naive,2}
\ = \ \Phi_{\mathbf{R}_{n}(\mathbf{Y}_{1}^{n})}(\textrm{$\boldsymbol{\Omega}$\hspace{-0.23cm}$\boldsymbol{\Omega}$}) \qquad \textrm{a.s.}
\label{brostu5:fo.BSmin.extended.improved.estim.lim.risk}
\end{equation}
\end{proposition}

\vspace{0.2cm}
\noindent
As the corresponding \textit{very natural naive (crude) estimator} of the minimizer-set 
$\mathcal{Q}^{\ast} := \argmin_{\mathds{Q} \in 
\textrm{$\boldsymbol{\Omega}$\hspace{-0.19cm}$\boldsymbol{\Omega}$}} 
\Phi_{\mathbf{R}_{n}(\mathbf{Y}_{1}^{n})}(\mathds{Q})$, 
we take 
\begin{equation}
\widehat{
\argmin_{\mathds{Q}\in \textrm{$\boldsymbol{\Omega}$\hspace{-0.19cm}$\boldsymbol{\Omega}$}} 
\Phi_{\mathbf{R}_{n}(\mathbf{Y}_{1}^{n})}(\mathds{Q})
}_{m,L}^{naive,2}
:= \argmin_{\bbnu \in \mathcal{V}_{m,L}} \Phi_{\mathbf{R}_{n}(\mathbf{Y}_{1}^{n})}(\bbnu)
\label{brostu5:fo.1555.risk.SBD}
\end{equation}
where $\mathcal{V}_{m,L} := \{\boldsymbol{\xi}_{m}^{w\mathbf{V}^{(\ell)}}:
\ell \in \{1,\ldots,L\} \,  \} \cap \textrm{$\boldsymbol{\Omega}$\hspace{-0.23cm}$\boldsymbol{\Omega}$}$.
In short, we take as minimizer-estimate
the (not necessarily unique) element $\boldsymbol{\xi}_{m}^{w\mathbf{V}^{L,\ast}}$
which minimizes $\Phi_{\mathbf{R}_{n}(\mathbf{Y}_{1}^{n})}(\cdot)$ amongst all values 
$\boldsymbol{\xi}_{m}^{w\mathbf{V}^{(\ell)}}$
at hand which fall into $\textrm{$\boldsymbol{\Omega}$\hspace{-0.23cm}$\boldsymbol{\Omega}$}$. 
We prove that if $L$ and $m$ tend to infinity, then 
$\boldsymbol{\xi}_{m}^{w\mathbf{V}^{L,\ast}}$
concentrates to the above-mentioned set $\mathcal{Q}^{\ast}$ 
of minimizers of $\Phi_{\mathbf{R}_{n}(\mathbf{Y}_{1}^{n})}(\cdot)$ on 
$\textrm{$\boldsymbol{\Omega}$\hspace{-0.23cm}$\boldsymbol{\Omega}$}$.
As usual in similar procedures, $L$ is assumed to be large enough in order to justify 
some approximation for fixed $m$, typically the substitution of empirical means by expectations,
since $L$ is at disposal.

\vspace{0.2cm}
\noindent
Next we derive that $\boldsymbol{\xi}_{m}^{w\mathbf{V}^{L,\ast}}$ is
a proxy minimizer of $\Phi_{\mathbf{R}_{n}(\mathbf{Y}_{1}^{n})}(\cdot)$ 
on $\textrm{$\boldsymbol{\Omega}$\hspace{-0.23cm}$\boldsymbol{\Omega}$}$, manifested by the following

\vspace{0.2cm}

\begin{proposition}
\label{brostu5:prop.generaldeterministic.minimizer.improved.risk}
There holds
\begin{equation}
\min_{\mathbf{Q}\in 
\textrm{$\boldsymbol{\Omega}$\hspace{-0.19cm}$\boldsymbol{\Omega}$}} 
\Phi_{\mathbf{R}_{n}(\mathbf{Y}_{1}^{n})}(\mathbf{Q}) 
\ \leq \ 
\Phi_{\mathbf{R}_{n}(\mathbf{Y}_{1}^{n})} \left( \boldsymbol{\xi}_{m}^{w\mathbf{V}^{L,\ast}} \right)
\ \leq \ 
\widehat{\Phi_{\mathbf{R}_{n}(\mathbf{Y}_{1}^{n})}(\textrm{$\boldsymbol{\Omega}$\hspace{-0.23cm}$\boldsymbol{\Omega}$})}_{m,\infty}^{naive,2}
\ + \ o_{\mathbb{\Pi}}(1)
\label{minim risk SBD}
\end{equation}
where $o_{\mathbb{\Pi}}(1)$ goes to $0$ as $L\rightarrow \infty$ and $m\rightarrow \infty$ under the
distribution $\mathbb{\Pi}$
(recall that 
$\mathbb{\Pi}[(V_{1},\ldots,V_{m}) \in \cdot \,]=
\bigotimes\limits_{k=1}^{K}U_{k}^{\otimes m_{k}}[\,\cdot \,]$).

\end{proposition}

\vspace{0.3cm}
\noindent
The proof of Proposition \ref{brostu5:prop.generaldeterministic.minimizer.improved.risk}
works analogously to the proof of Proposition \ref{brostu5:prop.generaldeterministic.minimizer.naive},
by replacing Theorem \ref{brostu5:thm.Fmin}(a) with \eqref{min.risk.5a.SBD.sample}.

\vspace{0.4cm}

\begin{proposition}
In the above set-up, one has
\begin{equation}
\lim_{m\rightarrow \infty} \lim_{L\rightarrow \infty} 
\Phi_{\mathbf{R}_{n}(\mathbf{Y}_{1}^{n})}\Big( \widehat{\argmin_{\mathds{Q} \in 
\textrm{$\boldsymbol{\Omega}$\hspace{-0.19cm}$\boldsymbol{\Omega}$}} 
\Phi_{\mathbf{R}_{n}(\mathbf{Y}_{1}^{n})}(\mathds{Q})}_{m,L}^{naive,2} \Big)
\ = \ \Phi_{\mathbf{R}_{n}(\mathbf{Y}_{1}^{n})}(\textrm{$\boldsymbol{\Omega}$\hspace{-0.23cm}$\boldsymbol{\Omega}$}) 
\qquad \textrm{a.s.},
\nonumber
\end{equation}
and thus the quantity 
$\Phi_{\mathbf{R}_{n}(\mathbf{Y}_{1}^{n})}\Big( \widehat{\argmin_{\mathds{Q} \in 
\textrm{$\boldsymbol{\Omega}$\hspace{-0.19cm}$\boldsymbol{\Omega}$}} 
\Phi_{\mathbf{R}_{n}(\mathbf{Y}_{1}^{n})}(\mathbf{Q})}_{m,L}^{naive,2} \Big)$
is a natural alternative to the estimate
$\widehat{\Phi_{\mathbf{R}_{n}(\mathbf{Y}_{1}^{n})}(\textrm{$\boldsymbol{\Omega}$\hspace{-0.23cm}$\boldsymbol{\Omega}$})}_{m,L}^{naive,2}$ 
given in 
\eqref{brostu5:fo.BSmin.extended.improved.estim.risk}.

\end{proposition}


\subsection{Naive estimators of max and argmax --- Base-Divergence-Method 2, compact case}
\label{SectEstimators.new.det.risk.meth2.compact.max}

\vspace{0.2cm}
\noindent
In the set-up of compact $\textrm{$\boldsymbol{\Omega}$\hspace{-0.23cm}$\boldsymbol{\Omega}$}$ 
--- with \eqref{regularity simplex} --- in the relative topology, we can handle
the maximizing problem completely analogously to the method in the previous Subsection 
\ref{SectEstimators.new.det.risk.meth2.compact.min}.
In fact, instead of \eqref{min.risk.5b.SBD.sample.var} we employ 
\eqref{max.risk.5b.SBD.sample}
--- for the random sample $\mathbf{Y}_{1}^{n} = (Y_{1},\ldots,Y_{n})$ ---
to obtain (for large $m$)
the approximation of the sample-dependent maximum
\begin{align}
& \Phi_{\mathbf{R}_{n}(\mathbf{Y}_{1}^{n})}(\textrm{$\boldsymbol{\Omega}$\hspace{-0.23cm}$\boldsymbol{\Omega}$}) := 
\max_{\mathds{Q}\in \textrm{$\boldsymbol{\Omega}$\hspace{-0.19cm}$\boldsymbol{\Omega}$}} 
\Phi_{\mathbf{R}_{n}(\mathbf{Y}_{1}^{n})}(\mathds{Q}) 
\nonumber \\
& \ \approx \  
\frac{1}{m} \log \negthinspace \left( \ 
\mathbb{E}_{\mathbb{\Pi}_{\mathbf{Y}_{1}^{n}}
}
\negthinspace \Big[
\exp\negthinspace\Big( m
\cdot \Big(
\breve{D}_{\widetilde{c} \cdot 
\varphi_{\gamma},\mathds{P}^{aux}}^{SBD}(\boldsymbol{\xi}_{m}^{w\mathbf{V}},\mathbf{Q}^{\ast\ast})  
+ \Phi_{\mathbf{R}_{n}(\mathbf{Y}_{1}^{n})}\big(\boldsymbol{\xi}_{m}^{w\mathbf{V}}\big)
\Big)
\Big)
\cdot \textfrak{1}_{
\textrm{$\boldsymbol{\Omega}$\hspace{-0.19cm}$\boldsymbol{\Omega}$}}\big(\boldsymbol{\xi}_{m}^{w\mathbf{V}}\big)
\, \Big] 
\right)
\nonumber
\end{align}
(where  where $\mathbf{Q}^{\ast\ast} \in \mathbb{R}_{>0}^{K}$ 
\textit{NEED NOT} be in $\textrm{$\boldsymbol{\Omega}$\hspace{-0.23cm}$\boldsymbol{\Omega}$}$),
from which we construct the corresponding \textit{naive BS-estimator} as
\begin{equation}
\widehat{
\Phi_{\mathbf{R}_{n}(\mathbf{Y}_{1}^{n})}(\textrm{$\boldsymbol{\Omega}$\hspace{-0.23cm}$\boldsymbol{\Omega}$})
}_{m,L}^{naive,2} \ := \ 
\frac{1}{m}\log \frac{1}{L}\sum_{\ell =1}^{L} 
\exp\negthinspace\Big(m \cdot \Big(
\breve{D}_{\widetilde{c} \cdot \varphi_{\gamma},\mathds{P}^{aux}}^{SBD}\negthinspace\left(
\boldsymbol{\xi}_{m}^{w\mathbf{V}^{(\ell)}},\mathbf{Q}^{\ast\ast}\right)
+ \Phi_{\mathbf{R}_{n}(\mathbf{Y}_{1}^{n})}\big(\boldsymbol{\xi}_{m}^{w\mathbf{V}^{(\ell)}}\big)
\Big)
\Big)
\cdot \textfrak{1}_{
\textrm{$\boldsymbol{\Omega}$\hspace{-0.19cm}$\boldsymbol{\Omega}$}}\big(\boldsymbol{\xi}_{m}^{w\mathbf{V}^{(\ell)}}\big)
\, ;
\label{brostu5:fo.BSmax.extended.improved.estim.risk}
\end{equation}
for this, we get
\begin{equation}
\lim_{m\rightarrow \infty} 
\widehat{\Phi_{\mathbf{R}_{n}(\mathbf{Y}_{1}^{n})}(\textrm{$\boldsymbol{\Omega}$\hspace{-0.23cm}$\boldsymbol{\Omega}$})}_{m,\infty}^{naive,2}
\ = \ 
\lim_{m\rightarrow \infty} \lim_{L\rightarrow \infty} 
\widehat{
\Phi_{\mathbf{R}_{n}(\mathbf{Y}_{1}^{n})}(\textrm{$\boldsymbol{\Omega}$\hspace{-0.23cm}$\boldsymbol{\Omega}$})
}_{m,L}^{naive,2}
\ = \ \Phi_{\mathbf{R}_{n}(\mathbf{Y}_{1}^{n})}(\textrm{$\boldsymbol{\Omega}$\hspace{-0.23cm}$\boldsymbol{\Omega}$}) \qquad \textrm{a.s.}
\nonumber
\end{equation}
instead of \eqref{brostu5:fo.BSmin.extended.improved.estim.lim.risk}
(notice the different definition of the involved quantities).
As the corresponding very natural naive BS-estimator of the 
maximizer(-set) \, 
$\argmax_{\mathds{Q} \in 
\textrm{$\boldsymbol{\Omega}$\hspace{-0.19cm}$\boldsymbol{\Omega}$}} 
\Phi_{\mathbf{R}_{n}(\mathbf{Y}_{1}^{n})}(\mathds{Q})$ \, 
we take 
\begin{equation}
\widehat{
\argmax_{\mathds{Q}\in \textrm{$\boldsymbol{\Omega}$\hspace{-0.19cm}$\boldsymbol{\Omega}$}} 
\Phi_{\mathbf{R}_{n}(\mathbf{Y}_{1}^{n})}(\mathds{Q})
}_{m,L}^{naive,2}
:= \argmax_{\bbnu \in \mathcal{V}_{m,L}} \Phi_{\mathbf{R}_{n}(\mathbf{Y}_{1}^{n})}(\bbnu)
\nonumber
\end{equation}
instead of 
\eqref{brostu5:fo.1555.risk.SBD};
in short, 
we take any $\boldsymbol{\xi}_{m}^{w\mathbf{V}^{L,\ast}}$
which maximizes $\Phi_{\mathbf{R}_{n}(\mathbf{Y}_{1}^{n})}(\cdot)$ amongst all values 
$\boldsymbol{\xi}_{m}^{w\mathbf{V}^{(\ell)}}$
at hand which fall into $\textrm{$\boldsymbol{\Omega}$\hspace{-0.23cm}$\boldsymbol{\Omega}$}$. 
For this, we obtain --- instead of \eqref{minim risk SBD} ---
the assertion
\begin{equation}
\max_{\mathbf{Q}\in 
\textrm{$\boldsymbol{\Omega}$\hspace{-0.19cm}$\boldsymbol{\Omega}$}} 
\Phi_{\mathbf{R}_{n}(\mathbf{Y}_{1}^{n})}(\mathbf{Q}) 
\ \geq \ 
\Phi_{\mathbf{R}_{n}(\mathbf{Y}_{1}^{n})} \left( \boldsymbol{\xi}_{m}^{w\mathbf{V}^{L,\ast}} \right)
\ \geq \ 
\widehat{\Phi_{\mathbf{R}_{n}(\mathbf{Y}_{1}^{n})}(\textrm{$\boldsymbol{\Omega}$\hspace{-0.23cm}$\boldsymbol{\Omega}$})}_{m,\infty}^{naive,2}
\ - \ o_{\mathbb{\Pi}}(1)
\nonumber
\end{equation}
where $o_{\mathbb{\Pi}}(1)$ goes to $0$ as $L\rightarrow \infty$ and $m\rightarrow \infty$ under the
distribution $\mathbb{\Pi}$.
Hence, as $L$ and $m$ tend to infinity,  
$\boldsymbol{\xi}_{m}^{w\mathbf{V}^{L,\ast}}$
concentrates to the 
set of maximizers of $\Phi(\cdot)$ on $\textrm{$\boldsymbol{\Omega}$\hspace{-0.23cm}$\boldsymbol{\Omega}$}$.

\vspace{0.2cm}

\begin{proposition}
In the above set-up, one has
\begin{equation}
\lim_{m\rightarrow \infty} \lim_{L\rightarrow \infty} 
\Phi_{\mathbf{R}_{n}(\mathbf{Y}_{1}^{n})}\Big( \widehat{\argmax_{\mathds{Q} \in 
\textrm{$\boldsymbol{\Omega}$\hspace{-0.19cm}$\boldsymbol{\Omega}$}} 
\Phi_{\mathbf{R}_{n}(\mathbf{Y}_{1}^{n})}(\mathds{Q})}_{m,L}^{naive,2} \Big)
\ = \ \Phi_{\mathbf{R}_{n}(\mathbf{Y}_{1}^{n})}(\textrm{$\boldsymbol{\Omega}$\hspace{-0.23cm}$\boldsymbol{\Omega}$}) 
\qquad \textrm{a.s.},
\nonumber
\end{equation}
and thus the quantity 
$\Phi_{\mathbf{R}_{n}(\mathbf{Y}_{1}^{n})}\Big( \widehat{\argmax_{\mathds{Q} \in 
\textrm{$\boldsymbol{\Omega}$\hspace{-0.19cm}$\boldsymbol{\Omega}$}} 
\Phi_{\mathbf{R}_{n}(\mathbf{Y}_{1}^{n})}(\mathbf{Q})}_{m,L}^{naive,2} \Big)$
is a natural alternative to the estimate
$\widehat{\Phi_{\mathbf{R}_{n}(\mathbf{Y}_{1}^{n})}(\textrm{$\boldsymbol{\Omega}$\hspace{-0.23cm}$\boldsymbol{\Omega}$})}_{m,L}^{naive,2}$ 
given in 
\eqref{brostu5:fo.BSmax.extended.improved.estim.risk}. 

\end{proposition}


\subsection{Improved/Speed-up estimators 
of min, argmin, max and argmax --- compact case}

\vspace{0.2cm}
\noindent
Recall from Subsection 
\ref{SectEstimators.new.det.simplex.improved.compact}
that --- for the case of compact constraint sets 
$A \cdot \textrm{$\boldsymbol{\Omega}$\hspace{-0.23cm}$\boldsymbol{\Omega}$}$
with \eqref{regularity simplex} ---
we have constructed improved/speed-up estimators of 
$\inf_{\mathbf{Q}\in A \cdot \textrm{$\boldsymbol{\Omega}$\hspace{-0.19cm}$\boldsymbol{\Omega}$}}\Phi (\mathbf{Q}) 
= \min_{\mathbf{Q}\in A \cdot \textrm{$\boldsymbol{\Omega}$\hspace{-0.19cm}$\boldsymbol{\Omega}$}} \Phi (\mathbf{Q})$ 
and of the corresponding
(set of) \textit{minimizers} \\
$\arg \inf_{\mathbf{Q}\in A \cdot 
\textrm{$\boldsymbol{\Omega}$\hspace{-0.19cm}$\boldsymbol{\Omega}$}}\Phi (\mathbf{Q}) 
=\arg \min_{\mathbf{Q}\in A \cdot \textrm{$\boldsymbol{\Omega}$\hspace{-0.19cm}$\boldsymbol{\Omega}$}} \Phi (\mathbf{Q})$,
by switching from the
base-divergence
$F_{\gamma,\widetilde{c},A}\Big(D_{\widetilde{c} 
\cdot \varphi_{\gamma}}(A \cdot \boldsymbol{\xi}_{n}^{w\mathbf{W}},\mathds{P})\Big)$
to the different base-divergence 
$\breve{D}_{\widetilde{c} \cdot \varphi_{\gamma},\mathbf{P}}^{SBD}\negthinspace\left(A 
\cdot \boldsymbol{\xi}_{n}^{w\mathbf{\widetilde{V}}},\mathbf{Q}^{\ast}\right)$
where $\mathbf{Q}^{\ast}$ is chosen to be in
$int(A \cdot \textrm{$\boldsymbol{\Omega}$\hspace{-0.23cm}$\boldsymbol{\Omega}$})$
such that by construction the involved random vector
$A \cdot \boldsymbol{\xi}_{n}^{w\mathbf{\widetilde{V}}}$
converges a.s. (as $n \rightarrow \infty$) to $\mathbf{Q}^{\ast}$
and consequently
$\lim_{n\rightarrow \infty} 
\mathbb{\Pi}[A \cdot \boldsymbol{\xi}_{n}^{w\mathbf{\widetilde{V}}} \in 
A \cdot \textrm{$\boldsymbol{\Omega}$\hspace{-0.23cm}$\boldsymbol{\Omega}$}]
\ = \ 1$.
Accordingly, we obtained --- for large enough approximation step $n$ --- the desired good hit-rate
needed for the construction of the improved/speed-up estimators.

\vspace{0.2cm}
\noindent
For the current \textit{risk-case} set-up on compact constraint sets 
$\textrm{$\boldsymbol{\Omega}$\hspace{-0.23cm}$\boldsymbol{\Omega}$} \in \mathbb{S}^{K}$
with \eqref{regularity simplex}, we can act similarly.
Indeed, for achieving improved/speed-up estimators of
of the \textit{minimum value}
$\inf_{\mathds{Q}\in \textrm{$\boldsymbol{\Omega}$\hspace{-0.19cm}$\boldsymbol{\Omega}$}} 
\Phi_{\mathbf{R}_{n}(\mathbf{Y}_{1}^{n})}(\mathds{Q}) 
= \min_{\mathds{Q}\in \textrm{$\boldsymbol{\Omega}$\hspace{-0.19cm}$\boldsymbol{\Omega}$}} 
\Phi_{\mathbf{R}_{n}(\mathbf{Y}_{1}^{n})}(\mathds{Q})$
and of the corresponding (set of) \textit{minimizers} 
$\arg \inf_{\mathds{Q}\in \textrm{$\boldsymbol{\Omega}$\hspace{-0.19cm}$\boldsymbol{\Omega}$}} 
\Phi_{\mathbf{R}_{n}(\mathbf{Y}_{1}^{n})}(\mathds{Q}) 
= \arg \min_{\mathds{Q}\in \textrm{$\boldsymbol{\Omega}$\hspace{-0.19cm}$\boldsymbol{\Omega}$}} 
\Phi_{\mathbf{R}_{n}(\mathbf{Y}_{1}^{n})}(\mathds{Q})$
we employ the base-divergence
$\breve{D}_{\widetilde{c} \cdot 
\varphi_{\gamma},\mathds{P}^{aux}}^{SBD}(\boldsymbol{\xi}_{m}^{w\mathbf{V}},\mathds{Q}^{\ast})$
where $\mathds{Q}^{\ast}$
is chosen to be in
$int(\textrm{$\boldsymbol{\Omega}$\hspace{-0.23cm}$\boldsymbol{\Omega}$})$
such that by construction the involved random vector
$\boldsymbol{\xi}_{m}^{w\mathbf{V}}$
converges --- with respect to the conditional distribution
$\mathbb{\Pi}_{\mathbf{Y}_{1}^{n}}$ --- a.s. (as $m \rightarrow \infty$) to $\mathds{Q}^{\ast}$
and consequently
$\lim_{m\rightarrow \infty} 
\mathbb{\Pi}_{\mathbf{Y}_{1}^{n}}[\boldsymbol{\xi}_{m}^{w\mathbf{V}} \in 
\textrm{$\boldsymbol{\Omega}$\hspace{-0.23cm}$\boldsymbol{\Omega}$}]
\ = \ 1$.
Accordingly, we obtain --- for large enough approximation step $m$ --- the desired good hit-rate
needed for the construction of the improved/speed-up estimators,
and proceed
analogously to Subsection \ref{SectEstimators.new.det.nonvoid.improved.compact}.1),
by appropriately applying the results of
the Subsections \ref{SectEstimators.new.det.risk.meth2.compact.min}
and \ref{SectEstimators.new.det.risk.meth2.compact.max}
(instead of Subsections \ref{SectEstimators.new.det.nonvoid.meth2.compact.min} 
and \ref{SectEstimators.new.det.nonvoid.meth2.compact.max}).
For the sake of brevity, the details are omitted.


\appendices
\section{Proofs}
\label{App.A}

\vspace{0.2cm}
\noindent
\textbf{Proof of Theorem \ref{brostu5:thm.BSnarrow.SBD}.}
We first prove that for all $k=1,\ldots,K$ the function 
$\widetilde{\varphi}_{k}(\cdot) : = M_{\mathbf{P}} \cdot \varphi_{k}(\cdot)$ (cf. \eqref{brostu5:fo.phi_k})
satisfies the representability
\begin{equation}
\widetilde{\varphi}_{k}(t) \ = \ 
\sup_{z \in \mathbb{R}} \Big( z\cdot t - \log \int_{\mathbb{R}} e^{z \cdot y} d\widetilde{U}_{k}(y) \Big),
\qquad t \in \mathbb{R}.  
\label{brostu5:fo.link.SBD.b}
\end{equation}
In other words, we have to show that for all $k=1,\ldots,K$ the function $ t \mapsto \widetilde{\varphi}_{k}(t)$
is the Fenchel-Legendre transform of the cumulant-generating function
(log-moment-generating function)
$\Lambda_{\widetilde{U}_{k}}(z):= 
\log \int_{\mathbb{R}} e^{z\cdot y} d\widetilde{U}_{k}(y)$ of the distribution $\widetilde{U}_{k}$.
To start with, let us fix a $k \in \{1,\ldots,K\}$ and recall that we have constructed
\begin{equation}
d\widetilde{U}_{k}(v) \ := \ 
\frac{\exp \left(\tau_{k} \cdot v\right)}{MGF_{\widetilde{\mathbb{\bbzeta}}}(\tau_{k})} 
\, d\widetilde{\mathbb{\bbzeta}}(v)
=
\frac{\exp \left(\tau_{k} \cdot v\right)}{
\int_{\mathbb{R}} e^{\tau_{k}\cdot y} d\widetilde{\mathbb{\bbzeta}}(y)
} \, d\widetilde{\mathbb{\bbzeta}}(v)
\qquad \textrm{(cf. \eqref{brostu5:Utilde_k_new})}
\nonumber
\end{equation}
where 
$\tau_{k} := 
\ M_{\mathbf{P}} \cdot \varphi^{\, \prime} \negthinspace
\left(\frac{q_{k}^{\ast\ast}}{p_{k}}\right)
=
\widetilde{\varphi}^{\, \prime} \negthinspace
\left(\frac{\widetilde{q}_{k}^{\ast\ast}}{\widetilde{p}_{k}}\right)
=
\widetilde{\varphi}^{\, \prime} \negthinspace
\left(t_{k}^{\ast\ast}\right)
< \infty$
with
$t_{k}^{\ast\ast} =  \frac{\widetilde{q}_{k}^{\ast\ast}}{\widetilde{p}_{k}} \in \, ]t_{-}^{sc},t_{+}^{sc}[$
(cf. \eqref{brostu5:fo.SBD.qstarstar}). From this, we get for all $t \in \mathbb{R}$
\begin{eqnarray}
&& \hspace{-1.4cm}
\sup_{z \in \mathbb{R}} \bigg( z\cdot t - 
\log \int_{\mathbb{R}} e^{z\cdot y} d\widetilde{U}_{k}(y) \bigg)
\ = \ \sup_{z \in \mathbb{R}} \left( z\cdot t 
- \Big(\Lambda_{\widetilde{\mathbb{\bbzeta}}}(z+\tau_{k}) 
- \Lambda_{\widetilde{\mathbb{\bbzeta}}}(\tau_{k}) \Big)
 \right)
\ = \ \sup_{z \in \mathbb{R}} \left( z\cdot t 
- \Lambda_{\widetilde{\mathbb{\bbzeta}}}(z) 
 \right)
+ \Lambda_{\widetilde{\mathbb{\bbzeta}}}(\tau_{k}) - t \cdot  \tau_{k}
\label{brostu5:fo.link.SBD.proof1}
\\
&& \hspace{-1.4cm}
= \ \widetilde{\varphi}(t) + \Lambda_{\widetilde{\mathbb{\bbzeta}}}(\tau_{k}) - t \cdot  \tau_{k}
\ = \ \widetilde{\varphi}(t) + 
\Lambda_{\widetilde{\mathbb{\bbzeta}}}\left(\widetilde{\varphi}^{\, \prime} \negthinspace
\left(t_{k}^{\ast\ast}\right)\right) 
- t \cdot \widetilde{\varphi}^{\, \prime} \negthinspace
\left(t_{k}^{\ast\ast}\right)
\ = \ 
\widetilde{\varphi}(t) +
t_{k}^{\ast\ast} \cdot \widetilde{\varphi}^{\, \prime}(t_{k}^{\ast\ast})
- \widetilde{\varphi}(t_{k}^{\ast\ast})
- t \cdot \widetilde{\varphi}^{\, \prime} \negthinspace
\left(t_{k}^{\ast\ast}\right)
= \widetilde{\varphi}_{k}(t),
\label{brostu5:fo.link.SBD.proof2}
\end{eqnarray}
where in the first equality of \eqref{brostu5:fo.link.SBD.proof2}
we have employed our basic representability assumption 
\begin{equation}
\widetilde{\varphi}(t) \ = \ 
\sup_{z \in \mathbb{R}} \Big( z\cdot t - \Lambda_{\widetilde{\mathbb{\bbzeta}}}(z) \Big),
\qquad t \in \mathbb{R},  
\qquad \textrm{(cf. \eqref{brostu5:fo.link.var})}
\nonumber
\end{equation}
and in the
third equality of \eqref{brostu5:fo.link.SBD.proof2}
that the corresponding explicit solution of \eqref{brostu5:fo.link.var}
for $t = t_{k}^{\ast\ast} \in \, ]t_{-}^{sc},t_{+}^{sc}[$
is given by \footnote{recall that $h^{\leftarrow}$ denotes the inverse of a function $h$}
\begin{equation}
\widetilde{\varphi}(t_{k}^{\ast\ast}) \ = \ 
t_{k}^{\ast\ast} \cdot \Lambda_{\widetilde{\mathbb{\bbzeta}}}^{\prime \, \leftarrow}(t_{k}^{\ast\ast})
- \Lambda_{\widetilde{\mathbb{\bbzeta}}}\left(
\Lambda_{\widetilde{\mathbb{\bbzeta}}}^{\prime \, \leftarrow}(t_{k}^{\ast\ast})
\right)
\ = \
t_{k}^{\ast\ast} \cdot \widetilde{\varphi}^{\, \prime}(t_{k}^{\ast\ast})
- \Lambda_{\widetilde{\mathbb{\bbzeta}}}\left(
\widetilde{\varphi}^{\, \prime}(t_{k}^{\ast\ast})
\right).  
\nonumber
\end{equation}
A comprehensive study of the properties of $\Lambda_{\widetilde{\mathbb{\bbzeta}}}$ 
is given in Section XI of Broniatowski \& Stummer \cite{Bro:23a} as well as 
in Broniatowski \& Stummer \cite{Bro:23b}.

\vspace{0.2cm}
\noindent
Summing up, so far we have proved the representability \eqref{brostu5:fo.link.SBD.b}.
Moreover, recall from Condition \ref{Condition  Fi Tilda in Minimization}
that the moment-generating function
$z \mapsto MGF_{\widetilde{\mathbb{\bbzeta}}}(z) = \int_{\mathbb{R}} e^{z \cdot y} d\widetilde{\mathbb{\bbzeta}}(y)$ 
(and equivalently, $z \mapsto  \Lambda_{\widetilde{\mathbb{\bbzeta}}}(z)$)
is supposed to be finite on some open interval containing zero.
This implies --- due to 
$\Lambda_{\widetilde{U}_{k}}(z) = 
\Lambda_{\widetilde{\mathbb{\bbzeta}}}(z+\tau_{k}) 
- \Lambda_{\widetilde{\mathbb{\bbzeta}}}(\tau_{k}) $ 
(cf. the first equality in \eqref{brostu5:fo.link.SBD.proof1}) 
and $\tau_{k} \in int(dom(\Lambda_{\widetilde{\mathbb{\bbzeta}}}))$ --- that 
also  $z \mapsto  \Lambda_{\widetilde{U}_{k}}(z)$ is finite on some open interval containing zero
($k=1,\ldots,K$).
Hence, one can prove for $D_{\varphi,\mathbf{P}}^{SBD}(\mathbf{Q},\mathbf{Q}^{\ast\ast}) = 
\sum_{k=1}^{K} p_{k} \cdot
\varphi_{k} \negthinspace\left( \frac{q_{k}}{p_{k}}\right)$ (cf. \eqref{brostu5:fo.SBD.smooth}) 
the narrow-sense BS-minimizability \eqref{LDP Minimization SBD} \textit{analogously} to the way
in which the 
narrow-sense BS-minimizability \eqref{LDP Minimization} was proved for 
$D_{\varphi}( \mathbf{Q}, \mathbf{P} ) = 
\sum_{k=1}^{K} p_{k} \cdot
\varphi \left( \frac{q_{k}}{p_{k}}\right)$ (cf. \eqref{brostu5:fo.div})
in Appendix A of Broniatowski \& Stummer \cite{Bro:23a}. For the sake of brevity,
we omit the details. 
\hspace{0.5cm}  $\blacksquare$  

\vspace{0.4cm}
\noindent
\textbf{Proof of Theorem \ref{brostu5:thm.Fmin}.}
It suffices to prove part (b); from there, the part (a) follows immediately as a special case.
Let us define the auxiliary function 
$\widetilde{F}(\mathbf{Q}) :=  D_{\varphi }(\mathbf{Q},\mathbf{P}) - \Phi(\mathbf{Q})$ which by   
our assumption \eqref{brostu5:fo.phibound} satisfies $\widetilde{F}(\mathbf{Q}) \leq c_{1}$
for all $\mathbf{Q} \in \mathbf{\Omega}$. Moreover, by rewriting
\eqref{LDP Minimization} of Theorem \ref{brostu5:thm.BSnarrow} we have
\begin{equation}
\lim_{n\rightarrow \infty }\frac{1}{n}\log \,\mathbb{\Pi }\negthinspace \left[
M_{\mathbf{P}} \cdot \boldsymbol{\xi }_{n}^{\mathbf{\widetilde{W}}}
\in \mathbf{\Omega} 
\right]
= - \inf_{\mathbf{Q}\in \mathbf{\Omega }}D_{\varphi }(\mathbf{Q},\mathbf{P}) \, .
\nonumber
\end{equation}
By applying the Varadhan Lemma (see e.g. the lecture notes of Franco \cite{Fra:15} and 
Swart \cite{Swa:23}, and the references therein) 
to the sequence $(\mu_{n})_{n\in \mathbb{N}}$ of probability distributions on $\mathbb{R}^{K}$
(equipped with the corresponding Borel $\sigma-$field) given by
$\mu_{n}[ \, \cdot \, ] := \mathbb{\Pi }\negthinspace \big[
M_{\mathbf{P}} \cdot \boldsymbol{\xi }_{n}^{\mathbf{\widetilde{W}}}
\in \, \cdot \, \big]$ we obtain (with a slight abuse of notation)
\begin{equation}
\lim_{n\rightarrow \infty }\frac{1}{n}\log \negthinspace \left( \ 
\int_{\mathbb{R}^{K}} 
\exp(n \cdot \widetilde{F}(\mathbf{Q})) \cdot \textfrak{1}_{\mathbf{\Omega}}(\mathbf{Q})
\, \mathrm{d}\mu_{n}(\mathbf{Q}) 
\right)
= \sup_{\mathbf{Q}\in \mathbf{\Omega}} \Big( \widetilde{F}(\mathbf{Q}) \, - \, D_{\varphi }(\mathbf{Q},\mathbf{P}) \Big) \, .
\label{LDP F Minimization}
\end{equation}
As usual, the left-hand side of \eqref{LDP F Minimization} can be equivalently rewritten as
\begin{equation}
\lim_{n\rightarrow \infty }\frac{1}{n}\log \negthinspace \left( \ 
\mathbb{E}_{\mathbb{\Pi}}\negthinspace \Big[
\exp\negthinspace\Big(n \cdot \widetilde{F}\big(M_{\mathbf{P}} \cdot \boldsymbol{\xi }_{n}^{\mathbf{\widetilde{W}}}\big) \Big)
\cdot \textfrak{1}_{\mathbf{\Omega}}\big(M_{\mathbf{P}} \cdot \boldsymbol{\xi }_{n}^{\mathbf{\widetilde{W}}}\big)
\, \Big] 
\right)
\nonumber
\end{equation}
whereas the right-hand side of \eqref{LDP F Minimization} is nothing else but
$\sup_{\mathbf{Q}\in \mathbf{\Omega}} \big( - \Phi(\mathbf{Q}) \big) = 
- \inf_{\mathbf{Q}\in \mathbf{\Omega}} \Phi(\mathbf{Q})$. 
\hspace{0.5cm}  $\blacksquare$  \\

\vspace{0.2cm}
\noindent
\textbf{Proof of bound \eqref{brostu5:fo.boundPOWdivmiss.altern2}.}
Clearly, due to \eqref{brostu5:fo.powdivgen} and \eqref{brostu5:fo.genLap3ba.equal}
it suffices to consider $\frac{\widetilde{c}}{\gamma} =1$. The case $t=0$ is clear:
\begin{eqnarray}
& & \varphi_{\beta,\beta,1}(0) 
\ = \ \beta \cdot \Big\{
\sqrt{2} \, - \, 1 
+ \log\Big(2 \cdot \Big(
\sqrt{2} \, - \, 1
\Big)
\Big)  \Big\}
\ < \ 1 \ = \ \varphi_{1}(0) \, .
\nonumber
\end{eqnarray}
So let $t \in \, ]0,\infty[$. For each $\beta \in \, ]0,\frac{8}{5}]$ we show the derivatives relations
\begin{eqnarray}
0 \ > \ \varphi_{\beta,\beta,1}^{\prime}(t) & > & \varphi_{1}^{\prime}(t)
\qquad \textrm{for all $t \in \, ]0,1[$,}
\label{brostu5:fo.proof.10a} \\
0 \ = \ \varphi_{\beta,\beta,1}^{\prime}(t) & = & \varphi_{1}^{\prime}(t)
\qquad \textrm{$t =1$,}
\label{brostu5:fo.proof.10b} \\
0 \ < \ \varphi_{\beta,\beta,1}^{\prime}(t) & < & \varphi_{1}^{\prime}(t)
\qquad \textrm{for all $t \in \, ]1,\infty[$,}
\label{brostu5:fo.proof.10c} 
\end{eqnarray}
Clearly, it suffices to consider $\beta = \frac{8}{5}$. By straightforward
calculations one gets from \eqref{brostu5:fo.genLap3ba.equal}
for all $t \in \mathbb{R}\backslash\{ 1 \}$
\begin{eqnarray}
& & \varphi_{\beta,\beta,1}^{\prime}(t) = \frac{\beta}{1-t} \cdot \Big(1 - \sqrt{1 + (1-t)^2}\Big)
\nonumber
\end{eqnarray}
(with $\lim_{t \rightarrow 1} \varphi_{\beta,\beta,1}^{\prime}(t) = 0$) and from \eqref{brostu5:fo.powdivgen} for all $t \in \, ]0,\infty[$
\begin{eqnarray}
& & \varphi_{1}^{\prime}(t) = \log(t) = \log(1 + (t-1)).
\nonumber
\end{eqnarray}
Moreover, for fixed $t \in \, ]0,1[$ its transform $x := t-1 \in \, ]-1,0[$
satisfies 
\begin{eqnarray}
& & \log(1 + x) \ < \ \frac{2 \cdot x}{2+x} \ < \ - \, \frac{8}{5 \cdot x} \cdot \Big(1 - \sqrt{1 + x^2}\Big);
\nonumber
\end{eqnarray}
indeed, the left-hand inequality is well-known (see e.g. Topsoe \cite{Top:07}) whereas the right-hand inequality 
follows from 
\begin{eqnarray}
& & \hspace{-1.0cm} \sqrt{1 + x^2} \ < \ 1 + \frac{5 \cdot x^2}{4\cdot (2+x)} 
\ \Longleftrightarrow \ x^2 \cdot (2+x)^2  \ < \ \frac{5 \cdot x^2 \cdot (2+x)}{2} + \frac{25 \cdot x^4}{16}
\ \Longleftrightarrow \ 0  \ < \  \Big(\frac{3 \cdot x}{4} - 1 \Big)^2 \, .
\nonumber
\end{eqnarray}
Analogously, one can show that for fixed $t \in \, ]1,\infty[$ its transform $x := t-1 \in \, ]0,\infty[$
satisfies 
\begin{eqnarray}
& & \log(1 + x) \ > \ \frac{2 \cdot x}{2+x} \ \geq \ - \, \frac{8}{5 \cdot x} \cdot \Big(1 - \sqrt{1 + x^2}\Big)
\nonumber
\end{eqnarray}
where the right-hand inequality turns into an equality if and only of $x=4/3$. Hence, we have shown 
\eqref{brostu5:fo.proof.10a},\eqref{brostu5:fo.proof.10b},\eqref{brostu5:fo.proof.10c} from which
\eqref{brostu5:fo.boundPOWdivmiss.altern2} follows immediately by deducing the monotonicity
properties of the function $t \mapsto g(t) := \varphi_{1}(t) - \varphi_{\beta,\beta,1}(t)$
which satisfies $g(1)=0$.
\hspace{0.5cm}  $\blacksquare$  \\

\vspace{0.1cm}
\noindent
\textbf{Proof of bound \eqref{brostu5:fo.boundPOWdivmiss}.}
Let us arbitrarily fix $\gamma \in \, ]1,\infty[$, $\beta \in \, ]0,1[$ and
$\widetilde{c} \in \, ]0,\infty[$.
For $t \in \, ]0,\infty[$ the bound \eqref{brostu5:fo.boundPOWdivmiss}
follows immediately from \eqref{brostu5:fo.boundPOWdivmiss.altern2} and \eqref{brostu5:fo.boundPOWdivmiss.altern1}.
Furthermore, for each $\gamma \in \, ]1,\infty[$ one gets from
\eqref{brostu5:fo.powdivgen} and \eqref{brostu5:fo.genLap3ba.equal} 
\begin{eqnarray}
& & \varphi_{\beta,\beta,\frac{\widetilde{c}}{\gamma}}(0) 
\ = \ 
\frac{\widetilde{c}}{\gamma} \cdot \beta \cdot \Big\{
\sqrt{2} \, - \, 1 
+ \log\Big(2 \cdot \Big(
\sqrt{2} \, - \, 1
\Big)
\Big)  \Big\}
\ < \ \frac{\widetilde{c}}{\gamma} \ = \ \widetilde{c} \cdot \varphi_{\gamma}(0) \, .
\nonumber
\end{eqnarray}
Moreover, by means of Example \ref{brostu5:ex.TV} one can even show for all $t \in \ ]-\infty,0]$ 
\begin{eqnarray}
& & \varphi_{\beta,\beta,\frac{\widetilde{c}}{\gamma}}(t) 
\ < \ \frac{\widetilde{c}}{\gamma} \cdot (1-t) \ \leq \ 
\frac{\widetilde{c}}{\gamma} \cdot (\gamma \cdot \varphi_{\gamma}(t)) \, 
\nonumber
\end{eqnarray}
where the right-hand inequality follows immediately from \eqref{brostu5:fo.powdivgen}.
\hspace{0.5cm}  $\blacksquare$  \\

\vspace{0.1cm}
\noindent
\textbf{Proof of Theorem \ref{brostu5:thm.Fmin.SBD}.}
We can proceed analogously as in the above proof of
Theorem \ref{brostu5:thm.Fmin},
by replacing the use of Theorem \ref{brostu5:thm.BSnarrow} with the
use of the analogous Theorem \ref{brostu5:thm.BSnarrow.SBD}.
For the sake of brevity, we omit the details.
\hspace{0.5cm}  $\blacksquare$ \\

\vspace{0.1cm}
\noindent
\textbf{Proof of Theorem \ref{brostu5:thm.Fmax}.}\ 
The assertions follow analogously to those of Theorem \ref{brostu5:thm.Fmin},
by instead taking the auxiliary function 
$\widetilde{F}(\mathbf{Q}) :=  D_{\varphi }(\mathbf{Q},\mathbf{P}) + \Phi(\mathbf{Q})$.
\hspace{0.5cm}  $\blacksquare$  \\

\vspace{0.1cm}
\noindent
\textbf{Proof of Theorem \ref{brostu5:thm.Fmax.SBD}.}\ 
We can proceed analogously as in the proof of
Theorem \ref{brostu5:thm.Fmax},
by replacing the use of Theorem \ref{brostu5:thm.BSnarrow} with the
use of the analogous Theorem \ref{brostu5:thm.BSnarrow.SBD}.
For the sake of brevity, we omit the details.
\hspace{0.5cm}  $\blacksquare$ \\

\vspace{0.1cm}
\noindent
\textbf{Proof of Theorem \ref{theorem inner min Bregman power general}.} 
From \eqref{brostu5:fo.scaledBregpow} we obtain straightforwardly
for arbitrary $A >0$, $\widetilde{c} >0$ and $\mathbf{P} \in \mathbb{R}_{>0}^{K}$
\begin{eqnarray}
D_{\widetilde{c} \cdot \varphi_{\gamma},\mathbf{P}}^{SBD}(m \cdot \mathbf{Q},\mathbf{Q}^{\ast\ast})
 \hspace{-0.2cm} &:=& \hspace{-0.2cm}
\begin{cases}
\widetilde{c} \cdot \Big\{\frac{ m^{\gamma} \cdot H_{\gamma}^{(1)}}{\gamma \cdot (\gamma-1)}
+ \frac{H_{\gamma}^{(2)}}{\gamma}
- \frac{m \cdot T_{\gamma}}{\gamma -1} 
\Big\}, 
\hspace{0.85cm}
\textrm{if }  \gamma \in \, ]-\infty,0[, \,  
\mathbf{Q}^{\ast\ast} \in \mathbb{R}_{>0}^{K}, \,
\mathbf{Q} \in A \cdot \mathbb{S}_{>0}^{K} 
\ \textrm{and } m>0, 
 \\
\widetilde{c} \cdot \Big\{ \breve{T}_{0} - M_{\mathbf{P}} \cdot \log m + m \cdot T_{0} - M_{\mathbf{P}} \Big\}, 
\hspace{0.5cm} \textrm{if }  \gamma = 0, \,  
\mathbf{Q}^{\ast\ast} \in \mathbb{R}_{>0}^{K}, \, 
\mathbf{Q} \in A \cdot \mathbb{S}_{>0}^{K} 
\ \textrm{and } m>0,
\\
\widetilde{c} \cdot \Big\{\frac{ m^{\gamma} \cdot H_{\gamma}^{(1)}}{\gamma \cdot (\gamma-1)}
+ \frac{H_{\gamma}^{(2)}}{\gamma}
- \frac{m \cdot T_{\gamma}}{\gamma -1} 
\Big\}, 
\hspace{1.55cm}
\textrm{if }  \gamma \in \, ]0,1[, \,  
\mathbf{Q}^{\ast\ast} \in \mathbb{R}_{>0}^{K}, \, \mathbf{Q} \in A \cdot \mathbb{S}^{K} 
\ \textrm{and } m \geq 0,  
 \\
\widetilde{c} \cdot \Big\{A \cdot m \cdot \log m + m \cdot (I-A) + M_{\mathbf{Q}^{\ast\ast}} \Big\}, 
\hspace{0.35cm} \textrm{if }  \gamma = 1, \,  
\mathbf{Q}^{\ast\ast} \in \mathbb{R}_{>0}^{K}, \, \mathbf{Q} \in  A \cdot \mathbb{S}^{K}
\ \textrm{and } m \geq 0,
\\
\widetilde{c} \cdot \Big\{\frac{ m^{\gamma} \cdot H_{\gamma}^{(1)}
\cdot \textfrak{1}_{[0,\infty[}(m)
}{\gamma \cdot (\gamma-1)}
+ \frac{H_{\gamma}^{(2)}}{\gamma}
- \frac{m \cdot T_{\gamma}}{\gamma -1} 
\Big\}, 
\\
\hspace{4.8cm} 
\textrm{if }  \gamma \in \, ]1,2[, \,  
\mathbf{Q}^{\ast\ast} \in \mathbb{R}_{>0}^{K}, \, \mathbf{Q} \in A \cdot \mathbb{S}^{K}
\ \textrm{and } m \in \, ]-\infty,\infty[,
\\
\widetilde{c} \cdot \Big\{\frac{ m^{2} \cdot H_{2}^{(1)}}{2}
+ \frac{H_{2}^{(2)}}{2}
- m \cdot T_{2}
\Big\}, 
\hspace{0.75cm}
\textrm{if }  \gamma = 2, \,  
\mathbf{Q}^{\ast\ast} \in \mathbb{R}^{K}, \, \mathbf{Q} \in A \cdot \mathbb{S}^{K}
\ \textrm{and } m \in \, ]-\infty,\infty[,
\\
\widetilde{c} \cdot \Big\{\frac{ m^{\gamma} \cdot H_{\gamma}^{(1)}
\cdot \textfrak{1}_{[0,\infty[}(m)
}{\gamma \cdot (\gamma-1)}
+ \frac{H_{\gamma}^{(2)}}{\gamma}
- \frac{m \cdot T_{\gamma}}{\gamma -1} 
\Big\}, 
\\
\hspace{4.6cm}   
\textrm{if }  \gamma \in \, ]2,\infty[, \,  
\mathbf{Q}^{\ast\ast} \in \mathbb{R}_{>0}^{K}, \, \mathbf{Q} \in A \cdot \mathbb{S}^{K}
\ \textrm{and } m \in \, ]-\infty,\infty[,
\\
\infty, \hspace{4.1cm} \textrm{else},
\end{cases}
\label{brostu5:fo.scaledBregpow.m} 
\end{eqnarray}
where we have employed the five $m-$independent abbreviations 
\begin{eqnarray}
&& H_{\gamma}^{(1)} := H_{\gamma}(\mathbf{Q},\mathbf{P}) =
\sum\displaylimits_{k=1}^{K} (q_{k})^{\gamma} \cdot (p_{k})^{1-\gamma},
\qquad
H_{\gamma}^{(2)} := H_{\gamma}(\mathbf{Q}^{\ast\ast},\mathbf{P}),
\qquad \text{(cf. \eqref{brostu5:fo.divpow.hellinger1.extended},\eqref{brostu5:fo.divpow.hellinger1.extended2})}
\nonumber
\\
&& T_{\gamma} := T_{\gamma}(\mathbf{Q},\mathbf{Q}^{\ast\ast},\mathbf{P}) := 
\sum\displaylimits_{k=1}^{K} q_{k} \cdot  (q_{k}^{\ast\ast})^{\gamma-1} \cdot (p_{k})^{1-\gamma},
\qquad \text{(cf. \eqref{brostu5:fo.triplepowersum})}
\nonumber
\\
&& I:= I(\mathbf{Q},\mathbf{Q}^{\ast\ast}) :=  
\sum\displaylimits_{k=1}^{K} q_{k} \cdot \log\left( \frac{q_{k}}{q_{k}^{\ast\ast}} \right),
\qquad \text{(cf. \eqref{brostu3:fo.divpow.Kull1.SBD})}
\nonumber
\\
&& \breve{T}_{0} := \breve{T}_{0}(\mathbf{Q},\mathbf{Q}^{\ast\ast},\mathbf{P}) := 
- \sum\displaylimits_{k=1}^{K} p_{k} \cdot \log\left( \frac{q_{k}}{q_{k}^{\ast\ast}} \right).
\qquad \text{(cf. \eqref{brostu5:fo.triplepowersum.log})}
\nonumber
\end{eqnarray}
\noindent
To proceed, let us fix an arbitrary constant $\widetilde{c} >0$.\\
(i) Case $\gamma \cdot (1-\gamma) \ne 0$. \\
\noindent
(ia) Let us start with the subcase $\gamma \in \, ]-\infty,0[$.
From the first and the last line of \eqref{brostu5:fo.scaledBregpow.m}, it is clear that
the corresponding $m-$infimum can not be achieved for $m \leq 0$;
since $H_{\gamma}^{(1)} > 0$ and $T_{\gamma} > 0$, 
one gets the unique minimizer $m_{min} = \Big(\frac{H_{\gamma}^{(1)}}{T_{\gamma}}\Big)^{1/(1-\gamma)} >0$
and the minimum 
\begin{eqnarray}
\inf_{m\neq 0} D_{\widetilde{c} \cdot \varphi_{1},\mathbf{P}}^{SBD}(m \cdot \mathbf{Q},\mathbf{Q}^{\ast\ast}) = 
D_{\widetilde{c} \cdot \varphi_{1},\mathbf{P}}^{SBD}(m_{min} \cdot \mathbf{Q},\mathbf{Q}^{\ast\ast})
=\frac{\widetilde{c}}{\gamma}
\cdot \left[H_{\gamma}^{(2)} - 
(T_{\gamma})^{\gamma/(\gamma-1)} \cdot 
(H_{\gamma}^{(1)})^{-1/\left( \gamma -1\right) }\right].
\nonumber
\end{eqnarray}
Thus, \eqref{brostu3:fo.676b.SBD} is established and from there, 
\eqref{LDP Normalized Vec BS2 nonnormalized SBD 2 simplex}
follows by means of \eqref{LDP Normalized Vec BS2 nonnormalized SBD 2}.\\
(ib) The subcase $\gamma \in \, ]0,1[$ (cf. the third line of \eqref{brostu5:fo.scaledBregpow.m})
works analogously to the subcase (ia) since  $H_{\gamma}^{(1)} > 0$ and $T_{\gamma} > 0$ due to the fact
that $\mathbf{Q} \in A \cdot \mathbb{S}^{K}$ implies that all the
components $q_{k}$ are nonnegative and that at least one component $q_{k}$ must be strictly positive.\\
(ic) In the subcase $\gamma \in \, ]1,2[ \, \cup \, ]2,\infty[$ (cf. the fifth and seventh
line of \eqref{brostu5:fo.scaledBregpow.m}) it is straightforward to see that the desired
infimum can not be achieved for $m < 0$. Hence, one can proceed analogously
to subcase (ia).\\
(id) Let $\gamma =2$, $\mathbf{Q}^{\ast\ast} \in \mathbb{R}^{K}$ and $\mathbf{Q} \in A \cdot \mathbb{S}^{K}$
for some $A \in \mathbb{R}\backslash\{0\}$.
The latter implies that 
either 
\textquotedblleft  all components $q_{k}$
are nonnegative and at least one component is strictly positive\textquotedblright\ 
(if $A>0$) or
\textquotedblleft  all components $q_{k}$
are nonpositive and at least one component is strictly negative\textquotedblright\
(if $A<0$). There holds $H_{2}^{(1)} > 0$, $H_{2}^{(2)} \geq 0$ and $T_{2} \in \, ]-\infty,\infty[$.
For $T_{\gamma} \ne 0$, we obtain 
the unique minimizer $m_{min} = \frac{T_{2}}{H_{2}^{(1)}} \ne 0$
and the minimum 
\begin{eqnarray}
\inf_{m\neq 0} D_{\widetilde{c} \cdot \varphi_{2},\mathbf{P}}^{SBD}(m \cdot \mathbf{Q},\mathbf{Q}^{\ast\ast}) = 
D_{\widetilde{c} \cdot \varphi_{2},\mathbf{P}}^{SBD}(m_{min} \cdot \mathbf{Q},\mathbf{Q}^{\ast\ast})
=\frac{\widetilde{c}}{2}
\cdot \left[H_{2}^{(2)} - 
(T_{2})^{2} \cdot 
(H_{2}^{(1)})^{-1}\right].
\label{brostu5:fo.gamma2}
\end{eqnarray}
In the subcase $T_{2}=0$, one arrives immediately at
\begin{eqnarray}
\inf_{m\neq 0} D_{\widetilde{c} \cdot \varphi_{2},\mathbf{P}}^{SBD}(m \cdot \mathbf{Q},\mathbf{Q}^{\ast\ast}) = 
\inf_{m\neq 0} \frac{\widetilde{c}}{2} \cdot \left[m^{2} \cdot H_{2}^{(1)} + H_{2}^{(2)} \right]
=\frac{\widetilde{c}}{2}
\cdot H_{2}^{(2)}
\nonumber
\end{eqnarray}
which is the corresponding special case of \eqref{brostu5:fo.gamma2}.\\
(ii) Case $\gamma =1$.
From the fourth and the last line of \eqref{brostu5:fo.scaledBregpow.m}, it is clear that
the corresponding $m-$infimum can not be achieved for $m \leq 0$.
We obtain the unique minimizer $m_{min} = \exp\{-I/A\} >0$
and the minimum 
\begin{eqnarray}
\inf_{m\neq 0} D_{\widetilde{c} \cdot \varphi_{1},\mathbf{P}}^{SBD}(m \cdot \mathbf{Q},\mathbf{Q}^{\ast\ast}) = 
D_{\widetilde{c} \cdot \varphi_{1},\mathbf{P}}^{SBD}(m_{min} \cdot \mathbf{Q},\mathbf{Q}^{\ast\ast})
=\widetilde{c}\cdot \Big[ M_{\mathbf{Q}^{\ast\ast}} - A \cdot \exp
\Big( -\frac{I}{A}\Big) \Big] . 
\nonumber
\end{eqnarray}
Thus, \eqref{brostu3:fo.677a.SBD} is established and from there, 
\eqref{LDP Normalized Vec BS2 nonnormalized SBD 2 simplex KL}
follows by means of \eqref{LDP Normalized Vec BS2 nonnormalized SBD 2}.\\
(iii) Case $\gamma =0$.
From the second and the last line of \eqref{brostu5:fo.scaledBregpow.m}, it is clear that
the corresponding $m-$infimum can not be achieved for $m \leq 0$.
By straightforward calculations, we deduce the unique minimizer $m_{min} = \frac{M_{\mathbf{P}}}{T_{0}} >0$
and the corresponding minimum 
\begin{eqnarray}
\inf_{m\neq 0} D_{\widetilde{c} \cdot \varphi_{0},\mathbf{P}}^{SBD}(m \cdot \mathbf{Q},\mathbf{Q}^{\ast\ast}) = 
D_{\widetilde{c} \cdot \varphi_{0},\mathbf{P}}^{SBD}(m_{min} \cdot \mathbf{Q},\mathbf{Q}^{\ast\ast})
= \ \widetilde{c} \cdot \Big[ 
M_{\mathbf{P}} \cdot \log(T_{0})
+ \breve{T}_{0}
-  M_{\mathbf{P}} \cdot \log( M_{\mathbf{P}}) \Big] .
\nonumber
\end{eqnarray}
Thus, \eqref{brostu3:fo.678.SBD} is proved and from there, 
\eqref{LDP Normalized Vec BS2 nonnormalized SBD 2 simplex RKL}
can be derived by means of \eqref{LDP Normalized Vec BS2 nonnormalized SBD 2}.
\hspace{0.5cm}  $\blacksquare$  \\

\vspace{0.1cm}
\noindent
\textbf{Proof of Theorem \ref{brostu5:thm.Fmin.simplex}.}\ 
We follow the lines of the above proof of Theorem \ref{brostu5:thm.Fmin}.
It suffices to prove part (b). Let us define the auxiliary function 
$\widetilde{F}(\mathbf{Q}) :=  F_{\gamma,\widetilde{c},A}\Big(D_{\widetilde{c} 
\cdot \varphi_{\gamma}}(\mathbf{Q},\mathds{P})\Big) - \Phi(\mathbf{Q})$
which by \eqref{brostu5:fo.phibound.simplex} satisfies $\widetilde{F}(\mathbf{Q}) \leq c_{1}$
for all $\mathbf{Q} \in A \cdot \textrm{$\boldsymbol{\Omega}$\hspace{-0.23cm}$\boldsymbol{\Omega}$}$. 
Moreover, by rewriting
\eqref{LDP Normalized Vec simplex} of Theorem 
\ref{brostu5:thm.divnormW.new} we have
\begin{equation}
\lim_{n\rightarrow \infty }\frac{1}{n}\log \, 
\mathbb{\Pi}\negthinspace \left[A \cdot \boldsymbol{\xi}_{n}^{w\mathbf{W}}\in 
A \cdot \textrm{$\boldsymbol{\Omega}$\hspace{-0.23cm}$\boldsymbol{\Omega}$}\right] 
= - \inf_{\mathbf{Q}\in A \cdot \textrm{$\boldsymbol{\Omega}$\hspace{-0.19cm}$\boldsymbol{\Omega}$} }
F_{\gamma,\widetilde{c},A}\Big(D_{\widetilde{c} \cdot \varphi_{\gamma}}(\mathbf{Q},\mathds{P})\Big)
\, .
\nonumber
\end{equation}
By applying the Varadhan Lemma to the sequence $(\mu_{n})_{n\in \mathbb{N}}$ 
of probability distributions on $\mathbb{R}^{K}$
(equipped with the corresponding Borel $\sigma-$field) given by
$\mu_{n}[ \, \cdot \, ] := \mathbb{\Pi }\negthinspace \big[
A \cdot \boldsymbol{\xi}_{n}^{w\mathbf{W}}
\in \, \cdot \, \big]$ we obtain (with a slight abuse of notation)
\begin{equation}
\lim_{n\rightarrow \infty }\frac{1}{n}\log \negthinspace \left( \ 
\int_{\mathbb{R}^{K}} 
\exp(n \cdot \widetilde{F}(\mathbf{Q})) \cdot \textfrak{1}_{A \cdot \textrm{$\boldsymbol{\Omega}$\hspace{-0.19cm}$\boldsymbol{\Omega}$}}(\mathbf{Q})
\, \mathrm{d}\mu_{n}(\mathbf{Q}) 
\right)
= \sup_{\mathbf{Q}\in A \cdot \textrm{$\boldsymbol{\Omega}$\hspace{-0.19cm}$\boldsymbol{\Omega}$}} 
\Big( \widetilde{F}(\mathbf{Q}) \, - \, F_{\gamma,\widetilde{c},A}\Big(D_{\widetilde{c} 
\cdot \varphi_{\gamma}}(\mathbf{Q},\mathds{P})\Big) \Big) \, .
\label{LDP F Minimization simplex}
\end{equation}
As usual, the left-hand side of \eqref{LDP F Minimization simplex} can be equivalently rewritten as
\begin{equation}
\lim_{n\rightarrow \infty }\frac{1}{n}\log \negthinspace \left( \ 
\mathbb{E}_{\mathbb{\Pi}}\negthinspace \Big[
\exp\negthinspace\Big(n \cdot \widetilde{F}\big(A \cdot \boldsymbol{\xi}_{n}^{w\mathbf{W}}\big) \Big)
\cdot \textfrak{1}_{A \cdot \textrm{$\boldsymbol{\Omega}$\hspace{-0.19cm}$\boldsymbol{\Omega}$}}\big(A \cdot \boldsymbol{\xi}_{n}^{w\mathbf{W}}\big)
\, \Big] 
\right)
\nonumber
\end{equation}
whereas the right-hand side of \eqref{LDP F Minimization simplex} is nothing else but
$\sup_{\mathbf{Q}\in A \cdot \textrm{$\boldsymbol{\Omega}$\hspace{-0.19cm}$\boldsymbol{\Omega}$}} 
\big( - \Phi(\mathbf{Q}) \big) = - \inf_{\mathbf{Q}\in A \cdot \textrm{$\boldsymbol{\Omega}$\hspace{-0.19cm}$\boldsymbol{\Omega}$}} \Phi(\mathbf{Q})$.
\hspace{0.5cm}  $\blacksquare$  \\

\vspace{0.1cm}
\noindent
\textbf{Proof of Theorem \ref{brostu5:thm.Fmax.simplex}.}\ 
The assertions follow analogously to those of Theorem \ref{brostu5:thm.Fmin.simplex},
by instead taking the auxiliary function 
$\widetilde{F}(\mathbf{Q}) :=  F_{\gamma,\widetilde{c},A}\Big(D_{\widetilde{c} 
\cdot \varphi_{\gamma}}(\mathbf{Q},\mathds{P})\Big) + \Phi(\mathbf{Q})$. 
\hspace{0.5cm}  $\blacksquare$  \\

\vspace{0.1cm}
\noindent
\textbf{Proof of Theorem \ref{brostu5:thm.Fmin.simplex.SBD}.}
We can proceed analogously as in the proof of
Theorem \ref{brostu5:thm.Fmin.simplex},
by replacing the use of \eqref{LDP Normalized Vec simplex} of Theorem \ref{brostu5:thm.divnormW.new} 
with the use of the analogous \eqref{LDP Normalized Vec BS2 nonnormalized SBD 2 simplex} 
(respectively \eqref{LDP Normalized Vec BS2 nonnormalized SBD 2 simplex KL} 
respectively \eqref{LDP Normalized Vec BS2 nonnormalized SBD 2 simplex RKL}) 
of Theorem \ref{theorem inner min Bregman power general}.
In particular, 
$F_{\gamma,\widetilde{c},A}\Big(D_{\widetilde{c} \cdot \varphi_{\gamma}}(\mathbf{Q},\mathds{P})\Big)$
is replaced by $\breve{D}_{\widetilde{c} \cdot \varphi_{\gamma},\mathbf{P}}^{SBD}( \mathbf{Q},\mathbf{Q}^{\ast\ast})$.
For the sake of brevity, we omit the details.
\hspace{0.5cm}  $\blacksquare$ \\

\vspace{0.1cm}
\noindent
\textbf{Proof of Theorem \ref{brostu5:thm.Fmax.simplex.SBD}.}
The assertions follow analogously to those of 
Theorem \ref{brostu5:thm.Fmax.simplex},
by employing the changes indicated in the above proof of Theorem \ref{brostu5:thm.Fmin.simplex.SBD}.
For the sake of brevity, we omit the details.
\hspace{0.5cm}  $\blacksquare$ \\

\vspace{0.1cm}
\noindent
\textbf{Proof of Theorem \ref{brostu3:thm.divnormW.new.det.SBD.risk}.} \ 
Recall from \eqref{brostu5:fo.phi_k} and \eqref{brostu5:fo.SBD.smooth} that we can represent 
the involved scaled Bregman distances as
$D_{\varphi,\mathds{P}}^{SBD}(\mathds{Q},\mathbf{Q}^{\ast\ast})= 
\sum_{k=1}^{K} p_{k} \cdot
\varphi_{k} \negthinspace\left( \frac{q_{k}}{p_{k}}\right)$, where 
for each $k=1,\ldots,K$ one has 
\begin{equation}
\varphi_{k}(t) \ = \ 
\sup_{z \in \mathbb{R}} \Big( z\cdot t - \log \int_{\mathbb{R}} e^{z \cdot y} dU_{k}(y) \Big),
\qquad t \in \mathbb{R},
\nonumber
\end{equation}
cf. \eqref{brostu5:fo.link.SBD.b} with omittance of the tildes since 
$\widetilde{\varphi}_{k}(\cdot) : = M_{\mathds{P}} \cdot \varphi_{k}(\cdot) = \varphi_{k}(\cdot)$
and $\widetilde{\mathbb{\bbzeta}}=\mathbb{\bbzeta}$; moreover,
let $V := (\mathbf{V}_{n})_{n \in \mathbb{N}}$ 
be a sequence of random vectors constructed (with the help of $\mathbb{\bbzeta}$) 
via \eqref{brostu5:V_new} and \eqref{brostu5:Utilde_k_new} (without tildes).
Moreover, recall the vector 
\begin{eqnarray}
\boldsymbol{\xi}_{n,\mathbf{X}}^{w\mathbf{V}} &:=&
\begin{cases}
\left(\frac{\sum_{i \in I_{1}^{(n)}(\mathbf{X}_{1}^{n})}V_{i}}{\sum_{k=1}^{K}
\sum_{i \in I_{k}^{(n)}(\mathbf{X}_{1}^{n})}V_{i}},
\ldots, \frac{\sum_{i \in I_{K}^{(n)}(\mathbf{X}_{1}^{n})}V_{i}}{\sum_{k=1}^{K}
\sum_{i \in I_{k}^{(n)}}V_{i}(\mathbf{X}_{1}^{n})} \right) ,
\qquad \textrm{if } \sum_{j=1}^{n} V_{j} \ne 0, \\
\ (\infty, \ldots, \infty) =: \boldsymbol{\infty}, \hspace{5.0cm} \textrm{if } \sum_{j=1}^{n} V_{j} = 0,
\end{cases}
\quad \textrm{(cf. \eqref{brostu5:fo.norweiemp.vec.det.SBD.risk})} 
\nonumber
\end{eqnarray}
and define
\begin{equation}
\boldsymbol{\xi}_{n,\mathbf{X}}^{\mathbf{V}}:=\Big(\frac{1}{n}\sum_{i\in
I_{1}^{(n)}(\mathbf{X}_{1}^{n})}V_{i},\ldots ,\frac{1}{n}\sum_{i\in I_{K}^{(n)}(\mathbf{X}_{1}^{n})}
V_{i}\Big)
\nonumber
\end{equation}
where the right-hand side structurally coincides with \eqref{Xi_n^W vector V new2} (without tildes) but
however --- as explained above --- the construction 
of the involved $I_{k}^{(n)}$ differs which is reflected in different indexing on the corresponding
left-hand sides. Recall that inside the block $I_{k}^{(n)}$, the $V_{i}$\textquoteright s
are independent with the same distribution 
$U_{k}$ (cf. \eqref{brostu5:Utilde_k_new}) (depending on the block).
With the help of these ingredients, to prove \eqref{LDP Normalized Vec BS2 SBD risk}
we can follow nearly verbatim our proof of Theorem 12 in Broniatowski \& Stummer \cite{Bro:23a};
in the following, for the sake of brevity we explain only major steps.
Firstly, we show for $m \ne 0$ that
\begin{equation}
\left\{ \boldsymbol{\xi}_{n,\mathbf{X}}^{\mathbf{V}} \in 
m\cdot \textrm{$\boldsymbol{\Omega}$\hspace{-0.23cm}$\boldsymbol{\Omega}$}\right\} \subset 
\left\{ \frac{1}{n}\sum_{i=1}^{n} V_{i} =m
\right\} .
\label{equ.00p.BS2}
\end{equation}
To see this, $\boldsymbol{\xi}_{n,\mathbf{X}}^{\mathbf{V}} \in 
m\cdot \textrm{$\boldsymbol{\Omega}$\hspace{-0.23cm}$\boldsymbol{\Omega}$}$ means
that there exists a probability vector $\mathds{Q} :=\{q_{1},\ldots,q_{K}\}$
such that $\boldsymbol{\xi}_{n,\mathbf{X}}^{\mathbf{V}} =  m \cdot \mathds{Q}$
and thus the sum of the components of $\boldsymbol{\xi}_{n,\mathbf{X}}^{\mathbf{V}}$
(which is equal to $\frac{1}{n}\sum_{i=1}^{n} V_{i}$) must be $m$.
To proceed, \eqref{equ.00p.BS2} implies
\begin{equation}
\frac{1}{n}\log \mathbb{\Pi}_{\mathbf{X}_{1}^{n}}\negthinspace \left[\boldsymbol{\xi}_{n,\mathbf{X}}^{w\mathbf{V}}
\in \textrm{$\boldsymbol{\Omega}$\hspace{-0.23cm}$\boldsymbol{\Omega}$} \right] =
\frac{1}{n}\log \mathbb{\Pi}_{\mathbf{X}_{1}^{n}}\negthinspace \bigg[ \boldsymbol{\xi}_{n,\mathbf{X}}^{\mathbf{V}} \in \bigcup\limits_{m\neq
0}m\cdot \textrm{$\boldsymbol{\Omega}$\hspace{-0.23cm}$\boldsymbol{\Omega}$} \bigg] 
\nonumber
\end{equation}
(analogously to the proof of equality (172) in \cite{Bro:23a}).
Secondly, in the present context the Lemma 42 of \cite{Bro:23a} still holds,
and the same is true for inequality (175) of \cite{Bro:23a} with $\Phi_{\mathds{P}}(\cdot) := 
D_{\varphi,\mathds{P}}^{SBD}(\cdot,\mathbf{Q}^{\ast\ast})$
as well as for inequality (173) of \cite{Bro:23a} with $\mathbf{V}$ instead of $\mathbf{W}$ 
(where we employ the corresponding part of the proof of Theorem \ref{brostu5:thm.BSnarrow.SBD} 
(of the current paper) instead of 
Proposition 39 of \cite{Bro:23a}).
From this, \eqref{LDP Normalized Vec BS2 SBD risk} (of the current paper) follows by considerations which are
analogous to those in the last paragraph of the proof of Theorem 12 in \cite{Bro:23a}. 
\hspace{0.5cm}  $\blacksquare$  \\

\vspace{0.1cm}
\noindent
\textbf{Proof of Theorem \ref{brostu3:thm.divnormW.new.det.SBD}.}
The corresponding assertions follow analogously to the above proof of 
Theorem \ref{brostu3:thm.divnormW.new.det.SBD.risk}, by taking
$\boldsymbol{\xi}_{n}^{w\mathbf{V}}$ instead of
$\boldsymbol{\xi}_{n,\mathbf{X}}^{w\mathbf{V}}$,
and $\mathbb{\Pi}$ instead of
$\mathbb{\Pi}_{\mathbf{X}_{1}^{n}}$. \hspace{0.5cm}  $\blacksquare$ \\

\vspace{0.1cm}
\noindent
\textbf{Proof of Proposition \ref{brostu5:prop.generaldeterministic.minimizer.naive}.}
We get, by the definition of 
$\Phi \left( M_{\mathbf{P}} \cdot \boldsymbol{\xi}_{n}^{\mathbf{\widetilde{W}}^{L,\ast}} \right)$ 
and \eqref{brostu5:fo.BSmin.extended.naive.estim} that
\begin{eqnarray}
\widehat{\Phi(\mathbf{\Omega})}_{n,L}^{naive,1} &=&
- \frac{1}{n}\log \frac{1}{L}\sum_{\ell =1}^{L} 
\exp\negthinspace\Big(n \cdot \Big(
D_{\varphi }\big(M_{\mathbf{P}} \cdot \boldsymbol{\xi }_{n}^{\mathbf{\widetilde{W}}^{(\ell)}},\mathbf{P}\big) 
- \Phi\big(M_{\mathbf{P}} \cdot \boldsymbol{\xi }_{n}^{\mathbf{\widetilde{W}}^{(\ell)}}\big)
\Big)
\Big)
\cdot
\mathbf{1}_{\mathbf{\Omega}}
\Big(M_{\mathbf{P}} \cdot \boldsymbol{\xi}_{n}^{\mathbf{\widetilde{W}}^{(\ell)}} \Big)
\nonumber
\\
&\geq&  
\Phi \left( M_{\mathbf{P}} \cdot \boldsymbol{\xi}_{n}^{\mathbf{\widetilde{W}}^{L,\ast}} \right) 
\ - \ 
\frac{1}{n}\log \frac{1}{L}\sum_{\ell =1}^{L} 
\exp\negthinspace\Big(n \cdot \Big(
D_{\varphi }\big(M_{\mathbf{P}} \cdot \boldsymbol{\xi }_{n}^{\mathbf{\widetilde{W}}^{(\ell)}},\mathbf{P}\big) 
\Big)
\Big)
\cdot
\mathbf{1}_{\mathbf{\Omega}}
\Big(M_{\mathbf{P}} \cdot \boldsymbol{\xi}_{n}^{\mathbf{\widetilde{W}}^{(\ell)}} \Big)
\ .  
\nonumber
\end{eqnarray}
For fixed $n$, by the strong law of large numbers there holds a.s. 
\begin{eqnarray*}
& & 
\lim_{L \rightarrow \infty}
\frac{1}{L}\sum_{\ell =1}^{L} 
\exp\negthinspace\Big(n \cdot \Big(
D_{\varphi }\big(M_{\mathbf{P}} \cdot \boldsymbol{\xi }_{n}^{\mathbf{\widetilde{W}}^{(\ell)}},\mathbf{P}\big) 
\Big)
\Big)
\cdot
\mathbf{1}_{\mathbf{\Omega}}
\Big(M_{\mathbf{P}} \cdot \boldsymbol{\xi}_{n}^{\mathbf{\widetilde{W}}^{(\ell)}} \Big)\\
&& = \ E_{\mathbb{\Pi}}\left[\exp\negthinspace\Big(n \cdot \Big(
D_{\varphi }\big(M_{\mathbf{P}} \cdot \boldsymbol{\xi }_{n}^{\mathbf{\widetilde{W}}},\mathbf{P}\big) 
\Big)
\Big)
\cdot
\mathbf{1}_{\mathbf{\Omega}}
\Big(M_{\mathbf{P}} \cdot \boldsymbol{\xi}_{n}^{\mathbf{\widetilde{W}}} \Big)
\right] .
\end{eqnarray*}
Since we have assumed that $\mathbf{\Omega}$ is compact, 
we get from the application of Theorem \ref{brostu5:thm.Fmin}(a) to the constant null function
$\breve{\phi}(\cdot) \equiv 0$ on $\mathbf{\Omega}$ that
\begin{equation}
\lim_{n\rightarrow \infty }\frac{1}{n}\log 
E_{\mathbb{\Pi}}\left[\exp\negthinspace\Big(n \cdot \Big(
D_{\varphi }\big(M_{\mathbf{P}} \cdot \boldsymbol{\xi }_{n}^{\mathbf{\widetilde{W}}},\mathbf{P}\big) 
\Big)
\Big)
\cdot
\mathbf{1}_{\mathbf{\Omega}}
\Big(M_{\mathbf{P}} \cdot \boldsymbol{\xi}_{n}^{\mathbf{\widetilde{W}}} \Big)
\right] \ = \ 0,
\nonumber
\end{equation}
which completes the proof of \eqref{minim}.
\hspace{0.5cm}  $\blacksquare$ \\

\vspace{0.1cm}
\noindent
\textbf{Proof of Proposition \ref{brostu5:prop.coincide}.}
First, we apply the following
representation 
\begin{equation}
\frac{\mathrm{d}\breve{\mathbb{\bbzeta}}}{\mathrm{d}\breve{\mathbb{S}}}
\negthinspace\left(M_{\mathbf{P}} \cdot \boldsymbol{\xi}_{n}^{\mathbf{\widetilde{V}}}\right)
\ = \
\exp \Big(\sum_{k=1}^{K} \, \Big( 
n_{k} \cdot \Lambda_{\widetilde{\mathbb{\bbzeta}}}(\tau_{k})
- \tau _{k} \cdot \sum_{i\in I_{k}^{(n)}} \mathbf{\widetilde{V}}_{i} \Big)  
\Big) 
\label{brostu5:Sdiffform}
\end{equation}
(cf. (114) in Subsection X-A of Broniatowski \& Stummer \cite{Bro:23a}).
By plugging \eqref{brostu5:Sdiffform} into 
\eqref{brostu5:fo.BSmin.extended.approx3}, the involved expectation becomes
\begin{equation}
\mathbb{E}_{\mathbb{\Pi}}\negthinspace \Big[ \, 
\textfrak{1}_{\mathbf{\Omega}}\big(M_{\mathbf{P}} \cdot \boldsymbol{\xi }_{n}^{\mathbf{\widetilde{V}}}\big)
\cdot
\exp\negthinspace\Big(n \cdot \Big(
D_{\varphi }\big(M_{\mathbf{P}} \cdot \boldsymbol{\xi }_{n}^{\mathbf{\widetilde{V}}},\mathbf{P}\big) 
- \Phi\big(M_{\mathbf{P}} \cdot \boldsymbol{\xi }_{n}^{\mathbf{\widetilde{V}}}\big)
\Big)
\Big)
\cdot 
\exp \Big(\sum_{k=1}^{K} \, \Big( 
n_{k} \cdot \Lambda_{\widetilde{\mathbb{\bbzeta}}}(\tau_{k})
- \tau _{k} \cdot \sum_{i\in I_{k}^{(n)}} \mathbf{\widetilde{V}}_{i} \Big) 
\Big) 
\, \Big] .
\nonumber
\end{equation}

\vspace{-0.3cm}
\noindent
Moreover, 
\begin{eqnarray}
& & \hspace{-0.5cm}
D_{\varphi}\big(M_{\mathbf{P}} \cdot \boldsymbol{\xi }_{n}^{\mathbf{\widetilde{V}}},\mathbf{P}\big)
+ \frac{1}{n} \cdot \sum_{k=1}^{K} \, \Big( 
n_{k} \cdot \Lambda_{\widetilde{\mathbb{\bbzeta}}}(\tau_{k})
- \tau _{k} \cdot \sum_{i\in I_{k}^{(n)}} \mathbf{\widetilde{V}}_{i} \Big) 
=  
D_{\widetilde{\varphi}}(\boldsymbol{\xi }_{n}^{\mathbf{\widetilde{V}}},\widetilde{\mathds{P}})
+ \sum_{k=1}^{K} \, \frac{n_{k}}{n} \cdot \Lambda_{\widetilde{\mathbb{\bbzeta}}}(\tau_{k})
- \sum_{k=1}^{K} \, \tau_{k} \cdot  \frac{1}{n} \sum_{i\in I_{k}^{(n)}} \mathbf{\widetilde{V}}_{i} 
\nonumber
\\
&& 
= \
D_{\widetilde{\varphi}}(\boldsymbol{\xi }_{n}^{\mathbf{\widetilde{V}}},\widetilde{\mathds{P}})
+ \sum_{k=1}^{K} \, \widetilde{p_{k}} \cdot \Lambda_{\widetilde{\mathbb{\bbzeta}}}(\tau_{k})
- \sum_{k=1}^{K} \, \widetilde{p_{k}} \cdot \tau_{k} \cdot 
\frac{1}{\widetilde{p_{k}}} \cdot \frac{1}{n} \sum_{i\in I_{k}^{(n)}} \mathbf{\widetilde{V}}_{i} 
\label{brostu5:fo.BSmin.extended.approx7}
\\
&& = \
\sum_{k=1}^{K} \, \widetilde{p_{k}} \cdot 
\left[ \,
\widetilde{\varphi} \left( \frac{\widetilde{x}_{k}}{\widetilde{p}_{k}}\right)
+
\Lambda_{\widetilde{\mathbb{\bbzeta}}}(\tau_{k})
-  \tau_{k} \cdot 
\frac{\widetilde{x}_{k}}{\widetilde{p}_{k}} \, \right]
\ = \ 
\sum_{k=1}^{K} \widetilde{p}_{k} \cdot
\widetilde{\varphi}_{k} \negthinspace\left( \frac{\widetilde{x}_{k}}{\widetilde{p}_{k}}\right)
\ = \ 
D_{\widetilde{\varphi},\widetilde{\mathds{P}}}^{SBD}(\widetilde{\mathbf{x}},\widetilde{\mathbf{Q}}^{\ast})
\ = \ D_{\varphi,\mathbf{P}}^{SBD}(\mathbf{x},\mathbf{Q}^{\ast}) 
\label{brostu5:fo.BSmin.extended.approx8}
\end{eqnarray}
which finishes the proof of Proposition \ref{brostu5:prop.coincide}.
In the above display,
in \eqref{brostu5:fo.BSmin.extended.approx7} we have used
the assumption $\frac{n_{k}}{n} = \widetilde{p}_{k}$ and
in the first equality of \eqref{brostu5:fo.BSmin.extended.approx8} we have employed 
the divergence definition \eqref{brostu5:fo.div} as well as the abbreviation
$\mathbf{\widetilde{x}} := \boldsymbol{\xi}_{n}^{\mathbf{\widetilde{V}}}$
which by construction leads to the corresponding components $\widetilde{x}_{k} =
\frac{1}{n} \sum_{i\in I_{k}^{(n)}} \mathbf{\widetilde{V}}_{i}$;
moreover, the second equality in \eqref{brostu5:fo.BSmin.extended.approx8}
follows --- in terms of  $\widetilde{\varphi}_{k} := M_{\mathbf{P}} \cdot \varphi_{k}$ --- 
from \eqref{brostu5:fo.link.SBD.proof2} in the 
above proof of Theorem \ref{brostu5:thm.BSnarrow.SBD},
the third equality comes from (the tilted version of) the divergence definition \eqref{brostu5:fo.SBD.smooth},
and the last equality 
is nothing but the divergence rewritability
\eqref{brostu5:fo.SBD.smooth.equality}. \hspace{0.5cm}  $\blacksquare$ \\


\vspace{-0.5cm}

\section*{Acknowledgment}

\vspace{0.2cm}
\noindent 
W. Stummer is grateful to the Sorbonne Universit\'{e} Paris 
for its multiple partial financial support and especially to the LPSM 
for its multiple great hospitality.
M. Broniatowski thanks very much the Friedrich-Alexander-Universit{\"a}t Erlangen-N{\"u}rnberg
(FAU) for its partial financial support and hospitality.

\enlargethispage{0.5cm}

%
%


\begin{thebibliography}{1}



\bibitem{Bro:23a}
M. Broniatowski and W. Stummer, 
\textquotedblleft A precise bare simulation approach to 
the minimization of some
distances. I. Foundations,\textquotedblright\ 
\emph{IEEE Trans. Inf. Theory}, Vol. 69, no. 5, pp. 3062--3120, 2023. 

\bibitem{Csi:63}
I. Csisz\'ar,
\textquotedblleft  Eine informationstheoretische Ungleichung und ihre Anwendung
auf den Beweis der Ergodizit\"at von Markoffschen Ketten,\textquotedblright\ 
\emph{Publ. Math. Inst. Hungar. Acad. Sci.}, Vol. 8, pp. 85--108, 1963. 

\bibitem{Ali:66}
M.S. Ali and D. Silvey, 
\textquotedblleft  A general class of coefficients of divergence of one
distribution from another,\textquotedblright\ 
\emph{J. Roy. Statist. Soc. Series B (Methodological)}, Vol. 28, no. 1, 
pp. 131--140, 1966. 
 
\bibitem{Mori:63}
T. Morimoto,
\textquotedblleft  Markov processes and the H-theorem,\textquotedblright\ 
\emph{J. Phys. Soc. Jpn.}, Vol. 18, no. 3, pp. 328--331, 1963. 
 
\bibitem{Cre:84}
N. Cressie and T.R.C. Read,
\textquotedblleft  Multinomial goodness-of-fit tests,\textquotedblright\ 
\emph{J. R. Statist. Soc. B}, Vol. 46, no. 3, pp. 440--464, 1984. 

\bibitem{Rea:88}
T.R.C. Read and N.A.C. Cressie, 
\emph{Goodness-of-Fit Statistics for Discrete Multivariate Data}. 
New York, USA: Springer, 1988.

\bibitem{Tsa:98}
C. Tsallis,
\textquotedblleft  Generalized entropy-based criterion for consistent testing,\textquotedblright\ 
\emph{Phys. Rev. E}, Vol. 58, No. 2,  pp. 1442--1445, 1998.

\bibitem{Ama:85} 
S.-I. Amari, 
\emph{Differential-Geometrical Methods in Statistics}.
Berlin, Germany: Springer, 1985.

\bibitem{Bha:43}
A. Bhattacharyya,
\textquotedblleft  On a measure of divergence between two statistical populations defined by their
probability distributions,\textquotedblright\ 
\emph{Bull. Calcutta Math. Soc.}, Vol. 35, pp. 99--109, 1943. 

\bibitem{Bha:46}
A. Bhattacharyya,
\textquotedblleft  On a measure of divergence between
two multinomial populations,\textquotedblright\ 
\emph{Sankhya}, Vol. 7, no. 4, pp. 401--406, 1946. 

\bibitem{Bha:47}
A. Bhattacharyya,
\textquotedblleft  On some analogues of the amount of information
and their use in statistical estimation (contd.),\textquotedblright\ 
\emph{Sankhya}, Vol. 8, no. 3, pp. 201--218, 1947.

\bibitem{Renyi:61}
A. Renyi,
\textquotedblleft On measures of entropy and information,\textquotedblright\ 
in: J. Neyman (ed.), \emph{Proc.\ 4th Berkeley Symp.\ Math.\ Stat.\ Probab. Vol.1}, 
pp. 547--561. 
Berkeley, CA, USA: Univ. of California Press, 1961. 

\bibitem{VanErv:14}
T. van Erven and P. Harremo{\"e}s,
\textquotedblleft  Renyi divergence and Kullback-Leibler divergence,\textquotedblright\ 
\emph{IEEE Trans. Inf. Theory}, Vol. 60, no. 7, pp. 3797--3820, 2014. 

\bibitem{Bur:82}
C. Burbea and C.R. Rao,
\textquotedblleft  On the convexity of some divergence measures
based on entropy functions,\textquotedblright\ 
\emph{IEEE Trans. Inf. Theory}, Vol. 28, no. 3, pp. 489--495, 1982.

\bibitem{Sha:48}
C.E. Shannon,
\textquotedblleft  A mathematical theory of communication,\textquotedblright\ 
\emph{Bell System Technical Journal}, Vol. 27, no. 3, pp. 379--423, 1948. 

\bibitem{Hav:67}
J. Havrda and F. Charvat,
\textquotedblleft  Quantification method of classification process,\textquotedblright\ 
\emph{Kybernetika}, Vol. 3, pp. 30--34, 1967.

\bibitem{Tsa:88}
C. Tsallis,
\textquotedblleft  Possible generalization of Boltzmann-Gibbs Statistics,\textquotedblright\ 
\emph{Journal of Statistical Physics}, Vol. 52, no. 1/2, pp. 479--487, 1988. 


\bibitem{Lie:87}
F. Liese and I. Vajda, 
\emph{Convex Statistical Distances}. 
Leipzig, Germany: Teubner, 1987.

\bibitem{Vaj:89} 
I. Vajda, 
\emph{Theory of Statistical Inference and Information}. 
Dordrecht, NL: Kluwer, 1989.

\bibitem{Csi:04} 
I. Csisz\'{a}r and P.C. Shields, 
\emph{Information Theory and Statistics: a Tutorial}. 
Hanover, MA, USA: now Publishers, 2004.

\bibitem{Stu:04a}
W. Stummer,
\emph{Exponentials, Diffusions, Finance, Entropy and Information}. 
Aachen, Germany: Shaker, 2004.

\bibitem{Par:06} 
L. Pardo, 
\emph{Statistical Inference Based on Divergence Measures}. 
Boca Raton, USA: Chapman \& Hall/CRC, 2006.

\bibitem{Lie:08}
F. Liese and K.J. Miescke,  
\emph{Statistical Decision Theory: Estimation, Testing, and Selection}.
New York, USA: Springer, 2008.

\bibitem{Bas:11} 
A. Basu, H. Shioya and C. Park,   
\emph{Statistical Inference: The Minimum Distance Approach}. 
Boca Raton, USA: CRC Press, 2011.

\bibitem{Lie:06}
F. Liese and I. Vajda,
\textquotedblleft  On divergences and informations in statistics and information theory,\textquotedblright\ 
\emph{IEEE Trans. Inf. Theory}, Vol. 52, no. 10, pp. 4394--4412, 2006. 

\bibitem{Vaj:10} 
I. Vajda and E.C. van der Meulen,
\textquotedblleft Goodness-of-fit criteria based on observations
quantized by hypothetical and empirical percentiles,\textquotedblright\ 
in: Z.A. Karian ZA and E.J. Dudewicz (eds.), 
\emph{Handbook of Fitting Statistical Distributions with R}, 
pp. 917 -- 994. 
Heidelberg, Germany: CRC, 2010. 

\bibitem{Reid:11}
M.D. Reid and R.C. Williamson,
\textquotedblleft  Information, divergence and risk for binary experiments,\textquotedblright\ 
\emph{J. Machine Learn. Res.}, Vol. 12, pp. 731--817, 2011.

\bibitem{Bass:13}
M. Basseville,
\textquotedblleft  Divergence measures for statistical data processing - an
annotated bibliography,\textquotedblright\ 
\emph{Signal Process.}, Vol. 93, pp. 621--633, 2013. 

\bibitem{Csi:72}
I. Csisz\'ar,
\textquotedblleft  A class of measures of informativity of
observation channels,\textquotedblright\ 
\emph{Periodica Mathem. Hungar.}, Vol. 2, no. 1-4, pp. 191--213, 1972.

\bibitem{BenB:78}
M. Ben-Bassat,
\textquotedblleft  f-entropies, probability of error, and feature selection,\textquotedblright\ 
\emph{Information and Control}, Vol. 39, pp. 227--242, 1978. 

\bibitem{BenT:86}
A. Ben-Tal and M. Teboulle,
\textquotedblleft  Rate-distortion theory with generalized information
measures via convex programming duality,\textquotedblright\ 
\emph{IEEE Trans. Inf. Theory}, Vol. 32, no. 5, pp. 630--641, 1986. 

\bibitem{Kes:89}
H.K. Kesavan and J.N. Kapur,
\textquotedblleft  The generalized maximum entropy principle,\textquotedblright\ 
\emph{IEEE Trans. Syst. Man Cyb.}, Vol. 19, no. 5, pp. 1042--1052, 1989. 

\bibitem{Dac:90}
D. Dacunha-Castelle and F. Gamboa,
\textquotedblleft  Maximum d'entropie et probleme des moments,\textquotedblright\ 
\emph{Ann. Inst. Henri Poincare}, Vol. 26, no. 4, pp. 567--596, 1990. 

\bibitem{Teb:93}
M. Teboulle and I. Vajda,
\textquotedblleft  Convergence of best $\phi-$entropy estimates,\textquotedblright\ 
\emph{IEEE Trans. Inf. Theory}, Vol. 39, no. 1, pp. 297--301, 1993. 

\bibitem{Gam:97}
F. Gamboa and E. Gassiat,
\textquotedblleft  Asymptotic distribution of $(h,\phi)-$entropies,\textquotedblright\ 
\emph{Ann. Stat.}, Vol. 25, no. 1, pp. 328--350, 1997. 

\bibitem{Vaj:07}
I. Vajda and J. Zvarova,
\textquotedblleft  On generalized entropies, Bayesian decisions 
and statistical diversity,\textquotedblright\ 
\emph{Kybernetika}, Vol. 43, no. 5, pp. 675--696, 2007. 

\bibitem{Sal:93}
M. Salicru, M.L. Menendez, D. Morales and L. Pardo,
\textquotedblleft  Asymptotic distribution of $(h,\phi)-$entropies,\textquotedblright\ 
\emph{Commun. Statist. - Theory Meth.}, Vol. 22, no. 7, pp. 2015--2031, 1993. 

\bibitem{Vaj:85}
I. Vajda and K. Vasek,
\textquotedblleft  Majorizations, concave entropies, and comparison
of experiments,\textquotedblright\ 
\emph{Problems of Control and Information Theory}, Vol. 14, no. 2, pp. 105--115, 1985. 

\bibitem{Breg:67}
L.M. Bregman,
\textquotedblleft  The relaxation method of finding the common point of convex sets and its
application to the solution of problems in convex programming,\textquotedblright\ 
\emph{USSR Comput. Math. Math. Phys.}, Vol. 7, no. 3, pp. 200--217, 167. 
 
\bibitem{Csi:91}
I. Csisz\'ar,
\textquotedblleft  Why least squares and maximum entropy? An axiomatic approach
to inference for linear inverse problems,\textquotedblright\  
\emph{Ann. Statist.}, Vol. 19, no. 4, pp. 2032--2066, 1991. 

\bibitem{Csi:94}
I. Csisz{\'a}r,
\textquotedblleft Maximum entropy and related methods,\textquotedblright\ 
in: \emph{Trans.\ 12th Prague Conf.\ 
Information Theory, Statistical Decision
Functions and Random Processes}, 
pp. 58--62.
Prague, Czech Republic: Czech Acad.\ Sci., 1994.
 
\bibitem{Csi:95}
I. Csisz{\'a}r,
\textquotedblleft  Generalized projections for non-negative functions,\textquotedblright\ 
\emph{Acta Math. Hung.}, Vol. 68, pp. 161--186, 1995. 
 
\bibitem{Par2:97}
M.C. Pardo and I. Vajda,
\textquotedblleft  About distances of discrete
distributions satisfying the data processing theorem of information theory,\textquotedblright\ 
\emph{IEEE Trans. Inf. Theory}, Vol. 43, no. 4, pp. 1288--1293, 1997. 

\bibitem{Par2:03}
M.C. Pardo and I. Vajda, 
\textquotedblleft  On asymptotic properties of information-theoretic divergences,\textquotedblright\ 
\emph{IEEE Trans. Inf. Theory}, Vol. 49, no. 7, pp. 1860--1868, 2003.
 
\bibitem{Stu:12}
W. Stummer and I. Vajda, 
\textquotedblleft  On Bregman distances and divergences of probability measures,\textquotedblright\ 
\emph{IEEE Trans. Inf. Theory}, Vol. 58, no. 3, pp. 1277--1288, 2012. 

\bibitem{Bro:19b}
M. Broniatowski and W. Stummer,
\textquotedblleft Some universal insights on divergences
for statistics, machine learning and
artificial intelligence,\textquotedblright\ 
in: F. Nielsen (ed.), \emph{Geometric Structures of Information}, 
Ser. Signals and Communications Technology, pp. 149--211. 
Cham, Switzerland: Springer Nature Switzerland, 2019.  

\bibitem{Bro:22}
M. Broniatowski and W. Stummer,
\textquotedblleft
A unifying framework for some directed distances in statistics,\textquotedblright\
in: F. Nielsen, A.S.R.S. Rao, C.R. Rao (eds.), \emph{Geometry and Statistics}, 
Handbook of Statistics, Vol. 46, pp. 145--223. 
Cambrigde MA, USA: Academic Press, 2022. 

\bibitem{Bas:98} 
A. Basu,  I.R. Harris,  N.L. Hjort and M.C. Jones,
\textquotedblleft  Robust and efficient estimation by minimizing a density
power divergence,\textquotedblright\ 
\emph{Biometrika}, Vol. 85, no. 3, pp. 549--559, 1998. 
 
\bibitem{Ban:05}
A. Banerjee, X. Guo and H. Wang,
\textquotedblleft  On the Optimality of Conditional Expectation
as a Bregman Predictor,\textquotedblright\ 
\emph{IEEE Trans. Inf. Theory}, Vol. 51, no. 7, pp. 2664--2669, 2005. 

\bibitem{Egu:01}
S. Eguchi and Y. Kano,
\textquotedblleft Robustifing Maximum Likelihood Estimation
by Psi-divergence,\textquotedblright\ 
\emph{Research Memorandum 802}, 2001. Tokyo: Institute of Statistical Mathematics.
 
\bibitem{Miho:02}
M. Mihoko and S. Eguchi,
\textquotedblleft Robust blind source separation by beta divergence,\textquotedblright\ 
\emph{Neural Comput.}, Vol. 14, 
pp. 1859--1886, 2002. 

\bibitem{Muk:19}
T. Mukherjee, A. Mandal and A. Basu,
\textquotedblleft  The B-exponential divergence and its generalizations with
applications to parametric estimation,\textquotedblright\ 
\emph{Stat. Methods Appl.}, Vol. 28, 
pp. 241–-257, 2019.

\bibitem{Basak:22}
S. Basak and A. Basu,
\textquotedblleft  The extended Bregman divergence and parametric estimation,\textquotedblright\ 
\emph{Statistics}, Vol. 56, no. 3,
pp. 699–-718, 2022.

\bibitem{Stu:07} 
W. Stummer,
\textquotedblleft  Some Bregman distances between financial diffusion processes,\textquotedblright\ 
\emph{Proc.\ Appl.\ Math.\ Mech.}, Vol. 7, no. 1, pp. 1050503--1050504, 2007. 
 
\bibitem{Kis:13}
A.-L. Ki{\ss}linger and W. Stummer, 
\textquotedblleft Some decision procedures based on scaled Bregman distance surfaces,\textquotedblright\  
in: F. Nielsen and F. Barbaresco (eds.),
\emph{Geometric Science of Information GSI 2013}, 
Lecture Notes in Computer Science, vol. 8085, pp. 479--486. 
Berlin, Germany: Springer, 2013.

\bibitem{Kis:15a}
A.-L. Ki{\ss}linger and W. Stummer, 
\textquotedblleft New model search for nonlinear recursive models, regressions 
and autoregressions,\textquotedblright\  
in: F. Nielsen and F. Barbaresco (eds.),
\emph{Geometric Science of Information GSI 2015}, 
Lecture Notes in Computer Science, vol. 9389, pp. 693--701. 
Berlin, Germany: Springer, 2015.

\bibitem{Kis:16}
A.-L. Ki{\ss}linger and W. Stummer,
\textquotedblleft Robust statistical engineering by means of scaled Bregman distances,\textquotedblright\ 
in: C. Agostinelli, A. Basu, P. Filzmoser and D. Mukherjee (eds.), 
\emph{Recent Advances in Robust Statistics -- Theory and Applications}, pp. 81--113.
New Delhi, India: Springer, 2016.

\bibitem{Kis:18}
A.-L. Ki{\ss}linger and W. Stummer,
\textquotedblleft A new toolkit for robust distributional change detection,\textquotedblright\ 
\emph{Appl. Stochastic Models Bus. Ind.}, Vol. 34, pp. 682--699, 2018.

\bibitem{Stu:17a}
W. Stummer and A.-L. Ki{\ss}linger,  
\textquotedblleft Some new flexibilizations of Bregman divergences and their asymptotics,\textquotedblright\  
In: F. Nielsen and F. Barbaresco (eds.), \emph{Geometric Science of Information GSI 2017}, 
Lecture Notes in Computer Science, vol. 10589, pp. 514--522. 
Cham, Switzerland: Springer International Publishing, 2017.

\bibitem{Liu8:10}
M. Liu, B.C. Vemuri, S.-I. Amari and F. Nielsen,
\textquotedblleft Total Bregman divergence and its applications to
shape retrieval,\textquotedblright\ 
in: \emph{Proc. 23rd IEEE CVPR}, 
pp. 3463--3468 , 2010. 
 
\bibitem{Liu8:12}
M. Liu, B.C. Vemuri, S.-I. Amari and F. Nielsen,
\textquotedblleft  Shape retrieval using hierarchical total Bregman soft
clustering,\textquotedblright\ 
\emph{IEEE Trans. Pattern Anal. Mach. Intell.}, Vol. 34, no. 12, pp. 2407--2419, 2012. 
 
\bibitem{Vem:11a}
B.C. Vemuri, M. Liu, S.-I. Amari and F. Nielsen,
\textquotedblleft  Total Bregman divergence and its applications to
DTI analysis,\textquotedblright\ 
\emph{IEEE Trans. Med. Imag.}, Vol. 30, no. 2, pp. 475--483, 2011. 
 
\bibitem{Noc:16sB}
R. Nock, A.K. Menon and C.S. Ong, 
\textquotedblleft A scaled Bregman theorem with applications,\textquotedblright\  
in: \emph{Advances in Neural Information Processing Systems 29 (NIPS 2016)},
2016, 9 pages.

\bibitem{Noc:16}
R. Nock, F. Nielsen and S.-I. Amari, 
\textquotedblleft  On conformal divergences and their population
minimizers,\textquotedblright\ 
\emph{IEEE Trans. Inf. Theory}, Vol. 62, no. 1, pp. 527--538, 2016. 


\bibitem{Maha:36}
P.C. Mahalanobis,
\textquotedblleft On the generalized distance in statistics,\textquotedblright\ 
\emph{Proc. Nat. Inst. Sci. Indida}, Vol. 2, no. 1, pp. 49--55, 1936. 
 



\bibitem{Stu:10}
W. Stummer and I. Vajda, 
\textquotedblleft  On divergences of finite measures and their
applicability in statistics and information theory,\textquotedblright\ 
\emph{Statistics}, Vol. 44, no. 2, pp. 169--187, 2010.

\bibitem{Gie:17}
C. Gietl and and F.P. Reffel,
\textquotedblleft  Continuity of $f-$projections and applications to the iterative proportional fitting procedure,\textquotedblright\ 
\emph{Statistics}, Vol. 51, no. 3, pp. 668--684, 2017. 

\bibitem{Bro:06} 
M. Broniatowski and A. Keziou,
\textquotedblleft Minimization of $\phi $-divergences on sets of 
signed measures,\textquotedblright\  
\emph{Stud. Scient. Math. Hungar.}, Vol. 43, pp. 403--442, 2006.


\bibitem{Bro:23b}
M. Broniatowski and W. Stummer,
\textquotedblleft On a cornerstone of bare-simulation
distance/divergence optimization,\textquotedblright\ 
in: F. Nielsen and F. Barbaresco (eds.), \emph{Geometric Science of Information GSI 2023,
Part I}, 
Lecture Notes in Computer Science, vol. 14071, pp. 105--116. 
Cham, Switzerland: Springer Nature Switzerland, 2023.

\bibitem{Csi:84}
I. Csisz\'ar,
\textquotedblleft  Sanov property, generalized I-projection and a conditional
limit theorem,\textquotedblright\ 
\emph{Ann. Probab.}, Vol. 12, no. 3, pp. 768--793, 1984. 


\bibitem{Tuc:13}
R.R. Tucci,
\textquotedblleft  Method for sampling probability distributions using
a quantum computer,\textquotedblright\ 
\emph{United States Patent}, Patent No. US 8543627 B1, 24th Sep. 2013.

\bibitem{Teh:15}
J.S. Teh, A. Samsudin,
M. Al-Mazrooie and A. Akhavan,
\textquotedblleft  GPUs and chaos: a new true random number generator,\textquotedblright\ 
\emph{Nonlinear Dyn.}, Vol. 82, pp. 1913--1922, 2015. 

\bibitem{Agh:17}
C. Aghamohammadi and J.P. Crutchfield,
\textquotedblleft  Thermodynamics of random number generation,\textquotedblright\ 
\emph{Phys. Rev. E}, Vol. 95, pp. 062139-1--062139-11, 2017. 

\bibitem{Herr:17}
M. Herrero-Collantes and J.C. Garcia-Escartin,
\textquotedblleft  Quantum random number generators,\textquotedblright\ 
\emph{Rev. Mod. Phys.}, Vol. 89, no. 1, pp. 015004-1 -- 015004-48, 2017. 

\bibitem{Bal:18}
K.A. Balygin, V.I. Zaitsev, A.N. Klimov, S.P. Kulik and S.N. Molotkov,
\textquotedblleft  A quantum random number generator based
on the 100-Mbit/s Poisson photocount statistics,\textquotedblright\ 
\emph{J. Exp. Theor. Phys.}, Vol. 126, no. 6, pp. 728--740, 2018.

\bibitem{Dang:19}
B. Dang, J. Sun, T. Zhang, S. Wang, M. Zhao, K. Liu, L. Xu,
J. Zhu , C. Cheng, L. Bao, Y. Yang, H. Wang,
Y. Hao and R. Huang,
\textquotedblleft  Physically transient true random number generators 
based on paired threshold switches enabling Monte Carlo method applications,\textquotedblright\ 
\emph{IEEE Electron. Device Lett.}, Vol. 40, no. 7, 
pp. 1096--1099, 2019.

\bibitem{Gon:19}
L. Gong, J. Zhang, H. Liu, L. Sang and Y. Wang,
\textquotedblleft  True random number generators using electrical noise,\textquotedblright\ 
\emph{IEEE Access}, Vol. 7, pp. 125796--125805, 2019. 

\bibitem{Chand:20}
S.T. Chandrasekaran, V.E.G. Karnam and A. Sanya,
\textquotedblleft  0.36-mW, 52-Mbps true random number generator based
on a stochastic delta-sigma modulator,\textquotedblright\ 
\emph{IEEE Solid-State Lett.}, Vol. 3, pp. 190--193, 2020.

\bibitem{Drah:20}
D. Drahi, N. Walk, M.J. Hoban, A.K. Fedorov, R. Shakhovoy, A. Feimov,
Y. Kurochkin, W.S. Kolthammer, J. Nunn, J. Barrett and I.A. Walmsley,
\textquotedblleft  Certified quantum random numbers from untrusted light,\textquotedblright\ 
\emph{Phys. Rev. X}, Vol. 10, pp. 041048-1 -- 041048-32, 2020. 

\bibitem{Kol:20} 
C. Kollmitzer, S. Schauer, S. Rass and B. Rainer (eds.), 
\emph{Quantum Random Number Generation}.
Cham, Switzerland: Springer Nature, 2020.

\bibitem{Liu3:20}
Y. Liu, C. Chen, D.D. Yang, Q. Li and X. Li,
\textquotedblleft  Fast true random number generator based on
chaotic oscillation in self-feedback weakly
coupled superlattices,\textquotedblright\ 
\emph{IEEE Access}, Vol. 8, 
pp. 182693--182703, 2020. 

\bibitem{Arc:21}
T. Arciuolo and K.M. Elleithy,
\textquotedblleft Parallel, true random number generator (P-TRNG):
using parallelism for fast true random number
generation in hardware,\textquotedblright\ 
in: \emph{Proc. 2021 IEEE 11th Ann. Comput. and Commun. Workshop Conf. (CCWC)}, 
0987--0992, 2021;
doi:10.1109/CCWC51732.2021.9375939. 

\bibitem{Awa:21}
A.M. Awaludin, D. Pratama and H. Kim,
\textquotedblleft AnyTRNG: generic, high-throughput,
low-area true random number generator based on synchronous edge
sampling,\textquotedblright\ 
in: H. Kim (ed.), \emph{Information Security Applications WISA 2021}, 
Lecture Notes in Computer Science, vol. 13009, pp. 157–-168. 
Cham, Switzerland: Springer Nature Switzerland, 2021. 

\bibitem{Bai:21}
B. Bai, J. Huang, G.-R. Qiao, Y.-Q. Nie,
W. Tang, T. Chu, J. Zhang and J.-W. Pan,
\textquotedblleft  18.8 Gbps real-time quantum random number generator with
a photonic integrated chip,\textquotedblright\ 
\emph{Appl. Phys. Lett.}, Vol. 118, No. 264001, 2021; doi:10.1063/5.0056027.

\bibitem{Cao15:21}
G. Cao, L. Zhang, X. Huang, W. Hu and X. Yang,
\textquotedblleft  16.8 Tb/s True Random Number Generator Based
on Amplified Spontaneous Emission,\textquotedblright\ 
\emph{IEEE Photonics Technol. Lett.}, Vol. 33, no. 14, pp. 699--702, 2021. 
 
\bibitem{Chan:21}
S.T. Chandrasekaran, A. Jayaraj,
N. Ramesh and A. Sanyal,
\textquotedblleft 33-200Mbps, 3pJ/bit true random number
generator based on CT Delta-Sigma modulator,\textquotedblright\ 
in: \emph{Proc. 2021 IEEE Intern. Symp. Circuits Systems (ISCAS)}, 
2021; doi:10.1109/ISCAS51556.2021.9401507. 
 
\bibitem{Deg:21}
A. Degada and H. Thapliyal,
\textquotedblleft  An integrated TRNG-PUF
architecture based on photovoltaic solar cells,\textquotedblright\ 
\emph{IEEE Consumer Electron. Magaz.}, Vol. 10, no. 4, pp. 99--105, 2021. 
 
\bibitem{Fis:21}
I. Fischer and D.J. Gauthier,
\textquotedblleft  High-speed harvesting of random numbers,\textquotedblright\ 
\emph{Science}, Vol. 371, 
26 February 2021,
pp. 889--890, 2021. 

\bibitem{Fu15:21}
Z. Fu, Y. Tang, X. Zhao, K. Lu, Y. Dong,
A. Shukla, Z. Zhu and Y. Yang,
\textquotedblleft  An overview of spintronic true random
number generator,\textquotedblright\ 
\emph{Front. Phys.}, Vol. 9, No. 638207, 2021; doi:10.3389/fphy.2021.638207.
 
\bibitem{Geh:21}
T. Gehring, C. Lupo, A. Kordts, D.S. Nikolic, 
N. Jain, T. Rydberg, T.B. Pedersen, S. Pirandola 
and U.L. Andersen,
\textquotedblleft  Homodyne-based quantum random number
generator at 2.9 Gbps secure against quantum
side-information,\textquotedblright\ 
\emph{Nature Comm.}, Vol. 12, No. 605, 2021; doi:10.1038/s41467-020-20813-w.
 
\bibitem{Gras:21}
G. Gras, A. Martin, J.W. Choi and F. Bussi\`eres,
\textquotedblleft  Quantum entropy model of an integrated
quantum-random-number-generator chip,\textquotedblright\ 
\emph{Phys. Rev. Applied}, Vol. 15, No. 054048, 2021; doi:10.1103/PhysRevApplied.15.054048.
 
\bibitem{Guo15:21}
Y. Guo, Q. Cai, P. Li, Z. Jia, B. Xu, Q. Zhang, Y. Zhang, R. Zhang, Z. Gao, K.A. Shore and Y. Wang,
\textquotedblleft  40 Gb/s quantum random number generation based on optically sampled amplified spontaneous emission,\textquotedblright\ 
\emph{APL Photonics}, Vol. 6, No. 066105, 2021; doi:10.1063/5.0040250.
 
\bibitem{Hof:21}
M. Hofert,
\textquotedblleft  Random number generators produce collisions: why, how many and more,\textquotedblright\ 
\emph{Americ. Statistician}, Vol. 75, no. 4, pp. 394--402, 2021.
 
\bibitem{Jac:21}
M.M. Jacak, P. J\'o\'zwiak, J. Niemczuk and J.E. Jacak,
\textquotedblleft  Quantum generators of random
numbers,\textquotedblright\ 
\emph{Sci. Rep.}, Vol. 11, No. 16108, 2021; doi:10.1038/s41598-021-95388-7.
 
\bibitem{Kim15:21}
K. Kim, S. Bittner, Y. Zeng, S. Guazzotti, O. Hess, Q.J. Wang and Hui Cao,
\textquotedblleft  Massively parallel ultrafast random bit generation
with a chip-scale laser,\textquotedblright\ 
\emph{Science}, Vol. 371, 
26 February 2021,
pp. 948--952, 2021. 

\bibitem{Kim16:21}
S. Kim, M.-S. KIM, Y. Lee, H.-D. KIM and S.-J. Choi,
\textquotedblleft  Low-power true random number generator
based on randomly distributed carbon
nanotube networks,\textquotedblright\ 
\emph{IEEE Access}, Vol. 9, 
pp. 91341--91346, 2021.

\bibitem{Li16:21a}
Y. Li, Y. Fei, W. Wang, X. Meng, H. Wang, Q. Duan
and Z. Ma,
\textquotedblleft Quantum random number generator using
a cloud superconducting quantum computer based
on source-independent protocol,\textquotedblright\ 
\emph{Sci. Rep.}, Vol. 11, 23873, 2021; doi:10.1038/s41598-021-03286-9. 
 
\bibitem{Li16:21b}
Y. Li, Y. Fei, W. Wang, X. Meng, H. Wang, Q. Duan
and Z. Ma,
\textquotedblleft Experimental study on the security of superluminescentvLED-based quantum random generator,\textquotedblright\ 
\emph{Opt. Eng.}, Vol. 60, No. 11,
116106-01--116106-22, 2021; doi:10.1117/1.OE.60.11.116106. 
 
\bibitem{Lu16:21}
Z. Lu, S. Yang, J. Liu, X. Wang and Y. Li
\textquotedblleft  Efficient FPGA implementation of high-speed
true random number generator,\textquotedblright\ 
\emph{Rev. Sci. Instrum.}, Vol. 92, No. 024706, 2021; doi:10.1063/5.0035519.
 
\bibitem{Luo15:21}
W. Luo, N. Takeuchi, O. Chen and N. Yoshikawa,
\textquotedblleft  Low-autocorrelation random number
generator based on adiabatic
quantum-flux-parametron logic,\textquotedblright\ 
\emph{IEEE Trans. Appl. Supercond.}, Vol. 31, no. 5, pp. 1302305, 2021. 
 
\bibitem{Luo16:21}
Y. Luo, S. Han, S. Zhang, Y. Wang and J. Liu,
\textquotedblleft High speed true random number generator
controlled by logistic map,\textquotedblright\ 
in: \emph{Proc. 2021 IEEE Int. Conf. High Perform. Comp. \& Commun.; 7th Int. Conf. Data Science \& Systems; 19th
Int. Conf. Smart City; 7th Int. Conf. Depend. Sensor, Cloud \& Big Data Systems \& Application}, pp. 57--62, 
2021;
doi:10.1109/HPCC-DSS-SMARTCITY-DEPENDSYS53884.2021.00035. 
 
\bibitem{Mal:21}
K. Malik, J. Pulikkotil and A. Sharma,
\textquotedblleft  Comparison of pseudorandom number generators and their
application for uncertainty estimation using Monte Carlo imulation,\textquotedblright\ 
\emph{MAPAN-J. Metrology Soc. India}, Vol. 36, no. 3, pp. 481--496, 2021. 
 
\bibitem{Mon:21}
F. Monet, J.-S. Boisvert and R. Kashyap,
\textquotedblleft  A simple high-speed random number generator with minimal
post-processing using a random Raman fiber laser,\textquotedblright\ 
\emph{Sci. Rep.}, Vol. 11, 13182, 2021; doi:10.1038/s41598-021-92668-0. 
 
\bibitem{Pat:21}
R. Patgiri,
\textquotedblleft Rando: a general-purpose true random number
generator for conventional computers,\textquotedblright\ 
in: \emph{Proc. 2021 IEEE 20th Intern. Conf. Trust, Security Privacy Comput. Commun.(TrustCom)}, 
pp. 107--113, 2021;
doi: 10.1109/TRUSTCOM53373.2021.00032. 
 
\bibitem{Serr:21}
R. Serrano, C. Duran, T.-T. Hoang, M. Sarmiento,
K.-D. Nguyen, K. Suzaki and C.-K. Pham,
\textquotedblleft  A fully digital true random number generator
with entropy source based in frequency collapse,\textquotedblright\ 
\emph{IEEE Access}, Vol. 9, 
pp. 105748--105755, 2021.

\bibitem{Sto:21}
S. Stoller and K.A. Campbell,
\textquotedblleft  Demonstration of three true random number generator
circuits using memristor created entropy and commercial
off-the-shelf components,\textquotedblright\ 
\emph{Entropy}, Vol. 23, No. 371, 2021; 
doi:10.3390/e23030371.

\bibitem{Tan:21}
S. Taneja and M. Alioto,
\textquotedblleft  Fully synthesizable unified true random number generator and cryptographic core,\textquotedblright\ 
\emph{IEEE J. Solid-State Circ.}, Vol. 56, no. 10, pp. 3049--3061, 2021. 

\bibitem{Tseng:21}
P.-H. Tseng, M.-H. Lee, Y.-H. Lin,
H.-L. Lung, K.-C. Wang and C.-Y. Lu,
\textquotedblleft  ReRAM-based pseudo-true random number
generator With high throughput and
unpredictability characteristics,\textquotedblright\ 
\emph{IEEE Trans. Electron. Devices}, Vol. 68, no. 4, pp. 1593--1597, 2021. 
 
\bibitem{Wang15:21}
X. Wang, H. Liang, Y. Wang, L. Yao, Y. Guo, M. Yi, Z. Huang, H. Qi
and Y. Lu,
\textquotedblleft  High-throughput portable true random number
generator based on jitter-latch structure,\textquotedblright\ 
\emph{IEEE Trans. Circ. Syst.-I: Reg. Papers}, Vol. 68, no. 2, pp. 741--750, 2021. 

\bibitem{Abr:22}
N. Abraham, K. Watanabe, T. Taniguchi and K. Majumdar,
\textquotedblleft  A high-quality entropy source using van der
Waals heterojunction for true random
number generation,\textquotedblright\ 
\emph{ACS Nano}, Vol. 16, 
pp. 5898--5908, 2022. 
 
\bibitem{Aka:22}
N. Akashi, K. Nakajima, M. Shibayama and Y. Kuniyoshi,
\textquotedblleft  A mechanical true random number generator,\textquotedblright\ 
\emph{New J. Phys.}, Vol. 24, No. 013019, 2022; doi:10.1088/1367-2630/ac45ca.
 
\bibitem{All:22}
Y. Alloun, M.S. Azzaz, A. Kifouche and R. Kaibou,
\textquotedblleft  Pseudo random number generator based on chaos
theory and artificial neural networks,\textquotedblright\ 
in: \emph{Proc. 2022 2nd Int. Conf. on Advanced Electr. Eng. (ICAEE)}, 
2022, 6 pages. 

\bibitem{Bae:22}
M.J. Bae,
\textquotedblleft 
Quantum walk random number generation:
memory-based models,\textquotedblright\ 
in: \emph{Proc. 2022 IEEE International Conference on Quantum Computing and Engineering (QCE)}, 
2022, 12 pages. 
 
\bibitem{Bha:22}
H. Bhatia, E. Tretschk, C. Theobalt and V. Golyanik,
\textquotedblleft  Generation of truly random numbers on a
quantum annealer,\textquotedblright\ 
\emph{IEEE Access}, Vol. 10, pp. 112832--112844, 2022. 

\bibitem{Cui15:22}
J. Cui, M. Yi, D. Cao, L. Yao, X. Wang, H. Liang, Z. Huang, H. Qi, T. Ni and Y. Lu,
\textquotedblleft  Design of True Random Number Generator Based
on Multi-Stage Feedback Ring Oscillator,\textquotedblright\ 
\emph{IEEE Trans. Circ. Syst.-II: Expr. Briefs}, Vol. 69, no. 3, pp. 1752--1756, 2022. 

\bibitem{Deb:22}
P. Debashis, H. Li, D. Nikonov and I. Young,
\textquotedblleft  Spin electronics
Gaussian random number generator with reconfigurable mean and variance
using stochastic magnetic tunnel junctions,\textquotedblright\ 
\emph{IEEE Magnetics Lett.}, Vol. 13, No. 4502905, 2022; doi:10.1109/LMAG.2022.3152991.
 
\bibitem{Dem:22}
K. Demir and S. Erg{\"u}n,
\textquotedblleft  A comparative analysis on chaos-based random number generation methods,\textquotedblright\ 
\emph{Eur. Phys. J. Plus}, Vol. 137, No. 591, 2022; doi:10.1140/epjp/s13360-022-02793-6.
 
\bibitem{Dut:22}
S. Dutta, A. Arunachalam and S. Misailovic,
\textquotedblleft To seed or not to seed?
An empirical analysis of usage of seeds for testing
in machine learning projects,\textquotedblright\ 
in: \emph{Proc. 2022 IEEE Conf. Softw. Test. Verif. Valid. (ICST)}, 
pp. 151--161, 2022; doi:10.1109/ICST53961.2022.00026.

\bibitem{Gar:22}
A.M. Garipcan and E. Erdem,
\textquotedblleft  A gigabit TRNG with novel lightweight post-processing method for cryptographic
applications,\textquotedblright\ 
\emph{Eur. Phys. J. Plus}, Vol. 137, No. 493, 2022; doi:10.1140/epjp/s13360-022-02679-7.
 
\bibitem{Gru:22}
M. Gruji\'c and I. Verbauwhede,
\textquotedblleft  TROT: a three-edge ring oscillator based
true random number generator with
time-to-digital vonversion,\textquotedblright\ 
\emph{IEEE Trans. Circ. Syst.-I: Reg. Papers}, Vol. 69, no. 6, pp. 2435--2448, 2022. 
 
\bibitem{Kim15:22}
K. Kim, S. Bittner, Y. Zeng, S. Guazzotti, O. Hess, Q.J. Wang and H. Cao,
\textquotedblleft Massively parallel generation of random numbers
using a semiconductor laser,\textquotedblright\ 
in: \emph{Proc. 2022 Conf. Lasers Electro-Optics (CLEO)}, 
AW4M.1, 2022. 
 
\bibitem{Kim17:22}
J. Kim and H. Chae,
\textquotedblleft A 10-Gbps, 0.121-pJ/bit, All-Digital True Random-Number
Generator using Middle Square Method,\textquotedblright\ 
in: \emph{Proc. 2022 IEEE Asian Solid-State Circ. Conf. (A-SSCC)}, 
Paper 9.4, 2022; doi:10.1109/A-SSCC56115.2022.9980831. 
 
\bibitem{Liao15:22}
T.L. Liao, P.Y. Wan and J.-J. Yan,
\textquotedblleft  Design and synchronization of chaos-based
true random number generators and
its FPGA implementation,\textquotedblright\ 
\emph{IEEE Access}, Vol. 10, pp. 8279--8286, 2022. 

\bibitem{Lv:22}
Y. Lv, B.R. Zink and J.-P. Wang,
\textquotedblleft  Bipolar random spike and bipolar random
number generation by two magnetic tunnel junctions,\textquotedblright\ 
\emph{IEEE Trans. Electron. Devices}, Vol. 69, no. 3, pp. 1582--1587, 2022. 

\bibitem{Naik:22}
R.B. Naik and U. Singh,
\textquotedblleft  A review on applications of chaotic maps in
pseudo-random number generators and encryption,\textquotedblright\ 
\emph{Ann. Data Science}, 18-01-2022, 26 pages; doi:10.1007/s40745-021-00364-7.

\bibitem{Park:22}
J. Park, B. Kim and J.-Y. Sim,
\textquotedblleft  A PVT-tolerant oscillation-collapse-based true
random number generator with an odd
number of inverter stages,\textquotedblright\ 
\emph{IEEE Trans. Circ. Syst.-II: Expr. Briefs}, Vol. 69, no. 10, pp. 4058--4062, 2022. 
 
\bibitem{Sala:22}
R. Della Sala, D. Bellizia and G. Scotti,
\textquotedblleft  High-throughput FPGA-compatible TRNG
architecture exploiting multistimuli metastable cells,\textquotedblright\ 
\emph{IEEE Trans. Circ. Syst.-I: Reg. Papers}, Vol. 69, no. 12, pp. 4886--4897, 2022. 
 
\bibitem{Zhang17:93}
R. Zhang, X. Wang, K. Liu and H. Shinohara,
\textquotedblleft  A 0.186-pJ per Bit Latch-Based True Random
Number Generator Featuring Mismatch
Compensation and Random Noise Enhancement,\textquotedblright\ 
\emph{IEEE J. Solid-State Circ.}, Vol. 57, no. 8, pp. 2498--2508, 2022. 
 
\bibitem{Zhou17:22}
X. Zhou, Z. Hu, Z. Chai, W. Zhang, S. Clima, R. Degraeve,
J.F. Zhang, A. Fantini, D. Garbin, R. Delhougne,
L. Goux and G. S. Kar,
\textquotedblleft  Impact of relaxation on the performance of
GeSe true random number generator based
on ovonic threshold switching,\textquotedblright\ 
\emph{IEEE Electron Device Lett.}, Vol. 43, no. 7, pp. 1061--1064, 2022. 
 
\bibitem{Zhu15:22}
Y. Zhu, Y. Bian, J. Yang, Y. Zhang and S. Yu,
\textquotedblleft 
21 Gbps source-independent quantum random number generator based on vacuum fluctuations,\textquotedblright\ 
in: \emph{Proc. 2022 Asia Commun. Photonics Conf. (ACP) and Intern. Conf. Information Photonics Optical Commun. (IPOC)}, 
2140--2142, 2022. 

\bibitem{Arg:23}
J. Argillander, A. Alarcon, C. Bao
C. Kuang, G. Lima, F. Gao and G.B. Xavier,
\textquotedblleft  Quantum random number generation based on a
perovskite light emitting diode,\textquotedblright\ 
\emph{Commun. Phys.}, Vol. 6, No. 157, 2023; doi:10.1038/s42005-023-01280-3.

\bibitem{Bruy:23}
C. Bruynsteen, T. Gehring, C. Lupo, J. Bauwelinck and X. Yin,
\textquotedblleft  100-Gbit/s Integrated Quantum Random Number Generator Based on Vacuum
Fluctuations
,\textquotedblright\ 
\emph{PRX QUANTUM}, Vol. 4, No. 010330, 2023; doi: 10.1103/PRXQuantum.4.010330.
 
\bibitem{Cai15:23}
Q. Cai, P. Li, Y. Shi, Z. Jia, L. Ma, B. Xu, 
X. Chen, K.A. Shore and Y. Wang,
\textquotedblleft  Tbps parallel random number generation based on a single quarter-wavelength-shifted DFB laser,\textquotedblright\ 
\emph{Optics \& Laser Technol.}, Vol. 162, No. 109273, 2023; doi:10.1016/j.optlastec.2023.109273.
 
\bibitem{Chat:23}
S. Chatterjee, N. Rangarajan, S. Patniak, D. Rajasekharan, O. Sinanoglu 
and Y.S. Chauhan,
\textquotedblleft  FerroCoin: ferroelectric tunnel junction-based
true random number generator,\textquotedblright\ 
\emph{IEEE Trans. Emerg. Topics Comput.}, Vol. 11, no. 2, pp. 541--547, 2023. 
 
\bibitem{Chen30:23}
C.-B. Chen, T. Chen, Y.-H. Huang and Y.-H. Huang,
\textquotedblleft  35.56-Gbits/sec chaos random number generator
supporting 1.2-GSamples/sec noise generation,\textquotedblright\ 
\emph{IEEE Trans. Circ. Syst.-II: Expr. Briefs}, Vol. 70, no. 10, pp. 3802--3806, 2023. 
 
\bibitem{Cheng15:23}
X. Cheng, Y. Zhang, H. Zhuand Y. Zhou,
\textquotedblleft  A true random number generator with high bit rate and
low energy efficiency,\textquotedblright\ 
\emph{Int. J.Circ. Theor. Appl.}, Vol. 51, 
pp. 3415--3431, 2023. 
 
\bibitem{Eat:23}
M. Eaton, A. Hossameldin, R.J. Birrittella,
P.M. Alsing, C.C. Gerry, H. Dong, C. Cuevas and
O. Pfister,
\textquotedblleft Resolution of 100 photons and quantum
generation of unbiased random numbers,\textquotedblright\ 
\emph{Nature Photonics}, Vol. 17, 
pp. 106--111, 2023.
 
\bibitem{Elm:23}
E. Elmitwalli and S. K{\"o}se,
\textquotedblleft  Bistable Josephson Junction-based true random
number generator without inductors,\textquotedblright\ 
\emph{IEEE Trans. Circ. Syst.-II: Expr. Briefs}, Vol. 70, no. 4, pp. 1615--1619, 2023. 
 
\bibitem{Equ:23}
M.S. Equbal, T. Ketkar and S. Sahay,
\textquotedblleft  Hybrid CMOS-RRAM true random number
generator exploiting coupled entropy sources,\textquotedblright\ 
\emph{IEEE Trans. Electron. Devices}, Vol. 70, no. 3, pp. 1061--1066, 2023. 
 
\bibitem{Fru:23}
F. Frustaci, F. Spagnolo, S. Perri and P. Corsonello,
\textquotedblleft  A high-speed FPGA-based true random number
generator using metastability with clock managers,\textquotedblright\ 
\emph{IEEE Trans. Circ. Syst.-II: Expr. Briefs}, Vol. 70, no. 2, pp. 756--760, 2023. 
 
\bibitem{Fu15:23}
Y. Fu, J. Wen, L. Wang, L. Yang, Q. Zhu, W. Zuo, P. Zhang,
Y. Li, H. Tong, G. Ma, H. Wang and X. Miao,
\textquotedblleft  A 2.22 Mb/s true random number generator
based on a GeTex ovonic threshold switching memristor,\textquotedblright\ 
\emph{IEEE Electron Device Lett.}, Vol. 44, no. 5, pp. 853--856, 2023. 
 
\bibitem{Gui:23}
O. Guillan-Lorenzo, M. Troncoso-Costas, D. Alvarez-Outarelo,
F.J. Diaz-Otero and J.C. Garcia-Escartin,
\textquotedblleft  Optical quantum random number generators: a comparative study,\textquotedblright\ 
\emph{Opt. Quant. Electron.}, Vol. 55, No. 185, 2023; doi:10.1007/s11082-022-04396-y.
 
\bibitem{Hanl:23}
J. Hanlon and S. Felix,
\textquotedblleft  A fast hardware pseudorandom number
generator based on xoroshiro128,\textquotedblright\ 
\emph{IEEE Trans. Computers}, Vol. 72, no. 5, pp. 1518--1524, 2023. 
 
\bibitem{Kes:23}
P. Keshavarzian, K. Ramu, D. Tang, C. Weill,
F. Gramuglia, S.S. Tan, M. Tng, L. Lim, E. Quek, D. Mandich, M. Stipcevi\'c and E. Charbon,
\textquotedblleft  A 3.3-Gb/s SPAD-Based Quantum Random
Number Generator,\textquotedblright\ 
\emph{IEEE J. Solid-State Circ.}, Vol. 58, no. 9, pp. 2632--2646, 2023. 
 
\bibitem{Kho:23}
F. Khodayari, A. Amirany, M.H. Moaiyeri and K. Jafari,
\textquotedblleft  A variation-aware ternary true random number generator
using magnetic tunnel junction at subcritical current regime,\textquotedblright\ 
\emph{IEEE Trans. Magnetics}, Vol. 59, no. 3, 3400208, 2023. 
 
\bibitem{Kumar3:23}
D. Kumar, L.L. Mankali, P.K. Misra and M. Goswami,
\textquotedblleft  A 0.7 pJ/bit, 1.5 Gbps energy-efficient
image-based true random number generator,\textquotedblright\ 
\emph{IETE J. Research}, Vol. 69, no. 3, pp. 1260--1270, 2023. 

\bibitem{Lee15:23}
S. Lee,
\textquotedblleft  Strategies for ultra-fast bit generation
of two-terminal threshold switch-based true random number generator using drift-free Ge-doped SiO2
threshold switch device,\textquotedblright\ 
\emph{Solid-State Electron.}, Vol. 201, No. 108609, 2023; doi:10.1016/j.sse.2023.108609.
 
\bibitem{Li30:23}
J. Li, Y. Li, Y. Dong, Y. Yuang, J. Xiao and Y. Huang,
\textquotedblleft  400 Gb/s physical random number generation based on deformed square self-chaotic lasers,\textquotedblright\ 
\emph{Chin. Optics Lett.}, Vol. 21, No. 6, 061901, 2023; doi:10.3788/COL202321.061901.
 
\bibitem{Li19:23}
J. Li, J. Liu, D. Liu, W. Tian, S. Jin, S. Hu, J. Guo
and Y. Jin,
\textquotedblleft  Ultrafast random number generation
based on random laser
,\textquotedblright\ 
\emph{J. Lightwave Technol.}, Vol. 41, no. 16, pp. 5233--5243, 2023. 
 
\bibitem{Li15:23}
S. Li, Y. Liu, F. Ren and Z. Yang,
\textquotedblleft  Design of a high throughput pseudorandom
number generator based on discrete
hyper-chaotic system,\textquotedblright\ 
\emph{IEEE Trans. Circ. Syst.-II: Expr. Briefs}, Vol. 70, no. 2, pp. 806--810, 2023. 
 
\bibitem{Li18:23}
X. Li, Y. Wang, Y. Yang, S. Lv, Q. Luo, X. Wang, X. Xu, 
D. Lei and F. Zhang,
\textquotedblleft  A 144-fJ/bit reliable and compact TRNG based
on the diffusive resistance of 3-D resistive
random access memory
,\textquotedblright\ 
\emph{IEEE Trans. Electron. Devices}, Vol. 70, no. 8, pp. 4139--4144, 2023. 
 
\bibitem{Luo16:23}
Y. Luo, J. Zhang, J. Hao and X. Zhao,
\textquotedblleft  A 2.5 pJ/bit PVT-tolerant true random number
generator based on native-NMOS-regulated
ring oscillator,\textquotedblright\ 
\emph{IEEE Trans. Circ. Syst.-II: Expr. Briefs}, Vol. 70, no. 10, pp. 3927--3931, 2023. 
 
\bibitem{Maity:23}
S. Maity, A. Prosad, H. Natarajan and V. Raghunathan
\textquotedblleft Comparison of high speed quantum random number generators based on
ASE-ASE and ASE-LASER beating,\textquotedblright\ 
in: P.R. Hemmer and A.L. Migdall (eds.), \emph{Quantum Computing, Communication, and Simulation III}, 
Proceedings of SPIE, vol. 12446, No. 1244610, 2023; 
doi:10.1117/12.2649916. 
 
\bibitem{Ming15:23}
H. Ming, H. Hu, F. Lv  and R. Yu,
\textquotedblleft  A high-performance hybrid random number 
generator based on a nondegenerate coupled chaos and its practical implementation,\textquotedblright\ 
\emph{Nonlinear Dyn.}, Vol. 111, 
pp. 847--869 2023. 
 
\bibitem{Monet:23}
F. Monet and R. Kashyap,
\textquotedblleft  On multiplexing in physical random
number generation, and conserved
total entropy content,\textquotedblright\ 
\emph{Sci. Rep.}, Vol. 13, 7892, 2023; doi:10.1038/s41598-023-35130-7. 
 
\bibitem{Nahar:23}
P. Nahar, P. Khandekar, M. Deshmukh, H.S. Jatana
and U. Khambete,
\textquotedblleft Survey of Stochastic Number Generators
and Optimizing Techniques,\textquotedblright\ 
in: A.J. Kulkarni et al. (eds.), \emph{Intelligent Systems and Applications}, 
Lecture Notes in Electrical Engineering, vol. 959, pp. 405–-415. 
Singapore: Springer Nature Singapore Pte Ltd., 2023. 
 
\bibitem{Peng15:23}
B. Peng, Q. Wu, Z. Wang and J. Yang,
\textquotedblleft  A RRAM-based true random number generator with 2T1R architecture for hardware security applications,\textquotedblright\ 
\emph{Micromachines}, Vol. 14, No. 1213, 2023; doi:10.3390/mi14061213.
 
\bibitem{Taniz:23}
K. Tanizawa, K. Kato and F. Futami,
\textquotedblleft Real-time 50-Gbit/s spatially multiplexed quantum
random number generator based on vacuum fluctuation,\textquotedblright\ 
in: \emph{Proc. 2023 Opt. Fiber Commun. Conf. Exh. (OCE)}, 
Th4A.8, 2023. 
 
\bibitem{Wang16:23}
X. Wang, R. Zhang, K. Liu and H. Shinohara,
\textquotedblleft  A 0.116 pJ/bit latch-based true random number
generator featuring static inverter selection
and noise enhancement,\textquotedblright\ 
\emph{IEEE Trans. Very Large Scale Integr. (VLSI) Systems}, to appear,
early access at doi:10.1109/TVLSI.2023.3328602, 2023. 
 
\bibitem{Yang15:23}
Z. Yang, Y. Liu, Y. Wu, Y. Qi, F. Ren and S. Li,
\textquotedblleft  A high speed pseudo-random bit generator
driven by 2D-discrete hyperchaos,\textquotedblright\ 
\emph{Chaos Solit. Fract.}, Vol. 167, No. 113039, 2023; doi:10.1016/j.chaos.2022.113039.

\bibitem{Yuan15:23}
F. Yuan, S. Li, Y. Deng, Y. Li and G. Chen,
\textquotedblleft  Cu-doped TiO2-x nanoscale memristive
applications in chaotic circuit and true
random number generator,\textquotedblright\ 
\emph{IEEE Trans. Ind. Electron.}, Vol. 70, no. 4, pp. 4120--4127, 2023. 
 
\bibitem{Zhang20:23}
W. Zhang,
\textquotedblleft  Analysis and Construction of Nonlinear Correctors
Used in True Random Number Generators,\textquotedblright\ 
\emph{IEEE Trans. Inf. Theory}, Vol. 69, no. 10, pp. 6671--6681, 2023. 

\bibitem{Zheng15:23}
J. Zheng and J. Li, 
\textquotedblleft An ultrafast cryptographically secure
pseudorandom number generator,\textquotedblright\  
in: S.-H. Seo and H. Seo (eds.),
\emph{Int. Conf. Inf. Security Crypt. ICISC 2022}, 
Lecture Notes in Computer Science, vol. 13849, pp. 267--291. 
Cham, Switzerland: Springer Nature Switzerland, 2023.



\bibitem{Bro:23c}
M. Broniatowski and W. Stummer,
\textquotedblleft Bare-Simulation optimization
of some distances between fuzzy sets respectively
basic belief assignments,\textquotedblright\ 
Preprint 2023.

\bibitem{Rat:72}
P.N. Rathie and P. Kannappan,
\textquotedblleft  A directed-divergence function of type $\beta$,\textquotedblright\ 
\emph{Information and Control}, Vol. 20, 
pp. 38--45, 1972. 

\bibitem{Shi:98}
M. Shiino,
\textquotedblleft  H-Theorem with generalized relative entropies and the 
Tsallis statistics,\textquotedblright\ 
\emph{J. Phys. Soc. Japan}, Vol. 67, no. 11, pp. 3658--3660, 1998. 

\bibitem{Mat:51}
K. Matusita,
\textquotedblleft  On the theory of statistical functions,\textquotedblright\ 
\emph{Ann. Inst. Statist. Math. Tokyo}, Vol. 3, 
pp. 17--35, 1951; \, corrections in
\emph{Ann. Inst. Statist. Math. Tokyo}, Vol. 4, 
pp. 51--53, 1952.

\bibitem{Dez:16} 
M.M. Deza and E. Deza, 
\emph{Encyclopedia of Distances}, 4th~ed. 
Berlin, Germany: Springer, 2016.










\bibitem{Roe:17}
B. Roensch and W. Stummer,  
\textquotedblleft 3D insights to some divergences for robust statistics and machine learning,\textquotedblright\  
In: F. Nielsen and F. Barbaresco (eds.), \emph{Geometric Science of Information GSI 2017}, 
Lecture Notes in Computer Science, vol. 10589, pp. 460--469. 
Cham, Switzerland: Springer International Publishing, 2017.

\bibitem{Roe:19a}
B. Roensch and W. Stummer,  
\textquotedblleft Robust estimation by means of scaled Bregman power distances; 
part I; non-homogeneous data,\textquotedblright\  
In: F. Nielsen and F. Barbaresco (eds.), \emph{Geometric Science of Information GSI 2019}, 
Lecture Notes in Computer Science, vol. 11712, pp. 319--330. 
Cham, Switzerland: Springer Nature, 2019.

\bibitem{Roe:19b}
B. Roensch and W. Stummer,  
\textquotedblleft Robust estimation by means of scaled Bregman power distances; 
part II; extreme values,\textquotedblright\  
In: F. Nielsen and F. Barbaresco (eds.), \emph{Geometric Science of Information GSI 2019}, 
Lecture Notes in Computer Science, vol. 11712, pp. 331--340. 
Cham, Switzerland: Springer Nature, 2019.

\bibitem{Kro:19}
S. Kr{\"o}mer and W. Stummer,
\textquotedblleft A new toolkit for mortality data analytics,\textquotedblright\ 
in: A. Steland, E. Rafajlowicz, and O. Okhrin (eds.), 
\emph{Stochastic Models, Statistics and Their Applications}, pp. 393--407.
Cham, Switzerland: Springer Nature Switzerland, 2019.
 
\bibitem{Henn:11}
R. Hennequin, B. David and R. Badeau,
\textquotedblleft Beta-Divergence as a Subclass
of Bregman Divergence,\textquotedblright\ 
\emph{IEEE Signal Proces. Lett.}, Vol. 18, no. 2, pp. 83--86, 2011. 
 
\bibitem{Itak:68}
F. Itakura and S. Saito,
\textquotedblleft Analysis synthesis telephony based on the
maximum likelihood method,\textquotedblright\ 
in: \emph{Proc. 6th Int. Congr. Acoust.}, 
1968, pp. C-17---C-20.
Los Alamitos, CA, USA.


\bibitem{Bas:15a}
A. Basu, A. Mandal, N. Martin and L. Pardo,
\textquotedblleft  Robust tests for the equality of two normal means
based on the density power divergence,\textquotedblright\ 
\emph{Metrika}, Vol. 78, 
pp. 611--634, 2015. 
 
\bibitem{Bas:16g}
A. Basu, A. Mandal, N. Martin and L. Pardo,
\textquotedblleft  Generalized Wald-type tests on minimum power divergence estimators,\textquotedblright\ 
\emph{Statistics}, Vol. 50, no. 1, 
pp. 1--26, 2016. 
 
\bibitem{Gho:16a}
A. Ghosh and A. Basu, 
\textquotedblleft  Robust Bayes estimation using the density power divergence,\textquotedblright\ 
\emph{Ann. Inst. Statist. Math.}, Vol. 68, 
pp. 413--437, 2016. 

\bibitem{Gho:16b}
A. Ghosh and A. Basu, 
\textquotedblleft  Robust estimation in generalized linear models:
the density power divergence approach,\textquotedblright\ 
\emph{TEST}, Vol. 25, 
pp. 269--290, 2016. 
 
\bibitem{Bas:17h}
A. Basu, A. Ghosh, A. Mandal, N. Martin and L. Pardo,
\textquotedblleft  A Wald-type test statistic for testing
linear hypothesis in logistic regression
models based on minimum density
power divergence estimator,\textquotedblright\ 
\emph{Electr. J. Statist.}, Vol. 11, 
pp. 2741--2772, 2017. 

\bibitem{Martin2:17}
F. Martin, J. Carballeira, L. Moreno, S. Garrido and P. Gonz{\'a}lez,
\textquotedblleft  Using the Jensen-Shannon, density power, and Itakura-Saito 
divergences to implement an evolutionary-based global localization filter 
for mobile robots,\textquotedblright\ 
\emph{IEEE Access}, Vol. 5, pp. 13922--13940, 2017. 

\bibitem{Bas:18f}
A. Basu, A. Mandal, N. Martin and L. Pardo,
\textquotedblleft  Testing composite hypothesis based on the density
power divergence,\textquotedblright\ 
\emph{Sankhya}, Vol. 80-B, Part 2, pp. 222--262, 2017. 
 
\bibitem{Balakri:19}
N. Balakrishnan, E. Castilla, N. Martin and L. Pardo,
\textquotedblleft  Robust estimators and test statistics for
one-shot device testing under
the exponential distribution,\textquotedblright\ 
\emph{IEEE Trans. Inf. Theory}, Vol. 65, no. 5, pp. 3080--3096, 2019. 

\bibitem{Balakri:20}
N. Balakrishnan, E. Castilla, N. Martin and L. Pardo,
\textquotedblleft  Robust inference for one-shot device testing data
under Weibull lifetime model,\textquotedblright\ 
\emph{IEEE Trans. Reliab.}, Vol. 69, no. 3, pp. 937--953, 2020. 
 
\bibitem{Ghosh:20d}
A. Ghosh and S. Majumdar,
\textquotedblleft  Ultrahigh-dimensional robust and efficient sparse
regression using non-concave penalized density power divergence,\textquotedblright\ 
\emph{IEEE Trans. Inf. Theory}, Vol. 66, no. 12, pp. 7812--7827, 2020. 
 
\bibitem{Leplat:20}
V. Leplat, N. Gillis and A.M.S. Ang,
\textquotedblleft  Blind audio source separation with
minimum-volume beta-divergence NMF,\textquotedblright\ 
\emph{IEEE Trans. Signal Proces.}, Vol. 68, 
pp. 3400--3410, 2020. 
 
\bibitem{Vandec:20}
M. Vandecappelle, N. Vervliet and L. De Lathauwer,
\textquotedblleft  A second-order method for fitting the canonical
polyadic decomposition with
non-least-squares cost,\textquotedblright\ 
\emph{IEEE Trans. Signal Proces.}, Vol. 68, 
pp. 4454--4465, 2020. 
 
\bibitem{Balakri:21}
N. Balakrishnan, E. Castilla, N. Martin and L. Pardo,
\textquotedblleft  Divergence-based robust inference under
proportional hazards model for one-shot device life-test,\textquotedblright\ 
\emph{IEEE Trans. Reliab.}, Vol. 70, no. 4, pp. 1355--1367, 2021. 
 
\bibitem{Basak:21}
S. Basak, A. Basu and M.C. Jones,
\textquotedblleft  On the \textquoteleft optimal\textquoteright\ density power divergence tuning parameter,\textquotedblright\ 
\emph{J. Appl. Statist.}, Vol. 48, no. 3, pp.536--556, 2021. 
 
\bibitem{Bas:21e}
A. Basu, A. Ghosh, A. Mandal, N. Martin and L. Pardo,
\textquotedblleft  Robust Wald-type tests in GLM with random design based
on minimum density power divergence estimators,\textquotedblright\ 
\emph{Statist. Meth. Appl.}, Vol. 30, 
pp.973--1005, 2021. 
 
\bibitem{Calv:21}
A. Calvino, N. Martin and L. Pardo,
\textquotedblleft  Robustness of minimum density power divergence
estimators and Wald-type test statistics in loglinear models with multinomial sampling,\textquotedblright\ 
\emph{J. Computat. Appl. Math.}, Vol. 386, No. 113214, 2021; doi:10.1016/j.cam.2020.113214.
 
\bibitem{Cast:21a}
E. Castilla, A. Ghosh, N. Martin and L. Pardo,
\textquotedblleft  Robust semiparametric inference for polytomous logistic
regression with complex survey design,\textquotedblright\ 
\emph{Adv. Data Anal. Classif.}, Vol. 15, 
pp. 701--734, 2021. 
 
\bibitem{Cast:21b}
E. Castilla, N. Martin, L. Pardo and K. Zografos,
\textquotedblleft  Composite likelihood methods: Rao-type tests based on
composite minimum density power divergence estimator,\textquotedblright\ 
\emph{Statist. Papers}, Vol. 62, 
pp. 1003--1041, 2021. 
 
\bibitem{Legros:21}
Q. Legros, J. Tachella, R. Tobin, A. McCarthy, S. Meignen, 
G.S. Buller, Y. Altmann, S. McLaughlinand M.E. Davies, 
\textquotedblleft  Robust 3D reconstruction of dynamic scenes
from single-photon Lidar using beta-divergences,\textquotedblright\ 
\emph{IEEE Trans. Image Process.}, Vol. 30, 
pp. 1716--1727, 2021. 
 
\bibitem{Pu11:22}
W. Pu, S. Ibrahim, X. Fu and M. Hong,
\textquotedblleft  Stochastic mirror descent for low-rank tensor
decomposition under non-Euclidean losses,\textquotedblright\ 
\emph{IEEE Trans. Signal Proces.}, Vol. 70, 
pp. 1803--1818, 2022. 
 
\bibitem{Rami:22}
M.A. Ramirez, W. Beccaro, D.Z. Rodriguez and R.L. Rosa,
\textquotedblleft  Differentiable measures for speech modeling\textquotedblright\ 
\emph{IEEE Access}, Vol. 10, pp. 17609--17618, 2022. 

\bibitem{Cast:23a}
E. Castilla and P.J. Chocano,
\textquotedblleft On the choice of the optimal tuning parameter in robust one-shot device testing analysis,\textquotedblright\ 
in: N. Balakrishnan et al. (eds.), 
\emph{Trends in Mathematical, Information and Data Sciences}, 
Studies in Systems, Decision and Control, vol. 445, pp. 169--180. 
Cham, Switzerland: Springer Nature Switzerland, 2023. 
 
\bibitem{Marmin:23}
A. Marmin, J.H. de Morais Goulart and C. F{\'e}votte,
\textquotedblleft  Majorization-minimization for sparse nonnegative
matrix factorization with the $\beta-$divergence,\textquotedblright\ 
\emph{IEEE Trans. Signal Proces.}, Vol. 71, 
pp. 1435--1447, 2023. 
 
\bibitem{Sara:23}
G. Saraceno, A. Ghosh, A. Basu and C. Agostinelli,
\textquotedblleft  Robust estimation of fixed effect parameters and variances
of linear mixed models: the minimum density power
divergence approach,\textquotedblright\ 
\emph{AStA Adv. Stat. Anal.},
2023; doi:10.1007/s10182-023-00473-z. 
 
\bibitem{Sharma2:23}
P. Sharma and P.M. Pradhan,
\textquotedblleft  A novel family of robust incremental adaptive
algorithms for distributed estimation based on Bregman divergence,\textquotedblright\ 
\emph{IEEE Sensor Lett.}, Vol. 7, No. 8, 7004104, 2023; doi:10.1109/LSENS.2023.3296396.
 
\bibitem{Zol:86} 
V.M. Zolotarev, 
\emph{One-dimensional Stable Distributions}. 
Providence, USA: American Mathematical Society, 1986.


\bibitem{Jeo:00}
D.-B. Jeong and S. Sarkar,
\textquotedblleft  Negative exponential disparity based family
of goodness-of-fit tests for multinomial models,\textquotedblright\ 
\emph{J. Statist. Comput. Simul.}, Vol. 65, 
pp. 43--61, 2000. 

\bibitem{Bhan:06}
S.K. Bhandari, A. Basu and S. Sarkar,
\textquotedblleft  Robust inference in parametric models
using the family of generalized negative exponential disparities,\textquotedblright\ 
\emph{Aust. N. Z. J. Stat.}, Vol. 48, no. 1, 
pp. 95--114, 2006. 

\bibitem{Lind:94}
B.G. Lindsay,
\textquotedblleft Efficiency versus robustness: the case for
Hellinger distance and related methods,\textquotedblright\ 
\emph{Ann. Stat.}, Vol. 22, no. 2, pp. 1081--1114, 1994. 

\bibitem{Bas:94}
A. Basu and S. Sarkar,
\textquotedblleft The trade-off between robustness and efficiency
and the effect of model smoothing
in minimum disparity inference,\textquotedblright\ 
\emph{J. Statist. Comput. Simul.}, Vol. 50, 
pp. 173--185, 1994. 


\bibitem{Berend:14}
D. Berend, P. Harremo{\"e}s and A. Kontorovich,
\textquotedblleft  Minimum KL-divergence on complements
of $L_{1}$ balls,\textquotedblright\ 
\emph{IEEE Trans. Inf. Theory}, Vol. 60, no. 6, pp. 3172--3177, 2014. 
 
\bibitem{Jiao7:14}
J. Jiao, T.A. Courtade, A. No, K. Venkat and T. Weissman,
\textquotedblleft  The curious case of the binary alphabet,\textquotedblright\ 
\emph{IEEE Trans. Inf. Theory}, Vol. 60, no. 12, pp. 7616--7626, 2014. 
 
\bibitem{Como:15}
G. Como and F. Fagnani,
\textquotedblleft  Robustness of large-scale stochastic
matrices to localized perturbations,\textquotedblright\ 
\emph{IEEE Trans. Netw. Sci. Eng.}, Vol. 2, no. 2, pp. 53--64, 2015. 
 
\bibitem{Han7:15}
Y. Han, J. Jiao and T. Weissman,
\textquotedblleft  Minimax Estimation of Discrete
Distributions Under $\ell_{1}$ loss,\textquotedblright\ 
\emph{IEEE Trans. Inf. Theory}, Vol. 61, no. 11, pp. 6343--6354, 2015. 
 
\bibitem{Sason:15}
I. Sason,
\textquotedblleft  Tight bounds for symmetric divergence measures
and a refined bound for lossless source coding,\textquotedblright\ 
\emph{IEEE Trans. Inf. Theory}, Vol. 61, no. 2, pp. 701--707, 2015. 
 
\bibitem{Bats:16}
A. Batsidis, N. Martin, L. Pardo and K. Zografos,
\textquotedblleft  $\phi-$Divergence based procedure for parametric
change-point problems,\textquotedblright\ 
\emph{Methodol. Comput. Appl. Probab.}, Vol. 18, 
pp. 21--35, 2016. 
 
\bibitem{Boech:16}
G. B{\"o}cherer and B.C. Geiger,
\textquotedblleft  Optimal quantization for distribution synthesis,\textquotedblright\ 
\emph{IEEE Trans. Inf. Theory}, Vol. 62, no. 11, pp. 6162--6172, 2016. 

\bibitem{Das:16}
N.V.A. Das and N. Kashyap,
\textquotedblleft  MCMC methods for drawing random samples
from the discrete-grains model
of a Magnetic Medium,\textquotedblright\ 
\emph{IEEE J. Sel. Areas Commun.}, Vol. 34, no. 9, pp. 2430--2438, 2016. 
 
\bibitem{Alonso:17}
J.M. Alonso-Revenga, N. Martin and L. Pardo,
\textquotedblleft  New improved estimators for overdispersion in models with
clustered multinomial data and unequal cluster sizes,\textquotedblright\ 
\emph{Stat. Comput.}, Vol. 27, 
pp. 193--217, 2017. 
 
\bibitem{Kez:17}
A. Keziou and P. Regnault,
\textquotedblleft  Semiparametric estimation of mutual information
and related criteria: optimal test of independence,\textquotedblright\ 
\emph{IEEE Trans. Inf. Theory}, Vol. 63, no. 1, pp. 57--71, 2017. 
 
\bibitem{LiuJingbo:17}
J. Liu, P. Cuff and S. Verd{\'u},
\textquotedblleft  $E_{\gamma}-$resolvability,\textquotedblright\ 
\emph{IEEE Trans. Inf. Theory}, Vol. 63, no. 5, pp. 2629--2658, 2017. 

\bibitem{Tzortzis:17}
I. Tzortzis, C.D. Charalambous, T. Charalambous,
C.N. Hadjicostis and M. Johansson,
\textquotedblleft  Approximation of Markov processes by lower
dimensional processes via total variation metrics,\textquotedblright\ 
\emph{IEEE Trans. Autom. Control}, Vol. 62, no. 3, pp. 1030--1045, 2017. 
 
\bibitem{Cast:18}
E. Castilla, N. Martin and L. Pardo,
\textquotedblleft  Minimum phi-divergence estimators for multinomial
logistic regression with complex sample design,\textquotedblright\ 
\emph{AStA Adv. Stat. Anal.}, Vol. 102, 
pp. 381--411, 2018. 
 
\bibitem{Csi:18}
I. Csisz{\'a}r and T. Breuer,
\textquotedblleft  Expected value minimization in information
theoretic multiple priors models,\textquotedblright\ 
\emph{IEEE Trans. Inf. Theory}, Vol. 64, no. 6, pp. 3957--3974, 2018. 
 
\bibitem{ElGheche:18}
M. El Gheche, G. Chierchia and J.-C. Pesquet,
\textquotedblleft  Proximity operators of discrete
information divergences,\textquotedblright\ 
\emph{IEEE Trans. Inf. Theory}, Vol. 64, no. 2, pp. 1092--1104, 2018. 

\bibitem{Feli:18}
A. Felipe, N. Martin, P. Miranda and L. Pardo,
\textquotedblleft  Statistical inference in constrained latent class models
for multinomial data based on $\phi-$divergence measures,\textquotedblright\ 
\emph{Adv. Data Anal. Classif.}, Vol. 12, 
pp. 605--636, 2018. 
 
\bibitem{Mark:18}
M. Markatou and Y. Chen,
\textquotedblleft  Non-quadratic distances in model assessment,\textquotedblright\ 
\emph{Entropy}, Vol. 20, No. 464, 2018; doi:10.3390/e20060464.
 
\bibitem{SunYujing:18}
Y. Sun, S. Schaefer and W. Wang,
\textquotedblleft  Image structure retrieval via $L_{0}$ minimization,\textquotedblright\ 
\emph{IEEE Trans. Visual. Comput. Graph.}, Vol. 24, no. 7, pp. 2129--2139, 2018. 
 
\bibitem{Asadi:19}
M. Asadi, N. Ebrahimi, O. Kharazmi and E.S. Soofi,
\textquotedblleft  Mixture models, Bayes Fisher information,
and divergence measures,\textquotedblright\ 
\emph{IEEE Trans. Inf. Theory}, Vol. 65, no. 4, pp. 2316--2321, 2019. 
 
\bibitem{Bro:19d}
M. Broniatowski, E. Miranda and W. Stummer,
\textquotedblleft Testing the number and the nature of the
components in a mixture distribution,\textquotedblright\ 
in: F. Nielsen and F. Barbaresco (eds.), \emph{Geometric Science of Information GSI 2019}, 
Lecture Notes in Computer Science, vol. 11712, pp. 309--318. 
Cham, Switzerland: Springer Nature Switzerland, 2019. 

\bibitem{Coll:19}
J.-F. Collet,
\textquotedblleft  An exact expression for the gap in the data
processing inequality for f-divergences,\textquotedblright\ 
\emph{IEEE Trans. Inf. Theory}, Vol. 65, no. 7, pp. 4387--4391, 2019. 
 
\bibitem{Sas:19}
I. Sason,
\textquotedblleft  On data-processing and majorization inequalities for
f-Divergences with applications,\textquotedblright\ 
\emph{Entropy}, Vol. 21, No. 1022, 2019; doi:10.3390/e21101022.

\bibitem{Yagli:19}
S. Yagli and P. Cuff,
\textquotedblleft  Exact exponent for soft covering,\textquotedblright\ 
\emph{IEEE Trans. Inf. Theory}, Vol. 65, no. 10, pp. 6234--6262, 2019. 
 
\bibitem{ZhaoKun:19}
K. Zhao, Y. Liu, H. Du and B. Yuan,
\textquotedblleft  Scaling analysis of multilevel polar coded modulation\textquotedblright\ 
\emph{IEEE Access}, Vol. 7, pp. 65129--65138, 2019.

\bibitem{DePonti:20}
N. De Ponti,
\textquotedblleft Metric properties of homogeneous and spatially
inhomogeneous F-Divergences,\textquotedblright\ 
\emph{IEEE Trans. Inf. Theory}, Vol. 66, no. 5, pp. 2872--2890, 2020. 
 
\bibitem{Kam:20}
N.B. Kammerer and W. Stummer,
\textquotedblleft Some dissimilarity measures of branching processes
and optimal decision making in the presence of potential pandemics,\textquotedblright\ 
\emph{Entropy}, Vol. 22(8), No. 874, 2020 (123 pages); doi:10.3390/e22080874.

\bibitem{Nish:20}
T. Nishiyama  and I. Sason,
\textquotedblleft  On relations between the relative entropy and
$\chi^{2}-$divergence, generalizations and applications,\textquotedblright\ 
\emph{Entropy}, Vol. 22, No. 563, 2020; doi:10.3390/e22050563.

\bibitem{Nom:20}
R. Nomura,
\textquotedblleft  Source resolvability and intrinsic randomness:
two random number generation problems With
respect to a subclass of f-divergences,\textquotedblright\ 
\emph{IEEE Trans. Inf. Theory}, Vol. 66, no. 12, pp. 7588--7601, 2020. 
 
\bibitem{Rassouli:20}
B. Rassouli and D. G{\"u}nd{\"u}z,
\textquotedblleft  Optimal utility-privacy trade-off with total
variation distance as a privacy measure,\textquotedblright\ 
\emph{IEEE Trans. Inf. Forensics Sec.}, Vol. 15, 
pp. 594--603, 2020. 
 
\bibitem{Espo:21}
A.R. Esposito, M. Gastpar and I. Issa,
\textquotedblleft  Generalization error bounds via R{\'e}nyi-,
f-divergences and maximal leakage,\textquotedblright\ 
\emph{IEEE Trans. Inf. Theory}, Vol. 67, no. 8, pp. 4986--5004, 2021. 
 
\bibitem{Mark:21}
M. Markatou, D. Karlis and Y. Ding,
\textquotedblleft  Distance-based statistical inference,\textquotedblright\ 
\emph{Annu. Rev. Stat. Appl.}, Vol. 8, 
pp. 301--327, 2021. 
 
\bibitem{Saleh:21}
S. Salehkalaibar, M.H. Yassaee, V.Y.F. Tan and M. Ahmadipour,
\textquotedblleft  State masking over a two-state
compound channel,\textquotedblright\ 
\emph{IEEE Trans. Inf. Theory}, Vol. 67, no. 9, pp. 5651--5673, 2021. 
 
\bibitem{Stu:21}
W. Stummer,
\textquotedblleft Optimal transport with some directed distances,\textquotedblright\ 
in: F. Nielsen and F. Barbaresco (eds.), \emph{Geometric Science of Information GSI 2021}, 
Lecture Notes in Computer Science, vol. 12829, pp. 829--840. 
Cham, Switzerland: Springer Nature Switzerland, 2021. 

\bibitem{Tzortzis:21}
I. Tzortzis, C.D. Charalambous and C.N. Hadjicostis,
\textquotedblleft  Jump LQR systems with unknown
transition probabilities,\textquotedblright\ 
\emph{IEEE Trans. Autom. Control}, Vol. 66, no. 6, pp. 2693--2708, 2021. 
 
\bibitem{Birr:22}
J. Birrell, M.A. Katsoulakis and Y. Pantazis,
\textquotedblleft  Optimizing variational representations of
divergences and accelerating their
statistical estimation,\textquotedblright\ 
\emph{IEEE Trans. Inf. Theory}, Vol. 68, no. 7, pp. 4553--4572, 2022. 
 
\bibitem{Cast:22}
E. Castilla and P.J. Chocano,
\textquotedblleft  A new robust approach for multinomial logistic
regression With complex design model,\textquotedblright\ 
\emph{IEEE Trans. Inf. Theory}, Vol. 68, no. 11, pp. 7379--7395, 2022. 
 
\bibitem{Dixit:22}
A. Dixit, M. Ahmadi and J.W. Burdick,
\textquotedblleft  Distributionally robust model predictive control
with total variation distance,\textquotedblright\ 
\emph{IEEE Control Syst. Lett.}, Vol. 6, 
pp. 3325--3330, 2022. 
 
\bibitem{HyunDongwoon:22}
D. Hyun, G.B. Kim, N. Bottenus  and J.J. Dahl,
\textquotedblleft  Ultrasound lesion detectability as a distance
between probability measures,\textquotedblright\ 
\emph{IEEE Trans. Ultrason. Ferr. Frequ. Control}, Vol. 69, no. 2, pp. 732--743, 2022. 
 
\bibitem{Melb:22}
J. Melbourne, S. Talukdar, S. Bhaban, M. Madiman and M.V. Salapaka,
\textquotedblleft  The differential entropy of mixtures: new bounds
and applications,\textquotedblright\ 
\emph{IEEE Trans. Inf. Theory}, Vol. 68, no. 4, pp. 2123--2146, 2022. 
 
\bibitem{Peng4:22}
J. Peng, G. Serrano, I.M. Traniello, M.E. Calleja-Cervantes,
U.V. Chembazhi, S. Bangru, T. Ezponda, J.R. Rodriguez-Madoz,
A. Kalsotra, F. Prosper, I. Ochoa and M. Hernaez,
\textquotedblleft  SimiC enables the inference of complex gene regulatory dynamics across cell phenotypes,\textquotedblright\ 
\emph{Commun. Biol.}, Vol. 5, No. 531, 2022; doi:10.1038/s42003-022-03319-7.
 
\bibitem{Tan11:22}
Z. Tan and X. Zhang,
\textquotedblleft  On loss functions and regret bounds for
multi-category classification,\textquotedblright\ 
\emph{IEEE Trans. Inf. Theory}, Vol. 68, no. 8, pp. 5295--5313, 2022. 
 
\bibitem{Zhang19:22}
F. Zhang, J. Li, and H.K.T. Ng,
\textquotedblleft  Minimum f-divergence estimation with
applications to degradation data analysis,\textquotedblright\ 
\emph{IEEE Trans. Inf. Theory}, Vol. 68, no. 10, pp. 6774--6789, 2022. 
 
\bibitem{Alba:23}
M.V. Alba-Fern{\'a}ndez and M.D. Jim{\'e}nez-Gamero,
\textquotedblleft Equivalence tests for multinomial data
based on $\phi-$divergences,\textquotedblright\ 
in: N. Balakrishnan et al. (eds.), 
\emph{Trends in Mathematical, Information and Data Sciences}, 
Studies in Systems, Decision and Control, vol. 445, pp. 121--129. 
Cham, Switzerland: Springer Nature Switzerland, 2023. 
 
\bibitem{Baudry:23}
J.-P. Baudry, M. Broniatowski and C. Thommeret,
\textquotedblleft Aggregated tests based on supremal
divergence estimators for non-regular
statistical models,\textquotedblright\ 
in: F. Nielsen and F. Barbaresco (eds.), \emph{Geometric Science of Information GSI 2023,
Part I}, 
Lecture Notes in Computer Science, vol. 14071, pp. 136--144. 
Cham, Switzerland: Springer Nature Switzerland, 2023.

\bibitem{Bouk:23}
M. Boukeloua and A. Keziou,
\textquotedblleft Empirical likelihood with censored data?,\textquotedblright\ 
in: F. Nielsen and F. Barbaresco (eds.), \emph{Geometric Science of Information GSI 2023,
Part I}, 
Lecture Notes in Computer Science, vol. 14071, pp. 125--135. 
Cham, Switzerland: Springer Nature Switzerland, 2023.

\bibitem{Cressie:23}
N. Cressie, A.R. Pearse and D. Gunawan,
\textquotedblleft Optimal spatial prediction
for non-negative spatial processes using
a phi-divergence loss function,\textquotedblright\ 
in: N. Balakrishnan et al. (eds.), 
\emph{Trends in Mathematical, Information and Data Sciences}, 
Studies in Systems, Decision and Control, vol. 445, pp. 181--197. 
Cham, Switzerland: Springer Nature Switzerland, 2023. 
 
\bibitem{Kateri:23}
M. Kateri,
\textquotedblleft Generalized Models for Binary and
Ordinal Responses,\textquotedblright\ 
in: N. Balakrishnan et al. (eds.), 
\emph{Trends in Mathematical, Information and Data Sciences}, 
Studies in Systems, Decision and Control, vol. 445, pp. 63--71. 
Cham, Switzerland: Springer Nature Switzerland, 2023. 
 
\bibitem{Mano:23}
T. Manole and A. Ramdas,
\textquotedblleft  Martingale methods for sequential estimation of
convex functionals and divergences,\textquotedblright\ 
\emph{IEEE Trans. Inf. Theory}, Vol. 69, no. 7, pp. 4641--4658, 2023. 
 
\bibitem{Mark:23}
M. Markatou and A. Liu,
\textquotedblleft Statistical distances in goodness-of-fit,\textquotedblright\ 
in: N. Balakrishnan et al. (eds.), 
\emph{Trends in Mathematical, Information and Data Sciences}, 
Studies in Systems, Decision and Control, vol. 445, pp. 213–-222. 
Cham, Switzerland: Springer Nature Switzerland, 2023. 
 
\bibitem{Masiha:23}
S. Masiha, A. Gohari and M.H. Yassaee,
\textquotedblleft  f-Divergences and their applications in lossy
compression and bounding generalization error,\textquotedblright\ 
\emph{IEEE Trans. Inf. Theory}, Vol. 69, no. 12, pp. 7538--7564, 2023. 
 
\bibitem{Miranda2:23}
P. Miranda, A. Felipe and N. Martin,
\textquotedblleft Phi-divergence test statistics applied
to latent class models for binary data,\textquotedblright\ 
in: N. Balakrishnan et al. (eds.), 
\emph{Trends in Mathematical, Information and Data Sciences}, 
Studies in Systems, Decision and Control, vol. 445, pp. 223–-231. 
Cham, Switzerland: Springer Nature Switzerland, 2023. 
 
\bibitem{Nielsen:23a}
F. Nielsen and K. Okamura,
\textquotedblleft On the f-divergences between
hyperboloid and Poincar{\'e} distributions,\textquotedblright\ 
in: F. Nielsen and F. Barbaresco (eds.), \emph{Geometric Science of Information GSI 2023,
Part I}, 
Lecture Notes in Computer Science, vol. 14071, pp. 176--185. 
Cham, Switzerland: Springer Nature Switzerland, 2023.

\bibitem{Perr:23}
P. Perrone,
\textquotedblleft Categorical information geometry,\textquotedblright\ 
in: F. Nielsen and F. Barbaresco (eds.), \emph{Geometric Science of Information GSI 2023,
Part I}, 
Lecture Notes in Computer Science, vol. 14071, pp. 268--277. 
Cham, Switzerland: Springer Nature Switzerland, 2023.

\bibitem{Vela:23}
S. Velasco-Forero,
\textquotedblleft Can Generalised Divergences Help
for Invariant Neural Networks?,\textquotedblright\ 
in: F. Nielsen and F. Barbaresco (eds.), \emph{Geometric Science of Information GSI 2023,
Part I}, 
Lecture Notes in Computer Science, vol. 14071, pp. 82--90. 
Cham, Switzerland: Springer Nature Switzerland, 2023.

\bibitem{Nielsen:24a}
F. Nielsen and K. Okamura,
\textquotedblleft  On the f-divergences between densities of a multivariate location or
scale family,\textquotedblright\ 
\emph{Statist. Comput.}, Vol. 34, No. 60, 2024; doi:10.1007/s11222-023-10373-6.
 

\bibitem{Csi:18c}
I. Csisz{\'a}r and T. Breuer,
\textquotedblleft  Expected value minimization in information
theoretic multiple priors models,\textquotedblright\ 
\emph{IEEE Trans. Inf. Theory}, Vol. 64, no. 6, pp. 3957--3974, 2018. 
 
\bibitem{Bas:19h}
S. Jana and A. Basu,
\textquotedblleft  A characterization of all single-integral,
non-kernel divergence estimators,\textquotedblright\ 
\emph{IEEE Trans. Inf. Theory}, Vol. 65, no. 12, pp. 7976--7984, 2019. 
 
\bibitem{Painsky:20}
A. Painsky and G.W. Wornell,
\textquotedblleft  Bregman divergence bounds and universality
properties of the logarithmic loss,\textquotedblright\ 
\emph{IEEE Trans. Inf. Theory}, Vol. 66, no. 3, pp. 1658--1673, 2020. 
 
\bibitem{Vial:21}
P.-H. Vial, P. Magron, T. Oberlin and C. F{\'e}votte,
\textquotedblleft  Phase retrieval with Bregman divergences and
application to audio signal recovery,\textquotedblright\ 
\emph{IEEE J. Sel. Topics Signal Proces.}, Vol. 15, no. 1, pp. 51--64, 2021. 
 
\bibitem{Lohit:23}
H. Lohit and D. Kumar,
\textquotedblleft  Modified total Bregman divergence driven picture
fuzzy clustering with local information for brain MRI image segmentation,\textquotedblright\ 
\emph{Appl. Soft Comput.}, Vol. 144, No. 110460, 2023; doi:10.1016/j.asoc.2023.110460.


\bibitem{Che:07}
B. Chen, J. Hu, L. Pu, and Z. Sun,
\textquotedblleft  Stochastic gradient algorithm under $(h,\phi)-$entropy,\textquotedblright\ 
\emph{Circuits Syst. Signal Process.}, Vol. 26, pp. 941--960, 2007. 

\bibitem{Girar:15}
V. Girardin and L. Lhote,
\textquotedblleft  Rescaling entropy and divergence Rates,\textquotedblright\ 
\emph{IEEE Trans. Inf. Theory}, Vol. 61, no. 11, pp. 5868--5882, 2015. 
 
\bibitem{Ren:15}
M. Ren, J. Zhang, M. Jiang, M. Yu, and J. Xu,
\textquotedblleft  Minimum $(h,\phi)-$entropy control
for non-Gaussian stochastic networked control systems 
and its application to a 
networked DC motor control system,\textquotedblright\ 
\emph{IEEE Trans. Control Syst. Technol.}, Vol. 23, no. 1, 
pp. 406--411, 2015.  

\bibitem{Girar:19}
V. Girardin, L. Lhote and P. Regnault,
\textquotedblleft  Different closed-form expressions for generalized entropy
rates of Markov chains,\textquotedblright\ 
\emph{Methodol. Comput. Appl. Probab.}, Vol. 21, 
pp. 1431--1452, 2019. 
 

\bibitem{Nielsen:11e}
F. Nielsen and S. Boltz,
\textquotedblleft The Burbea-Rao and Bhattacharyya centroids,\textquotedblright\ 
\emph{IEEE Trans. Inf. Theory}, Vol. 57, no. 8, pp. 5455--5466, 2011. 
 
\bibitem{Nielsen:17e}
F. Nielsen and R. Nock,
\textquotedblleft  Generalizing skew Jensen divergences and Bregman
divergences with comparative convexity,\textquotedblright\ 
\emph{IEEE Signal Proces. Lett.}, Vol. 24, no. 8, pp. 1123--5466, 2017. 

\bibitem{Xu18:13}
H. Xu, C. Caramanis and S. Mannor,
\textquotedblleft  Outlier-robust PCA: the high-dimensional case,\textquotedblright\ 
\emph{IEEE Trans. Inf. Theory}, Vol. 59, no. 1, pp. 546--572, 2013. 

\bibitem{Kim12:14}
S. Kim, B. Ham, B. Kim and K. Sohn,
\textquotedblleft Mahalanobis distance cross-correlation for
illumination-invariant stereo matching,\textquotedblright\ 
\emph{IEEE Trans. Circ. Syst. Video Techn.}, Vol. 24, no. 11, pp. 1844--1859, 2014. 
 
\bibitem{Mei1:16}
J. Mei, M. Liu, Y.-F. Wang and H. Gao,
\textquotedblleft  Learning a Mahalanobis distance-based dynamic
time warping measure for multivariate time series classification,\textquotedblright\ 
\emph{IEEE Trans. Cybernetics}, Vol. 46, no. 6, pp. 1363--1374, 2016. 
 
\bibitem{Zhang33:16}
Y. Zhang, B. Du, L. Zhang and S. Wang,
\textquotedblleft  A low-rank and sparse matrix
decomposition-based Mahalanobis distance method
for hyperspectral anomaly detection,\textquotedblright\ 
\emph{IEEE Trans. Geosci. Remote Sens.}, Vol. 54, no. 3, pp. 1376--1389, 2016. 
 
\bibitem{Li14:17}
Z. Li, D.P. Filev, I. Kolmanovsky, E. Atkins and J. Lu,
\textquotedblleft  A new clustering algorithm for processing
GPS-based road anomaly reports with a Mahalanobis distance,\textquotedblright\ 
\emph{IEEE Trans. Intell. Transp. Syst.}, Vol. 18, no. 7, pp. 1980--1988, 2017. 
 
\bibitem{Xu22:17}
Y. Xu, H. Min, Q. Wu, H. Song and B. Ye,
\textquotedblleft  Multi-instance metric transfer
learning for genome-wide protein
function prediction,\textquotedblright\ 
\emph{Sci. Rep.}, Vol. 7, 41831, 2017; doi:10.1038/srep41831. 
 
\bibitem{Mahony:18}
C.R. Mahony1 and A.J. Cannon,
\textquotedblleft  Wetter summers can intensify departures from
natural variability in a warming climate,\textquotedblright\ 
\emph{Nature Commun.}, Vol. 9, 783, 2018; doi:10.1038/s41467-018-03132-z. 

\bibitem{Xu33:18}
Y. Xu, Z. Wu, J. Chanussot and Z. Wei,
\textquotedblleft  Joint reconstruction and anomaly detection from
compressive hyperspectral images using Mahalanobis distance-regularized
tensor RPCA,\textquotedblright\ 
\emph{IEEE Trans. Geosci. Remote Sens.}, Vol. 56, no. 5, pp. 2919--2930, 2018. 
 
\bibitem{Ether:19}
T.R. Etherington,
\textquotedblleft  Mahalanobis distances and ecological niche modelling: 
orrecting a chi-squared probability error,\textquotedblright\ 
\emph{PeerJ}, Vol. 7, e6678, 2019; doi:10.7717/peerj.6678. 
 
\bibitem{Fitz:19}
M.C. Fitzpatrick and R.R. Dunn,
\textquotedblleft  Contemporary climatic analogs for 540 North
American urban areas in the late 21st century,\textquotedblright\ 
\emph{Nature Commun.}, Vol. 10, 614, 2019; doi:10.1038/s41467-019-08540-3. 
 
\bibitem{Kaka:19}
M. Kakavand, A. Mustapha, Z. Tan, S.F. Yazdani and L. Arulsamy,
\textquotedblleft  Using Mahalanobis distance map for web
service attacks classification\textquotedblright\ 
\emph{IEEE Access}, Vol. 7, pp. 167141--167156, 2019. 

\bibitem{Sun20:19}
S. Sun, H. Wang, Z. Chang, B. Mao and Y. Liu,
\textquotedblleft  On the Mahalanobis distance classification
criterion for a ventricular septal defect diagnosis system,\textquotedblright\ 
\emph{IEEE Sensors J.}, Vol. 19, no. 7, pp. 2665--2674, 2019. 

\bibitem{Bai:20}
Z. Bai, X.-L. Zhang and J. Chen,
\textquotedblleft  Speaker verification by partial AUC optimization
with Mahalanobis distance metric learning,\textquotedblright\ 
\emph{IEEE/ACM Trans. Audio Speech Lang. Proces.}, Vol. 28, 
pp. 1533--1548, 2020. 
 
\bibitem{Li22:20}
Q. Li, Z. Wu, L. Lin, J. Zeng, J. Zhang, H. Yan and S. Min,
\textquotedblleft  High-level fusion coupled with
Mahalanobis distance weighted
(MDW) method for multivariate calibration,\textquotedblright\ 
\emph{Sci. Rep.}, Vol. 10, 5478, 2020; doi:10.1038/s41598-020-62396-y. 
 
\bibitem{Naveed:20}
K. Naveed and N. ur Rehman,
\textquotedblleft  Wavelet based multivariate signal denoising using
Mahalanobis distance and EDF statistics,\textquotedblright\ 
\emph{IEEE Trans. Signal Proces.}, Vol. 68, 
pp. 5997--6010, 2020. 
 
\bibitem{Wang31:20}
R. Wang, J.-W. Chen, Y. Wang, L. Jiao and M. Wang,
\textquotedblleft SAR image change detection via spatial
metric learning with an improved Mahalanobis distance,\textquotedblright\ 
\emph{IEEE Geosci. Remote Sens. Lett.}, Vol. 17, no. 1, pp. 77--81, 2020. 
 
\bibitem{Winter:20}
A.J. Winter, J.M.D. Kruijssen, S.N. Longmore and M. Chevance,
\textquotedblleft Stellar clustering shapes the architecture of
planetary systems,\textquotedblright\ 
\emph{Nature}, Vol. 586, 
pp. 528--532, 2020,
and Supplement on Methods.
 
\bibitem{Bartlett:21}
T.E. Bartlett, P. Jia, S. Chandna and S. Roy,
\textquotedblleft  Inference of tissue relative
proportions of the breast epithelial
cell types luminal progenitor, basal,
and luminal mature,\textquotedblright\ 
\emph{Sci. Rep.}, Vol. 11, 23702, 2021; doi:10.1038/s41598-021-03161-7. 
 
\bibitem{Butter:21}
N.C. Butterfield, K.F. Curry, J. Steinberg , H. Dewhurst, D. Komla-Ebri,
N.S. Mannan, A.-T. Adoum, V.D. Leitch, J.G. Logan, J.A. Waung,
E. Ghirardello, L. Southam, S.E. Youlten, J.M. Wilkinson,
E.A. McAninch, V.E. Vancollie, F. Kussy, J.K. White,
C.J. Lelliott, D.J. Adams, R. Jacques, A.C. Bianco, A. Boyde,
E. Zeggini, P.I. Croucher, G.R. Williams  and J.H.D. Bassett,
\textquotedblleft  Accelerating functional gene discovery in
osteoarthritis,\textquotedblright\ 
\emph{Nature Commun.}, Vol. 12, 467, 2021; doi:0.1038/s41467-020-20761-5. 
 
\bibitem{Chamb:21}
M. Chamberland, S. Genc, C.M.W. Tax, D. Shastin, K. Koller,
E.P. Raven, A. Cunningham, J. Dohert, M.B.M. van den Bree,
G.D. Parker, K. Hamandi, W.P. Gray and D.K. Jones,
\textquotedblleft  Detecting microstructural deviations in
individuals with deep diffusion MRI tractometry,\textquotedblright\ 
\emph{Nature Commput. Sci.}, Vol. 1, 598--606, 2021; 
 
\bibitem{Kang9:21}
J.B. Kang , A. Nathan, K. Weinand, F. Zhang, N. Millard, L. Rumker,
D.B. Moody, I. Korsunsky and S. Raychaudhuri,
\textquotedblleft  Efficient and precise single-cell reference atlas
mapping with Symphony,\textquotedblright\ 
\emph{Nature Commun.}, Vol. 12, 5890, 2021; doi:10.1038/s41467-021-25957-x. 
 
\bibitem{Sato:21}
H. Sato, H. Ogihara, K. Takahashi, Y. Kawata, Y. Kojima,
K. Tominaga, J. Yokoyama, Y. Hamamoto and S. Terai,
\textquotedblleft  New cine magnetic resonance
imaging parameters
for the differential diagnosis
of chronic intestinal
pseudo-obstruction,\textquotedblright\ 
\emph{Sci. Rep.}, Vol. 11, 22974, 2021; doi:10.1038/s41598-021-02268-1. 
 
\bibitem{Zheng8:21}
J. Zheng, F. Shen and L. Ye,
\textquotedblleft  Improved Mahalanobis distance based JITL-LSTM
soft sensor for multiphase batch processes\textquotedblright\ 
\emph{IEEE Access}, Vol. 9, pp. 72172--72182, 2021.

\bibitem{A:22}
R. A, X. Mu and J. He,
\textquotedblleft  Enhance tensor RPCA-based Mahalanobis distance
method for hyperspectral anomaly detection,\textquotedblright\ 
\emph{IEEE Geosci. Remote Sens. Lett.}, Vol. 19, No. 6008305, 2022. 
 
\bibitem{Chakraborty:22}
A. Chakraborty, J.G. Wang and F. Ay,
\textquotedblleft  dcHiC detects differential compartments
across multiple Hi-C datasets,\textquotedblright\ 
\emph{Nature Commun.}, Vol. 13, 6827, 2022; doi:10.1038/10.1038/s41467-022-34626-6. 
 

\bibitem{Chen31:22}
Y. Chen, C. Han, J. He and G. Wang,
\textquotedblleft  A framework of Mahalanobis-distance metric with
supervised learning for clustering multipath
components in MIMO channel analysis,\textquotedblright\ 
\emph{IEEE Trans. Antenn. Propag.}, Vol. 70, no. 6, pp. 4069--4081, 2022. 
 
\bibitem{DosSantos:22}
R.F. dos Santos, M. Paraskevaidi, D.M.A. Mann, D. Allsop,
M.C.D. Santos, C.L.M. Morais and K.M.G. Lima,
\textquotedblleft  Alzheimer’s disease diagnosis
by blood plasma molecular
fluorescence spectroscopy (EEM),\textquotedblright\ 
\emph{Sci. Rep.}, Vol. 12, 16199, 2022; doi:10.1038/10.1038/s41598-022-20611-y. 
 
\bibitem{Guerra:22}
C.A. Guerra, M. Berdugo, D.J. Eldridge, N. Eisenhauer,
B.K. Singh, H. Cui, S. Abades, F.D. Alfaro,
A.R. Bamigboye, F. Bastida, J.L. Blanco-Pastor, A. de los Rios15,
J. Dur{\'a}n, T. Grebenc, J.G. Ill{\'a}n, Y.-R. Liu, T.P. Makhalanyane,
S. Mamet, M.A. Molina-Montenegro, J.L. Moreno, A. Mukherjee,
T.U. Nahberger, G.F. Penaloza-Bojac{\'a}, C{\'e}sar Plaza, Sergio Pic{\'o},
J.P. Verma, A. Rey, A. Rodriguez, L. Tedersoo,
A.L. Teixido, C. Torres-Diaz, P. Trivedi, J. Wang,
L. Wang, J. Wang, E. Zaady, X. Zhou, X.-Q. Zhou and M. Delgado-Baquerizo,
\textquotedblleft Global hotspots for soil nature conservation,\textquotedblright\ 
\emph{Nature}, Vol. 610, 
pp. 693--698, 2022,
and Supplement on Methods.
 
\bibitem{HuangHai:22}
H. Huang, J. Huang, X. Li, W. Zhuo, Y. Wu, Q. Niu, W. Su and W. Yuan,
\textquotedblleft  A dataset of winter wheat
aboveground biomass in China
during 2007-2015 based on data
assimilation,\textquotedblright\ 
\emph{Sci. Data}, Vol. 9, 200, 2022; doi:10.1038/s41597-022-01305-6. 
 
\bibitem{Nomoto:22}
M. Nomoto, E. Murayama, S. Ohno, R. Okubo-Suzuki, S.-i. Muramatsu and K. Inokuchi,
\textquotedblleft  Hippocampus as a sorter and reverberatory
integrator of sensory inputs,\textquotedblright\ 
\emph{Nature Commun.}, Vol. 13, 7413, 2022; doi:10.1038/s41467-022-35119-2. 
 
\bibitem{Reichen:22}
L. Reichen, A.-M. Burgdorf, S. Br{\"o}nnimann, J. Franke, R. Hand,
V. Valler, E. Samakinwa, Y. Brugnara and T. Rutishauser,
\textquotedblleft  A decade of cold Eurasian winters reconstructed
for the early 19th century,\textquotedblright\ 
\emph{Nature Commun.}, Vol. 13, 2116, 2022; doi:10.1038/s41467-022-29677-8. 
 
\bibitem{SunShuping:22}
S. Sun, T. Huang, B. Zhang, P. He, L. Yan, D. Fan, J. Zhang and J. Chen,
\textquotedblleft  A novel intelligent system based
on adjustable classifier models for diagnosing heart sounds,\textquotedblright\ 
\emph{Sci. Rep.}, Vol. 12, 1283, 2022; doi:10.1038/s41598-021-04136-4. 
 
\bibitem{Timm:22}
A. Timmermann, K.-S. Yu, P. Raia, J. Ruan,
A. Mondanaro, E. Zeller, C. Zollikofer, M.P. de Le{\'o}n,
D. Lemmon, M. Willeit and A. Ganopolski,
\textquotedblleft Climate effects on archaic human habitats and species successions,\textquotedblright\ 
\emph{Nature}, Vol. 604, 
pp. 495--501, 2022,
and Supplement on Methods.
 
\bibitem{Wauch:22}
H.S. Wauchope, J.P.G. Jones, J. Geldmann, B.I. Simmons,
T. Amano, D.E. Blanco, R.A. Fuller, A. Johnston,
T. Langendoen, T. Mundkur, S. Nagy and W.J. Sutherland,
\textquotedblleft Protected areas have a mixed impact on
waterbirds, but management helps,\textquotedblright\ 
\emph{Nature}, Vol. 605, 
pp. 103--107, 2022,
and Supplement on Methods.
 
\bibitem{Wein:22}
N. Weinberger,
\textquotedblleft  Generalization bounds and algorithms for learning
to communicate over additive noise channels,\textquotedblright\ 
\emph{IEEE Trans. Inf. Theory}, Vol. 68, no. 3, pp. 1886--1921, 2022. 
 
\bibitem{Wen6:22}
S. Wen, W. Guo, Y. Liu and R. Wu,
\textquotedblleft  Rotated object detection via scale-invariant
Mahalanobis distance in aerial images,\textquotedblright\ 
\emph{IEEE Geosci. Remote Sens. Lett.}, Vol. 19, No. 6514505, 2022. 
 
\bibitem{Yang19:22}
B. Yang, T. Karigo and D.J. Anderson,
\textquotedblleft Transformations of neural representations
in a social behaviour network,\textquotedblright\ 
\emph{Nature}, Vol. 608, 
pp. 741--749, 2022,
and Supplement on Methods.
 
\bibitem{Zhang34:22}
M. Zhang, Y. Zhang and G. Shen,
\textquotedblleft  PPDDS: a privacy-preserving disease diagnosis
scheme based on the secure Mahalanobis
distance evaluation model,\textquotedblright\ 
\emph{IEEE Systems J.}, Vol. 16, no. 3, pp. 4552--4562, 2022. 
 
\bibitem{Burss:23}
S. Burssens, D.M. Bowman, M. Michielsen,
S. Sim{\'o}n-Diaz, C. Aerts, V. Vanlaer, G. Banyard,
N. Nardetto, R.H.D. Townsend, G. Handler,
J.S.G. Mombarg, R. Vanderspek and G. Ricker,
\textquotedblleft  A calibration point for stellar evolution from
massive star asteroseismology\textquotedblright\ 
\emph{Nature Astronom.}, Vol. 7, pp. 913–-930, 2023; doi:10.1038/s41550-023-01978-y. 
 
\bibitem{Choi5:23}
K. Choi, S. Woo and J. Lee,
\textquotedblleft  Motor-effector dependent
modulation of sensory-motor
processes identified
by the multivariate pattern analysis
of EEG activity,\textquotedblright\ 
\emph{Sci. Rep.}, Vol. 13, 3161, 2023; doi:10.1038/s41598-023-30324-5. 
 
\bibitem{Choi7:23}
H.-K. Choi, P. Cong, C. Ge, A. Natarajan,
B. Liu, Y. Zhang, K. Li, M.N. Rushdi,
W. Chen, J. Lou, M. Krogsgaard and C. Zhu,
\textquotedblleft  Catch bond models may explain how force
amplifies TCR signaling and antigen discrimination,\textquotedblright\ 
\emph{Nature Commun.}, Vol. 14, 2616, 2023; doi:10.1038/s41467-023-38267-1. 
 
\bibitem{Dahlin:23}
J.L. Dahlin, B.K. Hua, B.E. Zucconi,
S.D. Nelson Jr, S. Singh, A.E. Carpenter,
J.H. Shrimp, E. Lima-Fernandes, M.J. Wawer,
L.P.W. Chung, A. Agrawal, M. O\'Reilly,
D. Barsyte-Lovejoy, M. Szewczyk, F. Li, P. Lak,
M. Cuellar, P.A. Cole, J.L. Meier, T. Thomas,
J.B. Baell, P.J. Brown, M.A. Walters,
P.A. Clemons, S.L. Schreiber and B.K. Wagner,
\textquotedblleft  Reference compounds for characterizing
cellular injury in high-content cellular morphology assays,\textquotedblright\ 
\emph{Nature Commun.}, Vol. 14, 1364, 2023; doi:10.1038/s41467-023-36829-x. 
 
\bibitem{Ebrahimi:23}
A.S. Ebrahimi, P. Orlowska-Feuer, Q. Huang, A.G. Zippo,
F.P. Martial, R.S. Petersen and R. Storchi,
\textquotedblleft  Three-dimensional unsupervised
probabilistic pose reconstruction (3D-UPPER) for freely moving
animals,\textquotedblright\ 
\emph{Sci. Rep.}, Vol. 13, 155, 2023; doi:10.1038/s41598-022-25087-4. 
 
\bibitem{Jeong7:23}
W. Jeong, S. Kim, J. Park and J. Lee,
\textquotedblleft  Multivariate EEG activity reflects the Bayesian
integration and the integrated Galilean relative
velocity of sensory motion during sensorimotor behavior,\textquotedblright\ 
\emph{Commun. Biol.}, Vol. 6, 113, 2023; doi:10.1038/s42003-023-04481-2. 
 
\bibitem{Kim18:23}
M. Kim, K.-R. Moon and B.-D. Lee,
\textquotedblleft  Motor-effector dependent
modulation of sensory-motor
processes identified
by the multivariate pattern analysis
of EEG activity,\textquotedblright\ 
\emph{Sci. Rep.}, Vol. 13, 3415, 2023; doi:10.1038/s41598-023-30589-w. 
 
\bibitem{Nowak:23}
A.J. Nowakowski, S.W.J. Canty, N.J. Bennett,
C.E. Cox, A. Valdivia, J.L. Deichmann, T.S. Akre,
S.E. Bonilla-Anariba, S. Costedoat and M. McField,
\textquotedblleft  Co-benefits of marine protected areas for
nature and people,\textquotedblright\ 
\emph{Nature Sustainab.}, Vol. 6, No. 10, pp. 1210--1218, 2023; doi:110.1038/s41893-023-01150-4. 
 
\bibitem{Qu4:23}
C. Qu, Y. Zhang, K. Huang, S. Wang and Y. Yang,
\textquotedblleft  Point clouds outlier removal method based on
improved Mahalanobis and completion,\textquotedblright\ 
\emph{IEEE Robot. Automat. Lett.}, Vol. 8, no. 1, pp. 17--24, 2023. 
 
\bibitem{Rabby:23}
Y.W. Rabby, Y. Li and H. Hilafu,
\textquotedblleft  An objective absence data
sampling method for landslide susceptibility mapping,\textquotedblright\ 
\emph{Sci. Rep.}, Vol. 13, 1740, 2023; doi:10.1038/10.1038/s41598-023-28991-5. 
 
\bibitem{Sarno:23}
J. Sarno, P. Domizi, Y. Liu, M. Merchant,
C.B. Pedersen, D. Jedoui, A. Jager, G.P. Nolan,
G. Gaipa, S.C. Bendall, F.-A. Bavaand K.L. Davis,
\textquotedblleft  Dasatinib overcomes glucocorticoid
resistance in B-cell acute lymphoblastic leukemia,\textquotedblright\ 
\emph{Nature Commun.}, Vol. 14, 2935, 2023; doi:10.1038/s41467-023-38456-y. 
 
\bibitem{Tang18:23}
Y. Tang, Y. Zhou, X. Ren, Y. Sun, Y. Huang and D. Zhou,
\textquotedblleft  A new basic probability assignment
generation and combination method for conflict data fusion
in the evidence theory,\textquotedblright\ 
\emph{Sci. Rep.}, Vol. 13, 8443, 2023; doi:0.1038/s41598-023-35195-4. 
 
\bibitem{Tsvi:23}
A. Tsvieli and N. Weinberger,
\textquotedblleft  Learning maximum margin channel decoders,\textquotedblright\ 
\emph{IEEE Trans. Inf. Theory}, Vol. 69, no. 6, pp. 3597--3626, 2023. 
 
\bibitem{Zhang27:23}
J. Zhang, Y. Feng, F.T. Maestre
M. Berdugo, J. Wang, C. Coleine,
T. S{\'a}ez-Sandino, L. Garcia-Vel{\'a}zquez, B.K. Singh and M. Delgado-Baquerizo,
\textquotedblleft Water availability creates global thresholds
in multidimensional soil biodiversity and functions,\textquotedblright\ 
\emph{Nature Ecol. Evol.}, Vol. 7, 
pp. 1002--1011, 2023,
and Supplement.
 
\bibitem{Zhou13:23}
M. Zhou, C. Shang, G. Li, L. Shen, N. Naik,
S. Jin, J. Peng and Q. Shen,
\textquotedblleft  Transformation-based fuzzy rule interpolation with
Mahalanobis distance measures supported by Choquet integral,\textquotedblright\ 
\emph{IEEE Trans. Fuzzy Syst.}, Vol. 31, no. 4, pp. 1083--1097, 2023. 

\bibitem{Varshney:14}
K.R. Varshney and L.R. Varshney,
\textquotedblleft  Optimal grouping for group minimax hypothesis testing,\textquotedblright\ 
\emph{IEEE Trans. Inf. Theory}, Vol. 60, no. 10, pp. 6511--6521, 2014. 
 
\bibitem{Hu14:15}
Z. Hu, Y. Zhu, J. Xu and Y. Yang,
\textquotedblleft  Bregman-based inexact excessive gap method
for multiservice resource allocation,\textquotedblright\ 
\emph{IEEE Trans. Wireless Commun.}, Vol. 14, no. 2, pp. 1115--1130, 2015. 
 
\bibitem{Noc:15f}
R. Nock, W. Bel Haj Ali, R. D'Ambrosio, F. Nielsen and M. Barlaud,
\textquotedblleft  Gentle nearest neighbors boosting
over proper scoring rules,\textquotedblright\ 
\emph{IEEE Trans. Pattern Anal. Mach. Intell.}, Vol. 37, no. 1, pp. 80--93, 2015. 
 
\bibitem{Rask:15}
G. Raskutti and S. Mukherjee,
\textquotedblleft  The information geometry of mirror descent,\textquotedblright\ 
\emph{IEEE Trans. Inf. Theory}, Vol. 61, no. 3, pp. 1451--1457, 2015. 
 
\bibitem{WangHuiwei:15}
H. Wang, X. Liao, T. Huang and C. Li,
\textquotedblleft  Cooperative distributed optimization in multiagent networks with delays,\textquotedblright\ 
\emph{IEEE Trans. Syst. Man Cyb.}, Vol. 45, no. 2, pp. 363--369, 2015. 
 
\bibitem{HeWenwu:17}
W. He, J.T.-Y. Kwok, J. Zhu and Y. Liu,
\textquotedblleft  A note on the unification of adaptive online learning,\textquotedblright\ 
\emph{IEEE Trans. Neural Netw. Learn. Syst.}, Vol. 28, no. 5, pp. 1178--1191, 2017. 
 
\bibitem{Li28:17}
X. Li, T. Pi, Z. Zhang, X. Zhao, M. Wang, X. Li and P.S. Yu,
\textquotedblleft  Learning Bregman distance functions
for structural learning to rank,\textquotedblright\ 
\emph{IEEE Trans. Knowledge Data Engin.}, Vol. 29, no. 9, pp. 1916--1927, 2017. 
 
\bibitem{Harre:18a}
P. Harremo{\"e}s,
\textquotedblleft Entropy on spin factors,\textquotedblright\ 
in: N. Ay et al. (eds.), \emph{Information Geometry and Its Applications}, 
Springer Proceedings in Mathematics \& Statistics, vol. 252, pp. 247–-278. 
Cham, Switzerland: Springer Nature Switzerland, 2018. 
 
\bibitem{Xu22:18}
J. Xu, S. Zhu, Y. C. Soh and L. Xie,
\textquotedblleft  A Bregman splitting scheme for distributed
optimization over networks,\textquotedblright\ 
\emph{IEEE Trans. Autom. Control}, Vol. 63, no. 11, pp. 3809--3824, 2018. 
 
\bibitem{Halder:19}
A. Halder,
\textquotedblleft  DeGroot–Friedkin map in opinion dynamics
is mirror descent,\textquotedblright\ 
\emph{IEEE Control Syst. Lett.}, Vol. 3, no. 2, pp. 463--468, 2019. 
 
\bibitem{ZhangQiping:19}
Q. Zhang, Y. Zhang, Y. Huang and Y. Zhang,
\textquotedblleft  Azimuth superresolution of forward-looking radar
imaging which relies on linearized Bregman,\textquotedblright\ 
\emph{IEEE J. Sel. Topics Appl. Earth Obs. Remote Sens.}, Vol. 12, no. 7, pp. 2032--2043, 2019. 
 
\bibitem{ShaoXiaodan:20}
X. Shao, X. Chen, D.W.K. Ng, C. Zhong and Z. Zhang,
\textquotedblleft  Cooperative activity detection: sourced and
unsourced massive random access paradigms,\textquotedblright\ 
\emph{IEEE Trans. Signal Proces.}, Vol. 68, 
pp. 6578--6593, 2020. 
 
\bibitem{Temb:20}
H. Tembine,
\textquotedblleft  Deep learning meets game theory:
Bregman-based algorithms for interactive deep
generative adversarial networks,\textquotedblright\ 
\emph{IEEE Trans. Cybernetics}, Vol. 50, no. 3, pp. 1132--1145, 2020. 
 
\bibitem{Brech:21}
C. Br{\'e}cheteau, A. Fischer and C. Levrard,
\textquotedblleft  Robust Bregman clustering,\textquotedblright\ 
\emph{Ann. Statist.}, Vol. 49, no. 3, pp. 1679--1701, 2021. 
 
\bibitem{Lin17:21}
C.-W. Lin, S.-H. Liao, H.-S. Huang, L.-M. Wang, J.-H. Chen, C.-H. Su and K.-L. Chen,
\textquotedblleft  Improvement of multisource localization of magnetic particles
in an animal,\textquotedblright\ 
\emph{Sci. Rep.}, Vol. 11, 9628, 2021; doi:110.1038/s41598-021-88847-8. 
 
\bibitem{YuanDeming:21}
D. Yuan, Y. Hong, D.W.C. Ho and S. Xu,
\textquotedblleft  Convergence of best $\phi-$entropy estimates,\textquotedblright\ 
\emph{IEEE Trans. Autom. Control}, Vol. 66, no. 2, pp. 714--729, 2021. 
 
\bibitem{Azizan:22}
N. Azizan, S. Lale and B. Hassibi,
\textquotedblleft  Stochastic mirror descent on overparameterized
nonlinear models,\textquotedblright\ 
\emph{IEEE Trans. Neural Netw. Learn. Syst.}, Vol. 33, no. 12, pp. 7717--7727, 2022. 
 
\bibitem{Dytso:22}
A. Dytso, M. Fau{\ss} and H.V. Poor, 
\textquotedblleft  Bayesian risk with Bregman loss: A Cram{\'e}r–Rao
type bound and linear estimation,\textquotedblright\ 
\emph{IEEE Trans. Inf. Theory}, Vol. 68, no. 3, pp. 1985--2000, 2022. 
 
\bibitem{Gruz:22}
T.V. Gruzdeva and A.V. Ushakov,
\textquotedblleft On a nonconvex distance-based
clustering problem,\textquotedblright\ 
in: P. Pardalos et al. (eds.), 
\emph{Mathematical Optimization Theory and Operations Research MOTOR 2022}, 
Lecture Notes in Computer Science, vol. 13367, pp. 139--152. 
Cham, Switzerland: Springer Nature Switzerland, 2022. 
 
\bibitem{Song6:22}
Y. Song, Y. Gu, R. Zhang and G. Yu,
\textquotedblleft  BrePartition: optimized high-dimensional kNN
search with Bregman distances,\textquotedblright\ 
\emph{IEEE Trans. Knowledge Data Engin.}, Vol. 34, no. 3, pp. 1053--1065, 2022. 
 
\bibitem{YuZhan:22}
Z. Yu, D.W.C. Ho and D. Yuan,
\textquotedblleft  Distributed randomized gradient-free mirror descent algorithm for
constrained optimization,\textquotedblright\ 
\emph{IEEE Trans. Autom. Control}, Vol. 67, no. 2, pp. 957--964, 2022. 
 
\bibitem{Capo:23}
M. Cap{\'o}, A. P{\'e}rez and J.A. Lozano,
\textquotedblleft  Fast computation of cluster validity measures
for Bregman divergences and benefits,\textquotedblright\ 
\emph{Pattern Recogn. Lett.}, Vol. 170, 
pp. 100--105, 2023. 
 
\bibitem{Chen32:23}
G. Chen, G. Xu, W. Li and Y. Hong,
\textquotedblleft  Distributed mirror descent algorithm With Bregman damping for
nonsmooth constrained optimization,\textquotedblright\ 
\emph{IEEE Trans. Autom. Control}, Vol. 68, no. 11, pp. 6921--6928, 2023. 
 
\bibitem{Fern:23}
J.D. Fern{\'a}ndez-Rodriguez, E.J. Palomo, J. Benito-Picazo, E. Dominguez,
E. L{\'o}opez-Rubio and F. Ortega-Zamorano,
\textquotedblleft  A convolutional autoencoder and a neural gas model based
on Bregman divergences for hierarchical color quantization,\textquotedblright\ 
\emph{Neurocomputing}, Vol. 544, No. 126288, 2023; doi:10.1016/j.neucom.2023.126288.
 
\bibitem{Haya:23}
M. Hayashi, 
\textquotedblleft  Bregman divergence based Em algorithm and its
application to classical and quantum rate distortion theory,\textquotedblright\ 
\emph{IEEE Trans. Inf. Theory}, Vol. 69, no. 6, pp. 3460--3492, 2023. 
 
\bibitem{Li33:23}
Z. Li and A. Ralescu,
\textquotedblleft  Generalized self-supervised contrastive learning with Bregman divergence
for image recognition,\textquotedblright\ 
\emph{Pattern Recogn. Lett.}, Vol. 171, 
pp. 155--161, 2023. 
 
\bibitem{XiongMenghui:23}
M. Xiong, B. Zhang, D.W.C. Ho, D. Yuan and S. Xu,
\textquotedblleft  Event-triggered distributed stochastic mirror
descent for convex optimization,\textquotedblright\ 
\emph{IEEE Trans. Neural Netw. Learn. Syst.},  Vol. 34, no. 9, pp. 6480--6491, 2023. 
 
\bibitem{LiuJie:24}
J. Liu, Z. Yu and D.W.C. Ho,
\textquotedblleft  Distributed constrained optimization with delayed
subgradient information over time-varying
network under adaptive quantization,\textquotedblright\ 
\emph{IEEE Trans. Neural Netw. Learn. Syst.}, Vol. 35, no. 1, pp. 143--156, 2024. 
 
\bibitem{Bau:97}
H.H. Bauschke and J.M. Borwein,
\textquotedblleft  Legendre functions and the method of random
Bregman projections,\textquotedblright\ 
\emph{J. Convex Anal.}, Vol. 4, no. 1, pp. 27--67, 1997. 

\bibitem{Dhi:07}
I.S. Dhillon and J.A. Tropp,
\textquotedblleft  Matrix nearness problems with Bregman divergences,\textquotedblright\ 
\emph{SIAM J. Matrix Anal. Appl.}, Vol. 29, no. 4, pp. 1120--1146, 2007. 

\bibitem{Kulis:09}
B. Kulis, M.A. Sustik and I.S. Dhillon,
\textquotedblleft  Low-rank kernel learning with Bregman matrix divergences,\textquotedblright\ 
\emph{J. Machine Learn. Res.}, Vol. 10, pp. 341--376, 2009.
 



\bibitem{Donoho:06a}
D.L. Donoho, M. Elad and V.N. Temlyakov,
\textquotedblleft  Stable recovery of sparse overcomplete
representations in the presence of noise,\textquotedblright\ 
\emph{IEEE Trans. Inf. Theory}, Vol. 52, no. 1, pp. 6--18, 2006. 
 
\bibitem{Can:06}
E.J. Cand{\`e}s, J.K. Romberg and T. Tao,
\textquotedblleft  Stable signal recovery
from incomplete and inaccurate measurements,\textquotedblright\ 
\emph{Commun. Pure Appl. Math.}, Vol. LIX, 
pp. 1207--1223, 2006. 
 
\bibitem{Lustig:07}
M. Lustig, D. Donoho and J.M. Pauly,
\textquotedblleft Sparse MRI: the application of compressed sensing
for rapid MR imaging,\textquotedblright\ 
\emph{Magnetic Reson. Med.}, Vol. 58, 
pp. 1182--1195, 2007. 
 
\bibitem{Can:08a}
E.J. Cand{\`e}s,
\textquotedblleft The restricted isometry property and its implications
for compressed sensing,\textquotedblright\ 
\emph{C. R. Acad. Sci. Paris, Ser. I}, Vol. 346, 
pp. 589--592, 2008.

\bibitem{Can:08b}
E.J. Cand{\`e}s, M.B. Wakin and S.P. Boyd,
\textquotedblleft  Enhancing sparsity by reweighted $\ell_{1}$ minimization,\textquotedblright\ 
\emph{J. Fourier Anal. Appl.}, Vol. 14, 
pp. 877--905, 2008. 
 
\bibitem{Goldst:09}
T. Goldstein and S. Osher,
\textquotedblleft  The split Bregman method for $L1-$regularized problems,\textquotedblright\ 
\emph{SIAM J. Imaging Sci.}, Vol. 2, no. 2, pp. 323--343, 2009. 
 
\bibitem{Zhang21:14}
Y. Zhang, B.S. Peterson, G. Ji and Z. Dong,
\textquotedblleft  Energy preserved sampling for compressed sensing MRI,\textquotedblright\ 
\emph{Comput. Mathem. Meth. Med.}, Vol. 2014, No. 546814, 2014; doi:10.1155/2014/546814.
 
\bibitem{Edgar:19}
M.P. Edgar, G.M. Gibson and M.J. Padgett,
\textquotedblleft Principles and prospects for single-pixel imaging,\textquotedblright\ 
\emph{Nature Photonics}, Vol. 13, 
pp. 13--20, 2019.
 
\bibitem{Foucart:09}
S. Foucart and M.-J. Lai,
\textquotedblleft  Sparsest solutions of underdetermined linear systems via
$\ell_{q}-$-minimization for $0 <q \leq 1$,\textquotedblright\ 
\emph{Appl. Comput. Harmon. Anal.}, Vol. 26, 
pp. 395--407, 2009. 
 
\bibitem{Liu8:15}
J. Liu, J. Jin and Y. Gu,
\textquotedblleft  Robustness of sparse recovery via F-minimization:
a topological viewpoint,\textquotedblright\ 
\emph{IEEE Trans. Inf. Theory}, Vol. 61, no. 7, pp. 3996--4014, 2015. 
 
\bibitem{Bruck:09}
A.M. Bruckstein, D.L. Donoho and M. Elad,
\textquotedblleft  From sparse solutions of
systems of equations to sparse
modeling of signals and images,\textquotedblright\ 
\emph{SIAM Rev.}, Vol. 51, no. 1, pp. 34--81, 2009. 




\bibitem{Lind:04}
B.G. Lindsay,
\textquotedblleft Statistical distances as loss functions in assessing model
adequacy,\textquotedblright\ 
in: M.P. Taper and S.R. Lele (eds.), \emph{The Nature of Scientific Evidence}, 
pp. 439--487. 
Chicago, IL, USA: The University of Chicago Press, 2004. 
This includes comments by D.R. Cox and S.P. Ellner as well as a rejoinder
by the author.

\bibitem{Lind:08}
B.G. Lindsay, M. Markatou, S. Ray, K. Yang, and S.-C. Chen,
\textquotedblleft  Quadratic distances on probabilities: a unified foundation,\textquotedblright\ 
\emph{Ann. Statist.}, Vol. 36, no. 2, pp. 983--1006, 2008.

\bibitem{Mark:19}
M. Markatou and E.M. Sofikitou,
\textquotedblleft  Statistical distances and the
construction of evidence functions for model adequacy,\textquotedblright\ 
\emph{Front. Ecol. Evol.}, Vol. 7, No. 447, 2019; doi: 10.3389/fevo.2019.00447. 

\bibitem{Bro:12}
M. Broniatowski and A. Keziou,
\textquotedblleft  Divergences and duality for estimation and test under moment
condition models,\textquotedblright\ 
\emph{J. Statistical Planning and Inference}, Vol. 142, pp. 2554--2573, 2012.

\bibitem{Bro:16}
M. Broniatowski and A. Decurninge,
\textquotedblleft  Estimation for models defined by conditions
on their L-moments,\textquotedblright\ 
\emph{IEEE Trans. Inf. Theory}, Vol. 62, no. 9, pp. 5181--5198, 2016. 








\bibitem{Vaa:98} 
A.W. Van der Vaart, 
\emph{Asymptotic Statistics, 8th Printing}.
Cambridge, UK: Cambridge University Press, 2007.

\bibitem{Kuc:17}
A.K. Kuchibhotla and A. Basu,
\textquotedblleft On the asymptotics of minimum disparity estimation,\textquotedblright\ 
\emph{TEST}, Vol. 26, pp. 481--502, 2017.
 



\bibitem{Fra:15}
T. Franco,
\textquotedblleft
A mini-course in large deviations,\textquotedblright\
\emph{https://w3.impa.br/$\sim$tertu/archives/LDP\_Notes\_2015.pdf}, 49 pages.
Salvador, Brazil: Universidade Federal da Bahia (UFBA), March 2015. 

\bibitem{Swa:23}
J. Swart,
\textquotedblleft
Large deviation theory, 6th edition,\textquotedblright\
\emph{http://staff.utia.cas.cz/swart/lecture\_notes/LDP23\_04\_15.pdf}, 179 pages.
Prague, Czech Republic: Institute of Information Theory
and Automation (UTIA), The Czech Academy of Sciences, April 2023. 

\bibitem{Top:07}
F. Topsoe, 
\textquotedblleft  Some bounds for the logarithmic function,\textquotedblright\ 
in: Y.J. Cho, J.K. Kim and S.S. Dragomir (eds.), \emph{Inequality Theory and Applications 4}, 
pp. 137–-151. 
New York, USA: Nova Sci. Publ., 2007. 



\end{thebibliography}
\end{document}